\pdfoutput=1
\documentclass [10pt, twoside] {uwthesis}
\usepackage{amsmath}
\usepackage{amssymb}

\usepackage{graphicx}
\usepackage{thophys}
\usepackage{array}
\usepackage{multirow}	
\usepackage{dcolumn}
\newcommand{\PreserveBackslash}[1]{\let\temp=\\#1\let\\=\temp}

\begin{document}

%  Macro for bold, hatted unit vectors.  -js
\newcommand{\bhat}[1]{\hat{\boldsymbol{#1}}}

% ==========   Preliminary pages
%
\prelimpages
%
% ----- title page
%
\Title{Single barium ion spectroscopy:  light shifts, hyperfine structure, and progress on an optical frequency standard and atomic parity violation}
\Author{Jeffrey A. Sherman}
\Year{2007}
\Program{Department of Physics}
\titlepage  

% ----- signature page %
\Chair{E. Norval Fortson}{Professor}{Physics}
\Signature{E. Norval Fortson}
\Signature{Warren Nagourney}
\Signature{Gerald Seidler}
\signaturepage

% ----- quoteslip

\setcounter{page}{-1}
% These are the real quote slips (choose one)
%  \thesisquoteslip
\doctoralquoteslip
%  \doctoralabstractquoteslip

%
% ----- abstract
%

\setcounter{page}{-1}
\abstract{Single trapped ions are ideal systems in which to test atomic physics at high precision:  they are effectively isolated atoms held at rest and largely free from perturbing interactions.  This thesis describes several projects developed to study the structure of singly-ionized barium and more fundamental physics.

First, we describe a spin-dependent `electron-shelving' scheme that allows us to perform single ion electron spin resonance experiments in both the ground $6S_{1/2}$ and metastable $5D_{3/2}$ states at precision levels of $10^{-5}$.  We employ this technique to measure the ratio of off-resonant light shifts (or ac-Stark effect) in these states to a precision of $10^{-3}$ at two different wavelengths.  These results constitute a new high precision test of heavy-atom atomic theory.  Such experimental tests in Ba$^+$ are in high demand since knowledge of key dipole matrix elements is currently limited to about 5\%.  Ba$^+$ has recently been the subject of theoretical interest towards a test of atomic parity violation for which knowledge of dipole matrix elements is an important prerequisite.  We summarize this parity violation experimental concept and describe new ideas. 

During the study of the nuclear spinless $(I=0)$ isotope of Ba$^+$, we discovered several worthwhile experimental goals for an isotope with nuclear spin, $^{137}$Ba$^+$ $(I = 3/2)$.  The hyperfine structure of the metastable $5D_{3/2}$ state is currently known to a precision $10^{-4}$.  We show how our rf spin-flip spectroscopy scheme could measure this structure to parts in $10^{-8}$ or better, allowing a determination of the nuclear magnetic dipole, electric quadrupole, and perhaps magnetic octopole moments.

Finally, the hyperfine structure of $^{137}$Ba$^+$ yields an optical transition with unique advantages in a single ion optical frequency reference.  Namely, the 2051~nm $6S_{1/2}, F = 2 \leftrightarrow 5D_{3/2}, F' = 0$ transition is effectively free of quadrupole (or gradient) Stark shifts which may plague competing ion frequency references at the $10^{-16}$ level.  We describe the performance and frequency narrowing of a diode-pumped solid state 2051~nm laser, and the observation of transitions in Ba$^+$.  We also estimate all known systematic effects on this transition and conclude that the realization of a frequency standard with long-term precision of $< 10^{-17}$ is possible at cryogenic temperatures.
}
 
%
% ----- contents & etc.
%
\tableofcontents
\listoffigures
\listoftables
 
%
% ----- glossary 
%
\chapter*{Glossary}      % starred form omits the `chapter x'
\addcontentsline{toc}{chapter}{Glossary}
\thispagestyle{plain}

\section*{Selected abbreviations}
\begin{glossary}
\item[AC]		Alternating-current, strictly.  Often used, in lowercase, to refer the oscillating part of a signal.
\item[AM]		Amplitude modulation 
\item[AOM]	Acousto-optic modulation, modulator
\item[BB]		Blackbody
\item[BNC]	Bayonet Neil-Concelman, a type of coaxial cable connector, named after its inventors.  Various (dubious?  true?) other names given by texts and internet authorities include Barrel Nut Connector, Bayonet Navy Connector, Baby N Connector, British Naval Connector, etc.
\item[BS]		Bloch-Siegert (shift), a component of the light shift (LS) which is often ignored at optical frequencies by treating an atomic system with the rotating wave approximation (RWA).
\item[BW]		Bandwidth
\item[CEO]	Carrier-envelope offset
\item[DBM]	Double-balanced mixer
\item[DC]		Direct current, strictly.  Often used, in lowercase, to indicate the limit of an oscillating signal at zero frequency. 
\item[E1, E2]	Electric dipole and electric quadrupole interactions, respectively
\item[ECDL]	External cavity diode laser
\item[EOM]	Electro-optic modulation, modulator
\item[FM]		Frequency modulation 
\item[FSR]		Free spectral range
\item[FWHM]	Full-width at half-maximum, a measure of spectral width
\item[GPS]	Global positioning system
\item[HCL]	Hollow cathode (gas discharge) lamp
\item[HF(S)]	Hyperfine (structure)
\item[HWP]	Optical half-wave plate
\item[IF]		Intermediate frequency, often the dc-coupled port of a double-balanced mixer (DBM)
\item[IR]		Infrared (light)
\item[LED]		Light emitting diode
\item[LO]		Local oscillator, often one of two ac-coupled ports on a double-balanced mixer (DBM)
\item[LS]		Light shift:  the energy shift due to the ac-Stark effect, or dynamic atomic polarization
\item[M1]		Magnetic dipole interaction
\item[PBS]		Polarizing beam splitter (cube)
\item[PM]		Polarization maintaining (fiber optics) or phase modulation.
\item[PMT]	Photo-multiplier tube
\item[PNC]	Parity non-conservation, non-conserving
\item[PPLN]	Periodically poled lithium niobate, a non-linear optical material designed for quasi phase matching (QPM) applications.
\item[PV]		Parity violation, violating
\item[PZT]		A particular piezo-electric material, lead zirconate titanate, but often referring to piezo-electric transducers in general.
\item[QED]	Quantum electrodynamics:  our present theoretical description of the interaction of electrically charged particles with photons
\item[QPM]	Quasi phase matching, a technique for efficient use of modular, poled nonlinear crystals. 
\item[QWP]	Optical quarter half-wave plate
\item[RAM]	Residual amplitude modulation, often referring to the erroneous amplitude modulation that often accompanies technical implementations of frequency or phase modulation
\item[RF]		Radio frequency, in general, and rendered in lower case letters.  Sometimes RF specifically refers to one of the ac-coupled ports on a double balanced mixer (DBM).
\item[RWA] 	Rotating wave approximation
\item[SHG]	Second harmonic generation
\item[SM]		Single mode (fiber optics)
\item[SMA]	SubMiniature, version A, a type of coaxial cable connector
\item[SWR]	Standing wave ratio
\item[TEC]		Thermo-electric cooler
\item[TEM]	Transverse electromagnetic, often referring to indexed radiation modes in an optical or microwave cavity.
\item[UHV]	Ultra-high vacuum
\item[ULE]		Ultra-low expansion, a brand name of glass/ceramic material
\item[UV]		Ultraviolet (light)
\item[VCO]	Voltage-controlled oscillator
\item[YAG,YLF]	 Yttrium aluminum garnet, yttrium lithium fluoride, two crystal hosts used in solid-state lasers
\end{glossary}
 
 \section*{Symbols, conventions}
\begin{glossary}
\item[$\alpha$]		The fine-structure constant
\item[$\alpha_0$, $\alpha_1$, $\alpha_2$]		Scalar, vector, and tensor polarizabilities, respectively
\item[$a_{m}$]		The probability to optically pump into a sub-level $m$, specific to Chapter~\ref{sec:lightShiftChapter}
\item[$c$]			The speed of light in vacuum
\item[$\Delta$]		In general, a difference of two frequencies
\item[$\boldsymbol{E}$]	Electric field, cartesian vector
\item[$E^{(1)}_q$]		Electric field, $q$ component of the rank-1 spherical vector
\item[$\Gamma$]		Atomic decay rate (angular frequency)
\item[$\boldsymbol{k}$]	Optical wavevector with magnitude $k = 2\pi / \lambda$
\item[$\lambda$]		Optical wavelength
\item[$\nu$]			Often a trapped ion secular oscillation frequency (in Hz)
\item[$\omega_\text{rf}$]	The trap radio frequency (angular frequency)
\item[$\omega_0$]		Frequency of an (unshifted) atomic transition (angular frequency)
\end{glossary}
 
%
% ----- acknowledgments
%
\acknowledgments{
To mix metaphors, it took a village to raise this lost idiot.  Foremost, thank you to my parents, Elaine and Jim Sherman, and grandparents, Sam and Anne Pink, for a lifetime of love and support.

Thanks, of course, to Professors E.\ Norval Fortson and Warren Nagourney, for showing me both the broad stokes, and nuts and bolts of atomic experimental physics.  Norval's knowledge of physics seems to be outdone only by his eagerness to pass it on.  Likewise, Warren's tenacity for getting dead argon lasers, circuits, ion traps, and graduate students running again demonstrates a deep reserve of patience and experience I only wish to inherit. So many thanks to colleague and good friend Timo K\"{o}rber and his wife Lydia.  I've had the pleasure of working with many talented undergraduate and graduate students on the barium ion project:  Ethan Clarke, Eryn Cook, S.\ Mayumi Fugami, Adam Kleczewski, Lauren Kost, Steven Metz, Edmund Meyer, Pete Morcos, Zach Simmons.  Thank you to several (past and present) colleagues in the Fortson and other atomic physics groups at the University of Washington:  Amar Andalkar, Boris Blinov (and his research group), Claire Cramer, W.\ Clark Griffith, Laura Kogler, Rob Lyman, Anna Markhotok, Reina Maruyama-Heeger, Chris Pearson, David Pinegar, M.\ David Swallows, William Trimble, and Roahn Wynar.

Many thanks to friends and fellow graduate students Conor Buechler (and spouse Alicia), Tom Butler,  Jon Chandra, Michael Endres, Andr\'{e} Walker-Loud, and Heather Zorn-Butler.  You all provided the (not little) things that kept me going:  bagels, coffee, and your conversation, support, and kindness.  A special thank you to S.\ Mayumi Fugami for so much love and support.

Our department's machine, glass, and electronic shops are staffed by several individuals worthy of recognition: Ted Ellis, Bob Morely, Ron Musgrave, Bryan Venema, and Mike Vinton.   Thank you to teachers that had an early impact on my life:  Roger Bengston, Sherwin Bennes, Phil Bombino.  I hope certain professors at the University of Washington helped shape my approach to physics during graduate school.  I probably don't know all that they think I ought to, but thank you to Eric Adelberger, Steve Ellis, Ann Nelson, Martin Savage, Gerald Seidler, and Larry Yaffe.

Finally, thank you to the NSF for funding this research.  Thank you to the University of Washington ARCS foundation for their generous fellowship.
}
%
% end of the preliminary pages

\textpages

\chapter{Introduction}
\section{A short story about a pale blue dot}
Billions of year ago, a star not yet a point of light in Earth's sky---because there was no sky, because there was no Earth---surrendered in a nuclear arms race with gravity.  If a white flag of truce was offered, it was a futile act because gravity's retribution was tremendous.  The star's death, violent and grand, is accompanied by a final desperate act of creation: a \emph{super nova} gives birth to a menagerie of the elements, including some of the first bits of matter heavier than iron that the universe had ever seen.

Some of these atoms are propelled outwards and embark on a lonely, silent, and cold journey lasting perhaps many millions of our lifetimes while the universe itself continues its morning stretches.  Guided only by faint signposts of gravitational attraction, they eventually coalesce into a swirling mass of dust and gas soon to be our home and its sun.  

Look closely: some of these atoms, each forty billion times smaller than you are tall, are carbon which now live inside you.  Some atoms are helium which will enliven a child's birthday party when she experiences what it might be like to talk as a squirrel.  And about four of every ten thousand atoms are barium or heavier elements destined to become barium through radioactive transformation.  \emph{These} are the subjects of this manuscript.

A few years ago under a sky now filled with stars, large amounts of this barium, named `heavy' after the greek \emph{barys}, are removed from Earth's crust by a mining firm eager to profit from an element so useful to those working in petroleum, medical imaging, and the rat poison business.  A student places a long distance telephone call to a company specializing in the sale of purified metals.  For just over twenty dollars, which are green pieces of paper equivalent in cost to two curry dishes at an exceptional restaurant located at Market St.\ and Ballard Ave.\ in a place called Seattle, he obtains from them a 100 gram chunk of barium, purified by a distillation process.

Quickly delivered, he shaves off a milligram or so of the barium (about four billion billion atoms) under an argon atmosphere and places these in a metal and glass contraption soon made devoid of air and placed on a smooth metal table in the basement of an attractive building near a waterway, bridge, and bicycle trail.

In the morning on one of several hundred days afterwards, the student gently warms the morsel of barium with a few Amperes of alternating current.  In the vacuum enclosure, the laws of kinetics dispassionately dictate the fate of one of these atoms.  Kicked by its neighbor, one atom leaves the accumulation of barium on another lonely ballistic trajectory but this time travels just ten millimeters to a ring-shaped metal electrode, where by chance, it encounters an energetic electron set on a collision course due to another human design.  

This crash, an isolated \emph{unit} of chemistry between one atom and one hot electron, ends with the barium atom losing a valance electron, one of the two left far away from its core as an unwanted couch left on the curb is easily spirited away in the night.  Immediately afterwards, our atom, now an \emph{ion}, is jerked into motion by a force unfelt just moments before.  Because it now carries a net electric charge, the barium ion is tugged into stable confined motion by a radio frequency electric field engineered on the ring-shaped electrode especially for this purpose.  

That's not all:  another electric field is now set upon our ion, an optical field oscillating some 100 million times faster than the radio frequency field.  This optical field, a blue laser beam, is specially crafted to be \emph{resonant} with the barium ion.  Now, likely for the first time in its existence, the barium ion spends a significant time outside of its quantum ground state.  About one million times per second it is excited by the blue laser, quivers, emits a blue photon and is excited again.  Because the laser is tuned just so, the ion is cooled by the laser, and within a millisecond or so is the coldest it has ever been and likely ever will be:  a quarter million times colder than room temperature.

A single barium ion traveled across the universe to be held fixed and cold in an enclosed vacuum on a table in a basement.  It sputters and shines, a pale blue dot, a one-atom star, artificially isolated in a small dark region of laboratory sky.

For about eight hours the student applies a premeditated, computer-controlled sequence of optical pulses and bursts of radio frequency energy in an effort to learn about the atomic structure of the ion.  After so long, a laser frequency drifts too far, or perhaps the student wants to go home and turns off the radio frequency trapping field; for one reason or another the ion is let go.  It flies out of the ring shaped electrode, and in just milliseconds collides with the metal vacuum enclosure, adheres, and remains even to this day.

\section{Overview}
\begin{figure}
\centering
\includegraphics[width=6in]{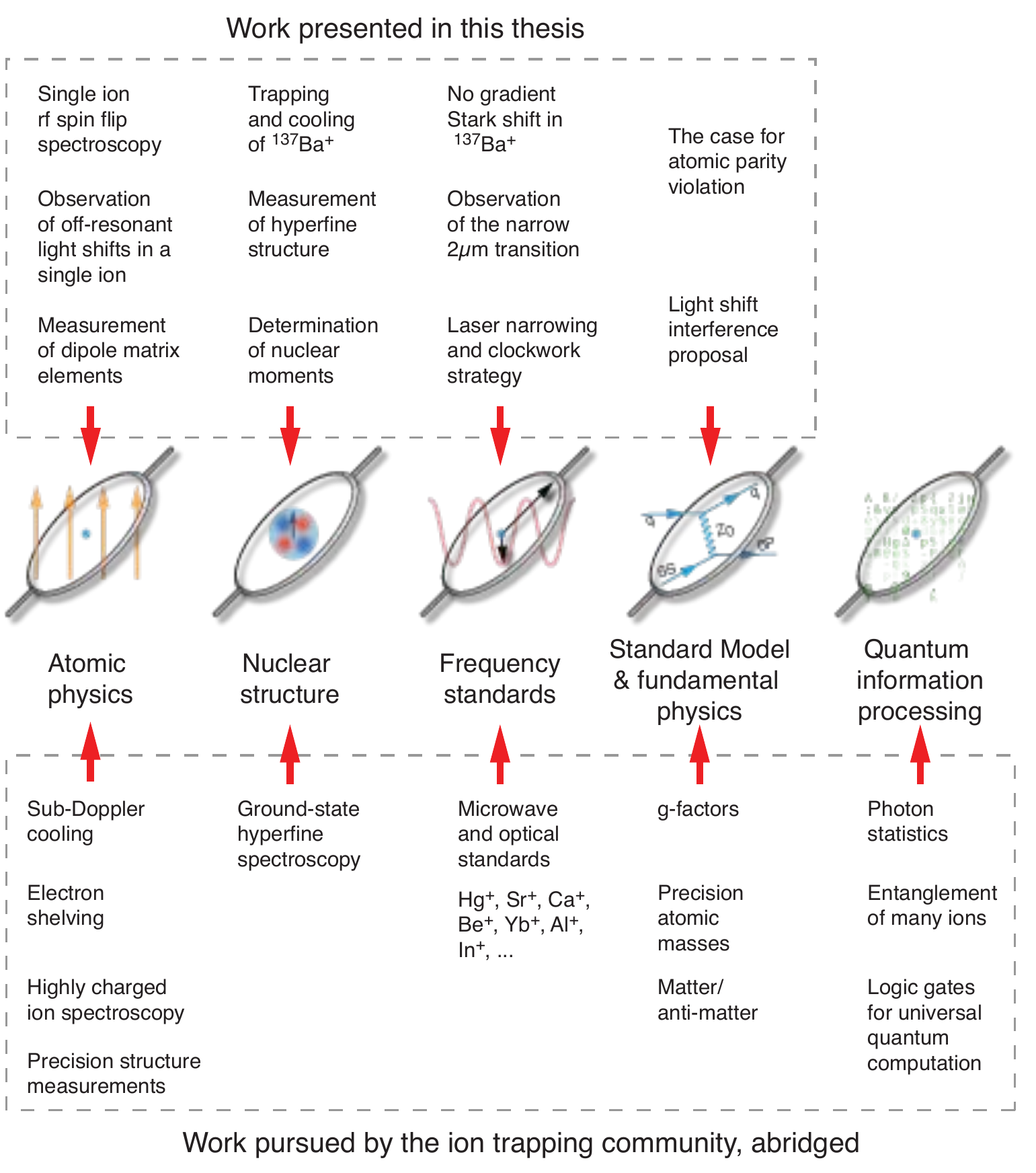}
\caption[Overview: physics pursued in this work and by the ion trap community]{The variety of experimental physics pursued by the ion trap community is staggering:  a lot of physics can come out of a ring shaped electrode holding a single charged particle.  Here we depict how aspects of our research projects fit into the bigger picture.}
\label{fig:summaryOfTrappedIonWork}
\end{figure}
We begin in Chapter~\ref{chap:AtomicPhysics} with a description of the interaction of electromagnetic fields and matter:  atomic physics.  Alongside the formalism, we illustrate realizations of many classic results in our trapped ion system.  In Chapter~\ref{sec:ionTrapping}, we describe a radio frequency quadrupole trap for charged particles and examine key similarities and differences between a free and trapped atom.  We make the ion trap and laser fields real in Chapter~\ref{sec:ApparatusChapter} by describing the apparatus constructed for our experimentation.

Trapping single ions of barium is an experimental effort that would please any armchair strategist:  it is a plan with many branches (see Figure~\ref{fig:summaryOfTrappedIonWork}).  The project began in 1993 as a proposal to measure atomic parity violation~\cite{fortson1993pmp};  progress along this line is documented in previous works~\cite{schacht2000thesis, koerber2003thesis,koerber2003rfs} and Chapter~\ref{sec:ParityChapter}.  Early on, our group realized that certain aspects of barium ion structure, namely dipole matrix elements such as $\langle 5D || er || nP \rangle$, were not known well enough, theoretically or experimentally, to properly interpret the parity violation measurement.  Therefore, we begun a long, eventually successful, project of measuring ratios of off-resonant vector light shifts in the single barium ion in order to precisely determine these matrix elements.  This work is fully documented in Chapter~\ref{sec:lightShiftChapter}.

Along the way, we discovered that the odd isotope $^{137}$Ba$^+$ offers unique, unrealized advantages as an optical atomic frequency standard.  As shown in Chapter~\ref{sec:clockChapter}, the ion lacks certain systematic field shifts that plague many other candidate ions of similar structure.  Further, we learned that radio frequency spectroscopy perfected during our measurements of light shifts could determine excited-state hyperfine structure in the odd-isotope to exacting precision.  As shown in Chapter~\ref{sec:hyperfineChapter}, such hyperfine splitting measurements can yield the details of the atom's nuclear structure by giving us the nuclear magnetic dipole, electric quadrupole, and magnetic octopole moments.
 \chapter{Atomic physics} \label{chap:AtomicPhysics}
\begin{quotation}
\noindent\small The map is not the territory. \\ \flushright{---Alfred Korzybski}
\end{quotation}
The properties of atoms are greatly determined by interactions of valence electrons with intra-atomic electric and magnetic fields that are hidden away from us yet gigantic compared to fields commonly experienced macroscopically.  For instance, the electrostatic field typically experienced by an atomic electron is a staggering $E \sim 10^{9}$~V/cm.  Manipulating matter by attempting to beat these forces in the lab is a biblical business fit for Goliath:  he'll need high energy collisions, large and expensive particle accelerators, and radiation safety training.  The rest of us have to employ a strategy David might find familiar:  applying tiny but \emph{resonant} laboratory fields to an atomic system allows us almost arbitrary manipulation and quantum control of atomic wavefunctions.

This chapter begins by describing an atomic system whose two levels are coupled by off, near, and on-resonance interactions.  We examine some basic behaviors of the system fundamental to our study of single trapped ion spectroscopy:  Rabi oscillations, the ac-Stark effect or light shift, spontaneous decay, and power broadening.  We then consider simple applications of these ideas: adiabatic rapid passage and laser cooling.  We briefly discuss atom/light interactions beyond the electric dipole approximation: selection rules and Rabi frequencies for magnetic dipole and electric quadrupole transitions.

The second half of the chapter treats systems of more than two levels directly relevant to our trapped ion work.  We begin by describing the density matrix formalism and immediately apply it to a three-level problem.  This discussion naturally leads to the technique of electron-shelving.  Since much of this work deals with electron spin resonance, we discuss the Zeeman effect and spin oscillations in $J \ge 1/2$ systems.  Finally we review the useful phenomena of optical pumping and the destabilization of dark states. 

\section{The two state coherent interaction} \label{sec:TwoStates}
\begin{figure}
\centering
\includegraphics{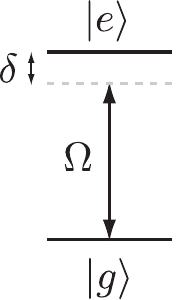}
\caption[Two states coupled by a Rabi frequency $\Omega$ detuned $\delta$ from resonance.]{A two state coupling diagram depicts two atomic levels: a ground state $|g\rangle$ and an excited state $|e \rangle$, connected by an interaction $\Omega = \tfrac{1}{\hbar}| \langle g | H_\text{int} | e \rangle|$.  The interaction is assumed to be near resonant with the atomic energy level splitting $\omega_0$. The resonant field responsible for $H_\text{int}$ is oscillatory at $\omega_L = \omega_0 + \delta$ where $\delta$ is called the \emph{detuning}.  Spontaneous decay of the excited state is ignored for now.} 
\label{fig:twoStateRabiLevels}
\end{figure}
 
\subsection{Rabi oscillations} \label{sec:rabiOscillations}
Following standard treatments (e.g., \cite{metcalf1999lct, foot2005ap}), we first posit an atomic Hamiltonian $H_0$ that includes all of the internal structure of the atom, such as the electrostatic and fine-structure energies, to have just two eigenstates we will label $|g \rangle$ and $| e \rangle$ for \emph{ground} and \emph{excited} states:
\begin{equation}
H_0 = \hbar \sum_{i=g,e} \omega_i | i \rangle \langle i |,
\end{equation}
where $\omega_i$ is the energy of state $i$.  We are free to set $\omega_g = 0$ and will sometimes make the identification $\omega_0 \equiv \omega_e - \omega_g$.  This atomic Hamiltonian is perturbed by a comparatively small time-varying interaction $H_\text{int}$ generated in the lab making the total Hamiltonian:
\begin{equation}
H = H_0 + H_\text{int}(t).
\end{equation}
Writing the atomic wavefunction as a linear combination of orthogonal eigenstates $\phi_k(\boldsymbol{r})$,
\begin{equation}
\Psi(\boldsymbol{r},t) = \sum_{k = g,e} c_k(t) \phi_k(\boldsymbol{r}) e^{-i \omega_k t},
\end{equation}
we then apply the time dependent Schr\"{o}dinger equation to see how it evolves in time:
\begin{align*}
H \Psi(\boldsymbol{r},t) &= i \hbar \frac{\partial}{\partial t} \Psi(\boldsymbol{r},t), \\
\left[H_0 + H_\text{int} \right]  \sum_{k = g,e} c_k(t) \phi_k(\boldsymbol{r}) e^{-i \omega_k t} &= \left( i \hbar \frac{\partial}{\partial t} \right)  \sum_{k = g,e} c_k(t) \phi_k(\boldsymbol{r}) e^{-i \omega_k t}.
\end{align*}
The diagonal terms of $H_\text{int}$ simply represent energy shifts and are absorbed into $H_0$ for now.  Multiplying both sides by $\phi_j(\boldsymbol{r})$ and integrating over $\boldsymbol{r}$ gives differential equations for the state amplitudes $c_g(t)$ and $c_e(t)$:
\begin{align}\label{eq:schrodingerNoDamping1}
i \hbar \frac{d c_g(t)}{dt} &= c_e(t) \langle g | H_\text{int}(t) | e \rangle e^{-i \omega_0 t}, \\
i \hbar \frac{d c_e(t)}{dt} &= c_g(t) \langle e | H_\text{int}(t) | g \rangle e^{i \omega_0 t}, \label{eq:schrodingerNoDamping2}
\end{align}
where we have introduced a \emph{matrix element}
\begin{equation}
 \langle i | H_\text{int}(t) | j \rangle \equiv \int \, \psi_i^*(\boldsymbol{r}) \, H_\text{int}(t) \,  \psi_j(\boldsymbol{r}) \, d^3 \boldsymbol{r}.
\end{equation}
We will first consider the interaction Hamiltonian $H_\text{int}(t)$ due to coherent electric field nearly resonant with this two state system split by $\hbar \omega_0$:
\begin{equation}
H_\text{int}(t) = - e \boldsymbol{E}(\boldsymbol{r},t) \cdot \boldsymbol{r},
\end{equation}
where $e$ is the electron charge.  The oscillating electric field is 
\begin{equation*}
\boldsymbol{E}(\boldsymbol{r},t) = \boldsymbol{E} \cos{(\omega_L t + \boldsymbol{k} \cdot \boldsymbol{r} + \phi)},
\end{equation*}
but we will find it convenient to express this as the sum of two (unphysical) complex exponentials,
\begin{equation}
\boldsymbol{E}(\boldsymbol{r},t) = \frac{1}{2} \boldsymbol{E} \left( \underbrace{e^{+i(\omega_L t + \boldsymbol{k} \cdot \boldsymbol{r} + \phi)}}_\text{co-rotating} + \underbrace{e^{-i(\omega_L t + \boldsymbol{k} \cdot \boldsymbol{r} + \phi)}}_\text{counter-rotating} \right), \label{eq:coAndCounterRotatingTerms}
\end{equation}
where we have assigned the names `co-rotating' and `counter-rotating' such that the co-rotating term has common time-dependence with the state phase factor $e^{-i \omega_0 t}$ due to the atomic Hamiltonian $H_0$ when on resonance.  The magnitude of the wavevector $\boldsymbol{k}$ is the inverse of the wavelength, reduced by $2\pi$: $k = |\boldsymbol{k}| = 2\pi/ \lambda$.  Choosing a constant phase $\phi = 0$ and expanding the exponentials in the interaction Hamiltonian yields
\begin{align}
H_\text{int} &= -\frac{e}{2} \boldsymbol{E} \left( e^{+i(\omega_L t + \boldsymbol{k} \cdot \boldsymbol{r} )} + e^{-i(\omega_L t + \boldsymbol{k} \cdot \boldsymbol{r})} \right)  \cdot \boldsymbol{r} \\
&= -\frac{e}{2} \boldsymbol{E} \left( \left[1 + i(\boldsymbol{k} \cdot \boldsymbol{r}) + \cdots \right] e^{+i\omega_L t }  +  \left[1 - i(\boldsymbol{k} \cdot \boldsymbol{r}) + \cdots \right]e^{-i \omega_L t} \right) \cdot \boldsymbol{r}. \label{eq:multipoleExansion}
\end{align}
Because atomic dimensions $a_0 \approx 0.05$~nm are much smaller than wavelengths of light often considered, the term $(\boldsymbol{k} \cdot \boldsymbol{r}) \sim a_0 / \lambda \ll 1$.  In other words, the electric field gradient across an atomic dimension is often relatively small:
\begin{equation*}
\left| \frac{1}{E} \frac{\partial E}{\partial z} \right| a_0 \ll 1. 
\end{equation*}
By ignoring the $(\boldsymbol{k} \cdot \boldsymbol{r})$ and higher order terms, we make the \emph{electric-dipole approximation}.  We will show later that each multipole component of the  $\boldsymbol{E}(\boldsymbol{r}) \cdot \boldsymbol{r}$ operator leads to unique selection rules and a hierarchy of transition strengths that allows them to be treated separately from the electric dipole coupling.  So, approximating $1 \pm i(\boldsymbol{k} \cdot \boldsymbol{r}) + \cdots \approx 1$, we have
\begin{equation}
H_\text{int} = \sum_{i,j} \frac{\hbar}{2} \big(\underbrace{\Omega_{ij} e^{-i \omega_L t}}_\text{co-rotating} + \underbrace{\Omega^*_{ij} e^{+i \omega_L t}}_\text{counter-rotating} \!\!\!\!\! \big), 
\end{equation}
where $i$ and $j$ each sum over the states $g$ and $e$ in this case, and we have introduced a Rabi-frequency $\Omega_{ij}$ that describes the strength of the coupling:
\begin{align} \label{eq:e1Rabi}
\Omega_{ij} &\equiv - \frac{e}{\hbar} \langle i | \boldsymbol{E} \cdot \boldsymbol{r} | j \rangle \\
&=  -\frac{e}{\hbar} \int \, \psi_i^*(\boldsymbol{r}) \,  \boldsymbol{E} \cdot \boldsymbol{r} \,\psi_j(\boldsymbol{r}) \, d^3 \boldsymbol{r}.
\end{align}
Diagrams like Figure~\ref{fig:twoStateRabiLevels} depict the internal atomic states coupled by a (perhaps detuned by $\delta$) interaction $\Omega$. It is worthwhile to keep in mind that the typical magnitude of an electric dipole matrix element $\sim e a_0$, which gives us~\cite{budker2004ape}
\begin{equation} \label{eq:RabiFrequencyBasic}
|\Omega_{ij}| \sim \frac{e a_0}{\hbar} E \approx \left[ 1.28 \,\, \frac{\text{MHz}}{\text{V/cm}} \right] \cdot E. 
\end{equation}

All that is left is to remove the time dependence from $H_\text{int}$, which in our context is oscillating quickly at optical frequencies.  One method is to apply a time dependent unitary transformation to the total Hamiltonian corresponding to rotation at $\omega_L$.  Under this transformation, the counter-rotating term is oscillating faster than any timescale in the problem.  We can make the \emph{rotating wave approximation}, which amounts to ignoring terms proportional to $1/\omega_L$ compared to those that scale with $1/\delta$, and neglect the counter-rotating term for now.  

\begin{figure}
\centering
\includegraphics[width=5 in]{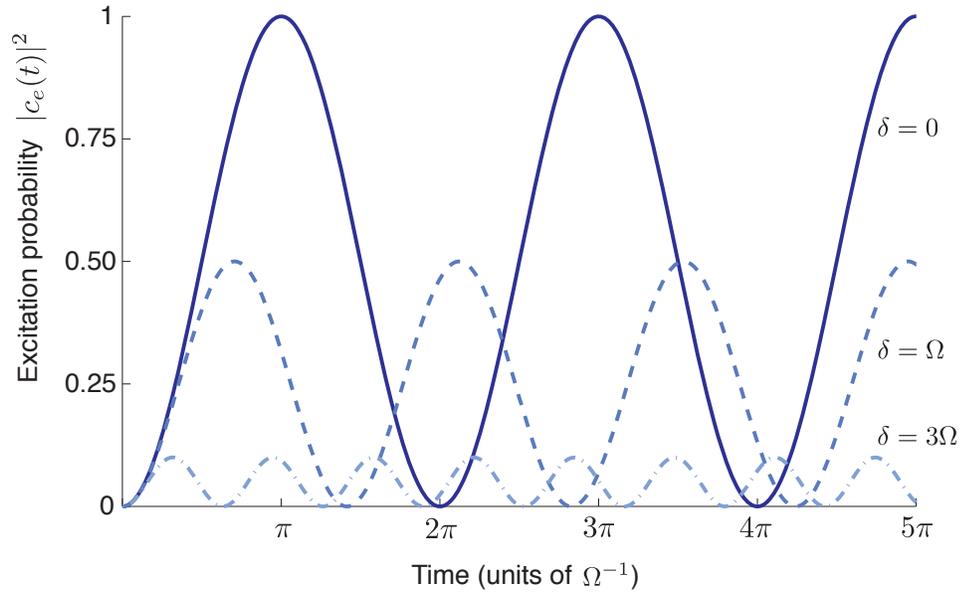}
\caption[Two-state Rabi oscillations: a time-domain picture]{The excitation probability of a two state system undergoing Rabi oscillation (with no damping) as a function of time at various detunings. $\delta = \omega_L - \omega_0 =0$ indicates an exactly resonant excitation field and we see full oscillation of the excited state probability with rate $\Omega$ called the Rabi oscillation frequency.  Faster but smaller amplitude oscillations are predicted for larger $\delta$ at a rate $\Omega' = \sqrt{\Omega^2 + \delta^2}$ sometimes called the generalized Rabi frequency.  Rabi oscillations are experimentally observed in an electron spin resonance in a single barium ion in section~\ref{sec:spinResonanceLS}.}
\label{fig:twoStateRabi}
\end{figure}
\begin{figure}
\centering
\includegraphics[width=5 in]{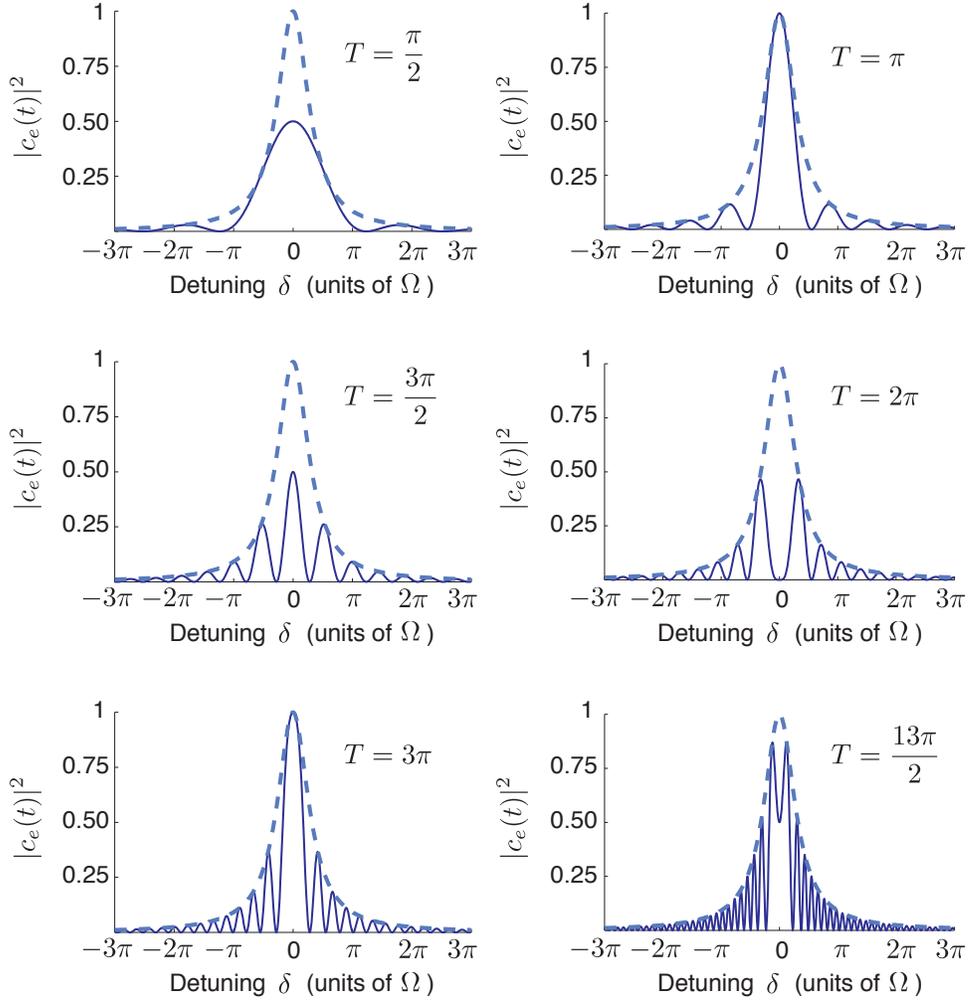}
\caption[Two-state Rabi oscillations: a frequency-domain picture]{The excitation probability of a two state system undergoing Rabi oscillation (with no damping) as a function of detuning away from resonance $\delta = \omega_L - \omega_0$.  The subplots show the state of the system after various excitation times $T$ in units of $\Omega^{-1}$, the inverse Rabi oscillation frequency. The dashed line is a Lorentzian of width $\Omega$ and represents the limiting shape as $T \to \infty$.  The narrowest features of the curves have widths governed by $T^{-1}$ due to the uncertainty principle. On-resonant pulses ($\delta =0$) of lengths $T = \pi/2$, $\pi$, $\ldots$ are often called $\pi/2$-pulses, $\pi$-pulses, etc.  Experimental spectra such as these are obtained from an electron spin resonance in a single barium ion in Section~\ref{sec:spinResonanceLS}.}
\label{fig:twoStateRabiFreq}
\end{figure}

Uncoupling the equations of motion Eq.~\ref{eq:schrodingerNoDamping1} and Eq.~\ref{eq:schrodingerNoDamping2} by differentiating again gives second-order differential equations~\cite{budker2004ape}
\begin{align}
\frac{d^2 c_g(t)}{dt^2} - i \delta \frac{d c_g(t)}{dt} + \frac{\Omega^2}{4} c_g(t) &= 0, \\
\frac{d^2 c_e(t)}{dt^2} + i \delta \frac{d c_e(t)}{dt} + \frac{\Omega^2}{4} c_e(t) &= 0,
\end{align}
where $\Omega = |\Omega_{ge}|$. Assuming initial conditions corresponding to a pure ground state at $t=0$,
\begin{equation*}
c_g(0) =1, \qquad c_e(0) =0,
\end{equation*}\
the time evolution is
\begin{align}
c_g(t) &= \left(\cos \frac{\Omega' t}{2} - i \frac{\delta}{\Omega'} \sin \frac{\Omega' t}{2} \right) e^{i \delta t/2}, \\
c_e(t) &= -i \frac{\Omega}{\Omega'} \sin \frac{\Omega' t}{2} e^{-i \delta t /2},
\end{align}
where 
\begin{equation} \label{eq:generalizedRabiFrequency}
\Omega' \equiv \sqrt{\Omega^2 + \delta^2}
\end{equation}
is sometimes called the \emph{generalized Rabi frequency}. Squaring to find the time evolution of the state probabilities $|c_g(t)|^2$ and $|c_e(t)|^2$, we find probability oscillations
\begin{align}
P( |g \rangle) &= |c_g(t)|^2 = 1- \frac{\Omega^2}{\Omega'^2} \sin^2 \frac{\Omega' t}{2}, \\
P( |e \rangle) &= |c_e(t)|^2 = \frac{\Omega^2}{\Omega'^2} \sin^2 \frac{\Omega' t}{2}.
\end{align}
We plot $ |c_e(t)|^2$ in Figure~\ref{fig:twoStateRabi} for various detunings $\delta$ of the laser field from resonance.  We see that an off-resonant laser ($|\delta| \gg \Omega$) field produces fast state oscillations that never acquire much probability in the excited state.  Full oscillation between the ground and excited states is possible when the interaction is on-resonance ($\delta = 0$) and occurs at the slower natural Rabi frequency $\Omega$.  

If we instead define a fixed interaction time $T$ and vary the detuning $\delta$ we observe the spectroscopic lineshapes shown in Figure~\ref{fig:twoStateRabiFreq}.  Though the lineshapes have sideband features (sometimes called \emph{Rabi sidebands}), they are bounded by Lorentzians, shown as dashed lines in Figure~\ref{fig:twoStateRabiFreq} with widths governed by $\Omega$.  These represent the maximum probability of excitation for an interaction time $T \to \infty$.  The narrowest structures in the resonances have widths that scale as $T^{-1}$ due to the uncertainty principle.  Resonant pulses with area $(\Omega \cdot T) = \pi/2$, $3 \pi/2$, $\ldots$ put an initial ground state into an equal superposition of $|g \rangle$ and $|e \rangle$. Pulses of $(\Omega \cdot T) = \pi$, $3 \pi$, $\ldots$ transfer an initial ground state into the excited state with ideally unit probability.  Pulses of $(\Omega \cdot T) = 2\pi$, $4 \pi$, $\ldots$ return an initial ground state back to the ground state.

In summary, a resonant interaction is characterized by a strength $\Omega$ which is proportional to an atomic matrix element and is numerically equal to the rate at which probability will flop between the two states coupled by the resonant interaction.  If the interaction is tuned away from resonance by $\delta$, the ground and excited state probabilities still flop, but with a lower maximum excitation probability and at a faster rate $\Omega' = \sqrt{\Omega^2 + \delta^2}$.

\subsection{The ac-Stark effect, or light shift} \label{sec:acStarkAtomic}
\begin{figure}
\centering
\includegraphics{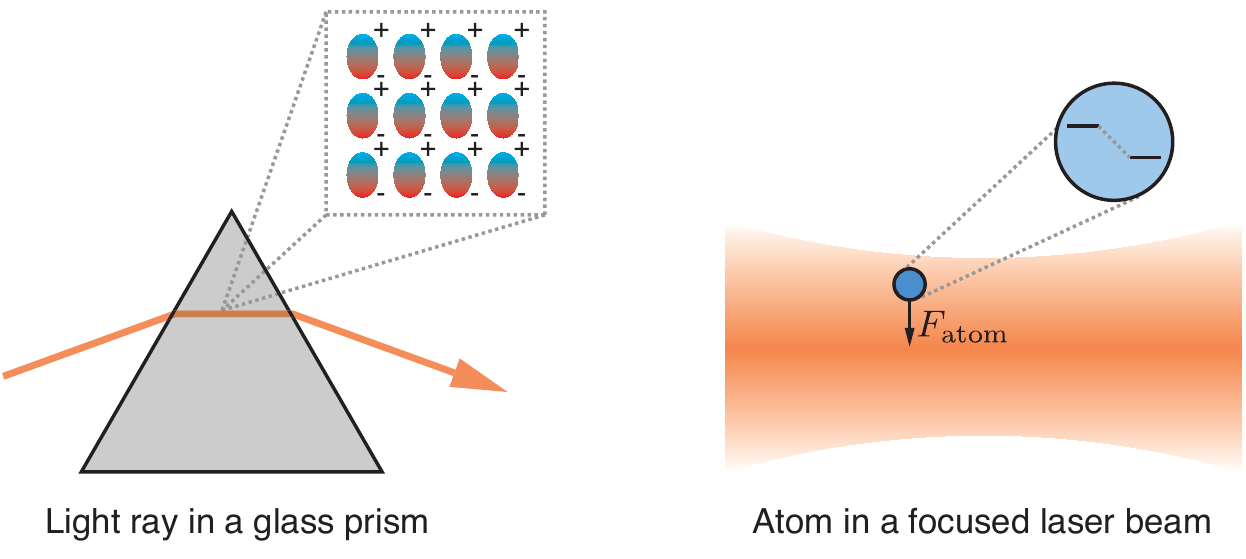}
\caption[Macroscopic refraction of light and the microscopic ac-Stark shift of atoms]{Macroscopic refraction of light is connected to the microscopic ac-Stark shift of atoms~\cite{foot2005ap}.  Refraction of light in a transmitting dielectric medium occurs because the light induces an atomic polarization which interacts with the light field.  This is complemented by an action of the light on the atoms called the ac-Stark effect.  In a beam of far off-resonant red-detuned light, a single atom finds its ground state energy reduced due to an oscillating polarization that is in phase with the light electric field.  If it lies in a gradient of light intensity (such as a focused laser beam), the atom feels a \emph{dipole force} and can be optically moved or trapped (see~\cite{miller1993for,chu1986eoo,stamperkurn1998ocb} for instance).}
\label{fig:lightShiftRefraction}
\end{figure}
Everyone knows something about refraction:  light beams are bent at the interface of macroscopic dielectric media.  We assign matter an index of refraction $n$ and often forget about its atomic origins.  If refraction is the action of atoms on off-resonant light, the ac-Stark shift is the back-action of the light on the atoms as suggested by Figure~\ref{fig:lightShiftRefraction}.  A more sophisticated representation of this idea is the Kramers-Kronig relation~\cite{jackson1998ced} which fixes the relationship between refraction and absorption (the imaginary part of a complex $n$). 

\begin{figure}
\centering
\includegraphics[scale=.75]{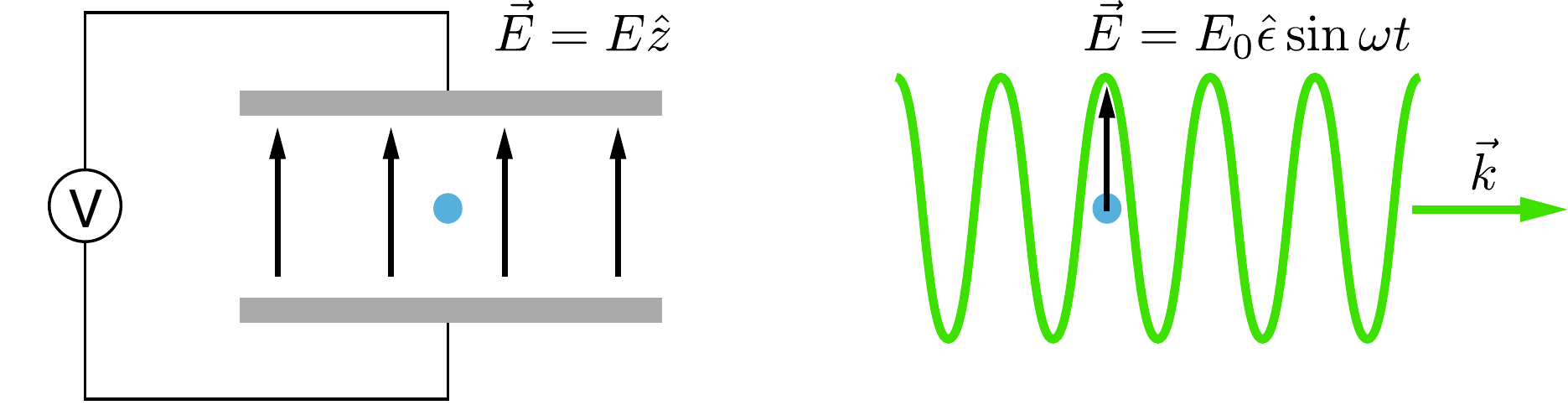}
\caption[Static and ac-Stark shift conceptual diagram]{It is useful to understand the ac-Stark shift as a dynamic polarizability in analogy to the well studied dc-Stark effect due to a static polarizability.  Here the electric field of a light wave $\boldsymbol{E} = E \hat{\boldsymbol{\epsilon}} \sin \omega t$ has polarization $ \hat{\boldsymbol{\epsilon}}$ and frequency $\omega$ compared with a static electric field.}
\label{fig:starkComparisonAtomic}
\end{figure}
To first guess the qualitative features of the ac-Stark effect, Figure~\ref{fig:starkComparisonAtomic} motivates the comparison with the action of a static electric field $\boldsymbol{E}$ on a non-degenerate atomic ground state: the dc-Stark shift.  The perturbation is second-order (one power of $E$ induces a polarization of the atom, another power of $E$ interacts with that polarization), and when the atomic polarization is aligned to $\boldsymbol{E}$, the energy of the state is reduced:
\begin{equation} \label{eq:dcStark}
\Delta E_{|g \rangle}^\text{dc-Stark} = \left(\frac{eE}{2}\right)^2 \sum_{e'} \frac{ | \langle g | z | e' \rangle|^2}{E_g - E_{e'}}. 
\end{equation}
Here we've explicitly allowed for additional excited states $| e' \rangle$ in the sum.  The static Stark shift analysis proves relevant for the ac-Stark shift due to an oscillating electric field tuned far \emph{below} an atomic resonance. This is not a surprise since, in some sense, a slowly oscillating field ought to have a similar effect as a dc field.  The induced dipole polarization in the atom oscillates \emph{in phase} with the applied field, and so the energy of the ground state is reduced.  Likewise, the atomic polarization induced by an oscillating electric field tuned above an atomic resonance will be 180$^\circ$ out of phase with that field, and so the ground state energy increases.

\begin{figure}
\centering
\includegraphics{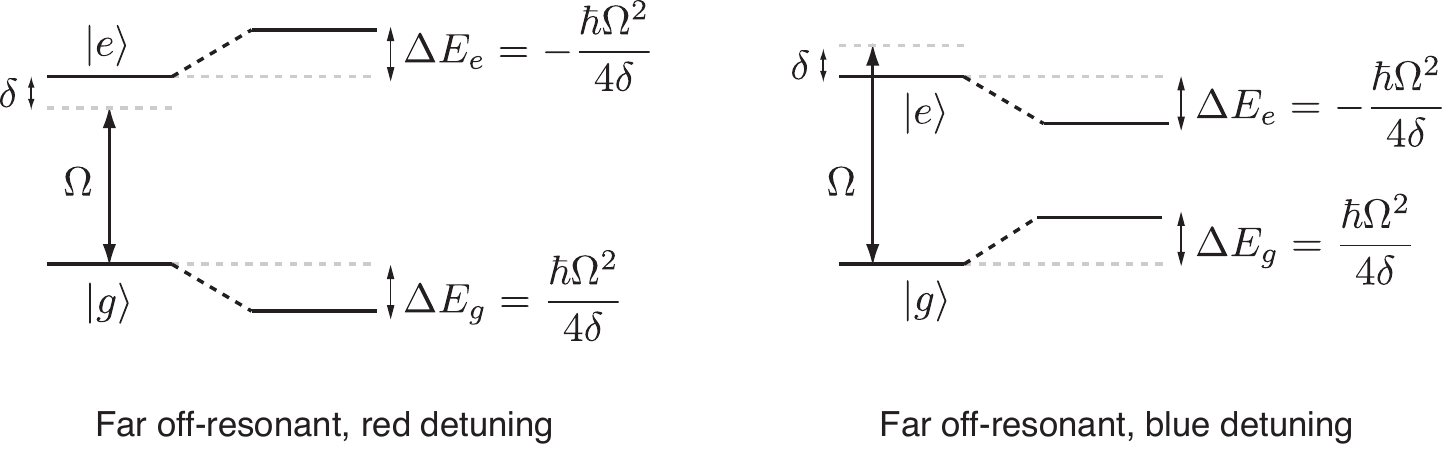}
\caption[The ac-Stark effect due to far detuned laser light on a two-level system]{Illustration of the ac-Stark effect or light shift on a two-level system due to far detuned laser light.  When the laser frequency is far below the two state resonance, the induced atomic polarization follows the electric field in phase, reducing the ground state energy.   Likewise, laser fields tuned far above the two state resonance will induce an atomic polarization out of phase with the electric field:  the ground state energy will be increased.}
\label{fig:twoStateRabiLightShift}
\end{figure}

We will eventually be able to write the ac-Stark energy shift in a form such as Eq.~\ref{eq:dcStark}, but we will begin with the treatment offered in the previous section.  There, we saw that the off-diagonal elements of $H_\text{int}$ lead to oscillation between the eigenstates of the system.  However, the energies of these states also change due to the application of the perturbing Hamiltonian.  To begin, we eliminate the time dependence in $H_\text{int}$ in order to find the new steady state energies.  This is accomplished by making the redefinitions~\cite{metcalf1999lct}
\begin{align}
c_g'(t) &\equiv c_g(t), \\
c_e'(t) &\equiv c_e(t) e^{-i \delta t},
\end{align}
which make the equations of motion
\begin{align}
i \hbar \frac{d c_g'(t)}{dt} &= c_e'(t) \frac{\hbar \Omega}{2}, \\
i \hbar \frac{d c_e'(t)}{dt} &= c_g'(t) \frac{\hbar \Omega}{2} - c_e'(t) \hbar \delta.
\end{align}
In the new basis, $H_\text{int}$ is
\begin{equation}
 H_\text{int} = \frac{\hbar}{2} \begin{pmatrix} -2 \delta &  \Omega \\ \Omega & 0 \end{pmatrix}.
\end{equation}
Using first order perturbation theory, the energy shifts are
\begin{equation}
\Delta  E_{e,g} = \frac{\hbar}{2} (-\delta \mp \Omega'),
\end{equation}
where the generalized Rabi frequency $\Omega' = \sqrt{\delta^2 + \Omega^2}$ as in the previous section.  In the far off-resonance limit $|\delta| \gg \Omega$, the shifts to the old eigenstates are
\begin{align} \label{eq:lightShiftsNoBlochS} 
\Delta E_g &= \frac{\hbar \Omega^2}{4 \delta} \\
\Delta E_e &= -\frac{\hbar \Omega^2}{4 \delta} \notag.
\end{align}
As illustrated in Figure~\ref{fig:twoStateRabiLightShift}, the sign of the shift depends on the detuning of the applied field from resonance.  Our comparison with the dc-Stark shift (Eq.~\ref{eq:dcStark}) had merit:  the ground state energy does decrease for light tuned below resonance, and by expanding $\Omega$ using Eq.~\ref{eq:e1Rabi} and $\delta = E_g - E_e - \hbar \omega_L$, the shift has the form
\begin{equation*}
\Delta E_g^\text{ac-Stark} = \frac{\hbar \Omega^2}{4 \delta} = \left(\frac{e E}{2} \right)^2 \frac{ | \langle g | \boldsymbol{\epsilon} \cdot \boldsymbol{r} | e' \rangle |^2}{E_g - E_{e'} - \hbar \omega_L}.
\end{equation*}
Since the intensity of an applied laser beam scales as $E^2$, another natural way to express the shift is 
\begin{equation}
\Delta E_g^\text{ac-Stark} = 2 \pi \alpha I \frac{ | \langle g | \boldsymbol{\epsilon} \cdot \boldsymbol{r} | e' \rangle|^2}{E_g - E_{e'} - \hbar \omega_L}.
\end{equation}
The matrix element in these expressions has angular dependence on the polarization of the light field, and in general, many levels $|e_i' \rangle$, including continuum states and core excitations, must be summed over in accordance with second-order perturbation theory.  These generalizations are discussed in Section~\ref{sec:lightShiftZeeman}.

A key application of the light shift, dipole force trapping, makes use of a ground state energy lowered by red-detuned light.  As shown in Figure~\ref{fig:lightShiftRefraction}, a focused laser beam red-detuned from an atomic resonance will have a spatial gradient of electric field strength which results in a net force on the atoms towards the center of the beam given by~\cite{budker2004ape}
\begin{equation}
(F_z)_\text{dipole} = - \frac{\partial (\Delta E_g^\text{ac-Stark})}{\partial z} = \frac{\hbar \Omega}{2 \delta} \frac{\partial \Omega}{\partial z}.
\end{equation}
If the atoms are sufficiently cold then this force can be made large enough to optically confine populations of many millions of atoms;  this serves as a key tool in modern experimental atomic physics (see~\cite{miller1993for,chu1986eoo,stamperkurn1998ocb} for instance).

The expression in Eq.~\ref{eq:lightShiftsNoBlochS} neglects an effect known as the Bloch-Siegert shift due to our early use of the rotating wave approximation.  Inclusion of the counter rotating term (see Eq.~\ref{eq:coAndCounterRotatingTerms}) tends to shift energy levels apart by
\begin{align} \label{eq:lightShiftsBlochS}
\Delta E^\text{Bloch-Siegert}_{g,e} &= \mp \frac{\Omega^2}{4 (\omega_0 + \omega_L)} \\
&\approx \frac{\Omega^2}{8 \omega_0} \notag
\end{align}
when $\omega_0 \approx \omega_L$.  The Bloch-Siegert term is small for optical fields where $\Omega \ll \omega_L$ is common but is comparable to the co-rotating light shift term for \emph{very} far off-resonance light $\delta \sim \omega_L$.  The Bloch-Siegert term is often important for radio frequency transitions where $\Omega \sim \omega_L$ is quite common.  See~\cite[pg.\ 460]{cohen-tannoudji1992api} for a complete treatment.

Section~\ref{sec:rabiOscillations} has shown us that the old states $| g \rangle$ and $|e \rangle$ are no longer stationary eigenstates.  However, we know from perturbation theory that the new states that diagonalize the total Hamiltonian can be written as admixtures of the old:
\begin{align} \label{eq:lightShiftStateMixture}
| 1 \rangle &= \cos \theta | g \rangle + \sin \theta | e \rangle, \\
| 2 \rangle &= -\sin \theta |g \rangle + \cos \theta | e \rangle, \\
\intertext{with} 
\tan 2\theta &= \frac{-\Omega}{\delta}.
\end{align}
For maximal coupling, $|\Omega /\delta| \to \infty$ (strong on-resonance interaction), we see that $\theta \to 45^\circ$ meaning that maximal mixing occurs.  In this limit, the approximation given in Eq.~\ref{eq:lightShiftsNoBlochS} is no longer valid, and the energy level shifts are not given by Eq.~\ref{eq:lightShiftsNoBlochS}.  The atomic state becomes \emph{dressed} by an infinite ladder of photon occupation number states $| N \rangle$.  The new eigenstates, Eq.~\ref{eq:lightShiftStateMixture}, have energy shifts~\cite{koerber2003thesis}
\begin{equation}
E_{1,2} = \frac{\hbar}{2} \left(- \delta \pm \sqrt{\Omega^2 + \delta^2} \right)
\end{equation}
so that on resonance, $\delta = 0$, three spectral frequencies, called the Mollow triplet, are observable:
\begin{center}
\begin{tabular}{c c}
Transition		&  Resonant frequency \\ \hline \hline
$|1, N \rangle \leftrightarrow |1, N-1 \rangle$		& $\omega_0$ \\
$|2, N \rangle \leftrightarrow |2, N-1 \rangle$		& $\omega_0$ \\
$|1, N \rangle \leftrightarrow |2, N-1 \rangle$		& $\omega_0 + \Omega$ \\
$|2, N \rangle \leftrightarrow |1, N-1 \rangle$		& $\omega_0 - \Omega$
\end{tabular}
\end{center}

\subsection{Spontaneous decay, power broadening}\label{sec:decayPowerBroadening}
\begin{figure}
\centering
\includegraphics{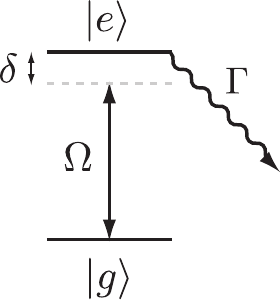}
\caption[An interaction Rabi frequency $\Omega$ detuned by $\delta$ with decay rate $\Gamma$]{Excited atomic states decay by spontaneous emission of radiation at a rate $\Gamma$ that depends on the strength of coupling to quantum vacuum fluctuations. Most phenomenology of atom and photon interactions can be understood by comparing the driving rate $\Omega$, the detuning frequency $\delta$, and the spontaneous decay rate $\Gamma$.}
\label{fig:twoStateRabiLevelsDecay}
\end{figure}
\begin{figure}
\centering
\includegraphics{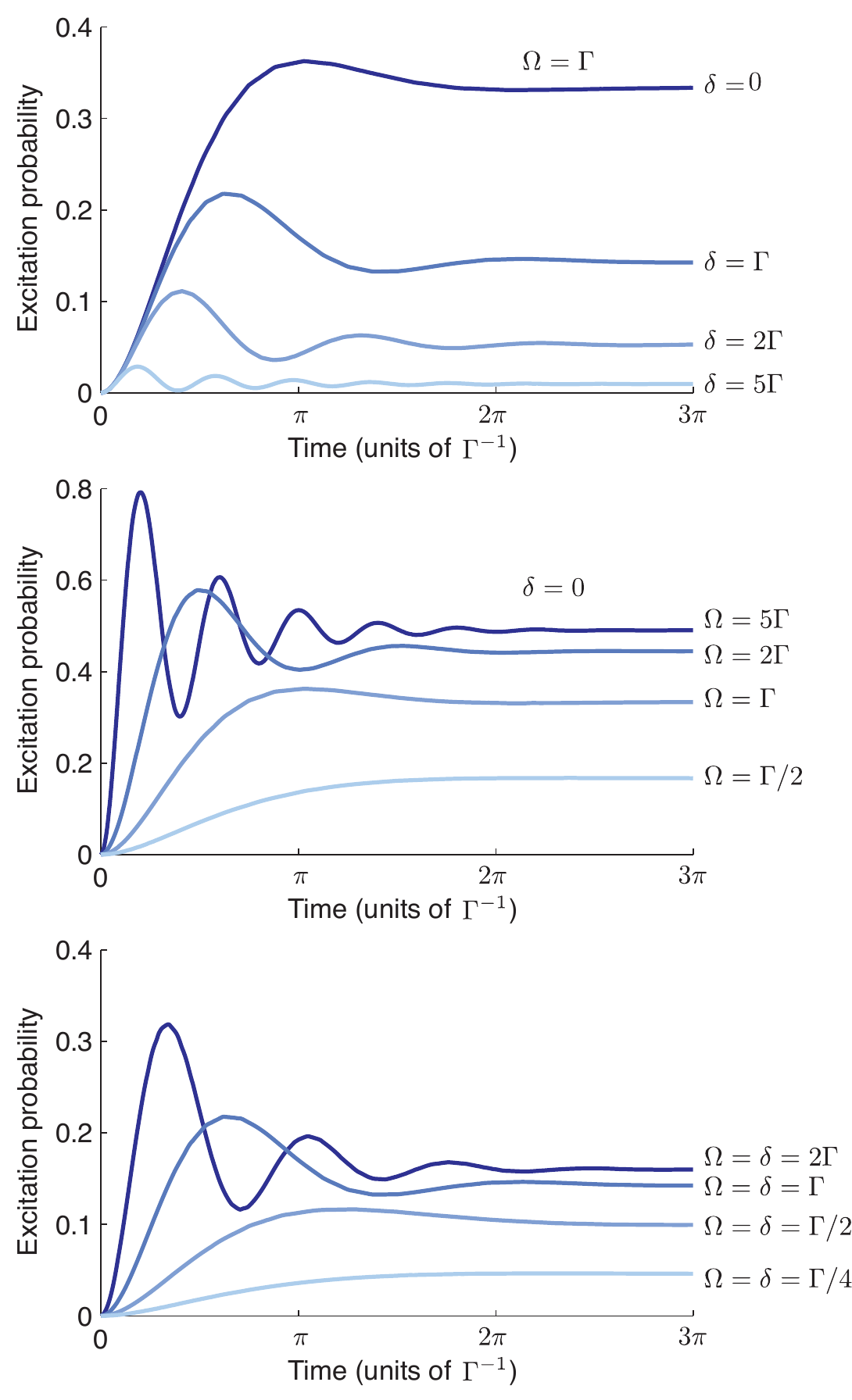}
\caption[Rabi oscillations modified by damping due to spontaneous decay]{Rabi oscillations are modified by damping due to spontaneous decay.  In the top plot, the Rabi frequency is set equal to the decay rate ($\Omega = \Gamma$) while the detuning is varied.  In the middle plot the interaction is made resonant ($\delta = 0$) while the strength of the coupling $\Omega$ is varied.  In the third plot, both $\delta$ and $\Omega$ are varied together. Note qualitative similarities to under-damped, critically damped, and over-damped classical oscillators.}
\label{fig:twoStateRabiDecayGraphs}
\end{figure}
All good things must come to an end---excited states are no exception. The process of \emph{spontaneous decay} is often depicted in diagrams such as Figure~\ref{fig:twoStateRabiLevelsDecay}.  If an electric dipole transition $|e \rangle \to |g \rangle$ is allowed, then an atom in the excited state $|e \rangle$ lives, on average, a lifetime $\tau = \Gamma^{-1}$, where
\begin{equation} \label{eq:spontaneousDecayAtomic}
\Gamma = \frac{\omega_0^3}{3 \pi \epsilon_0 \hbar c^3} | \langle e | e r | g \rangle |^2.
\end{equation}
This finite lifetime leads to a natural spectroscopic broadening of the state by $\Gamma$ due to the time/energy uncertainty principle.  As shown in Figure~\ref{fig:twoStateRabiDecayGraphs}, Rabi oscillations only appear when the driving interaction is at least comparable to the decay rate $\Omega \gtrsim \Gamma$. The process responsible for the state decay is coupling to the quantum vacuum~\cite{cohen-tannoudji1992api}.  As such, $\Gamma$ can be substantially modified by altering the boundary conditions of the vacuum with placement of dielectric surfaces, or high reflectance mirrors, techniques that fall under the name Cavity-QED\footnote{This fascinating and growing research area has demonstrated the modification of trapped ion decay lifetimes~\cite{kreuter2004sel} and level energies~\cite{wilson2003vfl}, cavity-enhanced cooling of a single trapped ion~\cite{bushev2006fcs,steixner2005qfc}, generation of single photon states using a coupled ion and cavity~\cite{keller2004cgs}, as well as coupled cavity/ion schemes for quantum information processing applications~\cite{steane2000qct}.  See \cite{buvzek1997cqc, mabuchi2002cqe,hinds1991cql,kimble1998sis} for reviews.}.  Decay via other multipole terms of the electromagnetic field are indeed possible but are much slower as discussed in Section~\ref{sec:multipoleInteractions}.

A rigorous inclusion of spontaneous decay into the treatment presented in Section~\ref{sec:rabiOscillations} would require state vectors accounting for all possible allowed photon modes in the vacuum, which is not straightforward.  Since we chiefly care about the quantum state of the atom, the effects of spontaneous decay can be included in our treatment by modifying Eq.~\ref{eq:schrodingerNoDamping2} with a term:
\begin{equation}
\frac{d c_e(t)}{dt} = - \frac{\Gamma}{2} c_e(t).
\label{eq:decayTerm}
\end{equation}

Spontaneous decay is largely how we detect atoms.  When a population is transferred from the excited state to the ground state via Rabi oscillation, the emitted photon goes \emph{into} the incident beam perfectly in phase and polarization.  Distinguishing the stimulated emission from the incident laser beam is possible~\cite{cohentannoudji1977dad} due to presence of the Mollow triplet (see Section~\ref{sec:acStarkAtomic}), but this is not straightforward.  When the excited state decays spontaneously, however, the emitted photon can be in any direction with a distribution of polarization states consistent with a semi-classical treatment of the atom as a dipole oscillator.  We say these photons are \emph{scattered} by the atom and are thus easily distinguished from the incident beam.  We discuss later that this random element of scattering via spontaneous decay is a key component of the laser cooling technique.

\begin{figure}
\centering
\includegraphics[width = 5 in]{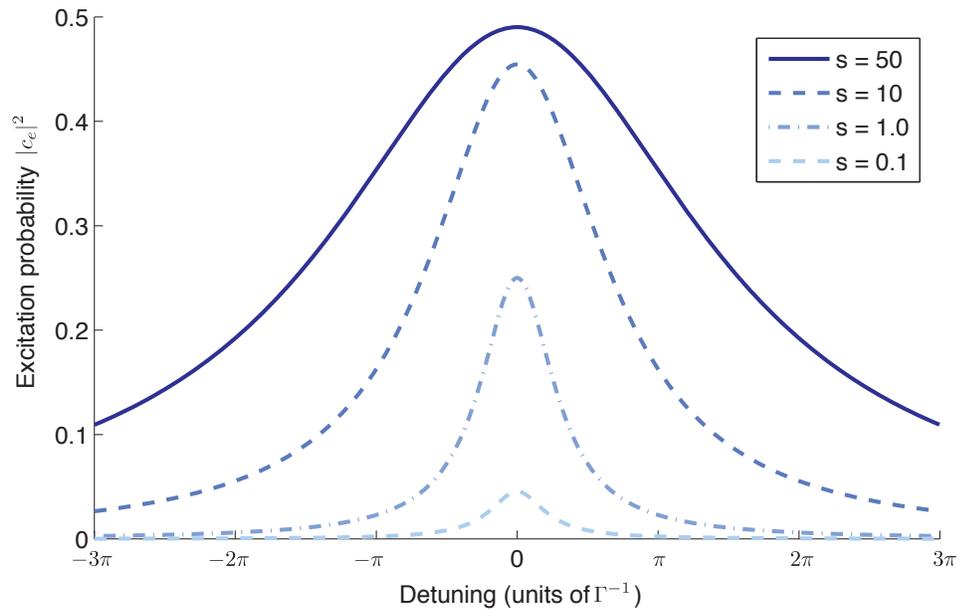}
\caption[Power broadening is important when $s = 2 \Omega^2/\Gamma \gg 1$.]{When the driving field $\Omega \gg \Gamma$, significant spectroscopic broadening occurs because the lifetimes of the ground and excited states are shortened by the coupling.  Notice that the excited state population $|c_e|^2$ (proportional to the photon scattering rate) peaks at 1/4 when the transition is saturated $s \equiv 2 \Omega^2 / \Gamma^2 = 1$.  As $s \to \infty$, the population tends to a maximum of 1/2 while the full width at half maximum scales as $\sqrt{1 + s}$.}
\label{fig:powerBroadening}
\end{figure}

From Eq.~\ref{eq:decayTerm} we see that each unit of time the atom spends in the excited state increases the probability it will decay and emit a photon.  So to make it scatter more photons per second, we should try and populate the excited state as often as possible.  Can we make our atom scatter ever faster, and shine ever brighter by simply increasing the driving interaction $\Omega$ that populates the excited state?  The answer is no.  Any interaction $\Omega$ that drives population from $|g \rangle \to |e \rangle$ also drives population back down, $|e \rangle \to |g \rangle$, via stimulated emission.  Eq.~\ref{eq:lightShiftStateMixture} shows that as $\Omega \to \infty$, the population of the excited state $|c_e(t)|^2 \to 1/2$.  This implies that the scattering rate has an upper limit:
\begin{equation}
R_\text{scattering} = \Gamma |c_e|^2  < 1/2 \Gamma.
\end{equation}

What happens as $\Omega$ is made larger than $\Gamma$ is called \emph{power broadening}.  As $\Omega \to \infty$, both the ground state and eventually (when $\Omega \gg \Gamma$) the excited state receive an effective lifetime given by $\Omega^{-1}$:  the resonant coupling shortens the lifetimes of the states and therefore broadens their energy uncertainties by the Heisenberg principle.  The spectroscopic result, shown in Figure~\ref{fig:powerBroadening}, is that the transition linewidth $\Delta \omega$ (FWHM) increases significantly from $\Gamma$ as the coupling strength is increased:
\begin{equation}
\Delta \omega = \Gamma \sqrt{1 + \frac{2 \Omega^2}{\Gamma^2}} = \Gamma \sqrt{1 + \frac{I}{I_\text{sat}}}
\end{equation}
where $I$ is the intensity of the driving field, and $I_\text{sat}$ is the intensity at which the Rabi oscillation frequency $\Omega$ becomes comparable with the spontaneous decay rate $\Gamma$.  In convenient laboratory units, this is~\cite{budker2004ape}
\begin{equation}
I_\text{sat} \, [ \text{mW}/\text{cm}^2]= \frac{1}{1.23} \left( \frac{\Gamma}{2\pi} \, [\text{MHz}]\right)^2 \left(\frac{1}{| \langle e | er | g \rangle | \, [ea_0]} \right)^2.
\end{equation}

The quantity $I / I_\text{sat} = 2 \Omega^2 / \Gamma^2$ is sometimes defined as the on-resonance saturation parameter; we define:
\begin{equation}
s \equiv \frac{2 \Omega^2}{\Gamma^2} = \frac{I}{I_\text{sat}}.
\end{equation}
For a given detuning $\delta$, the steady-state probability for the atom to be in the excited state is
\begin{equation} \label{eq: twoStateProbabilitySaturation}
P_e = \frac{s / 2}{1 + s + (2 \delta/\Gamma)^2},
\end{equation}
which is plotted in Figure~\ref{fig:powerBroadening} for various values of the saturation parameter $s$.  Notice an essential result:
\begin{equation} \label{eq:steadyStateExcitationProb}
P_e \to \frac{1}{2} \qquad \text{ as} \quad \frac{\Omega}{\Gamma} \to \infty.
\end{equation}

\subsection{Adiabatic rapid passage} \label{sec:atomicARP}
\begin{figure}
\centering
\includegraphics[width=5 in]{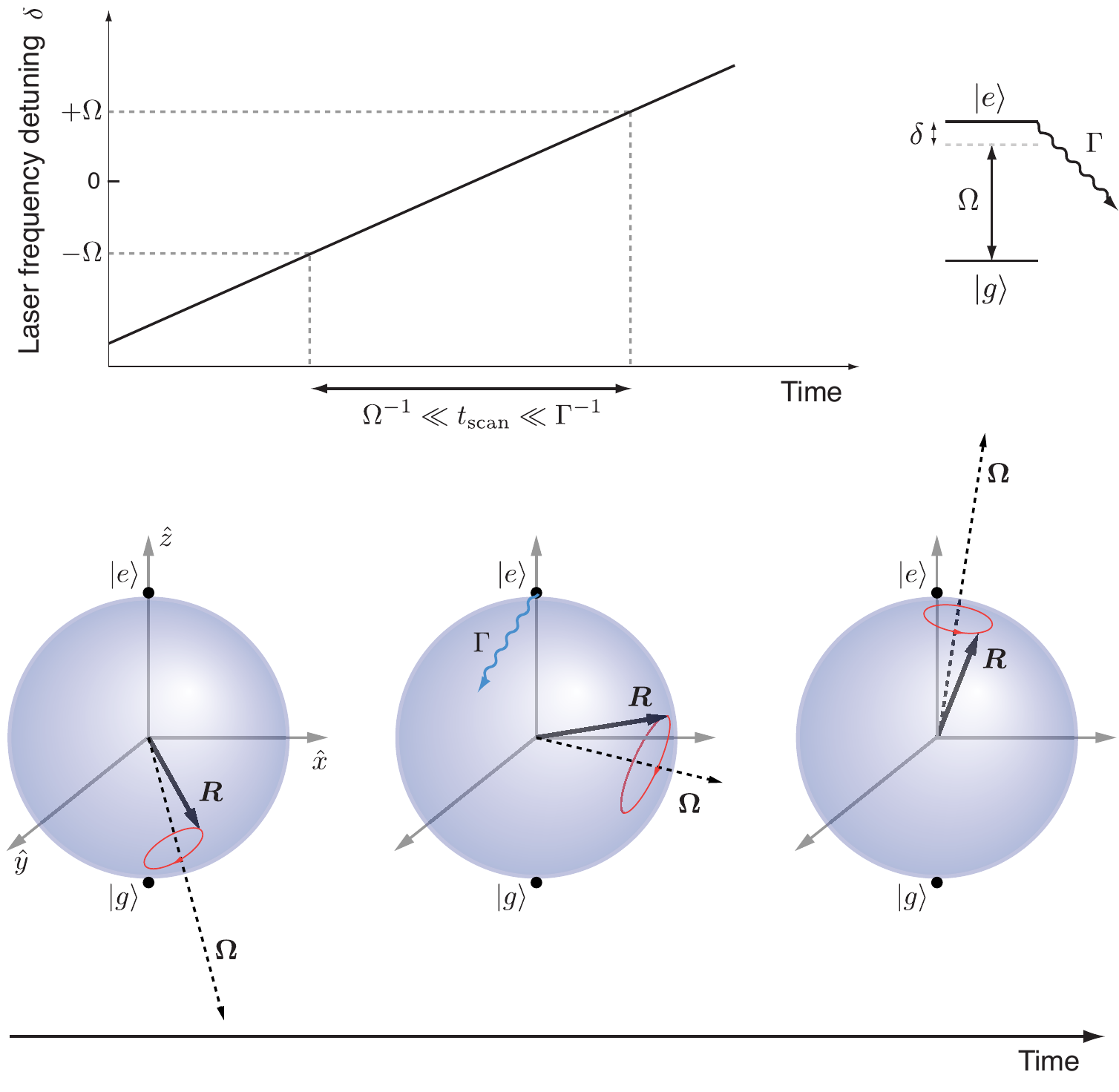}
\caption[Adiabatic rapid passage: an efficient method of state population transfer.]{Adiabatic rapid passage: an efficient method of state population transfer.  By modeling a two level atomic population as a Bloch vector $\boldsymbol{R}$ and expressing the interaction as a vector $\boldsymbol{\Omega}$ whose length is $\Omega' = \sqrt{\Omega^2 + \delta^2}$ and whose projection on the $\hat{\boldsymbol{z}}$ axis is the detuning $\delta$, we find that the Schr\"{o}dinger equation predicts that $\boldsymbol{R}$ undergoes precession around $\boldsymbol{\Omega}$ as long as variations in $\boldsymbol{\Omega}$ are made adiabatically relative to the precession rate.  A useful consequence is the discovery that scanning the detuning from $\delta \ll -\Omega'$ to $\delta \gg + \Omega'$ rapidly compared to any decay rate $\Gamma$ but adiabatically compared to $\Omega'$ one can efficiently transfer atomic population from the ground to excited state.}
\label{fig:adiabaticRapidPassage}
\end{figure}
\begin{figure}
\centering
\includegraphics{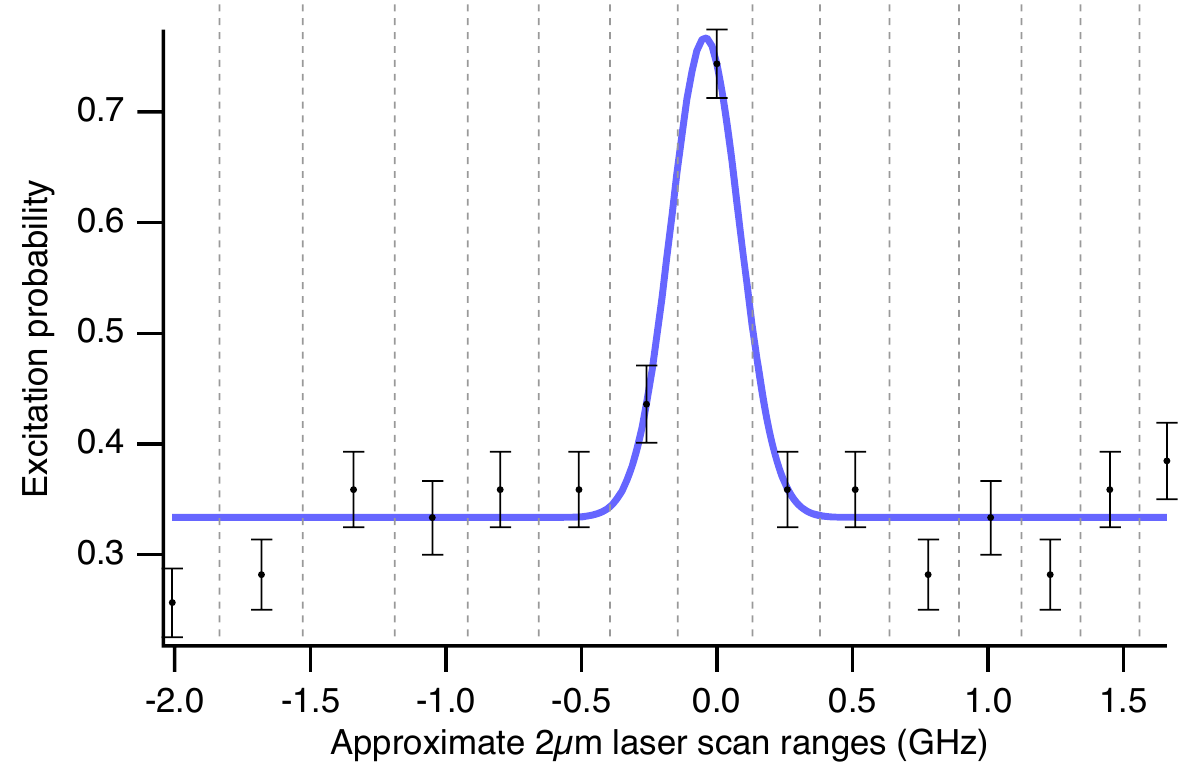}
\caption[Observation of adiabatic rapid passage on the $6S_{1/2} \leftrightarrow 5D_{3/2}$ transition]{Observation of adiabatic rapid passage on the $6S_{1/2} \leftrightarrow 5D_{3/2}$ transition.  One must scan a laser frequency quickly if the relevant decay modes are electric dipole allowed.  In this case, the long lifetime ($\sim$ 80~s) of the $5D_{3/2}$ state granted us easy requirements for adiabatic rapid passage on this 2051~nm transition.  The details of this experiment are given in Section~\ref{sec:ARPobservation}.}
\label{fig:2micronARP}
\end{figure}
Adiabatic rapid passage is a seemingly oxymoronic name for a highly efficient method of transferring state population with a simple laser sweep.  First we construct a Bloch vector representation of the state population $\boldsymbol{R}$ (following~\cite{metcalf1999lct}) that will lie on a unit sphere:
\begin{equation}
\boldsymbol{R} = \underbrace{(c_gc_e^* + c_g^*c_e) \hat{\boldsymbol{x}}
			+  i(c_gc_e^* - c_g^*c_e) \hat{\boldsymbol{y}}}_\text{coherences}
			+ \underbrace{( |c_e|^2 - |c_g|^2) \hat{\boldsymbol{z}}}_\text{state probabilities}. \\
\end{equation}
Note that the component $R_z$ is a measure of the probability for the state to be measured in the ground and excited states, whereas the other components $R_x$ and $R_y$ describe the phase relationship of the two eigenstates in the composite state.  Then we define an interaction vector $\boldsymbol{\Omega}$,
\begin{equation}
\boldsymbol{\Omega} = \text{Re }{ \Omega}  \hat{\boldsymbol{x}}
				   + \text{Im }{\Omega} \hat{\boldsymbol{y}}
				   + \delta \hat{\boldsymbol{z}},
\end{equation}
where, without loss of generality, the component $\Omega_y$ can be set to zero by a unitary transformation.  Note that the component $\Omega_z$ describes the detuning of the interaction from resonance which makes the length of $\boldsymbol{\Omega}$ equal to the generalized Rabi frequency $|\boldsymbol{\Omega}| = \Omega' = \sqrt{\delta^2 + \Omega^2}$ as defined in Section~\ref{sec:rabiOscillations}.  The equations of motion for the state amplitudes (Eq.~\ref{eq:schrodingerNoDamping1}, \ref{eq:schrodingerNoDamping2}) can be written~\cite{metcalf1999lct} in terms of the vectors $\boldsymbol{R}$ and $\boldsymbol{\Omega}$ in an intuitive way:
\begin{equation}
\frac{d \boldsymbol{R}}{dt} = \boldsymbol{\Omega} \times \boldsymbol{R}.
\end{equation}
This expression guarantees that the state vector $\boldsymbol{R}$ length does not vary, so the state amplitudes remain normalized.  Further, if $\boldsymbol{\Omega}$ is quasi-static then it also predicts that $\boldsymbol{R}$ undergoes precession around  $\boldsymbol{\Omega}$.

To see the utility in this formalism, follow Figure~\ref{fig:adiabaticRapidPassage}, and imagine the state starts in the ground state $\boldsymbol{R} = - \hat{\boldsymbol{z}}$ while the interaction is detuned far to the red: $|\Omega_z| \gg |\Omega_x|$ and $\delta < 0$.  As illustrated in the first Bloch sphere, the state vector precesses around $\boldsymbol{\Omega}$ but remains largely in the ground state.  Now suppose that the detuning $\delta$ is slowly swept through 0.  The precession of $\boldsymbol{R}$ follows $\boldsymbol{\Omega}$, very tightly as $\Omega_x \to \infty$, as expected in analogy to a spin and magnetic field adiabatically swept around the Bloch sphere.  The scan of $\delta$ proceeds until $|\Omega_z| \gg | \Omega_x|$ but with $\delta > 0$.  Now, as shown in the final Bloch sphere in Figure~\ref{fig:adiabaticRapidPassage}, the state vector points nearly along $+\hat{\boldsymbol{z}}$, meaning that the probability to find the system in the excited state is high.

This seems like a perfect way to create an excited state, but it comes with the caveat that spontaneous decay ruins the coherence of the process.  Therefore, the scan of $\delta$ (that must be slow with respect to the Rabi oscillation frequency $\Omega$) must be fast with respect to the decay rate $\Gamma$, which explains the seemingly oxymoronic name:  \emph{adiabatic} (with respect to $\Omega$) \emph{rapid} (with respect to $\Gamma$) \emph{passage}.  Because excited states commonly decay via electric dipole radiation, the necessary sweep time might be less than a few nanoseconds, meaning that the laser power required to make a fast Rabi oscillation frequency is often inconvenient or prohibitive.

However, the barium ion system offers a two level system where $\Gamma$ is several seconds:  the $6S_{1/2} \leftrightarrow 5D_{3/2}$ electric quadrupole transition at 2051~nm.  We implemented adiabatic rapid passage in this system by scanning a Tm,Ho:YLF laser over the resonance with $\Omega^{-1} \ll t_\text{scan} \ll \Gamma^{-1}$ and found we could indeed efficiently observe the resonance shown in Figure~\ref{fig:2micronARP}.  Detection of whether the ion was excited to $5D_{3/2}$ involved an \emph{electron shelving} technique discussed in a later Section;  more details of this experiment are given in section~\ref{sec:ARPobservation}.

\subsection{Semi-classical laser cooling}\label{sec:semiclassicalLaserCooling}
\begin{figure}
\centering
\includegraphics{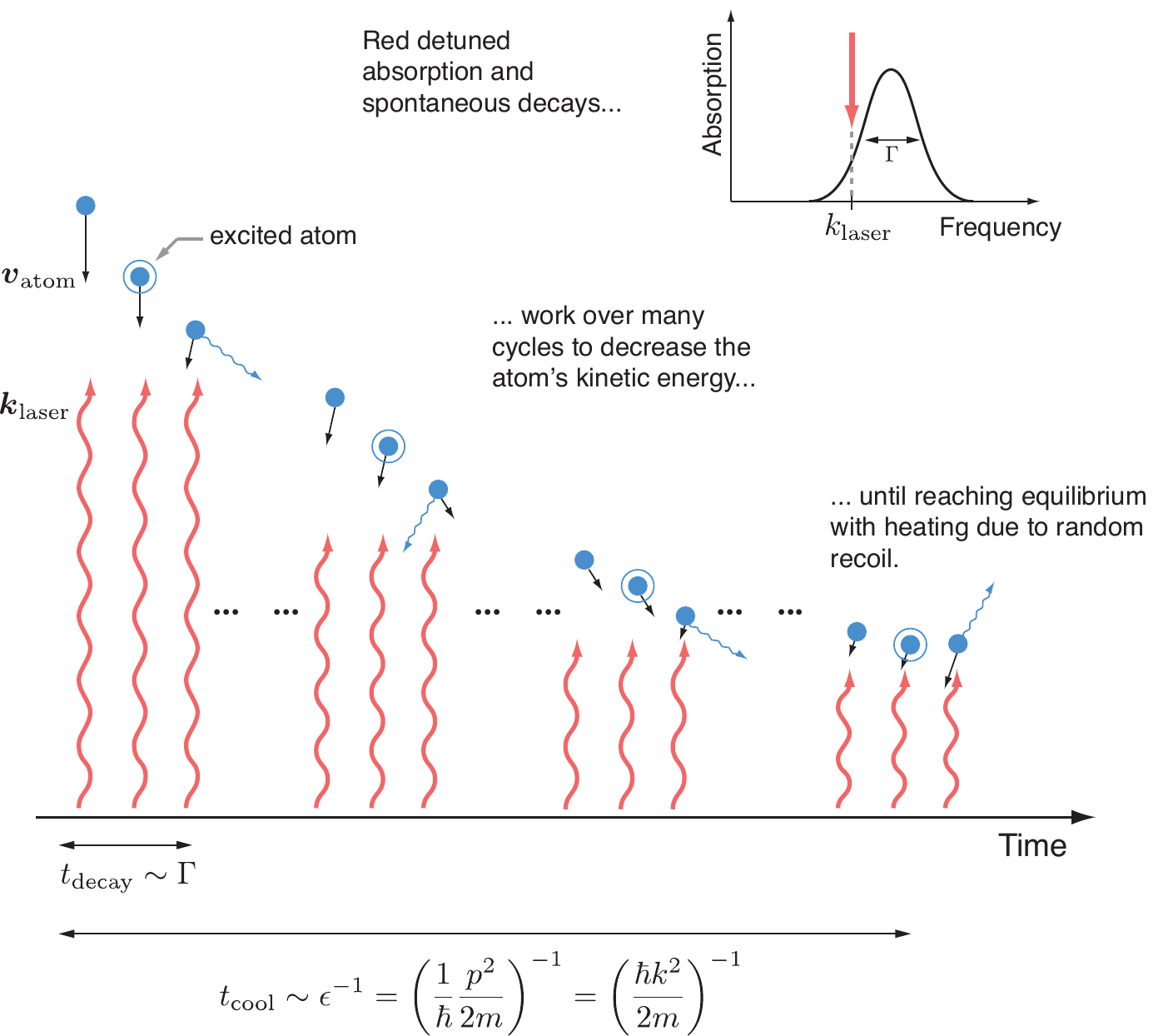}
\caption[Cartoon of the Doppler laser cooling process]{This cartoon of the Doppler laser cooling process shows a laser detuned to the red of an atomic resonance.  When the atom moves away from the laser, the frequency is red-shifted further out of resonance making the probability of absorption small.  However, when the atom moves towards the laser, the frequency is blue shifted closer to resonance making an interaction likely.  When a photon is absorbed, the atom loses momentum.  Since the spontaneous decay that follows is in a random direction, entropy is removed from the atom's motion and deposited into the light field.  The net result is a \emph{cooling} of the atomic motion.}
\label{fig:laserCooling}
\end{figure}
Consider an atom moving in a laser field tuned $\Delta_\text{laser}$ below resonance.  Because of the first-order Doppler shift, the effective detuning experienced by the atom depends on the atom's velocity $\boldsymbol{v}_\text{atom}$ with respect to the beam wavevector $\boldsymbol{k}_\text{laser}$:
\begin{equation}
\delta_\text{eff} = \Delta_\text{laser} - \boldsymbol{k}_\text{laser} \cdot \boldsymbol{v}_\text{atom}.
\end{equation}
The laser is tuned below the atomic resonance, so the atom is more likely to absorb from the beam when it is moving against the beam $\boldsymbol{k}_\text{laser}$.  Each absorption slows the atom by $\Delta p = \hbar k$; the subsequent spontaneous emission indeed results in an atomic recoil of the same impulse but in a random direction.  Over many absorptions one can define an average slowing force on the atom:
\begin{equation*}
\langle F_a \rangle = \underbrace{(\hbar k)}_{\Delta p \text{ per absorption}} \times \underbrace{(\Gamma)}_\text{Decay rate} \times \underbrace{\left( \frac{s/2}{1 + s + (2 \delta_\text{eff} / \Gamma)^2}\right)}_\text{Probability to absorb/decay},
\end{equation*}
where $s = 2 \Omega^2 / \Gamma^2$ is the saturation parameter, a measure of the strength of the resonant driving field relative to the natural decay rate, and the third term has its origin in Eq.~\ref{eq: twoStateProbabilitySaturation}.  Due to the velocity dependence of $\delta_\text{eff}$, an expansion of the denominator in $\langle F_a \rangle$ shows that it takes the form of a damping force when $\Delta_\text{laser} < 0$~\cite{leibfried2003qds}
\begin{align}
\langle F_a \rangle &\approx F_0(1 + \kappa v) \\
\intertext{with}
F_0 &= \hbar k \Gamma \frac{s/2}{1 + s + (2\Delta / \Gamma)^2}, \\
\intertext{and the damping coefficient}
\kappa &= \frac{8 k \Delta/ \Gamma^2}{1 + s + (2\Delta / \Gamma)^2}
\end{align}
in the limit of velocities small enough to make Doppler shifts smaller than $\Gamma$.

This cooling by radiation pressure, shown in Figure~\ref{fig:laserCooling}, is an approximation that holds until the heating that takes place due to the random recoil following spontaneous emission balances the cooling affected by the detuned laser beam.   At that point the minimum temperature theoretically achievable with ordinary Doppler cooling is
\begin{align}
T_\text{min}(s, \Gamma) &= \frac{\hbar \Gamma \sqrt{1 + s}}{4 k_B} (1 + \xi), \\
T_\text{min} &\approx \frac{\hbar \Gamma}{2 k_B},
\end{align}
where in the first expression, $\xi$ is a geometrical factor equal to 2/5 for dipole radiation, and the laser detuning is set to $\Delta_\text{laser} = \Gamma \sqrt{1+s}/2$.  In the second expression we assume the minimizing case of a laser detuned to $\Delta_\text{laser} = \Gamma/2$ and driving just at saturation ($s=1$).  For a sense of this limit, a barium ion cooled on the $6S_{1/2} \leftrightarrow 6P_{1/2}$ transition could reach a Doppler limited temperature of $T_\text{min} \approx 100 \mu$K.

To cool neutral atoms in all three dimensions, one employs three pairs of counter-propagating, red-detuned beams~\cite{phillips1998lct}.  As we will see in Chapter~\ref{sec:ionTrapping}, an ion trap stably confines our atom in all three directions, so Doppler cooling is achievable with only one laser beam as long as the beam wavevector does not accidentally lie along one of the trap's principle axes~\cite{wineland1979lca}.  Further, we show in Section~\ref{sec:coolingRegime} that the theory of Doppler laser cooling of free atoms largely applies to trapped ions when the time scale of photon emission (state lifetimes $\tau = \Gamma^{-1}$) is much shorter than the oscillation period of the trapped particle in its confining potential well.

\subsection{Multipole interactions} \label{sec:multipoleInteractions}
\begin{table}
\centering
\caption[Multipole transition selection rules]{Exact and approximate selection rules for electromagnetic transitions are derived from the Wigner-Eckhart theorem (see, for instance, \cite{sakurai1994mqm}).  The approximate rules are broken in cases where angular momenta $L$ and $S$ do not represent good quantum numbers.  For atoms with hyperfine structure, selection rules for $F$ and $m_F$ follow those of $J$ and $m_J$.  Derivations and discussion of the selection rules are found in many sources (e.g., \cite{corney1977aal}).  The relative strength estimates are discussed in the text.  Relationships between spontaneous decay rates and reduced matrix elements are available in~\cite{shore1990tca2}.}
\begin{tabular}{c|ccc}
	& Electric dipole (E1)	& Magnetic dipole (M1) & Electric quadrupole (E2) \\ \hline \hline
\multirow{4}{*}{\rotatebox{90}{Exact}} & $\Delta J = 0, \pm 1$ & $\Delta J = 0, \pm 1$ & $\Delta J = 0, \pm 1, \pm 2$ \\
& but $J = 0 \nleftrightarrow 0$&but $J = 0 \nleftrightarrow 0$ & but $J = 0 \nleftrightarrow 0, \tfrac{1}{2} \nleftrightarrow \tfrac{1}{2}, 0 \nleftrightarrow 1$ \\
 & $\Delta m_J = 0, \pm 1$ & $\Delta m_J = 0, \pm 1$ &$ \Delta m_J = 0, \pm 1, \pm 2$ \\
 & Parity changes & No parity change & No parity change \\ \hline
 \multirow{4}{*}{\rotatebox{90}{Approximate}} & $\Delta l = \pm 1$ & $\Delta l =0, \Delta n =0$ & $\Delta l = 0, \pm 2$ \\
 & $\Delta S = 0$ & $\Delta S = 0$ & $\Delta S = 0$ \\
 & $\Delta L = 0, \pm 1$ & $\Delta L = 0$ & $\Delta L = 0, \pm 1, \pm 2$ \\
 & but $L = 0 \nleftrightarrow 0$ & & but $L = 0 \nleftrightarrow 0, 0 \nleftrightarrow 1$ \\ \hline
 {\rotatebox{90}{\parbox[t]{0.75 in}{Relative strength}}} &  \mbox{1}	  & \mbox{$\sim  \alpha^2$}	& \mbox{$\displaystyle \sim \frac{a_0^2}{\lambda^2}$}
%{\rotatebox{90}{\parbox[t]{0.75 in}{Decay rate}}} &
%\mbox{$\displaystyle \Gamma_{E1} = \frac{4 c \alpha k^3}{3(2j' + 1)} |\langle \gamma, j || r || \rangle |^2$} &
%\mbox{$\displaystyle \Gamma_{M1} = \frac{4 c \alpha k^3}{3(2j' + 1)} |\langle \gamma, j || \boldsymbol{M} || \rangle |^2$} &
%\mbox{$\displaystyle \Gamma_{E2} = \frac{c \alpha k^5}{15(2j' + 1)} |\langle \gamma, j || r^2 \boldsymbol{C}^{(2)} || \rangle |^2$} 
\end{tabular}
\label{tab:selectionRules}
\end{table}

Recall from Eq.~\ref{eq:multipoleExansion} that the field of an electromagnetic wave, $\boldsymbol{E}(\boldsymbol{r},t) = \boldsymbol{E} e^{i (\boldsymbol{k} \cdot \boldsymbol{r} - \omega t)}$, can be expanded as:
\begin{equation}
\boldsymbol{E} e^{i (\boldsymbol{k} \cdot \boldsymbol{r} - \omega t)} = E \boldsymbol{\epsilon} (1 + i(\boldsymbol{k} \cdot \boldsymbol{r}) + \cdots) e^{-i \omega t},
\end{equation}
where $\boldsymbol{\epsilon}$ is a unit polarization vector.  Eq.~\ref{eq:e1Rabi} shows that the first term in the exponential expansion leads to a Rabi frequency that depends on dipole matrix elements:
\begin{align}
\Omega_\text{E1} &= \frac{e E_0}{\hbar} \left| \langle \gamma, j ,m  | \boldsymbol{\epsilon} \cdot \boldsymbol{r} | \gamma', j' ,m' \rangle \right|. \\
\intertext{We can similarly define a Rabi frequency for the next term which is proportional to a quadrupole matrix element:}
\Omega_\text{E2} &= \frac{e E_0}{2 \hbar} \left| \langle \gamma, j ,m | (\boldsymbol{\epsilon} \cdot \boldsymbol{r})(\boldsymbol{k} \cdot \boldsymbol{r}) | \gamma' j', m' \rangle \right|, \label{eq:e2Rabi}
\end{align}
where we have labeled states with their total angular momentum $j$, magnetic sublevel $m$ and other quantum numbers $\gamma$.  We can use the Wigner-Eckhart theorem to extract the angular dependence of these expressions.  Matrix elements of $T_q^{(k)}$, an irreducible spherical tensor of rank $k$ and index $q$, can be written in terms of a reduced matrix element and a Clebsch-Gordan coefficient:
\begin{equation}
\langle \gamma', j', m' | T_q^{(k)} | \gamma, j, m \rangle = \frac{1}{\sqrt{2j' + 1}} \underbrace{\langle j m ; k q | j' m' \rangle}_\text{Clebsch-Gordan}  \times  \underbrace{\langle \gamma', j' || T^{(k)} || \gamma, j \rangle}_\text{Reduced matrix element}.
\end{equation} 

For instance, to show the explicit angular dependence in the electric dipole Rabi frequency, we can write
\begin{equation}
\Omega_\text{E1} = \frac{e E_0}{\hbar} \left| \langle \gamma, j  || r ||  \gamma', j' \rangle \right| \sum_{q = -1}^{1} \langle j m; 1 q | j' m' \rangle E_q^{(1)},
\end{equation} \label{eq:e1RabiFrequency}
and in electric quadrupole transitions (from~\cite{roos2000thesis}),
\begin{align}
\Omega_\text{E2} &= \left| \frac{eE}{2 \hbar} \langle \gamma, j || r^2 \boldsymbol{C}^{(2)} || \gamma', j' \rangle \sum_{q = -2}^{q'} \begin{pmatrix} j & 2 & j' \\ -m & q & m' \end{pmatrix} \boldsymbol{\epsilon}_q^{(2)} \right| \\ 
&= \left| \frac{eE}{2 \hbar} \langle \gamma, j || r^2 \boldsymbol{C}^{(2)} || \gamma', j' \rangle \sum_{q = -2}^{q'} \langle j m; 2 q | j' m' \rangle \boldsymbol{\epsilon}_q^{(2)} \right|,
\label{eq:e2RabiFrequency}
\end{align}
where the quantity in parenthesis is a $3j$-symbol equivalent to Clebsch-Gordan coefficients.  The elements of the second rank polarization spherical tensor $\boldsymbol{\epsilon}_q^{(2)}$ can be written as
\begin{align}
\boldsymbol{\epsilon}_0^{(2)} &= \tfrac{1}{2} | \cos \gamma \sin 2 \phi |, \\
\boldsymbol{\epsilon}_{\pm 1}^{(2)} &= \tfrac{1}{\sqrt{6}} | \cos \gamma \cos 2 \phi + i \sin \gamma \cos \phi |, \\
\boldsymbol{\epsilon}_{\pm 2}^{(2)} &= \tfrac{1}{\sqrt{6}} | \tfrac{1}{2} \cos \gamma \sin 2 \phi + i \sin \gamma \sin \phi |,  
\end{align}
where we have assumed a quantization axis defined by a magnetic field $\boldsymbol{B}$, light wave-vector $\boldsymbol{k}$, and linear light polarization $\boldsymbol{\epsilon}$, parameterized in Cartesian coordinates
\begin{align}
\boldsymbol{B} &= B(0,0,1), \\
\boldsymbol{k} &= k(\sin \phi, 0, \cos \phi), \\
\boldsymbol{\epsilon} &= (\cos \gamma \cos \phi, \sin \gamma, - \cos \gamma \sin \phi).
\end{align}
\begin{figure}
\centering
\includegraphics{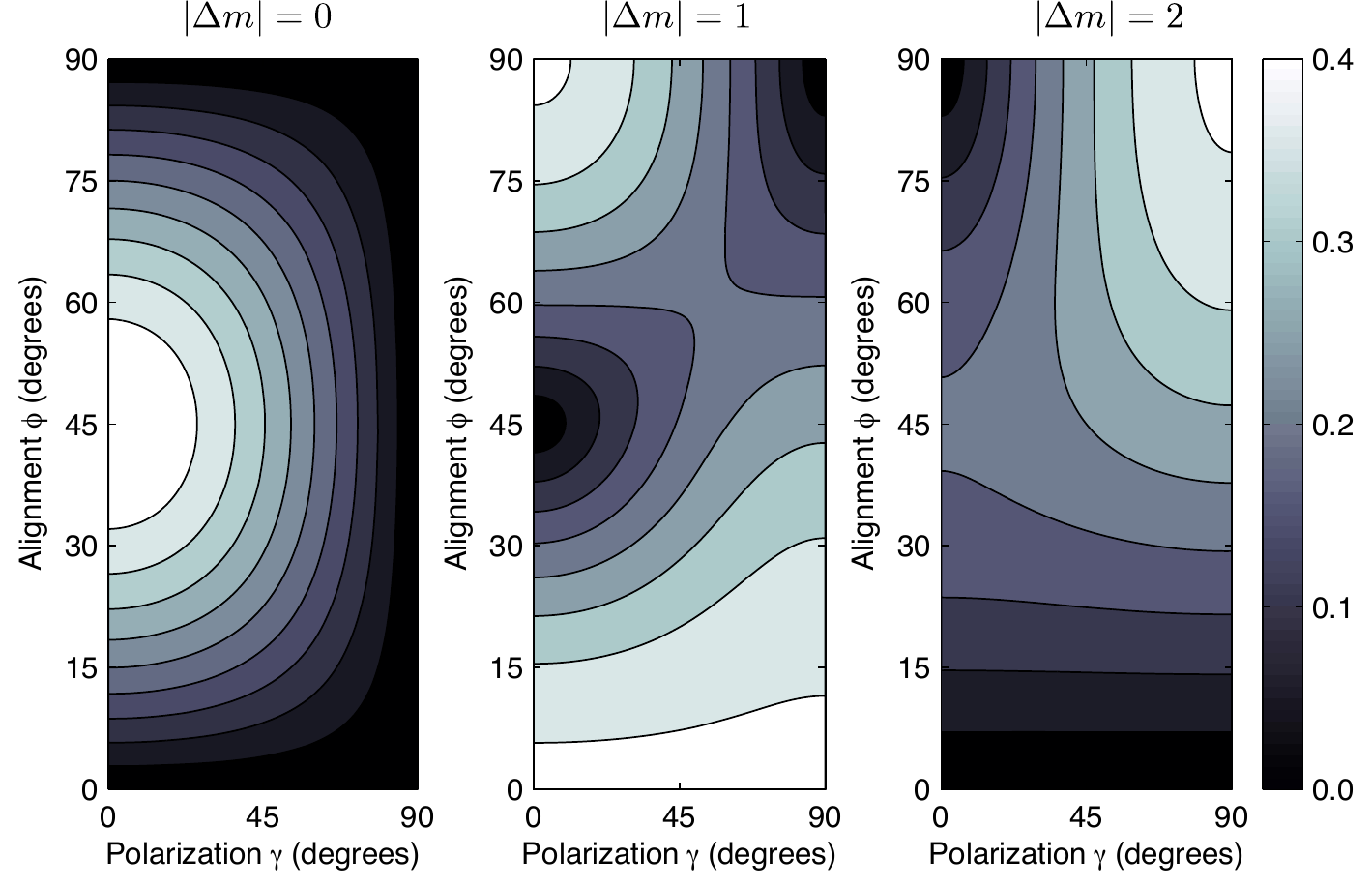}
\caption[Relative strengths of electric quadrupole transitions following~\cite{roos2000thesis}]{Relative strengths of electric quadrupole transitions given laser alignment  polarization angles defined in the text, independently reproduced from~\cite{roos2000thesis}.  From left to right, these are contour plots of the functions $\boldsymbol{\epsilon}_0^{(2)}$, $\boldsymbol{\epsilon}_1^{(2)}$, $\boldsymbol{\epsilon}_2^{(2)}$ assuming linear polarization whose projection is at an angle $\gamma$ with respect to the quantization axis $\boldsymbol{B}$ and  $\boldsymbol{k}_\text{laser}$, set at an angle $\phi$ with respect to $\boldsymbol{B}$.}
\label{fig:e2AngularContours}
\end{figure}
The $3j$-symbol in Eq.~\ref{eq:e2RabiFrequency} gives selection rules summarized in Table~\ref{tab:selectionRules}.  Additional selection rules on $\Delta m$ are present for particular laser alignment and polarization due to the structure of $\boldsymbol{\epsilon}^{(2)}$ and are illustrated in Figure~\ref{fig:e2AngularContours}, and tabulated for Cartesian components in Table~\ref{tab:parityPhaseFactors}. The rank-2 reduced matrix element is related to the electric quadrupole spontaneous decay rate~\cite{james1998qdc} with
\begin{equation}
\Gamma_\text{E2} = \frac{c \alpha k^5}{15(2j' +1)} \left|  \langle \gamma, j || r^2 \boldsymbol{C}^{(2)} || \gamma', j' \rangle \right|^2,
\end{equation}
which allows us to obtain the reduced matrix element without calculation if a decay rate is known.  The relative strength of electric quadrupole interactions to the typical electric dipole strength is
\begin{equation*}
\frac{ | \langle \gamma | (\boldsymbol{\epsilon} \cdot \boldsymbol{r})(\boldsymbol{k} \cdot \boldsymbol{r}) | \gamma' \rangle |^2}{ | \langle \gamma | \boldsymbol{r} \cdot \boldsymbol{\epsilon} | \gamma' \rangle |^2} \sim \frac{(e a_0^2 / \lambda)^2}{(e a_0)^2} \sim \frac{a_0^2}{\lambda^2}.
\end{equation*}

As a consequence of Maxwell's equation
\begin{equation}
\boldsymbol{\nabla} \times \boldsymbol{E} = -\frac{\partial \boldsymbol{B}}{\partial t},
\end{equation}
laser beams feature an oscillating magnetic field $B = E/c$ that couples to atomic magnetic moments by the Hamiltonian $-\boldsymbol{\mu} \cdot \boldsymbol{B}$.  A multipole expansion of $\boldsymbol{B}(\boldsymbol{r},t)$ similar to our treatment of the electric field is possible.  However, the magnetic dipole coupling due to a laser field is already in general much weaker than the electric dipole coupling~\cite{foot2005ap}:
\begin{equation}
\frac{ | \langle \gamma | \boldsymbol{\mu} \cdot \boldsymbol{B} | \gamma' \rangle |^2}{ | \langle \gamma | \boldsymbol{r} \cdot \boldsymbol{E} | \gamma'' \rangle |^2} \sim \left( \frac{\mu_B / c}{e a_0} \right)^2 \sim \alpha^2.
\end{equation}
The factor of $c$ is due to the ratio of magnetic to electric field strengths in an electro-magnetic wave.  Notice that the fine-structure constant $\alpha$ appears explicitly since it is a measure of the relative strength of magnetic to electric interactions in atoms~\cite{peskin1995iqf}.  A magnetic dipole coupling Rabi frequency can be written in analog to Eq.~\ref{eq:e1RabiFrequency},
\begin{equation}
\Omega_\text{M1} = \frac{B}{\hbar} \left| \langle \gamma' j' || \mu^{(1)} || \gamma j \rangle \sum_{q=-1}^1 \langle j' m' ; 1 q | j m \rangle \right| b_q^{(1)},Œ
\end{equation}
by defining a magnetic polarization vector $b_q^{(1)}$ in analog to the electric case.

\section{Relevant multi-state systems}
\subsection{The density matrix} \label{sec:densityMatrixFormalism}
Generalizing time-dependent perturbation theory to larger numbers of quantum states and to include `open sources' of decoherence~\cite{murao1998dnm} is made straightforward by the density matrix formalism~\cite{fano1957dsq}.  For an $n$-eigenstate system, the wavefunction is expressed as a linear combination of eigenfunctions $\phi_i$:
\begin{equation}
\psi = \sum_{i=1}^n c_i \phi_i.
\end{equation}
The Hermitian density matrix $\rho$ is then defined by
\begin{equation}
[\rho]_{ij} = c_i c_j^* \qquad \text{or } 
\rho = \begin{pmatrix} c_1 c_1^*	& c_1 c_2^* 	&\cdots 	& c_1 c_n^* \\
				   c_2 c_1^* 	& c_2 c_2^* 	& \cdots 	& c_2 c_n^* \\
				  \vdots	 	& \vdots		& \ddots 	&\vdots  \\
				  c_n c_1^* 	& c_n c_2^* 	& \cdots 	& c_n c_n^*
\end{pmatrix}.
\end{equation}
The dynamics of the density matrix $\rho$ are simply described by a rewriting of the Schr\"{o}dinger equation
\begin{equation}
i \hbar \frac{d \rho}{dt} = [\hat{H},\rho],
\end{equation}
where $\hat{H}$ is the Hamiltonian $H$ expressed as a matrix
\begin{equation}
[\hat{H}]_{ij} = \langle \phi_i | H | \phi_j \rangle  \qquad \text{or } 
\hat{H} = \begin{pmatrix}
\langle \phi_1 | H | \phi_1 \rangle	& \langle \phi_1 | H | \phi_2 \rangle &  \cdots &  \langle \phi_1 | H | \phi_n \rangle \\
\langle \phi_2 | H | \phi_1 \rangle	& \langle \phi_2 | H | \phi_2 \rangle &  \cdots &  \langle \phi_2 | H | \phi_n \rangle \\
\vdots						& \vdots				                 &  \ddots & \vdots \\
\langle \phi_n | H | \phi_1 \rangle	& \langle \phi_n | H | \phi_2 \rangle &  \cdots &  \langle \phi_n | H | \phi_n \rangle
\end{pmatrix}.
\end{equation}

\subsection{Three level spectroscopy}\label{sec:threeLevelSpectroscopy}
\begin{figure}
\centering
\includegraphics{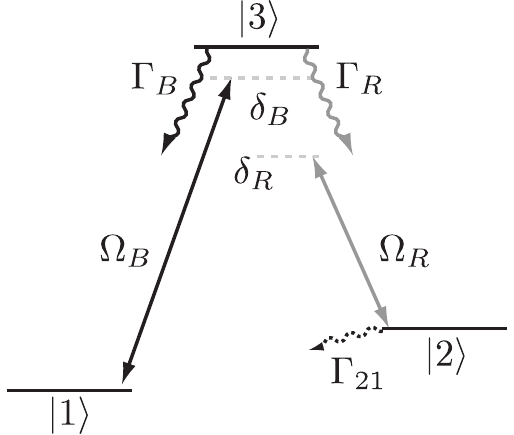}
\caption[The $\Lambda$-type, three level spectroscopy problem]{The so-called $\Lambda$-type three level spectroscopy problem matches the barium ion $6S_{1/2} \leftrightarrow 6P_{1/2} \leftrightarrow 5D_{3/2}$ system.  Here, the decay $\Gamma_{21}$ (which in the barium ion system is an electric-dipole forbidden decay) is presumed to be weak and is ignored in the density matrix treatment in this section.}
\label{fig:threeLevelDiagram}
\includegraphics[width = 5 in]{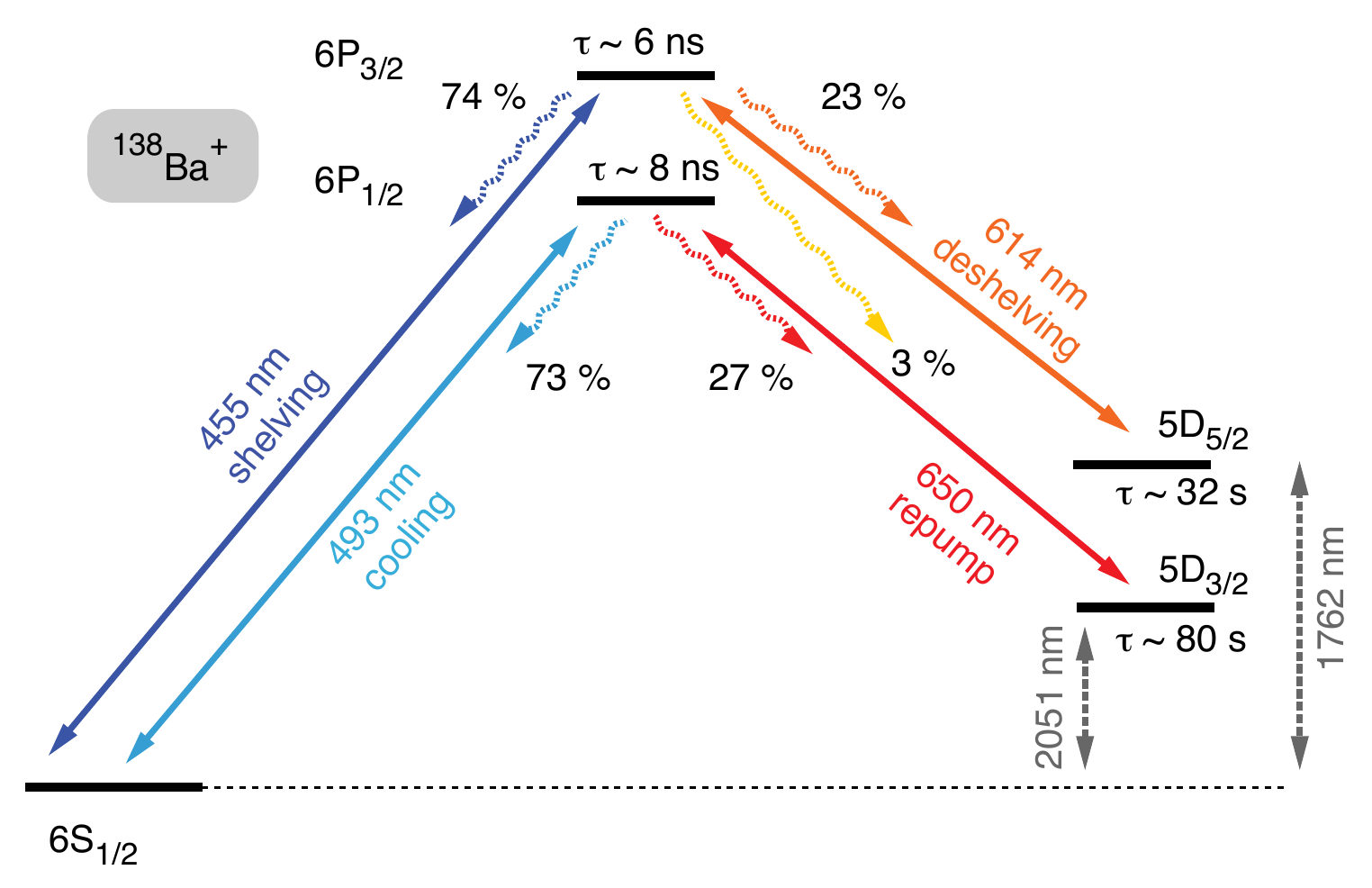}
\caption[Energy level diagram with decay rates of $^{138}$Ba$^+$]{Energy level diagram with decay rates of $^{138}$Ba$^+$.  Shown are the lowest lying energy levels of the outer valence electron in singly ionized barium, electric dipole transitions (solid lines), electric quadrupole transitions (dashed lines with arrows), and state lifetimes and decay branching rations (dashed wavy lines).}
\label{fig:bariumEnergyLevelsAtomic}
\end{figure}
The introduction of a third atomic level into the picture developed so far makes the system much richer.  We consider here the so-called $\Lambda$-arrangement of levels shown in Figure~\ref{fig:threeLevelDiagram} since states $|1\rangle$, $|2 \rangle$, and $|3 \rangle$ map exactly to the Ba$^+$ ground $6S_{1/2}$, metastable $5D_{3/2}$, and excited $6P_{1/2}$ states (see Figure~\ref{fig:bariumEnergyLevelsAtomic}).  Efficient laser cooling on the $6S_{1/2} \leftrightarrow 6P_{1/2}$ transition requires `repumping' on the $5D_{3/2} \leftrightarrow 6P_{1/2}$ transition.  We will find that the spectroscopic line shape of each of these transitions depends on \emph{both} couplings and detunings.  Yet more complication is introduced by considering a magnetic field that splits the Zeeman sublevels, but this case is completely treated in \cite{berkeland2002dds}.

In Figure~\ref{fig:threeLevelDiagram}, levels $|1\rangle$ and $|3\rangle$ are coupled with a Rabi frequency $\Omega_{13} = \Omega_B$, detuned from resonance by $\delta_B$.  Spontaneous decay $|3 \rangle \to |1 \rangle$ occurs with rate $\Gamma_B$.  Likewise, levels $|2 \rangle$ and $|3 \rangle$ are coupled with a Rabi frequency $\Omega_{23} = \Omega_R$, detuned from resonance by $\delta_R$.  Spontaneous decay $|3 \rangle \to |2 \rangle$ happens with rate $\Gamma_R$.  We could also account for the decay $|2 \rangle \to |1 \rangle$, which in the barium ion is weak, because it is electric-dipole forbidden.  We ignore it in the following treatment.

The density matrix equations, plus decay terms following Eq.~\ref{eq:decayTerm}, make up what are called the \emph{optical Bloch equations}~\cite{allen1987ora} and are given by~\cite{janik1985dfo} for this system:
\begin{align}
\frac{d \rho_{11}}{dt}	&= i \frac{\Omega_B}{2}(\rho_{13} - \rho_{31}) + \Gamma_B \rho_{33}, \\
\frac{d \rho_{22}}{dt} &= i \frac{\Omega_R}{2}(\rho_{23} - \rho_{32}) + \Gamma_R \rho_{33}, \\
\frac{d \rho_{12}}{dt} &= i \left[ (\Delta_R - \Delta_B) \rho_{12} + \frac{\Omega_R}{2} \rho_{13} - \frac{\Omega_B}{2} \rho_{32} \right], \\
\frac{d \rho_{13}}{dt} &= i \left[ \frac{\Omega_B}{2}(\rho_{11} - \rho_{33}) + \frac{\Omega_R}{2} \rho_{12} - \Delta_B \rho_{13} \right] - \frac{\Gamma}{2} \rho_{12}, \\
\frac{d \rho_{23}}{dt} &= i \left[ \frac{\Omega_R}{2} (\rho_{22} - \rho_{33} ) + \frac{\Omega_B}{2} \rho_{21} - \Delta_R \rho_{23} \right] - \frac{\Gamma}{2} \rho_{23}.
\end{align}
Here, $\Gamma_B$ and $\Gamma_R$ are the partial decay rates $6P_{1/2} \to 6S_{1/2}$ and $6P_{1/2} \to 5D_{3/2}$, respectively, with $\Gamma =\Gamma_B + \Gamma_R$ and $\Gamma_B / \Gamma_R = 2.7$ to match the branching ratio in Ba$^+$ (note that~\cite{janik1985dfo} sets this ratio to 2.88). By ensuring state normalization with $\rho_{11} + \rho_{22} + \rho_{33} = 1$, and demanding that $\rho = \rho^\dag$, the steady state solution for the excited state population is
\begin{equation} \label{eq:threeLevelSpecResult}
\rho_{33} = \frac{4 (\Delta_B - \Delta_R)^2 \Omega_B^2 \Omega_R^2 \Gamma}{Z},
\end{equation} 
where
\begin{align}
\begin{split}
Z &\equiv 	 8(\Delta_B - \Delta_R)^2 \Omega_B^2 \Omega_R^2 \Gamma 
	+ 4(\Delta_B - \Delta_R)^2 \Gamma^2 Y  \\
	&+ 16(\Delta_B - \Delta_R)^2[\Delta_B^2\Omega_R^2 \Gamma_B + \Delta_R^2 \Omega_B^2 \Gamma_R] \\
	&- 8\Delta_B( \Delta_B - \Delta_R) \Omega_R^4 \Gamma_B 
	+ 8\Delta_R(\Delta_B -\Delta_R) \Omega_B^4 \Gamma_R \\
	&+ (\Omega_B^2 + \Omega_R^2)^2 Y, \end{split} \\
\intertext{and,}
Y & \equiv \Omega_B^2 \Gamma_R + \Omega_R^2 \Gamma_B.
\end{align}
\begin{figure}
\centering
\includegraphics[width=2.9 in]{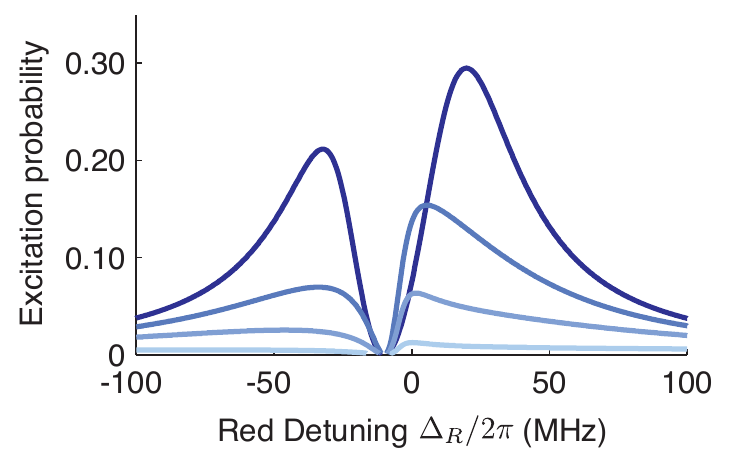} \includegraphics[width=2.9 in]{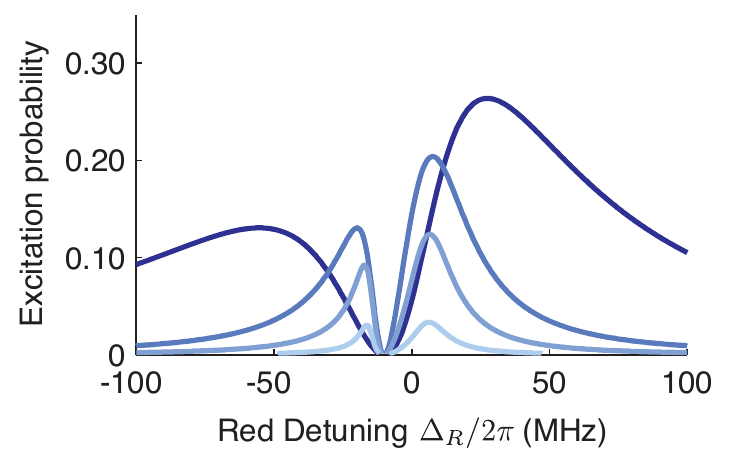} \\
\includegraphics[width=2.9 in]{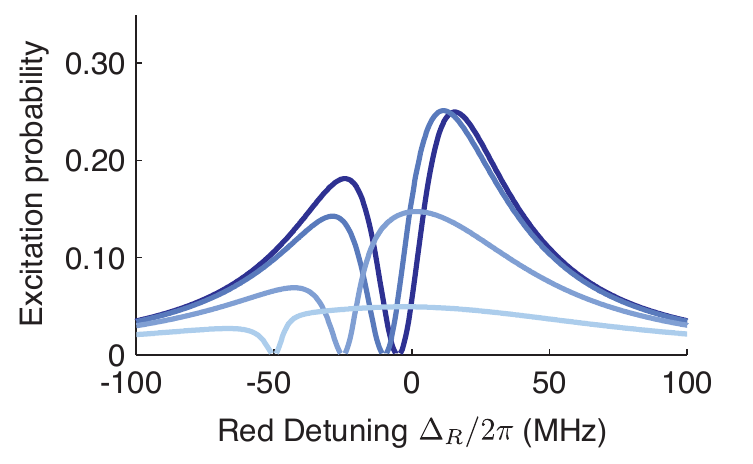} \includegraphics[width=2.9 in]{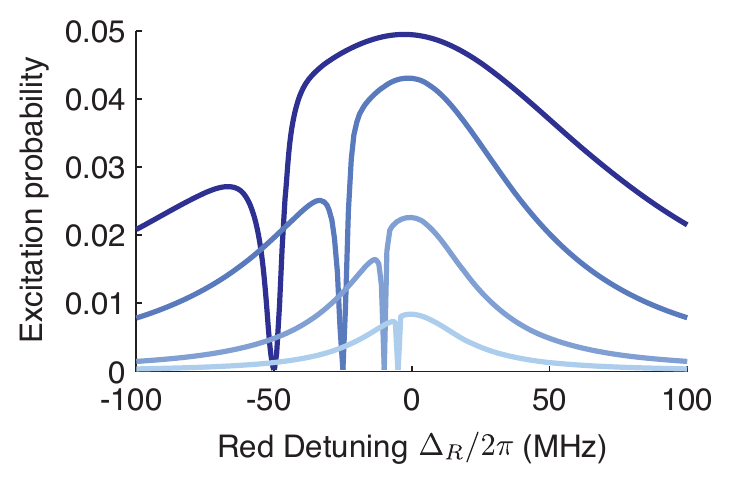} \\
\caption[Three level spectroscopy numerical simulations]{Several numerical simulations of Eq.~\ref{eq:threeLevelSpecResult} following~\cite{janik1985dfo} gives us important intuition about the three-level $\Lambda$-type spectroscopy described in this section and relevant to the two photon Ba$^+$ $6S_{1/2} \leftrightarrow 6P_{1/2} \leftrightarrow 5D_{3/2}$ coupled system. The following inset table shows the parameters for each graph: \\ \\
{\small \centering
\begin{tabular}{r|m{0.9 in}cl}
Plot...   	& ...shows & ...holds fixed & ...varies (darkest to lightest curve) \\ \hline \hline
Top-left 	& \raggedright{Blue power broadening}	 & $\Delta_B = \Gamma/2$, $\Omega_R = \Gamma$ & $\Omega_B/\Gamma = 2, 1/2, 1/4, 1/10$ \\ 
		&		&			& \\
Top-right	& \raggedright{Red power broadening}	& $\Delta_B = \Gamma/2$, $\Omega_B = \Gamma$ & $\Omega_R/\Gamma = 2, 1/2, 1/4, 1/10$ \\ 
		&		&			& \\
Bottom-left & \raggedright{Blue detuning}	& $\Omega_B = \Omega_R = \Gamma$ & $\Delta_B / 2\pi = -5, -10, -25, -50$~MHz \\ 
		&		&			& \\
\multirow{2}{*}{Bottom-right} & \multirow{2}{*}{\parbox{1.25 in}{\raggedright{The narrow dip width}}} & \multirow{2}{*}{---} & $\Omega_B/ \Gamma = \Omega_R/\Gamma = 1, 1/2, 1/5, 1/10$ \\
		&				&			&   $\Delta_B/ 2\pi = -50, -25, -10, -5$~MHz 
\end{tabular}} \\ \\
The important features of the resonance dip are that it occurs when $\Delta_B = \Delta_R$ and that the width is set by $\Omega_B$ and $\Omega_R$, not $\Gamma$.}
\label{fig:threeLevelModelFigs}
\end{figure}

\begin{figure}
\centering
\includegraphics{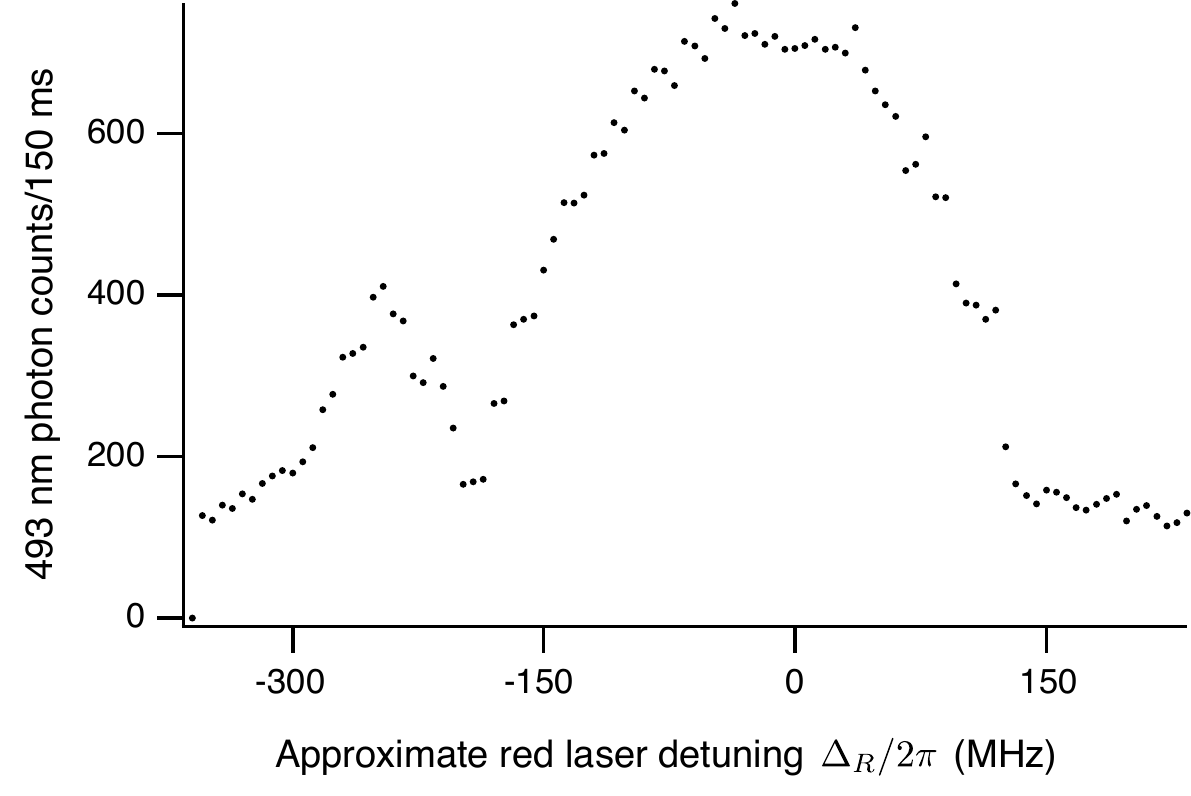}
\caption[Three level spectroscopy:  typical observation in Ba$^+$.]{Here, we scan a 650~nm red external-cavity diode laser across the Ba$^+$ $5D_{3/2} \leftrightarrow 6P_{1/2}$ while a blue laser is detuned $\approx 160$~MHz from the $6S_{1/2} \leftrightarrow 6P_{1/2}$ resonance.  Both laser amplitudes are far above saturation.  One should compare the qualitative features of the blue fluorescence spectrum to the theoretical predictions shown in Figure~\ref{fig:threeLevelModelFigs}.} 
\label{fig:threeLevelRedLaserScan}
\end{figure}
Some theoretical predictions of this model for the excited state population are shown in Figure~\ref{fig:threeLevelModelFigs}.  Notice that the $6P_{1/2}$ population approaches 1/3 when both lasers are driven above saturation, which is consistent with a simpler rate equation model. An independent and more complete treatment~\cite{berkeland2002dds} that includes the Zeeman effect shows that a more realistic maximum excited state population is roughly an order of magnitude lower.  Since the scattering rate of detectable photons is $\Gamma_B \rho_{33}$, this means our barium ion can emit $\sim 10^7$ photons/s.  An interesting qualitative feature in the theoretical curves is a fluorescence dip when the two laser detunings are equal $\Delta_B = \Delta_R$.  This \emph{dark state} is an example of \emph{coherent population trapping}~\cite{gray1978cta} formed by the two photon Raman interaction that allows population oscillation between $6S_{1/2}$ and $5D_{3/2}$ without significant probability of decay from $6P_{1/2}$ for strong interactions ($\Omega_B, \Omega_R \gg \Gamma$).

The qualitative features of the theoretical curves should be compared to Figure~\ref{fig:threeLevelRedLaserScan}, which shows experimental three level spectroscopy data of Ba$^+$ taken in typical operating conditions.

\subsection{Electron shelving}\label{sec:electronShelving}
\begin{figure}
\centering
\includegraphics{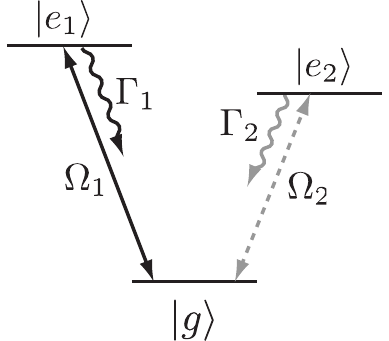}
\caption[The $V$-type, three level narrow spectroscopy problem]{Here we present an instance of a so-called $V$-type three level narrow spectroscopy problem where one of the decays is weak $\Gamma_2 \ll \Gamma_1$, perhaps because it is forbidden by electric-dipole interactions.  We intend to perform spectroscopy on the narrow resonance $|g \rangle \leftrightarrow |e_2 \rangle$ not by collecting photons scattered out of $|e_2\rangle$ but instead by driving both $\Omega_1$ and $\Omega_2$ and looking for modifications to the scattered light from $|e_1 \rangle$ when $\Omega_2$ goes into resonance.  This signal has earned names like \emph{quantum jumps}~\cite{cook1985pdo}, \emph{electron shelving} or detection via \emph{telegraph signals}~\cite{nagourney1986soe}.}
\label{fig:threeLevelShelvingDiagram}
\end{figure}
Examine the three level system diagramed in Figure~\ref{fig:threeLevelShelvingDiagram}.  One of the excited states $|e_2 \rangle$ is metastable, $\Gamma_2 \ll \Gamma_1$, perhaps because decays are forbidden by electric-dipole selection rules.  Nonetheless, suppose we wish to perform precision spectroscopy on the $|g\rangle \leftrightarrow |e_2 \rangle$ transition.  Because $|e_2 \rangle$ is so long-lived, it is difficult to imagine collecting many scattered photons:  recall from Section~\ref{sec:decayPowerBroadening} that increasing the coupling strength $\Omega_2$ power-broadens the state, but due to saturation, does not significantly increase the rate of scattering.
\begin{figure}
\centering
\includegraphics[width=5 in]{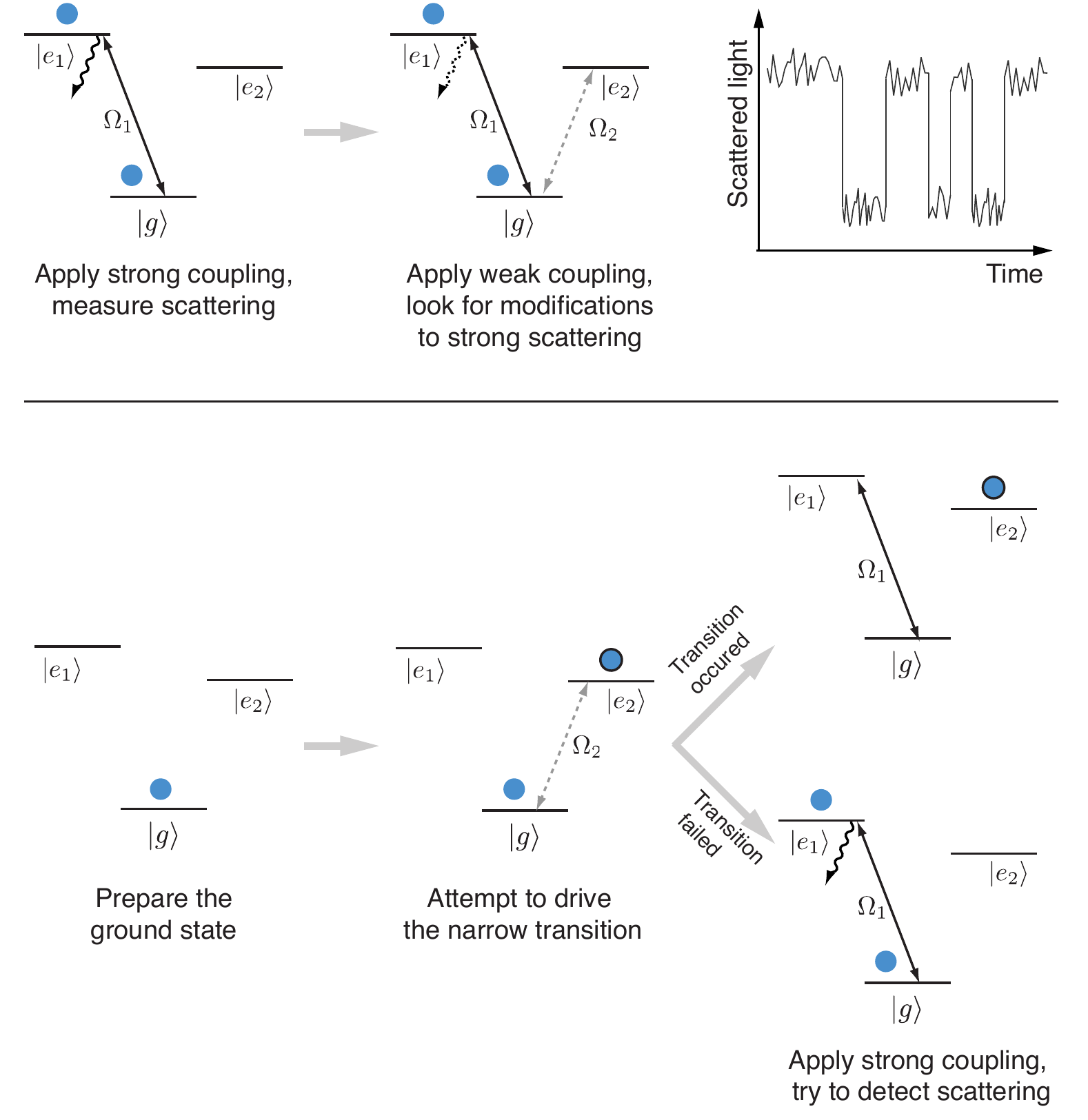}
\caption[Two methods of observing electron shelving]{Two methods of observing electron shelving. In the upper graph, we see that by applying a strong interaction $\Omega_1$ and a weak interaction $\Omega_2$ simultaneously we can detect transitions of the ion into $|e_2 \rangle$ by a disappearance of scattering radiation from $|e_1 \rangle$.  In the lower graph, we depict a `digital' test of whether $\Omega_2$ resulted in an ion transition by whether scattering is observed from $\Omega_1$ after the weak $\pi$-pulse excitation attempt.  Compare with actual data in Figure~\ref{fig:shelvingVariations}.}
\label{fig:electronShelvingExperiments}
\end{figure}
One way forward is to realize that photons are easy to collect on the strong $|g \rangle \leftrightarrow |e_1\rangle$ transition and that such a transition \emph{will not occur} when the atom is in the metastable or \emph{shelved} state $|e_2 \rangle$.  Figure~\ref{fig:electronShelvingExperiments} demonstrates that this disappearance experiment can be qualitatively run at least two ways:  random~\cite{berkeland2004tnq} interruptions in $\Omega_1$ scattering due to the application of a resonant $\Omega_2$~\cite{buhner2000rfs}, or a pump-with-$\Omega_2$ and probe-with-$\Omega_1$ approach.  These techniques have earned names in the literature like \emph{quantum jumps}~\cite{cook1985pdo}, \emph{electron shelving} or detection via \emph{telegraph signals}~\cite{nagourney1986soe}.

\begin{figure}
\centering
\includegraphics[width = 6 in]{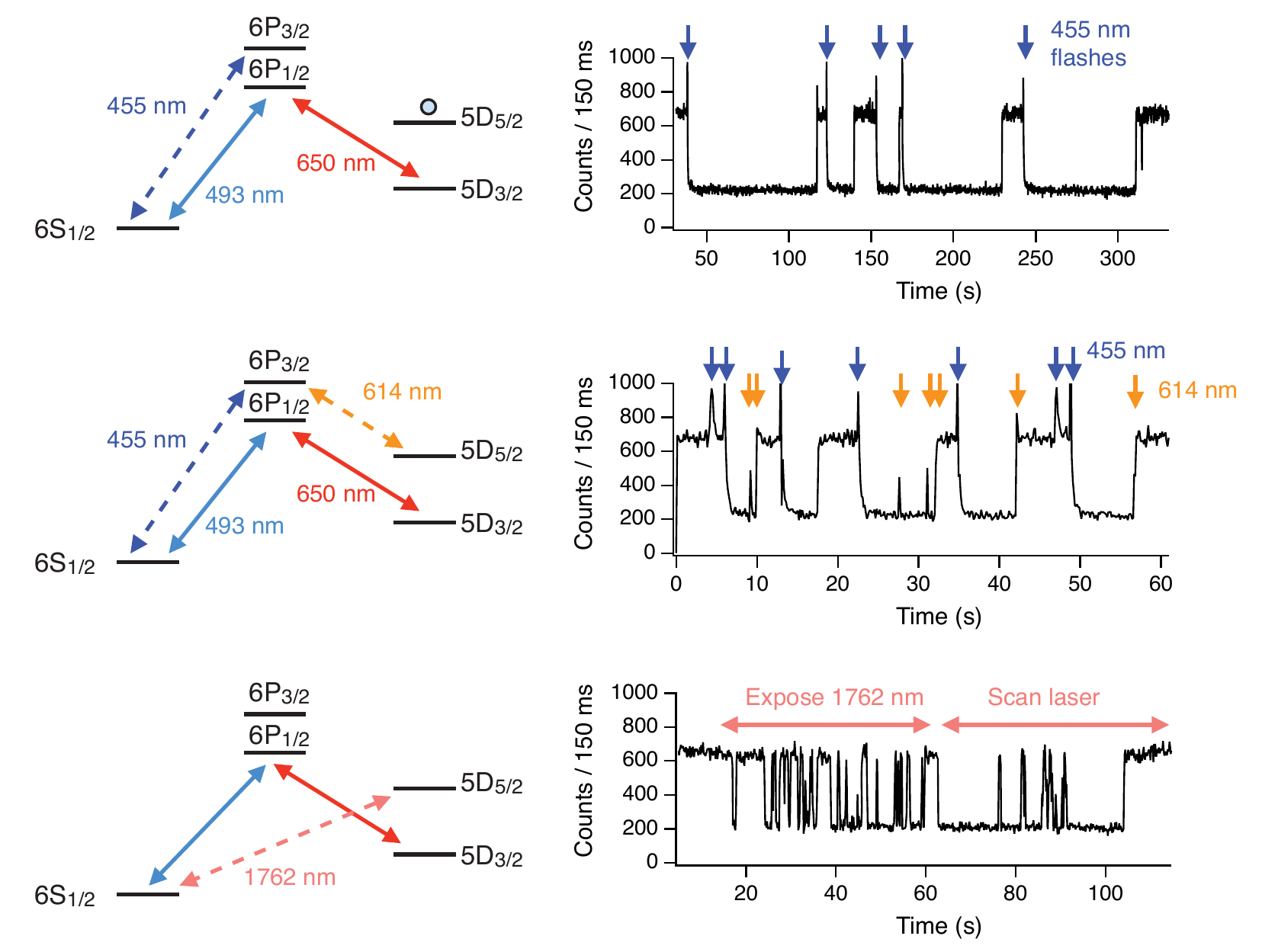}
\caption[Variations of observed electron shelving processes in a single barium ion]{Three variations of observed electron shelving in a single barium ion.  In the top graph, note the disappearance of 493~nm scattered fluorescence after a flash of 455~nm light (from an LED) excites the ion to $6P_{3/2}$ followed by a decay to meta-stable $5D_{5/2}$.  The fluorescence reappears after the decay lifetime $\tau \sim 30$s. In the middle plot we show how pulses of 455~nm and 614~nm move the ion into and out of the $5D_{5/2}$ at will.  In the bottom plot, we show recent observations of direct electron shelving caused by a 1762~nm fiber laser resonant with the $6S_{1/2} \leftrightarrow 5D_{5/2}$ transition.  These time series are recorded in data files {\tt 061201-1} through {\tt -5}.}
\label{fig:shelvingVariations}
\end{figure}
The energy level structure in Ba$^+$ shown in Figure~\ref{fig:bariumEnergyLevelsAtomic} is only slightly more complicated in terms of electron shelving.  Though the $5D_{3/2}$ state is metastable, it is strongly connected to the $6P_{1/2}$ with 650~nm radiation.  Therefore, when both 493~nm and 650~nm lasers are applied, strong scattering on the $6S_{1/2} \leftrightarrow 6P_{1/2} \leftrightarrow 5D_{3/2}$ closed cycle is observed.  However, application of 1762~nm light can move an ion from $6S_{1/2}$ to the shelved $5D_{5/2}$ state via an electric quadrupole transition~\cite{yu1994sss}.  Once the atom is in this metastable state, the application of the 493~nm and 650~nm lasers no longer produces fluorescence.

For the duration of this work, we achieved shelving in Ba$^+$ using an indirect method.  455~nm light excites an ion in the ground state to $6P_{3/2}$ and a subsequent decay to $5D_{5/2}$ is eventually likely, creating a shelved state.  An ion can then be deshelved with the application of 615~nm light, which excites the $5D_{3/2} \leftrightarrow 6P_{3/2}$ transition followed by a decay to the ground state.  Due to the very long $5D_{3/2}$ state lifetime ($\tau \sim 30$s), these `shelving' and `deshelving' transitions are easily observed.  Figure~\ref{fig:shelvingVariations} shows the action of both 455~nm and 614~nm light, along with very recent observation of direct shelving using light at 1762~nm.

\begin{figure}
\centering
\includegraphics[width = 5in]{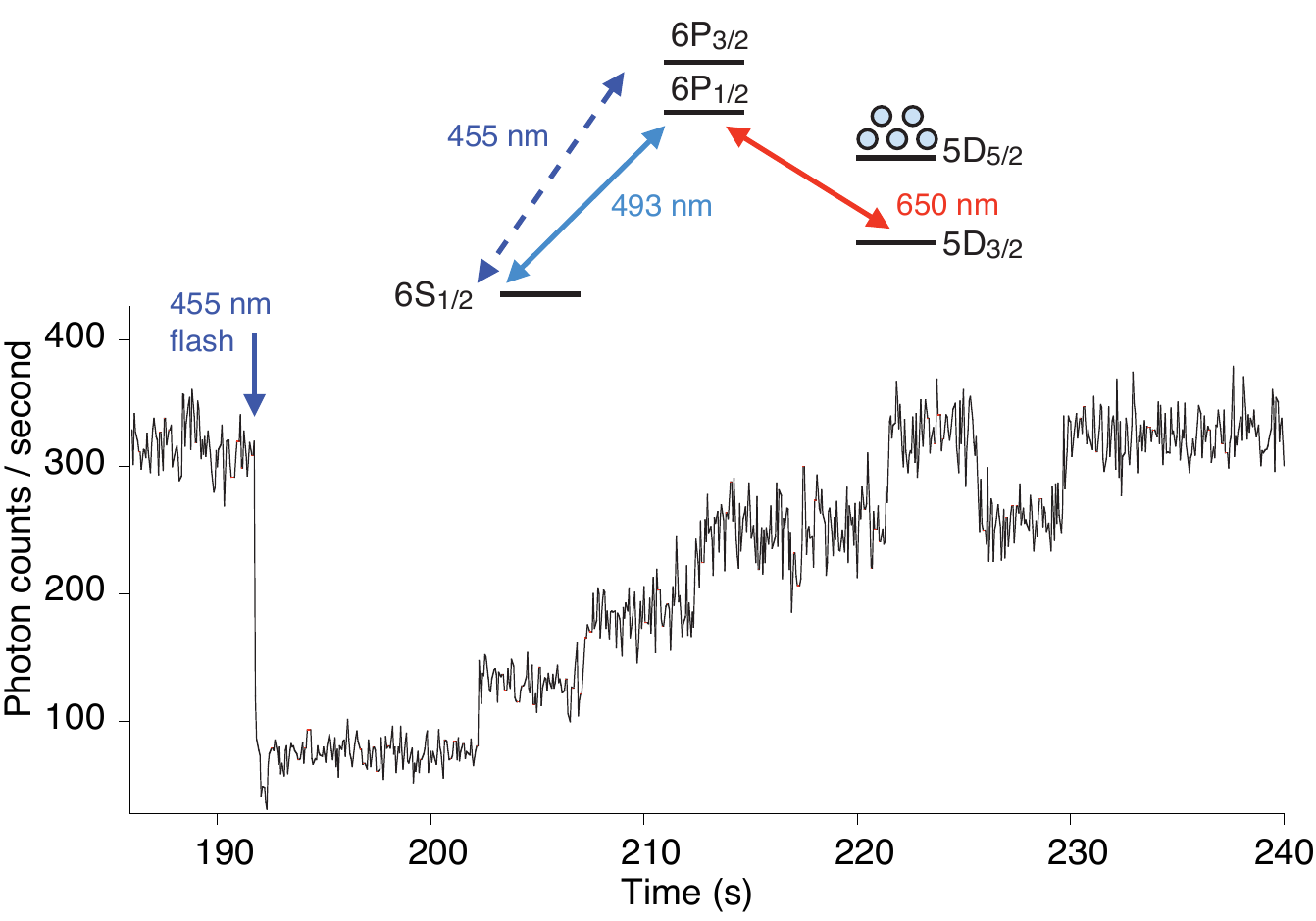}
\caption[Electron shelving allows us to count the number of trapped ions]{Electron shelving allows us to count the number of ions in our trap.  Here, a single pulse of 455~nm light shelves many trapped ions in the $5D_{5/2}$ state.  One by one, they decay to the ground state via the electric quadrupole interaction and produce an additional step in a staircase profile in the scattered fluorescence record.  In this example there are likely five or six ions present in the trap.}
\label{fig:shelvingFiveIons}
\end{figure}
In addition to finding utility in the precision experiments detailed in later chapters, electron shelving is also the most straightforward way to be sure one has trapped just one ion.  Figure~\ref{fig:shelvingFiveIons} depicts the observed fluorescence following a pulse of 455~nm shelving light when five or six ions were present in the trap.  The pulse succeeds in moving most or all of the ions to $5D_{5/2}$, and each decays to the ground state independently.  We observe each ion decay as another step in a staircase shape of the fluorescence record.  In practice, we would rid ourselves of the extra ions by either shutting of the trap and attempting to re-trap, or by lowering the trap depth (see Chapter~\ref{sec:ionTrapping}) with the cooling laser left off in order to let all but the coldest ion escape.

\subsection{Zeeman structure and spin resonance: systems with $J \geq 1/2$}\label{sec:spinDynamics}
\begin{table}
\centering
\caption[Electronic $g$-factors for several Ba$^+$ states.]{The predicted values $g_J(L,S,J)$ do not include high order QED corrections (see, for instance~\cite{lindroth1993aic}).  The high precision measurements above involved Penning trap spectroscopy of the ground and long-lived metastable state of clouds of barium ions.  The lower precision measurements involved level crossing spectroscopy on fast ion beams.}
\begin{tabular}{lcclr}
State 		& $|L,S,J \rangle$ 	 	& $g_J(L,S,J)$	 & \multicolumn{2}{c}{Measured $g_J$} \\ \hline \hline
$6S_{1/2}$	& $|0, 1/2, 1/2 \rangle$ 	& 2		 & 2.002 490 6(11)   & \cite{knoll1996egf} \\
$6P_{1/2}$	& $|1, 1/2, 1/2 \rangle$ 	& 2/3		 & 0.672(6)		&\cite{poulsen1976tdl} \\
$6P_{3/2}$	& $|1	, 1/2, 3/2 \rangle$	& 4/3 	& 1.328(8) 		&\cite{poulsen1976tdl}\\
$5D_{3/2}$	& $|2, 1/2, 3/2 \rangle$	& 5/4		& 0.799 327 8(3)	&\cite{knoll1996egf} \\
$5D_{5/2}$	& $|2, 1/2, 5/2 \rangle$	& 6/5		& \multicolumn{1}{c}{---}				& \multicolumn{1}{c}{---}
\end{tabular}
\label{tab:gfactors}
\end{table}
In $LS$-coupling, a weak magnetic field breaks the $(2J+1)$ degeneracy of atomic states through an interaction called the Zeeman effect:
\begin{align}
H_\text{zeeman} &= - \boldsymbol{\mu}_\text{eff} \cdot \boldsymbol{B}, \\
\intertext{where the atomic magnetic moment $\boldsymbol{\mu}_\text{eff}$ points along the total angular momentum $\boldsymbol{J}$}
\boldsymbol{\mu}_\text{eff} &= -g_J \mu_B \boldsymbol{J}.
\end{align}
Here $\mu_B$ is the Bohr magneton and $g_J$ is the known as the Land\'{e} $g$-factor:
\begin{equation}
g_J(L,S,J) = \frac{J(J+1) + L(L+1) - S(S+1)}{2J(J+1)} + g_s \frac{J(J+1) - L(L+1) + S(S+1)}{2J(J+1)}
\end{equation}
and $g_s \approx 2$ is the bound electron $g$-factor.  The predicted and precisely measured $g$-factors of several relevant states in Ba$^+$ are shown in Table~\ref{tab:gfactors}.  Deviations from the predictions of this formula are expected due to higher order corrections and are best studied using trapped ion techniques\footnote{Interest in single trapped ions was in some sense launched by precision measurements of the electron $g$-factor~\cite{vandyck1986emm,vandyck1987nhp,gabrielse2006ndf} which still constitute the most sensitive test of Quantum Electrodynamics~\cite{hughes1999agv} and atomic masses (\cite{vandyckjr1993thm}, for instance).  See~\cite{thompson1990pma, dehmelt1990eis} for reviews.}.

To first order in perturbation theory, using $\boldsymbol{B}$ to define the quantization axis, energy shifts due to the Zeeman Hamiltonian are linear in $m_J$, the projection of the spin along the magnetic field direction:
\begin{align}
\Delta E_{\gamma,J,m_J}^\text{Zeeman} &= \langle \gamma,J,m_J | g_J \mu_B \boldsymbol{J} \cdot \boldsymbol{B} | \gamma, J,m_J \rangle \notag \\
&= \langle \gamma,J,m_J | g_J \mu_B B J_z | \gamma, J,m_J \rangle \notag \\ 
&= g_J \mu_B B m_J.  \label{eq:ZeemanShiftJ}
\end{align}
Note that states with $m_J = 0$ have no linear Zeeman shift.  In general, shifts do occur in second-order perturbation theory but are often suppressed by large energy denominators:
\begin{equation}
\Delta E_{\gamma,J,m_J}^\text{Second-order Zeeman} = \sum_{\gamma',J',m_J'}
\frac{|\langle \gamma, J, m_J| g_J \mu_B B J_z| \gamma', J', m_J' \rangle |^2}{E_{\gamma,J,m_J} - E_{\gamma',J',m_J'}}.
\end{equation}
\begin{figure}
\centering
\includegraphics{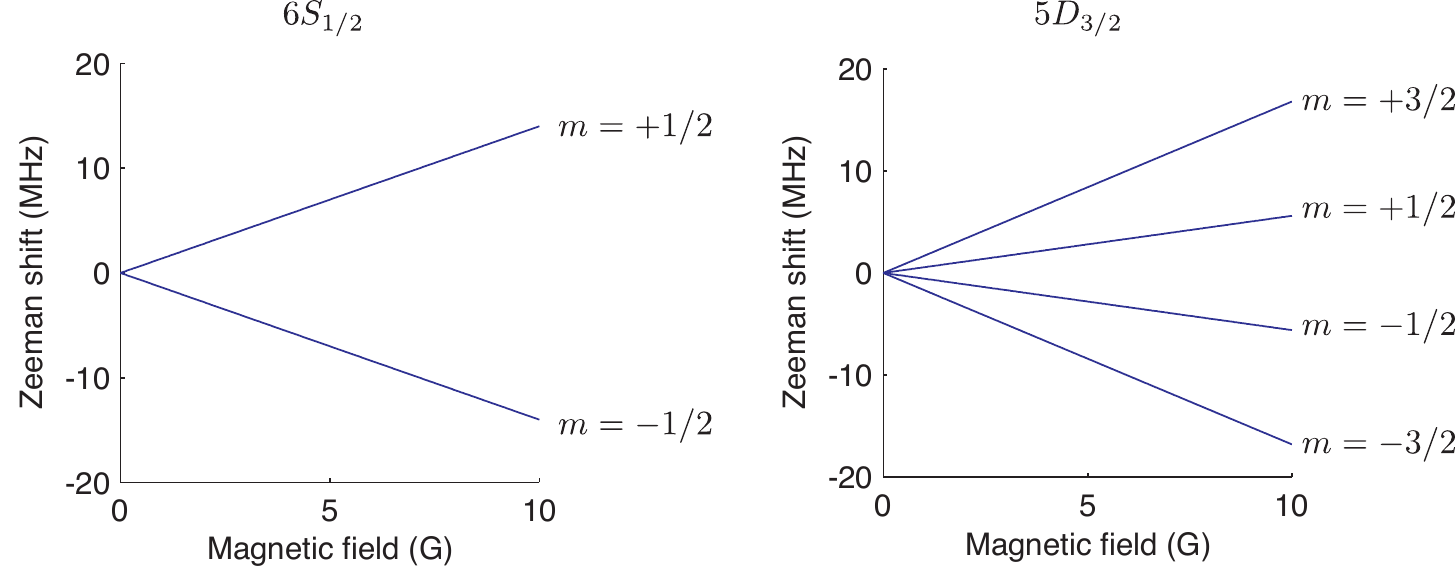}
\caption[Schematic of energy level shifts under the Zeeman effect]{Schematic of energy level shifts under the Zeeman effect in the single valence electron states $6S_{1/2}$ and $5D_{3/2}$.  The energy level shifts are linear in $|B|$ as long as the effect $|g_J \mu_B B m_J|$ is small compared to the fine-structure splittings.  Second-order Zeeman shifts, proportional to $B^2$, are highly suppressed for fields of several Gauss because the fine-structure splitting in heavy atoms such as Ba$^+$ is so large.}
\label{fig:zeemanEffectLevelShifts}
\end{figure}
Figure~\ref{fig:zeemanEffectLevelShifts} illustrates how magnetic substates shift under the Zeeman effect for small magnetic fields.

\begin{figure}
\centering
\includegraphics{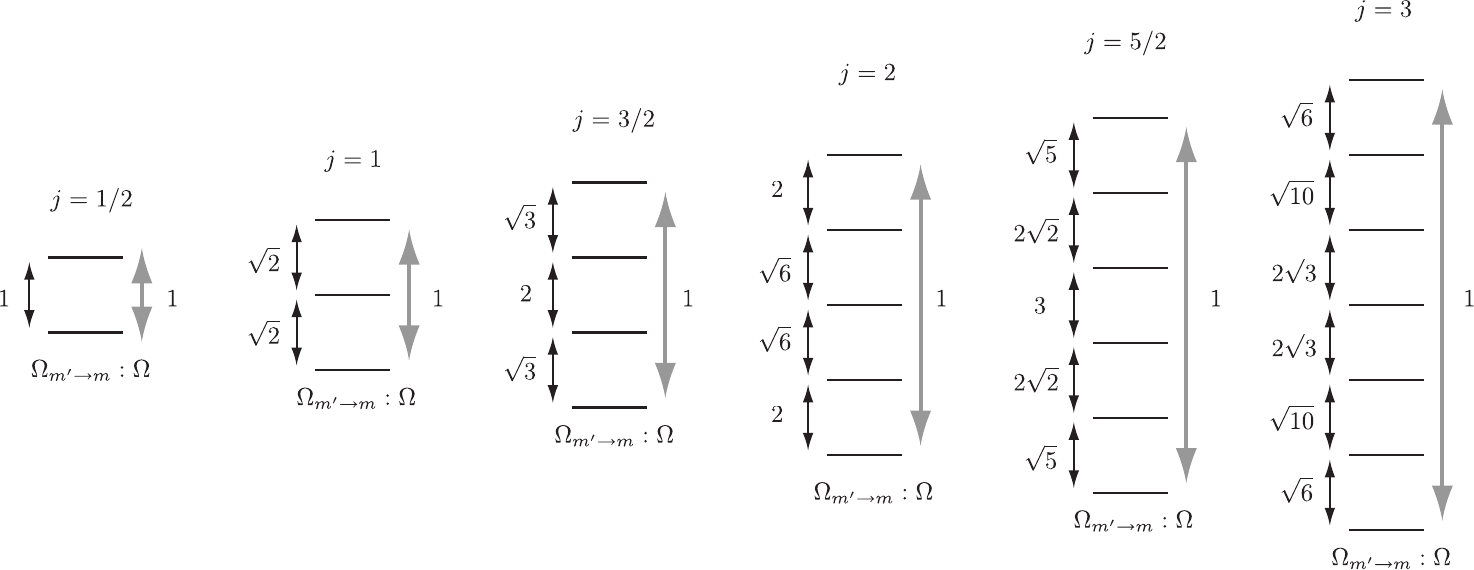}
\caption[For systems $J>1/2$, the spin resonance Rabi frequency is $m$-dependent.]{For systems $J>1/2$, the spin resonance Rabi frequency is $m$-dependent.  Here we show the ratio between the many effective `Rabi-frequencies' $\Omega_{m' \to m}$ in a spin $J>1/2$ system and the frequency $\Omega$ of total spin oscillation (e.g., for spin $j = 3/2$, a $m = 3/2 \to -3/2 \to 3/2$ cycle or `$2 \pi$-pulse.')}
\label{fig:rabiFrequenciesZeeman}
\end{figure}

We also must consider magnetic dipole transitions between Zeeman sublevels.  For a $J$-spin system, the interaction Hamiltonian is a $(2J+1) \times (2J+1)$ matrix with elements:
\begin{equation}
[H_\text{int}]_{m',m} = \frac{\hbar}{2} \Omega_{m' \to m},
\end{equation}
where we have defined `effective Rabi frequencies' between magnetic sublevels that in general have $m$-dependence:
\begin{align}
\Omega_{m' \to m} &= \frac{1}{\hbar} \langle j,m' | g_J \mu_B B \hat{J}_+ | j m \rangle \notag \\
&= g_J \mu_B B \sqrt{(j-m)(j+m+1)}.
\end{align}
Compare this expression with Eq.~\ref{eq:ZeemanShiftJ} and notice that this Rabi frequency formula is identical to the Zeeman energy shift expression for $J = 1/2$ systems only.  Figure~\ref{fig:rabiFrequenciesZeeman} shows these factors for several spin systems.  For the spin systems of interest in Chapter~\ref{sec:lightShiftChapter},
\begin{align*}
6S_{1/2}: \qquad \Omega_{1/2 \to -1/2} &= g(6S_{1/2}) \mu_B B, \\
5D_{3/2}: \qquad\Omega_{1/2 \to -1/2} &= 2 g(5D_{3/2}) \mu_B B, \\
5D_{3/2}: \qquad\Omega_{3/2 \to -1/2} &= \sqrt{3} g(5D_{3/2}) \mu_B B.
\end{align*}
For these spin systems, $J=1/2$ and $J=3/2$, an applied field detuned from by the Zeeman resonance frequency by $\delta$, yields the interaction Hamiltonians
\begin{align}
6S_{1/2}:  \qquad \hat{H} &= \frac{\hbar}{2}
\begin{pmatrix}
0			& g_J \mu_B B \\
g_J \mu_B B	& \delta
\end{pmatrix}, \\ \label{eq:dSpinFlipH}
5D_{3/2}:  \qquad \hat{H} &= \frac{\hbar}{2}
\begin{pmatrix}
0				& \sqrt{3} g_J \mu_B B	&	0 	                      &   0 \\
\sqrt{3} g_J \mu_B B	& \delta				& 2g_J \mu_B B            & 0 \\
0				& 2g_J \mu_B B 		& 2 \delta		             &\sqrt{3} g_J \mu_B B \\
0				& 0					& \sqrt{3} g_J \mu_B B & 3 \delta
\end{pmatrix}.
\end{align}
\begin{figure}
\centering
\includegraphics{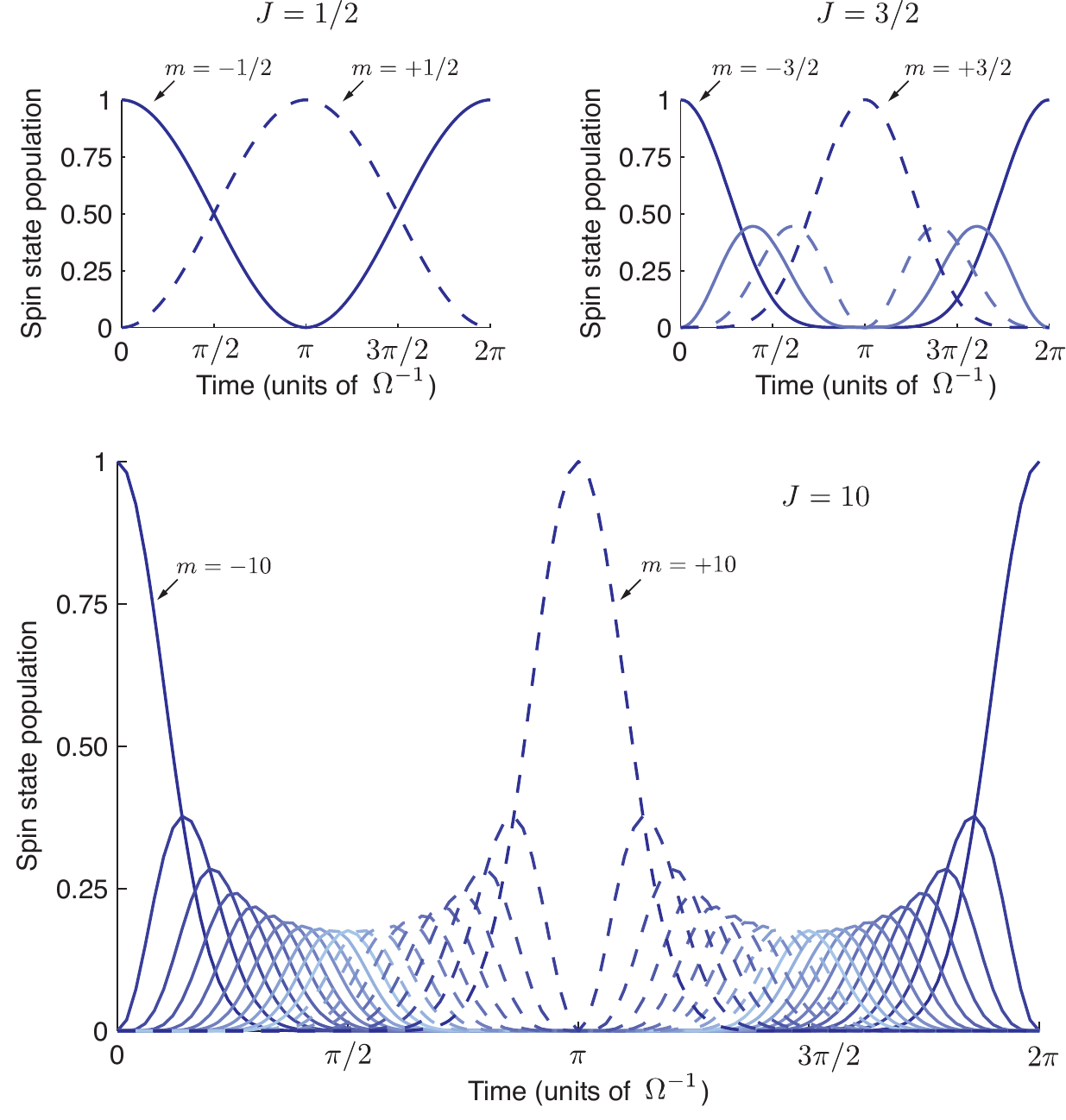}
\caption[Spin state evolution for $J \ge 1/2$ systems]{Spin state evolution for $J \ge 1/2$ systems.  Here we plot a full Rabi oscillation for the $J=1/2$ and $J=3/2$ spin state systems relevant for the spin resonance study in Chapter~\ref{sec:lightShiftChapter} and a high $J=10$ system to develop intuition.  In each plot, states $m \le 0$ are plotted with solid lines, states $m > 0$ are plotted with dashed lines, and the darkness of the line increases as $|m|$.  The states $m = \pm J$ are labeled.}
\label{fig:spinStateEvolution}
\end{figure}
Using the density matrix formalism presented in Section~\ref{sec:densityMatrixFormalism}, we can numerically integrate the time-dependent Schr\"{o}dinger equation
\begin{equation*}
i \hbar \frac{d \rho}{dt} = [\hat{H}, \rho]
\end{equation*}
and study the effect of near-resonant rf pulses on given initial state distributions.  It is illustrative to begin with a population completely in the lowest $m$-level:  that is $c_0(0) = 1$, and all other $c_i(0) = 0$, where $c_0$ labels the lowest $m$-level.  Figure~\ref{fig:spinStateEvolution} shows the time evolution of the spin-states for several different $J$ states. 

Another perspective is to treat the resonant magnetic dipole as a rotation of angle $\beta = \Omega t$ of the spin vector to some final state
\begin{equation}
| \psi(\beta) \rangle = e^{-2 i \hat{\boldsymbol{j}}_y \beta/\hbar} |j,m \rangle
\end{equation}
where $\hat{\boldsymbol{j}}_y$ is the generator of rotations for a spin-$J$ system.  Along the same axis that quantizes the initial state $|j m \rangle$ we find that amplitudes for the final state to be in some $|j m' \rangle$ are given by
\begin{equation}
\langle j m' | \psi(\beta) \rangle = d^{(j)}_{m' m}.
\end{equation}
The representation of rotations along $\hat{\boldsymbol{y}}$, $d^{(j)}$, is a matrix that is derived for particular spin systems of interest.  To see the generalization from $J =1/2$, in which SU(2) spin rotations are equivalent to ordinary rotations on a sphere, SO(3), we have
\begin{equation}
d^{(1/2)}_{m' m} = \begin{pmatrix}
\cos \beta/2	& -\sin \beta/2 \\
\sin \beta/2	& \cos \beta/2 \end{pmatrix}.
\end{equation}
The probability to rotate into the state $|1/2, +1/2 \rangle$ from an initial state $|1/2, -1/2 \rangle$ is
\begin{align*}
P(t) &= \left| \left\langle 1/2,+1/2 \left| e^{-2i \hat{\boldsymbol{j}}_y \Omega t /\hbar} \right| 1/2, -1/2 \right\rangle \right|^2 = \left|d^{(1/2)}_{1/2, -1/2} \right|^2 \\
&= \sin^2\left( \frac{\Omega t}{2} \right)
\end{align*}
which is equivalent to the derivation of two state Rabi oscillation given in Section~\ref{sec:rabiOscillations}.  See Figures~\ref{fig:twoStateRabi} and \ref{fig:twoStateRabiFreq} for explicit time and frequency dependence.  The lifetimes of Zeeman sub-levels are extremely long since the frequency splitting is so small, so damping can be completely neglected.  The representation for $J = 3/2$, is
\begin{equation}\label{eq:d32rotationmatrix} \small
d^{(3/2)}_{m' m} = \begin{pmatrix}
\cos^3 \tfrac{\beta}{2}					& 
-\sqrt{3} \sin \tfrac{\beta}{2} \cos^2 \tfrac{\beta}{2}	& 
\sqrt{3} \sin^2 \tfrac{\beta}{2} \cos \tfrac{\beta}{2}	& 
-\sin^3 \tfrac{\beta}{2} 					\\
\sqrt{3} \sin \tfrac{\beta}{2} \cos^2 \tfrac{\beta}{2} 			&
\cos \tfrac{\beta}{2} \left( 3\cos^2 \tfrac{\beta}{2} - 2\right)		&
\sin \tfrac{\beta}{2} \left( 3\sin^2 \tfrac{\beta}{2} - 2\right)		&
\sqrt{3} \sin^2 \tfrac{\beta}{2} \cos \tfrac{\beta}{2}			\\
\sqrt{3} \sin^2 \tfrac{\beta}{2} \cos \tfrac{\beta}{2} 			&
-\sin \tfrac{\beta}{2} \left( 3\sin^2 \tfrac{\beta}{2} - 2\right)		&
\cos \tfrac{\beta}{2} \left( 3\cos^2 \tfrac{\beta}{2} - 2\right)		&
-\sqrt{3} \sin \tfrac{\beta}{2} \cos^2 \tfrac{\beta}{2}			\\
\sin^3 \tfrac{\beta}{2}					&
\sqrt{3} \sin^2 \tfrac{\beta}{2} \cos \tfrac{\beta}{2}	&
\sqrt{3} \sin \tfrac{\beta}{2} \cos^2 \tfrac{\beta}{2}	&
\cos^3 \tfrac{\beta}{2} \end{pmatrix}.
\end{equation}
For arbitrary $j$, the $d^{(j)}_{m',m}$ operator is derived from the Wigner formula~\cite{sakurai1994mqm}
\begin{equation}
\begin{split}
d^{(j)}_{m',m} &= \sum_k (-1)^{k - m + m'} \frac{\sqrt{(j+m)! (j-m)! (j+m')! (j-m')!}}{(j+m-k)! k! (j-k-m')! (k-m+m')!} \\
& \qquad \qquad \times \left( \cos \frac{\beta}{2} \right)^{2j - 2k +m - m'} \left( \sin \frac{\beta}{2} \right)^{2k-m+m'}
\end{split}
\end{equation}
where $k$ is summed over all values that yield non-negative arguments in each denominator factorial term.

\subsection{Optical pumping} \label{sec:opticalPumping}
\begin{figure}
\centering
\includegraphics{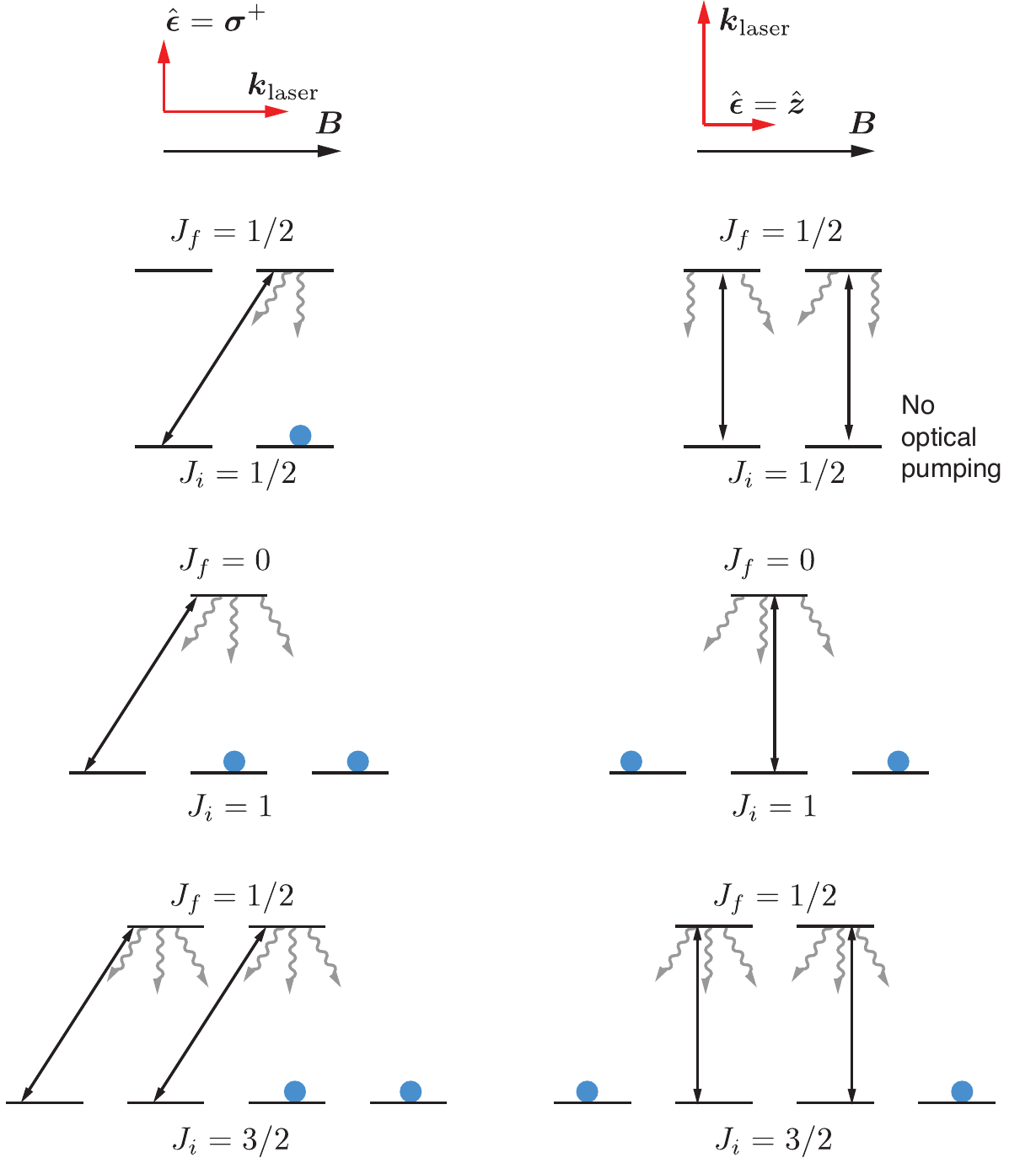}
\caption[Optical pumping examples]{A selection of optical pumping examples.  In the left column we depict a circularly polarized laser beam aligned with a magnetic field that serves are the quantization axis.  Because this configuration gives selection rules $\Delta m = 1$, the ground state population becomes polarized for $J_i \ge J_f$ since excited state decays are allowed for $\Delta m = 0,\pm1$.  In the right column we depict $\pi$-polarized light; that is, a linearly polarized laser aligned perpendicular to the magnetic field.  Here, only $\Delta m = 0$ excitations are allowed.  When $J_i > J_f$, optical pumping to the outer $m$-levels occurs.}
\label{fig:opticalPumpingExamples}
\end{figure}
Clever application of the transition angular selection rules implied by the Clebsch-Gordan coefficients in expressions such as:
\begin{equation*}
\Omega_\text{E1} = \frac{e E_0}{\hbar} \left| \langle \gamma, j  || r || | \gamma', j' \rangle \right| \sum_{q = -1}^{1} \langle j m; 1 q | j' m' \rangle E_q^{(1)}
\end{equation*}
yields a technique for optically directing atoms to occupy a chosen quantum state in the face of random spontaneous emission and even when many sublevels are degenerate or unresolved.  Figure~\ref{fig:opticalPumpingExamples} gives interesting configurations of an applied magnetic field and polarized light beam which leave certain ground state sublevels uncoupled and thus eventually occupied.

To demonstrate one example, take a system with initial and final angular momenta $J_i = J_f = 1/2$.  If a light beam is left circularly polarized and pointed along the quantization axis (magnetic field), then we can write see that only one coupling is non-zero
\begin{align*}
\left|\langle j m; 1 -1 | j' m' \rangle \cdot E_{-1}^{(1)}\right|^2 &= \frac{2}{3} \cdot (0) \\
\left|\langle j m; 1 0 | j' m' \rangle \cdot E_{0}^{(1)}\right|^2 &= \frac{1}{2} \cdot (0) \\
\left|\langle j m; 1 1 | j' m' \rangle \cdot E_{1}^{(1)}\right|^2 &= \frac{2}{3} \cdot (E)
\end{align*}
This means that only atoms in the $m=-1/2$ state can be coupled by the radiation (to excited state $m=+1/2$).  Since spontaneous decays out of the excited state can occur to either ground state $m = \pm 1/2$, this means that eventually atomic population will pile up in the $m= +1/2$ ground state.

A practical caveat to this technique for state preparation is imperfections in both laser alignment and polarization:  if the transition is highly saturated ($\Omega \gg \Gamma$) then small errors in either polarization or alignment leads to a large error in the final state population since the final state population when the laser is shut off is then determined by a single (random) decay.  Therefore in practice we attenuate optical pumping beams by a factor of $\sim 10^{-3}$ to ensure that any coupling due to errors in polarization and misalignment are far below saturated and that the intended pumping coupling is the only relevant interaction.  See Section~\ref{sec:opticalPumpingSD} for simulations of the optical pumping efficiency on the barium ion $6S_{1/2} \leftrightarrow 6P_{1/2}$ and $5D_{3/2} \leftrightarrow 6P_{1/2}$ transitions.

\subsection{Dark states and their destabilization}
\begin{table}
\let\PBS=\PreserveBackslash
\centering
\caption[Conditions for dark states in zero magnetic field (from \cite{berkeland2002dds})]{Reproduced from \cite{berkeland2002dds}, this table shows the conditions for dark state existence in systems with degenerate magnetic sublevels.  For optimal production of fluorescence and Doppler cooling, dark states must be destabilized by a proper choice of polarization, and, if necessary, a sufficiently strong magnetic field or modulation of laser polarization.}
\begin{tabular}{c|>{\PBS\centering}m{2.00 in}>{\PBS\centering}m{2.00in}}
\multicolumn{1}{c}{}	&\multicolumn{2}{c}{Lower level} \\ 
Upper level $J_f$	& Integer $J_i$	 	& Half-integer $J_i$ \\ \hline \hline
$J_i + 1$		& No dark state	 		& No dark state \\
$J_i$		& One dark state for any polarization & One dark state for circular polarization only \\
$J_i - 1$		& Two dark states for any polarization & Two dark states for any polarization
\end{tabular}
\label{tab:darkStates}
\end{table}
It is surprising to learn that in the absence of a magnetic field, \emph{any} polarization of light will optically pump a population of atoms if the excited state angular momentum $J_{f}$ is less than the ground state angular momentum $J_i$.  The states established by the pumping may not be the same eigenstates of the Zeeman operator but instead superpositions.  An intuitive way to understand is that in the absence of any other quantization axis, the light polarization vector defines the eigenstates.  These \emph{dark states} do not fluoresce, so cooling and observation of the ion is halted.

Since the dark states are in general coherent superpositions of magnetic $m$ levels, a sufficiently strong magnetic field can destabilize them.  Alternatively, when low magnetic fields are required, scrambling or modulating the polarization of light also destabilizes the dark states~\cite{devoe2002esa}.  A systematic theoretical study~\cite{berkeland2002dds} (from which Table~\ref{tab:darkStates} is reproduced) shows what one might expect:  the brightest fluorescence is obtained when the state destabilization rate $\delta$ is comparable to the driving laser field, $\delta \sim \Omega$.  

For $S_{1/2} \leftrightarrow P_{1/2} \leftrightarrow D_{3/2}$ systems such as Ba$^+$, the article~\cite{berkeland2002dds} specifically recommends a cooling and repumping intensities near but below saturation $\Omega_{SP} \sim \Gamma / 3$, $\Omega_{DP} \sim \Gamma/3$ to avoid power broadening.  Further, the repumping detuning should be positive $\Delta_{DP} = + \Gamma/2$ to place dark resonances on the blue side of the $S \leftrightarrow P$ transition.  Choosing the magnetic field such that $0.01 \Gamma < g_J \mu_B |\boldsymbol{B}| / \hbar < 0.1 \Gamma$ gives sufficient dark state destabilization without moving dark resonances by more than $\Gamma$ by the Zeeman interaction.

\chapter{Ion trapping} \label{sec:ionTrapping}
\begin{quotation}
\noindent\small I think it is a sad situation in all our chemistry that we are unable to suspend the constituents of matter free.\\ \flushright{---Georg Christoph Lichtenberg~\cite{paul1990emt}}
\end{quotation}

\section{A radio-frequency trap for single charged atoms}
It is impossible to confine charged particles with static electric fields alone.  This is so because Laplace's equation in a charge-free vacuum demands that an electrostatic potential have no hills or valleys: $\nabla^2 \phi = 0$, for any electric potential $\phi(x,y,z)$.  To illustrate this, we might imagine constructing a harmonic potential
\begin{equation*}
\phi(x,y,z) = ax^2 + by^2 + cz^2,
\end{equation*}
which may confine particles with a force of the form
\begin{equation*}
\boldsymbol{F} = - q (\boldsymbol{\nabla} \phi) = - 2 (ax \hat{\boldsymbol{x}} + by \hat{\boldsymbol{y}} + cz \hat{\boldsymbol{z}} )
\end{equation*}
that is restoring only if the coefficients $a$, $b$, and $c$ are all positive. However, Laplace's equation demands
\begin{equation*}
\nabla^2 \phi = -(2a + 2b + 2c) = 0,
\end{equation*}
which is only satisfied if $a = b = c = 0$ or if at least one of the coefficients is negative.

One solution, called a Penning trap (\cite{penning1936}, see~\cite{brown1986gtp} for a modern review), is to superimpose a strong magnetic field $B \hat{\boldsymbol{z}}$ onto an electric potential of the form $\phi(x,y,z) = ax^2 + by^2 - cz^2$ (which is precisely what results from an electrode shaped as a hyperboloid of revolution and held at constant electric potential).  The magnetic field curves ion trajectories into confining cyclotron orbits.  Another solution is to apply not a static electric voltage to a hyperbolic electrode pair but an \emph{oscillating} potential.  This solution, called a Paul trap~\cite{paul1990emt}, is employed in this work and described below.

\subsection{An oscillating potential on hyperbolic electrodes} \label{sec:paulTrap}
\begin{figure}
\centering
\includegraphics{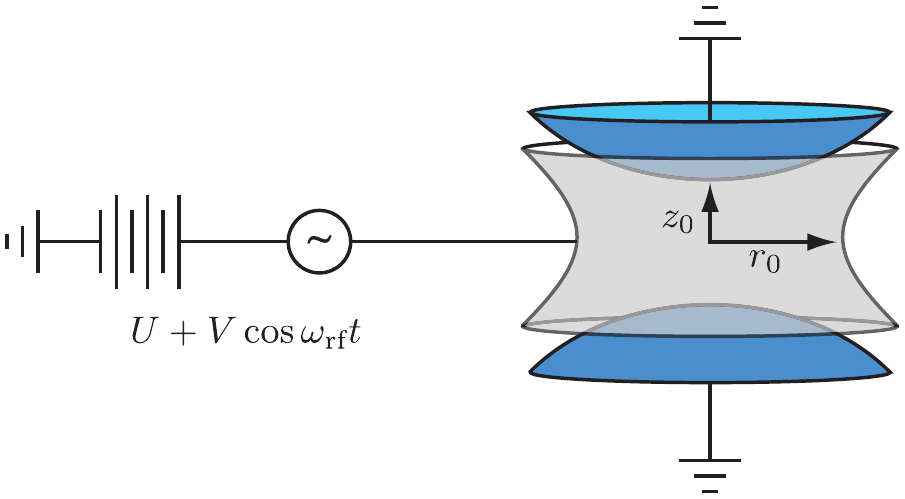}
\caption[The Paul trap:  a radio frequency trap for ions]{The Paul trap:  a radio frequency trap for ions.  Here $r_0$ and $z_0$, the distances to the ring and endcap electrodes, set the radial and axial distance scales.  In this treatment we consider the case $r_0 = \sqrt{2} z_0$.}
\label{fig:paulTrapVoltageDiagram}
\end{figure}
If we apply a voltage $U - V \cos \omega_\text{rf} t$ to the hyperbolic electrode structure shown in Figure~\ref{fig:paulTrapVoltageDiagram}, then the free space potential inside is
\begin{equation}
\phi(x,y,z) = (U - V \cos \omega_\text{rf} t) \left( \frac{x^2 + y^2 - 2z^2}{2r_0^2} \right)
\end{equation}
where $r_0$ is the central radius of the ring electrode and the distance to each endcap electrode is $z_0 = r_0 /\sqrt{2}$.  While the static voltage $U$ alone will not confine charged particles, the additional oscillating term $ V \cos \omega_\text{rf} t$ establishes a \emph{pseudo-potential} or force that is confining when averaged over oscillations of the fast frequency $\omega_\text{rf}$.

Given $\phi(x,y,z)$ above, the cylindrical symmetry of the potential, and the dynamical equation
\begin{equation*}
\boldsymbol{F} = m \frac{d^2 \boldsymbol{r}}{dt^2} = -e (\boldsymbol{\nabla} \phi),
\end{equation*}
we follow~\cite{ghosh1995it} and separate the equations of motion into radial and axial parts of motion
\begin{align}
\frac{d^2 r}{dt^2} + \left( \frac{e}{mr_0^2} \right) (U - V \cos \omega_\text{rf} t) r &= 0, \\
\frac{d^2 z}{dt^2} + \left( \frac{2e}{mr_0^2} \right) (U - V \cos \omega_\text{rf} t) z &= 0.
\end{align}
By defining dimensionless axial voltages $a_z$, $q_z$, radial voltages $a_r$, $q_r$, and a dimensionless time variable $\eta$,
\begin{align} \label{eq:trapParameters1}
a_z &= -2a_r \equiv - \frac{8eU}{mr_0^2 \omega_\text{rf}^2}, \\
q_z &= -2q_r \equiv - \frac{4eV}{mr_0^2 \omega_\text{rf}^2}, \label{eq:trapParameters2}\\
\eta &\equiv \frac{\omega_\text{rf} t}{2},
\end{align}
which are a set of universal parameters that contain all the physical scaling laws, we obtain the classical Mathieu equations:
\begin{align} \label{eq:mathieu}
\frac{d^2 r}{d\eta^2} + (a_r - 2q_r \cos 2 \eta ) r &= 0, \\
\frac{d^2 z}{d\eta^2} + (a_z - 2q_z \cos 2 \eta ) z &= 0. \notag
\end{align}
The dimensionless parameters $a_i$ and $q_i$ should be thought of physically as the strengths of the static and alternating electric field gradients.  Noticing that each $a_i$, $q_i$ takes the form
\begin{equation*}
(a,q)_i \sim \frac{\text{An electrostatic energy}}{\text{A kinetic energy}}
\end{equation*}
is not only an aid for remembering the expressions, but also suggests some of the limiting system behavior for very large or very small $a_i$ and $q_i$.  So far we have assumed a singly charged ion;  if $N$ electrons are removed, the $a$ and $q$ parameters are scaled up by $N$.

\begin{figure}
\centering
\includegraphics{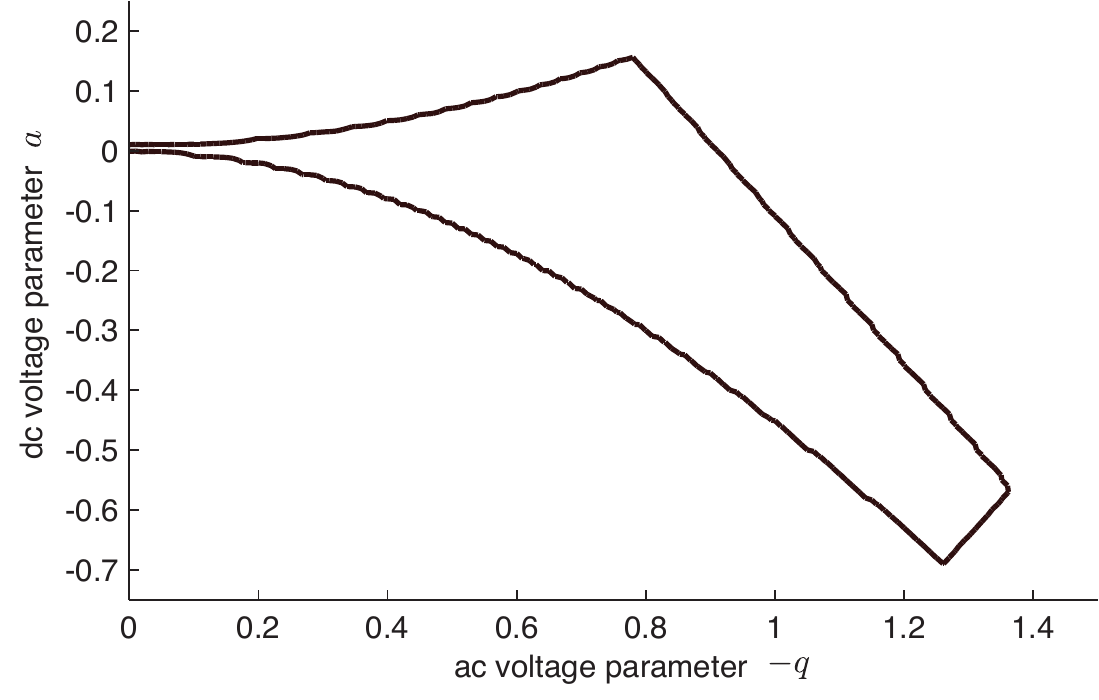}
\caption[Stability region of the trapping parameters $a$ and $q$]{This (first) stability region of the radial and axial dimensionless dc and ac trapping parameters $a_i$ and $q_i$ is obtained by numerically integrating the Mathieu equations for 100 rf cycles and plotting a contour for resulting ion orbits that remain confined to within the trap size $r_0$.  The plotted lines are somewhat rough due to the discrete set of $a_i$ and $q_i$ parameters chosen for simulation.  Other regions of stability exist in the parameter space, but this section is the most studied and the only one applicable to this work.}
\label{fig:stabilityDiagram}
\end{figure}
\begin{figure}[p]
\centering
\includegraphics[height = 7.0in]{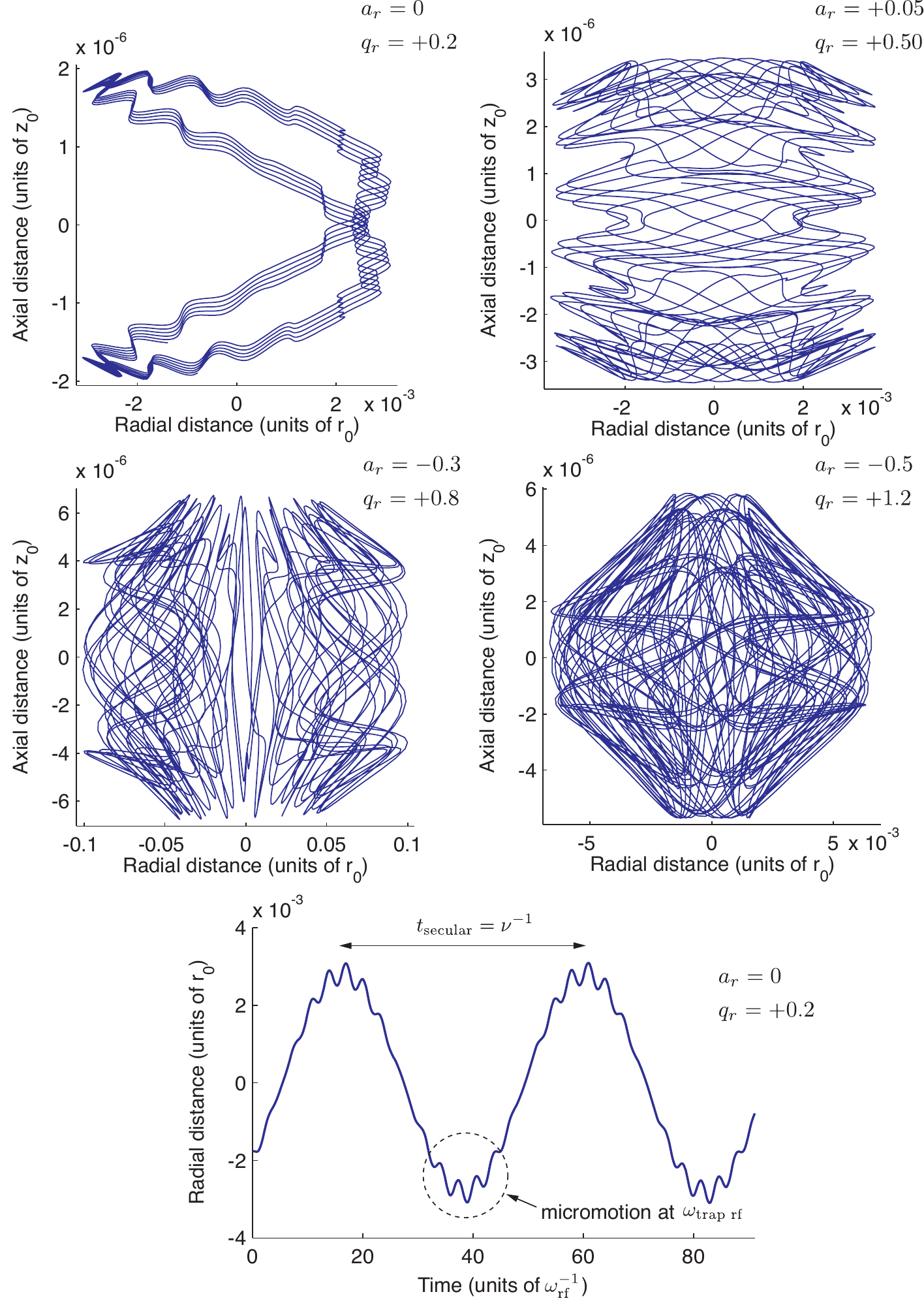}
\caption[Example trapped ion orbits]{Example trapped ion orbits.  We numerically integrate the Mathieu equations for $10^4$ cycles of $\omega_\text{rf}$ and plot the last 500 cycles.  In each case, the ion initial conditions are the same, $r(0) = r_0 \times 10^{-6}$, $z(0) = z_0 \times 10^{-6}$, $\dot{r}(0) =  r_0 \omega_\text{rf} \times 10^{-5}$, $\dot{z}(0) = 0$.  Notice in the last graph the two characteristic frequencies of motion:  secular motion at $2 \pi\nu = \beta \omega_\text{rf} / 2$ and micromotion at $\omega_\text{rf}$ with $\beta \ll 1$.}
\label{fig:ionOrbits}
\end{figure}

Mathieu equations (Eq.~\ref{eq:mathieu}) each have solutions expressible as power series.  Focusing on the radial equation,
\begin{equation}
u(\eta) = \alpha' e^{\mu \eta} \sum_{n = -\infty}^{n= +\infty} C_{2n} e^{2 i n \eta}
	    +\alpha'' e^{-\mu \eta} \sum_{n = -\infty}^{n= +\infty} C_{2n} e^{-2 i n \eta},
\end{equation}
where $\mu = \alpha \pm i \beta $ is a complex parameter expressed in pure real and imaginary parts $\alpha$ and $\beta$.  Unstable solutions exist for $\alpha \ne 0$ since one of the terms $e^{\mu \eta}$ or $e^{-\mu \eta}$ would increase without bound as the time variable $\eta \to \infty$.  Applying the stabilizing condition $\mu = \pm i \beta$ ($\alpha = 0$), we find the solution
\begin{equation}
u(\eta) = \alpha'  \sum_{n = -\infty}^{n= +\infty} C_{2n} e^{(2n \pm \beta)i \eta}
	   + \alpha''  \sum_{n = -\infty}^{n= +\infty} C_{2n} e^{-(2n \pm \beta)i \eta}.
\end{equation}
Here, $\alpha'$ and $\alpha''$ depend on the initial conditions of the ion while constants $C_{2n}$ and $\beta$ depend on the dimensionless voltages $a_i$ and $q_i$.  The so-called characteristic exponent $\beta$ is in principle expressed as a recursion relation~\cite{ghosh1995it, cirac1994lct}.  We will only be concerned with the region $0 < \beta < 1$, known as the first stability zone shown in Figure~\ref{fig:stabilityDiagram}.  In the lowest stability zone, which is the most commonly employed in practice, the lowest frequency of motion, termed the trap \emph{secular frequency} is
\begin{equation}
\nu = \frac{1}{2 \pi} \frac{\beta \omega_\text{rf}}{2}.
\end{equation}
Notice that when $\beta \ll 1$, the secular frequency $\nu \ll \omega_\text{rf}$.  This justifies a pseudo-potential approximation in understanding the oscillatory behavior of the trapped ion.  The radio-frequency field makes several sign changes during one secular oscillation in the trap.  The resulting ion orbit is therefore comprised of two pieces:  a harmonic secular oscillation at frequency $\nu$ and \emph{micromotion} at the trap frequency $\omega_\text{rf}$~\cite{berkeland1998mim}:
\begin{equation}
u(t) \sim \underbrace{u_0 \cos(2 \pi \nu t + \phi)}_\text{slow, secular motion}  \big[1 + \underbrace{\frac{q_r}{2} \cos \omega_\text{rf}}_\text{micromotion} \big].
\end{equation}
Refer to the final graph in Figure~\ref{fig:ionOrbits} for a graphical illustration of each component of the ion's oscillation.  In actuality there are three secular frequencies:  one for each degree of freedom.  Often, two are nearly degenerate due to the axial symmetry of the trap electrodes.  In the limit $\beta \ll 1$, we find the secular frequencies are
\begin{equation}
\nu_i \approx \frac{1}{2 \pi} \frac{\omega_\text{rf}}{2} \sqrt{a_i + \frac{1}{2}q_i^2}.
\end{equation}
A typical case is $a_i \ll q_i$ (low dc voltage) and $2\pi \nu_i \sim \omega_\text{rf}/10$ which gives $q_i \approx 0.28$.  The secular frequencies are easily measured by exciting the ion with an external rf source at $\nu_i$ and observing decreased fluorescence due to the ion heating that results.  A more robust method is the observation of motional sidebands on narrow optical transitions, discussed in Section~\ref{sec:coolingRegime}.

For much of the spectroscopy work in this thesis, the ion's micromotion at $\omega_\text{rf}$ is not carefully characterized and minimized.  The unavoidable micromotion that comes with the excursion of the ion away from the rf-null of the trap's center is responsible for about the same kinetic energy as the secular motion itself.  One treatment~\cite{berkeland1998mim} gives the radial kinetic energy as
\begin{align*}
E_{K,r} 	&= \frac{1}{2} m \langle \dot{r}^2 \rangle, \\
		&\simeq \frac{1}{4} m u_0^2 \left( (2\pi \nu)^2 + \frac{1}{8} q_r^2 \omega_\text{rf}^2 \right), \\
		&\simeq \underbrace{\frac{1}{4} m u_0^2 (2 \pi \nu)^2}_\text{secular} \Big( 1 + \underbrace{\frac{q_r^2}{q_r^2 + 2 a_r}}_\text{micromotion} \Big).
\end{align*}
We recognize that the factor $q_r^2 / (q_r^2 + 2 + a_r) \approx 1$ when we operate with low dc trap voltages $q_r \gg a_r$.  The ion's micromotion, heating rate, and mean kinetic energy can be significantly increased, however, by an uncompensated, stray electric field $\boldsymbol{E}_\text{dc}$.  The ion's equilibrium position must shift in the trap in response to such a field which in general means that it experiences larger oscillating fields and exhibits greater micromotion.  We can express the new orbit~\cite{berkeland1998mim} as
\begin{align*}
u(t) &\sim \underbrace{u_0' + u_0 \cos(2 \pi \nu t + \phi)}_\text{slow, secular motion}  \big[1 + \underbrace{\frac{q_r}{2} \cos \omega_\text{rf}}_\text{micromotion} \big] \\
\intertext{with the equilibrium displacement}
u_0' &\approx \frac{4 e \boldsymbol{E}_\text{dc} \cdot \hat{\boldsymbol{r}}}{m(a_r + \tfrac{1}{2} q_r)^2 \omega_\text{rf}^2} \approx \frac{e\boldsymbol{E}_\text{dc} \cdot \hat{\boldsymbol{r}} }{m (2 \pi \nu)^2}.
\end{align*}
While the term $u_0'$ does not increase the size of the secular orbit $u_0$ (in this approximate treatment), it does substantially increase the magnitude of the micromotion.  In an expression for the ion's kinetic energy~\cite{berkeland1998mim},
\begin{equation}
E_{K,r} \approx \underbrace{\frac{1}{4} m u_0^2 \left( (2 \pi \nu)^2 + \frac{1}{8} q_r \omega_\text{rf}^2 \right)}_\text{secular and accompanying micromotion} + \underbrace{\frac{4}{m} \left( \frac{e q_r \boldsymbol{E}_\text{dc} \cdot \hat{\boldsymbol{r}} }{(2 a_r + q_r^2) \omega_\text{rf}}  \right)^2}_\text{excess micromotion}.
\end{equation}
When converted to equivalent temperatures, the first term is the same order of magnitude of the $\sim 1$~mK Doppler cooling limit (see Section~\ref{sec:semiclassicalLaserCooling}), while the second term can be much higher.  For a barium trapped with $q_r = 0.25$ at $\omega_\text{rf} = 10$~MHz in a stray field of just 1~V/mm, the micromotion temperature is $\sim$~250 K.  Unlike secular motion which can be cooled, micromotion is driven by the trap's radio frequency field and cannot be cooled by Doppler cooling alone.  Fortunately, robust methods exist for the detection and minimization of micromotion~\cite{berkeland1998mim}.  The implication of micromotion and potential minimization techniques are discussed in the context of large second-order Doppler shifts of single ion optical frequency standards in Sections~\ref{sec:secondOrderDopplerShift} and~\ref{sec:clockMicromotion}.

\subsection{Real world trap shapes}\label{sec:realWorldTrapShapes}
\begin{figure}[p]
\centering
\includegraphics[width=6 in]{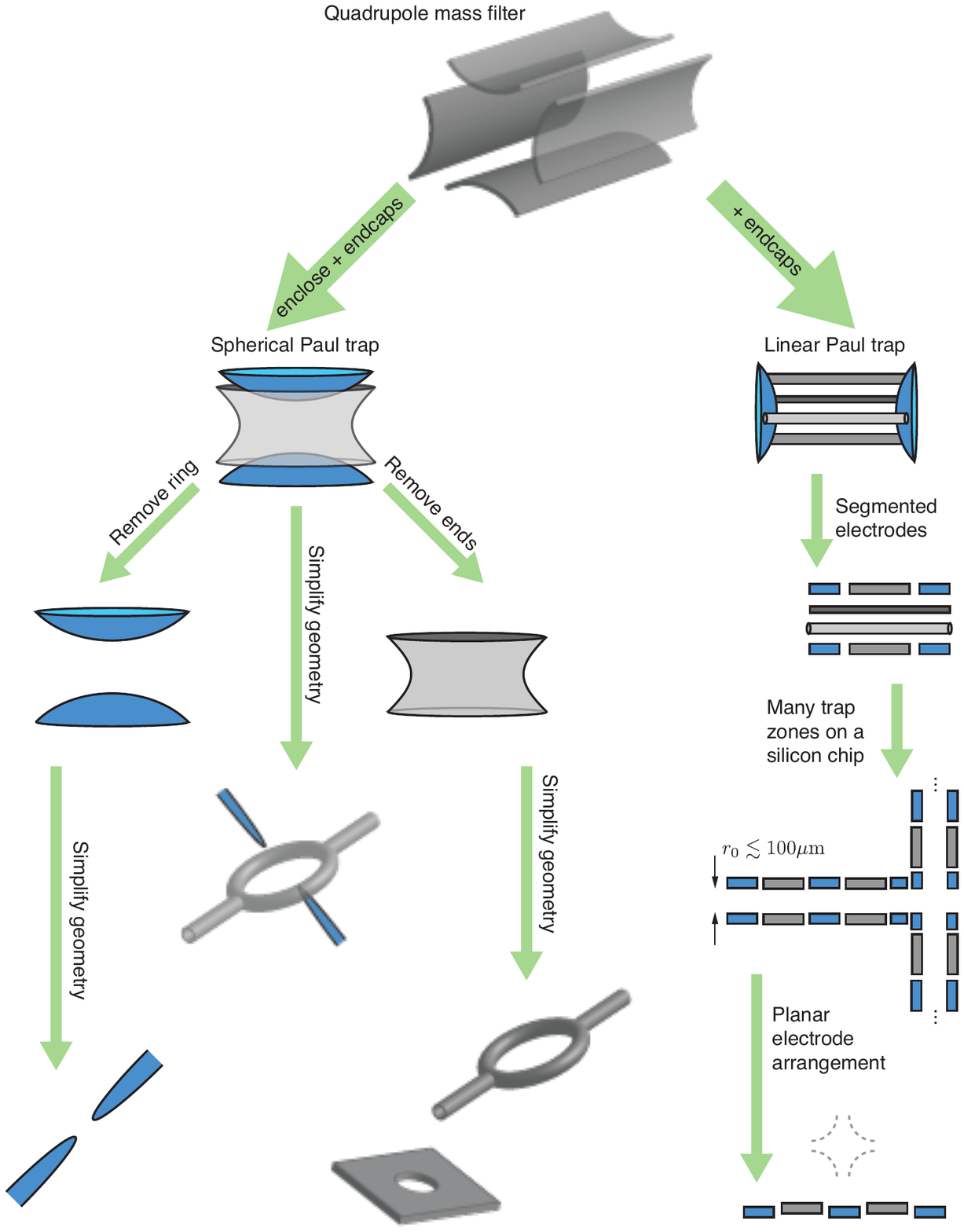}
\caption[Evolution of ion traps]{Evolution of ion traps from a quadrupole mass filter.  See text for a narrative.}
\label{fig:ionTrapEvolution}
\end{figure}
The hyperbolic ring and endcap electrodes shown in Figure~\ref{fig:paulTrapVoltageDiagram} do not offer easy optical access to the ion;  incident laser beams and collected fluorescence must be squeezed through tiny holes in the ring or endcaps, or through endcaps made of wire mesh.  Also, hyperbolic surfaces are difficult to manufacture, especially on miniaturized scales.  It is fortunate that almost any electrode shape with axial symmetry will provide a trapping center suitable for confining charged particles.  The cost of deviating from the ideal hyperbolic electrode shape can be understood with an ``inefficiency factor'' $f_\text{loss}$ that turns out to be of order 10 for many designs of interest.  The dimensionless trap strength parameters (Eq.~\ref{eq:trapParameters1} and~\ref{eq:trapParameters2}) are modified:
\begin{equation*}
a_r \to \frac{8e(U/f_\text{loss})}{mr_0^2 \omega_\text{rf}^2}, \qquad
q_r \to \frac{4e(V/f_\text{loss})}{mr_0^2 \omega_\text{rf}^2}, \qquad \text{etc.}
\end{equation*}
For a given trapping strength, more rf voltage is required on a non-hyperbolic trap because less pure gradient field is created per Volt applied.  As a corollary, the impure quadrupole field created by these electrodes contains higher order multipole moments which can lead to additional ion heating and deviations from the pseudo-potential model, especially for large ion clouds.  One systematic numerical study of the voltage loss factor of various electrode shapes is found in~\cite{yu1995aps}.  Another study~\cite{champenois2001cmp} experimentally determines the factor $f_\text{loss} \approx 7$ for a ring trap.

We see in Figure~\ref{fig:ionTrapEvolution} the origin of ion traps in an earlier revolutionary invention: the quadrupole mass filter.  This is a set of extruded hyperbolic electrodes that stably confine radially a beam of charged particles having a chosen range of charge-to-mass ratios~\cite{march1989qsm}.  After precision trapping of single electrons in hyperbolic Penning traps in 1987~\cite{vandyck1987nhp,dehmelt1990eis} the development of laser cooling techniques in the 1980's~\cite{stenholm1986stl}, ion trap designs evolved toward increasing simplicity and optical access in trap designs. \cite{yu1991dnp} is an early example of a trap without endcaps, also called a Paul-Straubel\footnote{Researcher H.\ Straubel demonstrated the trapping of charged dust particles in a ring shaped electrode in 1955~\cite{straubel1955}.} or ring trap.  The inverse, an endcap trap~\cite{schrama1993nmi}, does away with the ring and instead drives the endcaps, which are easily made out of thin metal rods or needles.  A ring can also be fashioned as a hole punched in a plane of metal~\cite{brewer1992pim}.  One can treat the removal of the ring or grounded electrodes as the limiting case of their placement farther away from the trapping center $z_0 \to \infty$ (or $r_0 \to \infty$ in the case of the endcap trap).  In practice, some grounded electrode in the apparatus outside the trapping region acts as the sink for electric field lines.

A similar road towards simplification has been traveled by linear ion trap designs, also shown in Figure~\ref{fig:ionTrapEvolution} after the groundbreaking work of Cirac and Zoller~\cite{cirac1995qcc} suggested a linear crystal of trapped cold ions as a vehicle for quantum information storage, manipulation, and general computation.  The rf electrodes confine ions radially;  endcap electrodes energized with a dc voltage axially confine the ions which mutually repel each other.  Early in the trap's evolution, we see the axial endcaps turned into segments aligned with the rf electrodes~\cite{berkeland1998mim}.  In the last 10 years, silicon etching and coating techniques native to the semi-conductor industry have yielded linear traps with $r_0 < 50$~$\mu$m, complex arrangements of segmented electrodes, and multiple independently operating trapping zones~\cite{madsen2004pit,stick2006its}.  Recently, at least one group has succeeded in driving trapped ions around $90^\circ$ bends~\cite{hensinger2006tji} which lends hope to someday realizing a complicated ion quantum information architecture~\cite{kielpinski2002als}.  At this writing, the state of the art in linear ion traps is a planar geometry that may support quiet and stable trapping of ions above a fabricated chip~\cite{chiaverini2005sea,pearson2006eip}.  The present challenge in these applications seems to be a matter of surface physics:  electric field noise experienced by ions in such small traps is far beyond the expected Johnson noise in the metal electrodes~\cite{deslauriers2006ssa}.   Dielectric patch-potentials~\cite{turchette2000hti} due to crystalline solid state zones or dirt on the metal seem to be implicated in worse than expected ion heating rates.  Careful study of the scaling behavior of the ion heating rate with the electrode/ion separation $d$~\cite{deslauriers2006ssa} demonstrates a $d^{-4}$ dependence, much stronger than the $d^{-2}$ scaling one expects from electric field fluctuations due to Johnson noise alone. Community-wide effort is underway to fully characterize and solve the problem.

\section{Ion/trap/laser interactions:  a tale of three rates}
Now we consider the rich behavior of three interacting physical systems:  a single trapped ion, a macroscopic ion trap, and a coherent laser beam.  Surprisingly, simply defining and comparing three rates describing the ion, trap, and ion/photon interactions yield valuable physical intuition about a cooled trapped ion.

\subsection{A decay rate, a secular oscillation rate, a recoil energy (disguised as a rate)}
\begin{table}
\centering
\caption[Common values:  decay rates, trap secular frequencies, recoil energies]{Three characteristic rates of the ion/trap/resonant photon system:  the spontaneous decay rate $\Gamma$, the trap secular oscillation frequency $\nu$ and the single photon atomic recoil $\epsilon$.  These rates often segregate into different orders of magnitude which makes their comparison a fruitful source of intuition.  The inset figures are convenient icons for remembering the definitions of the rates;  see the text for a full discussion.}
\begin{tabular}{m{2.0in}m{1.75in}m{1.75in}}
		& Typical minimum rates 			& Typical maximum rates \\ \hline \hline
\multicolumn{1}{c}{Decay rate $\Gamma$} & $< 1$ Hz & 100 MHz \\
\includegraphics{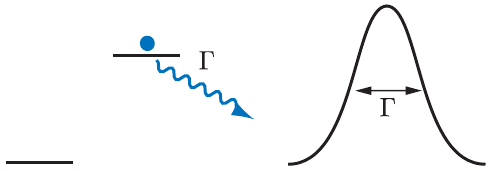}	& Electric-dipole forbidden or ``clock'' transitions & Fast electric-dipole cooling transitions \\ \hline
\multicolumn{1}{c}{Trap secular frequencies $\nu$} & 0.1 MHz & 10 MHz \\
\includegraphics{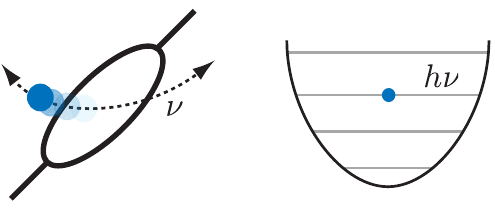}		&  Large, loose traps, slow rf drive $\omega_\text{rf}$, low voltage, heavy ions &  Small, stiff traps, fast rf drive $\omega_\text{rf}$, high voltage, light ions \\ \hline
\multicolumn{1}{c}{Recoil energies $\epsilon \equiv \hbar k^2/2m$} & 1 kHz & 500 kHz \\
\multicolumn{1}{c}{\includegraphics{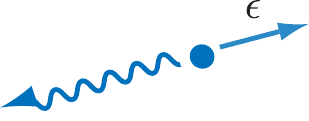}}			& IR and visible radiation, heavy ions &  UV radiation, light ions
\end{tabular}
\label{tab:threeRatesCommon}
\end{table}
Table~\ref{tab:threeRatesCommon} summarizes three characteristic rates describing the ion/trap/resonant photon system.  Any excited atomic state will have a finite lifetime $\tau$ (see Eq.~\ref{eq:spontaneousDecayAtomic}) meaning that transitions involving the state are spectroscopically broadened by at least the natural linewidth $\Gamma = 1 / \tau$, where $\Gamma$ is the decay rate of the excited state.  As shown in Section~\ref{sec:paulTrap}, the rf Paul trap creates a (time-averaged) harmonic potential that confines our ion.  If the energy level spacing in this potential is $h \nu$, then $\nu$ is identified as the trap secular frequency---it is the ion's classical natural frequency of oscillation.  Finally, due to conservation laws, the spontaneous emission of a photon with energy $\hbar k$ results in the emitting system to receive $(\hbar k)^2 / 2m$ of kinetic energy called \emph{recoil}.  By removing one $\hbar$ factor we can write this quantity as a rate $\epsilon \equiv \hbar k^2/2m$ and compare it with others.
		
See Table \ref{tab:threeRatesCommon} to become acquainted with the typical values encountered for these rates.  By analyzing dimensionless ratios of these rates we will gain at least qualitative understanding of the coupled ion/trap/laser system~\cite{stenholm1986stl}.

\subsection{The Lamb-Dicke parameter: $\epsilon / \nu$}\label{sec:threeRatesLambDicke}
\begin{table}
\centering
\caption{Two limiting regimes of trapping behavior: $\epsilon / \nu \ll 1$ and $\epsilon / \nu \gg 1$.}
\setlength{\extrarowheight}{4 pt}
\setlength{\tabcolsep}{20 pt}
\begin{tabular}{p{2.5 in} p{2.5 in}}
\multicolumn{1}{c}{{\Large Lamb-Dicke limit}} 	& \multicolumn{1}{c}{{\Large Poor confinement}} \\
\multicolumn{1}{c}{{\Large $\epsilon / \nu \ll 1$}} 		& \multicolumn{1}{c}{{\Large $\epsilon / \nu \gg 1$}} \\
\multicolumn{1}{c}{\includegraphics[width=2 in]{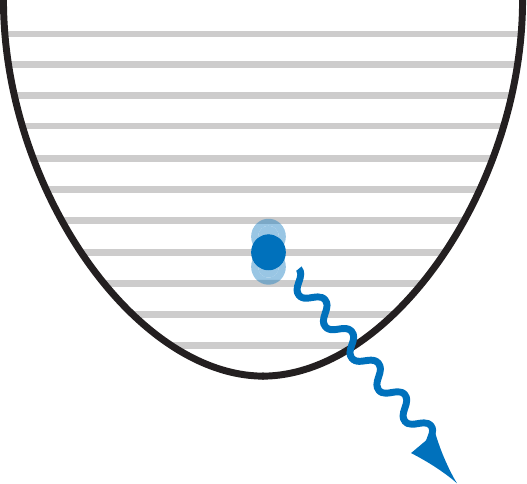}} &
\multicolumn{1}{c}{\includegraphics[width=2 in]{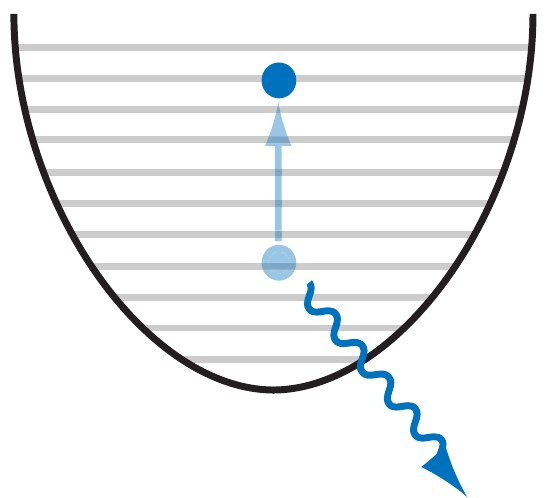}} \\
{\raggedright Atomic recoils are \emph{unlikely} to cause changes in the motional quantum number~$|n\rangle$.} &
{\raggedright  Atomic recoils are \emph{likely} to cause changes in the motional quantum number~$|n\rangle$.}\\
{\raggedright The recoil energy is absorbed by the macroscopic trap.}& 
{\raggedright The atom's kinetic energy increases with each recoil.} \\
{\raggedright $\epsilon / \nu \ll 1$ $\Rightarrow$ $\langle x^2 \rangle \ll \lambda$:  The atom is confined to a region much smaller than a wavelength of light (Dicke-narrowing).}& 
{\raggedright $\epsilon / \nu \gg 1$ $\Rightarrow$ $\langle x^2 \rangle \gg \lambda$:  The atom is not well confined.}  \end{tabular}
\label{tab:LambDicke}
\end{table}

The most important comparison to make is between the single photon emission recoil $\epsilon$ and the trap secular frequency $\nu$, illustrated in Table~\ref{tab:LambDicke}.  For even the lightest ions and ultra-violet emitted photons, the photon recoil energy $\epsilon = \hbar k^2/2m < 1$~MHz, a natural laboratory value for $\nu$, so the limit $\epsilon / \nu \ll 1$, called the Lamb-Dicke regime, is nearly always achievable.  Recall that the secular frequency $\nu$ also describes the energy level spacing $h \nu$ in the quantum harmonic potential well confining the ion.  If $\eta^2 \equiv \epsilon / \nu \ll 1$ then individual photon absorptions or emissions are unlikely to change the vibrational occupational number $|n\rangle$ of the ion in its potential well since the quantized ion motion can only increase in units of energy $\hbar \nu \gg \hbar \epsilon$.

Suppose an atom in the excited state decays without changing its vibrational quantum number $|n\rangle$, the most likely scenario in the Lamb-Dicke regime:  what happens to the recoil energy?  To conserve momentum, some mass must move in response to the emission of a photon, but it isn't likely to be the atom.  The answer is stunning:  in analogy with the M\"{o}ssbauer effect, the macroscopic trap absorbs the miniscule recoil energy~\cite{cirac1995qcc}.

A crucial consequence of this limit is that a trapped fluorescing ion is localized to dimensions smaller than the wavelength of light.  To see this, recall that for a harmonic potential well the dynamical variables $x$ and $p$ are related to raising and lowering operators
\begin{align}
a &= \sqrt{\frac{m \nu}{2 \hbar}} \left(x + i \frac{p}{m \nu} \right), \\
a^\dag &= \sqrt{\frac{m \nu}{2 \hbar}} \left(x - \frac{p}{m \nu} \right), \\
\intertext{which act on quantize vibrational states $| n \rangle$:}
a |n \rangle &= \sqrt{n} |n-1 \rangle, \\
a^\dag |n \rangle &= \sqrt{n+1} | n+1 \rangle.
\end{align}
One can show that the expected size of the wavefunction, $x_0 = ( \langle x^2 \rangle )^{1/2}$, is given by
\begin{align}
\langle n | x^2 | n \rangle &= \frac{\hbar}{2 m \nu} \langle n | (a + a^\dag) | n \rangle \\
&= \frac{\hbar}{2 m \nu} \left( 2n + 1 \right). 
\end{align}
Then, beginning with the condition for the Lamb-Dicke limit
\begin{align}
\frac{\hbar k^2}{2m} &\ll \nu, \\
\intertext{we see that}
k^2 &\ll \frac{2 m \nu}{\hbar}, \notag \\
\left( \frac{2 \pi}{\lambda} \right)^2 &\ll \frac{2n +1}{\langle x^2 \rangle}, \notag \\
\Rightarrow \langle x^2 \rangle &\ll \frac{2n + 1}{(2 \pi)^2} \lambda^2, \notag \\
\intertext{or}
x_0 = \sqrt{\langle x^2 \rangle} &\ll \frac{ \sqrt{2n +1}}{2 \pi} \lambda.
\end{align}
The condition $x_0 \ll \lambda$ holds true for $n \lesssim 30$ which is nearly always achievable with ordinary Doppler laser cooling techniques.  For instance, since laser cooling yields a thermal distribution of vibrational excitations, the expected value $\langle n \rangle$ with a secular motion temperature of 1~mK in a trap with $\nu / 2\pi = 1$~MHz is~\cite{stenholm1986stl, wineland1987lcl}
\begin{align} \label{eq:expectedNwithTemperature}
\langle n \rangle &= \frac{1}{e^{\hbar \nu / k_B T} -1} \\
		  	   &\approx 20. \notag
\end{align}

Finally, a vital consequence of the Lamb-Dicke limit is that first-order Doppler spectroscopic shifts on narrow transitions are negligible~\cite{wineland1987lcl}.  The second-order (relativistic) Doppler effect remains and is discussed in Section~\ref{sec:secondOrderDopplerShift}.

\subsection{The semi-classicalness of the interaction:  $\epsilon / \Gamma$}
\begin{table}
\centering
\caption[The applicability of semi-classical interactions: $\epsilon / \Gamma \ll 1$ and $\epsilon / \Gamma \gg 1$]{The applicability of semi-classical interactions: $\epsilon / \Gamma \ll 1$ and $\epsilon / \Gamma \gg 1$.}
\setlength{\extrarowheight}{4 pt}
\setlength{\tabcolsep}{20 pt}
\begin{tabular}{p{2.5 in} p{2.5 in}}
\multicolumn{1}{c}{{\Large Semiclassical cooling}} 		& \multicolumn{1}{c}{{\Large Non-classical interactions}} \\
\multicolumn{1}{c}{{\Large $\epsilon / \Gamma \ll 1$}} 	& \multicolumn{1}{c}{{\Large $\epsilon / \Gamma \gg 1$}} \\
\multicolumn{1}{c}{\includegraphics[width=2 in]{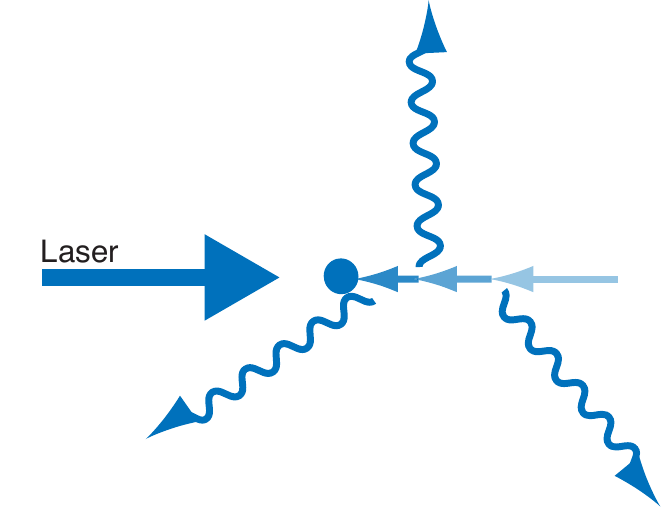}} &
\multicolumn{1}{c}{\includegraphics[width=2 in]{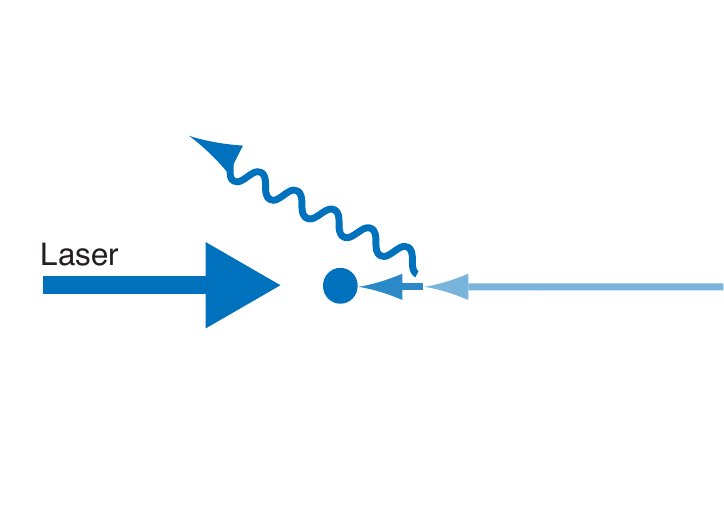}} \\
{\raggedright Significant cooling/heating is the result of \emph{many} scatterings.}  &
{\raggedright Significant cooling/heating can result from \emph{few} scatterings.} \\
{\raggedright The timescale of cooling is much larger than atomic state lifetimes $t_\text{cool} = \epsilon^{-1} \gg \tau$.} & 
{\raggedright The timescale of cooling is comparable or shorter than atomic state lifetimes $t_\text{cool} = \epsilon^{-1} \lesssim \tau$.} \\
{\raggedright Individual atom/photon interactions do not significantly change the atom's external degrees of freedom.}  &
{\raggedright  Each photon interaction is important. } \\
{\raggedright A semiclassical ``light-pressure'' model sufficiently explains the cooling process.} &
{\raggedright A full Hamiltonian~\cite{leibfried2003qds} including quantized photons and ion motion is needed to describe the system.}
\end{tabular}
\label{tab:semiclassicalCoolingLimits}
\end{table}
How quickly do resonant photons change atomic properties?  The timescale for changes in internal atomic states is the decay rate $\Gamma$.  However, the timescale for semi-classical changes in external properties like the atom's motion is $\epsilon^{-1} = (\hbar k^2/2m)^{-1}$.  How these two rates compare dictates whether the theory of semi-classical laser cooling can still apply to a trapped particle.  Table~\ref{tab:semiclassicalCoolingLimits} presents two limiting behaviors.  In the limit $\epsilon/ \Gamma \ll 1$, true for most electric-dipole allowed transitions, the absorption and emission of photons is fast compared to changes in the atom's motion;  we can think in terms of a semi-classical light-pressure on the atom because the timescale for changes in the internal degrees of freedom $\tau = \Gamma^{-1}$ is short compared the timescale for changes to external degrees of freedom $\tau_\text{cool} \sim \epsilon^{-1}$ (see Figure~\ref{fig:laserCooling}). 

In contrast, when atomic recoils are large or the state lifetimes long, $\epsilon /\Gamma \gg 1$ applies, and each absorption and emission process must be treated fully quantum-mechanically.

\subsection{The spectroscopy/cooling regime:  $\Gamma / \nu$} \label{sec:coolingRegime}
\begin{table}
\centering
\caption[Two limiting regimes of cooling behavior: $\Gamma /  \nu \gg 1$ and $\Gamma /  \nu \ll 1$]{Two limiting regimes of cooling behavior: $\Gamma /  \nu \gg 1$ and $\Gamma /  \nu \ll 1$.}
\setlength{\extrarowheight}{4 pt}
\setlength{\tabcolsep}{20 pt}
\begin{tabular}{p{2.5 in} p{2.5 in}}
\multicolumn{1}{c}{{\Large Heavy particle limit}} 		& \multicolumn{1}{c}{{\Large Resolved sideband limit}} \\
\multicolumn{1}{c}{{\Large $\Gamma /  \nu \gg 1$}} 	& \multicolumn{1}{c}{{\Large $\Gamma /  \nu \ll 1$}} \\
\multicolumn{1}{c}{\includegraphics[width=2 in]{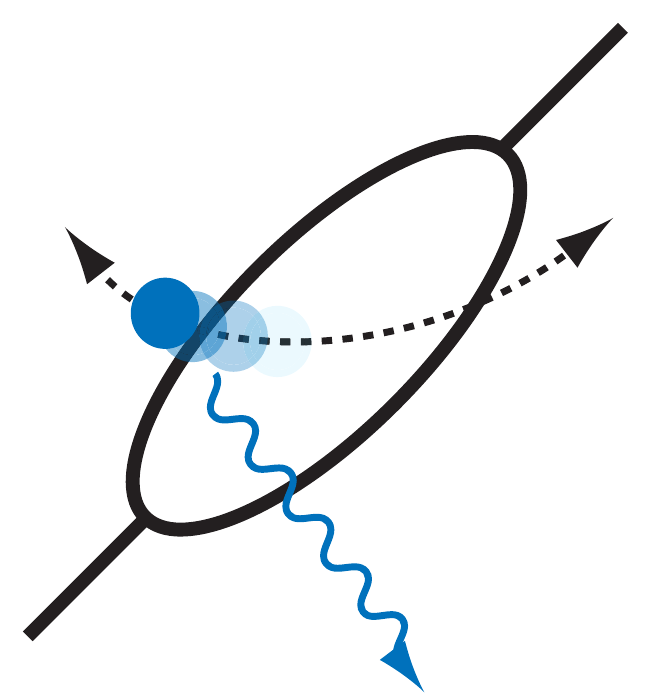}} &
\multicolumn{1}{c}{\includegraphics[width=2 in]{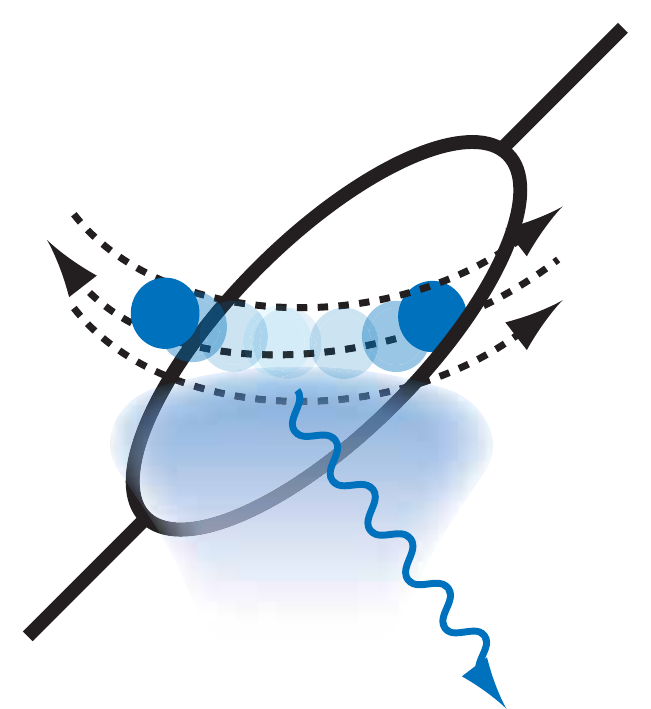}} \\
{\raggedright Spontaneous emission happens `instantaneously' with respect to the ion orbit in the trap.} & 
{\raggedright Spontaneous emission occurs over several trap oscillation cycles.} \\
{\raggedright No motional sidebands are resolved:  spectroscopy of the trapped particle is largely identical to that of a free particle.} &
{\raggedright Absorbed and emitted light is frequency modulated by the motion of the ion:  hence motional sidebands are observed separated by the trap secular frequencies $\nu$ with heights that scale with the Lamb-Dicke parameter $\eta^2 = \epsilon/\nu$.} \\  
{\raggedright The ion can be cooled to the Doppler limit $k_B T_\text{min} = \hbar \Gamma /2$.} &
{\raggedright The ion can be cooled to the $|n = 0\rangle$ motional state of the trap via absorptions of the red sideband: $k_B T_\text{min} = \hbar \nu /2$.}
\end{tabular}
\label{tab:coolingRegime}
\end{table}
How is the absorption spectra of a trapped particle different from that of a free particle?  We will find the answer by comparing the trap secular frequency and atomic state decay rates.  Typical trap secular frequencies obtained in the lab are $\nu/2\pi \sim 1$~MHz.  Common electric-dipole transitions (e.g.\ Doppler cooling transitions) have decay rates of $\Gamma \gg 1$~MHz while electric-dipole forbidden transitions (e.g.\ intercombination lines or ``clock'' transitions) are much slower, sometimes with $\Gamma \ll 1$~Hz.  We therefore often encounter practical examples of two limiting and very different behaviors.

As illustrated in Table~\ref{tab:coolingRegime}, when $\Gamma / \nu \gg 1$, the rate for spontaneous emission is fast compared to the time necessary to complete one oscillation in the trap.  Therefore, the processes of absorption and emission are little different for trapped particles in this regime than for free particles.  Given a sufficiently strongly trapped and cooled atom with a driving interaction below saturation, the spectroscopic linewidth of transitions with $\Gamma \gg \nu$ assume a Lorentzian profile with an unbroadened, natural linewidth.  The condition is sometimes known as the ``heavy-particle'' limit since more massive particles have slower oscillation frequencies and in some sense more `inertia' against the influence of the trapping field.

\begin{figure}
\centering
\includegraphics{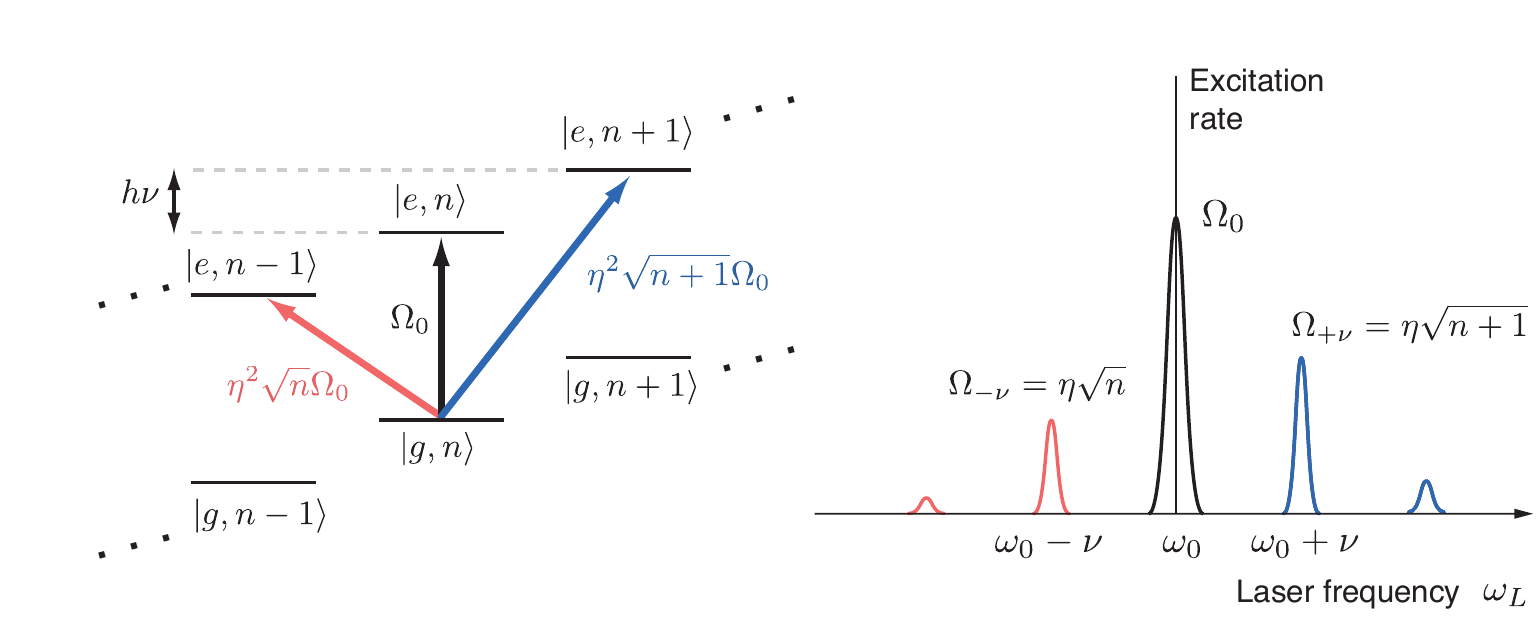}
\caption[Coupling diagram of motional sidebands in the Lamb-Dicke limit]{When the trap secular frequencies are much larger than transition linewidths $\nu \gg \Gamma$, one can spectroscopically resolve motional sidebands.  These sidebands represent absorptions and emissions that change the ion's motional state in addition to the internal energy state, visualized here as a ladder of states.  The so called \emph{blue sideband} is resolved at $\omega_0 + \nu$ and corresponds to adding one quantum of motion;  the \emph{red sideband} at $\omega_0 - \nu$ then corresponds to a removal of one motional quantum.  Their strengths are suppressed by the Lamb-Dicke factor $\eta^2 = \epsilon/\nu$ which is consistent with the reasoning that in the Lamb-Dicke limit ($\eta^2 \ll 1$) it is unlikely that atomic recoils due to spontaneous emission change the motional quantum number.  Often one refers to the atomic states being \emph{dressed} by the external degrees of freedom.}
\label{fig:lambDickeFluorescenceSpectrum}
\end{figure}

The case for narrow transitions, $\Gamma / \nu \ll 1$, is radically different;  the timescale for absorption and emission allows for many oscillations in the trap.  From the trapped atom's perspective, an interrogating laser beam or emitted photon seems frequency modulated at the oscillation frequency $\nu$.  In fact, the spectroscopic profile bears this intuition out:  one observes \emph{sidebands} at multiples of the trap secular frequencies $\nu$ with heights proportional to the extent of the ion's motion (which properly sets the modulation depth from the atom's perspective).  This regime is dubbed the ``fast particle'' limit, or in more recent discussions the ``resolved sideband'' limit.  See Section~\ref{sec:phaseModulation} for a discussion of the terminology and results of phase modulation that apply to these spectroscopic sidebands.

As shown in Figure~\ref{fig:lambDickeFluorescenceSpectrum}, the relative strengths of these sidebands are down with respect to the carrier transition by the Lamb-Dicke parameter $\eta = (\epsilon/\nu)^{1/2}$ and factors of $\sqrt{n}$ for red-sidebands and $\sqrt{n+1}$ for blue-sidebands where $n$ is the trap vibrational excitation quantum number.

\subsection{A quantum Hamiltonian}
To justify the assertions made in Sections~\ref{sec:threeRatesLambDicke}--\ref{sec:coolingRegime}, we will briefly examine a quantized Hamiltonian that treat a trapped ion immersed in a (classical) laser field.  This section is almost exclusively reproduced from~\cite{roos2000thesis,leibfried2003qds}.  First we write the total Hamiltonian in two parts:
\begin{equation}
H = H_\text{ion} + H_\text{interaction}.
\end{equation}
The $H_\text{ion}$ piece accounts for the ion's potential and kinetic energy in the trap potential, and the potential energy stored in a two-state internal degree of freedom.  The interaction term $H_\text{interaction}$ accounts for the addition or subtraction of ion motional and internal energy coupled with the disappearance or appearance of one or more (near) resonant photons at frequency $\omega_L = 1/k$.  We can write
\begin{align}
H_\text{ion} &= \underbrace{\frac{p^2}{2m} + \frac{1}{2} m \nu^2 x^2}_\text{motional} + \underbrace{\frac{1}{2} \hbar \omega_0 \sigma_z}_\text{internal}, \\
H_\text{interaction} &=  \frac{1}{2} \hbar \Omega (\sigma_+ + \sigma_-) \left( e^{i(kx - \omega_Lt + \phi)} + e^{-i(kx-\omega_Lt + \phi)}  \right),
\end{align}
where $\nu$ is the trap harmonic oscillator (secular) frequency (written in angular frequency units for this section only), $\sigma_z$, $\sigma_+$, and $\sigma_-$ are standard Pauli spin matrices acting on the internal $|g,e\rangle$ states of the ion, $\Omega$ is the Rabi frequency or strength of the laser field, defined in Eq.~\ref{eq:e1Rabi} for electric dipole coupling, Eq.~\ref{eq:e2Rabi} for electric quadrupole coupling.

Since the ion is harmonically trapped, the dynamical variables $x$ and $p$ can be described by the standard raising and lowering operators acting on excitation number states,
\begin{align}
a^\dag &= \sqrt{\frac{m \nu}{2 \hbar}}x + \frac{i}{\sqrt{2m \hbar \nu}} p, \\
a &= \sqrt{\frac{m \nu}{2 \hbar}}x - \frac{i}{\sqrt{2m \hbar \nu}} p, \\
a | n \rangle &= \sqrt{n}|n-1\rangle, \qquad a^\dag |n \rangle = (n+1)|n+1 \rangle.
\end{align}  
The Hamiltonian pieces are then rewritten as
\begin{align}
H_\text{ion} &=  \hbar \nu (a^\dag a + \tfrac {1}{2}) + \tfrac {1}{2} \hbar \omega_0 \sigma_z \\
H_\text{interaction} &= \frac{1}{2} \hbar \Omega \left( e^{i \eta(a + a^\dag)} \sigma_+ e^{-i \omega_L t} + e^{-i \eta(a + a^\dag)} \sigma_- e^{i \omega_L t} \right)
\end{align}
where we have again defined the Lamb-Dicke parameter as
\begin{equation*}
\eta = k \sqrt{ \frac{ \hbar}{2 m \nu}},
\end{equation*}
and a rotating wave approximation has been made~\cite{roos2000thesis,leibfried2003qds}.  Moving to an interaction picture by transforming the interaction Hamiltonian,
\begin{align}
U &= e^{i H_\text{ion} t /\hbar}, \\
H_I &= U^\dag H_\text{interaction} U \\
&= \frac{1}{2} \hbar \Omega \left( e^{i \eta(\hat{a} + \hat{a}^\dag)}\sigma^+ e^{-i \delta t} + e^{-i \eta (\hat{a} + \hat{a}^\dag)} \sigma^- e^{-i \delta t} \right),
\end{align}
where $\delta = \omega_L - \omega_0$ is the laser detuning and the raising/lowering operators have been transformed according to $\hat{a} = a e^{i \nu t}$.  Depending on the detuning $\delta$, an ion in the ground (internal) state and some harmonic oscillator motional state $|g, n\rangle$ can be coupled by this Hamiltonian to excited states in any harmonic oscillator state $|e, n' \rangle$;  one often hears that the atomic states are \emph{dressed} by the motional states.  Specifically, detunings that are multiples of the trap secular frequency $\delta \approx m \nu$ tend to couple $|g, n \rangle$ with $|e, n+m \rangle$.  In the limit of low laser power, $\Omega \ll \nu$, \emph{only} one such coupling is relevant, the others are sufficiently off-resonant to be ignored to first order.  

One finds that probability between the states flops with a Rabi frequency defined as
\begin{equation}
\Omega_{n+m,n} = \Omega \left\langle n + m \left| e^{i \eta(\hat{a} + \hat{a}^{\dag})} \right| n \right\rangle.
\end{equation}
In the Lamb-Dicke regime, $\eta (2n+1) \ll 1$, so the matrix element can be expanded,
\begin{equation}
\Omega_{n+m,n} = \Omega \left\langle n + m \left| 1 + i \eta(\hat{a}^{\dag}) + \hat{a} + O(\eta^2) \right| n \right\rangle.
\end{equation}
The most important interactions take place for $m = 0, \pm 1$:  spectroscopic features known as the carrier, and blue and red sidebands that correspond to the picture in Figure~\ref{fig:lambDickeFluorescenceSpectrum}.  Under the laser detunings of interest, the interaction Hamiltonians and effective Rabi frequencies are:
\begin{center} \footnotesize
\begin{tabular}{p{1in}ccc}
Feature				& Red sideband	&  Carrier		& Blue sideband \\ \hline \hline
Detuning $\delta$	& $-\nu$	& $0$		& $+ \nu$ \\
Interaction Hamiltonian $H_I$	& $\tfrac{1}{2}i \hbar \Omega_{n-1,n} (\hat{a} \sigma_+ - \hat{a}^\dag \sigma_-)$ & $\tfrac{1}{2} \hbar \Omega_{n,n}(\sigma_+ + \sigma_-)$ & $\tfrac{1}{2}i \hbar \Omega_{n+1,n} (\hat{a}^{\dag} \sigma_+ - \hat{a} \sigma_-) $ \\
Rabi frequency			& $\Omega_{n-1,n} = \eta \sqrt{n} \Omega$ & $\Omega_{n,n} = \Omega(1 - \eta^2 n) \approx \Omega$ &  $\Omega_{n+1,n} = \eta \sqrt{n+1} \Omega$
\end{tabular}
\vspace{0.25 in}
\end{center}

The form of the interaction on the red sideband
\begin{equation*}
H_I = \tfrac{1}{2}i \hbar \Omega_{n-1,n} (\hat{a} \sigma_+ - \hat{a}^\dag \sigma_-)
\end{equation*}
is the same as the Jaynes-Cummings Hamiltonian~\cite{vogel1995njc}, an important coupling in quantum optics, Cavity-QED, and many other theoretical systems.

\subsection{Application:  cooling to the ground state of motion}
\begin{figure}
\centering
\includegraphics{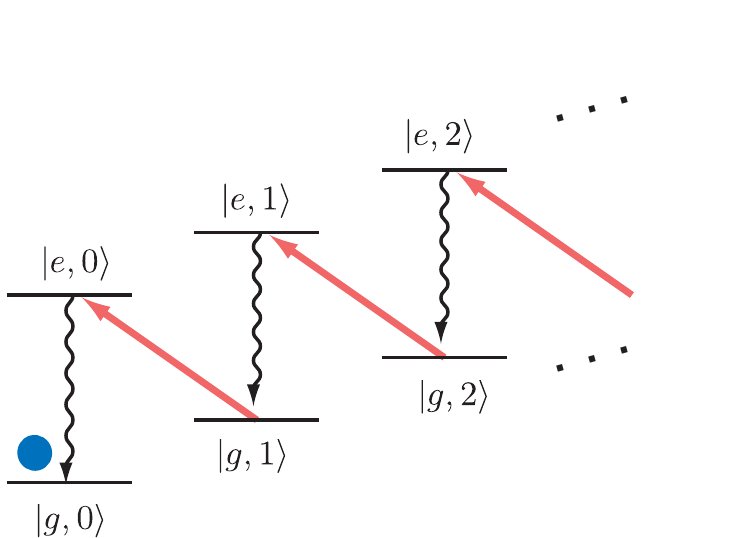}
\caption[Scheme for cooling to the quantum ground state]{Cooling to the quantum ground state of motion is possible~\cite{wineland1987lcl} by exciting the red sideband transition at $\omega_0 - \nu$ (see Figure~\ref{fig:lambDickeFluorescenceSpectrum}) in the Lamb-Dicke confinement regime.  In a sense we perform optical pumping in the dressed atomic states by only exiting the ion to an excited electronic state when doing so also subtracts a quantum of vibrational motion from the system.  Since decays conserving vibrational number $|n \rangle$, represented here as wavy lines, are most likely, one can efficiently pump the atom to occupy $\langle n \rangle \approx 0$~\cite{diedrich1989lcz}.}
\label{fig:groundStateCooling}
\end{figure}

Recall from Section~\ref{sec:semiclassicalLaserCooling} that Doppler cooling limits us to a temperature proportional to the natural linewidth $\Gamma$ of the cooling transition.  Typically this means about 1~mK, leaving our ion with an average vibrational occupation number of $\langle n \rangle \sim 20$ (see Eq.~\ref{eq:expectedNwithTemperature}).  While this is often sufficient to make first-order Doppler shifts negligible, significant micromotion still plagues the ion's motion.  The ultimate spectroscopic sensitivity, long-lived coherence, and robustness against systematic effects on a trapped ion requires cooling to the ground state of motion, $n = 0$.  Fortunately, one can achieve this in the resolved-sideband limit ($\Gamma / \nu \ll 1$) using a technique known as sideband cooling~\cite{diedrich1989lcz}.

Consider that for a narrow transition in this limit, the ground and excited internal states of the atom, $\ket{g}$ and $\ket{e}$, are dressed by the motional degrees of freedom, shown in Figure~\ref{fig:lambDickeFluorescenceSpectrum}.  It is now possible to resolve and independently drive transitions such as $\ket{g,n} \to \ket{e,n}$ at $\omega_0$ as well as $\ket{g,n} \to \ket{e,n \pm 1}$ at $\omega_0 \pm \nu$.  Likewise, spontaneous decays are possible that change the vibrational quantum number.  

What shall we make of the transition $\ket{g,n} \to \ket{e,n-1}$?  A photon with energy $\hbar (\omega_0 - \nu)$ has vanished, but it appears that the atomic system has gained $\hbar \omega_0$ in internal energy.  Where did the additional energy $\hbar \nu$ come from?  The ion has gained internal electronic energy but lost vibrational energy. Since a subsequent decay $\ket{e,n-1} \to \ket{g,n-1}$ is the most likely outcome, we can take advantage of this difference to cool the ion all the way to ground state by detuning our laser to $\omega_0 - \nu$ and driving only this \emph{red-sideband} transition as shown in Figure~\ref{fig:groundStateCooling}.  This allows one to reach a much colder temperature than was available with doppler cooling:
\begin{align*}
T_\text{min,doppler} &= \frac{\hbar \Gamma}{2 k_B}, \\
T_\text{min,sideband} &= \frac{\hbar \nu}{2 k_B}.
\end{align*}
Though the forms of these expressions for minimum temperatures look similar, the physical mechanisms behind them are quite different.  In Doppler cooling, the temperature limit results from equilibrium between the cooling that occurs due to differential absorption for atomic velocities directed against the laser beam and the random recoil that happens after spontaneous emission.  In sideband cooling, the limiting temperature is due to the zero-point temperature of the quantum harmonic potential well that confines the ion:  $E_0 = \tfrac{1}{2} h \nu$.  The spectroscopic red sideband disappears when $n=0$ (recall that $\Omega_{-v} = \eta \Omega_0 \sqrt{n}$), because there is no more vibrational energy to extract.

\subsection{Application:  single-ion temperature measurement}
Recall from Figure~\ref{fig:lambDickeFluorescenceSpectrum} that for a given laser intensity, the red and blue sidebands are driven more slowly than the carrier as their Rabi-frequencies are suppressed by the Lamb-Dicke factor.  Notice that the ratio of their effective Rabi-frequencies will be
\begin{equation}
\frac{\Omega_{+\nu}}{\Omega_{-\nu}} = \frac{\sqrt{n + 1}}{\sqrt{n}}.
\end{equation}
Therefore, for sufficiently small $n$, a ratio measurement of $\pi$-pulse times is a sensitive test of the average vibrational quantum number $\langle n \rangle$ of the ion.  If the measurement is made just after cooling for a long time on the red sideband, $n \to 0$, and one finds that the sideband \emph{vanishes}, ($\Omega_{-\nu} \to 0$).  If the measurement is made a controlled time after ending the sideband cooling routine, one can detect the ion heating rate in units of quanta/second (see~\cite{deslauriers2004zpc,devoe2002esa} for example).

\subsection{Application:  gates for quantum information processing}
Undoubtedly one of the chief reasons for interest in trapped ion systems is their utility in the storage and manipulation of quantum information.  One scheme suggested by Cirac and Zoller~\cite{cirac1995qcc} identifies long lived internal states (such as hyperfine spins or electronic metastable states) in a linear string of trapped ions as coherent storage mechanism of quantum information (dubbed \emph{qubits}) and the collective modes of oscillation shared by all the trapped ions as an bus for entanglement and execution of universal quantum `logic gates.'

Though a full explication is beyond the scope of this work, \cite{schmidtkaler2004qas} is an excellent demonstration of the realization of a quantum logic gate using sideband spectroscopy.
\chapter{Apparatus}\label{sec:ApparatusChapter}
\begin{quotation}
\noindent\small Once the last data is taken, a well designed experiment will utterly collapse into dust:  the duct-tape comes off, screws rattle loose, the vacuum pump burns out.  I say well designed because it means you didn't spend one moment or dollar over-engineering things.  That said, be damned sure it doesn't fall apart one minute too soon. \\ \flushright{---John Keto}
\end{quotation}

\section{Overview}
Briefly, our apparatus can be considered as three functional units:
\begin{enumerate}
\item A system to create, trap, and localize singly-ionized barium in vacuum.
\item Several light sources and laser systems resonant with transitions in Ba$^+$ and equipment to monitor and control the frequency and spectra of these lasers.
\item A computer to control the light incident on the ion and record the state of its fluorescence.
\end{enumerate}

%\begin{figure}
%\centering
%\includegraphics{figures/apparatusDiagram.pdf}
%\caption{A simplified schematic of the barium ion apparatus configured for $^{137}$Ba$^+$ clock experiments}
%\label{fig:apparatusOverview}
%\end{figure}

\section{The ion trap apparatus}
\begin{figure}
\centering
\includegraphics{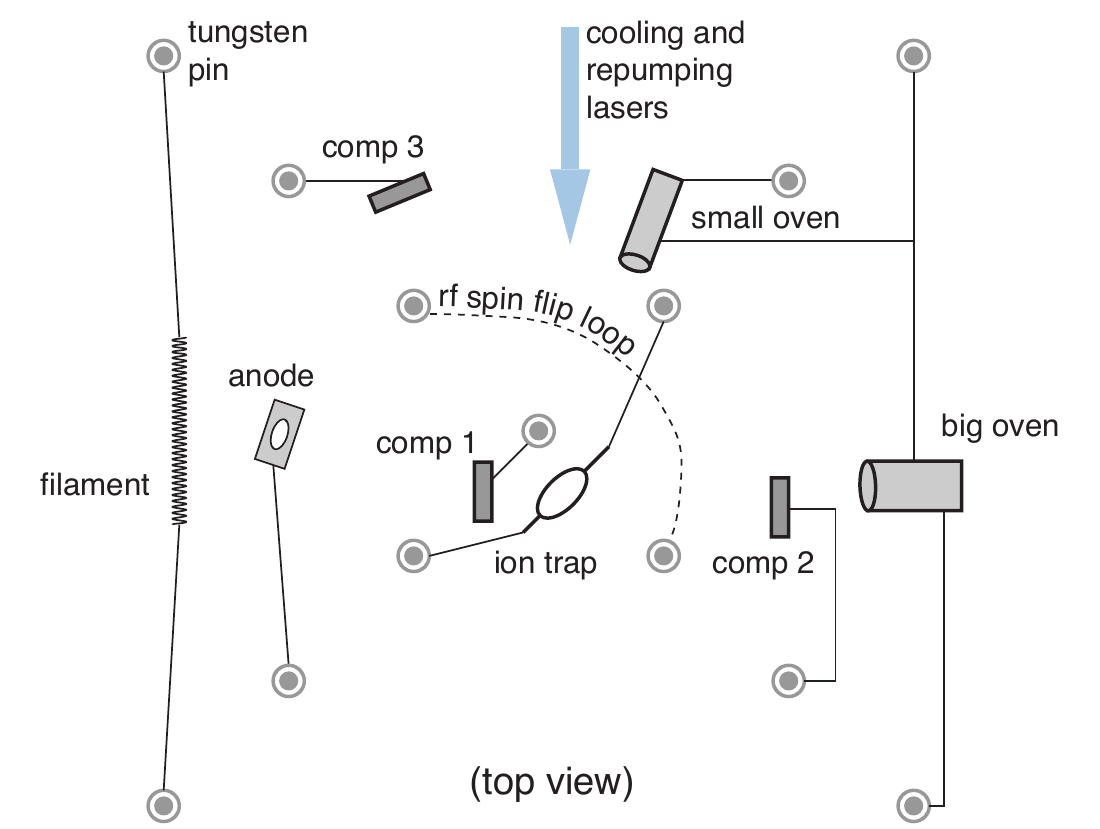}
\caption[Electrical pin-out of the ion trap vacuum header]{Electrical pin-out of the ion trap vacuum header:  birds' eye view, not to scale.  The `X' pattern of circles represent the tungsten vacuum feed-through pins. ``Comp'' is short for for dc compensation electrode.}
\label{fig:ionTrapPinout}
%\end{figure}
%\begin{figure}
%\centering
\includegraphics{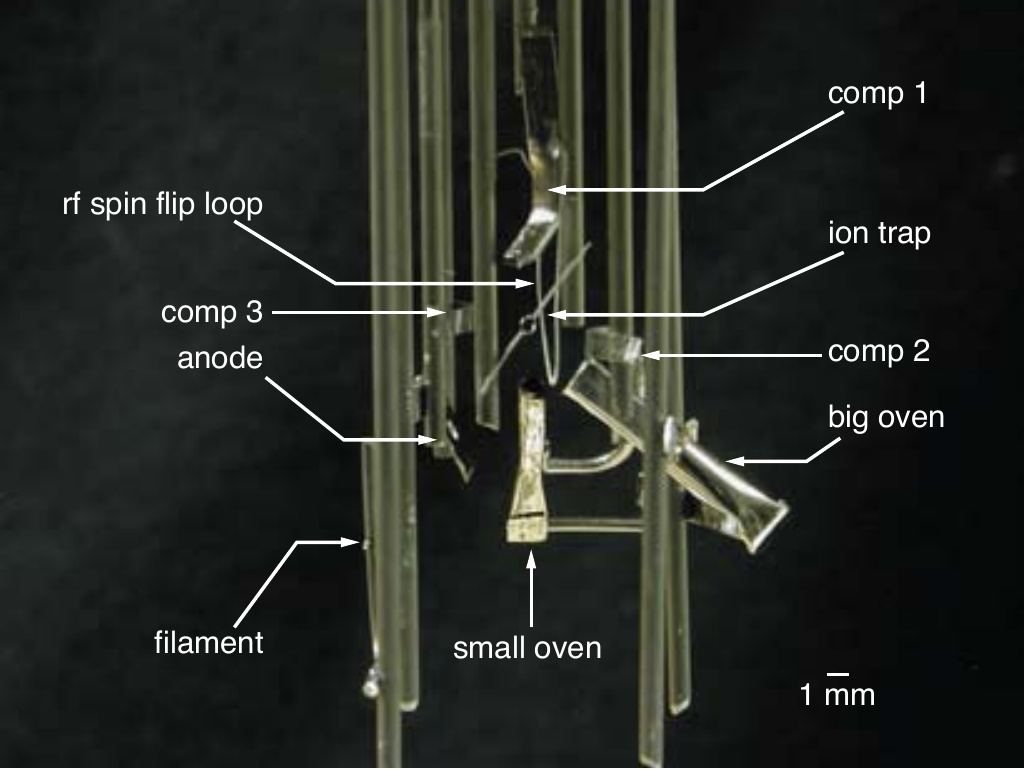}
\caption[Annotated photograph of the ion trap apparatus]{Annotated photograph of the ion trap apparatus.  In this view, blue and red cooling laser beams are oriented such that they would emerge from the page.}
\label{fig:ionTrapPhoto}
\end{figure}
\subsection{A twisted wire Paul-Straubel ring trap}
Of all the designs shown in Figure~\ref{fig:ionTrapEvolution}, perhaps the simplest ion trap one can construct is the twisted-wire Paul-Straubel ring trap.  One only has to twist two wires around a drill bit chosen to match the desired trap dimension $r_0$, trim the excess twisted wire, and spot weld them to vacuum feed-throughs.  In 2003, we rebuilt our trap apparatus completely using non-magnetic materials: see the electrical schematic in Figure~\ref{fig:ionTrapPinout}, and annotated photograph in Figure~\ref{fig:ionTrapPhoto}.  The trap ring (0.75~mm in diameter) is made from tantalum wire (0.2~mm thick).  Barium ovens are spot-welded rolls of 0.001 inch thick tantalum foil.  The electron source is a tungsten filament.  Micromotion compensation electrodes and the electron source anode are tantalum foil or flattened tantalum wire. Thirteen vacuum feed-throughs are made from tungsten rods cut to size with an electrical discharge machining (EDM) apparatus and embedded in a glass/metal seal by our department glass blower, Bob Morely.

\begin{figure}
\centering
\includegraphics{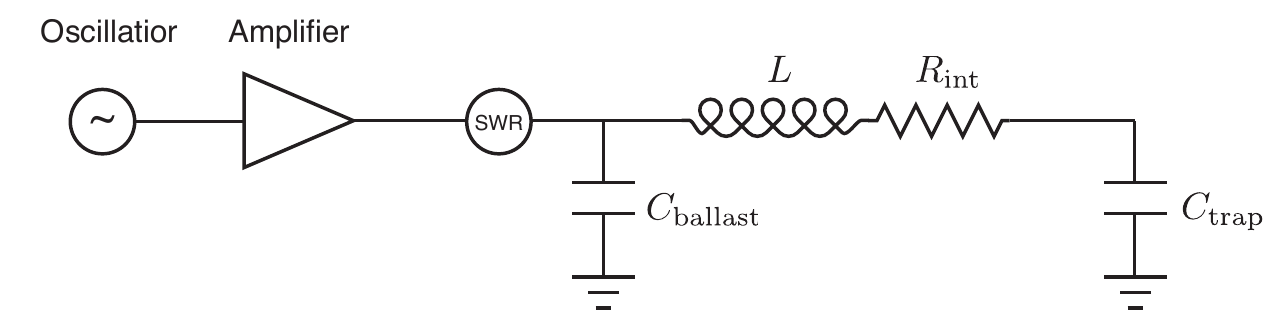}
\caption[Electrical schematic of the rf drive circuit powering the trap]{Electrical schematic of the rf drive circuit powering the trap.  The resonant circuit following the amplifier, sometimes called a $\pi$-network, matches the reactive impedance of the ion trap to the 50~$\Omega$ rf amplifier and increases the voltage seen on the trap.  The SWR (standing-wave ratio) meter is a passive device which measures the transmitted rf power as well as the fraction being reflected due to load mismatch.  See text for more detail.}
\label{fig:trapElectricalDrive}
\includegraphics{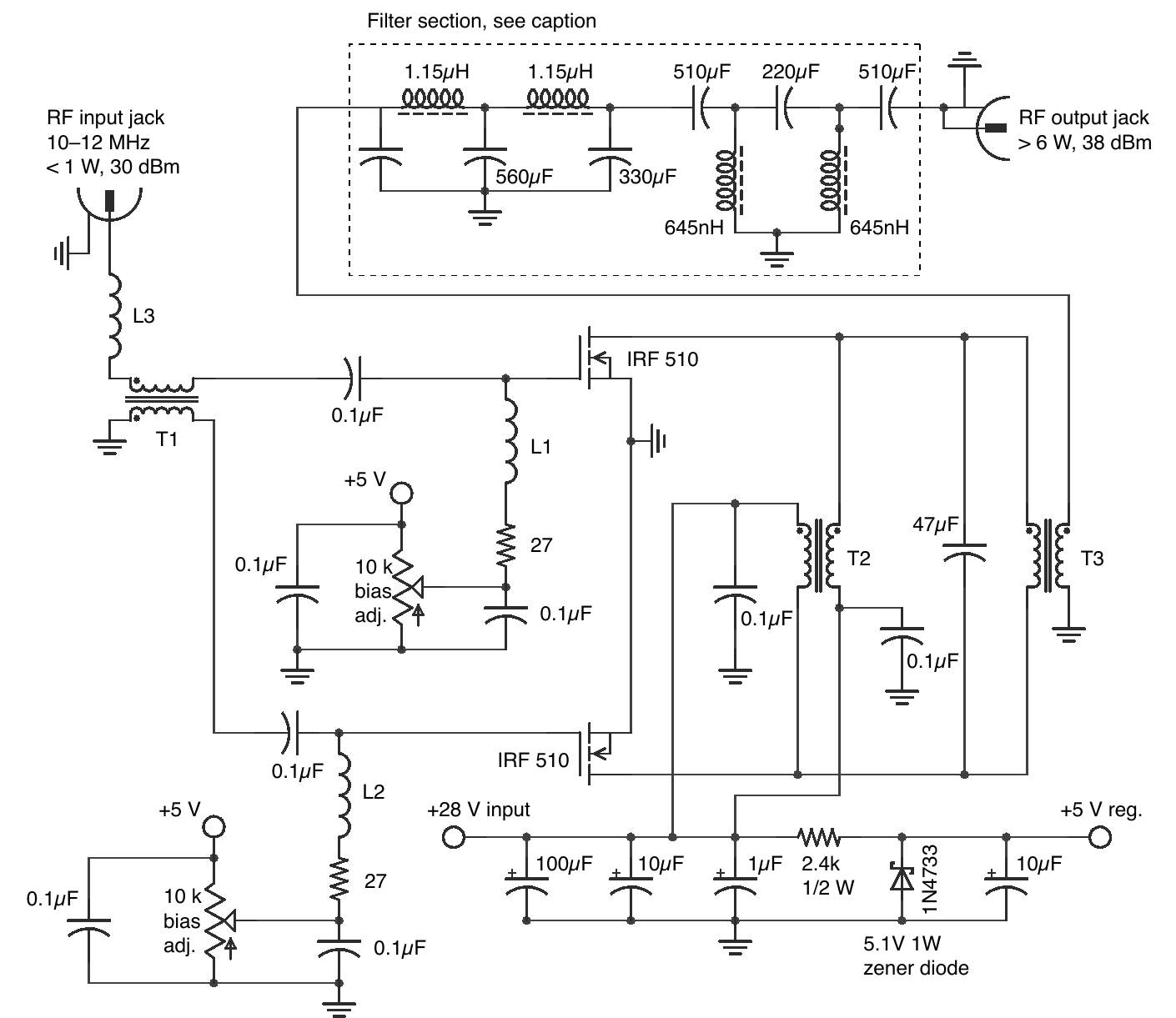}
\caption[Electrical schematic of a trap rf amplifier]{We adapted an amateur HF push-pull radio amplifier design~\cite{straw2005ARRL} to create several Watts of clean rf at trap frequencies of 10--12~MHz.  These amplifiers (or at least models we constructed) produce enough low frequency noise to quickly eject trapped ions through coupling with ion secular oscillations.  Therefore, we employ an aggressive band-pass filter, shown in the boxed section of the schematic.  See~\cite{straw2005ARRL} for inductor and transformer winding instructions.}
\label{fig:trapAmpSchematic}
\end{figure}

One encounters two practical problems when attempting to power rf ion traps:  how does one generate hundreds of Volts of rf necessary to stably trap ions, and how does one impedance-match the trap to commercial (likely 50~$\Omega$) amplifiers (see Figure~\ref{fig:trapAmpSchematic}) and transmission lines?  Both problems are solved by a resonant ``$\pi$-network'' step-up circuit shown in Figure~\ref{fig:trapElectricalDrive}.  The inductor $L$ forms a resonant circuit with the trap, modeled as a capacitance $C_\text{trap}$ at an angular frequency~\cite{straw2005ARRL}
\begin{equation}
\omega_\text{rf} \approx \frac{1}{\sqrt{L C_\text{trap}}}.
\end{equation}
The voltage on the trap is increased by (roughly) the $Q$ of the resonant circuit which scales as the ratio of the circuit reactance to losses coming from the internal resistance $R_\text{int}$
\begin{equation}
Q \sim \frac{|i \omega_\text{rf} L|}{R_\text{int}} \sim 80.
\end{equation}
which is determined by measuring the circuit bandwidth $\Delta \omega$ about the resonance frequency $\omega_\text{rf}$ and using $Q = \omega/\Delta \omega$.  A capacitance $C_\text{ballast} \gg C_\text{trap}$ helps impedance match the circuit to the 50 $\Omega$ radio frequency amplifier while only shifting the resonance frequency slightly.  In practice, we attach a short, open length of coaxial cable in parallel to the circuit to serve as $C_\text{ballast}$ since each foot of the cable yields a few dozen pF.  This practice is sometimes called \emph{stub tuning}~\cite{hobbs2000beo,pozar1998me}.  We adjust the length of the cable to minimize the measured \emph{standing wave ratio}
\begin{equation}
\text{SWR} = \frac{1 + | \Gamma|}{1 - |\Gamma|}
\end{equation}
where $\Gamma$ is the reflection coefficient of the circuit, proportional to the mismatch in the circuit impedance $Z_L$ to the characteristic impedance $Z_0 = 50 \, \Omega$,
\begin{equation}
\Gamma = \frac{Z_L - Z_0}{Z_L + Z_0}.
\end{equation}
We normally operate with $\text{SWR} < 2.0$ which translates into about 90 \% of the applied rf power bring transmitted to the resonant circuit and trap.

\subsection{Barium ovens and electron source}
\begin{figure}
\centering
\includegraphics{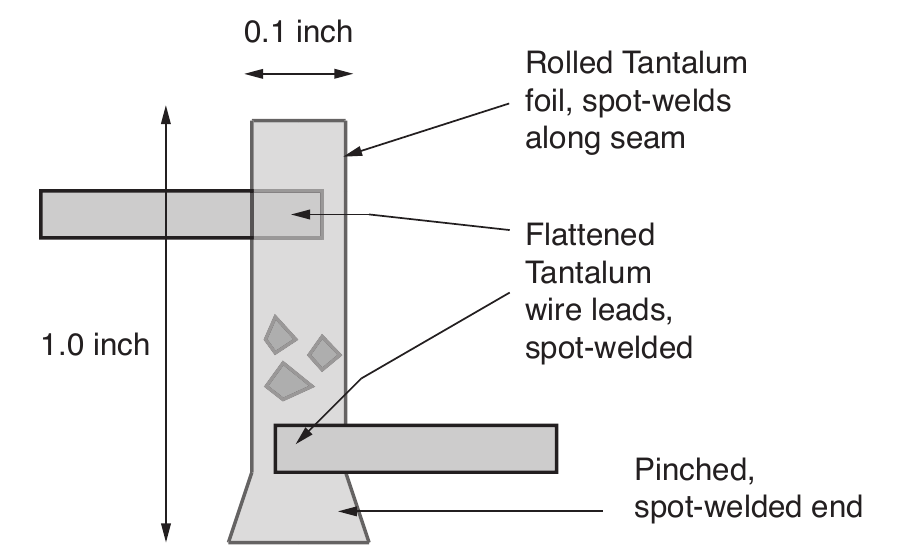}
\caption[Diagram of our barium ovens]{Diagram of our barium ovens, constructed by rolling and spot welding tantalum foil into tubes.  The tubes are crushed flat at the bottom and sealed by spot welding.  Tantalum wire electrodes, rolled flat, are spot welded to the tubes roughly 1/4 and 3/4 of the oven's length from the bottom.  Many identical ovens are made at one time, and several are tested in an evaporator apparatus before installation in the barium ion apparatus.  We achieve significant barium emission at about 4 A of oven current, as described in Figures~\ref{fig:filamentActivationTest} and~\ref{fig:ovenTestData}.}
\label{fig:bariumOvens}
\end{figure}

We abandoned using barium-aluminum alloy as a source of barium in 2003 due to evidence that either it or the previously used stainless-steel trap ring contained enough ferromagnetic impurities to corrupt our stable $\sim 1$~G magnetic field at the site of the ion at the $10^{-4}$ level.  Each time we heated the ovens to load an ion, we would find Zeeman transition frequencies shifted by several hundred Hz; these would continue to drift with hour-like time scales.  Switching to a non-magnetic trap and pure barium completely solved this problem.  As one might guess, the chief disadvantage of using pure barium is its fierce reactivity in air.  We have so far succeeded by purchasing, storing, and cutting the barium under a dry argon atmosphere and only exposing it to air for $< 10$~minutes while loading a few mg into each tantalum tube oven ($\sim$~3 minutes), bolting down the glass/metal vacuum feed-through header ($\sim$~5 minutes), performing final vacuum checks, arguing loudly about how white and therefore oxidized the unused barium chunks are beginning to look, and then pumping down to UHV ($\sim$~2 minutes).  Though the vapor pressure of barium at 500~$^\circ$C is $\sim 10^{-6}$~Torr~\cite{rudberg1935vpb}, we find that our ovens must be heated briefly to much hotter temperatures ($\sim$~800 $^\circ$C) before initially emitting barium.  We believe that some oxide layer is being `broken' by the procedure, but the facts are actually completely unknown.

\begin{figure}
\centering
\includegraphics{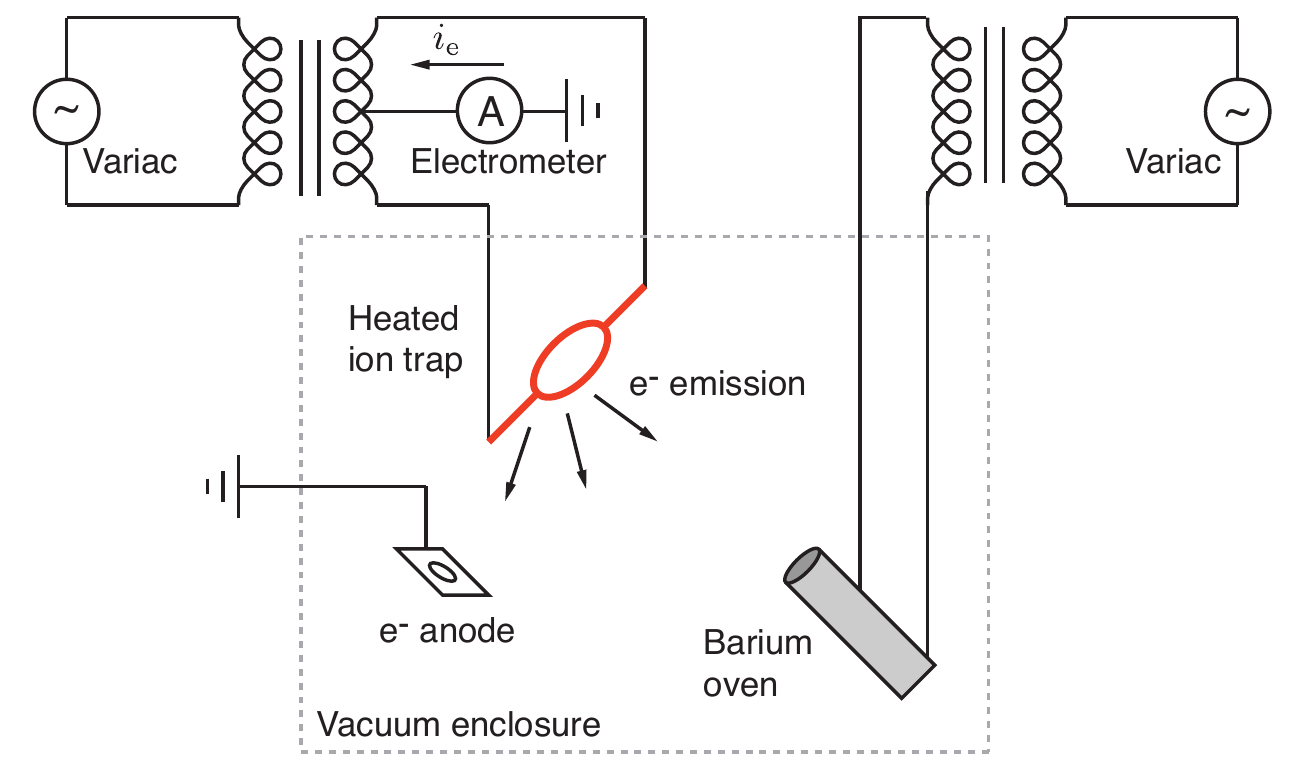}
\caption[Experimental schematic for confirming the ovens' barium emission]{The thermionic emission current of electrons, $i_e$, from a heated ion trap increases exponentially when the barium ovens are heated sufficiently to cause significant coverage of effused barium on the tantalum wire ion trap.  This allows us to set the barium oven temperature with confidence and confirms that un-oxidized barium is present in the oven.}
\label{fig:filamentActivationTest}
\vspace{0.5in}
\includegraphics{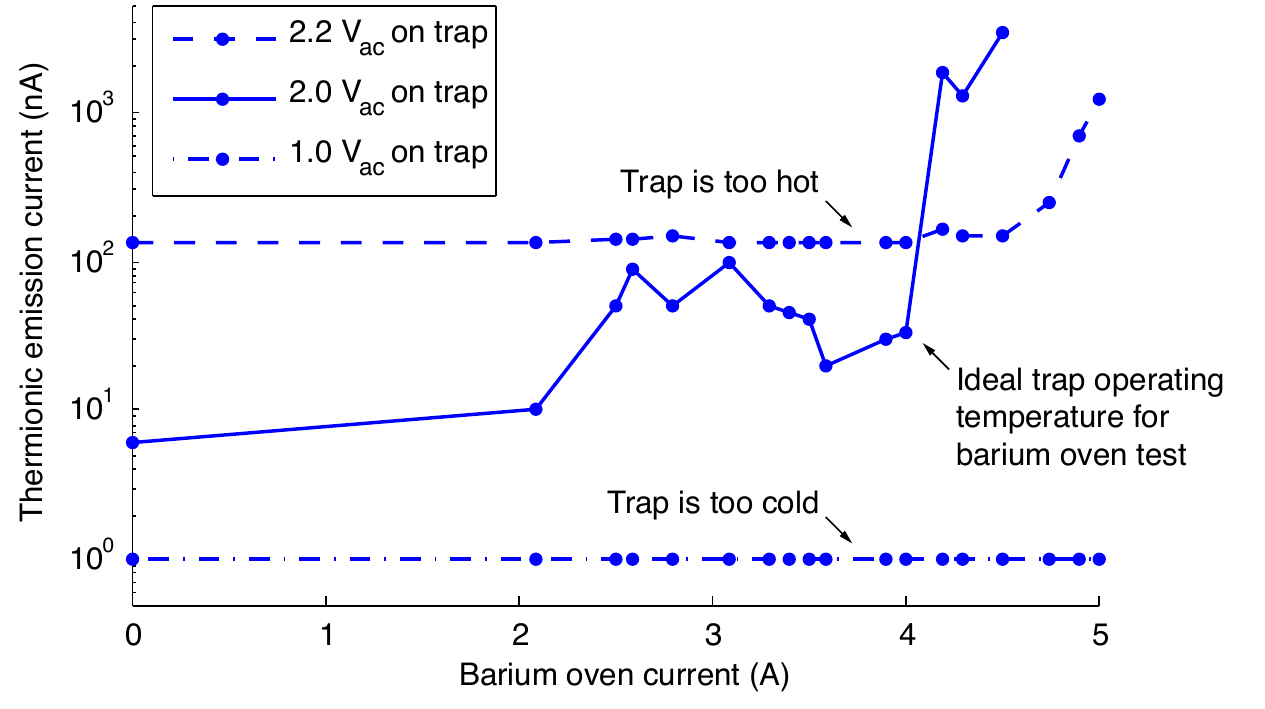}
\caption[Data confirming successful emission of barium from the ovens]{Here we plot data from the experimental test outlined in Figure~\ref{fig:filamentActivationTest}.  A critical increase in thermionic emission current is seen when the trap is heated to a dull orange glow (using 2.0~V of applied ac).  A less pronounced effect is present when the trap is heated to bright yellow (using 2.2~V of applied ac) and no effect is seen for a trap heated with far less power (1.0~V applied ac).}
\label{fig:ovenTestData}
\end{figure}

We \emph{do}, however, have an extremely robust and quantitative method for tuning up the oven temperature for optimal performance which we call the `filament activation' test.  Briefly, the work function of barium is so much lower than that of tantalum.  We turn our trap ring into a hot filament by passing ac current ($\sim$ 2.5 A) across it and look for greatly increased electron emission that results when the barium oven (see Figure~\ref{fig:bariumOvens}) is just hot enough to begin coating the trap surface with barium.  Figure \ref{fig:filamentActivationTest} shows the scheme:  the heating ac is passed through a transformer, while a precision electrometer (Keithley 6514) attached to the secondary center-tap measures electron emission to sub-nA sensitivity.  Typically, a red-hot trap ring will emit $\sim$~10 nA.  Current through the barium oven is slowly ramped up until it begins emitting barium.  The electron emission scales exponentially with the trap's metal workfunction $W$~\cite{melissinos1966emp},
\begin{equation*}
I_\text{emission} \propto T^2 e^{-eW/k_B T},
\end{equation*}
so when barium coats the tantalum trap, orders of magnitude more current is emitted.  Figure~\ref{fig:ovenTestData}) shows some results of this test.  The measured current jumps to $>$~1 $\mu$A just seconds after the ovens reach the critical temperature.  We begin trapping clouds of ions at this oven setting (about 4.5~A at this writing, though past ovens have required up to 6~A) and find that we must reduce it by 10~\% or so before we can stably trap single ions.  At the time of this writing, the larger oven is filled with enriched $^{137}$BaCO$_3$ and tantalum (Ta) powder while the smaller oven is filled with pure natural abundance barium metal (with some BaO).

Enriched barium is often sold in carbonate form since it is much easier to disassociate than BaO.  A recommended procedure~\cite{devoe2002esa} is a two stage loading scheme:  a large oven current heats the $^{137}$BaCO$_3$ and Ta mixture to a high temperature (1200 $^\circ$C) enabling a reduction reaction that frees Ba.  The emitted barium coats a plate which is then heated to $\sim 350^\circ$ in the second-stage of the recommended loading scheme.  For our initial work, we only implemented the first stage and expect to load large clouds of  $^{137}$Ba$^+$ from which a single ion can later be distilled.

We ionize barium atoms emitted from the ovens via collisions with electrons thermionically emitted from a hot tungsten filament.  About 2 A is passed over the filament, making it glow bright orange. We adjust the filament heating current so that the emission current is between 20~$\mu$A and 100~$\mu$A when the filament is biased by a -200~V source attached to the center tap of the secondary side of the transformer powering the filament.  Most of the accelerated electrons hit a grounded electrode near the filament which also blocks much of the filament glow from entering a photo-multiplier tube.  This electrode has a $\sim$ 1~mm hole punched in the center to allow some of the electrons to proceed to the trap in a beam.  We do believe that loading ions using electron bombardment increases ion micromotion and dirties the trap ring.  Periodically we clean the trap by disconnecting the rf and passing $\sim 3$~A of 60~Hz ac across it.  Afterwards, we find dramatic reduction in micromotion symptoms.

\subsection{rf spectroscopy loop}
Our trap apparatus features a half-loop of tantalum wire for driving rf spin flips in the single ion, shown in Figure~\ref{fig:ionTrapPhoto}.  The half-loop has a radius $R \approx 2.5$~mm, is situated about 1~mm from the trap loop, and is oriented so that the generated magnetic flux is roughly perpendicular to the main magnetic field.  Given oscillating current $I$, the magnetic field strength ought to be:
\begin{equation} \label{eq:rfLoopSpec}
B = \frac{\mu_0 I}{4 R} \approx 1 \cdot \frac{I}{1 [\text{mA}]} \text { mG}
\end{equation}

\subsection{Vacuum enclosure, magnetic shielding, magnetic coils}
\begin{figure}
\centering
\includegraphics{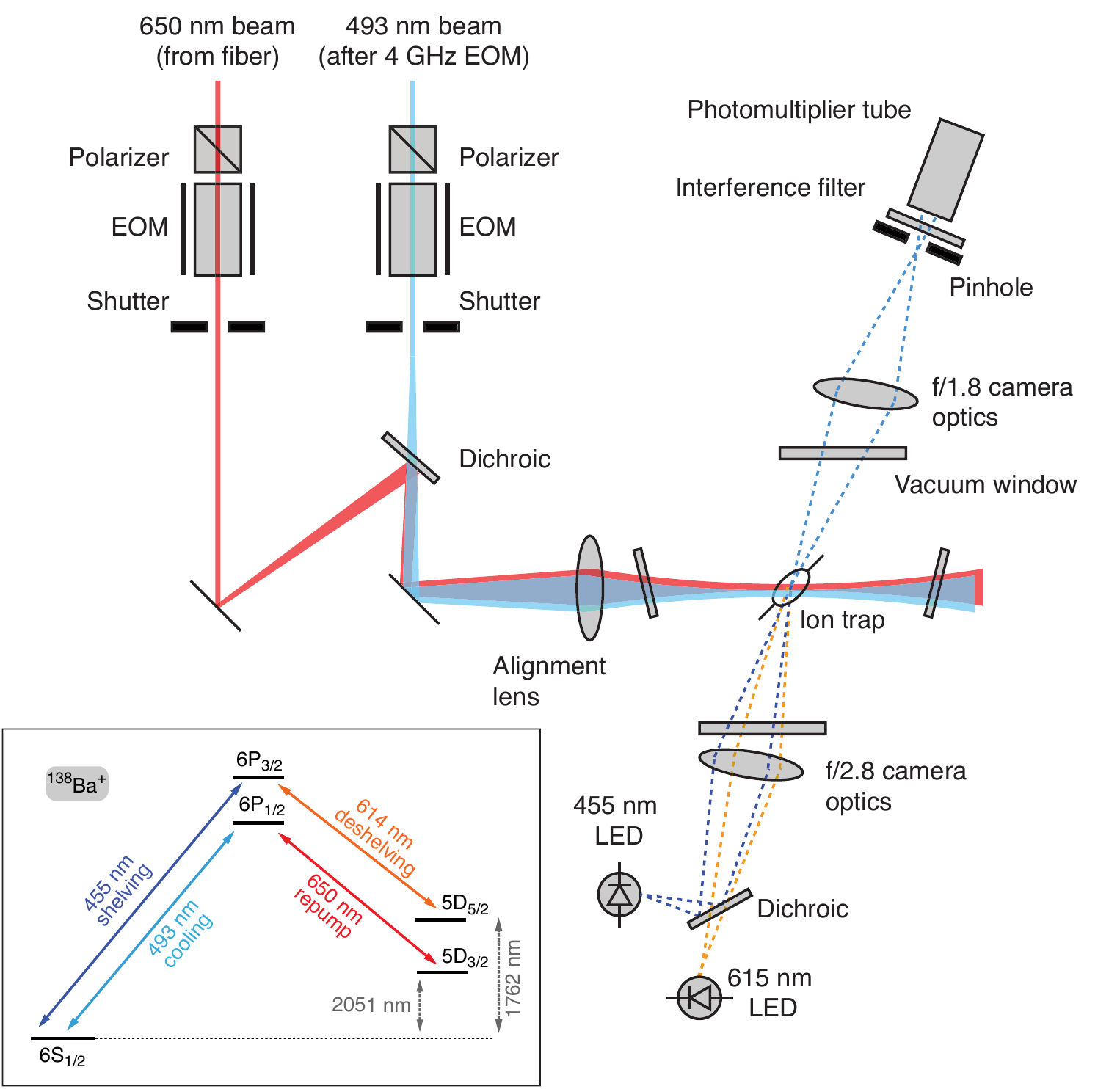}
\caption[Detailed schematic of optics near the ion trap, level diagram]{This detailed schematic (not to scale) of optics near the ion trap shows the blue cooling and red repumping laser being combined with a dichroic beamsplitter and focused into the center of the ion trap by a common alignment lens.  Independent lenses (not shown) in the blue and red beams allow the foci of each beam to occur at the ion trap.  The polarizer and EOM combination is configured so that the red and blue beam polarizations can be changed from linear to circular by switching between high voltages.  The diagram also depicts the ion fluorescence detection optics:  a camera lens followed by a 100~$\mu$m pinhole, 493~nm interference filter and photomultiplier tube.  Another dichroic beamsplitter combines the light from powerful orange and blue LEDs that are imaged onto the ion trap and provide 455~nm shelving and 615~nm deshelving radiation. Inset in the lower left is an energy level diagram of Ba$^+$ for reference.}
\label{fig:blueRedCombineOptics}
\end{figure}

%\begin{figure}
%\centering
%PLACEHOLDER FOR VACUUM SCHEMATIC
%\caption{Schematic of the ion trap vacuum enclosure and magnetic shielding}
%\label{fig:vacuumEnclosure}
%\end{figure}
A stainless steel block machined for six ConFlat attachments forms the core of the ultrahigh vacuum system.  In 2006, we redesigned the vacuum system somewhat, moving the ion pump further away from the trap;  for a detailed description of the apparatus during the light shift experiments, see~\cite{koerber2003thesis}.  Two angled window adapters form the cooling/repumping laser beam axis, and two flat windows located 3~cm away from the trap allow for photomultiplier tube viewing and access for the shelving and deshelving lamps (see Figure~\ref{fig:blueRedCombineOptics}).  The top port is used to secure the glass to metal vacuum feedthrough and ion trap apparatus.  We attached a 90$^\circ$ bend to the bottom port after which follows a cross containing an ion emission vacuum gauge, the copper vacuum pinch-off tube used during system pump-down and bake out, and a tube leading to a 20~l/s ion pump (Varian).

The system has been up to atmospheric pressure on only three occasions since 2001.  It is pumped down with an oil-free vacuum system.  First, two sorption pumps cooled with liquid nitrogen are opened to the system in series to reduce the pressure to 1~Torr and $\lesssim 1$~ mTorr, respectively.  Then, a 30 l/s ion pump is slowly opened to the system and it quickly becomes safe to turn on a vacuum ion gauge located close to this pump.  The system reaches $10^{-7}$~Torr within hours.  We degas trap wires, filaments, and the barium ovens by heating them with ac.  We then surround the system with clamshell ovens and bake it at 450~$^\circ$C under vacuum for about 7 days.  After cooling down to about 100~$^\circ$C, the copper umbilical tube connecting the system to the 30 l/s ion pump is pinched off, and the system's 20 l/s ion pump is switched on.  After various trap electrodes and the barium ovens are again degassed, the system reaches low $10^{-10}$~Torr within days.  Operating the barium ovens and the thermionic electron emission filament at optimal temperatures raises the pressure to mid $10^{-9}$~Torr, which falls back down within seconds of turning them off.

Two layers of box-shaped magnetic shielding surround main chamber of the vacuum enclosure and principle magnetic field coils.  When tested in isolation they shielded Earth's magnetic field by a factor of $\sim 2000$.  We have made no measurement of their shielding performance at oscillating frequencies such as 60~Hz.  The shields were last annealed by the manufacturer around 2001 and are regularly de-gaussed with coils providing a maximum of $\sim 70$ Ampere-turns of 60~Hz current.

Two magnetic field coils of 83 turns each are mounted 5~inches apart on vacuum flanges such that the central field is aligned with the intended blue/red laser beam path.  We pass about 1 A of current, generated by a home built precision current source with a relative day to day stability of $\sim 1 \times 10^{-5}$, through these coils in series to create a $\sim$1.5~G magnetic field at the site of the ion trap.  These coils likely saturate the first layer of magnetic shielding and reduce the overall shielding factor;  the current through the coils is kept constant and stable as much as possible, even during degaussing.

\section{Lasers and other bright lights}
\begin{figure}
\centering
\includegraphics{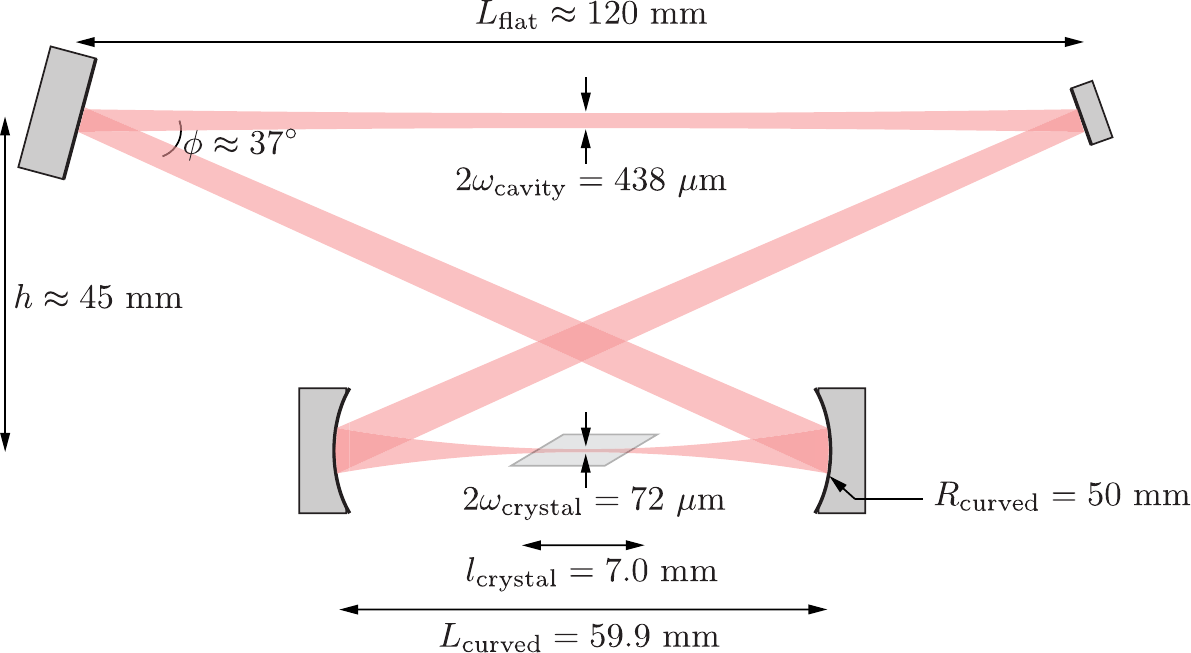}
\caption[Schematic diagram of the 986~nm doubling cavity]{This schematic diagram gives the important 986~nm doubling cavity physical dimensions.  The tilt of the (Brewster-cut) KNbO$_3$ crystal and resulting refraction of the cavity beam are not shown.}
\label{fig:blueCavitySchematic}
\includegraphics{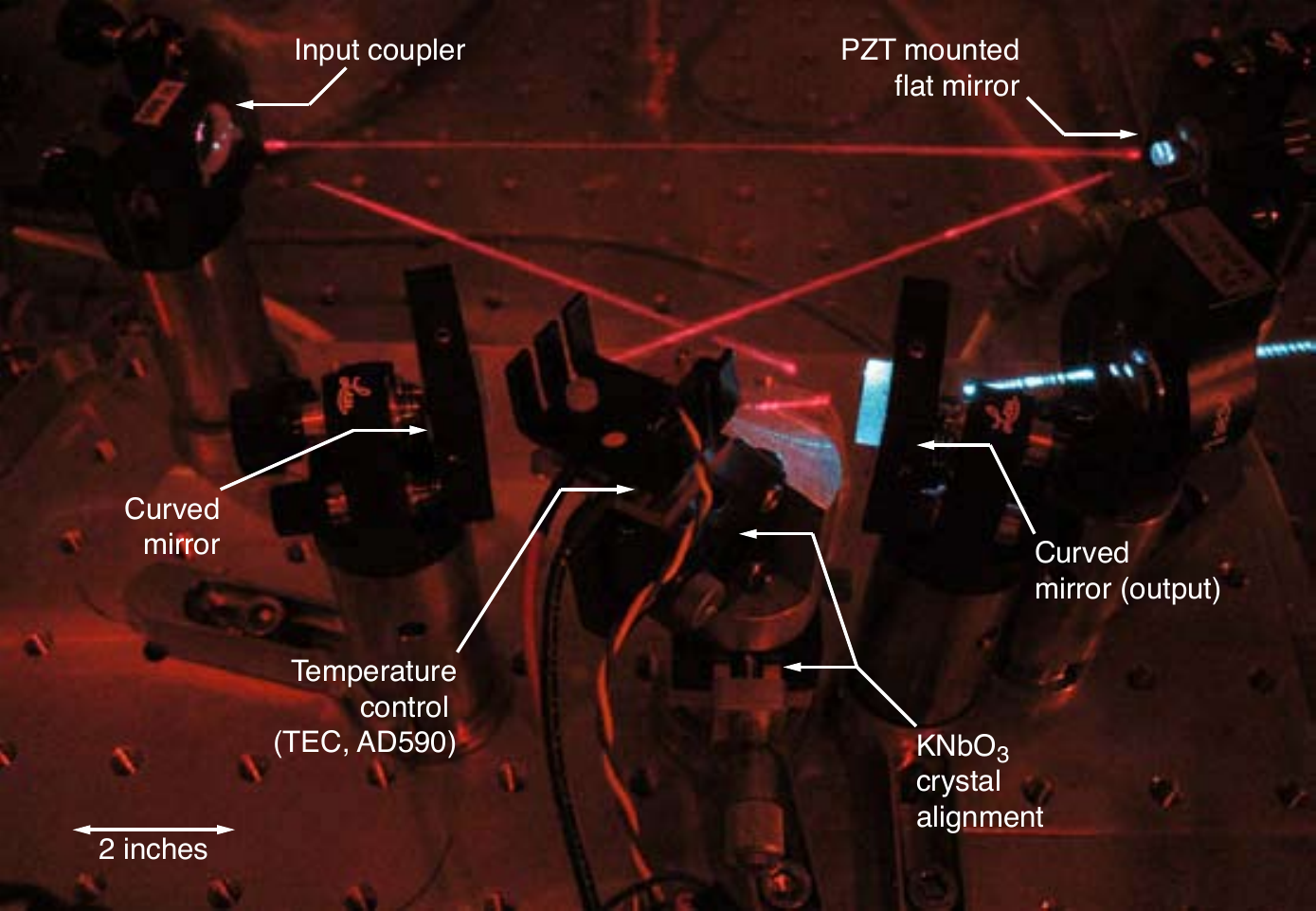}
\caption[Annotated photograph of the 986~nm doubling cavity.]{Annotated photograph of the 986~nm doubling cavity.  We acquired the photo with a 15 second digital camera exposure while sweeping the cavity PZT mounted mirror (to impede a saturating blue beam) and tracing the cavity and output beams with an IR fluorescent card.}
\label{fig:blueCavityPhoto}
\end{figure}

\subsection{The cooling laser:  493~nm}
\begin{figure}
\centering
\includegraphics{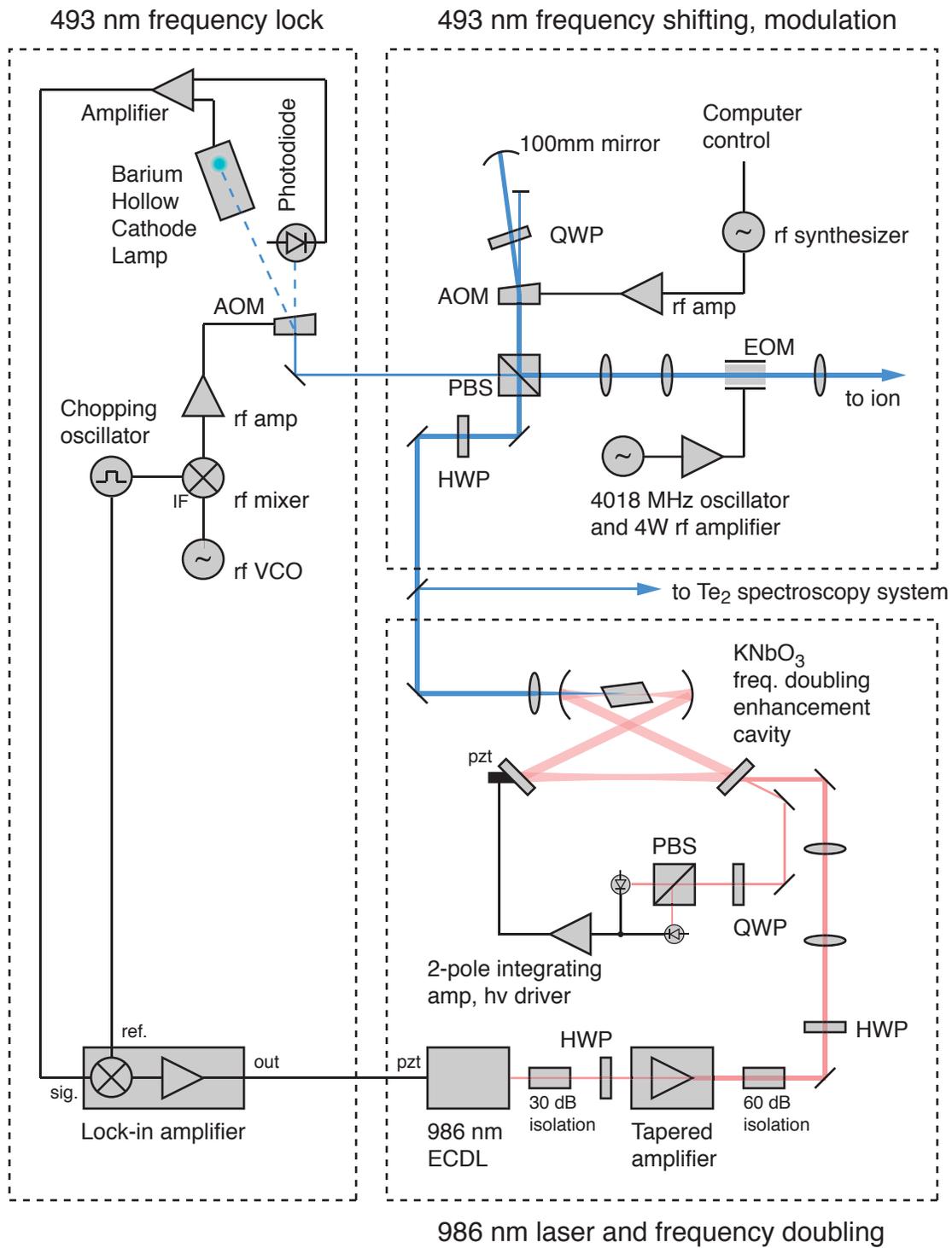}
\caption[Schematic of 493~nm laser system and frequency locking]{Schematic of 493~nm laser system and frequency locking.  See text for a full description.}
\label{fig:493nmLaserSchematic}
\end{figure}

\begin{figure}
\centering
\includegraphics{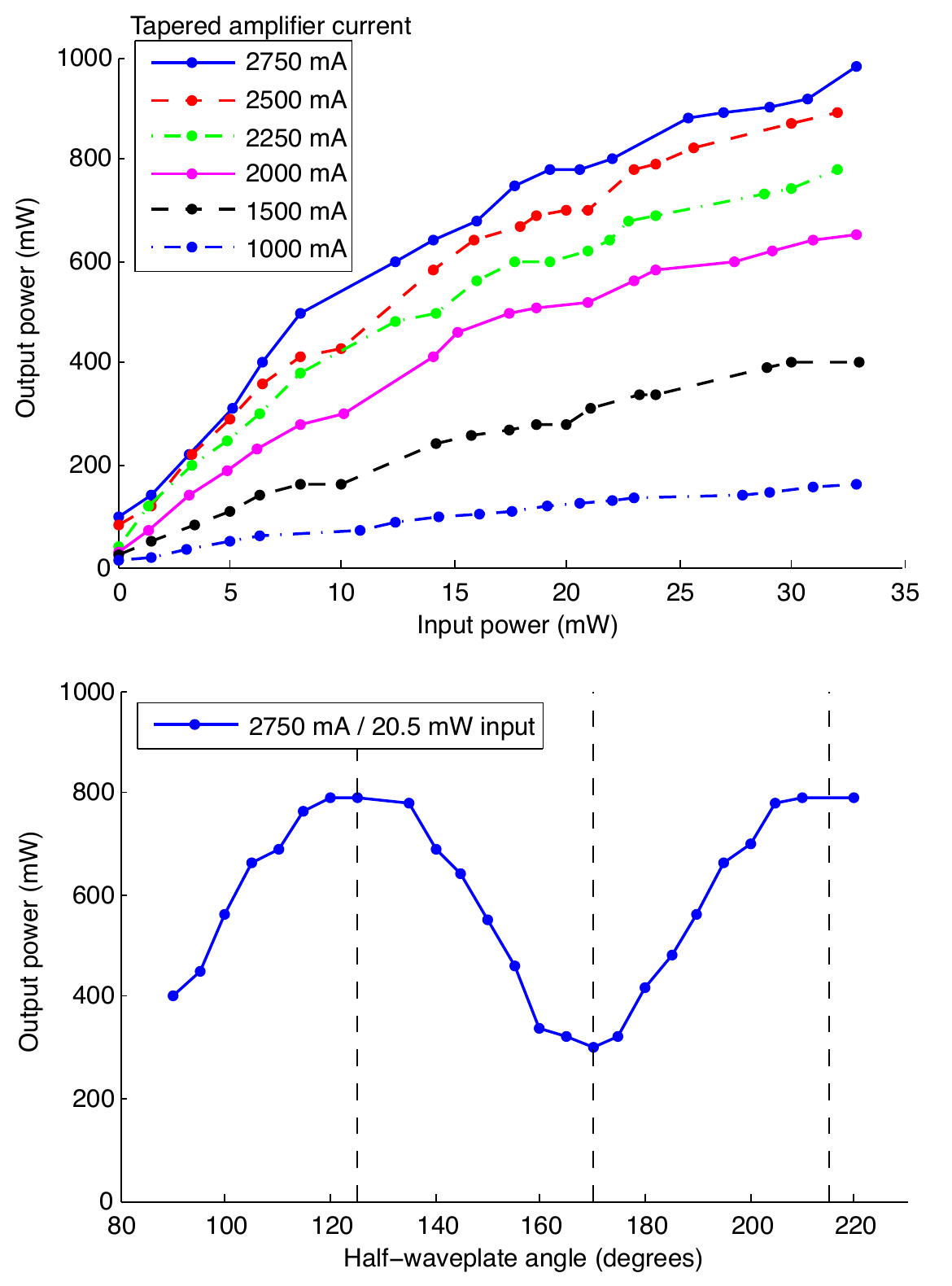}
\caption[Performance of the 986~nm tapered amplifier]{Performance of the 986~nm tapered amplifier.  The upper graph gives the output of the tapered amplifier as a function of input laser oscillator power under different amplifier current settings.  The lower graph shows how the tapered amplifier power varies as the polarization of the input beam is rotated by a half-wave plate.  The vertical dashed lines are spaced 45$^\circ$ apart.}
\label{fig:taperedAmpData}
\end{figure}

We use standard Doppler cooling methods on the barium ion $6S_{1/2} \leftrightarrow 6P_{1/2}$ transition at 493~nm.  This radiation is generated using the system shown in Figures~\ref{fig:blueCavitySchematic} and~\ref{fig:blueCavityPhoto}.  We begin with a commercial external cavity diode laser (New Focus) at 986~nm ($P \sim 30$~mW, linewidth $\sim 1$~MHz) which is amplified by a commercial tapered amplifier (Sacher) to about 800 mW (see Figure~\ref{fig:taperedAmpData} for detailed performance data).  After $> 60$~dB isolation (Optics for Research), the infrared power is frequency doubled to 493~nm using non-critically (type-I $ooe$~\cite{sutherland2003hno}) phase matched KNbO$_3$ in a ring enhancement cavity shown in Figure~\ref{fig:blueCavityPhoto}.  The cavity is of the Sandberg bow-tie design~\cite{freegarde2001ode}, with one of the flat mirrors used as an input coupler.  The curved mirrors have 50~mm radii of curvature and are located 59.9~mm apart to create a 36.0~$\mu$m waist in the 7~mm, Brewster-cut, nonlinear crystal.  Though we attempt to place the flat mirrors to minimize astigmatism in the 493~nm beam, we nonetheless observe a slightly aberrated beam which we attempt to correct with a tilted collimating lens immediately following the cavity.

The crystal is enclosed in a copper puck which can be either cooled or heated with an onboard heat-sunk TEC, an embedded AD590 sensor, and a commercial temperature controller (ILX Lightwave).  At this wavelength, the crystal should be a few degrees above room temperature for non-critical type-I phase matching~\cite{sutherland2003hno, smith2003snc}, but we find that we must cool the crystal mount slightly when full laser power is employed, presumably due to absorptive heating.

With minimal attempts at mode-matching, we easily achieve 40~mW of 493~nm output with 700~mW infrared input.  The mirrors are over 10 years old, have survived countless cleaning attempts, and a hairline scratch is visible on the small PZT mounted flat mirror.  We feel that this issue, in part, explains why we aren't seeing more output from the system.  40~mW is more than adequate for probing the ion, a barium hollow cathode lamp, and a saturated-absorption Te$_2$ system~\cite{raab1998dls} currently in development for use as a laser frequency lock.

\begin{figure}
\centering
\includegraphics{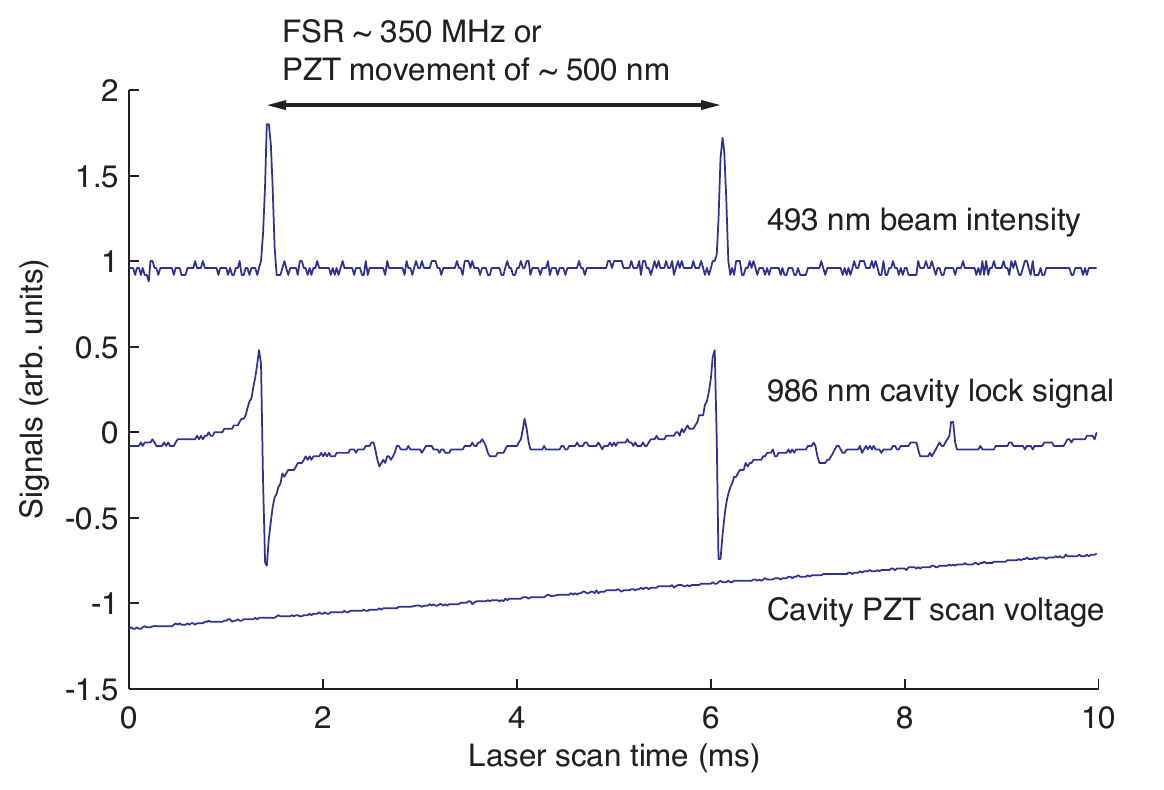}
\caption[Transmission and dispersion lock signals from a 986 nm doubling cavity]{493 nm transmission and 986~nm dispersion lock signals from a second harmonic generation cavity using the H\"{a}nsch-Couillaud scheme described in detail in Appendix~\ref{sec:opticalCavities}.  We obtained each curve with a digital sampling oscilloscope.}
\label{fig:doublingCavitySignals}
\end{figure}

The resonant doubling enhancement cavity is locked to the laser using the H\"{a}nsch-Couillaud polarization scheme~\cite{hansch1980lfs}, discussed in greater detail in Appendix~\ref{sec:opticalCavities}.  Briefly, the beam reflected by the input coupler is sent through a quarter-wave plate and polarization analyzer.  Near cavity resonances, the difference current seen by two photodiodes in the analyzer forms a dispersion shape suitable for input into a linear feedback loop;  see Figure~\ref{fig:doublingCavitySignals}.  We employed a custom two-integrator feedback loop circuit designed and modified by several former postdoctoral researchers and graduate students.

\begin{figure}
\centering
\includegraphics{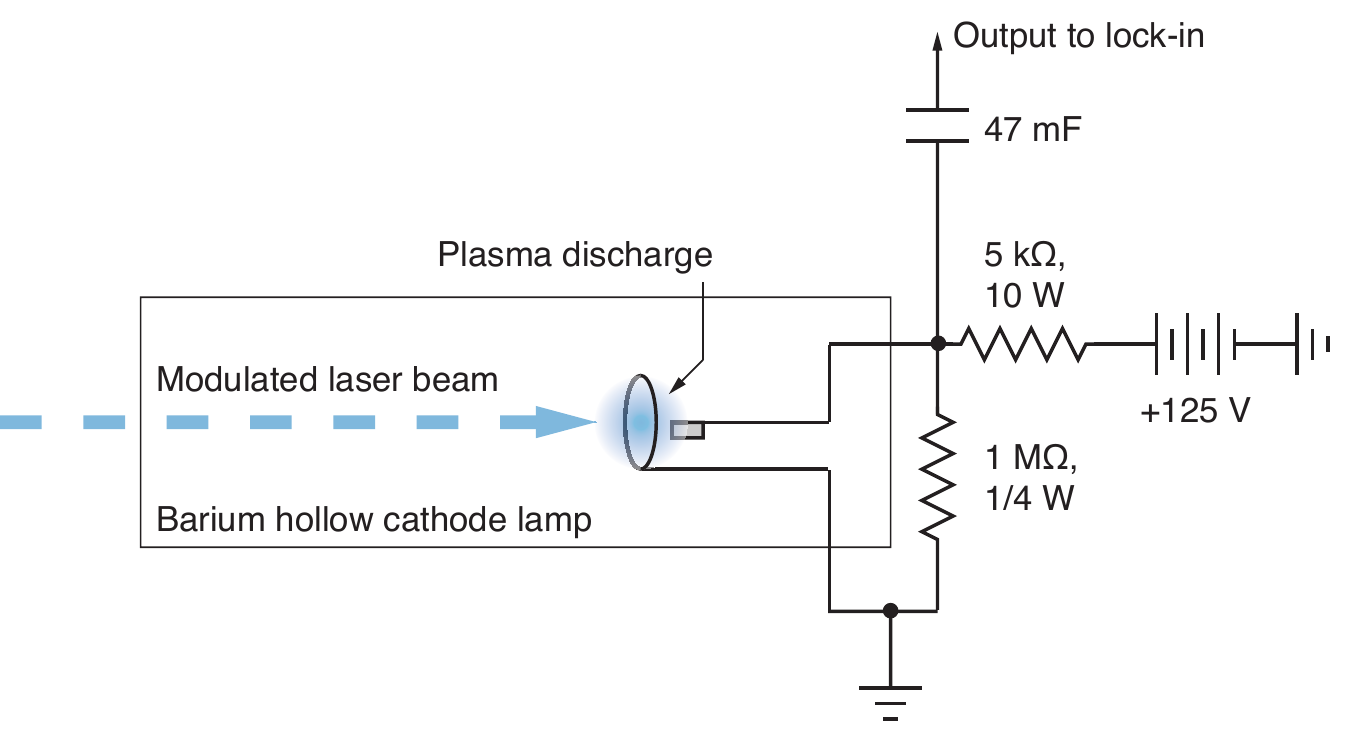}
\caption[A simplified schematic of the hollow-cathode lamp 493~nm laser lock]{A simplified schematic of the hollow-cathode lamp 493~nm laser lock.  It might seem odd that about 100 V from the regulated power supply is dropped across a 5~k$\Omega$ power resistor.  When the circuit is first turned on, the high voltage is necessary to start the hollow cathode tube discharge.  However, when the current increases to the rated tube operating value, 20~mA, the power resistor begins current limiting.  A near resonant laser beam can assist photo-ionization of some of the Ba$^+$ in the plasma discharge, an effect called an opto-galvanic resonance.  The increased current is very small---$< 1$~pA, perhaps---and buried in technical noise at dc.  But, by modulating the laser beam (i.e. chopping with an AOM) at tens of kHz, the signal can be ac-coupled and phase-sensitive detected in a lock-in amplifier.}
\label{fig:blueLaserLockSchematic}
\end{figure} 

The laser is nominally locked to the side of an optogalvanic resonance (see~\cite{bridges1978cog}, for instance) in a barium hollow cathode lamp shown in Figure~\ref{fig:blueLaserLockSchematic}.  The 125~V power supply starts a discharge in the tube;  once current flows, voltage is dropped on the 5~k$\Omega$ power resistor until the voltage across the tube is $\approx 25$~V.  Now, a glowing plasma (which includes Ba$^+$) exists in the tube;  one can identify prominent Ba$^+$ emission lines with a diffraction grating along with others belonging to the tube's buffer gas.  If 493~nm laser light resonant with the Doppler-broadened Ba$^+$ ions is introduced, many ions are excited out of the ground state to $6P_{1/2}$.  While there, there is a small probability that collisions in the plasma will ionize each to Ba$^{2+}$.  This weak process can, in principle, be detected by a slight increase in tube current. The effect is tiny and buried in awful $1/f$ noise at dc, but if we modulate the laser light at several kHz the signal is trivially detected by lock-in amplification, as shown in Figure~\ref{fig:493nmLaserSchematic}.

Therefore, the rf powering an AOM which deflects light into the tube is shuttered with a 10~kHz signal that also serves as the reference to a lock-in amplifier (Stanford Research Systems 510).  Another AOM, shown in its present double-pass configuration, implemented in 2006, raises the the frequency seen by the ion by a computer controlled frequency $f_\text{offset} \approx 200$~MHz.  This allows us to lock the laser to the side of the $\sim 1$~GHz where the \emph{slope} of the opto-galvanic response is high while sending the ion tunable light only $(\Gamma/2)/2 \pi \approx 10$~MHz away from resonance for optimal laser cooling.  

The lock-in output would also scale with 493~nm laser power, so we normalize the signal by detecting the intensity of the un-deflected beam through the dithered AOM and performing a weighted subtraction with the detected tube current.  A variable gain in this subtraction circuit accomplishes a frequency offset knob:  the proper setting zeros the lock-in output when the laser is in resonance with the ion.  Closing a feedback loop routes the lock-in output directly to the laser's PZT grating control.  The bandwidth and gain parameters of the lock-in amplifier are tuned up by observing the locked laser spectrum on a medium finesse scanning Fabry-Perot cavity.  The laser linewidth is undoubtedly \emph{increased} with this laser lock scheme, perhaps to 50~MHz.  The time constant of the lock-in amplifier is often set to its largest value, $\tau = 100$~s.  However, long term drift in the laser system is essentially eliminated for weeks at a time with only a few minutes of maintenance a day.  Ions are consistently trapped and laser cooled over many days; experiments could, in principle, be run for hundreds of hours uninterrupted without re-trapping an ion.  At the time of writing, discharge tubes cost approximately \$250 and last about 8 months of continuous operation for our purposes.  The very same tube can also be used to lock the red laser, as discussed below.

\subsection{The repump laser:  650~nm}
Red laser light at 650~nm resonant with the $5D_{3/2} \leftrightarrow 6P_{1/2}$ is required to empty out the metastable $5D_{3/2}$ state whose population would otherwise cease cooling and fluorescence on the $6S_{1/2} \leftrightarrow 6P_{1/2}$ transition.  Modulating it on and off also allows discrimination of background scattered 493~nm laser light from true ion fluorescence.  Our source of 650~nm light is a 5~mW commercial external cavity diode laser (New Focus Vortex) tunable over 80~GHz by a PZT-mounted mirror/grating system at rates of up to $\sim$~300~MHz via direct rf current injection.

\begin{figure}
\centering
\includegraphics{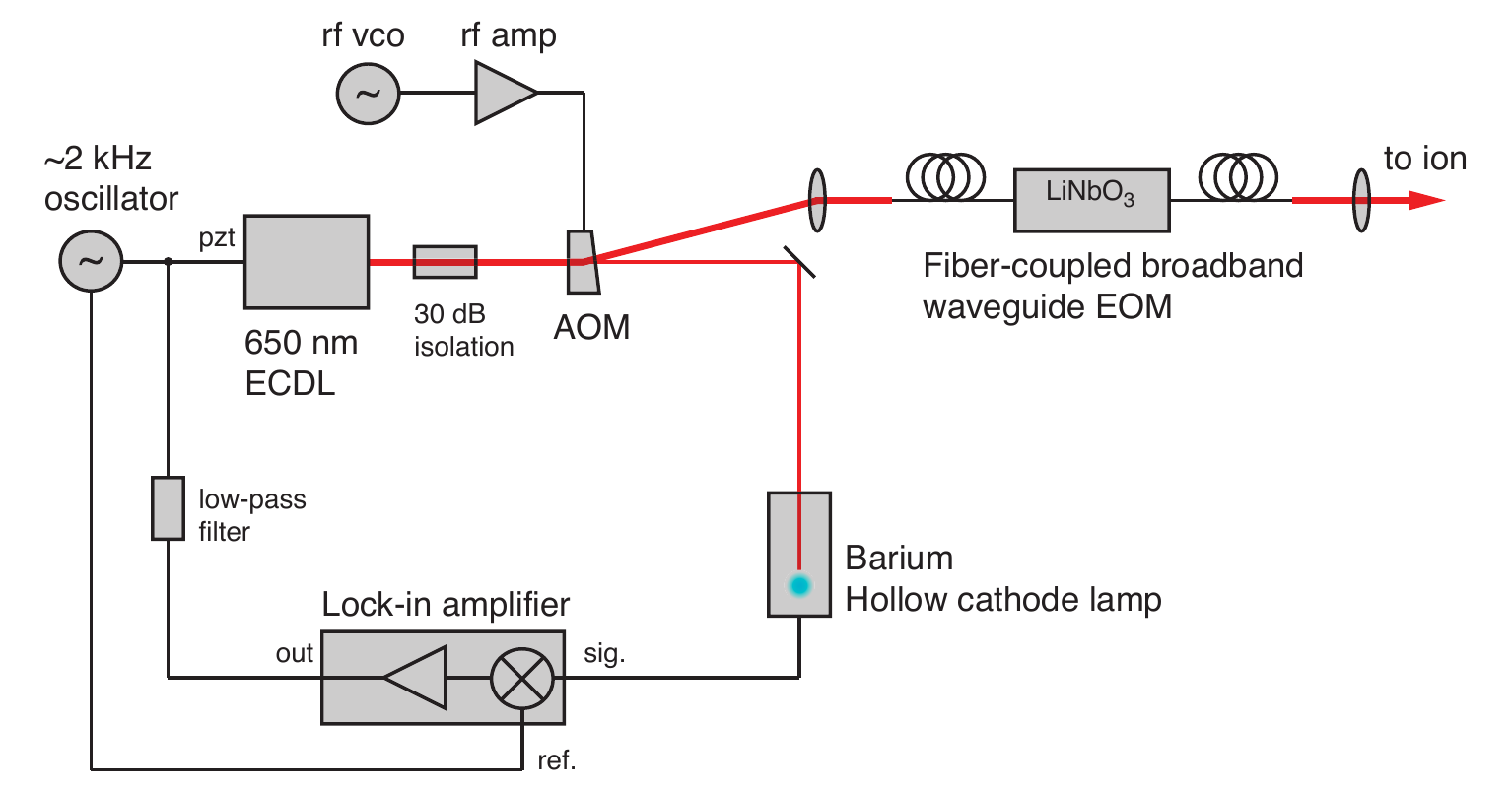}
\caption[A simplified schematic of the hollow-cathode lamp 650~nm laser lock]{A simplified schematic of the hollow-cathode lamp 650~nm laser lock.  Instead of modulating the amplitude of the 650~nm laser to produce an ac signal in the lamp current, we directly modulate its frequency at $\sim 2$~kHz to a depth of a few hundred 100~Mhz.  A dispersion shaped response (Figure~\ref{fig:redLaserLockData}, top) is detected at the output of a lock-in amplifier and serves as an error signal.}
\label{fig:redLaserLockSchematic}
\end{figure} 
\begin{figure}
\centering
\includegraphics{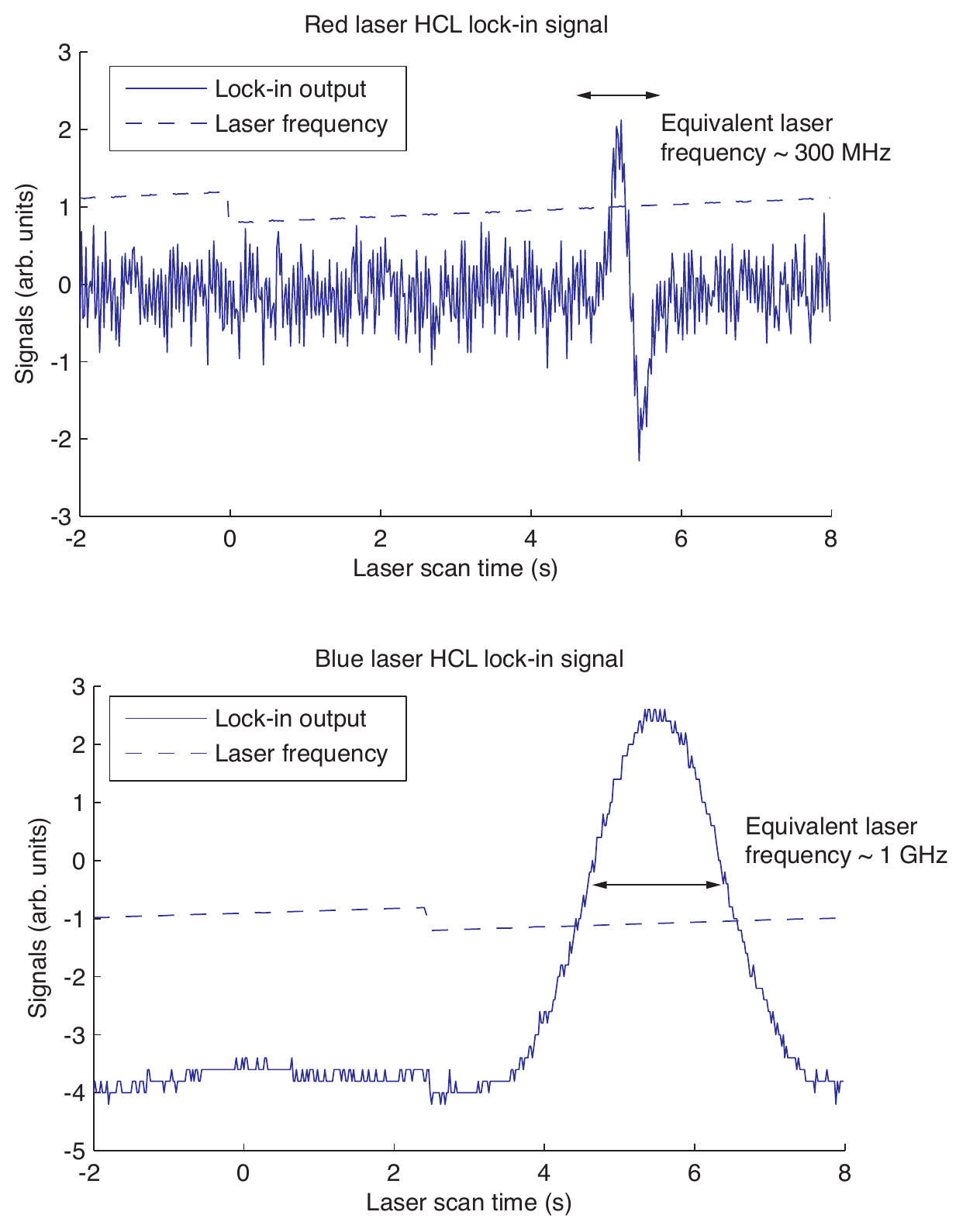}
\caption[Error signals from a hollow cathode lamp used for laser stabilization]{Lock-in amplifier output of a barium hollow cathode lamp as modulated 650~nm (top) and 493~nm (bottom) lasers are scanned across resonance.  For the red laser, the width of the dispersion feature scales with the depth of frequency modulation applied to the laser, $\sim 300$~MHz in this figure.  Since the blue laser beam into the discharge tube is amplitude modulated, it's hollow cathode lamp feature is a Gaussian whose width is set by Doppler broadening, $\sim 1$~GHz.}
\label{fig:redLaserLockData}
\end{figure}
The red repump laser, like the cooling laser, is frequency locked to an optogalvanic resonance in a barium hollow cathode lamp.  Since we do not need (and due to a narrow resonance feature, see Section~\ref{sec:threeLevelSpectroscopy}, we do not want) a narrow 650~nm laser, we choose a simpler scheme that locks to the top of the $\sim$1~GHz broad hollow cathode tube resonance.  We apply a 1.5~kHz signal to the laser's PZT-mounted grating with a sufficient amplitude to achieve $\sim 100$~MHz of laser frequency modulation.  The 1.5~kHz signal is used as the reference for lock-in amplification as shown in Figure~\ref{fig:redLaserLockSchematic}; a dispersion shaped signal results as the laser is tuned across resonance as shown in Figure~\ref{fig:redLaserLockData}.  The lock-in time-constant is often $\tau = 10$~s and the laser is not significantly broadened beyond the 100~MHz induced by direct modulation.

At times we have experimented with putting AOMs in various configurations between the red laser and discharge tube so that the laser could be kept narrow, increasing its spectral density.  So far, the limited bandwidths of available AOMs yield a poor dispersion curve signal-to-noise and thus an unstable lock. Another potential solution is implementing a sample/hold circuit with analog switches that toggle modulation to the laser while applying a sampled lock-in output to the laser during the short durations of time when a narrow red laser is necessary.

\subsection{Shelving/deshelving lamps:  455~nm and 615~nm}
\begin{figure}
\centering
\includegraphics{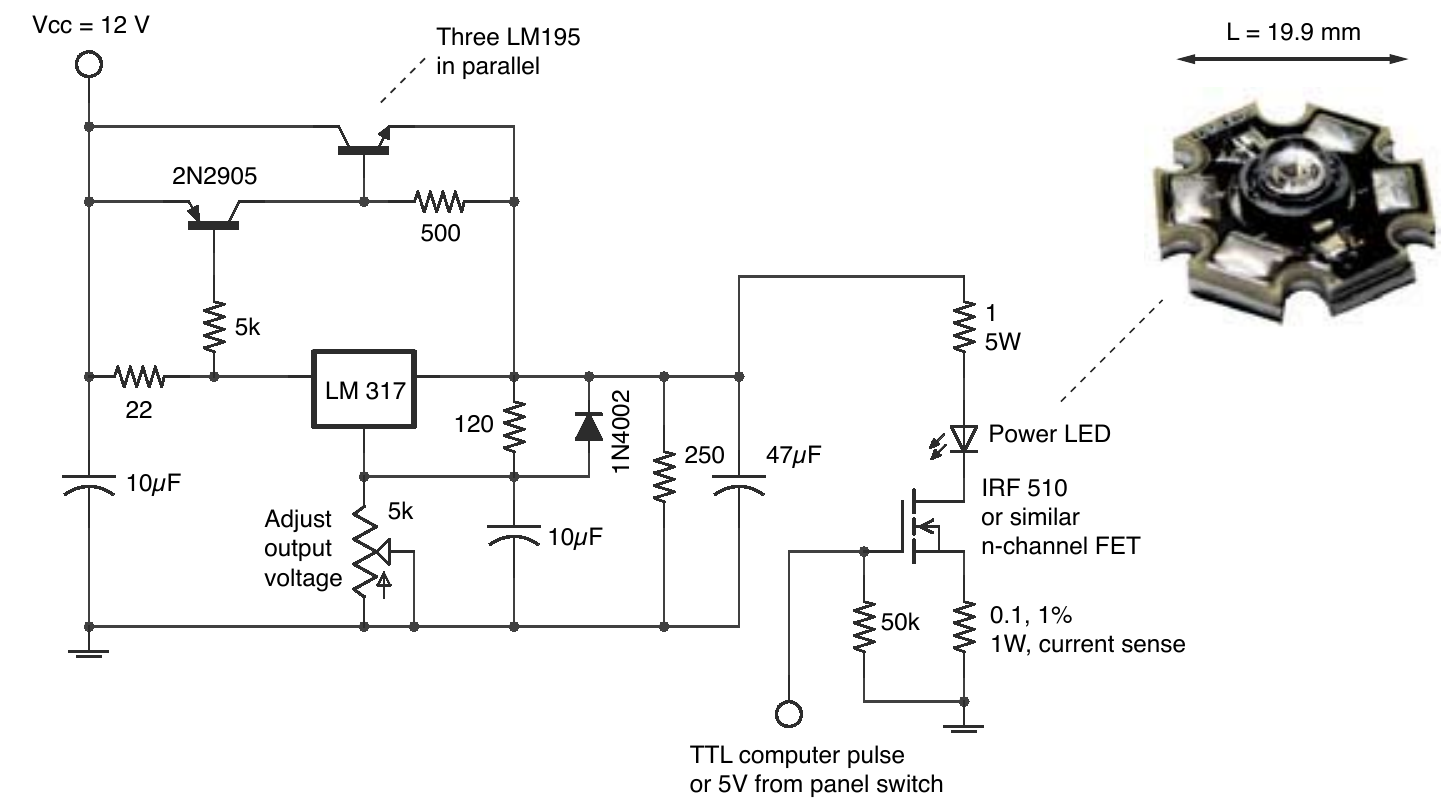}
\caption[High power LED and driver used for ion shelving/deshelving]{High power LED and driver used for ion shelving/deshelving.  The adjustable voltage regulator LM317 is power boosted with several power transistors LM195, a circuit copied from National's LM317 datasheet.  The output is tied to the drain of a power n-channel MOSFET, the IRF 510 or similar, whose gate is driven with a sufficiently high positive voltage to drive current across the power LED, a Lumileds Luxeon III.  An inset photo of the LED, copied from the Luxeon datasheet, also shows the integrated solder pads and aluminum heat spreader mount essential in conducting heat away from the LED to a large efficient aluminum heat sink.}
\label{fig:LEDs}
\end{figure}
This section describes the best blue and orange \emph{non-lasers} that \$8.00 can buy.  The metastable $5D_{5/2}$ `shelved' state can be reached via an excitation on the 455~nm $6S_{1/2} \leftrightarrow 6P_{3/2}$ transition followed by a decay with moderate probability.  Similarly, an ion in $5D_{5/2}$ can be `de-shelved' by an excitation on the 615~nm $5D_{5/2} \leftrightarrow 6P_{3/2}$ transition followed by a decay to the ground state.  Earlier in this experiment's history, researchers used barium hollow cathode lamps with colored glass or interference filters to excite these transitions.

With the aims of eliminating the need for additional hollow cathode lamps and their high voltage supplies, speeding up the shelving and deshelving optical pulses, and making the system more compact, we investigated bright light emitting diode (LED) sources.  Our current system uses Amber and Royal-blue Luxeon III diodes manufactured by Lumileds.  Each can be driven with over 1~A of current and are mounted with heatsinks.  We combine the light from the LEDs with a dichroic mirror and image them onto the ion trap with a 50~mm, f/2.8 camera lens. Though the spectral density of the LEDs is quite low---they each emit a 40~nm optical bandwidth---we drive shelving and deshelving transitions at rates exceeding 10~Hz.  The blue LED blinds the filtered PMT and therefore must be pulsed.  The orange LED has little effect and can be easily discriminated from the normal ion fluorescence signal.  Figure~\ref{fig:LEDs} contains a photograph of one LED and a power source schematic.

\subsection{Light shift lasers:  514~nm, and 1111~nm}
We employed a total of three argon-ion lasers, two of them already nearly dead from the start, for the light shift experiments at 514~nm.  Each ran with broadband mirrors and an intra-cavity prism that ensured single laser line operation.  The first was a Coherent Supergraphite CR-12 with a sagging cathode, disabled argon reservoir refill system, and power supply with a sporadic appetite for 70~A fuses and transient suppression diodes, the latter of which would vaporize in pairs nearly weekly leaving behind only wire stubs and black residue on the chassis door.  We had better luck with an aged Coherent Innova~400, though the water-cooled magnet jacket sprung a large leak while the laser was operating.  The tube survived unscathed, only to be flooded with water months later when it cracked.  We obtained the final data runs with a borrowed Innova tube which miraculously survived the curse of the light shift experiment.

Obtaining data at 1111~nm was a cake-walk by comparison.  A turn-key Yb fiber laser (Koheras Boostik) intended for another experiment became available for use in 2005.  It provides 1~W of light through a polarization-maintaining fiber and features long term intensity stability, a relatively narrow specified 100~kHz linewidth, and negligible frequency drift for our application.  More details about the measured performance and stabilization of both these lasers are provided in Section~\ref{sec:lightShiftBeamStabilization}.

\subsection{The clock laser: 2051~nm}\label{sec:clockLaserApparatusChapter}
\begin{figure}
\centering
\includegraphics{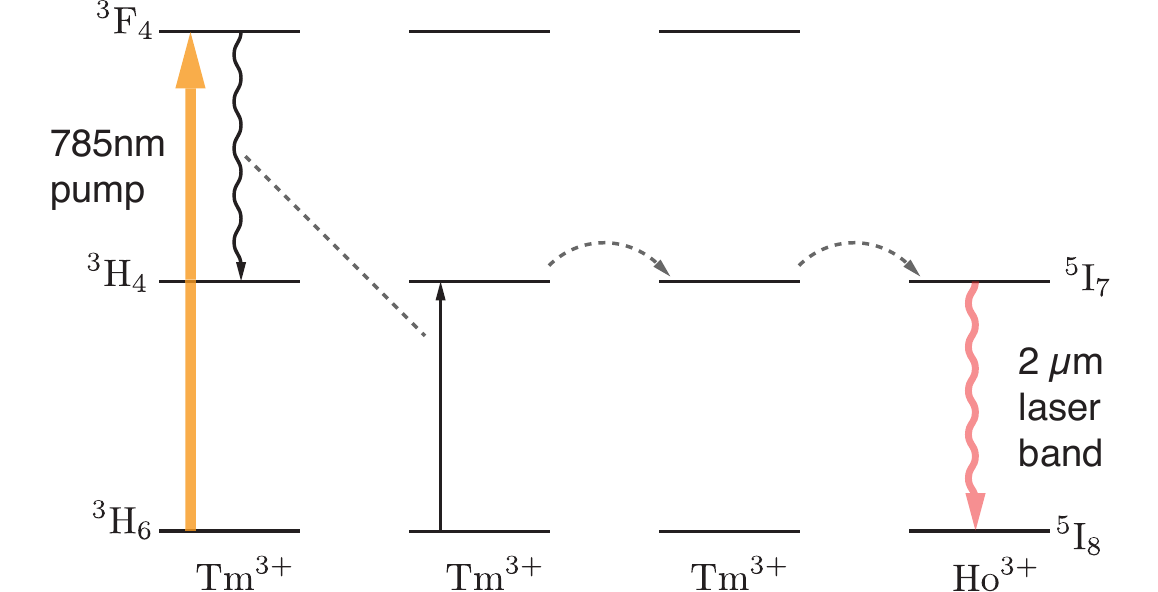}
\caption[The diode pumped Tm,Ho:YLF laser process at 2051~nm]{Though a full model is more complicated~\cite{bruneau1998mth}, the Tm,Ho:YLF laser process can be summarized by this diagram (reproduced with modifications from~\cite{fan1988sad}).  Pumping radiation excites Tm$^{3+}$ impurity ions in a YLF or YAG crystal matrix.  Radiative energy is transferred by a variety of mechanisms between Tm$^{3+}$ to eventually excite the metastable $^5$I$_7$ level of impurity Ho$^{3+}$ ions.  Laser action can be established on the $^5$I$_7 \leftrightarrow ^5$I$_8$ transition band covering 2.05 to 2.1 $\mu$m.} 
\label{fig:2micronLaserLevels}
\end{figure}

\begin{figure}
\centering
\includegraphics{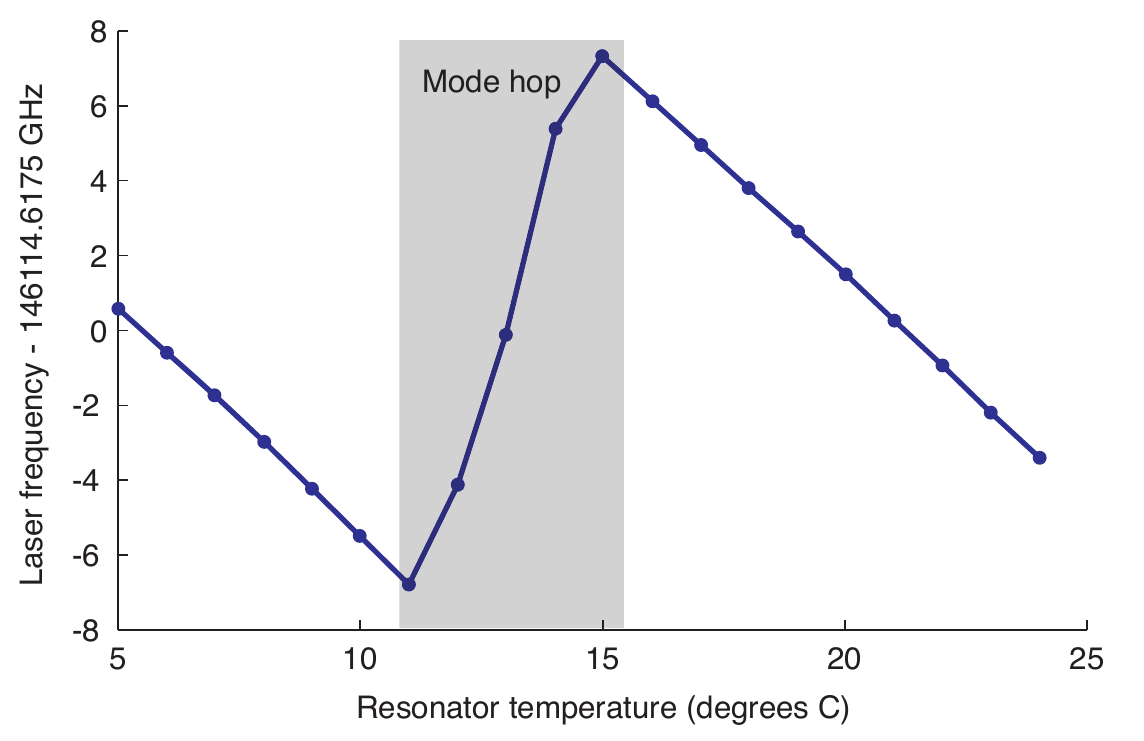}
\caption[Temperature tuning of the 2051~nm laser.]{This temperature tuning test of the 2051~nm laser, performed with a Burleigh WA-1500 wavemeter with long wavelength optics allows us to derive a $\approx -1.2$~GHz/$^\circ$C tuning curve outside the mode hop (shaded) region.}
\label{fig:2micronTempTune}
\end{figure}

\begin{table}
\centering
\caption[2051~nm laser PZT tuning characteristics]{2051~nm laser PZT tuning characteristics.  We used a commercial wavemeter and scanning Fabry-Perot cavity (see Table~\ref{tab:zerodurCavityProperties}) to determine the laser frequency tuning sensitivities in three PZT voltage ranges.}
\begin{tabular}{rr}
Voltage range		& Laser tuning rate \\ \hline \hline
0---133.7~V		& -50.0~MHz/V \\
133.7---243.3~V	& -75.0~MHz/V \\
243.3---454.0~V	& -81.2~MHz/V \\ \hline
\multicolumn{2}{l}{First PZT resonance $\approx 112$~kHz}
\end{tabular}
\label{tab:2micronPZTtune}
\end{table}

\begin{table}
\centering
\caption[Specifications of a medium-finesse Zerodur scanning cavity at 2051~nm]{Specifications of a medium-finesse Zerodur scanning cavity at 2051~nm constructed by previous graduate student K.\ Hendrickson~\cite{hendrickson1999Thesis}.}
\begin{tabular}{lll}
Property & & Measurement/Spec.  \\ \hline \hline
Zerodur spacer length & $L_\text{s}$ & ($8.83 \pm 0.1$)~in. \\
PZT length (specified) & $L_\text{pzt}$ & ($1.25 \pm 0.01$)~in. \\
Total cavity length & $L = L_\text{s} + L_\text{pzt}$ & 10.08~in. or 25.6~cm \\
Mirror radius of curvature & $R$ & 30~cm \\
PZT tube outer diameter & $d_\text{pzt}$& $(0.75 \pm 0.01)$~in. \\ \hline
Design free spectral range & $\nu_\text{FSR} = \frac{c}{2L}$ & 585.5~MHz \\
Design finesse &  $F = \frac{\pi \sqrt{R}}{1-R}$ & 500 \\
Design resolution & $\Delta \nu_{1/2} = \frac{c}{2L F}$ & 1.17~MHz \\
Design waist & $(\omega_0)_\text{conf} = \sqrt{\frac{\lambda L}{2 \pi}}$ & 0.289~mm \\
Design spot size at mirrors & $(\omega_1)_\text{conf} = \sqrt{2}(\omega_0)_\text{conf}$& 0.409~mm \\ \hline
Measured free spectral range & $\nu_\text{FSR}$ & $(576 \pm 21)$~MHz \\
Measured resolution & $\Delta \nu_{1/2}$ & $(1.7 \pm 0.2)$~MHz \\
Derived finesse & $F$ & $(339 \pm 50)$ 
\end{tabular}
\label{tab:zerodurCavityProperties}
\end{table}

The need for precision measurement and manipulation using the electric-dipole forbidden $6S_{1/2} \leftrightarrow 5D_{3/2}$ transition places stiff stability requirements on a 2051~nm laser system.  Fortunately, commercial development of narrow linewidth diode pumped Tm,Ho:YLF solid state lasers in the 2~$\mu$m wavelength region has been prompted by coherent lidar applications:  the eye-safe wavelength is useful for measuring atmospheric properties thanks to moderate absorption features of gasses like CO$_2$~\cite{henderson1993clr}.  Figure~\ref{fig:2micronLaserLevels} shows the laser process~\cite{bruneau1998mth}.  We have acquired a  80~mW, single longitudinal mode Tm,Ho:YLF laser manufactured by CLR Photonics (now owned by Lockheed-Martin) that is tunable by temperature, pump power, and a resonator-coupled PZT.

\begin{figure}
\centering
\includegraphics{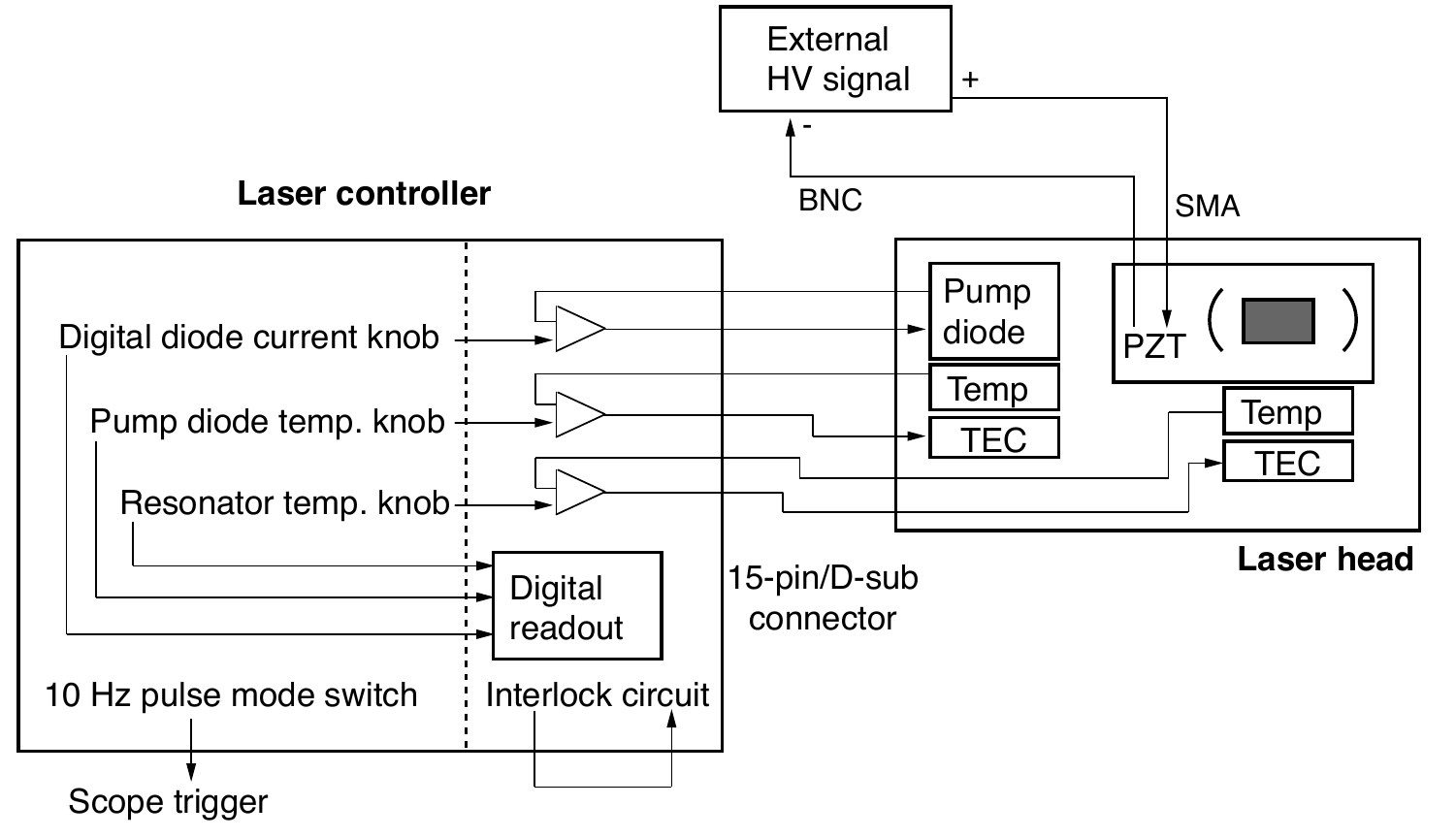}
\caption[Control schematic of the 2051~nm Tm,Ho:YLF laser]{Control schematic of the 2051~nm Tm,Ho:YLF laser.  Large range, slow tuning is provided by closed loop temperature control.  A PZT mounted on the laser cavity enables fast frequency tuning;  the specially designed PZT's first mechanical resonance appears to be at $\approx 112$~kHz.  A `pulse' mode is useful in detecting the laser with special IR photodiode.}
\label{fig:2micronControl}
\end{figure}
Changing its temperature tunes the laser by -1.2 GHz/$^\circ$C over a 14 GHz range between mode hops as shown in Figure~\ref{fig:2micronTempTune}.  We confirmed with a wavemeter (Burleigh WA-1500 with long IR optics) that the laser can hit all possible $^{138}$Ba$^+$ and $^{137}$Ba$^+$ transitions, shown in Table~\ref{tab:2micronTransitionsTable}.  We also explored the PZT tuning profile, shown in Table~\ref{tab:2micronPZTtune} using a wavemeter and scanning Fabry-Perot cavity measurements.  We employed a Zerodur spaced medium finesse scanning cavity (properties listed in Table~\ref{tab:zerodurCavityProperties}) to monitor the laser spectrum.  The control mechanisms are sketched in Figure~\ref{fig:2micronControl}.  A tedious detail about the laser control must be documented and emphasized since the user manual is insufficient:  the laser manufacturer employs a nonstandard wiring mechanism for the PZT.  A separate SMA connector on the laser head only provides a connection from the center-conductor of an attached cable to one side of the PZT.  The other side of the PZT is not grounded, connected to the laser chassis, or the shield of the SMA connector.  Instead, the PZT power return path is routed through an undocumented pin (pin 12) in the D-shaped 15-pin connector that carries the laser current and temperature control signals.  This is routed to the shield of a BNC connector that is spliced into the thick 15-conductor cable whose center pin is likewise routed to a dangling SMA cable at the other end of the bundle.

\begin{figure}
\centering
\includegraphics{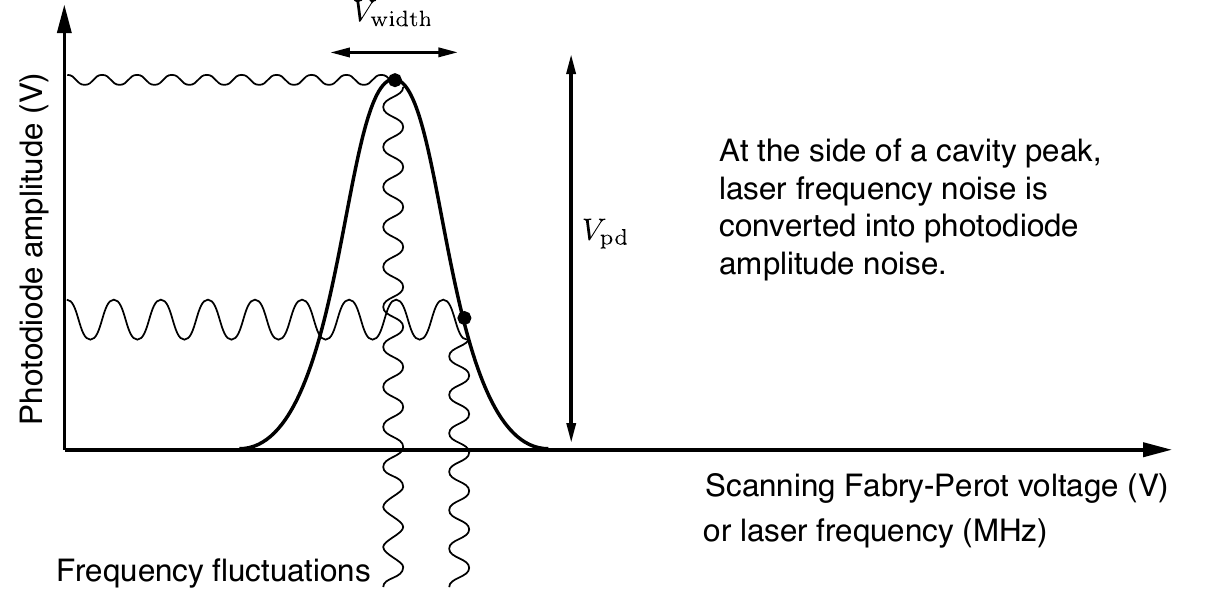}
\caption[The side of a Fabry-Perot peak converts laser noise into intensity fluctuations]{The side of a Fabry-Perot peak converts laser noise into intensity fluctuations.  We employed a medium finesse cavity to investigate the free-running 10~kHz linewidth of a 2051~nm laser.}
\label{fig:peakNoiseDiagram}
\end{figure}
\begin{figure}
\centering
\includegraphics[width=5.5 in]{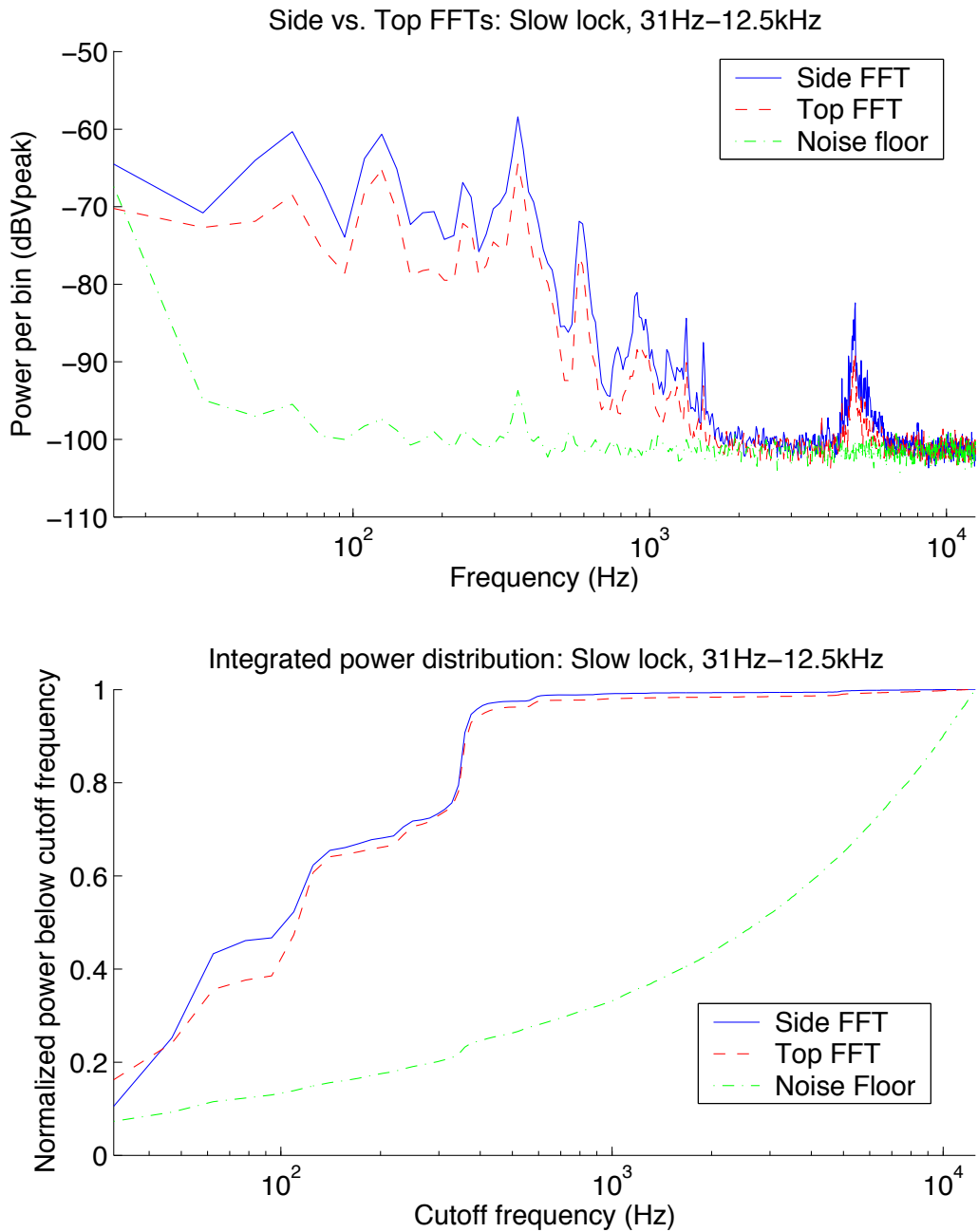}
\caption[A test of the 2051~nm laser free-running linewidth]{To test the 2051~nm laser's specified free-running linewidth of 10~kHz, we built a very low bandwidth laser frequency lock that would servo-lock the 2051~nm laser to the side of a medium finesse Fabry-Perot transmission peak (see Figure~\ref{fig:peakNoiseDiagram}).  Residual frequency (and amplitude) noise on the laser is then measured as intensity noise on a photodiode placed after the spectrum analyzer.  The frequency power spectrum shows that most of the noise occurs in a bandwidth of 1~kHz and that the integrated noise power confirms a linewidth of $\sim$ 10~kHz.  See text for additional details.} 
\label{fig:2micronLaserNoiseTest}
\end{figure}
The manufacturer specifies the free-running linewidth of the laser to be 10~kHz.  To confirm this using our scanning cavity, which has a linewidth much larger than 10~kHz, we measured the excess transmission noise of the laser locked to the side of a cavity fringe; see Figure~\ref{fig:peakNoiseDiagram} for a diagram.  With this rudimentary method, it is important to construct a rather poor laser frequency lock;  a good one would reduce the linewidth of the laser.  We built a simple integrator circuit with a variable lock time-constant and attempted to set the transmission lock-point to the 50\% maximum.  To normalize out amplitude noise on the laser, we also took measurements with the laser locked to the top of a cavity fringe.  

Sample data, recorded from a digital spectrum analyzer, are shown in Figure~\ref{fig:2micronLaserNoiseTest}.  The logarithmic units of power in these graphs are converted to linear (photodiode) voltages with
\begin{equation*}
V_i = 10^{(P_i [dB]/20)}.
\end{equation*}
Then we find the root mean squared value
\begin{equation*}
V_\text{total, rms} = \sqrt{\sum_{i=1}^{N_\text{bins}}V_i^2}.
\end{equation*}
To convert this value to a frequency (and an estimate of the linewidth of the laser), we perform a linear approximation of slope at the midpoint of a Fabry-Perot peak:
\begin{align*}
\text{Cavity peak width} &= 1.7 \text{ MHz} \\
\text{Cavity peak photodiode height} &= 0.275 \text{ V} \\
\Rightarrow \text{ Conversion factor} &= 6.2 \text{ MHz/V}.
\end{align*}
Therefore, we estimate the total rms frequency noise with
\begin{equation*}
\Delta \nu_\text{total,rms} =  \big( V_\text{total}(\text{SideFFT}) - V_\text{total}(\text{TopFFT}) \big) \cdot 6.2 \text{ [MHz/V]}.
\end{equation*}

\begin{table}
\centering
\caption[Table of data used in estimating the free-running 2051~nm laser linewidth]{Table of data used in estimating the free-running 2051~nm laser linewidth.  These data are obtained by converting the units and integrating digitally captured FFT curves such those in Figure~\ref{fig:2micronLaserNoiseTest}. The three sets shown here were obtained under different laser/cavity lock time constants.  The final set (`Twiddle lock'), was obtained with a very low bandwidth stabilization system:  a graduate student turning a knob.}
\label{tab:2micronLinewidthData}
\footnotesize
\begin{tabular}{ l| l| l| l| l| l| l}
\multicolumn{3}{c|}{} &  \multicolumn{3}{c|}{$V_\text{total,rms}$ (mV)} & \\
Dataset & Lock settings & Bandwidth & Side & Top &  Floor &  $\Delta \nu_\text{total,rms}$(kHz) \\ \hline \hline
Slow lock & (gain=10,$\tau$=30ms)&  31Hz-12.5kHz & 2.72 & 1.42 & 0.25 & 8.04 \\ \hline
Fast lock & (gain=10,$\tau$=600us)&  31Hz-12.5kHz & 2.33 & 1.11 & 0.25 & 7.53 \\ \hline
Twiddle lock & (student-knob-turn)&  31Hz-12.5kHz & 2.38 & 1.17 & 0.25 & 7.49
\end{tabular}
\end{table}
Table~\ref{tab:2micronLinewidthData} shows the results for three tests differing in lock time-constant.  Each gives a linewidth estimate of $\sim$ 8~kHz.  We checked these results by parking the laser on the side and top of a cavity Fabry-Perot peak and observing the peak-to-peak transmission noise on an oscilloscope.  Using our eyes to put bounds on the size of the noise signals, we estimated peak-to-peak noise while on the side of the peak as 12~mV, and noise while on the top of the peak as 8~mV.  By taking the difference (to cancel the effect of amplitude noise) and applying the conversion factor, this gives a peak-to-peak noise estimate of about 24~kHz.  This is consistent with data in Table~\ref{tab:2micronLinewidthData} if one assumes a peak-to-peak to rms conversion factor of 3, which is reasonable.

\section{A stable reference cavity at 1025~nm}
\begin{figure}
\centering
\includegraphics{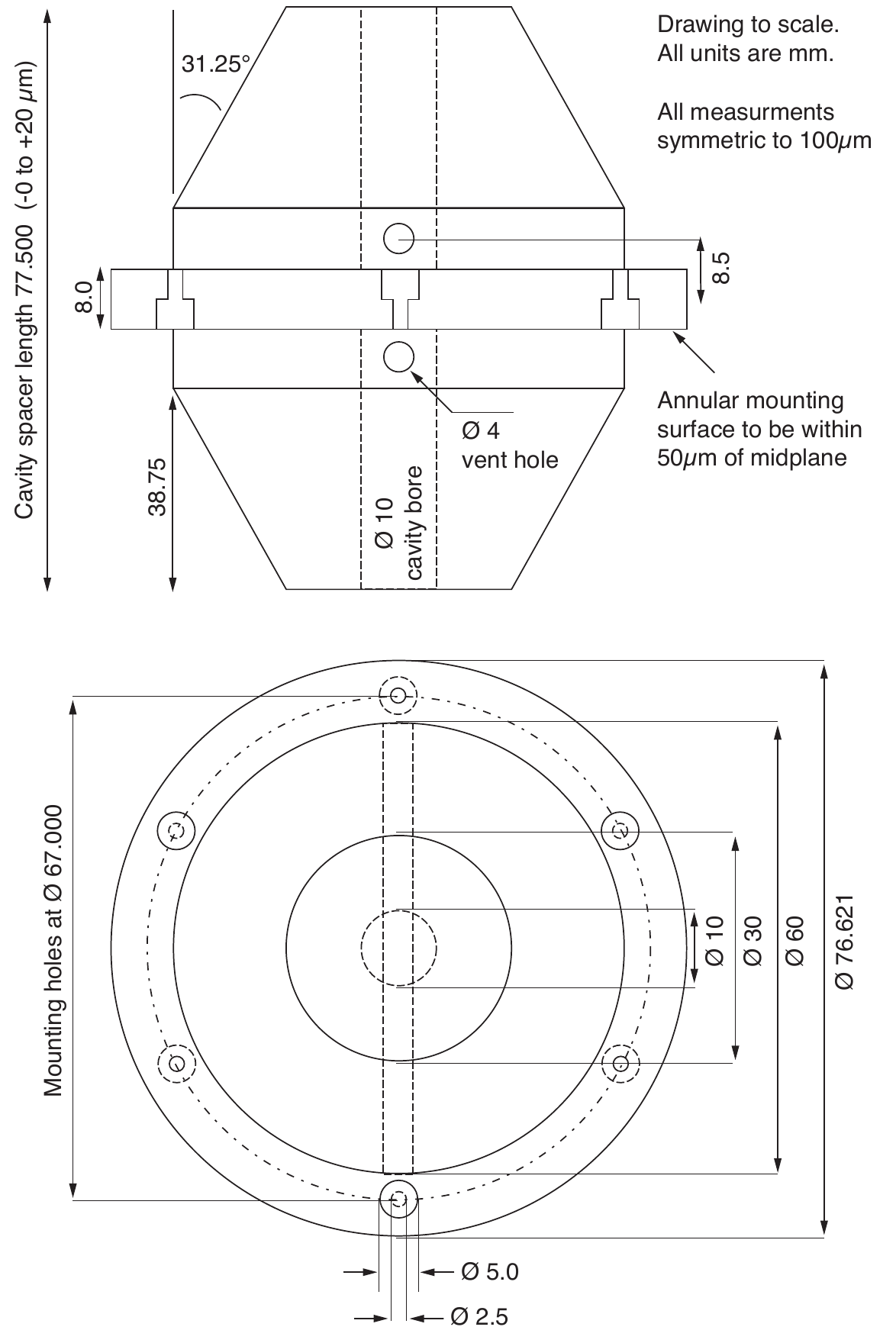}
\caption[Design of a stable, vertically-mounted ULE reference cavity]{This design of a stable vertically-mounted ULE reference cavity was pioneered by researchers Mark Notcutt, Jun Ye, and Jan Hall~\cite{notcutt2005sac} at JILA, manufactured by Advanced Thin Films, and is now in use at perhaps a dozen independent laboratories.  The key is the high degree of symmetry in the mass distribution about the midplane where the cavity is mounted on vertical support posts.  Notice that while only three mounting holes are used, three additional holes in opposite orientation are drilled to accomplish the mass distribution symmetry.}
\label{fig:referenceCavityDiagram}
\end{figure}

\begin{figure}
\centering
\includegraphics{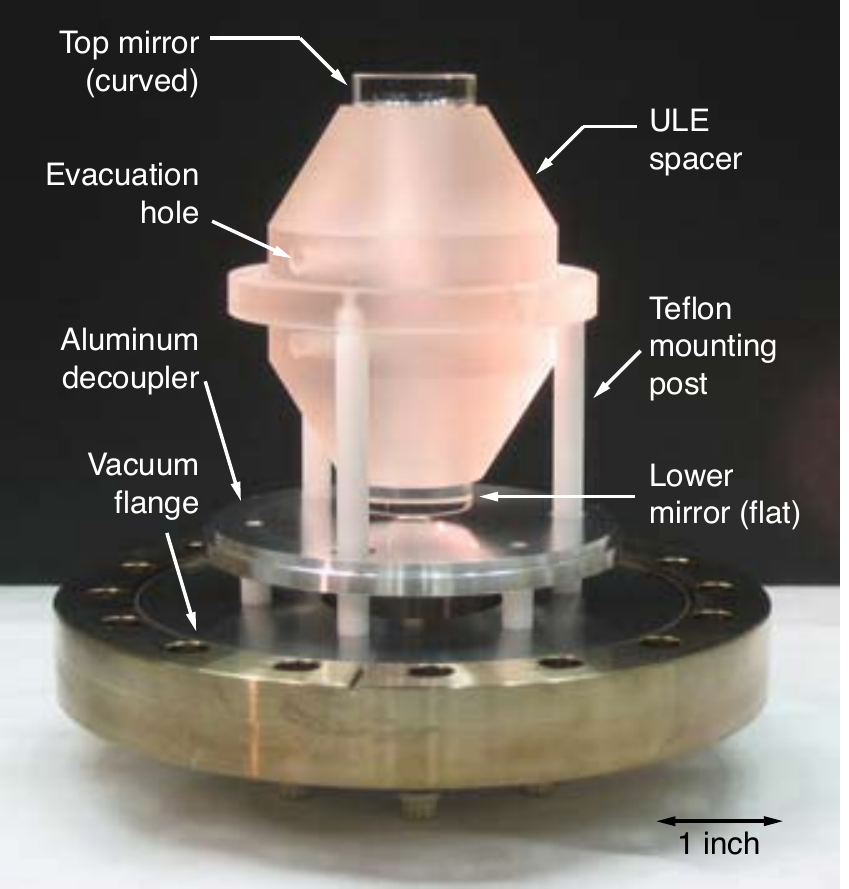}
\caption[Annotated photo of the vertically-mounted ULE reference cavity]{Annotated photo of the vertically-mounted ULE reference cavity.  High finesse mirrors are optically contacted to the ULE spacer described in Figure~\ref{fig:referenceCavityDiagram}.  We kinematically mount the cavity using three 4~mm Teflon ball bearings resting on 0.25~inch Teflon posts tapered to fit into the mounting annulus.  The posts are mounted on the base of a machined aluminum shield which itself is mounted onto a 6~inch stainless steel vacuum flange using short Teflon posts and stainless steel set screws.  All parts are treated and vented for operation at high vacuum.  A 2 l/s ion pump maintains a steady-state pressure known to be below $10^{-8}$ Torr after several days of 350 $^\circ$C baking treatment under vacuum.}
\label{fig:referenceCavityPhoto}
\end{figure}

\begin{figure}
\centering
\includegraphics[width=6in]{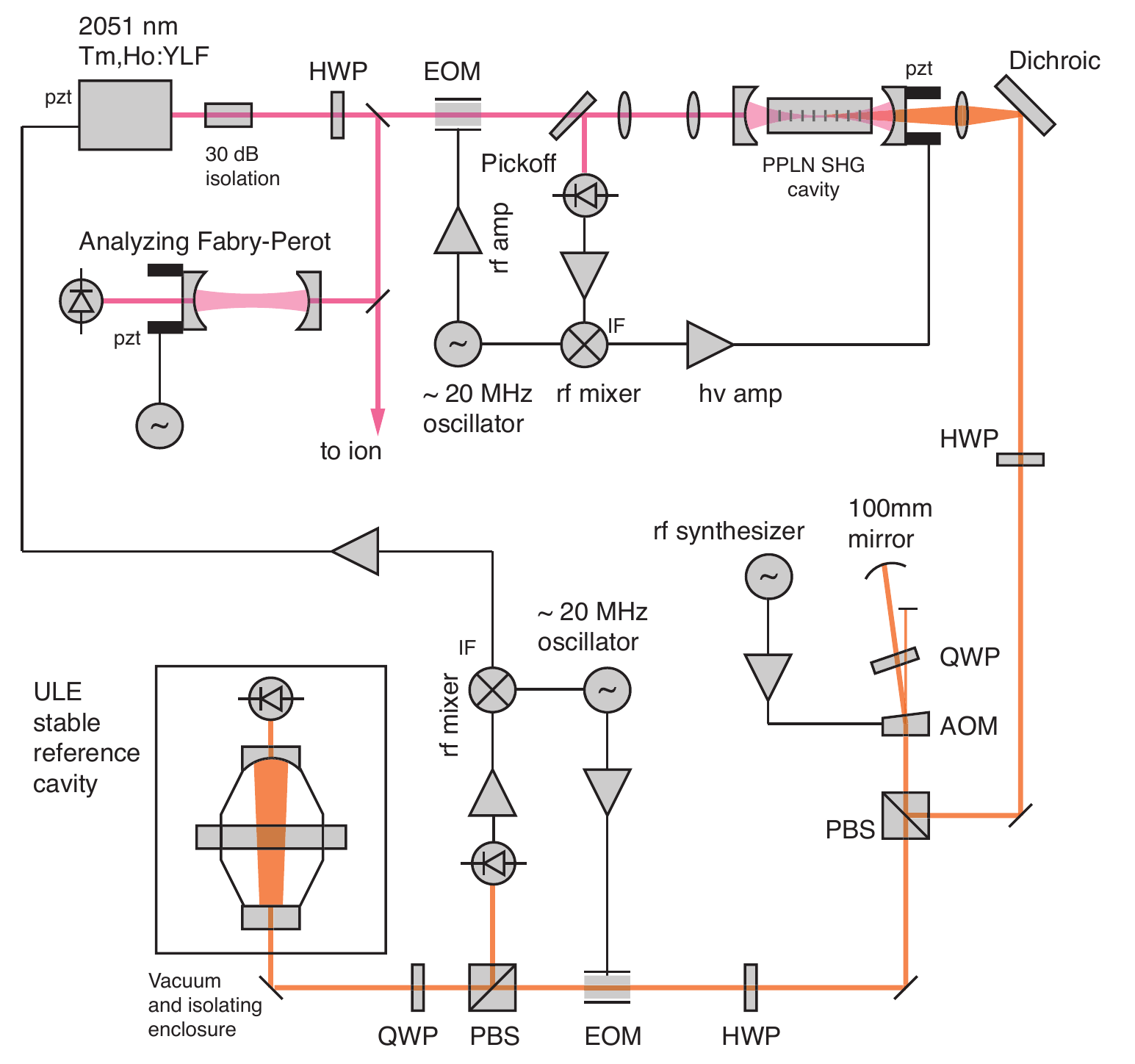}
\caption[Diagram of a stabilized 2051~nm clock laser apparatus]{This diagram of a stabilized 2051~nm clock laser apparatus shows a linear cavity enhanced frequency doubling stage using a PPLN nonlinear crystal to generate 1025~nm light.  This beam is frequency shifted by a wideband AOM in double-pass configuration and sent to the vertically mounted stable reference cavity to derive an error signal which is fed back to the 2051~nm in order to stabilize it.}
\label{fig:clockLaserApparatus}
\end{figure}
A stable, high-finesse reference Fabry-Perot optical cavity serves a crucial role in narrow spectroscopy.  Because a laser beam is only resonant with a laser cavity whose length is a half-integer number of wavelengths, a stable cavity length translates into a stable frequency reference.  Over short time scales (1 $\mu$s to 1 s) the cavity must serve as the frequency reference in a laser frequency noise cancellation feedback loop, disciplining lasers with free running linewidths as high as 1~MHz down to 1~Hz or even lower~\cite{young1999vls, webster2004sln}.  Such narrow laser linewidths allow one to resolve unbroadened, electric-dipole forbidden optical atomic resonances. Over longer time scales corresponding to many atomic interrogations (1 s to 30 s, perhaps), the cavity exhibits linear and small drift which can be calibrated.  As discussed in Section~\ref{sec:clockScheme}, an atomic clock scheme will use the optical atomic resonance to correct for drift in the optical cavity over yet longer timescales and thus establish its ultimate accuracy.

In a practical sense, good, short timescale stability means insensitivity of the cavity length to vibrations.  Researchers John Hall, Jun Ye, and Mark Notcutt are responsible for a noteworthy innovation in cavity design:  a highly symmetric vertical cavity mounted at its midplane~\cite{notcutt2005sac}, shown in Figures~\ref{fig:referenceCavityDiagram} and~\ref{fig:referenceCavityPhoto}.  In a midplane-mounted symmetric cavity, vertical accelerations that stretch the top portion of the cavity are compensated by compressions of the lower potion:  to first-order, the length of the cavity does not change, and thus the frequency reference markers are made insensitive to vibrations.  Finite-element modeling~\cite{chen2006vie} bears this assertion out and informs the designer about particular mounting methods which minimize bending stresses and reduce the second-order sensitivity to lab accelerations in all directions.  

Other approaches involve optimally mounting a typical cylindrical cavity~\cite{nazarova2006vir} or an algorithmically designed, horizontally-mounted cavity that optimizes the important vibrational modes to have nodes at the mounting points~\cite{webster2007vio}.  Future work by several groups will demonstrate the ultimate performance of these optical flywheels.  There is some evidence that a limiting factor will be fast temperature fluctuations/oscillations in the high finesse dielectric coatings~\cite{notcutt2006ctn}. 

We decided to have our ULE reference cavity coated for high finesse not at 2051~nm but at the second harmonic:  1025~nm.  Principally, the coating manufacturer (Advanced Thin Films) estimated they could obtain a finesse of only 20,000 at 2051~nm but could do ten times better at 1025~nm.  Also, we found commercial options for electro-optic and wideband acoustic-optic modulators at 2051~nm extremely lacking.  Long-wave InGaAs photodiodes appropriate for use at 2~$\mu$m are noisier at room temperature, have higher capacitance, and cannot be reversed biased by more than 1~V.  We concluded that all these technical challenges are better met at 1~$\mu$m wavelengths and focused on achieving second harmonic generation of the 2051~nm light. The details are discussed in Section~\ref{sec:frequencyDoubling2micron}.  The full apparatus diagram as it exists now is depicted in Figure~\ref{fig:clockLaserApparatus}.

\section{Radio frequencies and optical modulators}
A programmable digital function generator (SRS DS345) provided waveforms used in the rf Zeeman spectroscopy and light shift experiments discussed in Chapter~\ref{sec:lightShiftChapter}.  Several computer controlled coherent synthesizers (PTS 310, PTS 500) drive AOMs to shift critical laser frequencies, such as the doubled 2051~nm clock beam.  Many voltage controlled oscillators drive EOMs used in frequency modulation and Pound-Drever-Hall stabilization systems, and AOMs used to shift less crucial laser frequencies.  All synthesizers used 10~MHz oven-controlled crystal oscillators as frequency references rated for aging of $< 1 \times 10^{-6}$ per year.  At this writing we are investigating commercial Rb atomic standards which have rated stabilities of $\sim 10^{-11}$.  For an ultimate 2051~nm frequency standard effort, including the acquisition of an appropriate femtosecond laser frequency comb, the apparatus must be upgraded to include a better frequency reference:  perhaps a GPS referenced high quality atomic frequency standard.

All visible wavelength acousto-optic modulators (AOMs), present in Figures~\ref{fig:493nmLaserSchematic} and~\ref{fig:redLaserLockSchematic}, are made of PbMoO$_4$ (Isomet 1205 or very similar), with center frequencies and bandwidths of $(80 \pm40)$~MHz.  A wideband AOM (Brimrose) specifically designed for use at 1025~nm shifts the doubled 2051~nm beam in a double-pass configuration before it enters the stable ULE reference cavity (Figure~\ref{fig:clockLaserApparatus}).  A fiber-coupled transverse waveguide Mg:LiNbO$_3$ modulator (Guided Color Tech.), phase matched for red lasers, puts several frequency modulation\footnote{See Section~\ref{sec:phaseModulation} for a detailed discussion of laser modulation.} sidebands on the 650~nm repump laser to cover all the allowed transitions in $^{137}$Ba$^+$ (Figure~\ref{fig:redLaserLockSchematic}).  A transverse free-space Mg:LiNbO$_3$ modulator (New Focus Inc.), resonant at 4018~MHz creates sidebands on the 493~nm cooling transition to reach the relevant odd-isotope cooling transitions (Figure~\ref{fig:barium137Transitions}).  Two capacitive-type transverse modulators (Quantum Tech.\ and Conoptics) are used at switchable dc high voltage to change the polarization states of the cooling and repump laser beams for optical pumping purposes (Figure~\ref{fig:blueRedCombineOptics}).

\section{Computer control and data acquisition}
We initially employed a PC and later an Apple Macintosh to acquire data and control the experiment.  A National Instruments 6025E PCI card provides analog input (e.g.\ for photodiode and thermocouple measurements), analog output (for laser scanning), digital output (for shutter activation, AOM and EOM control, and rf switching), and buffered digital pulse counting (for ion fluorescence measurement and precision rf pulse timing).  In 2006 we added a National Instruments USB based digital I/O device for additional (non-critical) functionality.  A GPIB board and the computer's serial bus allow us to communicate with commercial rf synthesizers, laser control units, lock-in amplifiers, and digital oscilloscopes.

We used Igor Pro by Wavemetrics Inc. to program the data-acquisition devices.  Igor is very much a Frankenstein creation that blends much of the analysis and plotting functionality in programs such as MATLAB with the data acquisition features of programs such as CVI LabWindows.  Igor runs well on both Windows and Macintosh computers, allows easy creation of graphical user interfaces, and lets us analyze our data in the same program used to acquire it.  Software timing seems to be precise at the 5~ms level;  for shorter durations, or durations requiring higher precision timing, we used the hardware timing functionality of the NI devices.

\begin{figure}
\centering
\includegraphics[width=6in]{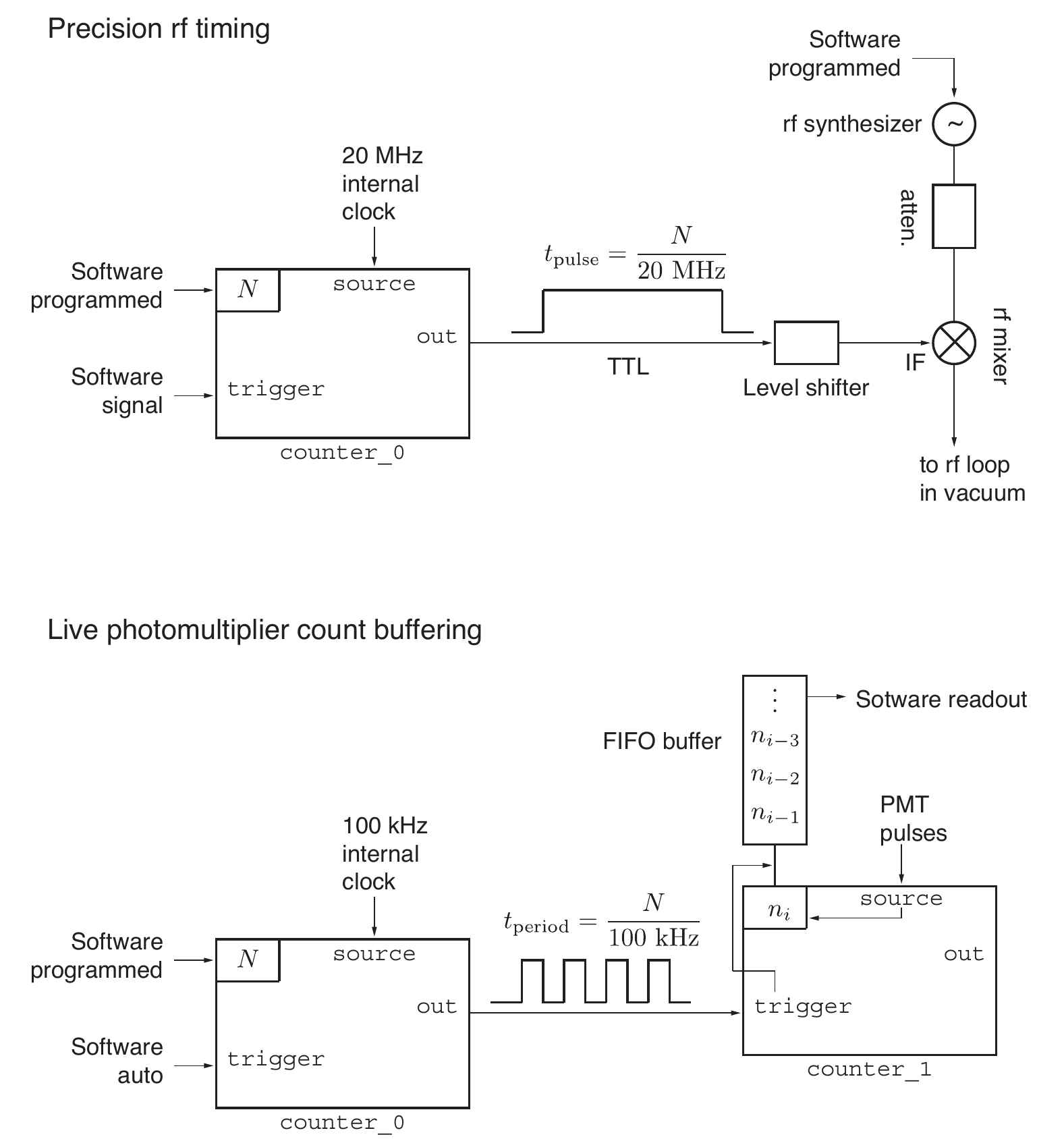}
\caption[Photon counting and precision rf pulse timing]{Photon counting and precision rf pulse timing with out data acquisition system; see National Instruments documentation for a precise description of terminology and hardware capabilities.}
\label{fig:daqTimingDiagrams}
\end{figure}

The data acquisition system configurations for photon counting and precision timing of radio frequency pulses is shown in Figure~\ref{fig:daqTimingDiagrams}.  Reprogramming the equipment in real time takes under 5~ms.  The labels and pseudocode shown use language in the NIDAQ and DAQmx Base architectures, the former of which we used on an early PC-based system, the latter of which we used on a Macintosh.

\section{Future apparatus upgrades}
\subsection*{Direct shelving via a 1762~nm laser}
\begin{figure}
\centering
\includegraphics{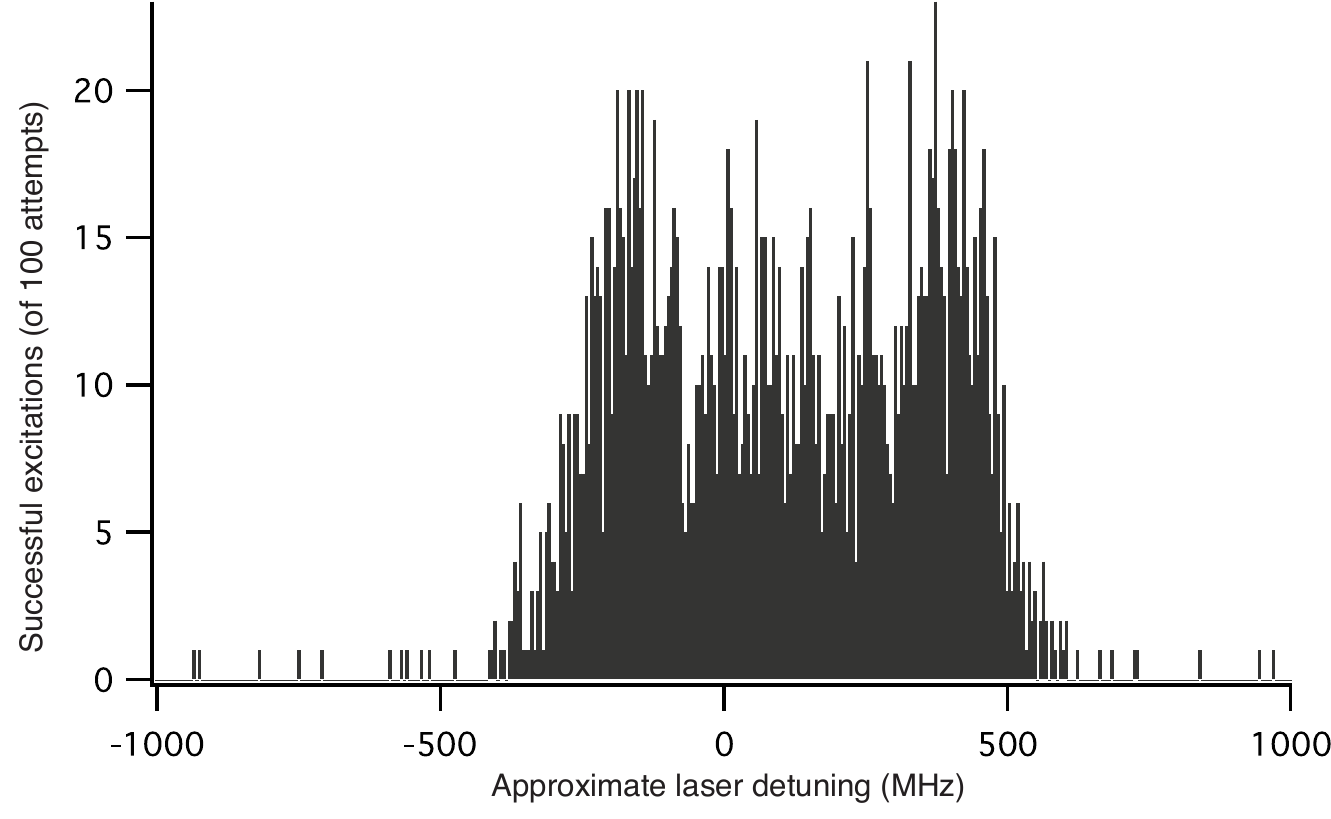}
\caption[Example of electron shelving with a 1762~nm fiber laser]{Example of electron shelving with a 1762~nm fiber laser.  Very recently, we borrowed the laser from UW researcher Dr.\ Boris Blinov and succeeded in performing adiabatic rapid passage (see Section~\ref{sec:atomicARP}) and shelving (see Section~\ref{sec:electronShelving}) on the $6S_{1/2} \leftrightarrow 5D_{5/2}$ transition.  Here, the laser takes 5~MHz steps through an approximate tuning range of 2~GHz;  100~ms of laser exposure shelves the ion when the laser is on resonance with moderate efficiency. We note that neither the laser or ion is well stabilized, and we believe the large width and character of the spectrum is due to extremely large modulation, or frequency modulation on the laser.}
\label{fig:1762sample}
\end{figure}
The narrow `direct shelving' transition $6S_{1/2} \leftrightarrow 5D_{5/2}$ was previously accessed at the University of Washington using a Nd:YAG pumped color-center laser at 1762~nm~\cite{yu1994sss}.  Though this liquid-nitrogen consuming monster still lurks in a corner of the lab, a new commercial solid state fiber laser solution is available (Koheras Adjustik) that outputs 20~mW of PZT-tunable 1762~nm light with a free-running linewidth of $\sim 50$~kHz.  University of Washington researcher Dr.\ Boris Blinov has acquired one of these lasers for use in quantum information experiments.  We have recently succeeded in driving transitions with this laser;  see the bottom graph in Figure~\ref{fig:shelvingVariations}, or some sample scanned laser shelving data in Figure~\ref{fig:1762sample}.

\subsection*{Photoionization loading schemes}
The ion-trapping community has verified that loading ions using e$^-$ bombardment is needlessly messy.  The electrons charge up nearby dielectrics (potentially disastrous when high-finesse optical cavities are located inside the vacuum) and therefore require continuous efforts at micromotion elimination.  Furthermore, since the electron-atom cross section is so low, we must make our ovens hot enough to create a large flux of atoms.  Both side-effects are not exhibited by photoionization, a scheme now employed by leading research groups.  A scheme for creating Ba$^+$ ought to be developed, perhaps involving two visible lasers.  A very recent publication by M.~Chapman's group~\cite{steele2007pip} shows success using lasers at 791~nm and 337~nm.

\subsection*{Lasers at 455~nm and 615~nm}
\begin{table}
\centering
\caption[Some parameters for SHG of 455~nm and 615~nm light using~\cite{smith2003snc}]{Some parameters for SHG of 455~nm and 615~nm light~\cite{smith2003snc}.  The slowest aspects of our apparatus during data taking are the LEDs used for the shelving and deshelving the ion via the $6P_{3/2}$ state and decays.  In the future, one may want to investigate a direct shelving scheme using the electric-quadrupole 1762~nm transition (see Figure~\ref{fig:1762sample}), or frequency doubling of infrared lasers if diode lasers at 455~nm and 615~nm are not inexpensively available.}
\begin{tabular}{m{1.5 in}l | cc | cc}
\multicolumn{2}{l}{Parameter}& 	\multicolumn{2}{c}{1228~nm $\to$ 614~nm}	& \multicolumn{2}{c}{911~nm $\to$ 455.5~nm} \\ \hline \hline
Crystal, (cut) &	& 	KNbO$_3$, ($xz$)	& LiNbO$_3$ & KNbO$_3$, ($xy$)	& LBO \\
Walkoff angle						& $\delta$	 (mrad)	&	60	& 30		& 40	 	& 13 \\
\raggedright Index of refraction (phase-matched)	& $n$& 2.210	& 2.230	& 2.272	& 1.608 \\
Phase-matching angles			& $\theta, \phi$ (deg.)\ & 29.6, 0	& 63.5, 0	&90, 66.3 & 90, 22 \\
Nonlinear coefficient				& $d_\text{eff}$ (pm/V)  & 7.58	& -4.75	& -12.5	& 0.802 \\
\raggedright Temperature sensitivity & $\Delta T$ (K$\cdot$cm) & 2.08 & 1.72	 &0.43	& 7.6
\end{tabular}
\label{tab:SHGshelvingLasers}
\end{table}
At this time, rare laser diodes at or near 455~nm are being produced by Toptica in extremely small quantities.  It does not seem that 615~nm are forthcoming, so it is natural to consider options for second-harmonic generation processes to get these blue and orange wavelengths at lower costs than a dye laser.  Powerful, inexpensive 910~nm laser diodes are common, and 1230~nm radiation is currently available from new diode lasers, Yb-doped fiber lasers, or high power pulsed Cr:Forsterite lasers.  Table~\ref{tab:SHGshelvingLasers} details good second-harmonic generation options.

\subsection*{Trap design}
Future experiments will require robust micromotion minimization, and a more rigid trap structure than the twisted wire ring currently provides.  Some new ideas relevant to the proposed atomic parity violation experiment require a linear trap geometry to take advantage of several ions whose spacings are controllable with dc endcap voltages.  The trap must be made compatible with one or more intra-vacuum mirrors or optical cavities, and present several non-orthogonal lines of access for the cooling laser in order to detect and cancel micromotion.  We have found it extremely useful to clean the trap by passing several amperes of ac across it;  the vacuum feedthrough structure of any future linear trap should be modular enough to enable this procedure.

\subsection*{A 685~nm laser for a $6P_{1/2}$ branching ratio measurement}
\begin{figure}
\centering
\includegraphics{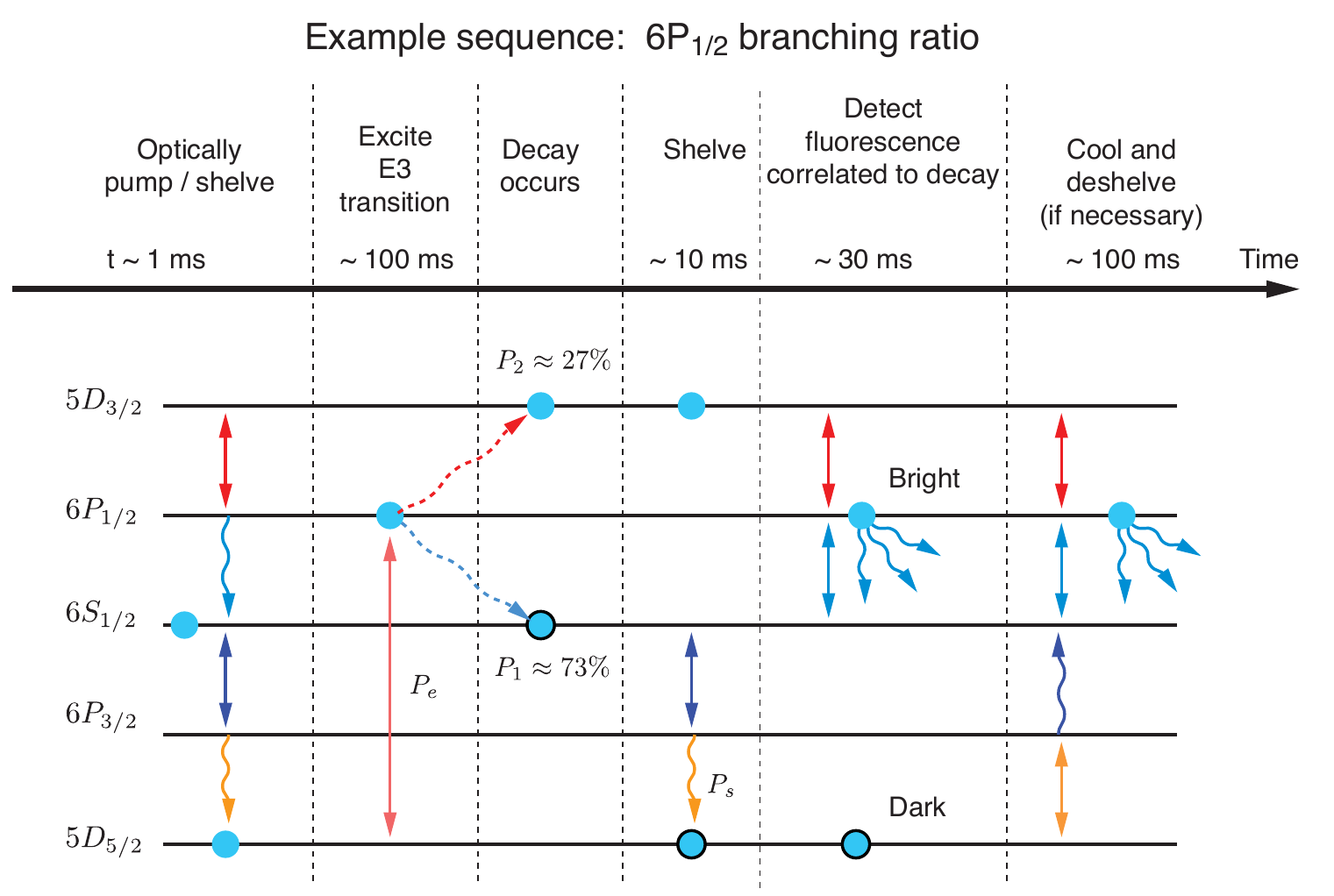}
\caption[A method of measuring the $6P_{1/2}$ decay branching ratio]{A precision measurement of the decay branching ratio of $6P_{1/2}$ would yield valuable information about the dipole matrix elements $\langle 6S_{1/2} || er || nP \rangle$ and $\langle 5D_{3/2} || er || nP \rangle$.  One way to accomplish a clean excitation to $6P_{1/2}$ is through an electric octopole transition $5D_{5/2} \leftrightarrow 6P_{1/2}$ at 685~nm.  The excitation probability $P_e$ needs to be controlled and measured during the experiment, as does the shelving probability $P_s$.}
\label{fig:branchingRatioSequence}
\end{figure}
Chapter~\ref{sec:lightShiftChapter} focuses on one attempt to precisely measure the ratio of $\langle 6S_{1/2} || er || nP \rangle$ and $\langle 5D_{3/2} || er || nP \rangle$ dipole matrix elements. Another scheme is the precision measurement of the branching ratio of electric dipole spontaneous decay out of $6P_{1/2}$ to the ground and metastable states $6S_{1/2}$ and $5D_{3/2}$.  A challenging but fruitful way of performing the controlled excitation to the $6P_{1/2}$ state is through the electric octopole transition $5D_{5/2} \leftrightarrow 6P_{1/2}$ at 685~nm.  We have modified an existing home made external cavity diode laser to function at this wavelength.  A sketch of the experimental sequence is proposed in Figure~\ref{fig:branchingRatioSequence}.

\chapter{Precision measurement of light-shifts} \label{sec:lightShiftChapter}
\begin{quotation}
\noindent\small Bishop, honey, bring the cart.  Don't run anyone over. \\ \flushright{---Anon., overheard}
\end{quotation}
\section{Motivation:  a key test of atomic structure}
\begin{table}
\centering
\caption[The hardships of precision atomic measurements, abridged]{The hardships of precision atomic measurements, abridged.  Most references are to trapped ion work, or measurements relevant to Ba$^+$.}
\begin{tabular}{cm{1 in}m{3.5 in}}
To test... & measure... & which is often hard because... \\ \hline \hline
$| \langle \psi_i | H_0 | \psi_i \rangle |$
& \raggedright State energies    
& \footnotesize The usual spectroscopic troubles:  Doppler and collisional shifts, broadening, need a spectrometer, one laser per state, or reliance on angle tuning a Doppler shifted beam. Fast decays limit precision.  Often, energy levels alone do not satisfy theorists' needs.  \\ \hline
$|\psi_i(0)|^2$
& \raggedright Hyperfine structure  
& \footnotesize  If an optical measurement of a metastable state, the usual broadening occurs but shifts are less crucial.  Rf probes of ground states are exquisite and straightforward, though. Compare laser experiment \cite{villemoes1993ish} with a microwave trapped ion \cite{trapp2000hsa} study. \\ 
\hline
\multirow{4}{*}[-0.75in]{$| \langle \psi_i | r | \psi_j \rangle |$}  
& \raggedright Transition rates 
& \footnotesize If direct, one must measure a laser intensity~\cite{kastberg1993mat,shoemaker2006dmt}, have fast and accurate timing. Signal is analog, so the detector must be linear.  Indirect double-resonance or avoided crossing experiments must be calibrated~\cite{gallagher1967osc}. \\ \cline{2-3}
& \raggedright State lifetimes
& \footnotesize  Short lifetimes $\sim 1$~ns require pulsed excitation scheme, accurate timing~\cite{moehring2006plm}.  Extraordinarily long lifetimes~\cite{yu1997rlm} demand proper treatment of collisional effects, background fields that may allow faster decay through admixture. \\ \cline{2-3}
& \raggedright Optical rotation
& \footnotesize Birefringent materials must be stabilized and calibrated, signal is totally analog so polarizer alignment and detector linearity are crucial.  Systematic effects scale with laser power, frequency, or the vapor optical thickness.  When the effect can be made differential, this technique can be extremely precise~\cite{meekhof1993hpm,lucas1998spn}. \\  \cline{2-3}
& \raggedright Polarizability, Stark effect, light shifts
& \footnotesize One must calibrate an electric field (see \cite{snow2005ddq} using Rydberg atoms); the measurement of the Stark shift must be systematic free itself.  Usually these experiments go the other way:  measure Stark shifts to determine an electric field.~\cite[Section 2.3]{budker2004ape}  \\ \hline
$| \langle \psi_i | H_\text{pnc} | \psi_j \rangle | $ 
& \raggedright  Parity violation or other exotic physics
& \footnotesize Effects are dwarfed by allowed processes that turn into systematic effects.  Interference experiments solve this but are complicated \cite{bouchiat1997pva}.
\end{tabular}
\label{tab:precisionMeasurementHardships}
\end{table}
Precision measurement of atomic physics is a tough business:  a slightly whimsical Table~\ref{tab:precisionMeasurementHardships} cites some reasons and contains references to a few selected experiments relevant to Ba$^+$ atomic structure.  Precision measurements on basic and heavy atoms are still vital for understanding the complicated, computer driven, many-body models used to explain the details of atomic structure.  Also, atomic physics techniques are essentially the only means to test the Standard Model at low energies---in order to do so, heavy atom atomic theory has to be confirmed to extraordinarily high accuracy as well.  Singly ionized barium is a somewhat ideal laboratory in which to test modern atomic theory:  it is a many electron atom but has only one valence electron, matching the case in Cesium which has already undergone extensive successful tests of Standard Model physics~\cite{noecker1988pmp, wood1997mpn} and corresponding theoretical study (\cite{derevianko2001cmb}, for instance).  At $Z = 56$, Ba$^+$ takes advantage of the heavy atom $Z^3$ scaling~\cite{bouchiat1997pva} (see also Chapter~\ref{sec:ParityChapter}) in the effects of atomic parity violation, and has large fine-structure energies which enhance the effects of potentially varying fundamental constants such as $\alpha$~\cite{dzuba2000aoc}.

However, at present, state of the art atomic theory for dipole matrix elements involving non-$s$ wavefunctions has been tested to only 5\% accuracy.  On the other hand, dipole matrix elements involving $s$ wavefunctions are known experimentally and theoretically to about 1\%.  Here we describe a method of improving the knowledge of some of the poorly known matrix elements in the barium ion which will help atomic theorists improve their models further.

As shown in Table~\ref{tab:precisionMeasurementHardships}, measurements of dipole reduced matrix elements often require exquisite calibration of either a laser intensity, an atomic decay time, a laser polarization, or an electric field.  To avoid such challenges one can instead measure \emph{ratios} of matrix elements sharing dependence on quantities hard to calibrate.  In this chapter we describe our success in measuring the vector ac-Stark or light shift ratios in two spin resonances of the barium ion.  We will show that by performing the measurement with multiple light shift wavelengths, one can extract atomic dipole matrix elements.  

As a historical note, work on single ion light shifts began in earnest in 2000.  By 2003, single ion spin flip experiments were routinely carried out by colleague Timo K\"{o}rber.  Over the next three years, we perfected the experimental method, and succeeded in resolving light shifts using off resonant lasers at 633~nm (a free-running dye laser), 514~nm (an Ar-ion laser), and at 1111~nm (a Yb-doped fiber laser).  The early data at 633~nm is presented in~\cite{koerber2003thesis} but not here because its accuracy is not acceptably known to us.  The power of this laser, 25~mW at the site of the ion, was not sufficient;  nor was the control of the beam polarization and alignment.

First, we describe the light shift phenomenon in multi-state atoms with non-degenerate magnetic sublevels.  We next describe a pioneering method for performing narrow electron spin resonance spectroscopy in the $6S_{1/2}$ and $5D_{3/2}$ states of Ba$^+$~\cite{koerber2002rss,koerber2003rfs}.  We then demonstrate precision measurement of off-resonance light shifts in these resonances at two wavelengths, 514~nm and 1111~nm.  We find that a quantity called the light shift ratio
\begin{equation}
R \equiv \frac{\Delta_S}{\Delta_D} \equiv \frac{\Delta E_{6S_{1/2}, m=1/2} - \Delta E_{6S_{1/2}, m=-1/2}}{\Delta E_{5D_{3/2}, m=1/2} - \Delta E_{5D_{3/2}, m=-1/2}} 
\end{equation}
is free to first order of many potentially troubling systematics, including the laser intensity, and errors in alignment and polarization.  At the two wavelengths sampled, we measure
\begin{align}
R( 514.531 \text{ nm} ) &= -11.494(13) \qquad (\text{0.11 \% precision}), \\
R( 1111.68 \text{ nm} ) &= +0.4176(8) \qquad   (\text{0.19 \% precision}).
\end{align}
We believe these form a new sensitive test of many body atomic models, including \emph{ab initio} methods and will lead to a better understanding of Ba$^+$ atomic structure.

\begin{figure}[p]
\centering
\includegraphics[width = 5.5 in]{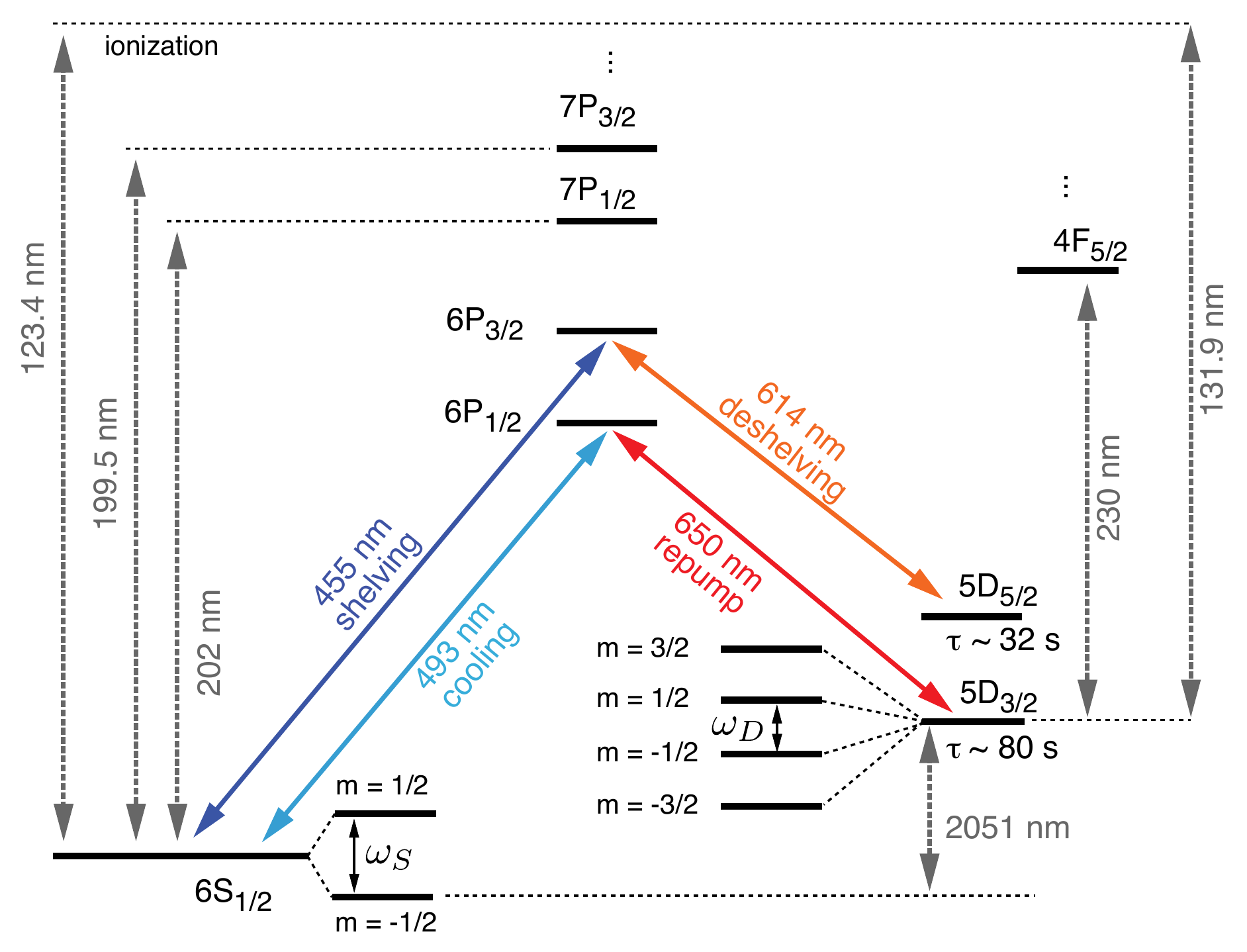}
\caption[Barium energy levels relevant to light shift experiments]{Lowest barium energy levels relevant to light shift experiments.  The real transitions performed in the experiments are labeled with solid arrows: cooling, repumping, shelving and deshelving optical transitions, and radio frequency Zeeman transitions $\omega_S$ and $\omega_D$.  The dashed line arrows indicate the most important energy level splittings necessary for the estimation of light shifts to spin resonances $\omega_S$ and $\omega_D$.}
\label{fig:bariumEnergyLevelsLightShifts}
\end{figure}

\section{Light shifts in non-degenerate Zeeman sublevels} \label{sec:lightShiftZeeman}
Recall from Section~\ref{sec:acStarkAtomic} that an off-resonant light beam shifts the energies of atomic states by
\begin{equation} \label{eq:lightShiftSimpleForm}
\Delta E_{g,e} = \pm \frac{\Omega^2}{4 \delta},
\end{equation}
where $\Omega$ is a Rabi frequency that scales linearly with the light electric field strength, and depends on an atomic dipole matrix element $\langle g | er | e \rangle$, and $\delta$ is the detuning of the light beam from a nearby electric dipole resonance.  When light shifts are kept small with respect to other energy scales in the atom (e.g.\ the fine structure splitting), Eq.~\ref{eq:lightShiftSimpleForm} is easily generalized to multi-state atoms.  The light shift of a state $|\gamma, j, m\rangle$ due to an oscillating field $\boldsymbol{E}(t) = \boldsymbol{E} \cos \omega_L t$ is, to second-order in perturbation theory \cite{stalnaker2006dse},
\begin{equation}
\Delta E_{\gamma, j, m} = \frac{(E)_i(E)_j}{4} \sum_{\gamma', j', m', \pm} \frac{\langle \gamma,j,m | e r_i | \gamma', j', m' \rangle \, \langle \gamma',j',m' | e r_j | \gamma, j, m \rangle}{E_{\gamma,j,m} - E_{\gamma',j',m'} \pm \hbar \omega_L},
\end{equation}
where $\boldsymbol{E}$ is the electric field strength times a polarization vector $\boldsymbol{\epsilon}$, $e$ is the electron charge, and summation over the indices $i$ and $j$ is implied.  The term containing $+ \hbar \omega_L$ in the denominator (termed the `Bloch-Siegert' shift in the context of radio frequency spectroscopy) is often ignored when the rotating wave approximation is made (see Eq.~\ref{eq:lightShiftsBlochS}) but must be included here.  Contributions from core excitations and coupling to continuum states must be included for an accurate calculation.  We have reasons to believe that corrections due to the continuum are small due to studies on the dc-Stark effect in Ba$^+$~\cite{stambulchik1997ehn} that imply the fraction of the dc shift due to high-$n$ and continuum levels is $\approx 1.4 \times 10^{-4}$ for $6S_{1/2}, m=\pm1/2$ states and $\approx 2.9 \times 10^{-2}$ for $5D_{3/2}, m=\pm 1/2$ states\footnote{The dc-Stark effect does not have a vector component like our ac-Stark effect with circularly polarized light, so this argument should be considered very loose}.  The next order in perturbation theory, called the \emph{hyperpolarizability} is unimportant unless the supposedly off-resonant laser is accidentally resonant with a two-photon transition (see~\cite{brusch2006hes}, for instance).

It is also possible to write the light shift Hamiltonian in terms of irreducible tensors of rank 2 and lower \cite{stalnaker2006dse}:
\begin{equation}
H = \underbrace{(\epsilon_i \epsilon_j^* \delta_{ij})}_\text{scalar} T_{ji}^{(0)} + 
\underbrace{(\epsilon_i \epsilon_j^* - \epsilon_j \epsilon_i^*)}_\text{vector}  T_{ji}^{(1)} +
\underbrace{(\epsilon_i \epsilon_j^* - \tfrac{1}{3} \epsilon_i \epsilon_j^* \delta_{ij}) }_\text{tensor} T_{ji}^{(2)}.
\end{equation}
The braced terms are the unique irreducible spherical tensors formed by the polarization vector $\boldsymbol{\epsilon}$, and each term is contracted with the associated tensor operator $T_{ji}^{(k)}$. 

\begin{figure}
\centering
\includegraphics{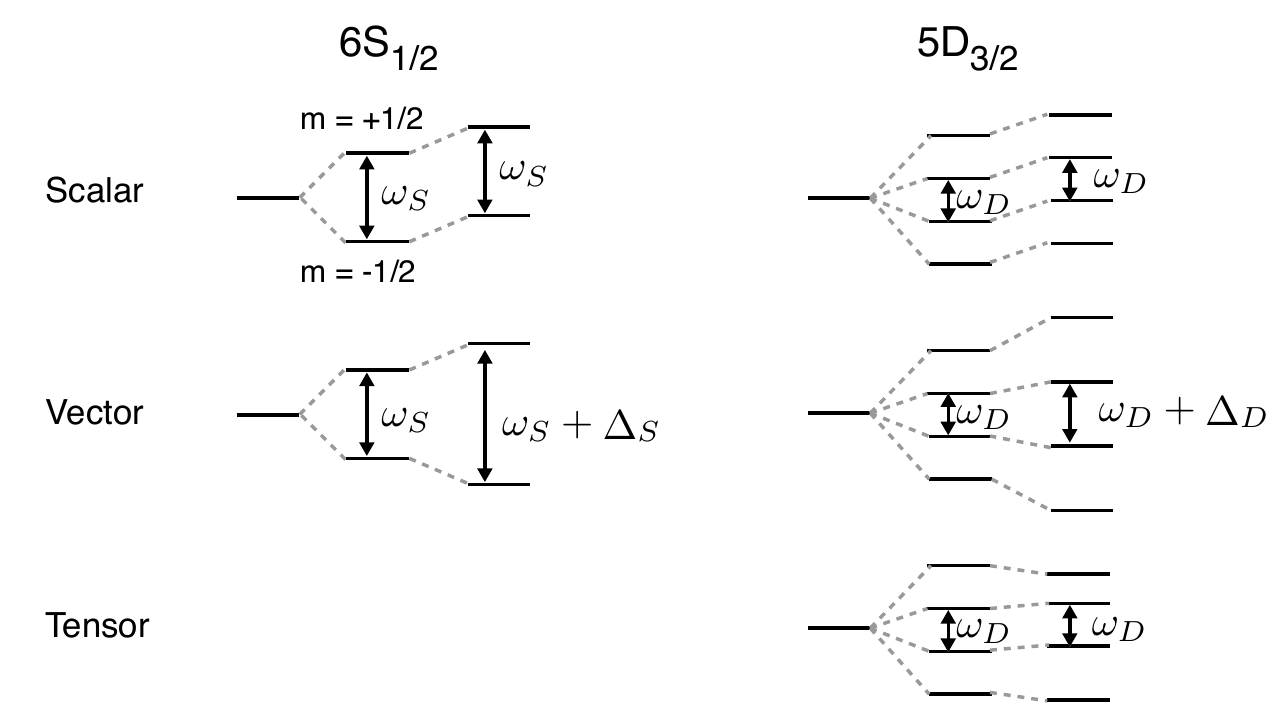}
\caption[Structure of the multipole light shifts in the $6S_{1/2}$ and $5D_{3/2}$ states.]{Structure of the multipole light shifts in the $6S_{1/2}$ and $5D_{3/2}$ states.  In this schematic, we imagine a $\sim 1$ G magnetic field splitting the $6S_{1/2}$ and $5D_{3/2}$ sublevels by $\omega_S$ and $\omega_D$ in the vector structure of the Zeeman interaction.  Further (assumed small) perturbations due to the ac-Stark effect add scalar, vector, and tensor-like shifts.}
\label{fig:lightShiftMultipoleCartoon}
\end{figure}

If the light shifts are much smaller than the Zeeman energies, the resulting energy shift also can be written in terms of tensor ranked polarizabilities  $\alpha_0$, $\alpha_1$, and $\alpha_2$ \cite{stalnaker2006dse}:
\begin{equation}
\Delta E_{\gamma, j, m} = - \underbrace{\frac{\alpha_0}{2} | \boldsymbol{E} |^2}_\text{scalar}
- \underbrace{i \frac{\alpha_1}{2} \frac{m}{j} (i |\boldsymbol{E} \times \boldsymbol{E^*}|) }_\text{vector}
- \underbrace{\frac{\alpha_2}{2} \left( \frac{3m^2 - j(j+1)}{j(2j-1)}\right) \frac{3 |E_z|^2 - |\boldsymbol{E}|^2}{2}}_\text{quadrupole/tensor},
\end{equation}
which should be compared to the result for static electric fields:
\begin{equation}
\Delta E^\text{static stark}_{\gamma,j,m} = - \underbrace{\frac{\alpha_0}{2} |\boldsymbol{E}|^2}_\text{scalar} - \underbrace{\frac{\alpha_2}{2} \left( \frac{3m^2 - j(j+1)}{j(2j-1)}\right) \frac{3 |E_z|^2 - |\boldsymbol{E}|^2}{2}}_\text{quadrupole/tensor}.
\end{equation}
Unlike the dc-Stark effect (due to static electric fields), the ac-Stark effect can result in vector-like shifts.  In general, the vector shift term is maximal for pure circularly polarized light aligned with any existing magnetic field.  Other researchers call this arrangement the `Zeeman-like' ac-Stark shift~\cite{park2001mzl} or a `fictitious' magnetic field~\cite{haroche1970mzh}, and have employed the shift as a practical replacement for a magnetic field:  e.g., an ac-Stark optical Stern-Gerlach effect~\cite{park2002osg}, an optically induced Faraday effect~\cite{cho2005oif}, an optical Feshbach resonance~\cite{dickerscheid2005fro}, an ac-Stark phase shift compensator akin to a spin-echo~\cite{haffner2003pma}, even a technique to enable highly forbidden $J = 0 \leftrightarrow 0$ optical `clock' transitions in optical lattice trapped atoms~\cite{ovsiannikov2007mwi}.

\begin{table}
\centering
\caption[Scalar, vector, and tensor coefficients used in the light shift ratio estimate]{Scalar, vector, and tensor coefficients used in estimating the light shift ratio shift magnitudes in the $m = \pm 1/2$ splittings in the $6S_{1/2}$ and $5D_{3/2}$.  In particular, the constants $v(j,j')$ are necessary in the computation of our measured light shift ratios in Eq.~\ref{eq:lightshiftRatioEstimate}.}
\begin{tabular}{c | c  c  c}
	   & \multicolumn{3}{c}{Relevant light shift coefficients} \\
$(j, j')$ & Scalar $s(j,j')$ & Vector $v(j,j')$ & Tensor $t(j,j')$ \\ \hline \hline
(1/2, 1/2) & -1/6 & -1/3 & 0 \\
(1/2, 3/2) & -1/2 & 1/6 & 0 \\
(3/2, 1/2) & -1/12 & -1/12 & 1/12 \\
(3/2, 3/2) & -1/12 & -1/30 & -1/15 \\
(3/2, 5/2) & -1/12 & 1/20 & 1/60
\end{tabular}
\label{tab:lightShiftCoefficients}
\end{table}
\begin{table*}
\centering
\caption[Collection of calculated and measured $5D_{3/2}$ and $6S_{1/2}$ dipole matrix elements]{Collection of calculated and measured $5D_{3/2}$ and $6S_{1/2}$ dipole matrix elements.  Data from \cite{dzuba2001cpn} are derived from radial integrals.  Data from~\cite{kastberg1993mat} are derived from absolute transition rates with signs inferred from other references.  The light shift ratio $R$ estimates use data from~\cite{gopakumar2002edq} when it is otherwise not available.}
%\longtable*   --  spans whole page
%\footnotemark
%\footnotetext
%dcolumn.sty -- align d columns along decimal point
\small 
\begin{tabular}{cl | lll | ll}
					&					& \multicolumn{5}{c}{Dipole matrix elements (a.u)} \\
					&					& \multicolumn{3}{c}{Theory} & \multicolumn{2}{c}{Experiment} \\
Transition				& Splitting (cm$^{-1}$)  	& \multicolumn{1}{c}{Ref.~\cite{gopakumar2002edq}}  &   \multicolumn{1}{c}{Ref.~\cite{dzuba2001cpn}} &   \multicolumn{1}{c}{Ref.~\cite{guet1991rmb}}  &  \multicolumn{1}{|r}{Ref.~\cite{kastberg1993mat}} &  \multicolumn{1}{r}{Ref.~\cite{gallagher1967osc}} \\ \hline \hline
$6S_{1/2}$--$6P_{1/2}$	& 20261.561			& 3.3266            &  3.310        & 3.300        & 3.36(16)  & 3.36(12) \\
$6S_{1/2}$--$6P_{3/2}$	& 21952.404			& -4.6982           & -4.674       & (-)4.658     & 4.45(19) &  4.69(16) \\
$6S_{1/2}$--$7P_{1/2}$	& 49389.822			& 0.1193            & -0.099       & & &\\
$6S_{1/2}$--$7P_{3/2}$	& 50011.340			& -0.3610           & -0.035       & & &\\
$6S_{1/2}$--$8P_{1/2}$	& 61306				& -0.4696	          & -0.115       & & &\\
$6S_{1/2}$--$8P_{3/2}$	& 61613				& 0.5710             & 0.073        & & &\\ \hline
$5D_{3/2}$--$6P_{1/2}$	& 15387.708			& -2.9449            & 3.055       & 3.009        & 3.03(9)  & 2.99(18) \\
$5D_{3/2}$--$6P_{3/2}$	& 17078.552			& -1.2836            & -1.334      & (-)1.312    & 1.36(4) & 1.38(9) \\
$5D_{3/2}$--$7P_{1/2}$	& 44515.970			& -0.3050            & 0.261       & & &\\
$5D_{3/2}$--$7P_{3/2}$	& 45137.488			& -0.1645            & -1.472      & & &\\
$5D_{3/2}$--$8P_{1/2}$	& 56432					& -0.1121            & 0.119       & & & \\
$5D_{3/2}$--$8P_{3/2}$	& 56739					& -0.0650            & -0.070      & & &\\ \hline
$5D_{3/2}$--$4F_{5/2}$	& 43384.765			& -3.69 \cite{das2004pc}	            &                  & & & \\
$5D_{3/2}$--$5F_{5/2}$	& 52517.070			& 1.59  \cite{das2004pc}                 &                  & & &\\
$5D_{3/2}$--$6F_{5/2}$	& 64337.85 					& 0.44 \cite{das2004pc}                  &                  &  & &\\ \hline \hline
\multicolumn{2}{l|}{$R_{\lambda = 514 \text{ nm}}$ prediction} & -13.4056 & -12.5517 & -12.75  & -13.2099 & -13.3477 \\
\multicolumn{2}{l|}{$R_{\lambda = 1111.6 \text{ nm}}$ prediction} & 0.4444 & 0.4146 & 0.4168 &  0.5001 & 0.5110
\end{tabular}
\label{tab:dipoleMatrixElementTable}
\end{table*}

\begin{figure}
\centering
\includegraphics{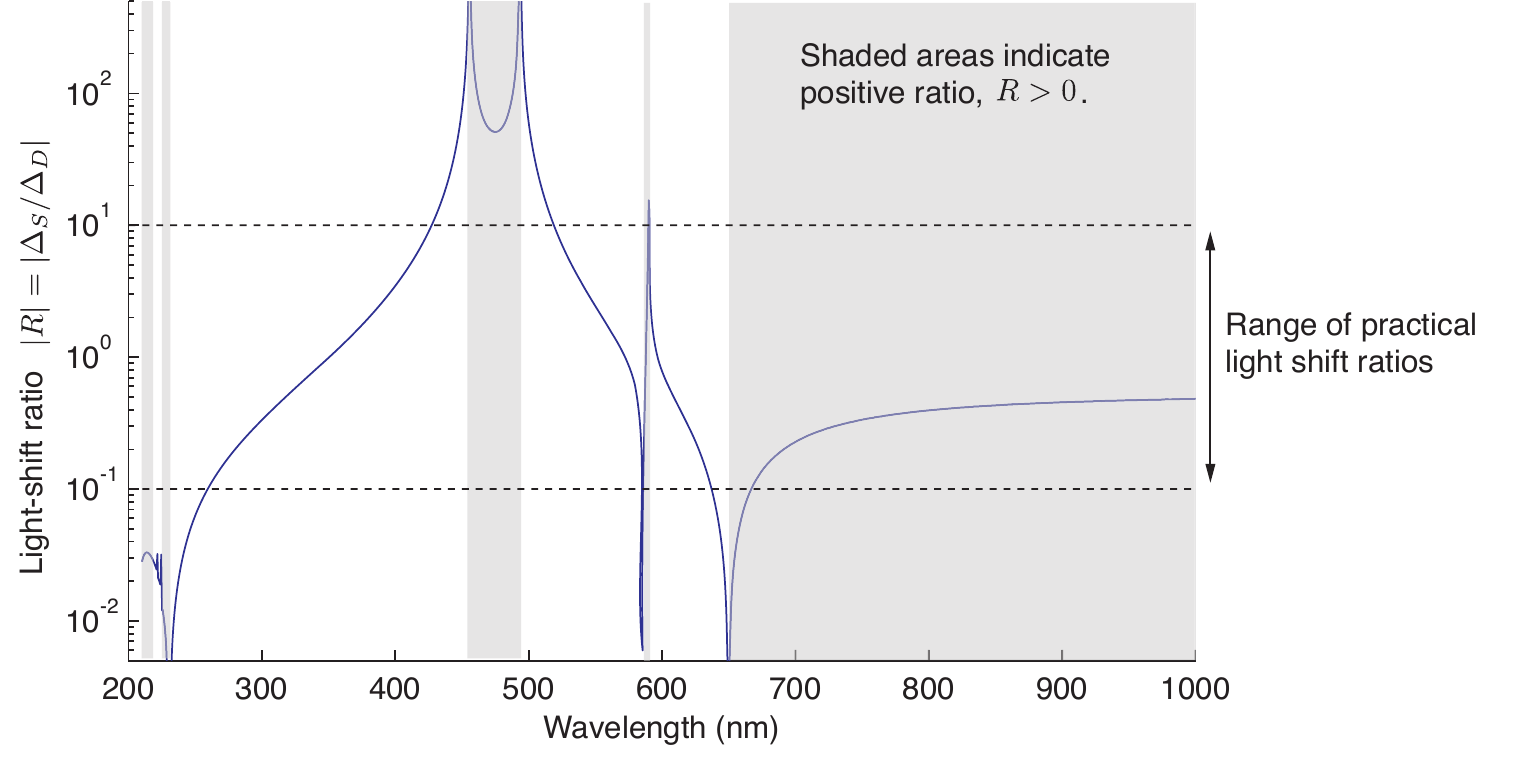}
\caption[Logarithmic plot of $|R| = |\Delta_S / \Delta_D|$ and practical guidelines]{A logarithmic plot of $|R| = |\Delta_S / \Delta_D|$ allows us to asses the feasibility of a light shift ratio experiment at various off-resonant wavelengths.  A reasonable guideline is to ensure $0.1 < |R| < 10$ so that one can make both $\Delta_D$ and $\Delta_S$ large enough to obtain good sensitivity against drift while keeping both $\Delta_D$ and $\Delta_S$ smaller than Zeeman shifts $\omega_D$ and $\omega_S$ in order to avoid misalignment and polarization systematic effects.  One must also consider the stability of various lasers in this regime.}
\label{fig:lightShiftRatioLog}
\end{figure}

\begin{figure}
\centering
\includegraphics{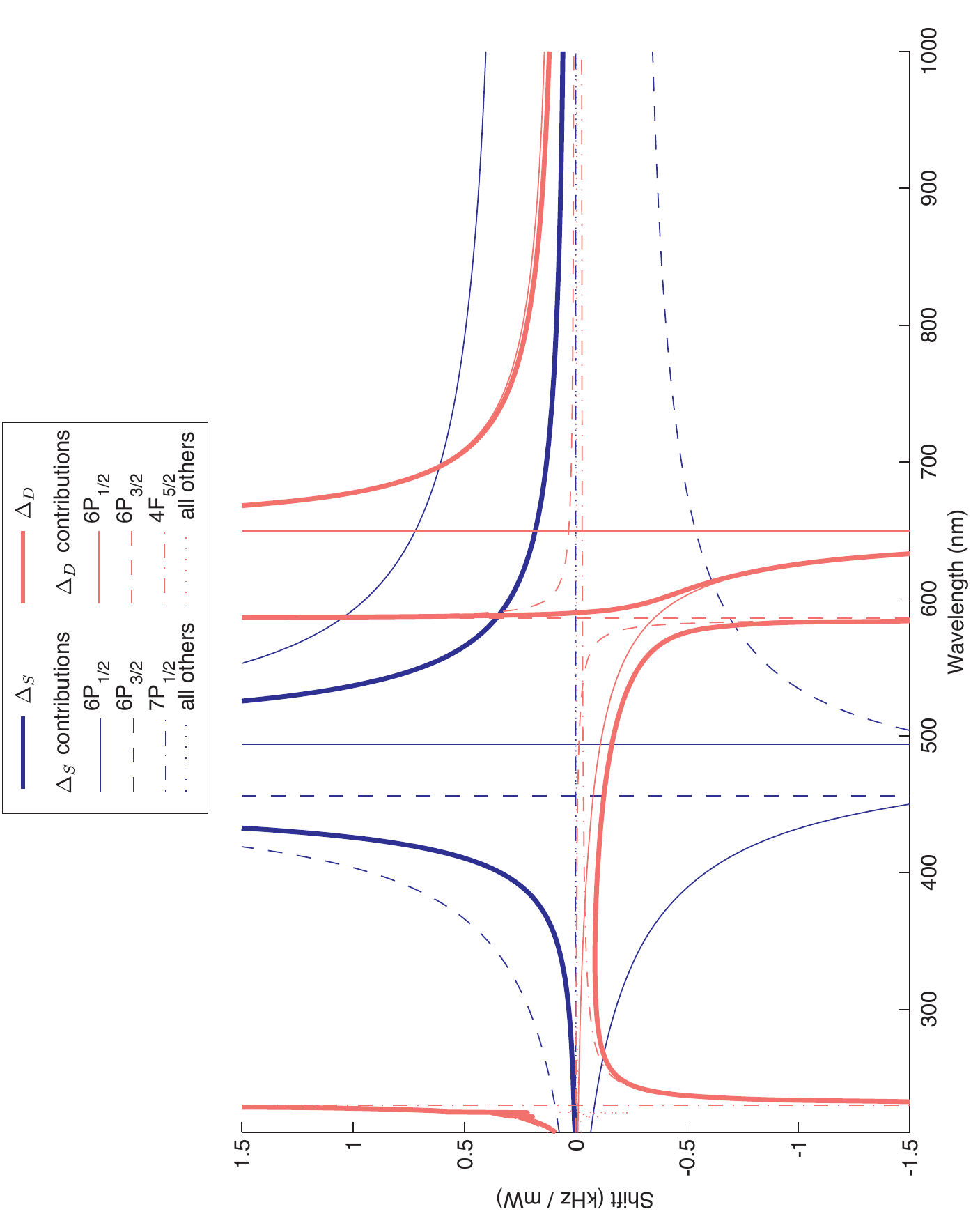}
\caption[Estimates of light shifts $\Delta_S$ and $\Delta_D$ (kHz/mw) over the visible spectrum]{Here we show estimates of light shifts $\Delta_S$ and $\Delta_D$ (kHz/mw) over the visible spectrum.  In each case we assume a 20~$\mu$m spot with perfect circular polarization.  All states listed in Table~\ref{tab:dipoleMatrixElementTable} are included in the calculation, and the contributions of the most important are shown here.}
\label{fig:lightShiftEstimate}
\end{figure}

\begin{figure}
\centering
\includegraphics{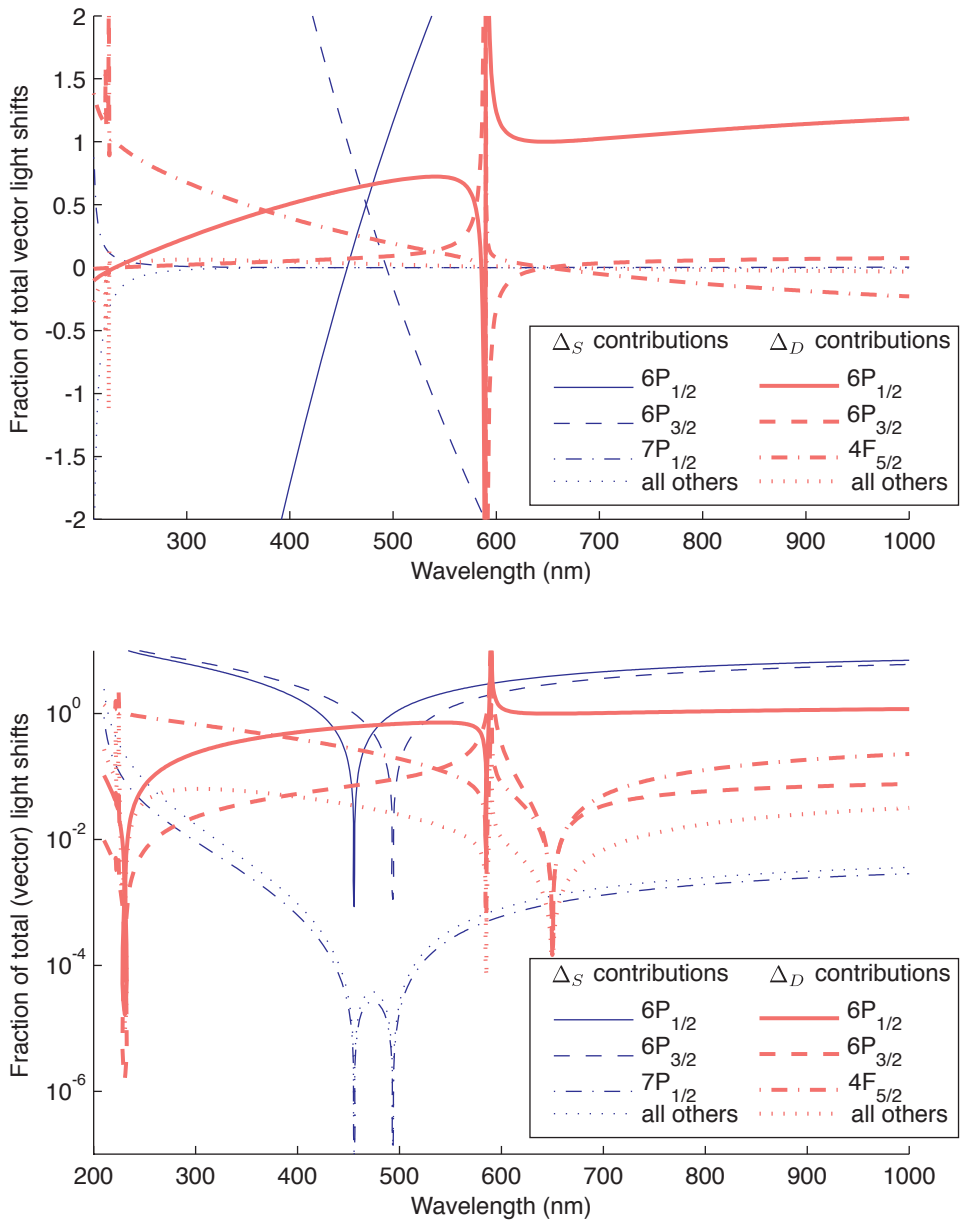}
\caption[Fractional contributions to light shifts $\Delta_S$ and $\Delta_D$]{Plotted are the fractional contributions of several excited states to the (vector) light shifts $\Delta_S$ and $\Delta_D$.  For clarity, the top graph shows a linear scale, including the \emph{sign} of each contribution.  A key observation is that the contributions of $6P_{1/2}$ and $6P_{3/2}$ to $\Delta_S$ largely cancel outside of the fine-structure splitting.  The bottom graph demonstrates how each contribution changes in relative weight over many orders of magnitude across the visible spectrum.}
\label{fig:lightShiftEstimateFractional}
\end{figure}

We wish to focus on vector light shifts $\Delta_S$ and $\Delta_D$ to the $6S_{1/2}, m= \pm 1/2$ and $5D_{3/2}, m= \pm 1/2$ splittings, $\omega_S$ and $\omega_D$ (refer to Figure~\ref{fig:bariumEnergyLevelsLightShifts}).  As shown in Figure~\ref{fig:lightShiftMultipoleCartoon}, only vector pieces of the ac-Stark effect can accomplish such shifts.  Therefore, scalar and tensor light shifts are not direct systematic effects (though we will see that the tensor piece in the $5D_{3/2}$ state plays a role in small systematic $\omega_D$ resonance line-shape distortions).  If we define shifted resonances
\begin{align}
\omega_S^{LS} &= \omega_S + \Delta_S, \\
\omega_D^{LS} &= \omega_D + \Delta_D, \label{eq:deltaDdef}
\end{align}
then we can form a light shift ratio which results from the measurement of two shifted and two unshifted resonance measurements
\begin{equation}
R \equiv \frac{\Delta_S}{\Delta_D} = \frac{\omega_S^{LS} - \omega_S}{\omega_D^{LS} - \omega_D}.
\end{equation}
In terms of atomic dipole reduced matrix elements, the light shift ratio is
\begin{align}
R = \frac{\Delta_S}{\Delta_D} &= \frac{\Delta E_{6S_{1/2}, m=1/2} - \Delta E_{6S_{1/2}, m=-1/2}}{\Delta E_{5D_{3/2}, m=1/2} - \Delta E_{5D_{3/2}, m=-1/2}} \\
	&= \frac{\displaystyle \sum_{\gamma', j', \pm} v(\tfrac{1}{2},j') \frac{ | \langle 6S_{1/2} || r || \gamma', j' \rangle |^2}{(W_{\gamma',j'} - W_{6S_{1/2}} \pm \hbar \omega)}}{ \displaystyle \sum_{\gamma', j', \pm} v(\tfrac{3}{2},j')  \frac{| \langle 5D_{3/2} || r || \gamma' j' \rangle |^2}{(W_{\gamma', j'} - W_{5D_{3/2}} \pm \hbar \omega)}}, \label{eq:lightshiftRatioEstimate}
\end{align}
where $E_{\gamma', j'}$ is the energy of state $|\gamma', j' \rangle$, $\omega$ is off-resonant light shift laser frequency, and $v(j,j')$ is a vector light shift coefficient calculated from Clebsch-Gordan coefficients and tabulated with similar scalar and tensor shift coefficients in Table~\ref{tab:lightShiftCoefficients}.  The dipole reduced matrix elements $\langle 6S_{1/2} || r || \gamma', j' \rangle$ and $\langle 5D_{3/2} || r || \gamma', j' \rangle$ are the subject of study.  Current theoretical calculations and previous measurements of these are shown in Table~\ref{tab:dipoleMatrixElementTable}.  Using these data and Eq.~\ref{eq:lightshiftRatioEstimate}, we compute an estimate of the light shift ratio $R$ as a function of laser wavelengths and plot the result in Figure~\ref{fig:lightShiftRatioLog}.

Figures~\ref{fig:lightShiftEstimate} and \ref{fig:lightShiftEstimateFractional}  show the relative contributions of each excited state coupling to the shifts $\Delta_S$ and $\Delta_D$.  We see that the light shifts $\Delta_D$ are, in general, large near dipole resonances involving the $5D_{3/2}$ state while $\Delta_S$ light shifts are large near transition wavelengths of resonances with the ground state $6S_{1/2}$.  The expected $1/\delta$ character of the light shifts is modified by the presence of fine-structure.  For instance, the contributions of $6P_{1/2}$ and $6P_{3/2}$ to the \emph{vector} light shift $\Delta_S$ are opposite in sign and therefore tend to suppress the magnitude of the shift by $1/\delta^2$ at wavelengths far away from the fine structure splitting.  This is understood by remembering that the difference in light shift between $m =-1/2$ and $m = +1/2$ depends only on the electron spin.

\section{Single ion spin-flip spectroscopy} \label{sec:spinResonanceLS}
The core of the experiment is the measurement of electron spin resonances $\omega_S$ and $\omega_D$ (refer to Figure~\ref{fig:bariumEnergyLevelsLightShifts}) with and without an applied light shift laser.  But how does one measure an electron spin resonance in a single ion?  We have developed a technique that employs a spin-dependent version of the electron shelving mechanism discussed in Section~\ref{sec:electronShelving} to detect whether pulses of rf are resonant with the transitions $\omega_S$ and $\omega_D$.  This section describes the method, which is not totally straightforward, so we will give an overview here:
\begin{enumerate}
\item Using circularly polarized light, we prepare a cold trapped ion in a chosen atomic spin state in either $6S_{1/2}$ or $5D_{3/2}$.
\item With all resonant lasers turned off, we subject the ion to a pulse of radio frequency energy designed to be a $\pi$-pulse (see Figure~\ref{fig:twoStateRabiFreq} for a review) when on resonance, causing a spin flip.
\item Using a brief, dim pulse of polarized laser light, we probe whether the spin flip has occurred by translating the spin information into state information via a decay from $6P_{1/2}$.
\item A pulse of 455~nm shelving light moves any ion population still in the $6S_{1/2}$ state to the $5D_{5/2}$ level, where it is held safely due to the long metastable lifetime of $5D_{3/2}$ ($\tau \sim 30$~s).
\item Resonant cooling and re-pumping lasers detect if the ion is in the shelved state, which by design is correlated to whether an rf spin flip occurred.
\item The process is repeated hundreds of times to build up statistics and at dozens of spin-flip frequencies to form the resonance curves shown in this chapter.
\end{enumerate}
We will walk through each of these stages, beginning with the state preparation via optical pumping.

\subsection{Spin state preparation via optical pumping} \label{sec:opticalPumpingSD}
\begin{figure}
\centering
\includegraphics{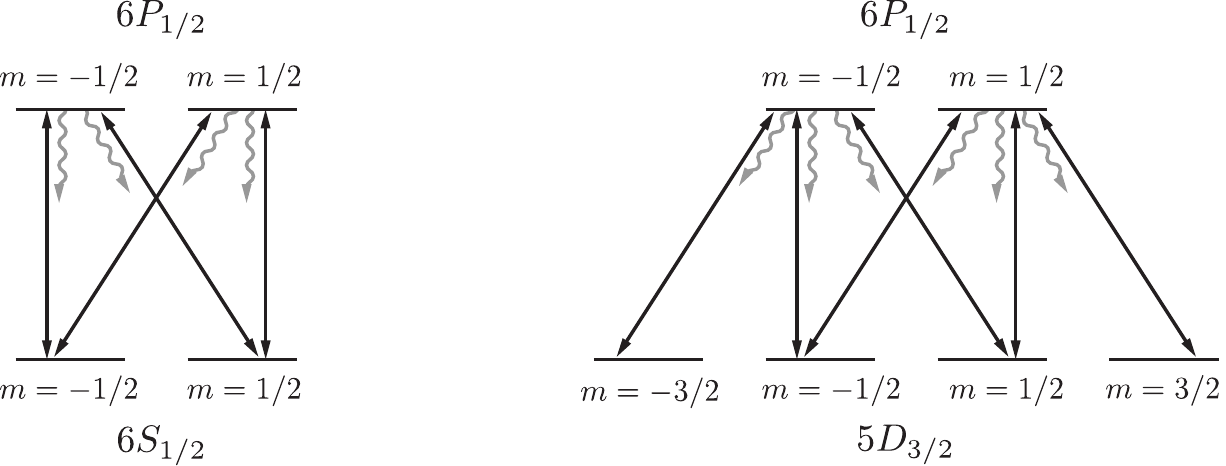}
\caption[A rate equation model for optical pumping]{A rate equation model for optical pumping allows each of the excitations and decays shown, weighted by the appropriate Clebsch-Gordan coefficients given a laser polarization and alignment.  As long as the excitation rate is much slower than the decays, $\Omega \ll \Gamma$, the steady state solution to rate equations Eq.~\ref{eq:sOpticalPumping}--\ref{eq:dOpticalPumping} give the atomic polarizations plotted in Figures~\ref{fig:opticalPumpingS} and~\ref{fig:opticalPumpingD}.}
\label{fig:rateEquationOpticalPumping}
\end{figure}
For the $6S_{1/2}$ and $5D_{3/2}$ spin resonance measurements, the essential first step is a reliable technique for initializing the spin state of the ion.  We employ the optical pumping technique discussed in section~\ref{sec:opticalPumping}.  Earlier work \cite{koerber2003thesis} shows that the steady state optical pumping efficiency for the $6S_{1/2}$ state by applying 493~nm light with circular polarization is
\begin{equation}\label{eq:opticalPumpingSPrediction}
P_{m \to 1/2}(t = \infty) = \frac{1}{2} + \frac{2 \sigma \cos \theta}{3 + 2 \sigma^2 - \cos 2\theta}.
\end{equation}
Here, $\sigma$ (defined in Appendix~\ref{sec:conventions}) is unity for perfect circular polarization;  perfect alignment of the beam with the magnetic field corresponds to $\theta = 0$.  We verified Eq.~\ref{eq:opticalPumpingSPrediction} by numerically integrating a rate equation model described below.

Since, in practice, we perform optical pumping with lasers attenuated far below saturation, we can model the atomic population with rate equations that ignore the off diagonal coherence terms in the density matrix.  We will treat the pumping processes separately:  when pumping into $6S_{1/2}$, we assume the $5D_{3/2}$ is being ideally pumped out and \emph{vice versa}.  The following differential equations describe the time varying probabilities $s_m$, and $p_m$ to be in each magnetic sublevel of $6S_{1/2}$ and $6P_{1/2}$ when pumping on the 493~nm transition:
\begin{align} \label{eq:sOpticalPumping}
\frac{ds_m}{dt} &= \sum_{m'} \left( \Gamma_B |\langle \tfrac{1}{2},m; 1,(m'-m) | \tfrac{1}{2}, m' \rangle|^2  p_{m'} - \Omega \sum_{q=-1}^{1} |\langle \tfrac{1}{2},m ; 1 q | \tfrac{1}{2}, m' \rangle|^2 E_q^2 s_m \right), \\
\frac{d p_m}{dt} &= \sum_{m'} \left( - \Gamma_B |\langle \tfrac{1}{2}, m; 1 (m'-m) | \tfrac{1}{2}, m' \rangle|^2  p_{m'} + \Omega \sum_{q=-1}^1 |\langle \tfrac{1}{2}, m; 1 q | \tfrac{1}{2}, m' \rangle|^2 E_q^2 s_m \right),
\end{align}
and with probabilities $d_m$ to be in the $5D_{3/2}$ substates when pumping on the 650~nm transition:
\begin{align} \label{eq:dOpticalPumping}
\frac{d d_m}{dt} &= \sum_{m'} \left( \Gamma_R |\langle \tfrac{1}{2},m; 1,(m'-m) | \tfrac{3}{2}, m' \rangle|^2  p_{m'} - \Omega \sum_{q=-1}^{1} |\langle \tfrac{3}{2},m ; 1 q | \tfrac{1}{2}, m' \rangle|^2 E_q^2 d_m \right), \\
\frac{d p_m}{dt} &= \sum_{m'} \left( - \Gamma_R |\langle \tfrac{1}{2}, m; 1 (m'-m) | \tfrac{3}{2}, m' \rangle|^2  p_{m'} + \Omega \sum_{q=-1}^1 |\langle \tfrac{3}{2}, m; 1 q | \tfrac{1}{2}, m' \rangle|^2 E_q^2 d_m \right),
\end{align}
where $\Gamma_B$ and $\Gamma_R$ are the partial decay rates for the two blue and red transitions from $6P_{1/2}$, and $\Omega$ sets an average excitation rate. $E_q$ describes the magnitude of polarization alignment with the principle types of transitions $q = (m' - m) = \pm 1,$ 0 with
\begin{align}
E_{\pm 1} &= \frac{\cos \theta \pm \sigma}{\sqrt{2} \sqrt{1 + \sigma^2}}, \\
E_0		&= \frac{\sin \theta}{\sqrt{1 + \sigma^2}},
\end{align}
where $\theta$ is the angle between the laser wavevector $\boldsymbol{k}$ and the quantization axis $\boldsymbol{B}$ and $\sigma$ is the magnitude of circular polarization.

\begin{figure}
\centering
\includegraphics[width= 5in]{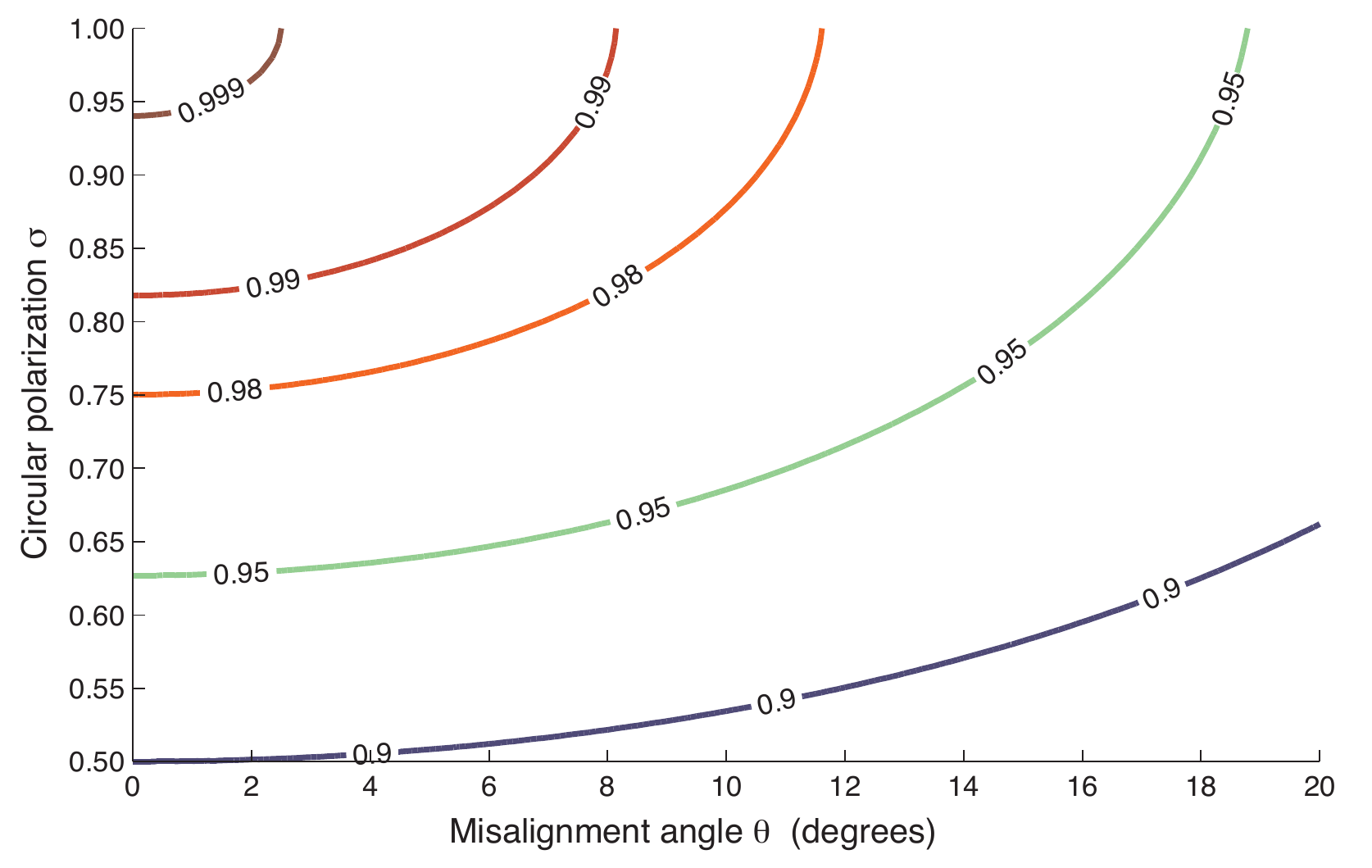}
\caption[Contour plot of optical pumping efficiency in the $6S_{1/2}$ state]{Contour plot of the steady-state efficiency of optical pumping in the $6S_{1/2}, m=-1/2$ state given a misalignment $\theta$ (degrees) of the laser with respect to magnetic field and circular polarization strength $\sigma$. In this simulation, $\Omega = \Gamma_B/10$ (see Eq.~\ref{eq:sOpticalPumping}). }
\label{fig:opticalPumpingS}
\end{figure}
\begin{figure}
\centering
\includegraphics[width= 5in]{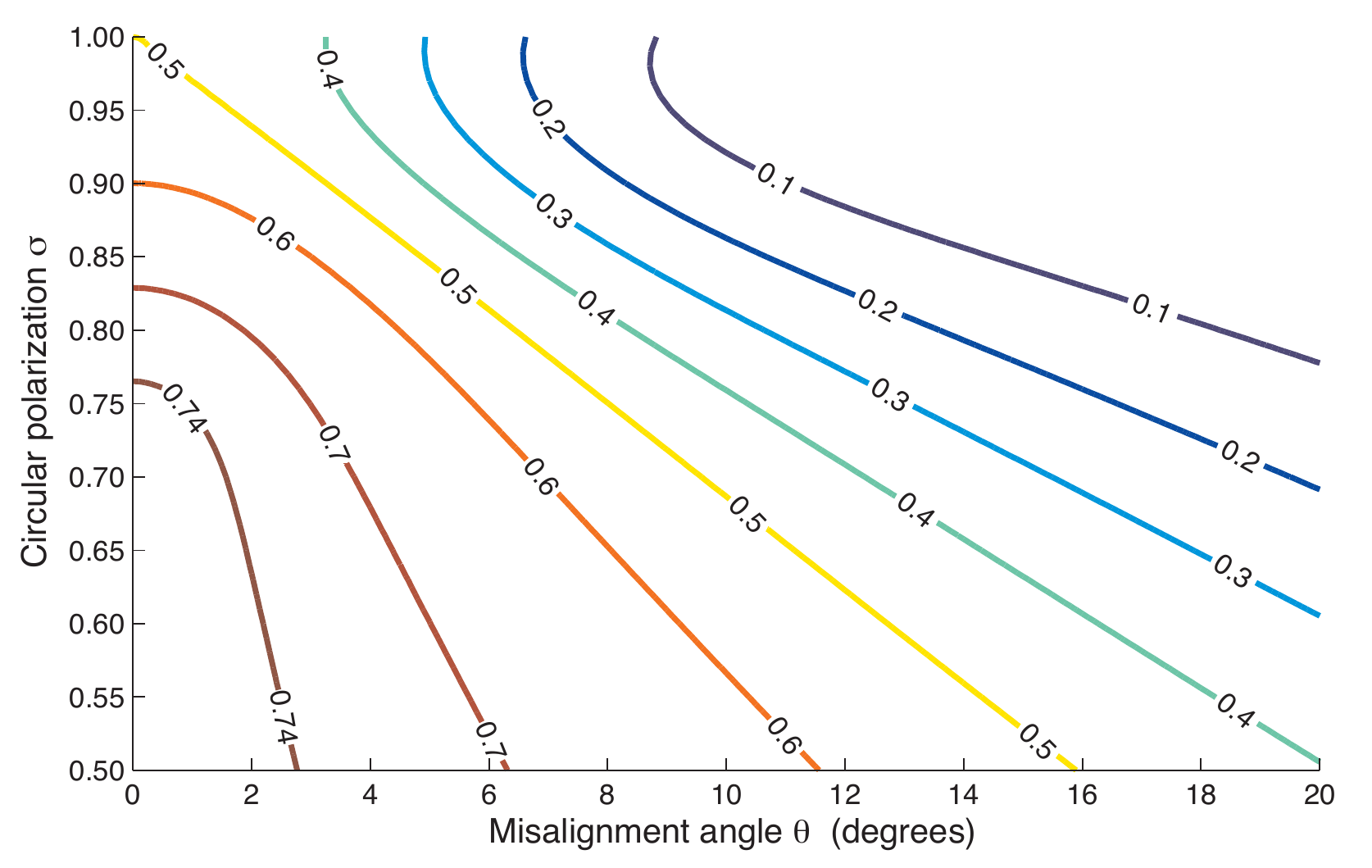}
\caption[Contour plot of optical pumping efficiency in the $5D_{3/2}$ state]{Contour plot of the steady-state efficiency of optical pumping in the $5D_{3/2}, m = -1/2$ state given a misalignment $\theta$ (degrees) of the laser with respect to magnetic field and circular polarization strength $\sigma$.  In this simulation, $\Omega = \Gamma_R/2$ (see Eq.~\ref{eq:dOpticalPumping}).}
\label{fig:opticalPumpingD}
\end{figure}

Figures~\ref{fig:opticalPumpingS} and~\ref{fig:opticalPumpingD} show the results of a numerical integration of the rate equation models for a range of misalignment angles $\theta$ and polarization errors $|\sigma| < 1$.  In Figure~\ref{fig:opticalPumpingS}, we plot the fraction of the ground state population pumped into the $m = -1/2$ state:
\begin{equation}
6S_{1/2} \text{ pumping fraction}(t \to \infty) = \frac{s_{-1/2}}{s_{1/2} + s_{-1/2}}.
\end{equation}
We are most interested in the $5D_{3/2} m = -1/2 \to 1/2$ splitting and therefore we are most interested in pumping into the $m = -1/2$ state, not $m = -3/2$ (though both levels are significantly populated by even the most efficient optical pumping process). We define the pumping fraction $a_{-1/2}$ as
\begin{equation}
a_{-1/2} \equiv 5D_{3/2} \text{ pumping fraction}(t \to \infty) = \frac{d_{-1/2}}{d_{3/2} + d_{1/2} + d_{-1/2} + d_{-3/2}}.
\end{equation}
Simulations of $a_{-1/2}$ are plotted in Figure~\ref{fig:opticalPumpingD}, and we note that it becomes an important parameter in a resonance curve lineshape systematic effect.  Both pumping fraction metrics do not include the small excited state probability which will, in practice, randomly decay into any of the available ground states.  Our approximations will therefore match reality well when $\Omega \ll \Gamma$.

The simulations indicate that achieving good pumping efficiency in the $6S_{1/2}$ is relatively easy;  we can suffer large laser misalignments $\theta < 10^\circ$ while $\sigma \approx 0.95$ (typically quickly achievable in the lab) and still expect 99~\% pumping efficiency.  Pumping into the $5D_{3/2}, m =-1/2$ state depends more critically on alignment and polarization.  From Figure~\ref{fig:opticalPumpingD}, we find maximizing $|\sigma|$ does not maximize the pumping fraction $a_{-1/2}$ even for perfect alignment ($\theta = 0$).  A maximum fraction of $a_{1/2} \approx 3/4$ seems to be available with good alignment and purposeful error in the circular polarization $|\sigma| \approx 0.70$.  During light shift data taking, the 650~nm laser circular polarization was in practice anywhere between $0.70 < |\sigma | < 1$ due to temperature drift in the transverse electro-optic modulator used to create the circular polarization; therefore, from Figure~\ref{fig:opticalPumpingD}, the pumping fraction varied from $0.5 \lesssim a_{-1/2} \lesssim 0.75$ in an uncontrolled way for most data runs.  We discovered the implications of this uncontrolled variation midway through data taking, and discuss its ramifications in Section~\ref{sec:lightShiftLineShapes}.  In pumping either to specific spin states of $6S_{1/2}$ and $5D_{3/2}$, an important practical tip is to shutter the circularly polarized laser before turning off the repumping laser.  

\subsection{Spin state probing with optical pumping}
\begin{figure}
\centering
\includegraphics[width=5in]{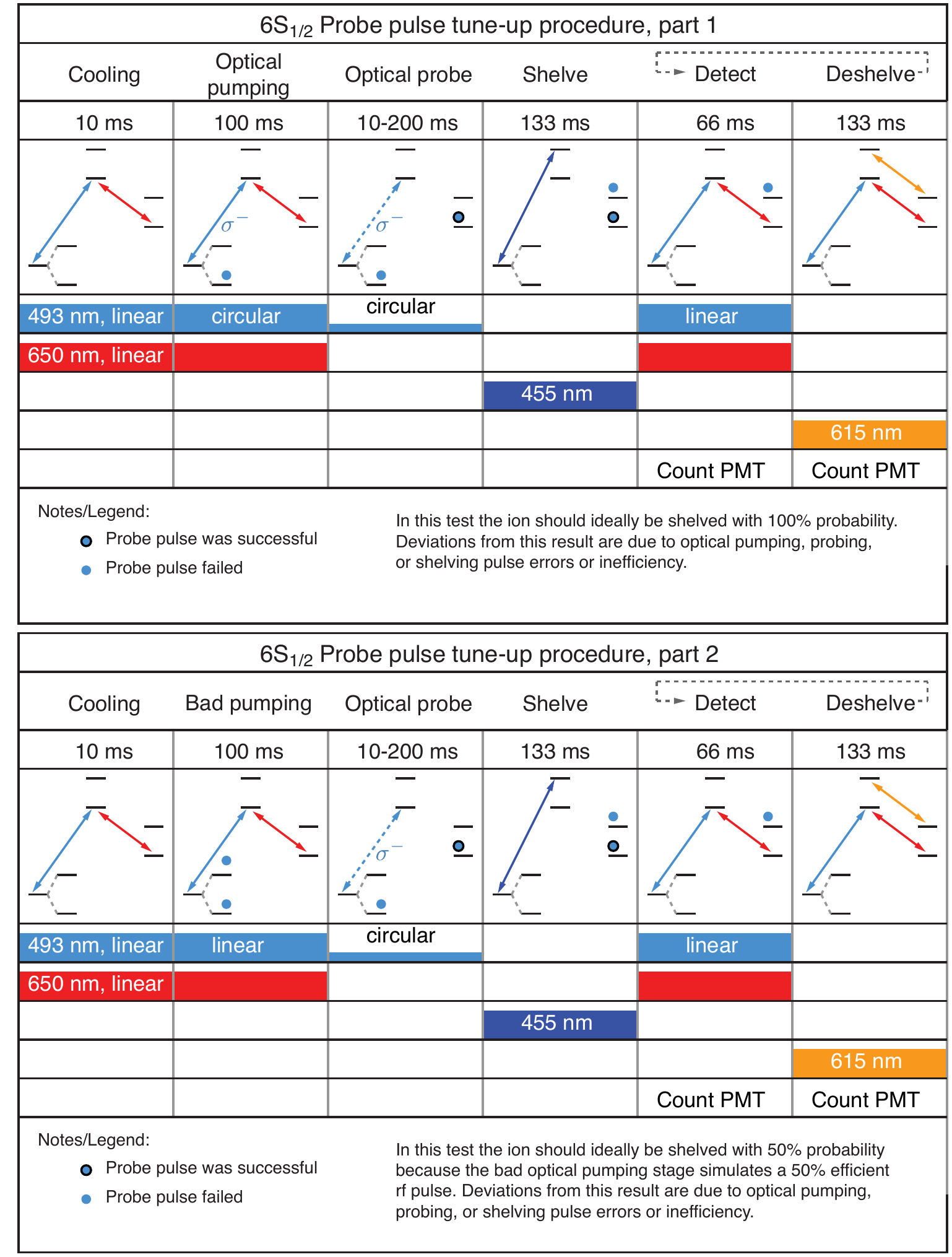}
\caption[Experimental method for tuning up probe pulse timing]{Experimental method for tuning up probe pulse timing.  Part 1 of the sequence is essentially an rf spin flip experiment (without any actual rf) and should ideally measure a shelved ion 100\% of the time.  Part 2 of the sequence simulates a 50\% efficient spin flip pulse and so ideally 50 \% of the time the ion should be shelved.  Deviations from these probabilities indicate inefficiency in the optical pumping, probing, and shelving stages;  we commonly use these sequences, and a similar test for the $5D_{3/2}$ resonance, to tune up these parameters.}
\label{fig:probePulseTuneup}
\end{figure}
After spin state preparation, we expose the ion to a pulse of resonant rf through a half loop antenna near the ion trap that may cause a spin flip (see Section~\ref{sec:spinDynamics}).  By applying a dim `probe' pulse of circularly polarized light connecting the state to $6P_{1/2}$, we can `readout' whether the spin flip occurred by causing a spin-dependent electronic transition.  Using the same rate equation model as the last section, we also perform simulations to determine the optimal probe pulse exposure time that will move out a spin-flipped ion without a great probability of emptying out the state completely even in the case of laser misalignments and polarization errors. In practice, the probe pulse exposure times are tuned up experimentally with experiment sequences illustrated in Figure~\ref{fig:probePulseTuneup}.

\subsection{The $6S_{1/2}$ state measurement procedure}
\begin{figure}[p]
\centering
\includegraphics[angle=90]{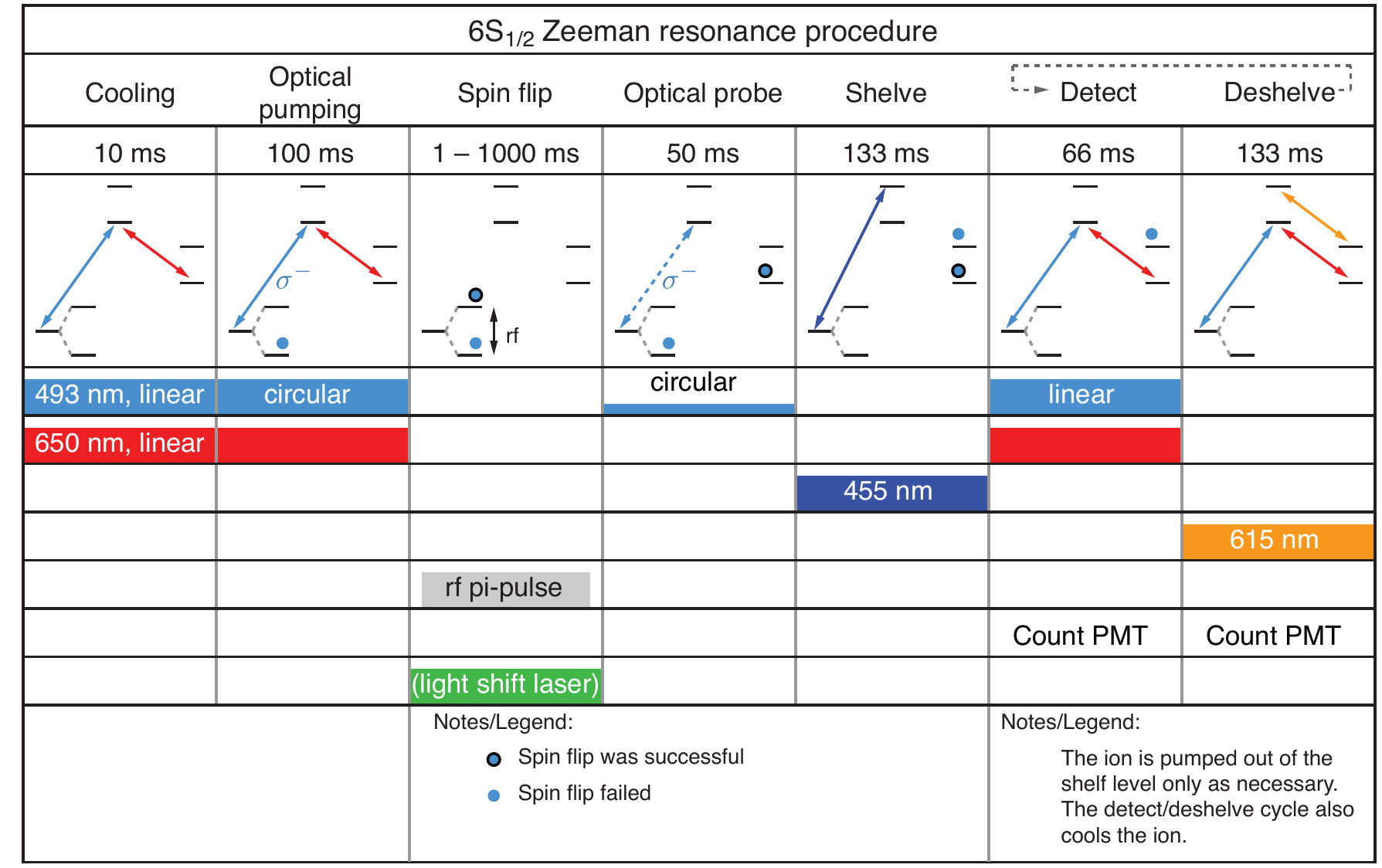}
\caption[Spin-dependent shelving technique for measuring $6S_{1/2}$ Zeeman splitting]{Spin-dependent shelving technique for measuring $6S_{1/2}$ Zeeman splitting.  As described in the text, this procedure detects if an rf pulse is resonant with the $6S_{1/2}$ Zeeman splitting by a circularly polarized optical probe pulse followed by a shelving transition.  The procedure is repeated many times using different spin flip frequencies to build a resonance curve.}
\label{fig:zeemanResonanceSstateProc}
\end{figure}
\begin{figure}[p]
\centering
\includegraphics[angle=90]{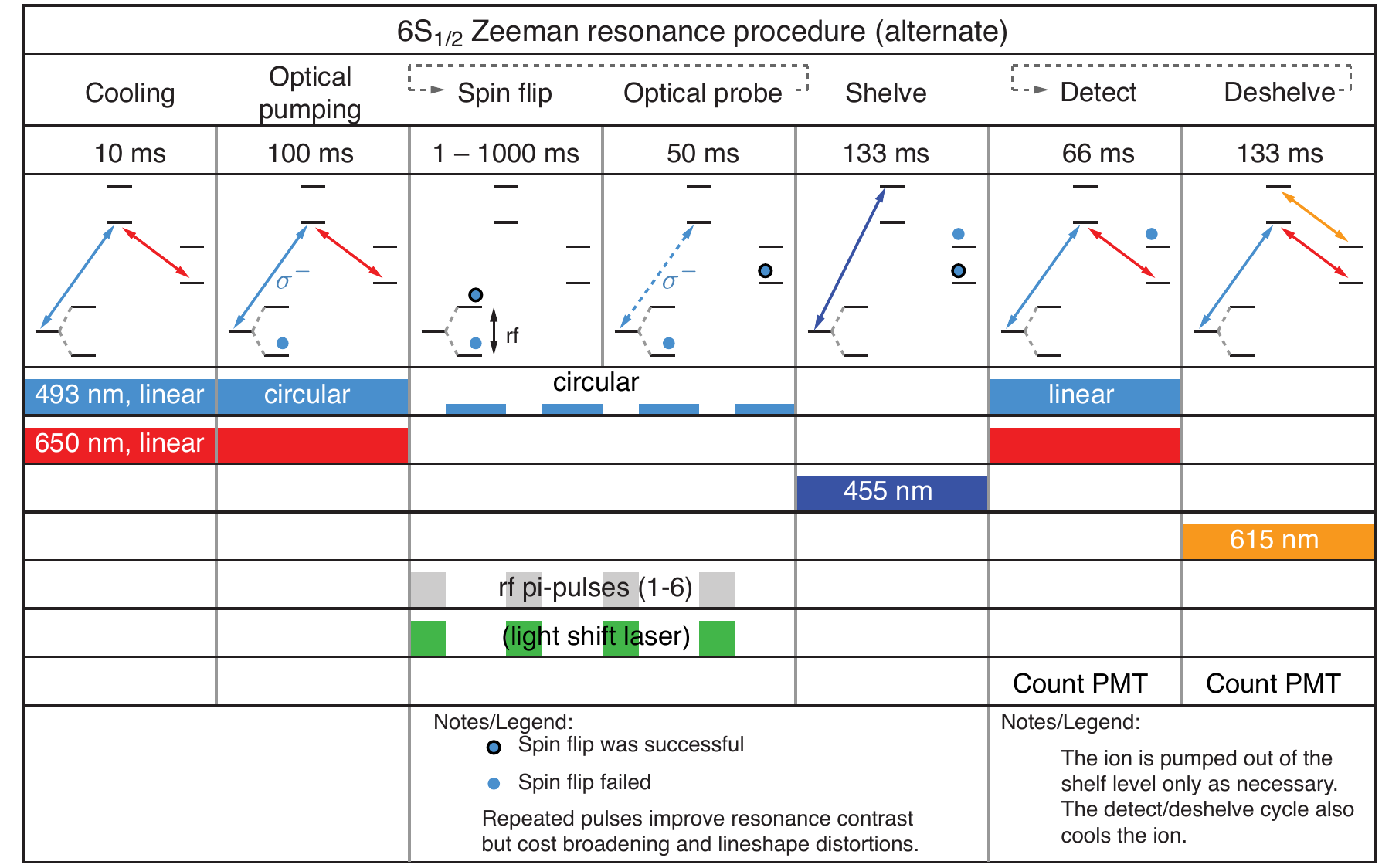}
\caption[Improved pulsing technique for measuring $6S_{1/2}$ Zeeman splitting]{The spin-dependent shelving technique for measuring $6S_{1/2}$ Zeeman splitting can be improved by introducing multiple pulses of rf $\pi$-pulses interleaved with circularly polarized optical probe pulses.  This improves the resonance contrast by giving the ion many chances to be excited to $6P_{1/2}$ and follow the lower probability decay to $5D_{3/2}$.  It broadens the resonance curve somewhat and heavily distorts Rabi-oscillation curves, as seen in Figure~\ref{fig:multiplePulseSExamples}.}
\label{fig:zeemanResonanceSstatePulsesProc}
\end{figure}

As illustrated Figure~\ref{fig:zeemanResonanceSstateProc}, we begin by initializing the ion in (suppose) the $m = -1/2$ sublevel of $6S_{1/2}$ via optical pumping by applying a weak 493~nm $\sigma^-$ polarized beam along with linearly polarized 650~nm repumping radiation.  Then, with all the lasers off, we expose the ion to a hardware-timed, mixer gated (see Figure~\ref{fig:daqTimingDiagrams}) pulse of rf intended to be resonant with the $m = -1/2 \to 1/2$ spin-flip transition frequency $\omega_S$.  To detect whether the spin flip has occurred, we now apply another dim pulse of $\sigma^-$ polarized 493~nm light.  The probability that excitation to the $6P_{1/2}$ occurs is proportional to the probability that the rf pulse succeeded in changing the spin of the ground state to $m = +1/2$.  Due to the branching ratio of decays from $6P_{1/2}$, roughly 30\% of the time a successfully probed ion ends up decaying to $5D_{3/2}$.  In these cases, we have successfully stored the spin information of the ion into state information that is readily extracted by an electron shelving technique.  We apply a pulse of 455~nm light designed to move any ion population in $6S_{1/2}$ to the `shelved' level $5D_{5/2}$ via a decay from $6P_{3/2}$.  Then, we apply bright, linearly polarized 493~nm cooling and 650~nm repump light to the ion and watch the PMT for fluorescence.  If the ion is in the shelved level, correlated to the rf spin-flip being unsuccessful, then the scattered 493~nm radiation is low.  If the ion is in the $5D_{3/2}$ level, correlated to the rf spin-flip and circular-polarized 493~nm probe beam being successful, then the scattered 493~nm radiation is high.

Though just one bit of information is gleaned from this procedure---a `1' if the ion was shelved in $5D_{5/2}$, a `0' if the ion was not---upon repetition it yields an ensemble measurement of `shelving probability' that is correlated to the success of the rf spin flip:
\begin{equation*}
P(\text{ion shelved}) = 1 - P(\text{spin-flip}) \cdot \underbrace{P(\text{probe}) \cdot P(\text{shelve})}_\text{efficiency factor} - \underbrace{P(\text{probe-error})  \cdot P(\text{shelve})}_\text{constant bias}, \\
\end{equation*}
where
\begin{align*}
P(\text{probe}) &\approx 0.30, \\
P(\text{probe-error}) &\approx 0.30 \pm 0.20, \\
P(\text{shelve}) &> 0.90.
\end{align*}

\begin{figure}
\centering
\includegraphics{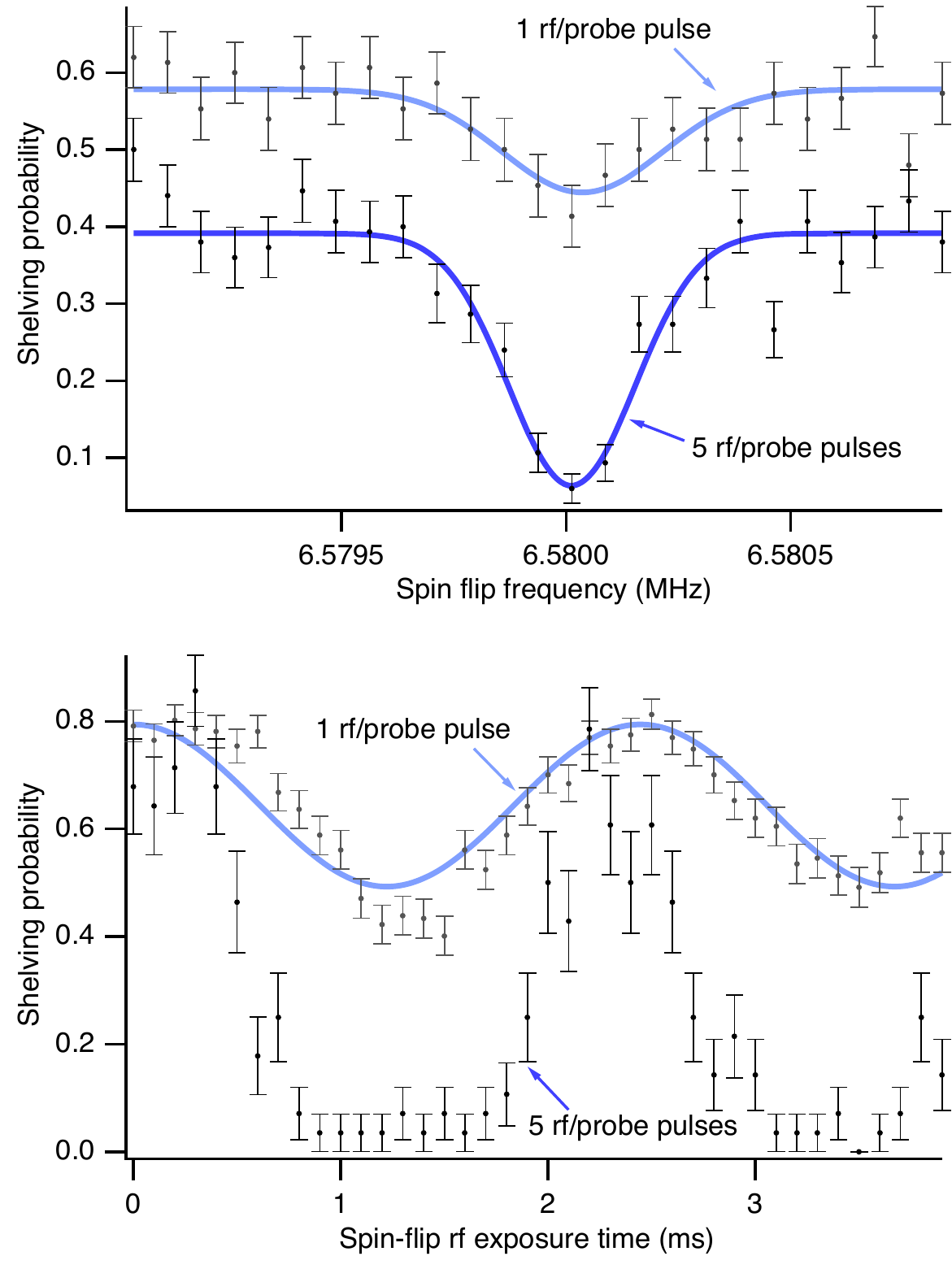}
\caption[$6S_{1/2}$ measurement with multiple rf/probe pulses]{$6S_{1/2}$ spin resonance profiles with multiple rf/probe pulses increase curve contrasts.  In the top plot we examine scan {\tt 040303-1} in which we performed two $\omega_S$ scans, identical in all ways except one featured a sequence of 5 rf $\pi$-pulses followed by optical probe pulses (compare Figures~\ref{fig:zeemanResonanceSstateProc} and~\ref{fig:zeemanResonanceSstatePulsesProc}.  The multiple-pulse procedure performs better in all respects:  the curve contrast is better by about a factor of 2, Gaussian curve fits show no resolvable shifts in the center frequency, and the center frequency error (1 standard deviation) derived from the fits is three times smaller in the multiple pulse case.  However, notice in the bottom plot time domain scans {\tt 041217-7} and {\tt -8} (which show the $\sin^2(\Omega/2 t)$ Rabi flopping behavior obeyed in the case of a single rf/probe pulse) destroyed when 5 pulses are applied.  We observe a larger resonance contrast but decreased sensitivity to the $\pi$-pulse time which is not at all undesirable from a practical viewpoint.}
\label{fig:multiplePulseSExamples}
\end{figure}
Because the probe pulse efficiency is so low due to the branching ratio of decays from $6P_{1/2}$ that put a probed ion back into the $6S_{1/2}$ state rather than the $5D_{3/2}$, the resonance contrast is similarly lousy.  We discovered that by applying multiple cycles of rf $\pi$-pulses followed by optical probe pulses, we could increase the contrast of the curves, lose sensitivity on the $\pi$-pulse exposure time (which isn't necessarily bad), and destroy the coherent Rabi sidebands present on the resonance line shape.  The modified experimental sequence is detailed in Figure~\ref{fig:zeemanResonanceSstatePulsesProc} and some example curves are shown in Figure~\ref{fig:multiplePulseSExamples}.  In practice, we limited the number of rf/probe repetitions to five or six.  Additional pulses resulted in broadening, and fewer pulses yielded odd lineshapes, presumably because it then poorly mimics Ramsey spectroscopy~\cite{townes1956ms}. 

\subsection{The $5D_{3/2}$ state measurement procedure}
\begin{figure}[p]
\centering
\includegraphics[angle=90]{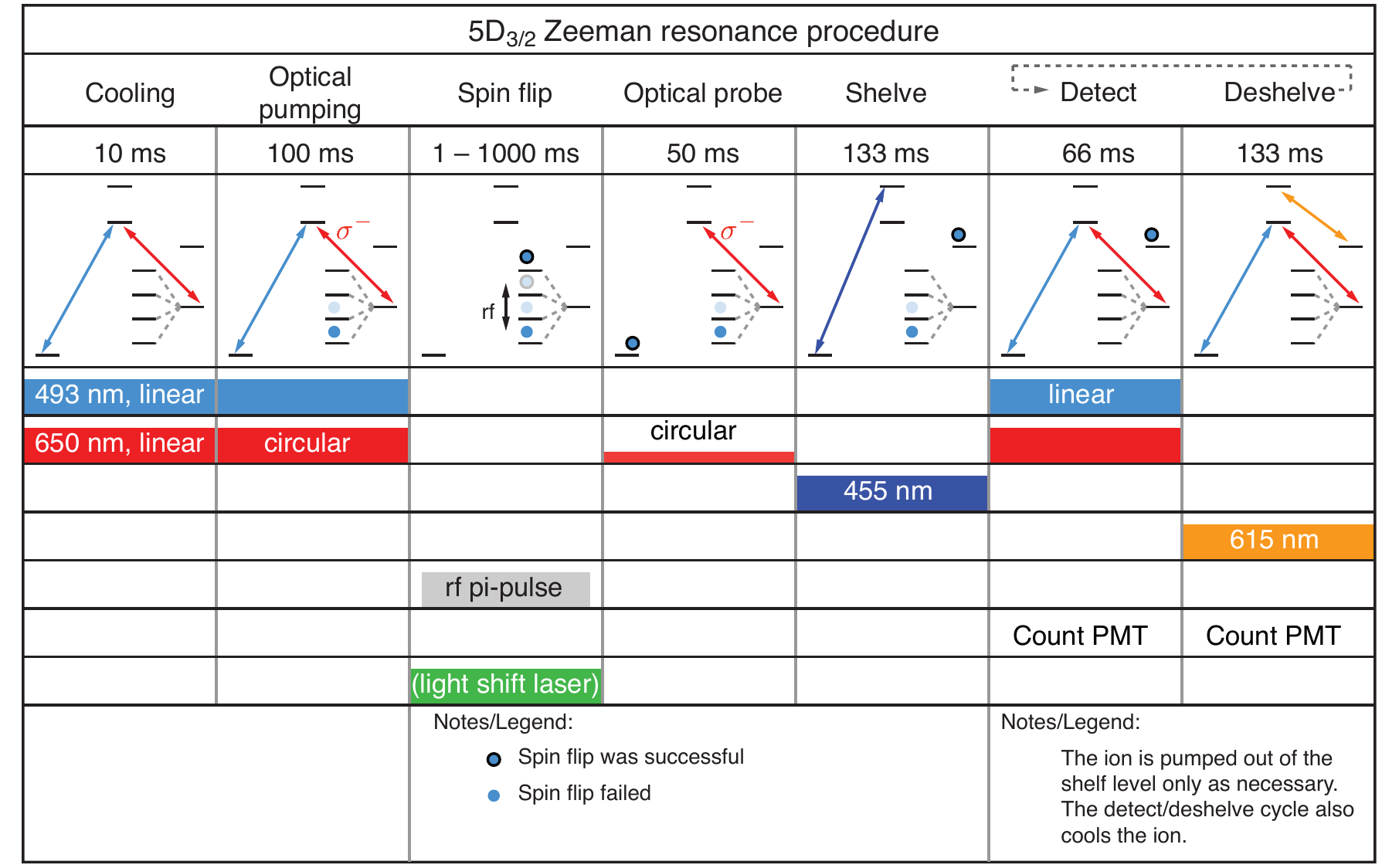}
\caption[Spin-dependent shelving technique for measuring $5D_{3/2}$ Zeeman splitting]{Spin-dependent shelving technique for measuring $5D_{3/2}$ Zeeman splitting.  The procedure is similar to that of $6S_{1/2}$ state, except now the ion is shelved when an rf spin flip is successful.}
\label{fig:zeemanResonanceDstateProc}
\end{figure}
\begin{figure}
\centering
\includegraphics{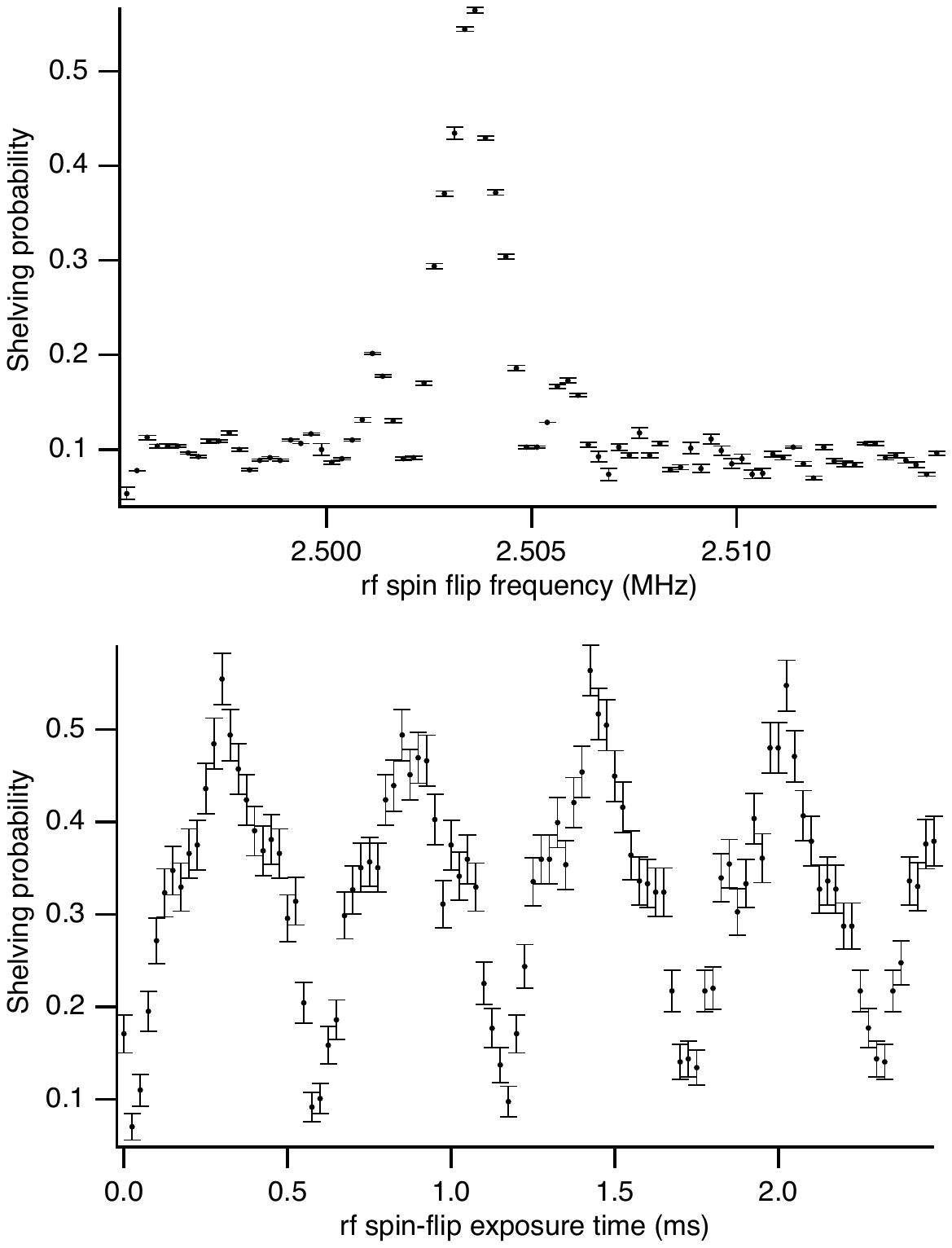}
\caption[Sample $5D_{3/2}$ resonance and Rabi flopping curves]{The upper $5D_{3/2}$ resonance curve (data file {\tt 041214-1}) shows a high contrast resonance with Rabi sideband features.  The lower trace (data file {\tt 031215-1}) shows Rabi oscillations characteristic of the four state $J = 3/2$ system.}
\label{fig:dResonanceSample}
\end{figure}

We also measure the $m = -1/2 \to 1/2$ splitting the in the $5D_{3/2}$ state with a similar technique.  Using circularly polarized light, we cannot initialize the ion in a single $5D_{3/2}$ sublevel;  instead, we can pump into a statistical mixture of $m = -3/2,-1/2$ with $\sigma^-$ or $m = +3/2,+1/2$ with $\sigma^+$ 650~nm light.  Supposing we choose to pump into the lower spin manifold, we apply a radio frequency pulse designed to move the population to the upper manifold considering the $J > 1/2$ spin dynamics presented in Section~\ref{sec:spinDynamics}.  In analogy to the $6S_{1/2}$ spin resonance procedure, we convert this spin information into state information with the application of a dim probe pulse of $\sigma^-$ 650~nm light which transfers the $5D_{3/2},m = +3/2,+1/2$ population to $6P_{1/2}$ that decays $\sim$~70\% of the time to $6S_{1/2}$.  We then shelve this population to $5D_{5/2}$ with a pulse of 455~nm light.  Such a probe pulse leaves any population still in $5D_{3/2}$ alone.  Now, application of bright linearly polarized 493~nm and 650~nm light distinguishes between an ion in the shelved state and one left in the $5D_{3/2}$ state and hence whether a spin-flip was successful or not.  Here,
\begin{equation*}
P(\text{ion shelved}) = P(\text{spin-flip}) \cdot  \underbrace{P(\text{probe}) \cdot P(\text{shelve})}_\text{efficiency factor} + \underbrace{P(\text{probe-error})  \cdot P(\text{shelve})}_\text{constant bias}, \\
\end{equation*}
where
\begin{align*}
P(\text{probe}) &\approx 0.70, \\
P(\text{probe-error}) &\approx 0.10, \\
P(\text{shelve}) &> 0.90.
\end{align*}
Due to the favorable branching ratio of decays from $6P_{1/2}$, the contrast of $5D_{3/2}$ spin resonance curves is far better than those obtained for $6S_{1/2}$ (examine Figure~\ref{fig:dResonanceSample}).  However, the four-state system with impure initial conditions yields complicated resonance curve shapes.

In both the $6S_{1/2}$ and $5D_{3/2}$ measurement procedures, we assign statistical error bars to each spin-flip frequency shelving probability $p_s$ using
\begin{equation}
\sigma_p = \sqrt{ \frac{p_s (1-p_s)}{N} }.
\end{equation}
where $N$ is the number of spin-flip trials.  This has the right $N^{-1/2}$ scaling, and derives from the variance in a binomial distribution.

\section{Light shift ratio measurement technique}
\begin{figure}
\centering
\includegraphics{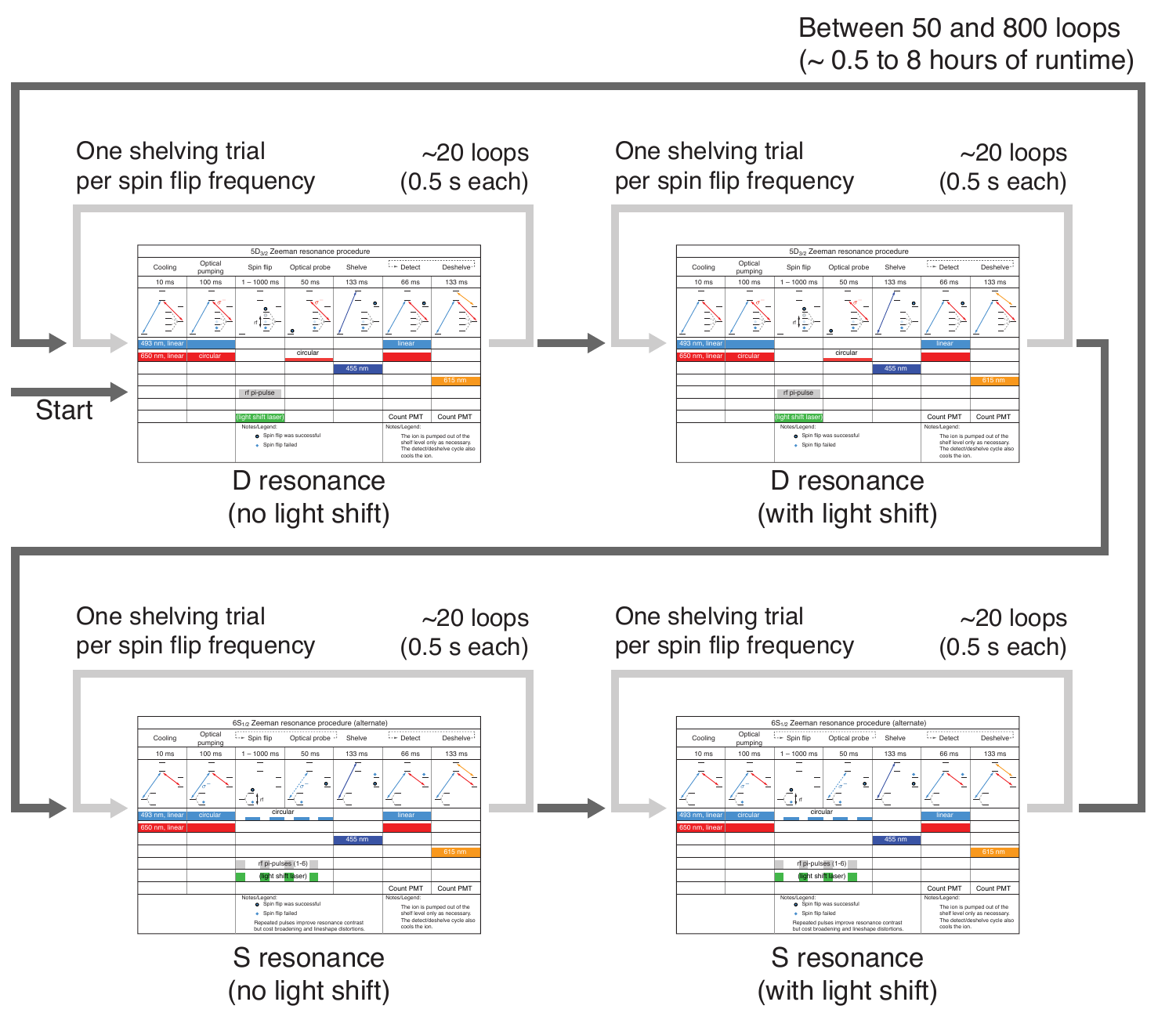}
\caption[Interleaved measurement routine for light shift ratio data]{Interleaved measurement routine for light shift ratio data. The miniature timing diagrams are full-size in Figures~\ref{fig:zeemanResonanceSstatePulsesProc} and~\ref{fig:zeemanResonanceDstateProc}. We choose the order of the loops (the grey-arrow loops through trial spin flip frequencies, and the black-arrowed loops to gather statistics) to make drifts in the magnetic field and light shift laser parameters as common-mode as possible without introducing additional overhead (e.g.\ additional rf synthesizer switching time when large frequency changes are programmed).}
\label{fig:measurementLoop}
\end{figure}

\begin{figure}
\centering
\includegraphics[width = 5 in]{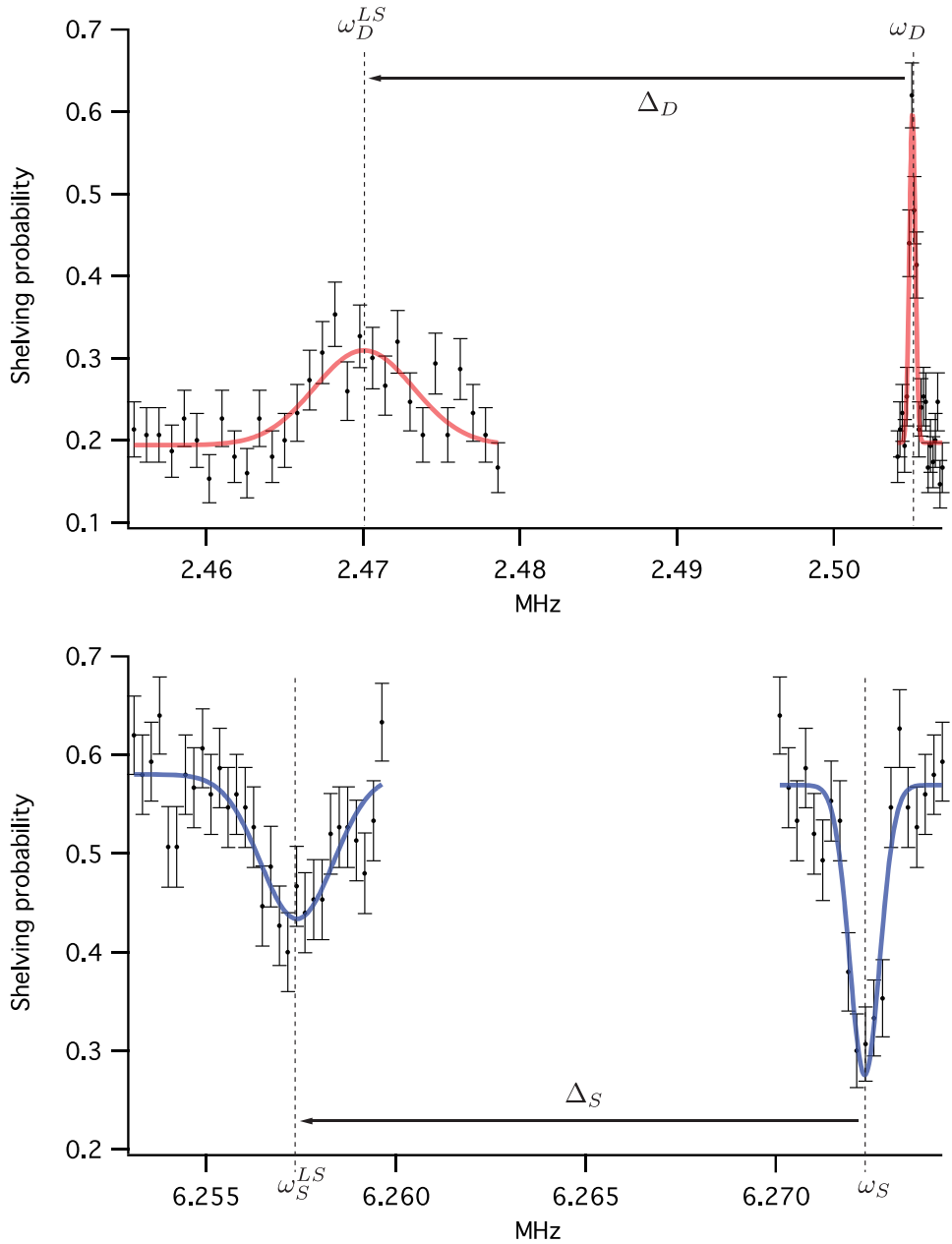}
\caption[An example light shift ratio measurement]{An example light shift ratio measurement at 1111~nm:  data file {\tt 050810-1}.  Measurements of the shifted resonances $\omega_D^{LS}$, $\omega_S^{LS}$ and unshifted resonances $\omega_D$, $\omega_S$ are interleaved for several hours.  In this case, we attempted about 150 spin-flip and shelving sequences at each trial frequency for each of the four resonances before the light shift laser power began to drift.  The light shift ratio is derived from curve fits to these four resonances $R = \Delta_S / \Delta_D = (\omega_S^{LS} - \omega_S)/(\omega_D^{LS} - \omega_D)$.}
\label{fig:fourResonanceSample}
\end{figure}

To measure the light shift ratio
\begin{equation}   \label{eq:lsrDefinition}
R = \frac{\Delta_S}{\Delta_D} = \frac{\omega_S^\text{LS} - \omega_S}{\omega_D^\text{LS} - \omega_D}
\end{equation}
we interleave measurements of the two ion spin resonances $\omega_S$, $\omega_D$ without application of the light shift beam and $\omega_S^{LS}$, $\omega_D^{LS}$, with application of the light shift beam.  By `interleave' I mean the nested measurement cycle shown in Figure~\ref{fig:measurementLoop}.  In order for all resonances to share drifts in the magnetic field, light intensity, pointing, and polarization, etc., we perform just one shelving measurement at each trial spin flip frequency for a given resonance before moving on to the next resonance.  In this way, equal amounts of data are taken for each resonance, roughly equal time is spent taking data for each resonance, and many systematic effects explored later are made common mode and cancel out in $R$.  An example of four interleaved resonance measurements is shown in Figure~\ref{fig:fourResonanceSample}.

\subsection{Light shift beam stabilization} \label{sec:lightShiftBeamStabilization}
\begin{figure}
\centering
\includegraphics{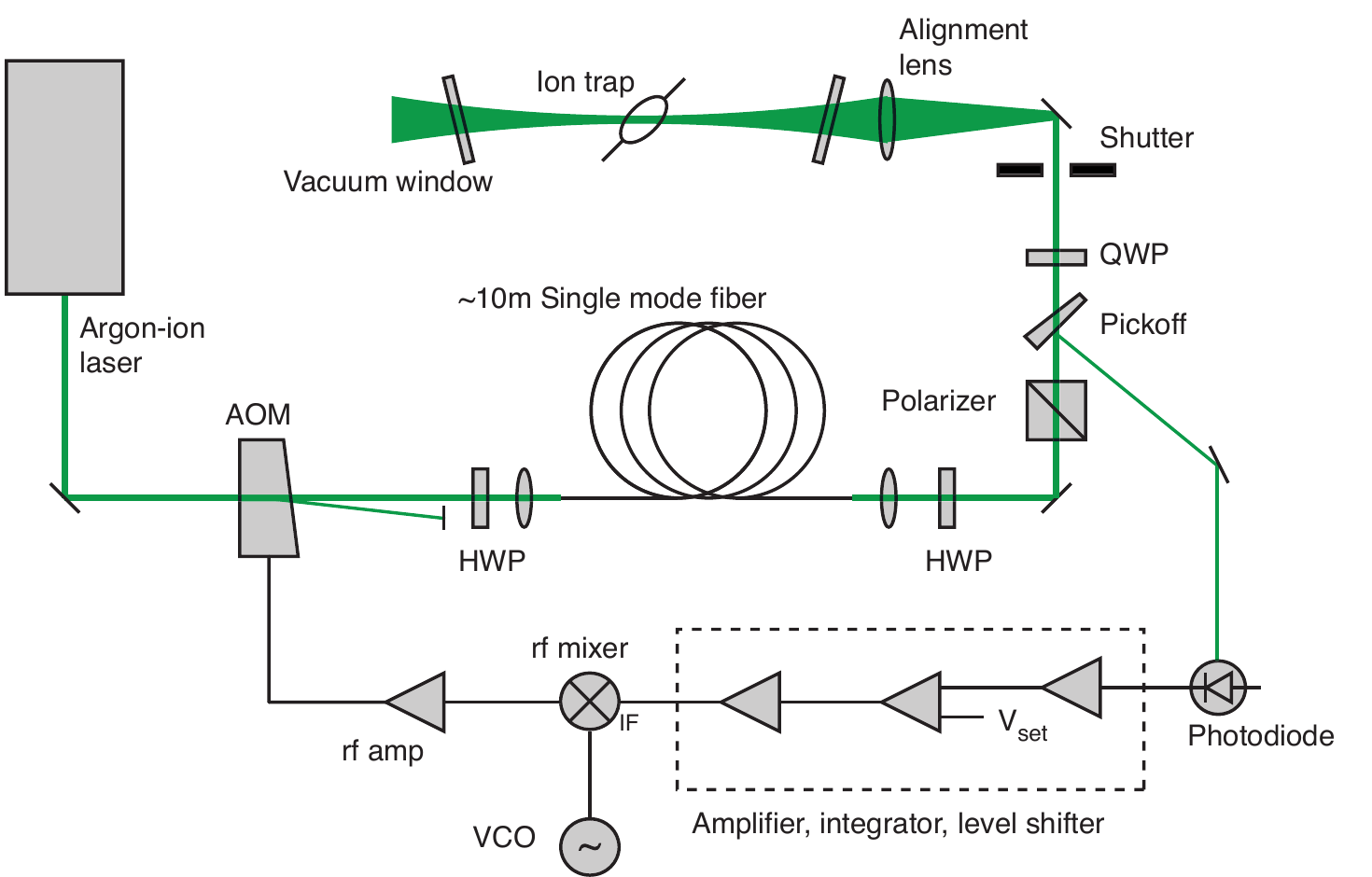}
\caption[Stabilization of the light shift beam power, polarization, and pointing]{We stabilize the light shift beam power, polarization, and pointing with the apparatus shown above.  See the text for a full description of the technique.  Briefly, in this 514~nm light shift laser subsystem, argon-ion laser pointing and polarization noise are converted into intensity noise by the polarization maintaining fiber followed by a polarizer.  We control the power of the laser with a feedback loop to maintain a constant intensity at a photodiode after the polarizer.  A quarter-wave plate converts the beam polarization to circular.  An achromatic lens focuses the beam to $20$~$\mu$m at the site of the ion.}
\label{fig:lightshiftStabilization}
\end{figure}

One remaining challenge is the stabilization of the light shift laser over the many hours each experiment takes.  Temporal variation of the off-resonant light intensity (and polarization) at the site of the ion is especially problematic:  $\omega_S^\text{LS}$ and $\omega_D^\text{LS}$ resonances become smeared and require larger spin-flip driving fields to power-broaden them.  To minimize these fluctuations we have taken the following steps.  We arrange the light shift beam spot size at the site of the ion to be 20~$\mu$m which in turn is much greater than the $\lesssim$ 1 $\mu$m presumed ion orbit and trap motion with respect to the optical table.  During the 514~nm light shift measurements, we controlled the light shift beam power, polarization, and pointing using the subsystem diagramed in Figure~\ref{fig:lightshiftStabilization}.  Laser power fluctuations are detected on a photodiode and corrected using an acousto-optic modulator (AOM).  Pointing noise out of the laser is converted into power noise by the single-mode fiber which is thus servo corrected.  Laser polarization noise is indeed transmitted over the fiber, which is a polarization maintaining type, but since the photodiode is placed after a polarizer, this noise source too is converted into power noise and corrected by the AOM servo.

It is vital that the AOM be placed \emph{before} the fiber.  We found that the servo AOM itself introduced substantial pointing noise ($\sim$ 10~$\mu$m spot movement at the site of the ion) when placed after the fiber coupling perhaps due to fluctuating thermal effects.  Referring again to Figure~\ref{fig:lightshiftStabilization}, the half-wave plate (HWP) is tuned to align the polarization of the laser beam with either the fast or the slow axis of the single mode, polarization maintaining fiber (see \cite{hobbs2000beo} for one technique).  Another half-wave plate and polarizer following the fiber are adjusted for maximum output.  The quarter-wave plate (QWP) is then adjusted to give maximum circular polarization as measured by an inserted polarization analyzer and checked later using the measured light shifts.  We achieve polarizations of $|\sigma| \gtrsim 0.95$ measured just in front of the final alignment lens.  A wedged glass pickoff sends a sample of the beam to a large area silicon photodiode, masked to minimize stray light and other reflections.  A feedback circuit compares the transimpedance-amplified photocurrent to a set point voltage and adjusts the rf amplitude in the AOM via an rf mixer.

The Yb-doped fiber light shift laser at 1111~nm featured a single-mode, polarization maintaining fiber output.  Since an AOM placed after this laser would introduce large pointing error, and since the laser features no electronic current control, we simply operated the laser at constant current.  This sufficed since the free-running power stability of the laser over many hours is better than a few percent.  We acquired about half of this data with circular polarization created by an achromatic IR quarter-wave plate ($\sigma > 0.90$) and half with a quartz Babinet-Solei compensator  ($\sigma > 0.95$).

\section{Systematic error analysis}
A leading virtue of the experiment design is the lack of significant systematic effects.  Light shifts are proportional to the laser intensity $I$, which in general is difficult to measure at the site of the ion with any precision.  By measuring the light shifts in two states, however, the intensity drops out completely in the ratio.  By choosing to measure the $m = \pm 1/2$ splittings in both the $6S_{1/2}$ and $5D_{3/2}$, errors in laser polarization and alignment affect the light shift ratio only in second order.  By interleaving the measurements, as shown in Figure~\ref{fig:measurementLoop}, noise and temporal drift in the magnetic field, laser pointing, polarization, and intensity is shared by both shifted and unshifted resonances and thus do not contribute to the ratio.

The effects that remain are smaller than the statistical precision of the data.  The first effect documented in this section is the most important:  line-shape effects that plague the light-shifted $5D_{3/2}$ resonance.  We then consider laser errors:  misalignments, polarization errors, fluctuations, and spectral purity.  Next, we analyze ac-Zeeman effects from the trapping and spin-flip fields, and finally, systematic fluctuations in the magnetic field and ion position.

\subsection{$5D_{3/2}$ resonance line shape effects} \label{sec:lightShiftLineShapes}
\begin{figure}
\centering
\includegraphics{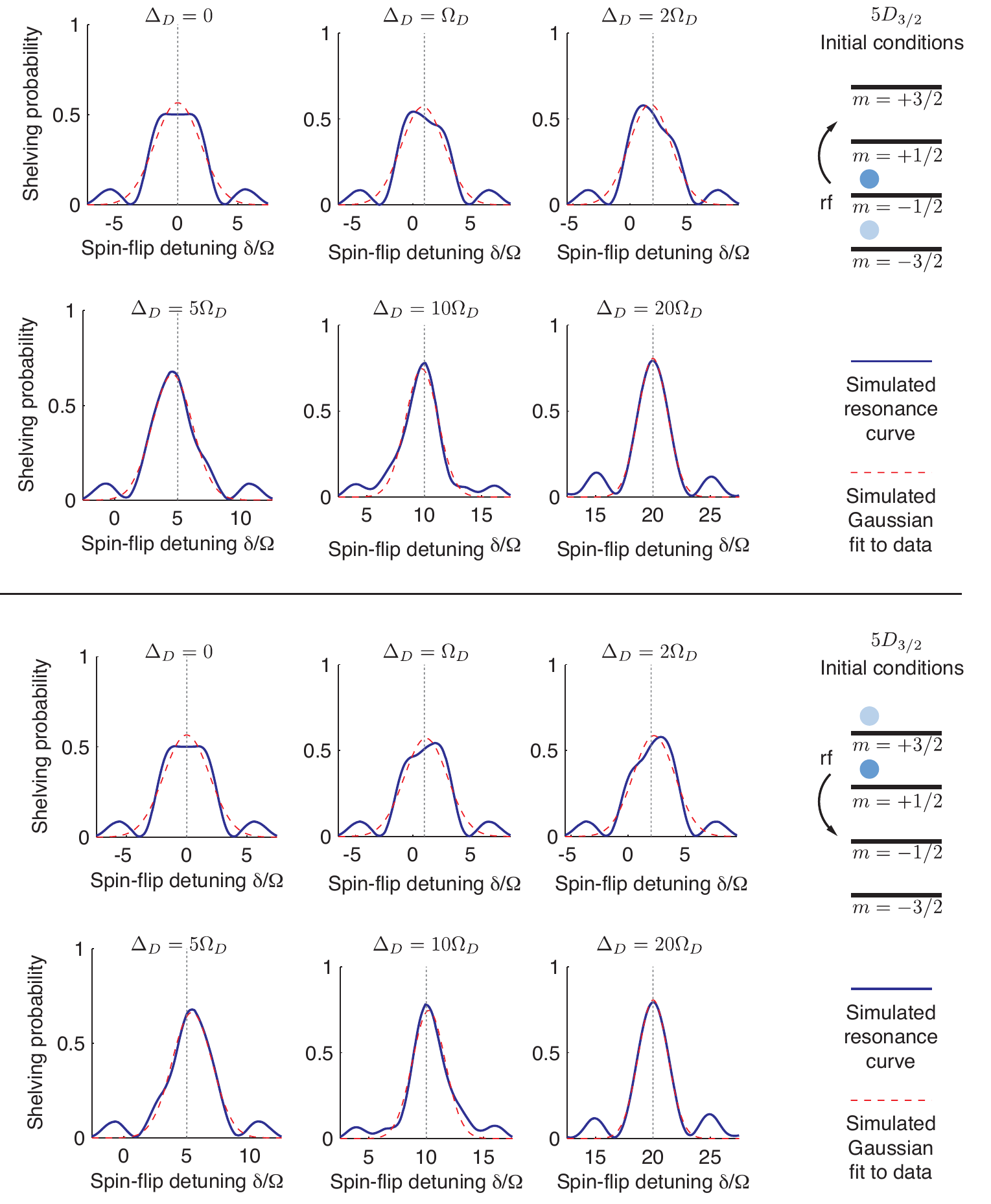}
\caption[Sensitivity of the $5D_{3/2}$ light-shifted resonance lineshape to initial conditions]{Sensitivity of the $5D_{3/2}$ light-shifted resonance lineshape to initial conditions.  These simulations show a systematic error in the $\omega_D^{LS}$ resonances worst for small values of $\Delta_D / \Omega_D$ and that changes sign when the relative sense of circular polarization of the 650~nm laser compared to the light shift laser are reversed.}
\label{fig:dLineShapeModels}
\end{figure}
\begin{figure}
\centering
\includegraphics{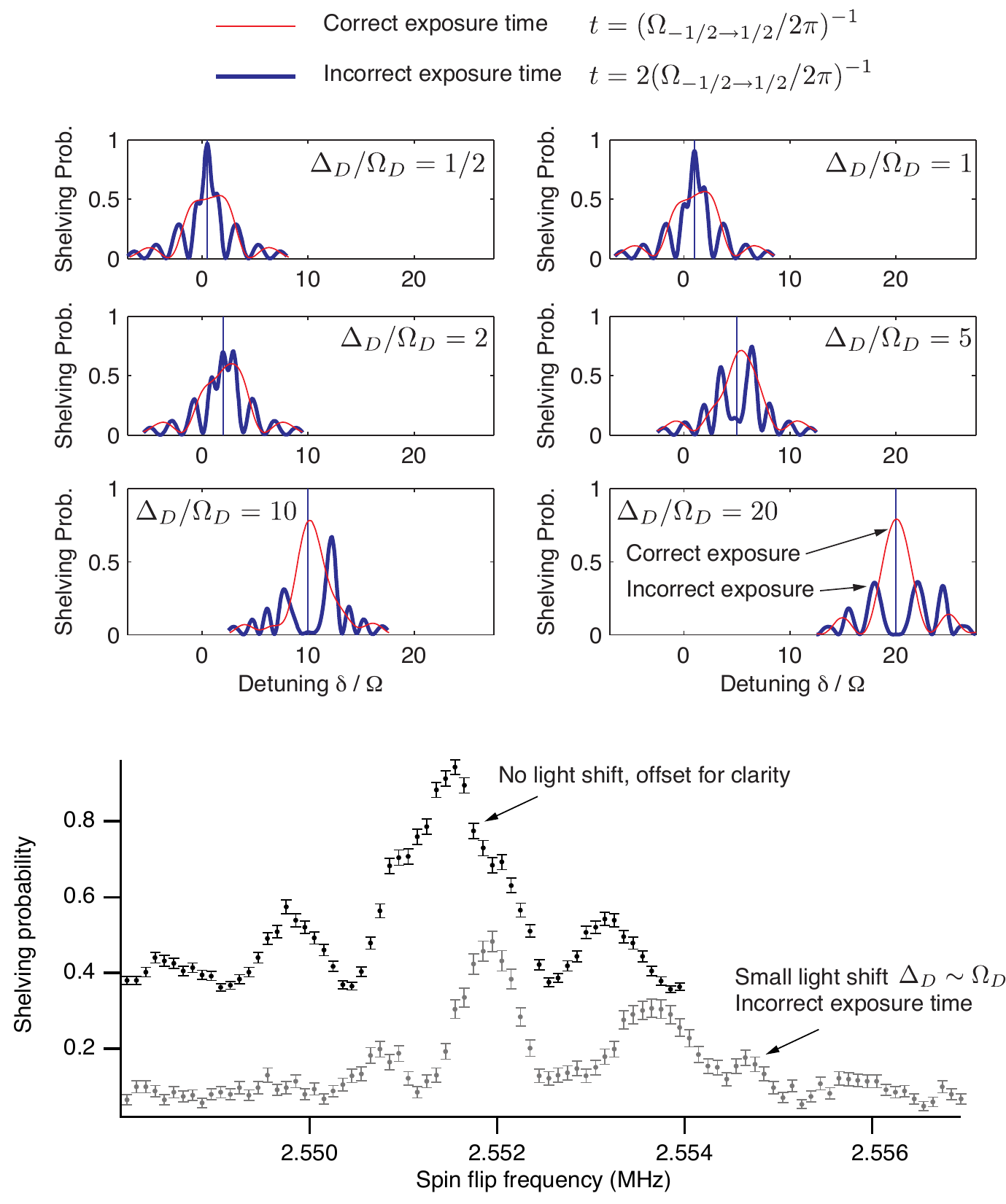}
\caption[Distortion in light shift data from incorrect exposure timing]{These data demonstrate theoretically (top) and experimentally (bottom, {\tt 030403-6}) distortions in light shifted spin resonance curves resulting from incorrect spin-flip exposure time.  We encountered this in very early data by mistakenly assuming a spin-1/2 Rabi frequency instead of the correct $\Omega_{-1/2 \to 1/2}$ Rabi frequency for a $J = 3/2$ system (see Section~\ref{sec:spinDynamics}).}
\label{fig:distortedLightShiftExample}
\end{figure}

Though we expect to measure the $m = \pm 1/2$ splitting in $5D_{3/2}$, the outer two splittings
\begin{align}
\Delta_\text{upper} &\equiv \Delta E_{5D_{3/2}, m=3/2} - \Delta E_{5D_{3/2}, m=1/2},\\
\Delta_\text{lower} &\equiv \Delta E_{5D_{3/2}, m=-1/2} - \Delta E_{5D_{3/2}, m=-3/2}
\end{align}
nevertheless play a role in shifting the $\omega_D^{LS}$ lineshape through distortions of the resonance profile.  Only the vector part of the light shift Hamiltonian shifts magnetic sublevels an amount proportional to $m$.  While the scalar shifts are completely undetectable, the tensor shift makes  $\Delta_\text{upper}$ and  $\Delta_\text{lower}$ in general different than $\Delta_D = \Delta_\text{middle}$.  If the light shifts are comparable in size to the spin-flip Rabi frequency $\Delta_D \sim \Omega$, then all the shifted $5D_{3/2}$ resonances are quasi-resonant with the spin-flip field.  Combine with this the result from Section~\ref{sec:opticalPumpingSD}---that in general, through optical pumping, the $5D_{3/2}$ begins the experiment in a statistical mixture of the $m = -3/2$ and $m= -1/2$, levels and predicted resonance curves such as those in Figure~\ref{fig:dLineShapeModels} begin to make sense.  Example data taken in the regime $\Delta_D \sim \Omega$ are shown in Figure~\ref{fig:distortedLightShiftExample}.

\begin{table}
\centering
\caption[Estimated light shift magnitudes at various wavelengths]{When choosing a light shift wavelength, it is important to minimize the $5D_{3/2}$ resonance line shape effects.  This table shows various candidate lasers, an estimate of the light shift ratio $R_\text{est}$, and observable vector light shifts $\Delta_S$, $\Delta_D$ defined the in the next as well as the shifts of each of the individual $5D_{3/2}$ sublevels.}
\footnotesize
\begin{tabular}{lr|r|rrllll}
& & & \multicolumn{6}{c}{Light shift estimates (kHz/mW), $20 \mu$m spot} \\
\multicolumn{1}{c}{Laser type} & \multicolumn{1}{c|}{$\lambda$~(nm)} &
\multicolumn{1}{c|}{$R_\text{est}$} & $\Delta_S$ & $\Delta_D$ &
$\Delta_D^{m=-3/2}$ & $\Delta_D^{m=-1/2}$ & $\Delta_D^{m=+1/2}$ & $\Delta_D^{m=+3/2}$ \\ \hline \hline
Diode ECDL & 399.000 & -3.37 & 0.328 & -0.097 & -0.147 & -0.025 & 0.072 & 0.146 \\
Argon-ion & 487.986 & 99.20 & -15.436 & -0.156 & -0.341 & -0.120 & 0.035 & 0.126 \\
Argon-ion & 514.687 & -13.09 & 2.524 & -0.193 & -0.456 & -0.180 & 0.013 & 0.122 \\
Nd:YAG & 532.000 & -5.08 & 1.161 & -0.229 & -0.565 & -0.240 & -0.012 & 0.121 \\
HeNe/Dye & 632.816 & -0.14 & 0.209 & -1.477 & -4.318 & -1.285 & 0.192 & 0.113 \\
Diode ECDL & 635.000 & -0.12 & 0.205 & -1.705 & -5.003 & -1.519 & 0.187 & 0.113 \\
Diode ECDL & 660.000 & 0.06 & 0.168 & 2.656 & 8.081 & 2.807 & 0.151 & 0.112 \\
Diode ECDL & 685.000 & 0.18 & 0.143 & 0.801 & 2.512 & 0.933 & 0.133 & 0.111 \\
Nd:YAG & 946.000 & 0.47 & 0.062 & 0.132 & 0.501 & 0.224 & 0.092 & 0.105 \\
Nd:YAG & 1064.000 & 0.49 & 0.053 & 0.107 & 0.425 & 0.195 & 0.088 & 0.103 \\
Yb:Fiber & 1111.600 & 0.50 & 0.050 & 0.101 & 0.406 & 0.187 & 0.087 & 0.103 \\
Tm,Ho:YLF & 2051.000 & 0.53 & 0.035 & 0.067 & 0.301 & 0.146 & 0.079 & 0.100
\end{tabular}
\label{tab:lightShiftComponents}
\end{table}

\begin{figure}
\centering
\includegraphics{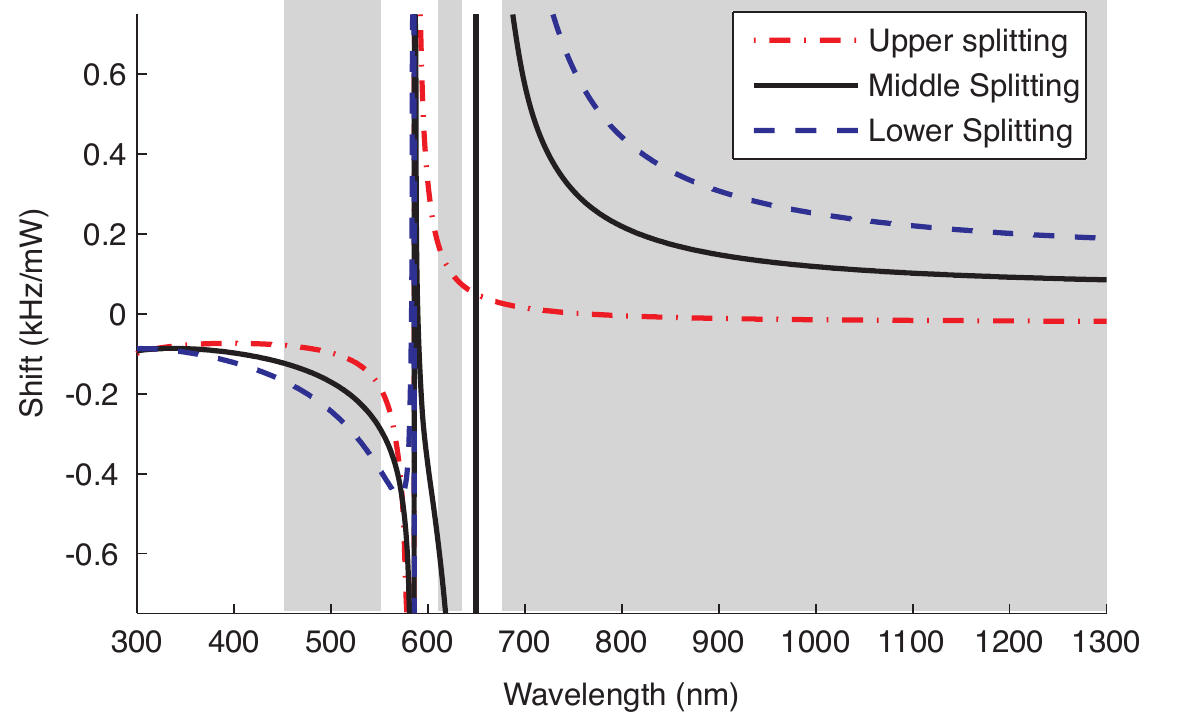}
\caption[Suitable wavelengths, given the $5D_{3/2}$ light shift distribution]{The $5D_{3/2}$ line fitting problem is worst when $\Delta_\text{upper}$ or $\Delta_\text{lower}$ are too similar to $\Delta_D$.  In this plot, the shaded regions indicate where the splittings are sufficiently non-degenerate at wavelengths not too close to dipole resonances (vertical lines).}
\label{fig:dStateLSSplittings}
\end{figure}

\begin{figure}
\centering
\includegraphics[width=5 in]{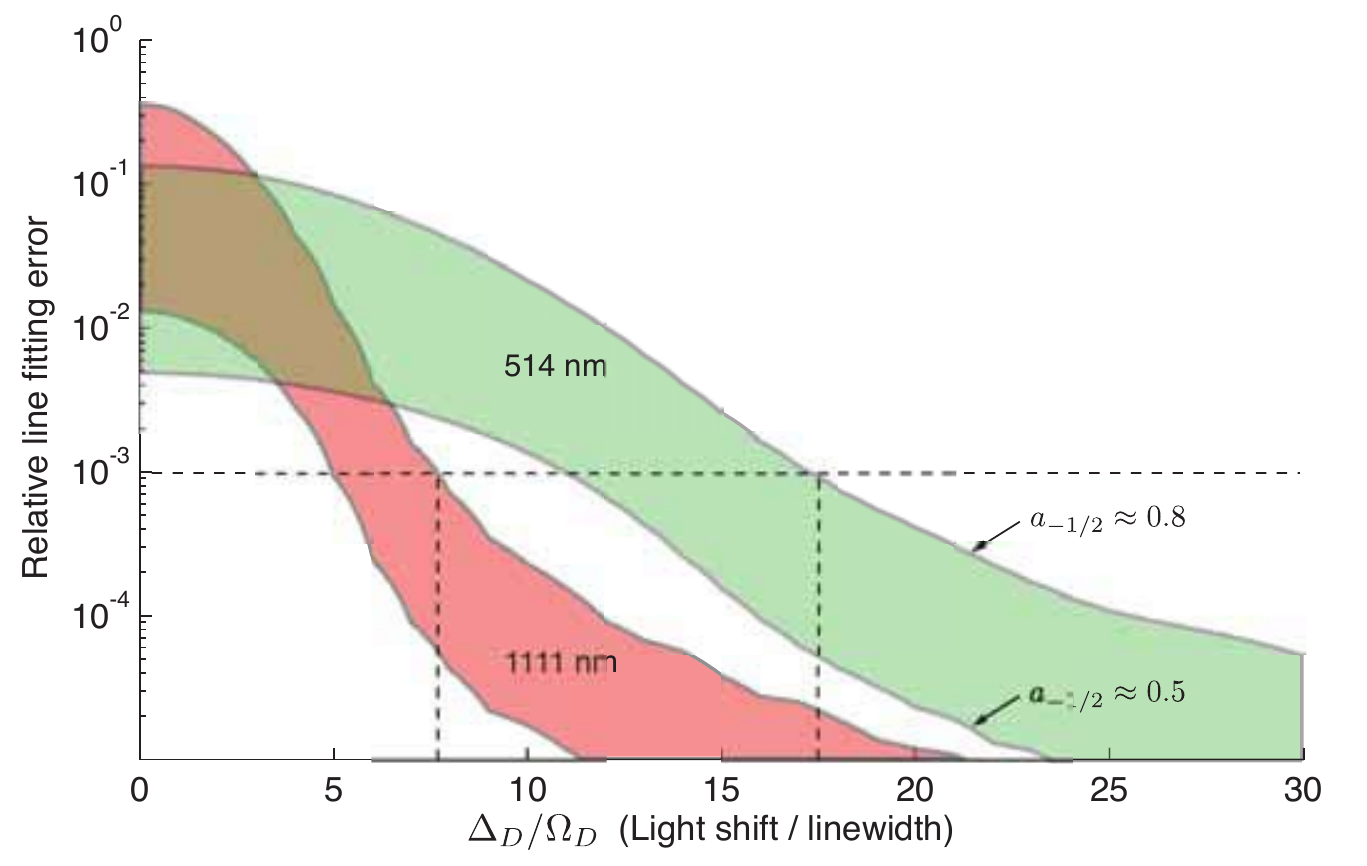}
\includegraphics[width=5 in]{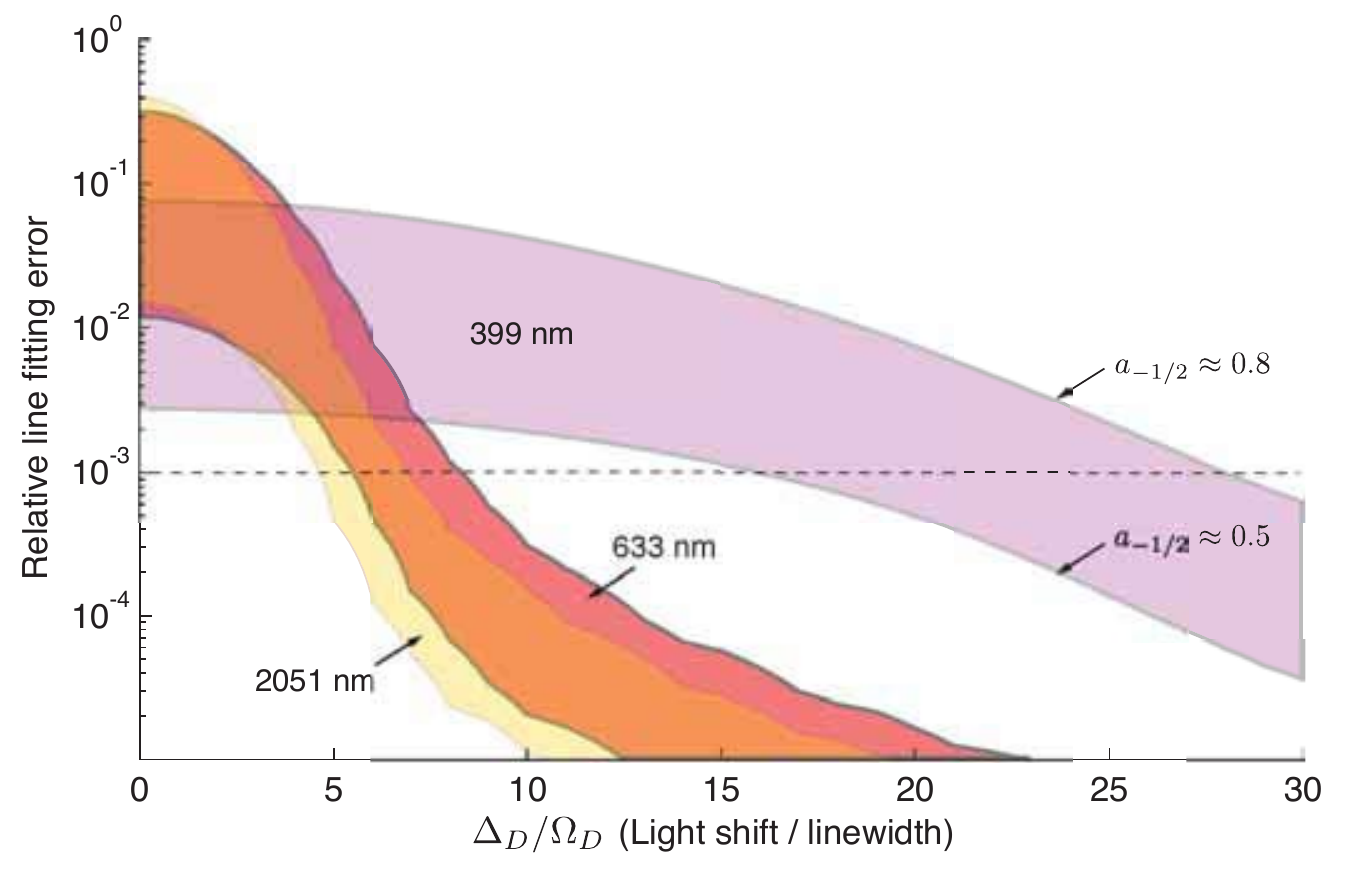}
\caption[Models of the $5D_{3/2}$ spin resonance line-fitting error]{Models of the $5D_{3/2}$ spin resonance line-fitting error.  The top graph compares the typical error magnitude for 514~nm and 1111~nm light shifts considered here.  The bottom graph shows the error profiles for some other candidate light shift wavelengths.  For each wavelength, the top curve assumes initial $5D_{3/2}$ optical pumping fraction $a_{-1/2} = 0.8$, the maximum typically observed while the bottom curve assumes $a_{-1/2} = 0.51$ which minimizes lineshape error.  Dashes lines illustrate approximate bounds on $\Omega_D / \Delta_D$ that keep the line fitting relative error below 0.1\%.}
\label{fig:linefittingModel}
\end{figure}

We observe that the \emph{size} of the distortion depends sensitively on the optical pumping conditions that prepare the ion, while the \emph{sign} depends on the sense of light shift circular polarization $\sigma < 0$ or $\sigma >0$, and whether the ion is initially prepared in the $m = -1/2,-3/2$ or $m = +1/2,+3/2$ manifolds by choosing a $\sigma^-$ or $\sigma^+$ polarized 650~nm beam during optical pumping.  We modeled the optical pumping, light shift Hamiltonian, rf spin-flip, and detection-by-shelving aspects of the experiment by numerically integrating the Bloch equations with the spin flip Hamiltonian in Eq.~\ref{eq:dSpinFlipH} using Runge-Kutta routines~\cite{press1992nrc} for all conceivable experimental conditions at several candidate light shift wavelengths.  The expected relative $\Delta_D$ shift error is plotted in Figure~\ref{fig:linefittingModel} against the relevant scaling parameter:  the ratio of light shift to spin-flip Rabi frequency $(\Delta_D / \Omega_D)$.  Experimentally, for $\pi$-pulses of rf, $(\Delta_D / \Omega_D)$ is also a measure of the vector light shift divided by the spin-flip linewidth.

Since the sign of the error depends on both the sign of the light shift laser polarization and the sign of the circular polarization used for state preparation, in principle the error can be cancelled by performing reversals of these parameters.  While we did implement such reversals in the 1111~nm data set, the earlier runs using 514~nm were not systematically controlled in this way.  Therefore, for a consistent treatment, we have decided to cut data which our model suggests has a line fitting error of more than $0.1~\%$.  In both data sets, this cut procedure does not end up shifting the reported $R$ values by more than the statistical precision.

\subsection{Laser misalignments, polarization errors, and hyperpolarizability}
Our experiment is designed to ideally measure only \emph{vector} light shifts which are maximized in the case of pure circular polarization aligned along the magnetic field.  We will show in this section that the light shift ratio $R$ is insensitive to first order to misalignments and errors in polarization as long as the light shifts are much smaller than the Zeeman shifts,
\begin{align*}
\Delta_S &\ll \omega_S, \\
\Delta_D &\ll \omega_D.
\end{align*}
In order for the light shift ratio $R = \Delta_S / \Delta_D$ to be independent of the light shift laser intensity $I$, we also must guarantee that the next-order of perturbation due the the light shift, scaling with $I^2$ and called hyperpolarizability, is small.

First we will treat laser misalignments and polarization errors following~\cite{koerber2003thesis,schacht2000thesis}.  If a (positive) circularly polarized laser beam $\boldsymbol{k}$ is misaligned by a small angle $\alpha$ with respect to the magnetic field $\boldsymbol{B}$, the polarization in a spherical basis becomes
\begin{equation*}
\boldsymbol{\epsilon} = \frac{1}{\sqrt{2}} \begin{pmatrix} \cos \alpha \\ 0 \\ \sin \alpha \end{pmatrix}.
\end{equation*}
If we further allow a small relative phase $\delta$ between the orthogonal components $\epsilon_x$ and $\epsilon_y$, the light becomes elliptically polarized,
\begin{equation*}
\boldsymbol{\epsilon} = \frac{1}{\sqrt{2}} \begin{pmatrix} \cos \alpha \\ ie^{i \delta} \\ \sin \alpha \end{pmatrix}.
\end{equation*}
The size of the vector part of the light shift is proportional to what we will term the circular polarization strength
\begin{align}
\sigma &= | \boldsymbol{\sigma} | = | \boldsymbol{\epsilon}^* \times \boldsymbol{\epsilon} |, \\
& = |(-\cos \delta \sin \alpha, 0, \cos \delta \cos \alpha) |.
\end{align}
Some refer to the vector part of the light shift as an effective magnetic field along the vector $\boldsymbol{\sigma}$.  When $\alpha = 0$, $\boldsymbol{\sigma}$ points along the direction of the main magnetic field $\bhat{z}$, and the quantity $\Delta_D$ defined in Eq.~\ref{eq:deltaDdef} exactly measures the vector light shift.  But for $\alpha \ne 0$, the light shift perturbation does not commute with the Zeeman Hamiltonian.  Specifically, the total Hamiltonian in the $5D_{3/2}$ state is
\begin{align}
H &= H_\text{Zeeman} + H_\text{Light shift} \\
&= \scriptsize \begin{pmatrix}
-\tfrac{3}{2}(\omega_D + \Delta_D' \cos \alpha \cos \delta)  & -\tfrac{\sqrt{3}}{2} \Delta_D' \cos \delta \sin \alpha & 0 & 0 \\
-\tfrac{\sqrt{3}}{2} \Delta_D' \cos \delta \sin \alpha & -\tfrac{1}{2}(\omega_D + \Delta_D' \cos \alpha \cos \delta) & - \Delta_D' \cos \delta \sin \alpha & 0 \\
0	& - \Delta_D' \cos \delta \sin \alpha & +\tfrac{1}{2}(\omega_D + \Delta_D' \cos \alpha \cos \delta) & -\tfrac{\sqrt{3}}{2} \Delta_D' \cos \delta \sin \alpha \\
0	&	0 	& -\tfrac{\sqrt{3}}{2} \Delta_D' \cos \delta \sin \alpha	& +\tfrac{3}{2}(\omega_D + \Delta_D' \cos \alpha \cos \delta) 
\end{pmatrix}.
\end{align}
One can diagonalize this matrix;  the eigenvalues correspond to the shifted state energies.  The difference between adjacent $m$-levels, the observed vector-like shift, is
\begin{align}
\Delta_D &= \sqrt{\omega_D^2 + \Delta_D' \cos \delta(2\omega_D \cos \alpha + \Delta_D' \cos \delta)} - \omega_D \\
\intertext{which, for small shifts $\Delta_D' \ll \omega_D$, can be expanded into}
 \Delta_D &=- \cos \alpha \cos \delta \Delta_D' + \frac{\Delta_D'^2}{2\omega_D}(\cos \delta \sin \alpha)^2 + O(\Delta_D'^3). \label{eq:lightshiftAlignError}
\end{align}
We see that the observed energy shift $\Delta_D$ is reduced given finite misalignments and polarization errors.  The first term, linear in $\Delta_D'$ however, is common to both $\Delta_D$ and $\Delta_S$ and therefore cancels in the ratio.  We mechanically constrain misalignment angles to well below $10^\circ$, verified by the maximum efficiency of optical pumping.  Deviations from true perfect circular polarization is almost always less than 5\%.  For the typical $\Delta_D \approx 10$~kHz light shift on a background magnetic field splitting $\omega_D \approx 2.5$~MHz, the relative error due to Eq.~\ref{eq:lightshiftAlignError} is less than $1 \times 10^{-4}$.  A similar treatment of the tensor shift mixed into $\Delta_D$ by a misaligned beam gives a similarly small error.  The relative systematic shift is
\begin{equation}
\frac{\Delta_{D,\text{tensor}}}{\Delta_D} \sim \frac{3}{4} \frac{\Delta_D}{\omega_D} \left(\sin^2 \delta - 2 \sin^2 \alpha \right).
\end{equation}

\subsection{Laser frequency error, spectral impurity, residual fluctuations}
\begin{figure}
\centering
\includegraphics{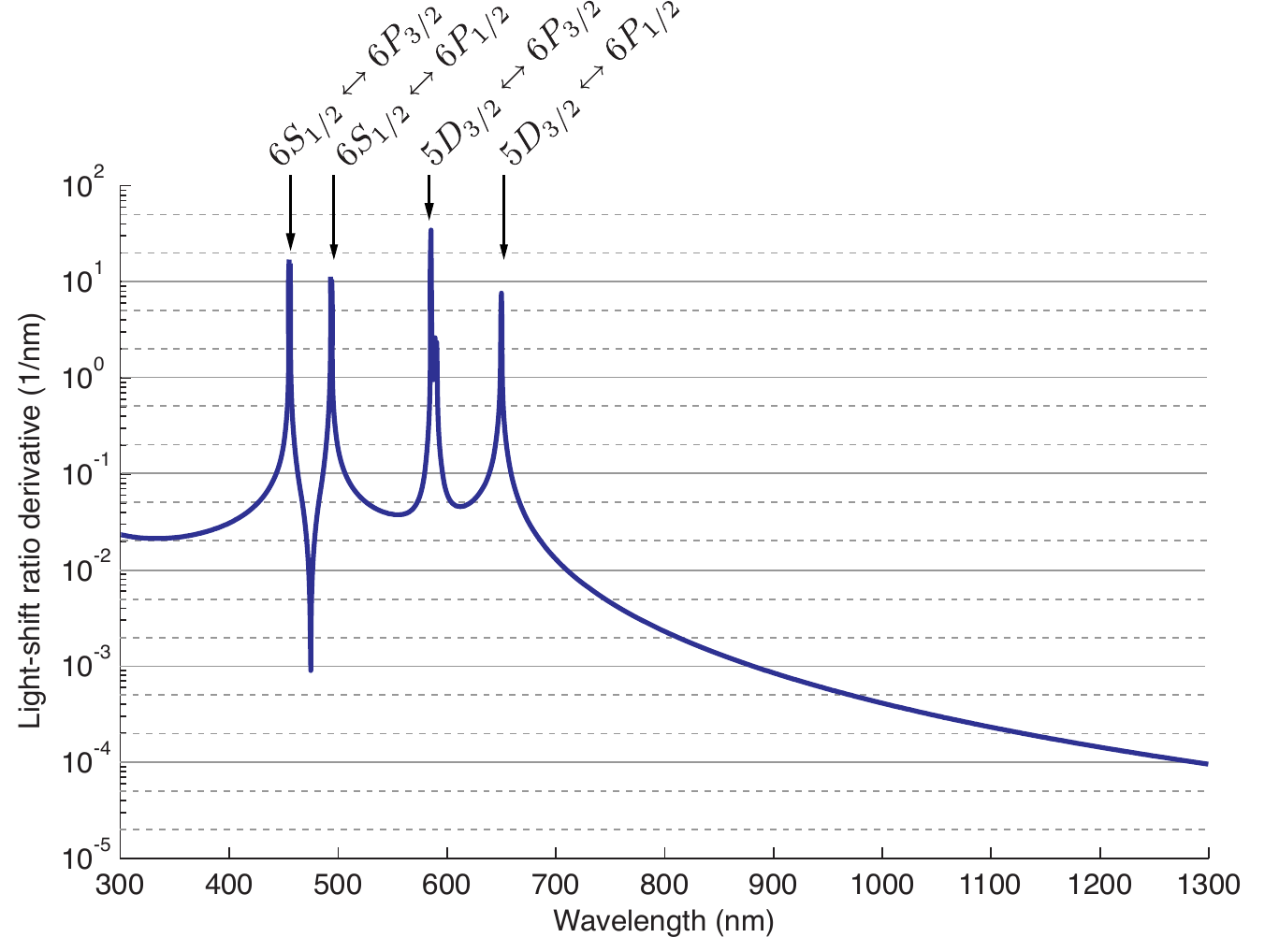}
\caption[An estimate of the light shift ratio \emph{slope}.]{An estimate of the light shift ratio \emph{slope} $|R'|$ with respect to wavelength gives us an idea of how sensitive the measured light shift ratio will be to fluctuations in the laser frequency.  The slope diverges near atomic dipole resonances.  In regions with an expected relative error $|R'| < 0.1$ nm$^{-1}$, a laser with free running stability of $\Delta \lambda = 0.01$~nm contributes a relative systematic error to the light shift ratio of $10^{-3}$.  The region with low $|R'|$ around 470~nm is in-between the $6S_{1/2} \leftrightarrow 6P_{1/2}$ and $6S_{1/2} \leftrightarrow 6P_{3/2}$ transitions.}
\label{fig:LSRslope}
\end{figure}
How accurately must we measure and control the light shift beam wavelength? Consider a quantity $R'$ the relative light shift ratio \emph{slope} with respect to wavelength:
\begin{equation}
R'(\lambda_0) \equiv \left. \frac{1}{R(\lambda_0)} \frac{d \, R(\lambda)}{d \lambda} \right|_{\lambda_0}.
\end{equation}
We expect $|R'|$ to become large only near Ba$^+$ dipole transitions.  Using existing literature values for the barium reduced matrix elements, we plot this function in Figure~\ref{fig:LSRslope}.  Examining it helps us determine the level of laser stability required at particular wavelengths.  If a given laser can be stabilized to within $\Delta \lambda$ over the course of the experiment then the contribution to the expected relative error in the reported light shift ratio is
\begin{equation}
\frac{\sigma_{\lambda}}{R} =  |R'(\lambda_0)| \Delta \lambda.
\end{equation}

\begin{table}
\centering
\caption[Estimated errors in the light shift ratio due to laser frequency fluctuations]{Estimated errors in the light shift ratio due to laser frequency fluctuations for a selection of several candidate light shift lasers.  Here we pick $\Delta \lambda$ to be characteristic of the free-running stability of the candidate laser.  Then we can estimate the relative systematic error of the light shift ratio measurement due to wavelength errors $\sigma_\lambda / R = R' \Delta \lambda$}
\small
\begin{tabular}{rll|rrrr}
Laser type & \multicolumn{1}{c}{$\lambda$ (nm)} & \multicolumn{1}{c|}{$\Delta \lambda$ (nm)} & \multicolumn{1}{c}{$R_\text{est}$} & \multicolumn{1}{c}{$R'$ (nm)$^{-1}$} & \multicolumn{1}{c}{$|\sigma_\lambda/R|$} \\ \hline \hline
Diode ECDL 	 & 399.000 & 0.01 & -3.37 & 0.030 			& $3 \times 10^{-4}$ \\
Argon-ion		 & 487.986 & 0.001 & 99.20 & 0.148 		& $2 \times 10^{-4}$  \\
Argon-ion		 & 514.687 & 0.001 & -13.09 & -0.069		& $7 \times 10^{-5}$  \\
Nd:YAG		 & 532.000 & 0.001 & -5.08 & -0.045 		& $5 \times 10^{-5}$  \\
He-Ne/Dye	 & 632.816 & 0.01 & -0.14 & -0.071 			& $4 \times 10^{-4}$  \\
Diode ECDL 	 & 635.000 & 0.01 & -0.12 & -0.079 			& $7 \times 10^{-4}$  \\
Diode ECDL 	 & 660.000 & 0.01 & 0.06 & 0.090 			& $8 \times 10^{-4}$  \\
Diode ECDL 	 & 685.000 & 0.01 & 0.18 & 0.021 			& $2 \times 10^{-4}$  \\
Nd:YAG		 & 946.000 & 0.001 & 0.47 & 0.001 			& $1 \times 10^{-6}$  \\
Nd:YAG 		 & 1064.000 & 0.001 & 0.49 & $\sim 10^{-4}$ 	& $<1 \times 10^{-6}$  \\
Yb:Fiber		 & 1111.600 & 0.01 & 0.50 & $\sim 10^{-4}$	& $<1 \times 10^{-6}$  \\
Tm,Ho:YLF	 & 2051.000 & 0.01 & 0.53 & $\sim 10^{-4}$ 	& $<1 \times 10^{-6}$ 
\end{tabular}
\label{tab:LSRslopeErrors}
\end{table}
Table~\ref{tab:LSRslopeErrors} lists the expected $R$ slope and expected free-running stability $\Delta \lambda$ at several candidate wavelengths of interest.  The $\Delta \lambda$ values shown for 514~nm and 1111~nm are the maximum drifts observed using a wavemeter over the months spent collecting data; for other wavelengths, $\Delta \lambda$ are speculations.  We adjusted the wavelength of the Yb-fiber laser at the beginning of each data run but made no attempt to actively frequency stabilize it afterwards.

We considered the possibility of poor spectral purity in each the light shift lasers.  For the 514~nm data using a single-frequency argon-ion laser we found that $\lesssim 20$~$\mu$W of background fluorescence emerges from the fiber coupling when laser is made non-lasing at but driven at full current.  The argon-ion fluorescence line at 454.2~nm is close enough to the barium ion $6S_{1/2} \leftrightarrow 6P_{3/2}$ transition to conceivably cause a worrisome shift in the $S_{1/2}$ Zeeman resonance.  Scans \verb.040220-9. and \verb.-8., are direct searches for light shifts caused just by the fluorescence and showed no such shift at the 20~Hz level.  During the latter half of data collection, we inserted an interference filter at 514~nm to remove all possibility of additional light shifts.

The Yb-fiber laser at 1111~nm is pumped with high power 795~nm diode lasers.  Using a prism we separated out and observed $< 1$~mW of pump light emitting from the fiber.  Any residual pump light that did end up focused on the ion would not likely be well circularly polarized since the polarizing elements we employ are wavelength dependent.

The closed-loop power servo detailed in Section~\ref{sec:lightShiftBeamStabilization} kept the laser power fluctuations below 0.1~\% in a dc-10~kHz bandwidth.  Slow drifts in the intensity seen by the ion, caused either by motion of the trapping center or ion trap itself, or by temperature drift in the servo electronics should not show up as a systematic effect on the light shift ratio since the four resonance measurements $\omega_S$, $\omega_S^\text{LS}$, $\omega_D$, and $\omega_D^\text{LS}$ are interleaved.  We programmed the data acquisition system to monitor the light shift beam power with a photodiode operating in photovoltaic mode and halt data taking in the case that the light shift power drifted 5\% above or below a software setpoint.  After hearing an audible alarm triggered by this event, we would discover the cause---a railing power control servo, laser interlock shutdown, the smoking remains of an argon-ion laser, etc.---and were usually able to fix the problem and resume data taking without trapping a new ion.

\subsection{Off-resonant, rf ac-Zeeman shifts (i.e.\ from trapping currents)} \label{sec:acZeemanLS}
\begin{figure}
\centering
\includegraphics[width = 5 in]{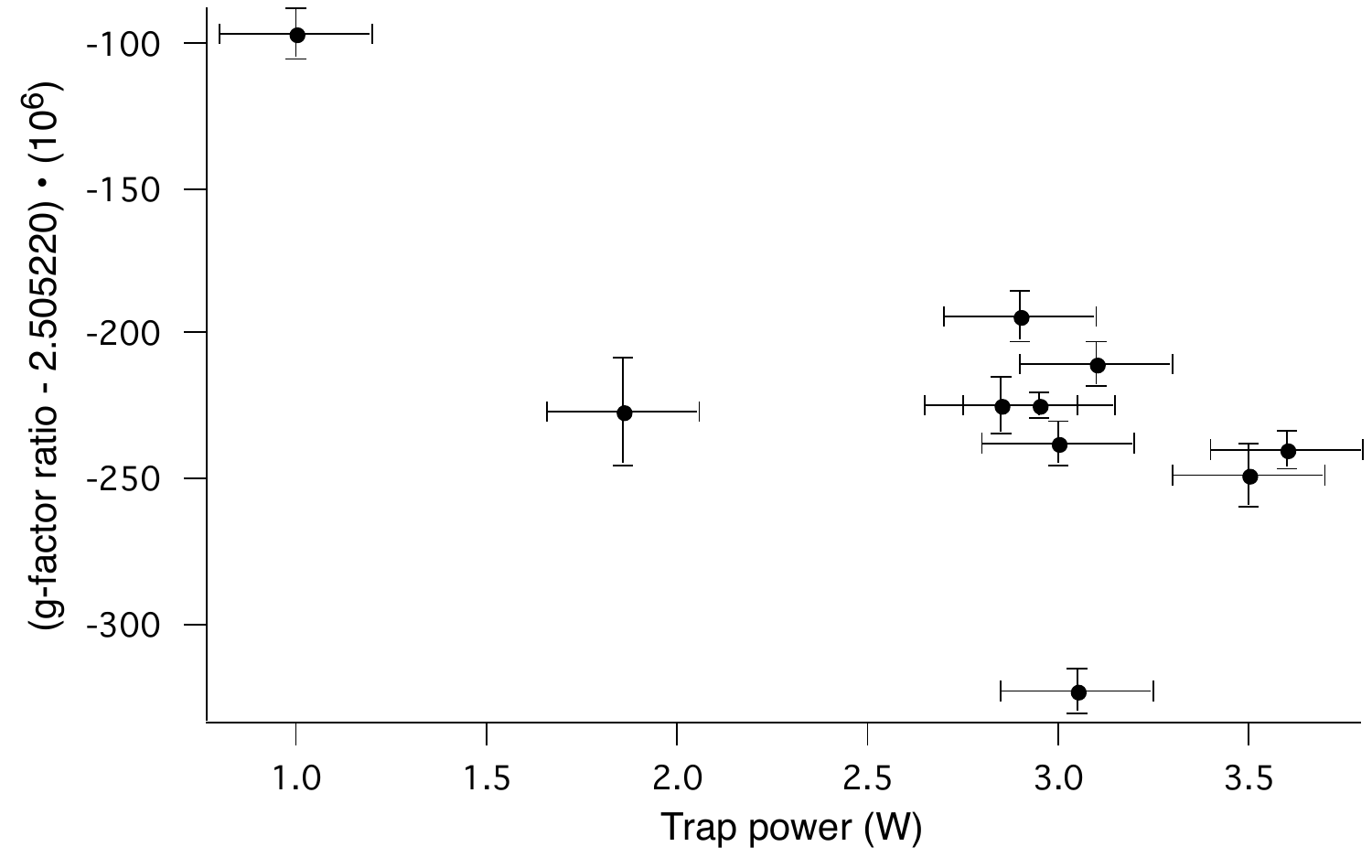}
\caption[Data showing the measured $g$-factor ratio deviation with trap strength]{Data showing the measured $g$-factor ratio deviation with trap strength.  In the text we argue that currents generated by the trapping rf create an oscillating magnetic field not very far off-resonant from relevant Zeeman transitions.  These are data {\tt 041112-1} through {\tt 041112-9} and {\tt 041215-2}.}
\label{fig:gFactorTrapShiftData}
\includegraphics{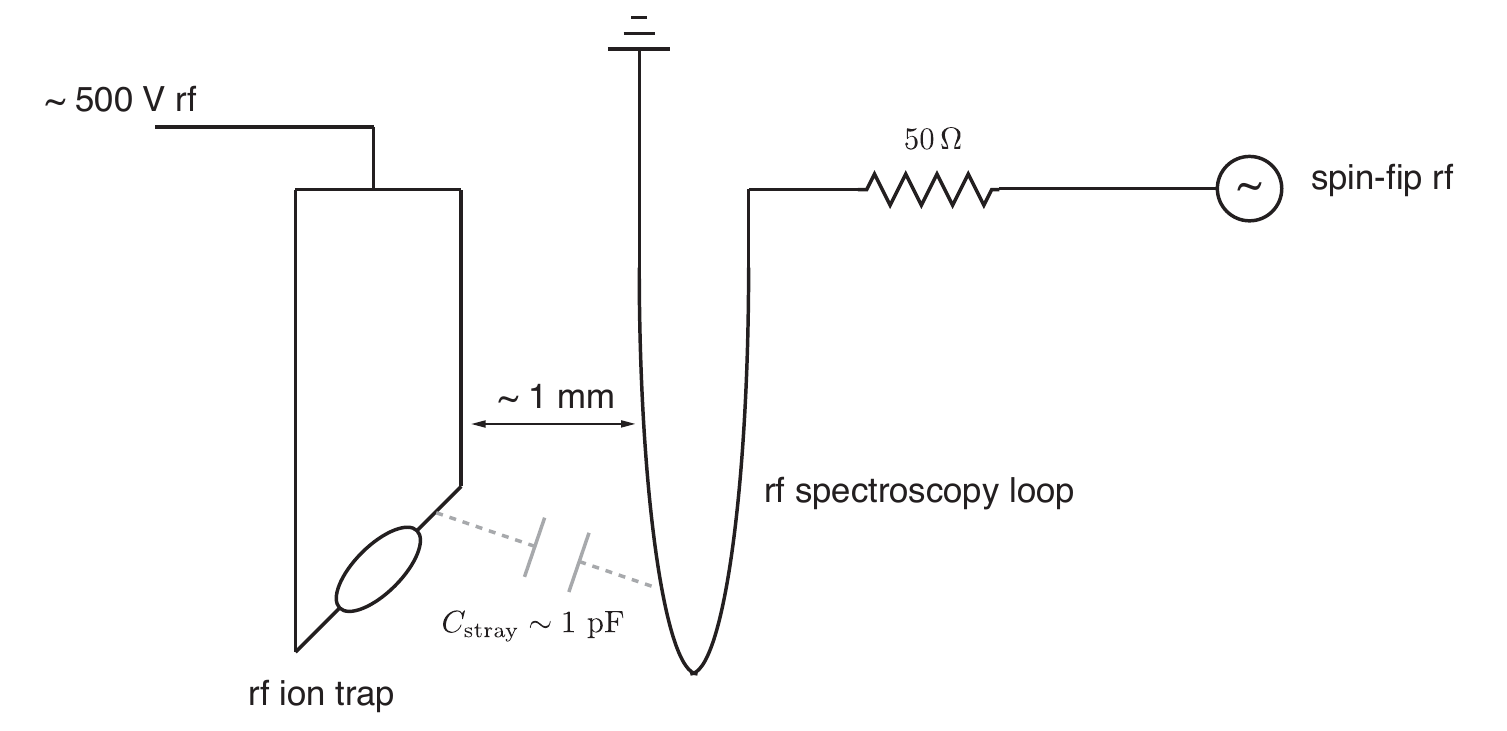}
\caption[An electrical model explaining one possible mechanism for a trap rf shift]{An electrical model explaining one possible mechanism for a trap rf shift on Zeeman resonances:  stray capacitance between the ion trap and rf spectroscopy look.}
\label{fig:trapLoopCoupling}
\end{figure}

\begin{table}
\centering
\caption[Evidence of ac-Zeeman shifts due to rf trapping fields.]{Reproduced from \cite{koerber2003thesis}, these data are evidence of off-resonance ac-Zeeman shifts due to rf fields at the trapping frequency $\omega_\text{trap} \approx 10$~MHz.  Notice that the measured error in the $g$-factor changes sign when the $6S_{1/2}$ resonance $\omega_S$ is placed above $\omega_\text{trap}$ by raising the magnetic field.}
\begin{tabular}{ccD{.}{.}{10}D{.}{.}{8}D{.}{.}{10}}
Data & Trap rf power & \multicolumn{1}{c}{$\omega_S$ (MHz)} & \multicolumn{1}{c}{$\omega_D$ (MHz)} &\multicolumn{1}{c}{$\omega_S / \omega_D - G_0$~\cite{knoll1996egf}} \\ \hline \hline
\verb.030314-5. & 2.0 W & 6.381877(24) 		& 2.547755(7)	& -0.000318(12) \\
\verb.030311-1. & 1.0 W & 6.340522(7) 		& 2.531105(7)	& -0.000178(7) \\ \hline
\verb.030313-1. & 0.8 W & 12.375745(20) 	& 4.939907(4)	& +0.000039(5) \\
\verb.030314-1. & 1.9 W & 12.376323(14)	& 4.939902(9)	& +0.000158(5) \\
\verb.030313-2. & 3.0 W & 12.376350(200) 	& 4.939737(14)& +0.000248(43)
\end{tabular}
\label{tab:gFactorTrapShiftData}
\end{table}
We began to believe our spin resonance measurements were infected with off-resonant \emph{magnetic} light shifts (the ac-Zeeman effect) because measurements of the g-factor ratio
\begin{equation}
G_\text{meas} \equiv \frac{g(6S_{1/2})}{g(5D_{3/2})} = \omega_S / \omega_D
\end{equation}
consistently deviated from the precisely measured value $G = 2.505220(2)$~\cite{knoll1996egf}.  Table~\ref{tab:gFactorTrapShiftData} offers strong evidence:  the sign of the deviation changes as we move the resonances above the trap frequency by increasing $B$.  Further, as shown in Figure~\ref{fig:gFactorTrapShiftData}, the deviation scaled with the trapping rf strength which suggested that rf currents at the trapping frequency $\omega_\text{trap} \approx 9.5$~MHz generate an oscillating magnetic field of sufficient size to light-shift our single ion resonances.  Such an off-resonant shift can be written in terms of the magnetic dipole Rabi frequency $\Omega_\text{trap rf}$ and detuning from resonance.  Including the Bloch-Siegert term we have shifts in the $6S_{1/2}$ and $5D_{3/2}$ spin resonances
\begin{align} \label{eq:trapRfZeemanResShift}
\Delta \omega_S^\text{trap rf} &= -\frac{\Omega_\text{trap rf}^2}{2(\omega_\text{trap} - \omega_S)} + \frac{\Omega_\text{trap rf}^2}{2(\omega_\text{trap} + \omega_S)}, \\
\Delta \omega_D^\text{trap rf} &= -\frac{\Omega_\text{trap rf}^2}{2(\omega_\text{trap} - \omega_D)} + \frac{\Omega_\text{trap rf}^2}{2(\omega_\text{trap} + \omega_D)}.
\end{align}
Such a shift modifies the $g$-factor ratio $G$ as measured:
\begin{equation}
G_\text{meas} = \frac{\omega_S + \Delta \omega_S^\text{trap rf}}{\omega_D + \Delta \omega_D^\text{trap rf}}.
\end{equation}
Because in our case $\omega_D < \omega_S < \omega_\text{trap}$, and $g(6S_{1/2}) \approx 2.5 g(5D_{3/2})$ we can begin by ignoring $\Delta \omega_D$ term and expect that the measured $g$-factor ratio will be \emph{decreased} by the light shift.  Then,
\begin{align}
G_\text{meas} &\approx \frac{\omega_S}{\omega_D} + \frac{\Omega_\text{trap rf}^2}{2 \omega_D} \left( \frac{1}{\omega_\text{trap} + \omega_S} - \frac{1}{\omega_\text{trap} - \omega_S} \right) \\
&= G \left(1 - \frac{\Omega_\text{trap rf}^2}{(\omega_\text{trap} - \omega_S)(\omega_\text{trap} + \omega_S)} \right) 
\end{align}
We can invert this expression and instead express the off-resonant trap rf Rabi frequency in terms of a measured g-factor ratio, the known value, and resonance frequencies:
\begin{equation}\label{eq:trapRFShift}
\Omega_\text{trap rf}^2 = \left(1 - \frac{G_\text{meas}}{G} \right)(\omega_\text{trap} + \omega_S)(\omega_\text{trap} - \omega_S)
\end{equation}

Representative data (\verb.041215-1.) taken at moderate trap power (2W) gives:
\begin{align*}
(\omega_S)_\text{meas} &= 6.270566(20) \text{ MHz} \\ 
(\omega_D)_\text{meas} &= 2.503226(11) \text{ MHz} \\ 
\Rightarrow G_\text{meas} &= \frac{(\omega_S)_\text{meas}}{(\omega_D)_\text{meas}}  = 2.504994(14).
\end{align*}
Using Eq.~\ref{eq:trapRFShift} above, we can estimate the induced trap rf Rabi frequency as
\begin{align*}
\Omega_\text{trap rf}^2 &= \left(1 - \frac{2.504994}{2.505220} \right)(9.6 + 6.3)(9.6 - 6.3)(10^6 \text{ Hz})^2 \\
\Rightarrow \Omega_\text{trap rf} &\approx 70 \text{ kHz}.
\end{align*}
According to Eq.~\ref{eq:trapRfZeemanResShift}, this means the ac-Zeeman shift of the $S_{1/2}$ resonance is about 600 Hz, or a part in $10^{4}$!

Is this even a plausible model?  Is $ \Omega_\text{trap rf} \approx 70$~kHz within reason?  A simplified electrical schematic shown in Figure~\ref{fig:trapLoopCoupling} shows the trap loop capacitively coupled to its nearest grounded electrode, the spin-flip resonance antenna.  If we assume a capacitance of 1~pF between them and a 500~V trapping voltage, the impedance $Z$ and induced rf current $I$ is
\begin{align*}
Z &= \frac{1}{2 \pi (9.5 \text{ MHz} )(1 \text{ pF})} \approx 1.7 \times 10^{4} \, \Omega \\
\Rightarrow I &\approx \frac{500 \text{ V}}{1.7 \times 10^{4} \, \Omega} = 30 \text{ mA}.
\end{align*}
Since such currents flow through the rf spectroscopy loop to ground, we can use the information in Eq.~\ref{eq:rfLoopSpec} to infer that an oscillating magnetic field of nearly $B_\text{rf} = 30$~mG results.  For the $S_{1/2}$ state, this would imply a Rabi frequency
\begin{equation*}
\Omega^S_\text{trap rf} = g_S \mu_B B_\text{rf} \approx 84 \text{ kHz}
\end{equation*}
which seems to be just the right order of magnitude to explain the $g$-factor anomaly in the data shown above.  Yet another mechanism for the generation of an oscillating magnetic field at the site of the ion is micromotion within a magnetic field gradient.  This does not seem plausible however:  the required magnetic field gradient of about 10~mG/$\mu$m is too large given our coil arrangement.  

Having established a plausible model for the apparent $g$-factor ratio deviations, we can ask what effects might be present on the light shift ratio measurement?  It is clear that the Zeeman resonances $\omega_S$ and $\omega_D$ must be far from the trapping frequency $\omega_\text{trap}$ and light shifts $\Delta_S$ and $\Delta_D$ must be kept small compared to the detunings $\omega_\text{trap} - \omega_S$ and $\omega_\text{trap} - \omega_D$ to minimize the effect on the light shift ratio.  Expressing each resonance shift in the light shift ratio:
\begin{equation}
R_\text{meas} = \frac{\omega_S^{LS} - \omega_S + (\Delta \omega_S^{LS} - \Delta \omega_S)^\text{trap rf}}{\omega_D^{LS} - \omega_D + (\Delta \omega_D^{LS} - \Delta \omega_D)^\text{trap rf}}.
\end{equation}
For light shift wavelengths with $\Delta_S \gg \Delta_D$ (such as 514~nm) it suffices to consider only the trap rf shift on the $S_{1/2}$ resonances:
\begin{align}
R_\text{meas} &= \frac{\omega_S^{LS} - \omega_S + (\Delta \omega_S^{LS} - \Delta \omega_S)^\text{trap rf}}{{\omega_D^{LS} - \omega_D}} \\
&= R + \frac{ (\Delta \omega_S^{LS} - \Delta \omega_S)^\text{trap rf}}{\omega_D^{LS} - \omega_D}, \\
\intertext{with}
\begin{split}
(\Delta \omega_S^{LS} - \Delta \omega_S) &= -\frac{\Omega^2_\text{trap rf}}{2} \left(\frac{1}{\omega_\text{trap} - \omega_S^{LS}} - \frac{1}{\omega_\text{trap} + \omega_S^{LS}} \right. \\
&\left. - \frac{1}{\omega_\text{trap} - \omega_S} + \frac{1}{\omega_\text{trap} + \omega_S} \right)
\end{split}
\end{align}
where $\Omega^2_\text{trap rf}$ can be derived from a data set using Eq.~\ref{eq:trapRFShift}.  So, for a representative data run, \verb.041130-2., we have
\begin{alignat*}{2}
\omega_D &= 2.629673(9) \text{ MHz} & \qquad \omega_D^{LS} &= 2.643767(151) \text{ MHz} \\
\omega_S &= 6.587177(5) \text{ MHz} & \qquad \omega_S^{LS} &= 6.423016(463) \text{ MHz} \\
\end{alignat*}
which gives
\begin{align*}
R_\text{meas} &= -11.56(13) \\
\Omega^S_\text{trap rf} &\approx 74 \text{ kHz} \\
\Rightarrow (\Delta \omega_S^{LS} - \Delta \omega_S) &= -58 \text{ Hz} \\
\Rightarrow R_0 &= R_\text{meas} - \frac{-58 \text{ Hz}}{14.1 \text{ kHz}} \\
\end{align*}
which is a relative shift of about $0.04 \%$.  This is below the overall statistical error but not by much, and thus is included in the final analysis along with an analogous treatment for the $5D_{3/2}$ resonance more relevant for the 1111~nm data.  It is also clear that the systematic effect could be much bigger had either of the magnetic resonances been closer to the trap frequency.

\subsection{On-resonant, rf Bloch-Siegert shift (i.e.\ from spin-flipping field)}
\begin{figure}
\centering
\includegraphics{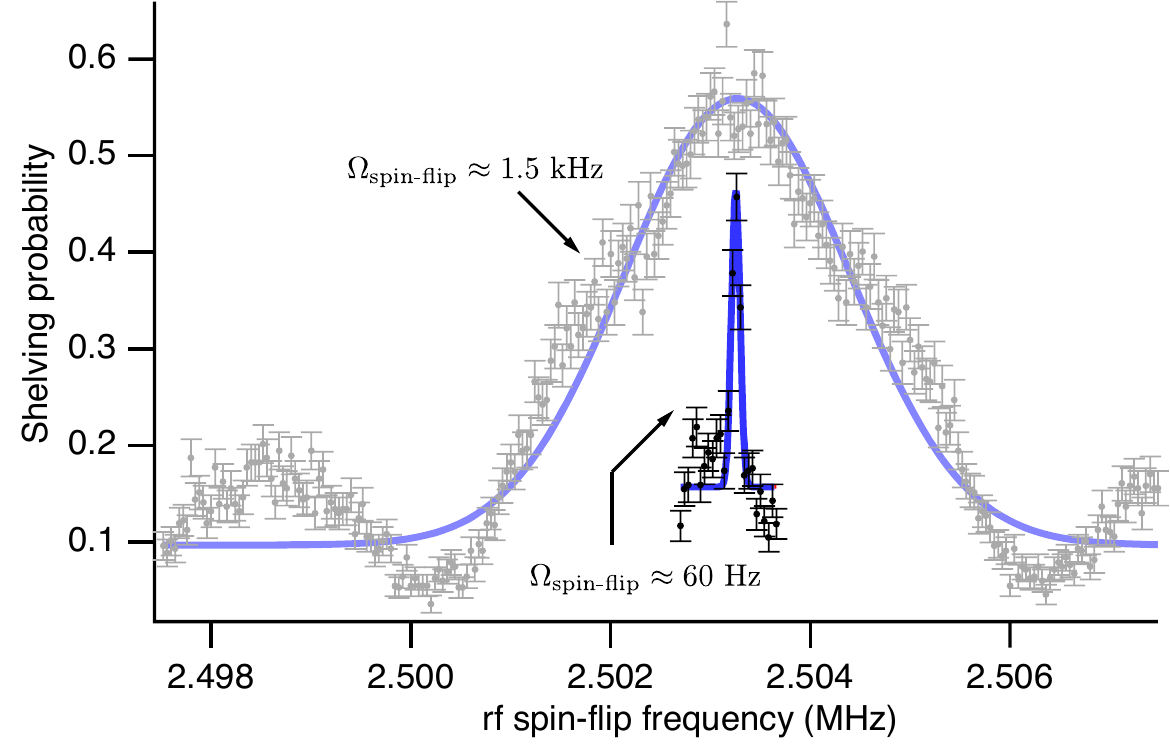}
\caption[An experimental bound of resonance shifts scaling with spin-flip rf power.]{A data run demonstrating the lack of resonance shifts scaling with spin-flip power. Scan {\tt 041216-1} features two interleaved $\omega_D$ resonance scans differing only in spin-flip amplitude by a factor of $\approx 25$.  The $\sim 1$~Hz shift we expect to see is not resolved due to statistical uncertainty, but the data nonetheless constrain the amount of resonance shift scaling with spin-flip amplitude.}
\label{fig:spinFlipShiftRun}
\end{figure}
The resonant spin-flip field is itself responsible for an ac-Zeeman light shift through the Bloch-Siegert effect (see Eq.~\ref{eq:lightShiftsBlochS}):
\begin{equation}
\Delta \omega_0^{BS} = +\frac{\Omega_\text{spin flip}^2}{4 \omega_0}.
\end{equation}
The effect would cancel in the light shift ratio \emph{if} we probed the light-shifted and unshifted resonances with the same rf power.  In practice we do not since the light-shifted resonances undergo a substantial amount of temporal noise, from fluctuating laser intensity or pointing, that require broadening the resonances with larger spin-flip fields.  $\Omega_\text{spin flip}^D < 2$~kHz and  $\Omega_\text{spin flip}^S < 4$~kHz during our experiments, so we can bound the effect on the resonances as:
\begin{alignat*}{2}
\Delta \omega_D^{BS} &\le \frac{ \Omega_\text{spin flip}^2}{4 \omega_D} &\qquad  \Delta \omega_S^{BS} &\le \frac{ \Omega_\text{spin flip}^2}{4 \omega_S} \\
&\le \frac{(2 \text{ kHz})^2}{4 \cdot 2.5 \text{ MHz}} & \qquad &\le \frac{(4 \text{ kHz})^2}{4 \cdot 6.3 \text{ MHz}} \quad ,\\
&\le 0.4 \text{ Hz} & \qquad &\le 0.6 \text{ Hz}
\end{alignat*}
which might amount to a light shift error of $\sim \Delta \omega_D^{BS} / \Delta D \approx 10^{-5}$ given a 10~kHz $5D_{3/2}$ light shift.  Though this effect is well below the statistical sensitivity and dwarfed by other systematics, we attempted to search for any shift that scaled with $\Omega_\text{spin flip}$ by taking special interleaved data runs differing only in spin-flip magnitude by a factor of $\approx 25$.  One such data run, \verb.041216-1., is shown in Figure~\ref{fig:spinFlipShiftRun}.

\subsection{Correlated magnetic field shifts}
Any change in the magnetic field correlated with the measurement of $6S_{1/2}$ versus $5D_{3/2}$ resonances vanishes entirely in the light-shift ratio.  However, a change in $B$ correlated with the application of the light shift laser does not cancel out.  The only conceivable mechanism is a fluctuation of the magnetic field due to the actuation of the stepper-motor shutter gating the light shift laser.  We try to suppress any such effect with two layers of magnetic shielding, and by placing the shutter more than 1~m from the trap with the solenoid aligned perpendicular to the quantization axis.  A flux-gate magnetometer placed just outside the magnetic shielding showed no change larger than 1~mG correlated with the light shift beam shutter.  Testing for any shift using the ion itself is straightforward:  leaving the light shift laser turned off, we performed a typical light shift ratio data run with narrow $\omega_D$ resonances.  During several dedicated scans, we find that there are no apparent shutter-correlated shifts at the $\pm 2$~Hz level.  The error would be correlated in both $6S_{1/2}$ and $5D_{3/2}$ resonances and larger in the former by the $g$-factor ratio of 2.5.  We put limits on the error in $R$ due to this effect to $\sim 2 \times 10^{-4}$ for 10~kHz type shifts.  Despite being below the statistical sensitivity, we assign this error estimate to each light shift run.

\subsection{Systematic ion displacements}
Any systematic shift in the ion's position due to the application of the light shift laser will lead to a light shift ratio error:  a displaced ion experiences a different light intensity and magnetic field.  Also, the dipole force on an ion in the $6S_{1/2}$ will be in principle different than when in the $5D_{3/2}$ state.  Fortunately the rf trapping force dwarfs these effects by many orders of magnitude.  Previous work \cite{koerber2003thesis} also treats systematic movement of the ion trap loop and fluctuations of the pseudo-potential due to application of the spin-flip rf:  such mechanisms contribute to a systematic shift at a level far below the other effects considered in this chapter.

\section{Data and summary}
\begin{figure}
\centering
\includegraphics[width=5in]{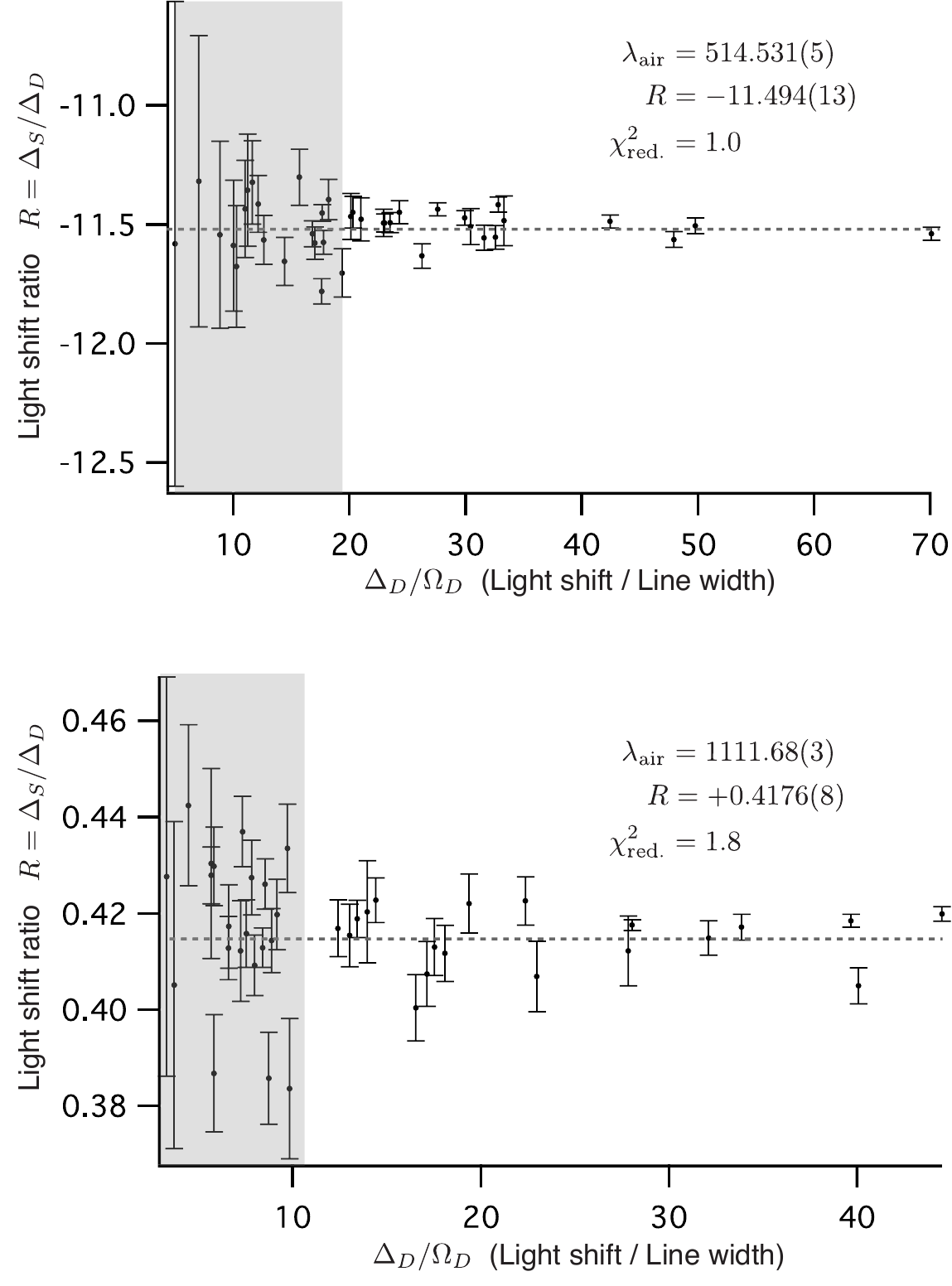}
\caption[Light shift ratio reduced data at 514~nm and 1111~nm]{Light shift ratio data at 514~nm and 1111~nm plotted against the most important parameter for systematic effects, $\Delta_D/\Omega_D$ (the light shift magnitude over the $5D_{3/2}$ resonance linewidth).  Data in the shaded regions are cut because the size of the lineshape fitting error is expected to be $> 10^{-3}$ (see Figure~\ref{fig:linefittingModel}) and the sign is uncontrolled.  In each case, this data cut does not change the fitted values for $R$ beyond the quoted errors. No statistically significant slopes are derived from  these plots which might imply an effect due to hyperpolarizability.  To account for the larger than expected scatter in the 1111~nm data, we have $\chi^2$-corrected the reported value.}
\label{fig:lightShiftRatioData}
\end{figure}

At both wavelengths surveyed, 514~nm and 1111~nm, we determined the light shift ratio to about 0.1~\% accuracy,
\begin{align}
R_\text{measured}(514.531 \text{ nm}) &= -11.494(13), \\
R_\text{measured}(1111.68 \text{ nm}) &= +0.4176(8).
\end{align}
Both values are weighted averages of all light shift data thought to be free (at the 0.1\% level) from the resonance lineshape fitting issue as discussed in Section~\ref{sec:lightShiftLineShapes}.  Inclusion of all data did not change these reported values by more than the quoted error.  Each reported error is the combination of the weighted total statistical error and systematic errors added in quadrature.  Figure~\ref{fig:lightShiftRatioData} shows that the reduced-$\chi^2$ is 1.0 for the 514~nm light shift data, and 1.8 for the 1111~nm data;  a value of 1.0 indicates that the amount of scatter in the data is well characterized statistically by the assigned error.  Omission of one strong outlier in the 1111~nm data set reduces this value to 1.3.  Nevertheless, we have decided to $\chi^2$-correct the reported errors, increasing them by a factor of $\sqrt{\chi^2_\text{red}}$, to account for the unexplained additional scatter.

Figure~\ref{fig:lightShiftRatioData} shows the light shift ratio data at both 514~nm and 1111~nm wavelengths plotted as a function of the most important systematic effect parameter, $\Delta_D / \Omega_D$, which is thought of as the strength of the light shift compared to the rf spin flip power or $5D_{3/2}$ resonance linewidth.  We found earlier (see Figure~\ref{fig:linefittingModel}) that a line fitting error becomes very large for small values of $\Delta_D / \Omega_D$.  At each wavelength, we searched for and found no statistically significant correlation of $R$ with light shift laser intensity $I$, the trap rf power, either light shift magnitude $\Delta_S$ or $\Delta_D$, the parameter $\Delta_D / \Omega_D$, the length or dates of the experimental runs, or the least-squares fit quality ($\chi^2$) of each of the four resonances. No statistically significant slopes are derived from plots of $R$ against the size of the light shift $\Delta_D$, which might imply an effect due to hyperpolarizability.

To turn our results into values for reduce dipole matrix elements, one could solve for two desired quantities in Eq.~\ref{eq:lightshiftRatioEstimate}, using other literature values as `known' quantities.  Performing this operation for the most significant matrix elements involving $5D_{3/2}$,
\begin{align*}
| \langle 5D_{3/2} || e r || 6P_{1/2} \rangle | &= 3.295(37) \\
| \langle 5D_{3/2} || e r || 4F_{5/2} \rangle | &= 3.898(160).
\end{align*}  
These values should not be taken too seriously, however.  We have not yet included the effects of core excitations or continuum states.  We believe a better approach is to ask atomic theorists to generate an \emph{ab initio} model of Ba$^+$ and calculate $R$ directly by simulating the dynamic polarization at our experimentally tested laser wavelengths.  Collaboration with theorists is ongoing towards this end.  

A more accurate determination of $\langle 5D_{3/2} || e r || 6P_{1/2} \rangle$ using our light shift method is suggested by Figure~\ref{fig:lightShiftEstimate}:  one should pick a light shift wavelength closer to the $5D_{3/2} \leftrightarrow 6P_{1/2}$ transition wavelength of 650~nm.  In fact, we spent nearly a year (circa 2002) attempting to take data at 633~nm using a dye laser;  some meager results are presented in~\cite{koerber2003thesis}.  Due to the poor performance of its aging pump laser, we simply could not regularly generate enough power to take accurate light shift data.  Also, at the time we were unaware of many experimental systematic effects and a proper laser power stabilization technique.  At this wavelength, the light shift ratio is approximately $R_\text{measured}(632.8) \text{ nm}) \approx -0.14$ but I cannot report the accuracy level.  As solid state laser sources near 650~nm improve, another light shift experiment that picks out the $\langle 5D_{3/2} || e r || 6P_{1/2} \rangle$ coupling should be attempted if theorists are not satisfied with the data at present.

For reference, our light shift data in Figure~\ref{fig:lightShiftRatioData} is presented in three tables per light shift wavelength:
\begin{center}
\begin{tabular}{lll}
	&  514~nm	& 1111~nm \\
Reduced spin-resonance data	  & Table \ref{tab:514lightShiftRawData} & Table \ref{tab:1111lightShiftRawData} \\
Derived light shift ratios		& Table \ref{tab:514lightShiftAnalysis} & Table \ref{tab:1111lightShiftAnalysis} \\
Error analysis				& Table \ref{tab:514lightShiftRatioAnalysis} & Table \ref{tab:1111lightShiftRatioAnalysis}
\end{tabular}
\end{center}
The reduced light shift data tables contain the spin resonance frequencies $\omega_S$, $\omega_S^{LS}$, $\omega_D$, and $\omega_D^{LS}$, errors in these values $\sigma$ (1 standard deviation), and linewidths $\Gamma$ extracted from least-squares Gaussian curve fits of the shelving probabilities.  In these tables $N$ is the number of shelving experiments attempted at each rf spin flip frequency for each of the four resonances.  The rest of the symbols (not previously discussed in the text) are
\renewcommand{\descriptionlabel}[1]%
	{\hspace{\labelsep}\textrm{#1}}
\begin{description}
\item[$\sigma_{LS}$]	  Sense of circular polarization of the light shift beam with respect to the magnetic field.  Either $+$ or $-$.
\item[$\sigma_\text{red}$]	 Sense of circular polarization of the 650~nm optical pumping beam.  Determined whether we began the $5D_{3/2}$ spin flip experiment in the upper or lower magnetic spin manifold.  This choice couples to the sign of the lineshape fitting error.  Either $+$, $-$, or n if not controlled. 
\item[$\omega_i$] Center frequency from a resonance line fit
\item[$\sigma_{\omega_i}$] Estimated 1-standard deviation of statistical error in a resonance line fit
\item[$\Gamma_{\omega_i}$] Gaussian width from a resonance line fit
\item[$\Delta_D, \Delta_S$]  Measured vector light shifts (Hz) 
\item[$\sigma_{\Delta_D}, \sigma_{\Delta_S}$]  Absolute 1-standard deviation error (Hz) in the light shifts obtained through error propagation 
\item[$\sigma_\text{line}/R$] Relative 1-standard deviation error  assigned to the light shift ratio due to our lineshape fitting error model and the measured parameter $\Delta_D / \Omega_D$
\item[$\sigma_{\theta,\sigma}/R$]  Relative 1-standard deviation error assigned to the light shift ratio due to possible misalignments of the laser beam with respect to the magnetic field, and errors in polarization
\item[$\sigma_\lambda / R$]  Relative 1-standard deviation error assigned to the light shift ratio due to laser frequency fluctuations 
\item[$\sigma_\text{rf} / R$]  Relative 1-standard deviation error assigned to the light shift ratio due to presumed ac-Zeeman effects inferred from a g-factor ratio $\omega_S / \omega_D$ that deviates from the known value. 
\item[$\sigma_R$]  Absolute 1-standard deviation of the light shift ratio from propagated statistical errors 
\item[$\sigma_\text{tot}/R$] Relative 1-standard deviation of the light shift ratio from all effects added in quadrature.
\end{description}

\begin{table}[p]
\centering
\caption[514~nm reduced light shift data]{514~nm reduced light shift data.  See text for a description of the column labels.  All frequency units are Hz.}
\scriptsize
\setlength{\tabcolsep}{3 pt}
\newcolumntype{V}{>{\tt}c<{\tt}}
\begin{tabular}{cVc|ccc|ccc|ccc|ccc}
\# & File & $N$ & $\omega_D$ & $\sigma_{\omega_D}$ & $\Gamma_D$ & $\omega_D^{LS} $ & $\sigma_{\omega_D^{LS}}$ & $\Gamma_D^{LS}$ & $\omega_S$ & $\sigma_{\omega_S}$ & $\Gamma_S$ & $\omega_S^{LS} $ & $\sigma_{\omega_S^{LS}}$ & $\Gamma_S^{LS}$ \\ \hline \hline
1 & 030713-4 & 64 & 2,551,742 & 11 & 176 & 2,546,743 & 40 & 453 & 6,392,182 & 34 & 328 & 6,449,342 & 286 & 3,097\\
2 & 030714-3 & 350 & 2,551,674 & 5 & 191 & 2,546,592 & 18 & 402 & 6,392,022 & 17 & 355 & 6,450,791 & 88 & 2,479\\
3 & 030716-1 & 327 & 2,551,639 & 4 & 137 & 2,545,372 & 11 & 258 & 6,391,946 & 8 & 207 & 6,463,698 & 255 & 3,731\\
4 & 030716-4 & 306 & 2,551,777 & 4 & 143 & 2,544,699 & 10 & 216 & 6,392,238 & 12 & 274 & 6,473,050 & 157 & 3,590\\
5 & 030716-5 & 223 & 2,551,772 & 4 & 135 & 2,545,198 & 14 & 285 & 6,392,220 & 12 & 256 & 6,467,780 & 156 & 2,556\\
6 & 030719-2 & 147 & 2,551,697 & 7 & 153 & 2,544,760 & 9 & 232 & 6,392,234 & 12 & 240 & 6,471,816 & 113 & 2,442\\
7 & 030719-3 & 410 & 2,551,690 & 8 & 224 & 2,548,500 & 8 & 263 & 6,392,234 & 13 & 415 & 6,428,654 & 57 & 1,909\\
8 & 030720-1 & 216 & 2,551,698 & 7 & 167 & 2,549,964 & 7 & 247 & 6,392,242 & 11 & 287 & 6,411,865 & 20 & 729\\
9 & 030720-3 & 302 & 2,551,728 & 6 & 164 & 2,544,695 & 8 & 255 & 6,392,272 & 10 & 266 & 6,472,706 & 99 & 2,545\\
10 & 030721-1 & 27 & 2,551,454 & 17 & 160 & 2,544,040 & 35 & 365 & 6,391,752 & 17 & 159 & 6,476,628 & 179 & 1,990\\
11 & 030722-2 & 44 & 2,551,577 & 14 & 158 & 2,542,535 & 32 & 538 & 6,392,086 & 25 & 305 & 6,496,430 & 196 & 2,880\\
12 & 030723-1 & 139 & 2,551,755 & 12 & 215 & 2,542,366 & 18 & 532 & 6,392,388 & 17 & 309 & 6,499,901 & 145 & 3,300\\
13 & 030723-3 & 30 & 2,551,646 & 19 & 151 & 2,546,860 & 32 & 305 & 6,392,172 & 33 & 318 & 6,446,264 & 347 & 2,998\\
14 & 030729-1 & 118 & 2,551,698 & 9 & 153 & 2,546,236 & 16 & 310 & 6,392,371 & 15 & 214 & 6,456,702 & 166 & 2,380\\
15 & 030729-3 & 174 & 2,551,812 & 9 & 193 & 2,546,345 & 14 & 307 & 6,392,504 & 13 & 243 & 6,455,796 & 157 & 3,109\\
16 & 030730-3 & 142 & 2,551,678 & 10 & 179 & 2,541,808 & 15 & 206 & 6,392,267 & 20 & 331 & 6,506,400 & 198 & 2,701\\
17 & 030730-5 & 106 & 2,551,789 & 7 & 126 & 2,542,681 & 25 & 288 & 6,392,496 & 15 & 208 & 6,497,740 & 351 & 4,667\\
18 & 030731-2 & 79 & 2,551,770 & 10 & 107 & 2,543,386 & 30 & 492 & 6,392,378 & 18 & 259 & 6,489,450 & 403 & 6,254\\
19 & 030801-2 & 101 & 2,551,715 & 9 & 124 & 2,543,838 & 19 & 300 & 6,392,296 & 24 & 414 & 6,483,926 & 303 & 4,584\\
20 & 030801-4 & 69 & 2,551,750 & 12 & 169 & 2,545,487 & 35 & 323 & 6,392,357 & 21 & 247 & 6,465,664 & 448 & 4,172\\
21 & 030801-5 & 98 & 2,551,618 & 10 & 158 & 2,545,374 & 35 & 432 & 6,392,063 & 22 & 298 & 6,464,830 & 389 & 6,213\\
22 & 030804-4 & 84 & 2,551,868 & 11 & 153 & 2,541,679 & 12 & 145 & 6,392,674 & 17 & 288 & 6,510,254 & 114 & 1,823\\
23 & 030805-3 & 104 & 2,551,644 & 8 & 149 & 2,541,767 & 14 & 198 & 6,392,170 & 25 & 357 & 6,505,817 & 201 & 2,680\\
24 & 030805-7 & 144 & 2,551,630 & 8 & 133 & 2,544,043 & 13 & 233 & 6,392,118 & 16 & 242 & 6,479,774 & 313 & 4,301\\
25 & 030806-2 & 112 & 2,551,643 & 9 & 144 & 2,546,408 & 23 & 260 & 6,392,132 & 20 & 273 & 6,452,156 & 408 & 4,453\\
26 & 030806-3 & 96 & 2,551,659 & 13 & 210 & 2,546,497 & 22 & 246 & 6,392,126 & 30 & 346 & 6,451,376 & 350 & 4,179\\
27 & 040302-1 & 114 & 2,626,925 & 2 & 47 & 2,615,353 & 37 & 380 & 6,580,564 & 12 & 157 & 6,713,748 & 727 & 7,167\\
28 & 040302-2 & 111 & 2,626,927 & 4 & 68 & 2,616,120 & 88 & 961 & 6,580,589 & 16 & 144 & 6,703,324 & 1,599 & 5,322\\
29 & 040303-1 & 105 & 2,626,880 & 2 & 39 & 2,612,610 & 66 & 428 & 6,580,478 & 21 & 182 & 6,744,374 & 1,244 & 4,686\\
30 & 040304-3 & 237 & 2,626,638 & 4 & 66 & 2,619,666 & 46 & 694 & 6,579,934 & 11 & 190 & 6,660,738 & 621 & 9,644\\
31 & 040312-1 & 108 & 2,627,345 & 6 & 84 & 2,622,266 & 68 & 1,017 & 6,581,400 & 10 & 179 & 6,640,222 & 1,037 & 11,661\\
32 & 040312-2 & 116 & 2,627,346 & 7 & 102 & 2,623,070 & 41 & 482 & 6,581,376 & 6 & 153 & 6,630,736 & 339 & 4,876\\
33 & 040312-3 & 198 & 2,627,335 & 6 & 111 & 2,603,358 & 31 & 565 & 6,581,313 & 7 & 191 & 6,856,756 & 251 & 4,767\\
34 & 040625-1 & 73 & 2,622,100 & 32 & 302 & 2,637,262 & 63 & 661 & 6,567,372 & 39 & 472 & 6,393,096 & 206 & 1,363\\
35 & 040713-1 & 92 & 2,627,938 & 5 & 102 & 2,618,474 & 21 & 403 & 6,582,569 & 16 & 142 & 6,691,334 & 277 & 9,855\\
36 & 041129-3 & 72 & 2,629,744 & 17 & 218 & 2,644,743 & 117 & 1,458 & 6,587,345 & 28 & 596 & 6,412,202 & 675 & 9,336\\
37 & 041130-1 & 61 & 2,629,728 & 9 & 116 & 2,644,944 & 109 & 1,307 & 6,587,236 & 7 & 155 & 6,414,926 & 1,136 & 15,367\\
38 & 041207-2 & 95 & 2,629,771 & 10 & 44 & 2,646,582 & 69 & 924 & 6,587,386 & 7 & 137 & 6,395,798 & 1,127 & 8,731
\end{tabular}
\label{tab:514lightShiftRawData}
\end{table}

\begin{table}[p]
\centering
\caption[514~nm light shift analysis]{514~nm light shift analysis.  See text for a description of the column labels.  All frequency units are Hz.}
\scriptsize
\setlength{\tabcolsep}{3 pt}
\newcolumntype{M}{>{$}c<{$}}
\begin{tabular}{cc|cccc|cMc}
\#   & $\sigma_\text{LS}, \sigma_\text{red}$ & $\Delta_D$ & $\sigma_{\Delta_D}$ & $\Delta_S$ & $\sigma_{\Delta_S}$ & $\Delta_D / \Omega_D$ & G_\text{meas} - G & $\Omega_\text{trap rf}$ (kHz) \\ \hline \hline
1 & (+,n) & -4998 & 41 & 57,160 & 288 & 11.0 & -1.9 \times 10^{-4} & 62\\
2 & (+,n) & -5082 & 19 & 58,769 & 90 & 12.7 & -1.9 \times 10^{-4} & 61\\
3 & (+,n) & -6267 & 12 & 71,753 & 255 & 24.3 & -1.8 \times 10^{-4} & 60\\
4 & (+,n) & -7078 & 11 & 80,812 & 157 & 32.8 & -2.1 \times 10^{-4} & 64\\
5 & (+,n) & -6574 & 15 & 75,559 & 156 & 23.1 & -2.1 \times 10^{-4} & 64\\
6 & (+,n) & -6936 & 12 & 79,582 & 114 & 30.0 & -1.3 \times 10^{-4} & 50\\
7 & (+,n) & -3191 & 11 & 36,419 & 59 & 12.2 & -1.2 \times 10^{-4} & 49\\
8 & (+,n) & -1734 & 9 & 19,623 & 23 & 7.0 & -1.3 \times 10^{-4} & 50\\
9 & (+,n) & -7033 & 10 & 80,433 & 100 & 27.6 & -1.4 \times 10^{-4} & 53\\
10 & (+,n) & -7415 & 39 & 84,876 & 179 & 20.3 & -7.9 \times 10^{-5} & 39\\
11 &(+,n) & -9042 & 35 & 104,344 & 198 & 16.8 & -6.9 \times 10^{-5} & 37\\
12 & (+,n) & -9389 & 21 & 107,513 & 146 & 17.6 & -1.3 \times 10^{-4} & 50\\
13 & (+,n) & -4786 & 38 & 54,092 & 348 & 15.7 & -1.0 \times 10^{-4} & 45\\
14 & (+,n) & -5461 & 19 & 64,332 & 166 & 17.6 & -7.5 \times 10^{-5} & 39\\
15 & (+,n) & -5468 & 17 & 63,291 & 158 & 17.8 & -1.4 \times 10^{-4} & 52\\
16 & (+,n) & -9870 & 18 & 114,133 & 199 & 47.9 & -9.8 \times 10^{-5} & 44\\
17 & (+,n) & -9108 & 26 & 105,244 & 351 & 31.6 & -1.2 \times 10^{-4} & 48\\
18 & (+,n) & -8384 & 32 & 97,072 & 403 & 17.0 & -1.4 \times 10^{-4} & 53\\
19 & (+,n) & -7877 & 21 & 91,630 & 304 & 26.3 & -1.2 \times 10^{-4} & 49\\
20 & (+,n) & -6264 & 37 & 73,307 & 449 & 19.4 & -1.3 \times 10^{-4} & 51\\
21 & (+,n) & -6243 & 37 & 72,767 & 390 & 14.4 & -1.2 \times 10^{-4} & 48\\
22 & (+,n) & -10189 & 16 & 117,580 & 115 & 70.1 & -1.2 \times 10^{-4} & 49\\
23 & (+,n) & -9877 & 17 & 113,647 & 202 & 49.8 & -1.0 \times 10^{-4} & 45\\
24 & (+,n) & -7587 & 15 & 87,656 & 314 & 32.6 & -1.1 \times 10^{-4} & 46\\
25 & (+,n) & -5235 & 24 & 60,024 & 408 & 20.1 & -1.2 \times 10^{-4} & 48\\
26 & (+,n) & -5162 & 26 & 59,250 & 351 & 21.0 & -1.3 \times 10^{-4} & 51\\
27 & (+,n) & -11572 & 37 & 133,184 & 727 & 30.5 & -1.8 \times 10^{-4} & 57\\
28 & (+,n) & -10808 & 88 & 122,735 & 1599 & 11.2 & -1.7 \times 10^{-4} & 56\\
29 & (+,n) & -14270 & 66 & 163,896 & 1244 & 33.3 & -1.7 \times 10^{-4} & 56\\
30 & (+,n) & -6972 & 46 & 80,804 & 621 & 10.1 & -1.4 \times 10^{-4} & 51\\
31 & (+,n) & -5079 & 68 & 58,823 & 1037 & 5.0 & -2.6 \times 10^{-4} & 70\\
32 & (+,n) & -4276 & 42 & 49,361 & 339 & 8.9 & -2.7 \times 10^{-4} & 71\\
33 & (+,n) & -23978 & 31 & 275,443 & 251 & 42.5 & -2.8 \times 10^{-4} & 73\\
34 & (-,n) & 15161 & 70 & -174,275 & 209 & 22.9 & -6.0 \times 10^{-4} & 110\\
35 & (+,n) & -9464 & 21 & 108,765 & 277 & 23.5 & -3.8 \times 10^{-4} & 84\\
36 & (-,n) & 14999 & 118 & -175,143 & 676 & 10.3 & -2.8 \times 10^{-4} & 73\\
37 & (-,n) & 15216 & 109 & -172,310 & 1136 & 11.6 & -3.1 \times 10^{-4} & 76\\
38 & (-,n) & 16811 & 69 & -191,588 & 1127 & 18.2 & -2.9 \times 10^{-4} & 74
\end{tabular}
\label{tab:514lightShiftAnalysis}
\end{table}

\begin{table}[p]
\centering
\caption[514~nm light shift ratio error analysis]{514~nm light shift ratio error analysis.  See text for a description of the column labels.  All frequency units are Hz.}
\scriptsize
\newcolumntype{M}{>{$}c<{$}}
\newcolumntype{V}{>{\tt}c<{\tt}}
\setlength{\tabcolsep}{3 pt}
\begin{tabular}{cV|MMMM|ccM}
\#   & File &  \sigma_\text{line} / R & \sigma_{\theta, \sigma}/R & \sigma_\lambda/R & \sigma_\text{rf}/R & $R$ & $\sigma_R$ & \sigma_\text{tot}/R \\ \hline \hline
1 & 030713-4 & 1.5\times 10^{-2} & 2.7\times 10^{-4} & 7\times 10^{-5} & -9.8\times 10^{-5} & -11.44 & 0.20 & 1.8\times 10^{-2}\\
2 & 030714-3 & 7.8\times 10^{-3} & 2.8\times 10^{-4} & 7\times 10^{-5} & -9.6\times 10^{-5} & -11.56 & 0.10 & 8.8\times 10^{-3}\\
3 & 030716-1 & 5.0\times 10^{-4} & 3.4\times 10^{-4} & 7\times 10^{-5} & -9.4\times 10^{-5} & -11.45 & 0.05 & 4.3\times 10^{-3}\\
4 & 030716-4 & 5.0\times 10^{-4} & 3.8\times 10^{-4} & 7\times 10^{-5} & -1.0\times 10^{-4} & -11.42 & 0.03 & 2.8\times 10^{-3}\\
5 & 030716-5 & 5.0\times 10^{-4} & 3.6\times 10^{-4} & 7\times 10^{-5} & -1.1\times 10^{-4} & -11.49 & 0.04 & 3.3\times 10^{-3}\\
6 & 030719-2 & 5.0\times 10^{-4} & 3.8\times 10^{-4} & 7\times 10^{-5} & -6.5\times 10^{-5} & -11.47 & 0.03 & 2.6\times 10^{-3}\\
7 & 030719-3 & 9.6\times 10^{-3} & 1.7\times 10^{-4} & 7\times 10^{-5} & -6.2\times 10^{-5} & -11.41 & 0.12 & 1.0\times 10^{-2}\\
8 & 030720-1 & 5.4\times 10^{-2} & 9.3\times 10^{-5} & 7\times 10^{-5} & -6.5\times 10^{-5} & -11.32 & 0.61 & 5.4\times 10^{-2}\\
9 & 030720-3 & 5.0\times 10^{-4} & 3.8\times 10^{-4} & 7\times 10^{-5} & -7.3\times 10^{-5} & -11.44 & 0.02 & 2.3\times 10^{-3}\\
10 & 030721-1 & 5.0\times 10^{-4} & 4.0\times 10^{-4} & 7\times 10^{-5} & -4.0\times 10^{-5} & -11.45 & 0.07 & 5.8\times 10^{-3}\\
11 & 030722-2 & 1.2\times 10^{-3} & 4.9\times 10^{-4} & 7\times 10^{-5} & -3.5\times 10^{-5} & -11.54 & 0.05 & 4.7\times 10^{-3}\\
12 & 030723-1 & 9.1\times 10^{-4} & 5.1\times 10^{-4} & 7\times 10^{-5} & -6.3\times 10^{-5} & -11.45 & 0.03 & 3.1\times 10^{-3}\\
13 & 030723-3 & 1.9\times 10^{-3} & 2.6\times 10^{-4} & 7\times 10^{-5} & -5.3\times 10^{-5} & -11.30 & 0.12 & 1.0\times 10^{-2}\\
14 & 030729-1 & 9.1\times 10^{-4} & 3.0\times 10^{-4} & 7\times 10^{-5} & -3.8\times 10^{-5} & -11.78 & 0.05 & 4.5\times 10^{-3}\\
15 & 030729-3 & 8.6\times 10^{-4} & 3.0\times 10^{-4} & 7\times 10^{-5} & -6.9\times 10^{-5} & -11.58 & 0.05 & 4.3\times 10^{-3}\\
16 & 030730-3 & 5.0\times 10^{-4} & 5.4\times 10^{-4} & 7\times 10^{-5} & -4.9\times 10^{-5} & -11.56 & 0.03 & 2.9\times 10^{-3}\\
17 & 030730-5 & 5.0\times 10^{-4} & 5.0\times 10^{-4} & 7\times 10^{-5} & -5.9\times 10^{-5} & -11.56 & 0.05 & 4.6\times 10^{-3}\\
18 & 030731-2 & 1.1\times 10^{-3} & 4.6\times 10^{-4} & 7\times 10^{-5} & -7.3\times 10^{-5} & -11.58 & 0.07 & 5.9\times 10^{-3}\\
19 & 030801-2 & 5.0\times 10^{-4} & 4.3\times 10^{-4} & 7\times 10^{-5} & -6.2\times 10^{-5} & -11.63 & 0.05 & 4.4\times 10^{-3}\\
20 & 030801-4 & 5.6\times 10^{-4} & 3.5\times 10^{-4} & 7\times 10^{-5} & -6.8\times 10^{-5} & -11.70 & 0.10 & 8.6\times 10^{-3}\\
21 & 030801-5 & 3.4\times 10^{-3} & 3.4\times 10^{-4} & 7\times 10^{-5} & -6.0\times 10^{-5} & -11.66 & 0.10 & 8.7\times 10^{-3}\\
22 & 030804-4 & 5.0\times 10^{-4} & 5.5\times 10^{-4} & 7\times 10^{-5} & -6.3\times 10^{-5} & -11.54 & 0.02 & 2.3\times 10^{-3}\\
23 & 030805-3 & 5.0\times 10^{-4} & 5.4\times 10^{-4} & 7\times 10^{-5} & -5.1\times 10^{-5} & -11.51 & 0.03 & 2.8\times 10^{-3}\\
24 & 030805-7 & 5.0\times 10^{-4} & 4.1\times 10^{-4} & 7\times 10^{-5} & -5.5\times 10^{-5} & -11.55 & 0.05 & 4.3\times 10^{-3}\\
25 & 030806-2 & 5.0\times 10^{-4} & 2.8\times 10^{-4} & 7\times 10^{-5} & -5.9\times 10^{-5} & -11.47 & 0.09 & 8.4\times 10^{-3}\\
26 & 030806-3 & 5.0\times 10^{-4} & 2.8\times 10^{-4} & 7\times 10^{-5} & -6.8\times 10^{-5} & -11.48 & 0.09 & 7.8\times 10^{-3}\\
27 & 040302-1 & 5.0\times 10^{-4} & 6.1\times 10^{-4} & 7\times 10^{-5} & -7.9\times 10^{-5} & -11.51 & 0.07 & 6.5\times 10^{-3}\\
28 & 040302-2 & 1.4\times 10^{-2} & 5.6\times 10^{-4} & 7\times 10^{-5} & -7.6\times 10^{-5} & -11.36 & 0.23 & 2.1\times 10^{-2}\\
29 & 040303-1 & 5.0\times 10^{-4} & 7.5\times 10^{-4} & 7\times 10^{-5} & -7.4\times 10^{-5} & -11.49 & 0.10 & 9.0\times 10^{-3}\\
30 & 040304-3 & 2.1\times 10^{-2} & 3.7\times 10^{-4} & 7\times 10^{-5} & -6.4\times 10^{-5} & -11.59 & 0.27 & 2.4\times 10^{-2}\\
31 & 040312-1 & 8.5\times 10^{-2} & 2.7\times 10^{-4} & 7\times 10^{-5} & -1.2\times 10^{-4} & -11.58 & 1.02 & 8.8\times 10^{-2}\\
32 & 040312-2 & 3.2\times 10^{-2} & 2.3\times 10^{-4} & 7\times 10^{-5} & -1.2\times 10^{-4} & -11.54 & 0.39 & 3.4\times 10^{-2}\\
33 & 040312-3 & 5.0\times 10^{-4} & 1.3\times 10^{-3} & 7\times 10^{-5} & -1.2\times 10^{-4} & -11.49 & 0.02 & 2.4\times 10^{-3}\\
34 & 040625-1 & 5.0\times 10^{-4} & -8.0\times 10^{-4} & 7\times 10^{-5} & 2.9\times 10^{-4} & -11.49 & 0.06 & 5.0\times 10^{-3}\\
35 & 040713-1 & 5.0\times 10^{-4} & 5.0\times 10^{-4} & 7\times 10^{-5} & -1.7\times 10^{-4} & -11.49 & 0.04 & 3.7\times 10^{-3}\\
36 & 041129-3 & 2.0\times 10^{-2} & -8.0\times 10^{-4} & 7\times 10^{-5} & 1.3\times 10^{-4} & -11.68 & 0.25 & 2.2\times 10^{-2}\\
37 & 041130-1 & 1.2\times 10^{-2} & -7.9\times 10^{-4} & 7\times 10^{-5} & 1.5\times 10^{-4} & -11.32 & 0.17 & 1.5\times 10^{-2}\\
38 & 041207-2 & 7.6\times 10^{-4} & -8.8\times 10^{-4} & 7\times 10^{-5} & 1.4\times 10^{-4} & -11.40 & 0.08 & 7.4\times 10^{-3}
\end{tabular}
\label{tab:514lightShiftRatioAnalysis}
\end{table}

\begin{table}[p]
\centering
\caption[1111~nm reduced light shift data]{1111~nm reduced light shift data. See text for a description of the column labels.  All frequency units are Hz.}
\scriptsize
\setlength{\tabcolsep}{3 pt}
\newcolumntype{V}{>{\tt}c<{\tt}}
\begin{tabular}{cVc|ccc|ccc|ccc|ccc}
\# & File & $N$ & $\omega_D$ & $\sigma_{\omega_D}$ & $\Gamma_D$ & $\omega_D^{LS} $ & $\sigma_{\omega_D^{LS}}$ & $\Gamma_D^{LS}$ & $\omega_S$ & $\sigma_{\omega_S}$ & $\Gamma_S$ & $\omega_S^{LS} $ & $\sigma_{\omega_S^{LS}}$ & $\Gamma_S^{LS}$ \\ \hline \hline
1 &  050801-1 & 127 & 2,500,306 & 29 & 508 & 2,529,750 & 112 & 1,519 & 6,263,612 & 125 & 2,384 & 6,276,040 & 119 & 1,937\\ 
2 &  050802-1 & 172 & 2,500,311 & 15 & 226 & 2,528,853 & 80 & 712 & 6,263,431 & 35 & 416 & 6,274,990 & 95 & 1,199\\ 
3 &  050802-3 & 82 & 2,501,048 & 19 & 213 & 2,521,333 & 98 & 1,557 & 6,265,430 & 63 & 851 & 6,273,856 & 108 & 1,596\\ 
4 &  050803-2 & 111 & 2,501,226 & 11 & 130 & 2,515,493 & 107 & 1,467 & 6,265,858 & 72 & 575 & 6,272,042 & 98 & 985\\ 
5 &  050803-4 & 82 & 2,501,426 & 21 & 115 & 2,530,976 & 223 & 1,685 & 6,266,288 & 59 & 475 & 6,278,493 & 136 & 1,390\\ 
6 &  050804-1 & 47 & 2,501,859 & 32 & 219 & 2,527,649 & 193 & 926 & 6,267,370 & 56 & 461 & 6,278,002 & 159 & 986\\ 
7 &  050805-1 & 203 & 2,504,727 & 42 & 519 & 2,474,438 & 296 & 1,831 & 6,271,471 & 107 & 1,175 & 6,259,344 & 131 & 1,089\\ 
8 &  050808-2 & 373 & 2,504,800 & 32 & 553 & 2,473,006 & 399 & 3,465 & 6,271,130 & 72 & 1,109 & 6,257,784 & 143 & 1,900\\ 
9 &  050809-2 & 51 & 2,504,705 & 47 & 307 & 2,473,502 & 307 & 1,725 & 6,271,232 & 91 & 646 & 6,258,386 & 93 & 914\\ 
10 &  050809-3 & 60 & 2,504,626 & 38 & -190 & 2,483,358 & 337 & 4,768 & 6,271,184 & 80 & 540 & 6,261,774 & 118 & 1,237\\ 
11 &  050809-4 & 60 & 2,504,710 & 27 & 227 & 2,487,245 & 117 & 1,020 & 6,271,024 & 68 & 549 & 6,263,908 & 82 & 548\\ 
12 &  050809-5 & 52 & 2,504,647 & 27 & 227 & 2,492,885 & 102 & 842 & 6,270,946 & 93 & 554 & 6,266,002 & 69 & 482\\ 
13 &  050809-7 & 60 & 2,504,833 & 36 & 402 & 2,499,272 & 110 & 980 & 6,271,676 & 64 & 555 & 6,269,282 & 72 & 623\\ 
14 &  050809-8 & 60 & 2,504,974 & 37 & 264 & 2,499,166 & 141 & 1,571 & 6,272,057 & 71 & 582 & 6,269,704 & 87 & 466\\ 
15 &  050810-1 & 333 & 2,505,018 & 18 & 268 & 2,470,054 & 529 & 4,460 & 6,272,344 & 44 & 561 & 6,257,400 & 141 & 1,364\\ 
16 &  050812-1 & 59 & 2,502,723 & 91 & 473 & 2,478,033 & 510 & 2,827 & 6,268,277 & 65 & 649 & 6,258,752 & 109 & 504\\ 
17 &  050822-1 & 73 & 2,500,235 & 56 & 497 & 2,474,941 & 195 & 1,883 & 6,262,733 & 24 & 500 & 6,252,139 & 41 & 695\\ 
18 &  050823-1 & 500 & 2,500,152 & 18 & 459 & 2,473,099 & 55 & 682 & 6,262,748 & 9 & 526 & 6,251,426 & 22 & 524\\ 
19 &  050823-2 & 98 & 2,500,033 & 45 & 377 & 2,470,448 & 423 & 3,332 & 6,262,678 & 25 & 488 & 6,250,417 & 81 & 1,241\\ 
20 &  050823-3 & 159 & 2,500,077 & 36 & 416 & 2,476,724 & 334 & 3,548 & 6,262,768 & 19 & 434 & 6,253,128 & 57 & 1,273\\ 
21 &  050823-4 & 127 & 2,500,084 & 32 & 350 & 2,483,499 & 211 & 2,079 & 6,262,865 & 24 & 492 & 6,256,078 & 51 & 1,044\\ 
22 &  050823-5 & 97 & 2,500,252 & 45 & 441 & 2,491,888 & 270 & 2,533 & 6,262,926 & 29 & 419 & 6,259,348 & 53 & 725\\ 
23 &  050824-1 & 184 & 2,500,160 & 30 & 439 & 2,491,594 & 81 & 1,003 & 6,262,943 & 18 & 504 & 6,259,293 & 19 & 520\\ 
24 &  050825-2 & 117 & 2,500,152 & 17 & 196 & 2,521,563 & 304 & 3,673 & 6,262,982 & 12 & 228 & 6,272,184 & 100 & 1,485\\ 
25 &  050825-3 & 97 & 2,500,226 & 17 & 209 & 2,533,495 & 493 & 4,538 & 6,263,143 & 11 & 265 & 6,277,681 & 112 & 1,826\\ 
26 &  050825-4 & 75 & 2,500,288 & 25 & -221 & 2,532,640 & 311 & 2,243 & 6,263,185 & 13 & 308 & 6,276,860 & 66 & 1,159\\ 
27 &  050826-1 & 182 & 2,500,072 & 10 & 190 & 2,505,032 & 113 & 851 & 6,262,824 & 12 & 290 & 6,264,742 & 38 & 825\\ 
28 &  050826-2 & 112 & 2,500,198 & 23 & 213 & 2,508,258 & 113 & 1,220 & 6,262,997 & 19 & 271 & 6,266,360 & 46 & 585\\ 
29 &  050826-3 & 139 & 2,500,270 & 19 & 243 & 2,519,242 & 192 & 1,528 & 6,263,248 & 11 & 249 & 6,271,158 & 76 & 1,658\\ 
30 &  050829-2 & 106 & 2,500,095 & 21 & 245 & 2,519,667 & 400 & -2,703 & 6,262,733 & 27 & 330 & 6,270,802 & 116 & 1,033\\ 
31 &  050829-3 & 143 & 2,500,199 & 23 & 290 & 2,519,523 & 200 & 863 & 6,262,910 & 14 & 343 & 6,271,076 & 44 & 888\\ 
32 &  050829-4 & 107 & 2,500,232 & 20 & 235 & 2,529,976 & 452 & 3,938 & 6,262,964 & 14 & 311 & 6,275,331 & 88 & 1,581\\ 
33 &  050829-5 & 152 & 2,500,203 & 17 & 241 & 2,528,674 & 853 & 2,894 & 6,263,076 & 12 & 295 & 6,273,997 & 254 & 3,986\\ 
34 &  050830-1 & 500 & 2,500,250 & 9 & 229 & 2,527,709 & 252 & 4,844 & 6,263,110 & 7 & 262 & 6,274,860 & 81 & 2,208\\ 
35 &  050830-2 & 474 & 2,500,283 & 9 & 230 & 2,524,348 & 208 & 2,864 & 6,263,250 & 7 & 252 & 6,273,186 & 48 & 1,446\\ 
36 &  050831-1 & 159 & 2,500,351 & 17 & 253 & 2,530,219 & 374 & 1,299 & 6,263,344 & 11 & 268 & 6,275,496 & 157 & 2,335\\ 
37 &  050907-1 & 994 & 2,500,278 & 8 & 258 & 2,531,536 & 64 & 1,114 & 6,263,094 & 4 & 351 & 6,276,148 & 15 & 675\\ 
38 &  050907-2 & 90 & 2,500,288 & 18 & 237 & 2,538,298 & 107 & 854 & 6,263,436 & 11 & 220 & 6,279,398 & 25 & 441\\ 
39 &  050908-1 & 84 & 2,500,267 & 22 & 211 & 2,538,504 & 190 & 1,129 & 6,263,438 & 12 & 216 & 6,279,389 & 63 & 684\\ 
40 &  050908-2 & 282 & 2,500,221 & 8 & 190 & 2,537,213 & 257 & 1,152 & 6,263,166 & 10 & 228 & 6,278,515 & 74 & 837
\end{tabular}
\label{tab:1111lightShiftRawData}
\end{table}

\begin{table}[p]
\centering
\caption[1111~nm light shift analysis]{1111~nm light shift analysis. See text for a description of the column labels.  All frequency units are Hz.}
\scriptsize
\setlength{\tabcolsep}{3 pt}
\newcolumntype{M}{>{$}c<{$}}
\begin{tabular}{cc|cccc|cMc}
\#   & $\sigma_\text{LS}, \sigma_\text{red}$ & $\Delta_D$ & $\sigma_{\Delta_D}$ & $\Delta_S$ & $\sigma_{\Delta_S}$ & $\Delta_D / \Omega_D$ & G_\text{meas} - G & $\Omega_\text{trap rf}$ (kHz) \\ \hline \hline
1 &  (+,+) & 29443 & 115 & 12,428 & 172 & 19.4 & -8.3 \times 10^{-5} & 41\\ 
2 &  (+,+) & 28542 & 82 & 11,559 & 101 & 40.1 & -1.6 \times 10^{-4} & 57\\  
3 & (+,+) & 20285 & 100 & 8,427 & 125 & 13.0 & -9.8 \times 10^{-5} & 45\\  
4 &  (+,+) & 14267 & 107 & 6,184 & 122 & 9.7 & -1.0 \times 10^{-4} & 46\\ 
5 & (+,+) & 29550 & 224 & 12,205 & 148 & 17.5 & -1.3 \times 10^{-4} & 52\\  
6 &  (+,+) & 25790 & 195 & 10,632 & 169 & 27.8 & -1.3 \times 10^{-4} & 52\\  
7 &  (-,+) & -30289 & 299 & -12,127 & 169 & 16.5 & -1.4 \times 10^{-3} & 170\\ 
8 &  (-,+) & -31794 & 400 & -13,346 & 160 & 9.2 & -1.6 \times 10^{-3} & 180\\  
9 & (-,+) & -31203 & 311 & -12,845 & 130 & 18.1 & -1.4 \times 10^{-3} & 170\\  
10 & (-,+) & -21268 & 339 & -9,410 & 143 & 4.5 & -1.4 \times 10^{-3} & 170\\  
11 &  (-,+) & -17465 & 120 & -7,116 & 107 & 17.1 & -1.5 \times 10^{-3} & 180\\  
12 &  (-,+) & -11762 & 106 & -4,944 & 116 & 14.0 & -1.5 \times 10^{-3} & 170\\  
13 &  (-,+) & -5561 & 116 & -2,393 & 96 & 5.7 & -1.4 \times 10^{-3} & 170\\ 
14 &  (-,+) & -5807 & 145 & -2,353 & 112 & 3.7 & -1.4 \times 10^{-3} & 170\\  
15 &  (-,+) & -34964 & 530 & -14,945 & 147 & 7.8 & -1.3 \times 10^{-3} & 160\\ 
16 &  (-,-) & -24690 & 518 & -9,525 & 126 & 8.7 & -6.4 \times 10^{-4} & 110\\  
17 &  (-,-) & -25294 & 203 & -10,594 & 47 & 13.4 & -3.6 \times 10^{-4} & 86\\  
18 &  (-,-) & -27054 & 58 & -11,321 & 24 & 39.7 & -2.7 \times 10^{-4} & 75\\  
19 &  (-,-) & -29585 & 425 & -12,260 & 85 & 8.9 & -1.8 \times 10^{-4} & 61\\  
20 &  (-,-) & -23353 & 336 & -9,641 & 60 & 6.6 & -1.9 \times 10^{-4} & 62\\  
21 &  (-,-) & -16584 & 214 & -6,787 & 56 & 8.0 & -1.6 \times 10^{-4} & 57\\  
22 &  (-,-) & -8364 & 273 & -3,577 & 60 & 3.3 & -3.0 \times 10^{-4} & 78\\  
23 &  (-,-) & -8566 & 86 & -3,650 & 26 & 8.5 & -2.0 \times 10^{-4} & 64\\  
24 &  (+,+) & 21411 & 305 & 9,201 & 101 & 5.8 & -1.8 \times 10^{-4} & 60\\  
25 &  (+,+) & 33269 & 493 & 14,538 & 113 & 7.3 & -1.9 \times 10^{-4} & 62\\  
26 &  (+,+) & 32352 & 312 & 13,676 & 67 & 14.4 & -2.3 \times 10^{-4} & 69\\  
27 &  (+,+) & 4960 & 114 & 1,919 & 40 & 5.8 & -1.6 \times 10^{-4} & 58\\  
28 &  (+,+) & 8061 & 115 & 3,364 & 50 & 6.6 & -2.2 \times 10^{-4} & 67\\  
29 &  (+,+) & 18971 & 193 & 7,909 & 77 & 12.4 & -1.9 \times 10^{-4} & 62\\  
30 &  (+,+) & 19572 & 401 & 8,068 & 119 & 7.2 & -2.2 \times 10^{-4} & 67\\  
31 &  (+,+) & 19324 & 201 & 8,166 & 46 & 22.4 & -2.6 \times 10^{-4} & 72\\  
32 &  (+,+) & 29743 & 452 & 12,367 & 89 & 7.6 & -2.7 \times 10^{-4} & 74\\  
33 &  (+,+) & 28471 & 853 & 10,921 & 254 & 9.8 & -1.9 \times 10^{-4} & 63\\  
34 &  (+,+) & 27459 & 252 & 11,750 & 81 & 5.7 & -2.3 \times 10^{-4} & 68\\  
35 &  (+,+) & 24065 & 209 & 9,936 & 48 & 8.4 & -2.0 \times 10^{-4} & 64\\  
36 &  (+,+) & 29868 & 374 & 12,153 & 157 & 23.0 & -2.3 \times 10^{-4} & 69\\ 
37 &  (+,+) & 31259 & 65 & 13,054 & 15 & 28.1 & -2.6 \times 10^{-4} & 73\\  
38 &  (+,-) & 38010 & 109 & 15,962 & 27 & 44.5 & -1.3 \times 10^{-4} & 52\\ 
39 &  (+,-) & 38237 & 191 & 15,951 & 64 & 33.9 & -1.1 \times 10^{-4} & 48\\  
40 &  (+,-) & 36992 & 257 & 15,349 & 75 & 32.1 & -1.8 \times 10^{-4} & 60
\end{tabular}
\label{tab:1111lightShiftAnalysis}
\end{table}

\begin{table}[p]
\centering
\caption[1111~nm light shift ratio error analysis]{1111~nm light shift ratio error analysis. See text for a description of the column labels.  All frequency units are Hz.}
\scriptsize
\newcolumntype{M}{>{$}c<{$}}
\newcolumntype{V}{>{\tt}c<{\tt}}
\setlength{\tabcolsep}{3 pt}
\begin{tabular}{cV|MMMM|ccM}
\#   & File &  \sigma_\text{line} / R & \sigma_{\theta, \sigma}/R & \sigma_\lambda/R & \sigma_\text{rf}/R & $R$ & $\sigma_R$ & \sigma_\text{tot}/R \\ \hline \hline
1 &  050801-1 & 1.0 \times 10^{-4} & 6.0 \times 10^{-5} & 1 \times 10^{-6} & -4.6 \times 10^{-5} & 0.422 & 0.006 & 1.4 \times 10^{-2}\\ 
2 &  050802-1 & 1.0 \times 10^{-4} & 5.6 \times 10^{-5} & 1 \times 10^{-6} & -8.8 \times 10^{-5} & 0.405 & 0.004 & 9.3 \times 10^{-3}\\ 
3 &  050802-3 & 1.0 \times 10^{-4} & 4.1 \times 10^{-5} & 1 \times 10^{-6} & -5.5 \times 10^{-5} & 0.415 & 0.006 & 1.6 \times 10^{-2}\\ 
4 &  050803-2 & 2.3 \times 10^{-4} & 3.0 \times 10^{-5} & 1 \times 10^{-6} & -5.8 \times 10^{-5} & 0.433 & 0.009 & 2.1 \times 10^{-2}\\  
5 &  050803-4 & 1.0 \times 10^{-4} & 5.9 \times 10^{-5} & 1 \times 10^{-6} & -7.4 \times 10^{-5} & 0.413 & 0.006 & 1.4 \times 10^{-2}\\ 
6 &  050804-1 & 1.0 \times 10^{-4} & 5.1 \times 10^{-5} & 1 \times 10^{-6} & -7.5 \times 10^{-5} & 0.412 & 0.007 & 1.8 \times 10^{-2}\\ 
7 &  050805-1 & 1.0 \times 10^{-4} & -5.8 \times 10^{-5} & 1 \times 10^{-6} & 7.6 \times 10^{-4} & 0.400 & 0.007 & 1.7 \times 10^{-2}\\  
8 &  050808-2 & 2.8 \times 10^{-4} & -6.4 \times 10^{-5} & 1 \times 10^{-6} & 8.7 \times 10^{-4} & 0.420 & 0.007 & 1.7 \times 10^{-2}\\  
9 &  050809-2 & 1.0 \times 10^{-4} & -6.2 \times 10^{-5} & 1 \times 10^{-6} & 8.0 \times 10^{-4} & 0.412 & 0.006 & 1.4 \times 10^{-2}\\  
10 &  050809-3 & 3.1 \times 10^{-2} & -4.5 \times 10^{-5} & 1 \times 10^{-6} & 7.7 \times 10^{-4} & 0.442 & 0.017 & 3.8 \times 10^{-2}\\  
11 &  050809-4 & 1.0 \times 10^{-4} & -3.4 \times 10^{-5} & 1 \times 10^{-6} & 8.5 \times 10^{-4} & 0.407 & 0.007 & 1.7 \times 10^{-2}\\  
12 &  050809-5 & 1.0 \times 10^{-4} & -2.4 \times 10^{-5} & 1 \times 10^{-6} & 8.3 \times 10^{-4} & 0.420 & 0.011 & 2.5 \times 10^{-2}\\  
13 &  050809-7 & 7.5 \times 10^{-3} & -1.2 \times 10^{-5} & 1 \times 10^{-6} & 7.7 \times 10^{-4} & 0.430 & 0.020 & 4.6 \times 10^{-2}\\  
14 &  050809-8 & 6.4 \times 10^{-2} & -1.1 \times 10^{-5} & 1 \times 10^{-6} & 7.6 \times 10^{-4} & 0.405 & 0.034 & 8.4 \times 10^{-2}\\  
15 &  050810-1 & 8.4 \times 10^{-4} & -7.2 \times 10^{-5} & 1 \times 10^{-6} & 7.3 \times 10^{-4} & 0.427 & 0.008 & 1.8 \times 10^{-2}\\  
16 &  050812-1 & 4.1 \times 10^{-4} & -4.6 \times 10^{-5} & 1 \times 10^{-6} & 3.5 \times 10^{-4} & 0.386 & 0.010 & 2.5 \times 10^{-2}\\  
17 &  050822-1 & 1.0 \times 10^{-4} & -5.1 \times 10^{-5} & 1 \times 10^{-6} & 2.0 \times 10^{-4} & 0.419 & 0.004 & 9.2 \times 10^{-3}\\  
18 &  050823-1 & 1.0 \times 10^{-4} & -5.5 \times 10^{-5} & 1 \times 10^{-6} & 1.5 \times 10^{-4} & 0.418 & 0.001 & 3.2 \times 10^{-3}\\  
19 &  050823-2 & 3.5 \times 10^{-4} & -5.9 \times 10^{-5} & 1 \times 10^{-6} & 1.0 \times 10^{-4} & 0.414 & 0.007 & 1.6 \times 10^{-2}\\  
20 &  050823-3 & 2.7 \times 10^{-3} & -4.6 \times 10^{-5} & 1 \times 10^{-6} & 1.1 \times 10^{-4} & 0.413 & 0.007 & 1.6 \times 10^{-2}\\  
21 &  050823-4 & 7.2 \times 10^{-4} & -3.3 \times 10^{-5} & 1 \times 10^{-6} & 8.8 \times 10^{-5} & 0.409 & 0.006 & 1.5 \times 10^{-2}\\  
22 &  050823-5 & 9.0 \times 10^{-2} & -1.1 \times 10^{-6} & 1 \times 10^{-6} & 1.7 \times 10^{-4} & 0.428 & 0.041 & 9.7 \times 10^{-2}\\  
23 &  050824-1 & 4.8 \times 10^{-4} & -1.8 \times 10^{-5} & 1 \times 10^{-6} & 1.1 \times 10^{-4} & 0.426 & 0.005 & 1.2 \times 10^{-2}\\  
24 &  050825-2 & 5.9 \times 10^{-3} & 4.4 \times 10^{-5} & 1 \times 10^{-6} & -1.0 \times 10^{-4} & 0.430 & 0.008 & 1.9 \times 10^{-2}\\  
25 &  050825-3 & 1.3 \times 10^{-3} & 7.0 \times 10^{-5} & 1 \times 10^{-6} & -1.1 \times 10^{-4} & 0.437 & 0.007 & 1.7 \times 10^{-2}\\ 
26 &  050825-4 & 1.0 \times 10^{-4} & 6.6 \times 10^{-5} & 1 \times 10^{-6} & -1.3 \times 10^{-4} & 0.423 & 0.005 & 1.1 \times 10^{-2}\\  
27 &  050826-1 & 5.9 \times 10^{-3} & 9.2 \times 10^{-6} & 1 \times 10^{-6} & -9.0 \times 10^{-5} & 0.387 & 0.012 & 3.1 \times 10^{-2}\\  
28 &  050826-2 & 2.6 \times 10^{-3} & 1.6 \times 10^{-5} & 1 \times 10^{-6} & -1.2 \times 10^{-4} & 0.417 & 0.009 & 2.1 \times 10^{-2}\\  
29 &  050826-3 & 1.0 \times 10^{-4} & 3.8 \times 10^{-5} & 1 \times 10^{-6} & -1.1 \times 10^{-4} & 0.417 & 0.006 & 1.4 \times 10^{-2}\\  
30 &  050829-2 & 1.4 \times 10^{-3} & 3.9 \times 10^{-5} & 1 \times 10^{-6} & -1.2 \times 10^{-4} & 0.412 & 0.010 & 2.5 \times 10^{-2}\\  
31 &  050829-3 & 1.0 \times 10^{-4} & 3.9 \times 10^{-5} & 1 \times 10^{-6} & -1.4 \times 10^{-4} & 0.423 & 0.005 & 1.2 \times 10^{-2}\\  
32 &  050829-4 & 1.1 \times 10^{-3} & 6.0 \times 10^{-5} & 1 \times 10^{-6} & -1.5 \times 10^{-4} & 0.416 & 0.007 & 1.7 \times 10^{-2}\\  
33 &  050829-5 & 2.2 \times 10^{-4} & 5.3 \times 10^{-5} & 1 \times 10^{-6} & -1.1 \times 10^{-4} & 0.384 & 0.015 & 3.8 \times 10^{-2}\\  
34 &  050830-1 & 7.6 \times 10^{-3} & 5.1 \times 10^{-6} & 1 \times 10^{-6} & -1.3 \times 10^{-4} & 0.428 & 0.006 & 1.4 \times 10^{-2}\\ 
35 &  050830-2 & 5.4 \times 10^{-4} & 4.8 \times 10^{-5} & 1 \times 10^{-6} & -1.1 \times 10^{-4} & 0.413 & 0.004 & 1.0 \times 10^{-2}\\  
36 &  050831-1 & 1.0 \times 10^{-4} & 5.9 \times 10^{-5} & 1 \times 10^{-6} & -1.3 \times 10^{-4} & 0.407 & 0.007 & 1.8 \times 10^{-2}\\  
37 &  050907-1 & 1.0 \times 10^{-4} & 6.3 \times 10^{-5} & 1 \times 10^{-6} & -1.4 \times 10^{-4} & 0.418 & 0.001 & 2.7 \times 10^{-3}\\  
38 &  050907-2 & 1.0 \times 10^{-4} & 7.1 \times 10^{-6} & 1 \times 10^{-6} & -7.5 \times 10^{-5} & 0.420 & 0.001 & 3.6 \times 10^{-3}\\ 
39 &  050908-1 & 1.0 \times 10^{-4} & 7.1 \times 10^{-6} & 1 \times 10^{-6} & -6.2 \times 10^{-5} & 0.417 & 0.003 & 6.5 \times 10^{-3}\\ 
40 &  050908-2 & 1.0 \times 10^{-4} & 7.4 \times 10^{-5} & 1 \times 10^{-6} & -9.1 \times 10^{-6} & 0.415 & 0.004 & 8.6 \times 10^{-3} 
\end{tabular}
\label{tab:1111lightShiftRatioAnalysis}
\end{table}

\chapter{Experiments with the odd-isotope $^{137}$B\lowercase{a}$^+$}\label{sec:hyperfineChapter}
\begin{quotation}
\noindent\small You will reach high levels of intelligence. \\ \flushright{---Fortune cookie, received by the author \emph{after} writing most of this manuscript.}
\end{quotation}
\section{New levels, hyperfine splittings, isotope shifts}
\begin{figure}
\centering
\includegraphics{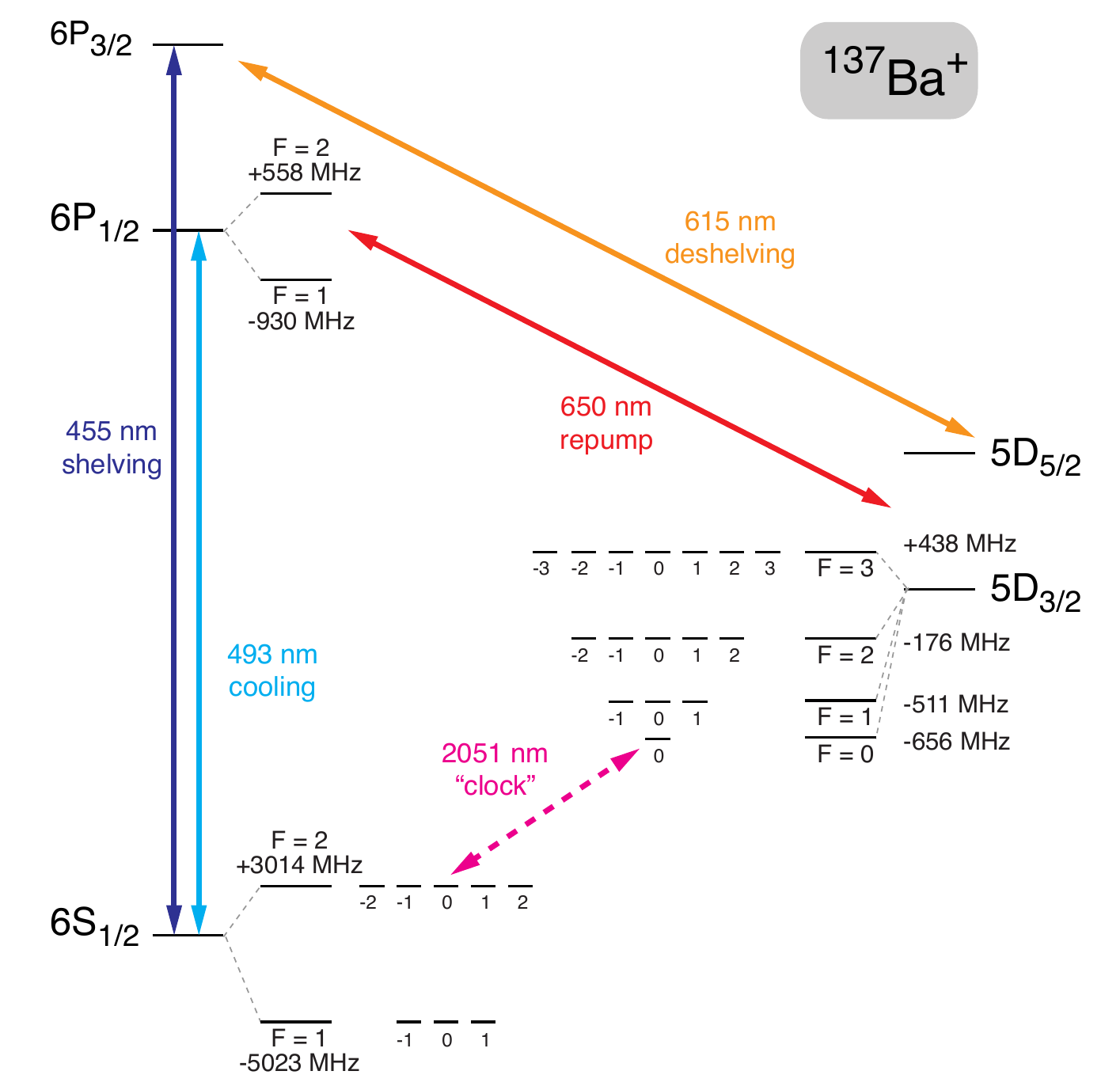}
\caption[$^{137}$Ba$^+$ energy level diagram]{This $^{137}$Ba$^+$ energy level diagram shows the hyperfine structure of the $6S_{1/2}$, $6P_{1/2}$ and $5D_{3/2}$ states.  The 2051~nm ``clock'' transition $6S_{1/2}, F = 2, m_F = 0 \leftrightarrow 5D_{3/2}, F=0, m_F = 0$ is also highlighted.  While the hyperfine splitting the ground state is extremely well known, the excited state hyperfine structure has only been experimentally measured to perhaps 0.1\%.  Precision knowledge of the $5D_{3/2}$ splittings at to a level of $10^{-8}$ are possible using the rf spin-flip spectroscopy detailed in Chapter~\ref{sec:lightShiftChapter}.}
\label{fig:bariumEnergyLevelsHyperfine}
\end{figure}
\subsection{The hyperfine dipole Hamiltonian}
$^{137}$Ba$^+$ has something its even isotope cousin $^{138}$Ba$^+$ does not:  nuclear spin.  The odd-isotope nucleus has spin $I = 3/2$ and thus a magnetic moment.  The electronic wavefunctions are modified, in part, by the interaction of this magnetic moment with the magnetic field generated by the electrons~\cite{woodgate1980eas}
\begin{equation}
H_\text{hf dipole} = - \boldsymbol{\mu}_I \cdot \boldsymbol{B}_\text{el}.
\end{equation}
In the approximation that $J$ and $I$ are both good quantum numbers, which is to say the hyperfine interaction energy scale is much smaller than the fine-structure scale, states will have total angular momentum
\begin{equation}
\boldsymbol{F} = \boldsymbol{I} + \boldsymbol{J}.
\end{equation}
Each state $|F (IJ) \rangle$ is $(2F + 1)$ degenerate (each sublevel is assigned a magnetic quantum number $m_F$) and for a given fine-structure level $J$, $F$ can range over
\begin{align}
F &\in \left\{ |J - I|, |J - I| + 1, \ldots, J + I \right \},
\intertext{while in a given $F$ level, the magnetic quantum number}
m_F &\in \left\{-F, -F+1, \ldots, F \right\}.
\end{align}
These levels are shown for many relevant states in Figure~\ref{fig:bariumEnergyLevelsHyperfine}.  Matrix elements of $\boldsymbol{F}$ follow rules similar to $\boldsymbol{J}$ and $\boldsymbol{I}$:
\begin{align}
\langle (IJ) F m_F | \boldsymbol{F}^2 | (IJ) F m_F \rangle &= \hbar^2 F(F+1) \\
\langle (IJ) F m_F | F_z | (IJ) F m_F \rangle &= \hbar m_F.
\end{align}

\begin{table}
\centering
\caption[Hyperfine magnetic dipole and electric quadrupole coefficients in $^{137}$Ba$^+$.]{Hyperfine magnetic dipole and electric quadrupole coefficients in $^{137}$Ba$^+$.  Measurements are followed by reported errors in parenthesis.  Note that two theoretical predictions of $A(5D_{3/2})$ differ in sign and that the electric quadrupole interaction $B(J)$ is larger than the dipole term.  Notice also the exceptional precision of the ground state hyperfine splitting measurement compared to several measurements in the excited states.  Indeed, ground state hyperfine splittings serve us as microwave frequency standards~\cite{fisk1997tia}.}
\begin{tabular}{ccll}
State 		& Reference 		& $A(J)$ (MHz) 		& $B(J)$ (MHz) \\ \hline \hline
\multirow{2}{*}{$6S_{1/2}$}
			&\cite{trapp2000hsa}& 4018.87083385(18) 	& --- \\
			&\cite{wendt1984rjd}& 4020.3(2.3) 			& --- \\ \hline
\multirow{2}{*}{$6P_{1/2}$}	
			&\cite{wendt1984rjd}& 744.1(1.6) 			& --- \\
			&\cite{villemoes1993ish}& 743.7(0.3)		& --- \\  \hline
\multirow{2}{*}{$6P_{3/2}$}	
			&\cite{wendt1984rjd}& 126.7(1.1)			& 95.0(3.7) \\
			& \cite{villemoes1993ish}	& 127.2(0.2)		& 92.5(0.2) \\ \hline
\multirow{3}{*}{$5D_{3/2}$}	
			& \cite{hove1985}	& 189.8277(6)			& 44.5417(16) \\
			& \cite{sahoo2006rcc} & 190.89			& 46.82 \\
			& \cite{itano2006qmh} & 192.99			& 51.21 \\ \hline
\multirow{3}{*}{$5D_{5/2}$}	
			& \cite{silverans1986hss}  & -12.028(11)		& 59.533(43)  \\
			& \cite{sahoo2006rcc} & -11.99			& 62.27  \\
			& \cite{itano2006qmh} & 9.39 				& 68.16  \\
\end{tabular}
\label{tab:hfAvalues}
%\begin{table}
\vspace{0.5in}
\caption[Isotope shift data for $^{137}$Ba$^+$]{Isotope shift data for $^{137}$Ba$^+$.  These experimental data are quoted for transitions rather than levels since relative spectral shifts between isotopes is directly observable.  In general, states of low orbital angular momentum account for the bulk of the shift due to the increased wavefunction penetration into the nucleus.}
\begin{tabular}{llrc}
Transition		& $\lambda$~[nm]	& $\delta \nu^{137\text{--}138}$~[MHz]	& Reference \\ \hline \hline
$6S_{1/2} \leftrightarrow 6P_{1/2}$	& 493.4	& 271.1(1.7) & \cite{wendt1984rjd} \\
$6S_{1/2} \leftrightarrow 6P_{3/2}$	& 455.4	& 279.0(2.6) & \cite{wendt1984rjd} \\ \hline
$5D_{3/2} \leftrightarrow 6P_{1/2}$	& 649.7	& 13.0(0.4) & \cite{villemoes1993ish} \\
$5D_{3/2} \leftrightarrow 6P_{3/2}$	& 614.2	& 2.3(0.4)	& \cite{villemoes1993ish} \\ \hline
$5D_{5/2} \leftrightarrow 6P_{3/2}$	& 585.4	& 5.3(0.5) & \cite{villemoes1993ish}
\end{tabular}
\label{tab:isotopeShifts}
%\end{table}
\end{table}

The magnetic dipole hyperfine Hamiltonian can be written~\cite{woodgate1980eas}
\begin{equation}
H_\text{hf dipole} = A(J) \boldsymbol{I} \cdot \boldsymbol{J}
\end{equation}
which means that first order energy shifts are
\begin{align}
\Delta E^\text{hf dipole}_{|(IJ)F \rangle} &= \langle (IJ) F | A(J)  \boldsymbol{I} \cdot \boldsymbol{J} | (IJ) F \rangle \\
&= \frac{A(J)}{2}\left\langle (IJ) F \left| (\boldsymbol{F}^2 - \boldsymbol{I}^2 - \boldsymbol{J}^2) \right| (IJ) F \right\rangle \\
&=  A(J) \frac{F(F+1) - J(J+1) - I(I+1)}{2}.
\end{align}
This leads to the well known interval rule describing the splittings of adjacent hyperfine states:
\begin{equation}
\Delta E^\text{hf dipole}(F) - \Delta E^\text{hf dipole}(F-1) = A(J) F.
\end{equation}
This rule is not exact; we expect and observe deviations due to higher order multipole terms of the hyperfine interaction discussed later.  In general these deviations are small for $s$-wavefunctions but can be comparable or even dominate the dipole term for higher angular momentum states.  Indeed, Table~\ref{tab:hfAvalues} shows disagreement in the literature over the \emph{sign} of the dipole hyperfine term for $5D_{5/2}$ which is known to be smaller than the electric quadrupole piece.

\subsection{Isotope shifts}
The subtraction of a neutron from $^{138}$Ba$^+$ has two small effects of comparable size on atomic wavefunctions and therefore on transition frequencies.  The first is called the \emph{mass shift} and the second is called the \emph{volume shift}.  It is spectroscopically convenient to specify the isotope shifts in transition frequencies rather than individual energy levels:  Table~\ref{tab:isotopeShifts} gives the barium ion isotope shift data present in the literature.  Of course, in an odd isotope no real transition lies just an isotope shift away from a transition in an even isotope due to the hyperfine splitting.  Instead, the isotope shift measures the deviation of the \emph{centroid} or \emph{center of gravity} of the hyperfine structure.  That is, where the transition would be if hyperfine interaction terms like $A(J) \to 0$.

Often transition frequencies are calculated for simple atoms by assuming an infinitely heavy nucleus compared to the mass of the electron, a feature of the Bohr atomic model.  Following~\cite{foot2005ap}, if the atomic mass is $M_A$ (the atomic mass $A$ should not to be confused with the dipole hyperfine coupling constant $A(J)$), then the application of a `reduced-mass' term can correct this idealization $\nu_\infty$:
\begin{equation}
\nu_A = \nu_\infty \left( \frac{M_A}{m_e + M_A} \right).
\end{equation}
When the mass is changed to $M_{A'}$ by removing (or adding) neutrons, we can account for observed transition frequency changes due to the alteration of nuclear mass by
\begin{align}
\Delta \nu^\text{mass shift} & = \nu_{A'} - \nu_A \\
&\simeq \frac{m_e}{M_p} \frac{(A' - A)}{A' A} \nu_\infty
\end{align}
For our case, this \emph{normal mass shift} has a relative size $\sim 3 \times (10)^{-8}$ which implies $\sim 10$~MHz shifts in visible transitions.  Another part of the mass shift, the \emph{specific mass shift} is due to the modifications to all other non-valence electrons, the calculation of which is beyond the scope of this work~\cite{berengut2003isc}.

The second component of the isotope shift accounts for the decreased nuclear volume with which atomic wavefunctions can interact.  This piece of the isotope shift is large for $s$-wavefunctions compared to ones of higher angular momentum because $s$-wavefunctions tend to have significant amplitude $|\psi(0)|$ near the nucleus.  An estimate of the shift is given as~\cite{foot2005ap}
\begin{align}
\Delta \nu^\text{volume shift} & = \nu_{A'} - \nu_A \\
&\simeq \frac{ \langle r_N^2 \rangle}{a_0^2} \frac{(A' - A)}{A} \frac{Z^2}{(n^*)^3} R_\infty ,
\end{align}
where $r_N \sim 1.2 \times A^{1/3}$~fm is the nuclear size and $n^*$ is computed using the quantum defect method.

\section{Cooling and trapping $^{137}$Ba$^+$}
The laser cooling of $^{137}$Ba$^+$ is complicated due to the presence of hyperfine structure.  Laser frequency lock points that work for trapping the even isotope must be modified due to the isotope shift.  Also, as seen in Figure~\ref{fig:barium137Transitions} two 493~nm laser frequencies must be applied to empty and cycle the $F = 1$ and $F = 2$ ground states.  Further, at least four 650~nm laser frequencies are required to empty the $F =$ 0, 1, 2 and 3  metastable $5D_{3/2}$ states.  Instead of shopping for and frequency locking a large quantity of blue and red laser systems, we choose to generate spectral sidebands on single blue and red lasers that are resonant with the additional transitions.

What follows is a section of the general principles of modulation, and applications of laser modulation in the cooling, repumping, and optical pumping of $^{137}$Ba$^+$ in Section~\ref{sec:modulationApparatus}.

\subsection{Phase modulation}\label{sec:phaseModulation}
\begin{figure}
\centering
\includegraphics[width=5.5in]{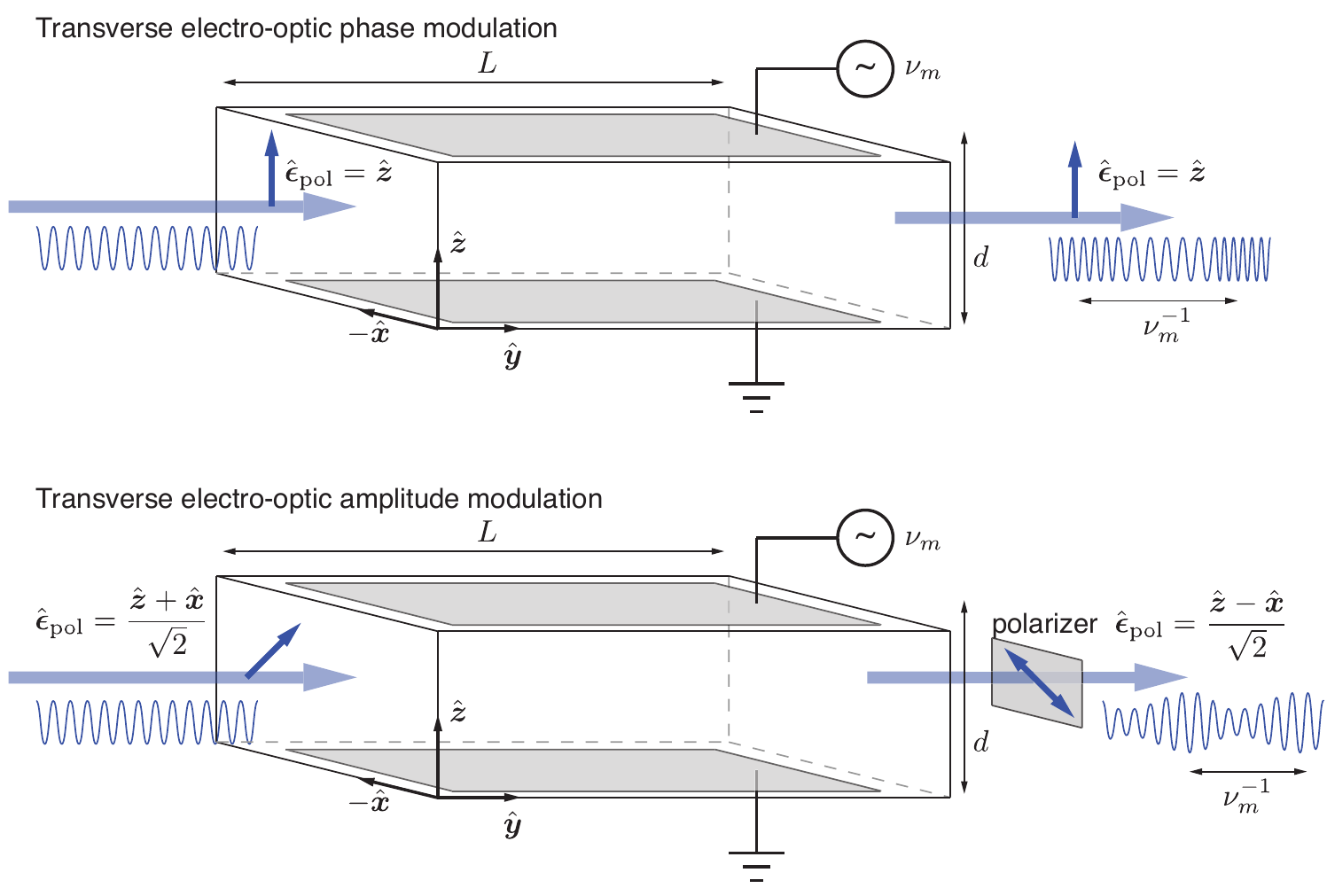}
\caption[Diagram of transverse electro-optic phase and amplitude modulators]{Diagram of transverse phase and amplitude electro optic modulators, following \cite{yariv2006poe}.  The the phase modulator configuration, the polarization of the incident light beam is not rotated, }
\label{fig:transverseEOM}
%\begin{figure}
%\centering
\includegraphics{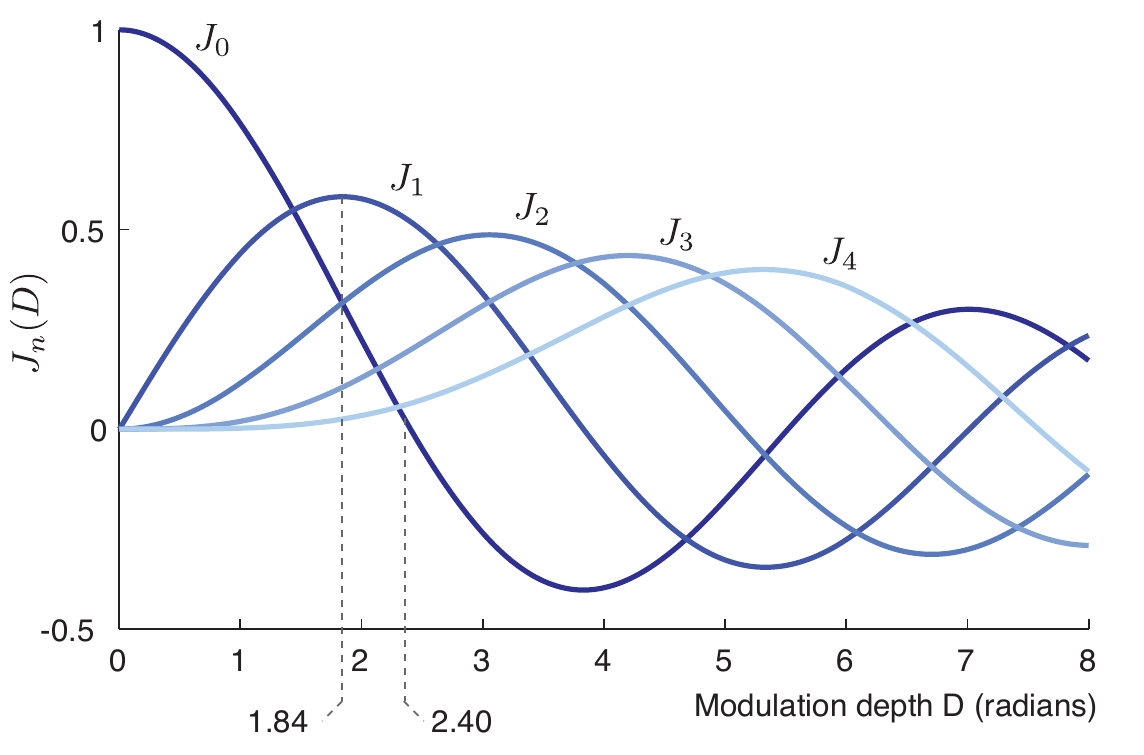}
\caption[Bessel functions describe phase modulation]{The asymptotic behavior of the Bessel functions describes phase modulation.  Here we plot the first few bessel functions $J_n(D)$ for small values of the modulation depth $D$.  Notice the special values of $D=1.84$ that maximizes the first-order sideband and $D=2.40$ that suppresses the carrier ($n=0$).  Note also that for $D \ll 1$, the asymptotic behavior of the Bessel functions is $J_n(D) \sim x^n$}
\label{fig:besselFunctions}
%\end{figure}
\end{figure}

A key application of electro-optic materials is phase modulation of optical waves at radio frequencies.  Briefly, a transmissive material exhibits a large linear electro-optic or \emph{Pockels} effect when the application of an electric field significantly changes the indices of refraction of that material.  Consummate reviews are found in~\cite{yariv2006poe, kaminow1966elm}.  In the transverse configuration, the polarization of an incident laser beam is aligned along the $c$-axis of the crystal (direction with highest index of refraction), as shown in Figure~\ref{fig:transverseEOM}.  If the nonlinear electro-optic coefficient $r_{33}$ is non-zero, then the index of refraction experienced by the light beam is a function of an applied electric field $E$~\cite{yariv2006poe},
\begin{equation}
n_z = n_e - \frac{1}{2} n_e^3 r_{33} E,
\end{equation}
where $n_e$ is the extraordinary index of refraction of the material.  Given an applied voltage $V$ to a crystal of length $l$ and thickness $d$, the phase shift of the optical wave upon exiting the crystal is
\begin{equation} \label{eq:transverseEOMPhaseShift}
\phi = \frac{2 \pi}{\lambda} \left(n_e - \frac{1}{2} n_e^3 r_{33} \frac{V}{d} \right) l.
\end{equation}
$V_\pi$ is a commonly used metric signifying the voltage necessary to achieve $\pi$-radians of phase modulation:
\begin{equation}
V_\pi \equiv \frac{\lambda}{n_e^3 r_{33}} \frac{d}{l}.
\end{equation}
For the common electro-optic material LiNbO$_{3}$, the coefficient $r_{33} \approx 30$~pm/V making $V_\pi \sim 150$~V for a 1~cm long, 1~mm thick crystal.  While this magnitude of voltage is not trivial to produce at radio frequencies, resonant circuits can both step up the applied voltage and transform the capacitive load of an EOM to match rf amplifiers (see Figure~\ref{fig:trapElectricalDrive}).  A chief disadvantage of transverse phase modulators is that Eq.~\ref{eq:transverseEOMPhaseShift} contains a term that does not vary directly with $V$ but is temperature dependent.  Modulators of this type often must be thermally insulated or compensated by orthogonally mounting two matched modulators in series, each tipped $\pm$45$^\circ$ with respect to the incident light~\cite{enscoe2005eom}.

A small constant phase shift on an optical signal isn't so breathtaking, but phase modulators become very interesting tools when the applied voltage is made time varying.  Suppose that
\begin{equation}
V(t) = V_m \sin \omega_m t
\end{equation}
where $\omega_m$ is called the \emph{modulation frequency}.  Then, an optical field of frequency $\omega_0$ (called the \emph{carrier frequency}) gains a time varying phase:
\begin{equation}
E(t) = E_0 e^{i (\omega_0 t - D \sin \omega_m t)}
\end{equation}
where $D = \pi (V_m / V_\pi)$ is called the \emph{modulation depth}. The constant phase offset has been dropped.  A Bessel function identity
\begin{equation}
e^{-i D \sin \omega_m t} = \sum_{n = -\infty}^{\infty} J_n(D) e^{-i n \omega_m t}
\end{equation}
tells us that we can write the resulting frequency spectrum of the optical wave as a sum of pure spectral components.
\begin{equation} \label{eq:besselFunctionExpansion}
E(t) = E_0 \sum_{n = -\infty}^{\infty} J_n(D) e^{i(\omega_0 - n \omega_m)t}.
\end{equation}
There appear to be infinitely many pure spectral components, but at least their magnitudes are weighted by Bessel functions which have simple asymptotic behavior for either $D \to 0$ or $D \to \infty$.  In particular, for small modulation depths $D \ll 1$, the Bessel functions $J_n(D) \sim x^n$ which mean that to first order we can consider just the terms $n = 0, \pm 1$.  The $n = 0$ term is called the \emph{carrier} while the $n = \pm 1$ terms are first order sidebands:  they have opposite phases due to the Bessel function identity $J_{-n}(D) = (-1)^n J_n(D)$.  From Figure~\ref{fig:besselFunctions} we see that the first order sideband spectral power is maximized when $D \approx 1.84$ radians, the carrier is suppressed when $D \approx 2.40$ radians.

\begin{figure}
\centering
\includegraphics{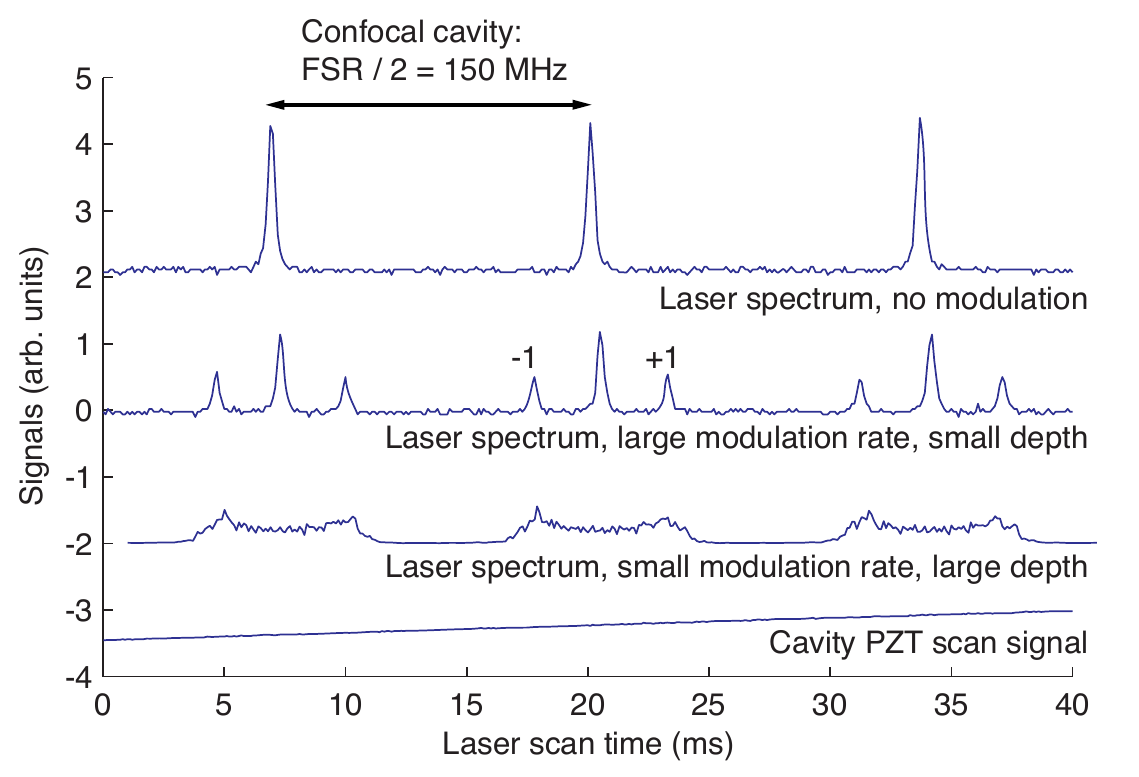}
\caption[Two limits of modulation:  sidebands, and a classical oscillator]{These digitally recorded oscilloscope traces are spectra of a red laser incident on a scanning confocal optical resonator ($\nu_\text{fsr} \approx 300$~MHz) with and without phase modulation.  In the first case of modulation we apply fast, small depth phase modulation at roughly 25~MHz to a fiber-coupled traveling wave EOM.  Two sidebands, labeled $\pm 1$, appear while the carrier is reduced in intensity.  In the second case we apply a slow, large depth modulation by sending several volts of a 2~kHz oscillation to the laser's PZT-mounted intra-cavity tuning element (a diffraction grating).  We increased the modulation to achieve roughly 25~MHz of depth. The laser spectrum (averaged over 50 samples) seems consistent with a classical oscillator:  small peaks at the ``turning points'' are actually a conspiracy of very many Bessel function orders in Eq.~\ref{eq:besselFunctionExpansion}.}
\label{fig:redLaserModulationExamples}
\end{figure}
As $D$ increases, more and more sidebands become relevant since the asymptotic behavior of the Bessel functions as $D \to \infty$ is damped and oscillatory. Cross terms between the sidebands develop in the intensity $I(t) = E(t)^2$ that are proportional to different Bessel function orders, a phenomenon called \emph{intermodulation}. These two limits are depicted in actual laser spectra in Figure~\ref{fig:redLaserModulationExamples}.

\subsection{Modulation apparatus}\label{sec:modulationApparatus}
\begin{figure}
\centering
\includegraphics{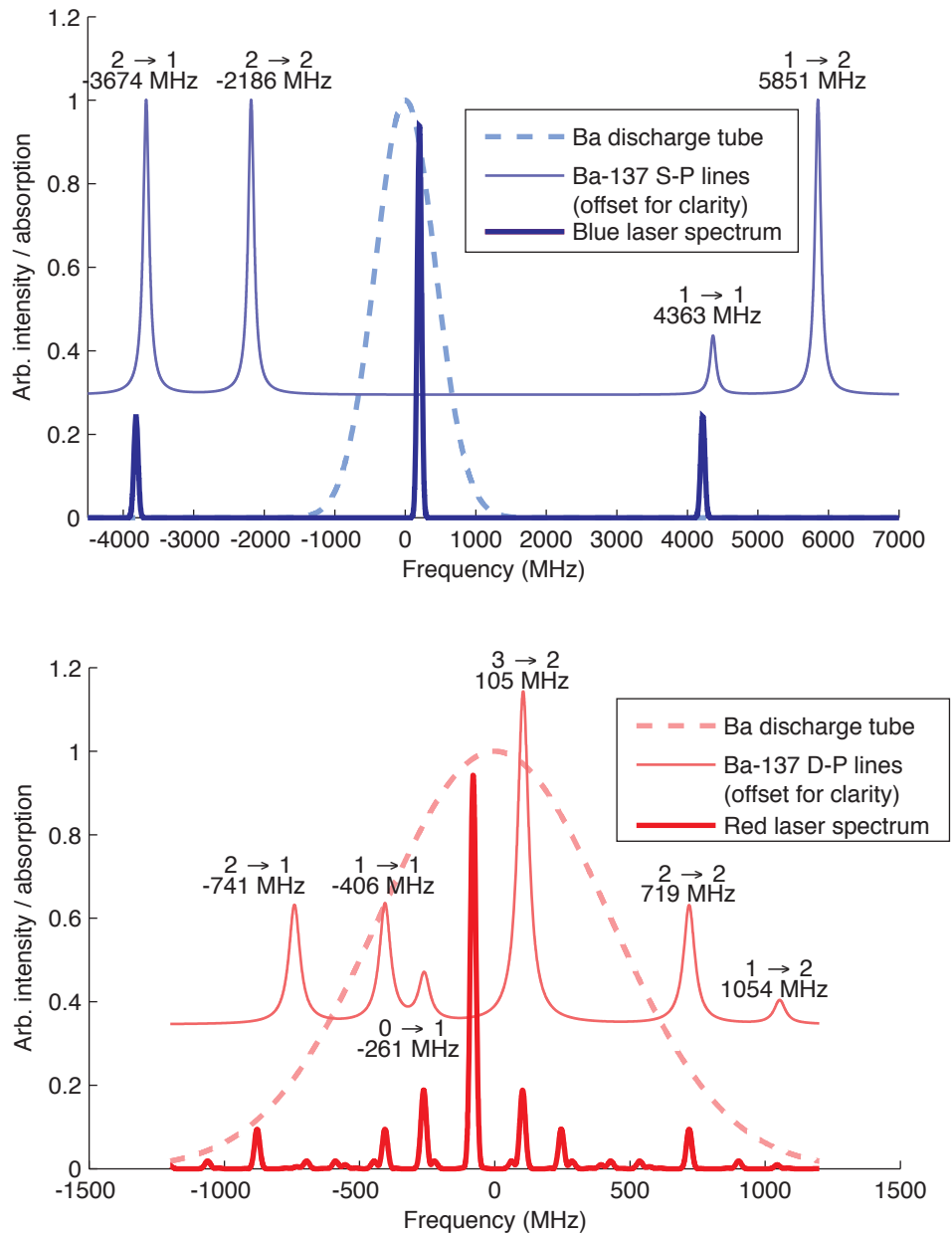}
\caption[Locations and strengths of the 493~nm and 650~nm transitions in $^{137}$Ba$^+$]{Locations and strengths of the 493~nm and 650~nm transitions in $^{137}$Ba$^+$.  The relative transition strengths are depicted graphically.  Also shown are modulated blue and red laser spectra designed to close the cooling and repumping transitions.  Here the blue laser is phase modulated at 4018~MHz at a depth of 0.5~radian while the red laser is modulated at four different frequencies; the details of which follow in the text.}
\label{fig:barium137Transitions}
\end{figure}

\begin{figure}
\centering
\includegraphics{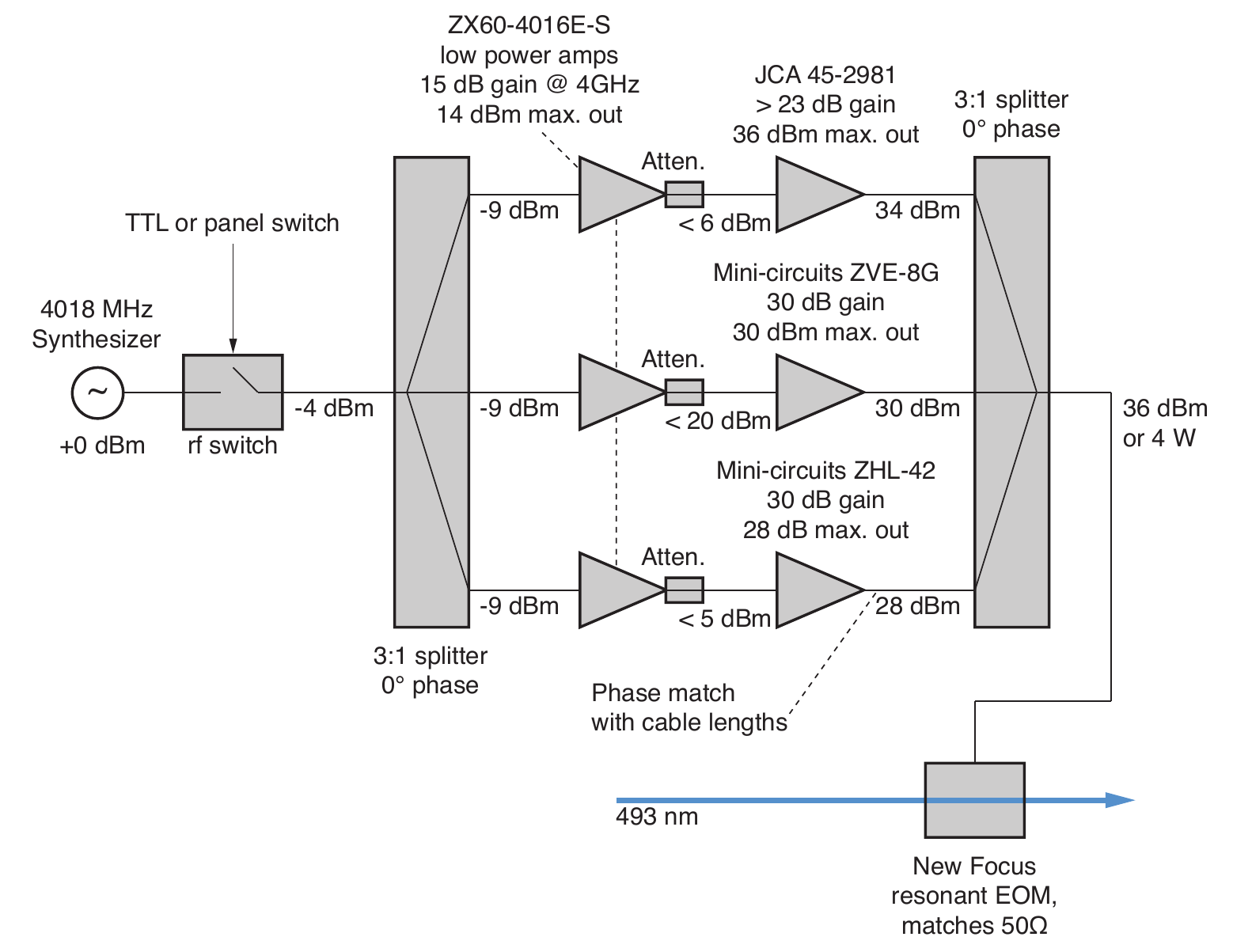}
\caption[Radio-frequency apparatus for modulation of the 493~nm laser]{Radio-frequency apparatus for modulation of the 493~nm laser.  This somewhat roundabout design came about because the three high power amplifiers and splitter/combiners were already purchased.  The resonant Mg:LiNbO$_3$ modulator (New Focus 4431) presents a matched load at its resonant frequency of 4018~MHz, mechanically tunable via a metal slug in the top of the case.  A maximum of 4W (36 dBm) of rf can be applied.  At 500~nm, the modulation depth is specified as 0.1~rad/V, making $V_\pi \approx 32~V$. }
\label{fig:blueModulationRFApparatus}
\end{figure}

\begin{figure}
\centering
\includegraphics{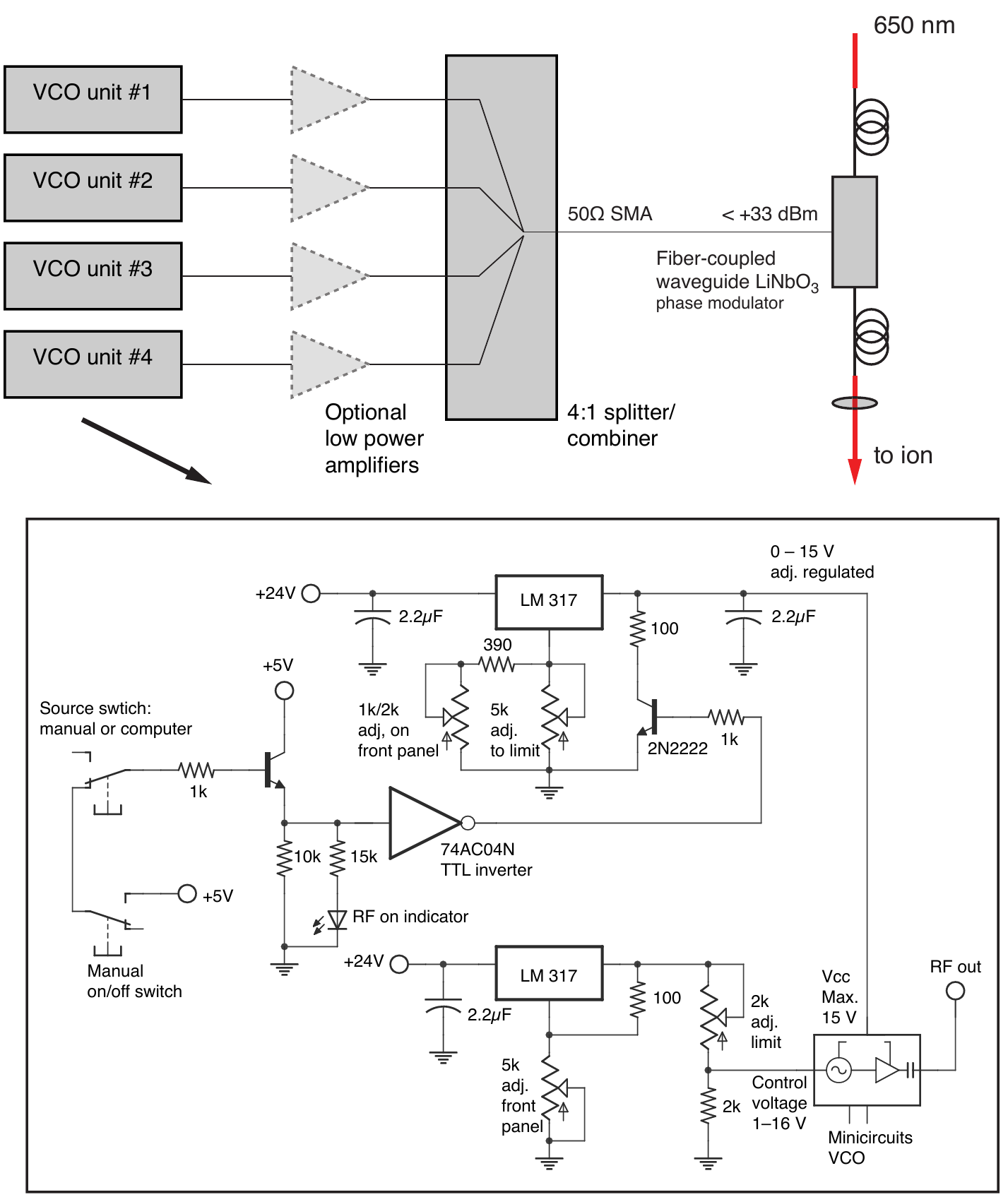}
\caption[Apparatus for many frequency phase modulation of the 650~nm laser]{Apparatus for many frequency phase modulation of the 650~nm laser.  Each voltage-controlled oscillator (VCO) frequency can be tuned to correspond to a particular $5D_{3/2} F \leftrightarrow 6P_{1/2} F'$ repumping transition.  Either front panel controls or TTL computer signals can turn individual sidebands on or off, allowing for specific $F$ level optical pumping.}
\label{fig:redLaserVCOSchematic}
\end{figure}

\begin{figure}
\centering
\includegraphics{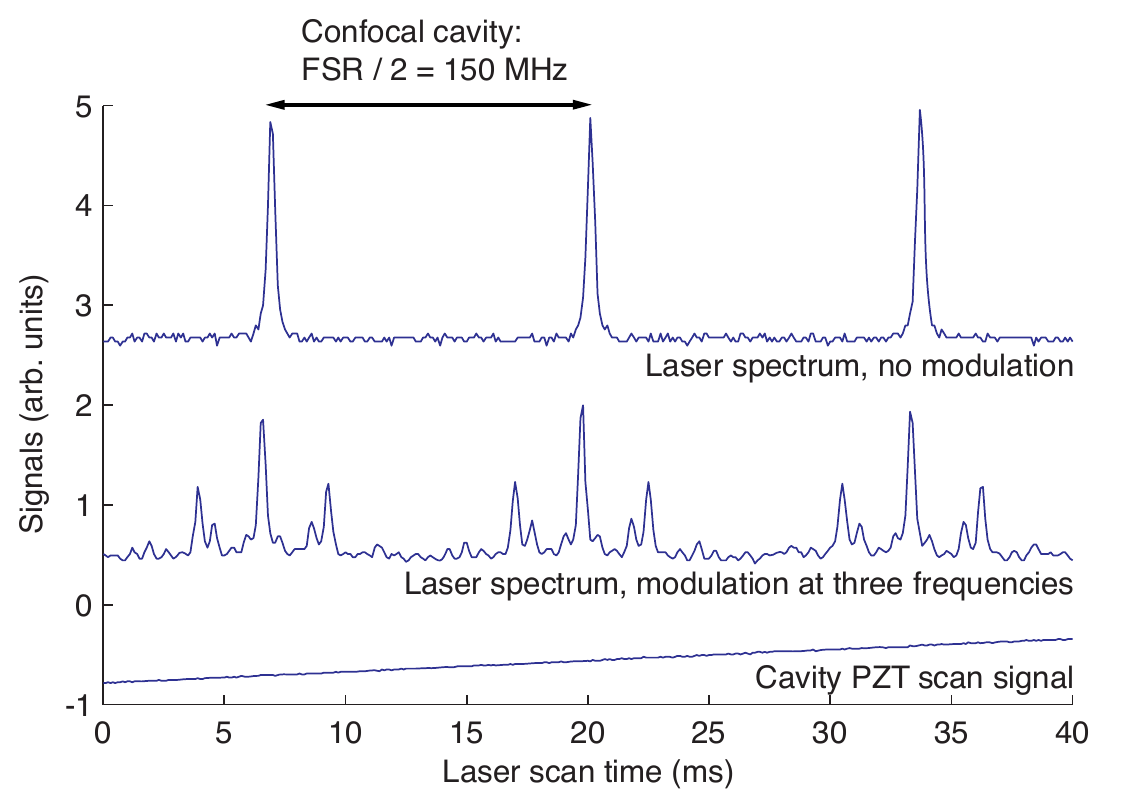}
\caption[Laser spectra with phase modulation at three frequencies]{Laser spectra with phase modulation at three frequencies chosen to repump particular $5D_{3/2}, F \to 6P_{1/2}, F'$ transitions as described in the text.  As long as all modulation depths are kept small, peaks corresponding to intermodulation between the frequencies are also small.}
\label{fig:redModulationThreeFreq}
\end{figure}

To create the necessary cooling sidebands on the 493~nm laser, we decided to purchase a resonant Mg:LiNbO$_3$ modulator (New Focus 4431) that presents an impedance matched load at a (slightly tunable) frequency of $(4018\pm200)$~MHz.  At 500~nm, the modulation depth of this unit is specified as 0.1~rad/V, making $V_\pi \approx 32~V$.  Our modulation scheme is to use both the $\pm 1$ phase modulation sidebands to excite the $6S_{1/2}, F = 2 \leftrightarrow 6P_{1/2}, F = 1$ and $6S_{1/2}, F = 1 \leftrightarrow 6P_{1/2}, F = 1$ transitions.  While this has the advantage of leaving the carrier of the laser near the Ba$^+$ hollow cathode tube resonance for frequency locking, there are two disadvantages.  First, as shown in Figure~\ref{fig:barium137Transitions}, the $6S_{1/2}, F = 1 \leftrightarrow 6P_{1/2}, F = 1$ is substantially weaker than the rest (see the coupling strength diagram in Figure~\ref{fig:oddIsotopeRelativeStrengths}).  Secondly, while the carrier is not resonant with any transitions in $^{137}$Ba$^+$, and due to the isotope shift is placed somewhat \emph{above} the $^{138}$Ba$^+$ resonance, we have experienced simultaneous loading and cooling of  $^{138}$Ba$^+$ and $^{137}$Ba$^+$, presumably through sympathetic cooling.  We are currently developing techniques for either lowering the trap depth to let the poorly cooled $^{138}$Ba$^+$ leave the trap~\cite{devoe2002esa}, selective excitation of the $^{138}$Ba$^+$ secular frequency~\cite{hasegawa2000rii}, or for otherwise preventing the ion to be trapped at all.

The high $V_\pi$ for this modulator means that we need to generate several Watts of rf at 4~GHz in order to put substantial amounts of power in the $\pm 1$ sidebands.  Suppression of the carrier is not possible while adhering to the maximum drive limits recommended by the manufacturer.  To that end, using many parts that were available in the laboratory, we have constructed a parallel amplification scheme to generate up to 4~W of rf, shown in Figure~\ref{fig:blueModulationRFApparatus}.

Frequency selective optical pumping with this modulation scheme is possible.  By simultaneously shifting the carrier downwards by $f' \sim 200$~Hz and increasing the modulation frequency by the same amount, one keeps the upper transition $6S_{1/2}, F = 1 \leftrightarrow 6P_{1/2}, F = 1$ in resonance while the lower transition is suppressed.  Likewise, by reversing the sign of the shifts, the lower transition $6S_{1/2}, F = 2 \leftrightarrow 6P_{1/2}, F = 1$ can be enhanced.  One might also employ a somewhat low finesse cavity with $\nu_\text{fsr}$ made non commensurate with 4018~MHz as a `sideband selector'~\cite{devoe2002esa}.

Partially due to our experience with the resonant modulator for 493~nm, we opted for a widely tunable, low-voltage, transverse phase modulator to generate the sidebands on the 650~nm repumping laser.  The unit is fiber-coupled with polarization maintaining single mode fiber.  The waveguide device only modulates one sense of polarization and in fact acts like a very good polarizer.  Over a bandwidth of 1~GHz, $V_\pi \approx 6$~V, which is accomplished with very minor rf amplification.  Since we expect to achieve selective optical pumping amongst all the $5D_{3/2}$ $F$-levels, we designed a modular, many frequency, rf scheme shown in Figure~\ref{fig:redLaserVCOSchematic}.  Several voltage-controlled oscillators are programmed to make sidebands resonant with one or more of the $5D_{3/2} \leftrightarrow 6P_{1/2}$ repumping lines.  Then, by digitally or manually controlling which of these frequencies are combined and routed to the modulator, one can turn on and off individual transitions for selective optical pumping.  The system has been well calibrated (see Figure~\ref{fig:redModulationThreeFreq} for an example spectrum), but has not been tested on a trapped ion for optical pumping purposes yet.

\subsection{Optical pumping strategies}
\begin{figure}
\centering
\includegraphics{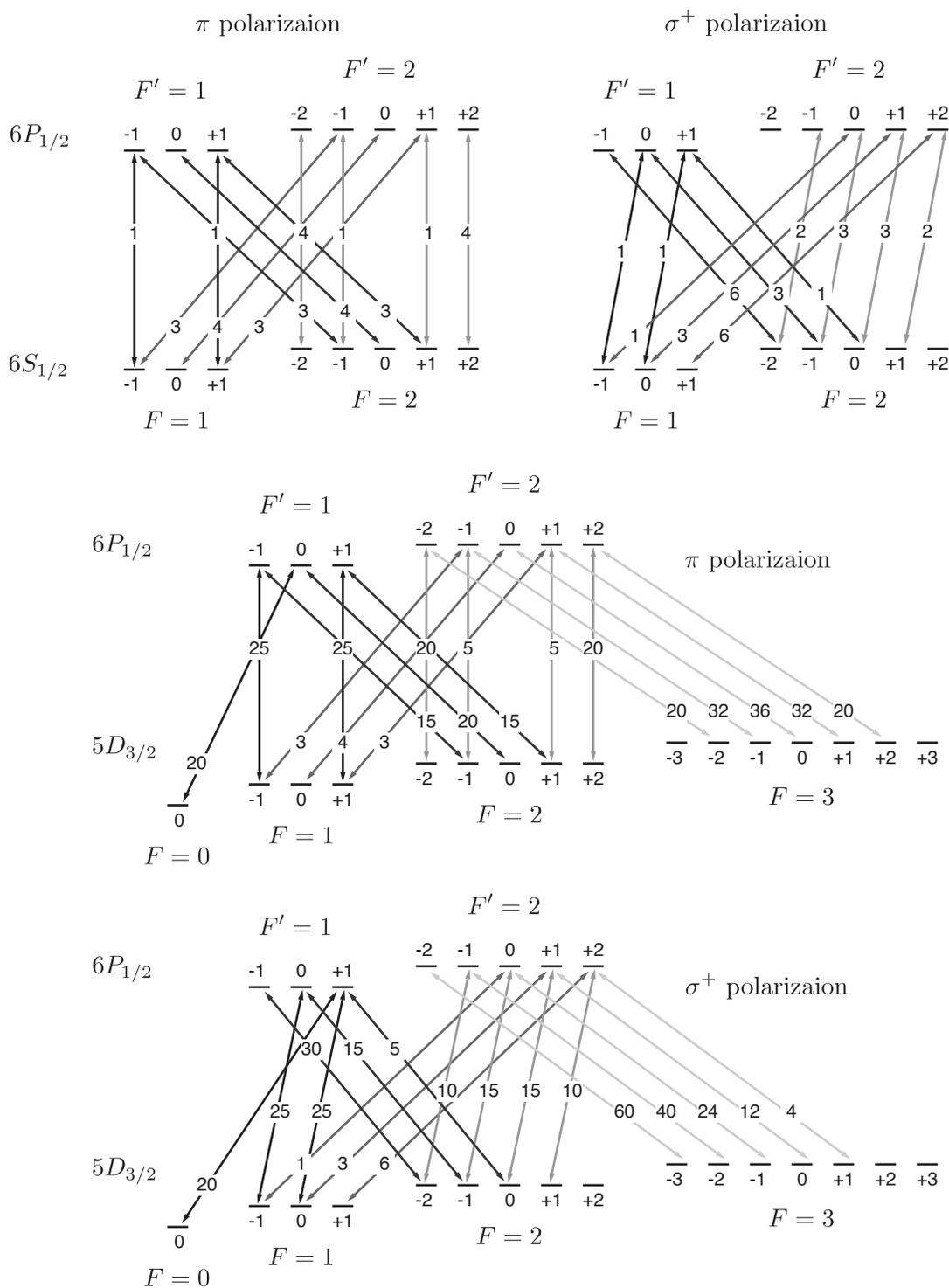}
\caption[Relative strengths of the 493~nm and 650~nm transitions in $^{137}$Ba$^+$.]{Relative strengths of the $6S_{1/2} \leftrightarrow 6P_{1/2}$ and $5D_{3/2} \leftrightarrow 6P_{1/2}$ transitions in $^{137}$Ba$^+$ following~\cite{metcalf1999lct}.  Each coupling strength is normalized to the weakest in each set.  This chart allows us to craft potential optical pumping strategies.}
\label{fig:oddIsotopeRelativeStrengths}
\end{figure}
Besides the selective frequency optical pumping schemes described in the laser section, we can also make use of the lasers' polarization states to achieve traditional optical pumping (see Section~\ref{sec:opticalPumping}). The same electric dipole selection rules that govern $|J, m \rangle \to |J', m' \rangle$ also apply to atoms with hyperfine structure when $F$ and $m_F$ are substituted for $J$ and $m_J$.  Matrix elements can be written in terms of the same dipole reduced matrix elements~\cite{metcalf1999lct},
\begin{equation}
\begin{split}
\langle \gamma', F', m' | \boldsymbol{\epsilon} \cdot \boldsymbol{r} | \gamma, F, m \rangle &= (-1)^{1+L'+S+J+J'+I-m'} \langle \gamma' , L' || r || \gamma, L \rangle \\
 &\times \sqrt{(2J + 1)(2J' +1)(2F+1)(2F'+1)} \\
&\times \left\{ \begin{array}{ccc} L' & J' & S' \\ J & L & 1\end{array} \right\}
 \left\{ \begin{array}{ccc} J' & F' & I \\ F & J & 1 \end{array}\right\}
 \left( \begin{array}{ccc} F & 1 & F' \\ m & q & -m' \end{array} \right) 
\end{split}
\end{equation}
when $\boldsymbol{\epsilon}$ picks out a single tensor index $q$ allowed by selection rules.  Figure~\ref{fig:oddIsotopeRelativeStrengths} shows a computation of the relative strengths of the Zeeman components of the $6S_{1/2} \leftrightarrow 6P_{1/2}$ and $5D_{3/2} \leftrightarrow 6P_{1/2}$ transitions, normalized by the weakest transition among each set.

\section{Precision measurement of $5D_{3/2}$ hyperfine structure}
We are motivated to look beyond the magnetic dipole approximation of the hyperfine interaction by careful studies such as~\cite{gerginov2003onm} which resolved the effects of the next two multipole terms allowed by parity:  the electric quadrupole and magnetic octopole hyperfine interactions. These measurements yield precious information on the details of many-nucleon nuclear structure often unavailable to atomic physics techniques.  We will show in this chapter that by employing the spin resonance technique successfully used to measure light shifts, a measurement of the hyperfine splittings in the $^{137}$Ba$^+$ metastable $5D_{3/2}$ is possible to a relative precision approaching $10^{-9}$.  

\subsection{The multipole hyperfine interaction}
The hyperfine interaction is usually characterized as the interaction of the nuclear and electronic angular momentum:  $H_\text{hf} \propto \boldsymbol{I} \cdot \boldsymbol{J}$.  This, however, only captures the dominant magnetic dipole interaction.  For nuclear spins $I > 1$, the electric-quadrupole and magnetic octopole interactions are also allowed.  For atoms with a single valance electron, the Hamiltonian can be expressed as in these multipole parts:
\begin{equation} \label{eq:hyperfineMultipoleHamiltonian}
H_\text{hf} = A \boldsymbol{I} \cdot \boldsymbol{J} +B f_\text{E2}(\boldsymbol{I},\boldsymbol{J}) + C f_\text{M3}(\boldsymbol{I},\boldsymbol{J})
\end{equation}
where  $f_\text{E2}$ and  $f_\text{M3}$ are geometrical weighting functions for the electric quadrupole and magnetic octopole interactions \cite{armstrong1971ths}:
\begin{align}
f_\text{E2}(\boldsymbol{I},\boldsymbol{J}) &= \frac{3(\boldsymbol{I} \cdot \boldsymbol{J})^2 + \tfrac{3}{2}(\boldsymbol{I} \cdot \boldsymbol{J}) - I(I+1)J(J+1)}{2I(2I-1)J(2J-1)} \\
\begin{split}
f_\text{M3}(\boldsymbol{I},\boldsymbol{J}) &= 
\left\{ 10(\boldsymbol{I} \cdot \boldsymbol{J})^3 + 20(\boldsymbol{I} \cdot \boldsymbol{J})^2 \right. \\
& + 2(\boldsymbol{I} \cdot \boldsymbol{J})[-3I(I+1)J(J+1) + I(I+1) + J(J+1) +3]   \\
& \left. -5I(I+1)J(J+1) \right\} / [I(I-1)(2I-1)J(J-1)(2J-1)].
\end{split}
\end{align}
For a given $|F,I,J \rangle$ state, we can write the product
\begin{equation}
\boldsymbol{I} \cdot \boldsymbol{J} = \frac{F(F+1) - I(I+1) - J(J+1)}{2}.
\end{equation}

\begin{table}
\centering
\caption[Hyperfine multipole geometrical coefficients for several $^{137}$Ba$^+$ states.]{Hyperfine multipole geometrical coefficients for several $^{137}$Ba$^+$ states.  These geometrical coefficients weight the nuclear magnetic dipole, electric quadrupole, and magnetic octopole moments $A$, $B$, and $C$ in Eq.~\ref{eq:hyperfineMultipoleHamiltonian}. }
\begin{tabular}{cc|ccc}
\multicolumn{2}{c}{State} 	& $ \boldsymbol{I} \cdot \boldsymbol{J}$ & $ f_\text{E2}$ & $ f_\text{M3}$ \\ \hline \hline
\multirow{2}{*}{$6S_{1/2}$ and $6P_{1/2}$} & $F=1$	& -5/4 & ---  & --- \\
								    & $F=2$         & 3/4  & ---  & --- \\ \hline
\multirow{4}{*}{$5D_{3/2}$ and $6P_{3/2}$}& $F = 0$ 	& -15/4 & 5/4 & -35 \\
								    & $F =1$	& -11/4 & 1/4 & 21 \\
								    & $F =2$	&  -3/4  & -3/4 & -7 \\
								    & $F =3$	&  9/4    & 1/4  & 1 \\ \hline
\multirow{4}{*}{$5D_{5/2}$}			    & $F = 1$ 	&-21/4 & 7/10 & -42/5 \\
								    & $F = 2$	& -13/4 & -1/10 & 54/5 \\
								    & $F = 3$	&  -1/4  & -11/20 & -27/5 \\
								    & $F = 4$	&  15/4 & 1/4  & 1
\end{tabular}
\label{tab:hyperfineMultipoleCoeff}
\end{table}

From Table~\ref{tab:hyperfineMultipoleCoeff} we can compute the hyperfine splittings in terms of the multipole interaction strengths $A$, $B$, and $C$ given in Eq.~\ref{eq:hyperfineMultipoleHamiltonian}:
\begin{align*}
\Delta_{F=0 \leftrightarrow 1} &= H_\text{hf}(F=1) - H_\text{hf}(F=0) =  -A + B-56C +\text{ (corrections)} \\
\Delta_{F=1 \leftrightarrow 2} &= H_\text{hf}(F=2) - H_\text{hf}(F=1) = -2A + B -28C +\text{ (corrections)} \\
\Delta_{F=2 \leftrightarrow 3} &= H_\text{hf}(F=3) - H_\text{hf}(F=1) = -3A - B -8C+\text{ (corrections)}
\end{align*}
Not yet included in these expressions are corrections due to the second-order perturbation terms of the dipole term as well as other perturbations such as the linear and quadratic Zeeman shifts.

In the case of neutral Cs~\cite{gerginov2003onm}, the nuclear magnetic octopole term is found to be $C = 0.56(7)$~kHz, which implies a nuclear magnetic octopole moment term some 40 times larger than one expects from a nuclear shell model.  If this moment has a comparable size in $^{137}$Ba$^+$, we will easily be able to resolve it using rf spin-slip spectroscopy;  an interesting question is whether there is a general principle that makes the moments so large.  If it is much smaller, however, it will not be resolvable due to other energy shifts and corrections.

\subsection{The second-order dipole hyperfine corrections}
\begin{figure}
\centering
\includegraphics{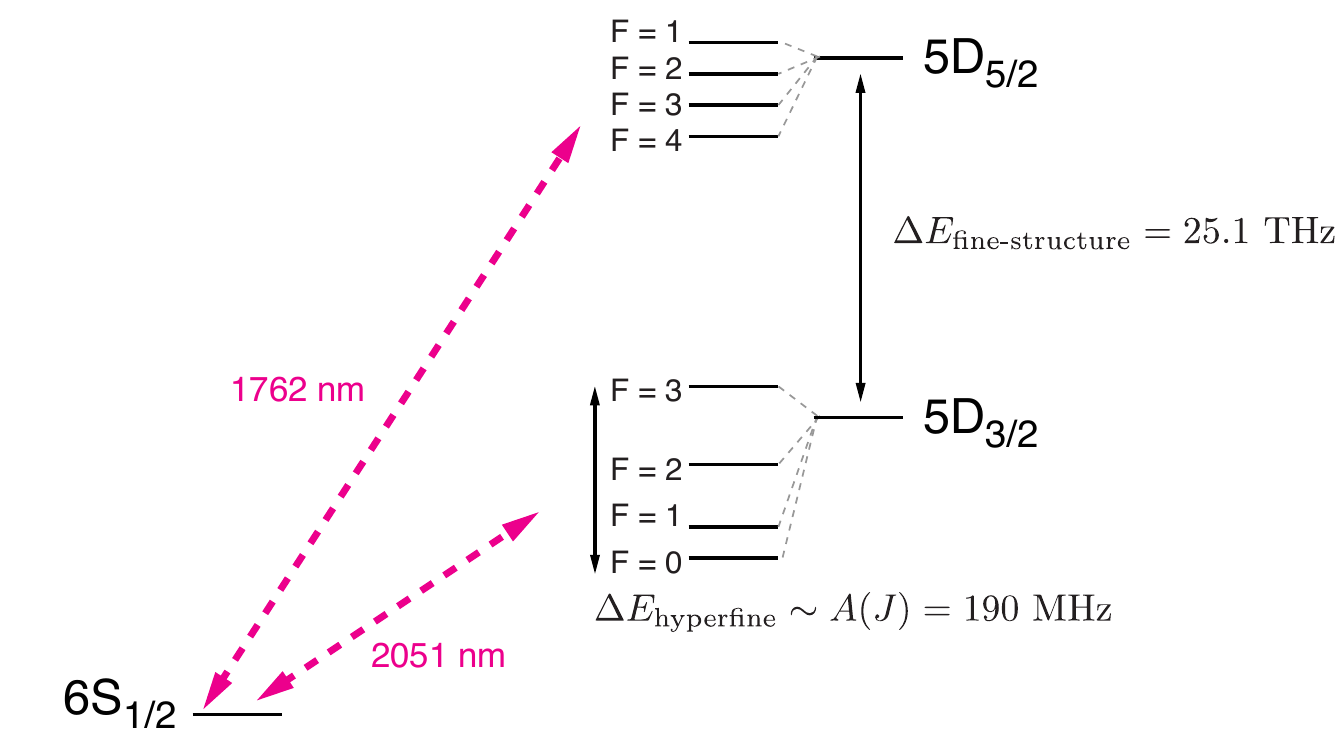}
\caption[Energy scales for the second-order dipole hyperfine shift]{Energy scales for the second-order dipole hyperfine shift.  We must estimate the second-order contribution of the nuclear magnetic dipole part of the hyperfine interaction since we anticipate that it will be as large or larger than any magnetic octopole term.  Naively, such contributions are suppressed by the ratio of the hyperfine to fine-structure energy scales;  see Eq.~\ref{eq:secondOrderDipole}.}
\label{fig:secondOrderDipoleHF}
\end{figure}
Anticipating that the magnetic octopole contribution to hyperfine splittings is small, we must calculate the magnetic dipole ($\boldsymbol{I } \cdot \boldsymbol{J}$) term out to second order in perturbation theory.  Following \cite[(Section 1.11)]{budker2004ape}, we expect the $5D_{3/2}$ states to gain an admixture of $5D_{5/2}$ states.  Since the dipole hyperfine Hamiltonian only mixes states with the same $F$, we can expect:
\begin{equation*}
| 5D_{3/2}, F \rangle \to | 5D_{3/2}, F \rangle_0 + O(1) \frac{\Delta E_\text{hyperfine}}{\Delta E_\text{fine-structure}} | 5D_{5/2}, F \rangle
\end{equation*}
which means that the energy shifts will be of the scale
\begin{equation*}
\Delta E_{5D_{3/2}, F}^\text{Second-order hyperfine} \sim \frac{\Delta E_\text{hyperfine}^2}{\Delta E_\text{fine-structure}} \approx \frac{(150 \text{ MHz})^2}{25 \text{ THz}} = 0.9 \text{ kHz}.
\end{equation*}

More precisely, using second-order perturbation theory, we write
\begin{equation}
\Delta E_{5D_{3/2}, F} = \sum_{F'} \frac{|\langle 5D_{3/2}, F | A(J) \boldsymbol{I} \cdot \boldsymbol{J} | 5D_{5/2}, F' \rangle|^2}{E_{5D_{3/2}, F} - E_{5D_{5/2, F'}}}
\end{equation}
We proceed by decomposing the $|F \rangle$ wavefunctions into the constituent $|I \rangle |J \rangle$ pieces, valid in the $IJ$-coupling approximation that both $I$ and $J$ are good separate quantum numbers. Since the dipole Hamiltonian can be broken up into pieces that independently couple electronic orbital angular momentum $\boldsymbol{l}$ and spin angular momentum $\boldsymbol{s}$~\cite{budker2004ape}, we write
\begin{equation*}
A(J) \boldsymbol{I} \cdot \boldsymbol{J} = \boldsymbol{I} \cdot (a_l \boldsymbol{L} + a_s \boldsymbol{S})
\end{equation*}
noting that often it is the $ a_s \boldsymbol{I} \cdot  \boldsymbol{S}$ piece that is dominant.  In this case, the second-order energy shifts are
\begin{equation} \label{eq:secondOrderDipole}
\Delta E_{5D_{3/2}, F} = \sum_{F'} \frac{|\langle 5D_{3/2}, F | a_s \boldsymbol{I} \cdot \boldsymbol{S} | 5D_{5/2}, F' \rangle|^2}{E_{5D_{3/2}, F} - E_{5D_{5/2, F'}}}.
\end{equation}
Writing
\begin{equation}
\boldsymbol{I} \cdot \boldsymbol{s} = \frac{1}{2}(I_+ S_- + I_- S_-) + I_z S_z,
\end{equation}
and using the operators
\begin{align}
I_\pm | I, m_I \rangle &= \sqrt{I(I+1) - m_I(m_I \pm 1)} |I, m_I \pm 1 \rangle, \\
I_z | I, m_I \rangle &= m_I  | I, m_I \rangle, \\
S_\pm | S, m_S \rangle &= \sqrt{S(S+1) - m_S(m_S \pm 1)} |S, m_S \pm 1 \rangle, \\
S_z | S, m_S \rangle &= m_S  | S, m_S \rangle
\end{align}
we can calculate the matrix elements
\begin{align}
\langle 5D_{3/2}, F = 0 | \boldsymbol{I} \cdot \boldsymbol{S} | 5D_{5/2}, F = 0 \rangle &= 0, \\
\langle 5D_{3/2}, F = 1 | \boldsymbol{I} \cdot \boldsymbol{S} | 5D_{5/2}, F = 1 \rangle &= \frac{6}{50}, \\
\langle 5D_{3/2}, F = 2 | \boldsymbol{I} \cdot \boldsymbol{S} | 5D_{5/2}, F = 2 \rangle &= \sqrt{\frac{3}{7}} \frac{22}{30}, \\
\langle 5D_{3/2}, F = 3 | \boldsymbol{I} \cdot \boldsymbol{S} | 5D_{5/2}, F = 3 \rangle &= \sqrt{\frac{3}{2}}\frac{4}{5},
\end{align}
with all other matrix elements vanishing.  Note that the operator $\boldsymbol{I} \cdot \boldsymbol{S}$ only couples states with $F = F'$.  These lead to frequency shifts in the $5D_{3/2}$ state of
\begin{align}
\Delta \omega_{F = 0 \leftrightarrow 1}^\text{second-order dipole} & \approx +0.1 \text{ kHz} \\
\Delta \omega_{F = 1 \leftrightarrow 2}^\text{second-order dipole} & \approx +1.3 \text{ kHz} \\
\Delta \omega_{F = 2 \leftrightarrow 3}^\text{second-order dipole} & \approx +5.8 \text{ kHz}.
\end{align}
These shifts could all be comparable or larger than those due to the nuclear magnetic octopole term.  Therefore a more accurate calculation should be completed before deriving values of hyperfine multipole moments $A(J)$, $B(J)$, and $C(J)$ from measurements of the hyperfine splittings.

\subsection{The Zeeman effect in $^{137}$Ba$^+$ $5D_{3/2}$}
\begin{figure}
\centering
\includegraphics{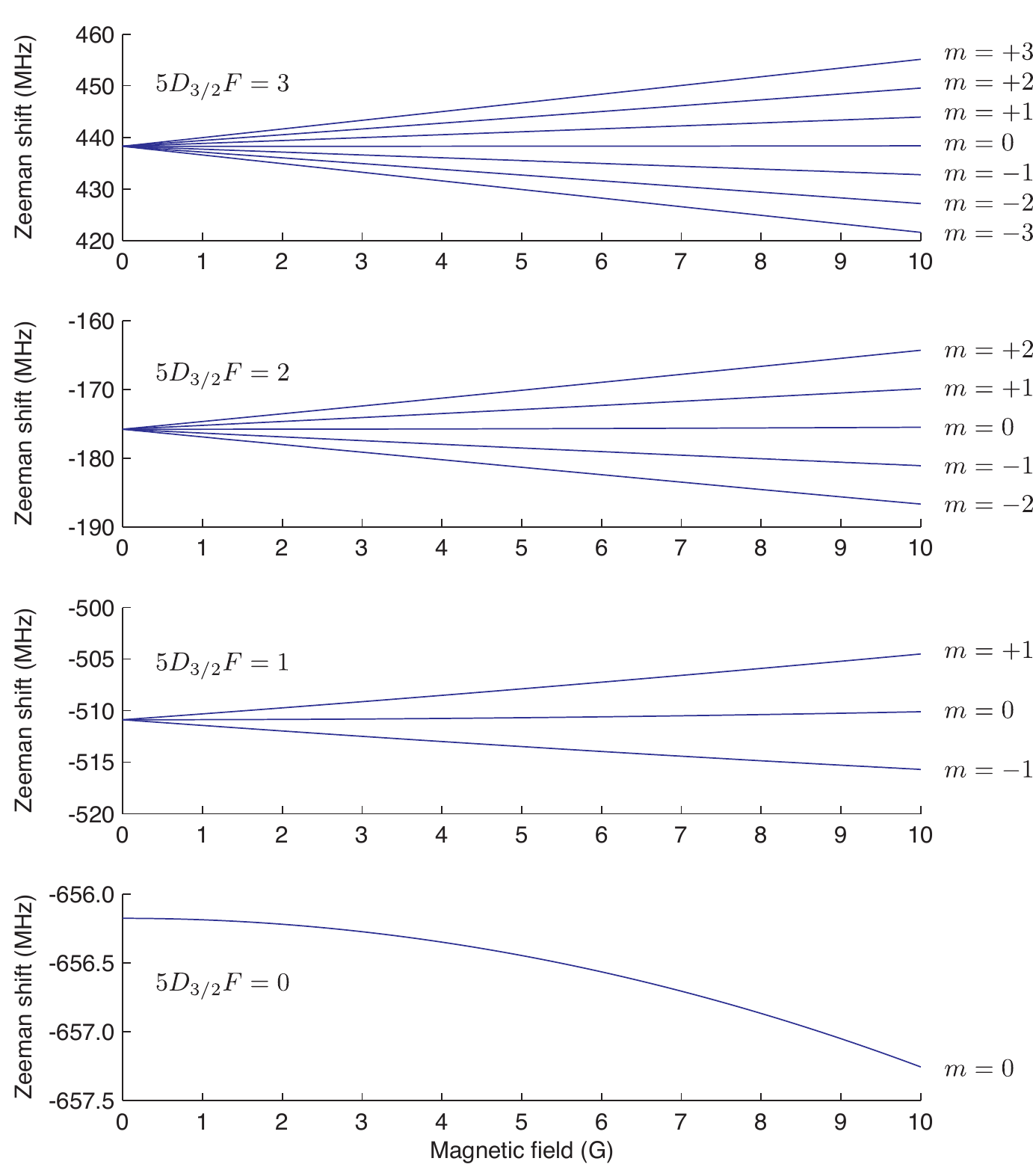}
\caption[The Zeeman effect in the $^{137}$Ba$^+$ $5D_{3/2}$ hyperfine states]{The Zeeman effect in the $^{137}$Ba$^+$ $5D_{3/2}$ hyperfine states for magnetic fields small compared to the hyperfine energy scale $g_J \mu_B B \ll A$.  To first order in perturbation theory, $m_F = 0$ states are free of Zeeman shifts.  Second-order terms allow admixture between $F$ states, and shifts proportional to $B^2$ result.}
\label{fig:hyperfineZeemanFigure}
\end{figure}
A special cancellation occurs in the Zeeman effect for the odd isotope metastable $5D_{3/2}$ state:  the splittings are independent of $F$ level because $I = J = 3/2$.  In general, the linear Zeeman shifts are:
\begin{equation*}
\Delta_{|F, m_F \rangle}^\text{Zeeman} = g_F \mu_B B m_F
\end{equation*}
where the hyperfine state $g$-factor is expressed in terms of the electronic and nuclear $g$-factors $g_J$ and $g_I'$:
\begin{equation*}
g_F \equiv g_J \frac{F(F+1) + J(J+1) - I(I+1)}{2F(F+1)} - g_I' \frac{F(F+1) - J(J+1) + I(I+1)}{2F(F+1)}
\end{equation*}
and when $I=J$ as is the case for $5D_{3/2}$, this simply reduces to
\begin{equation}
g_F = \frac{g_J - g_I'}{2}
\end{equation}
which is independent of $F$.  

The second-order Zeeman effect is relevant at lower magnetic field magnitudes in the odd-isotope because the relevant scale at which admixture occurs is the hyperfine splitting rather than the fine-structure.  Section~\ref{sec:clockZeeman} shows the calculation.  Both the linear and Zeeman effects are plotted for the hyperfine states in Figure~\ref{fig:hyperfineZeemanFigure}.

\subsection{Experimental method}
\begin{figure}
\centering
\includegraphics{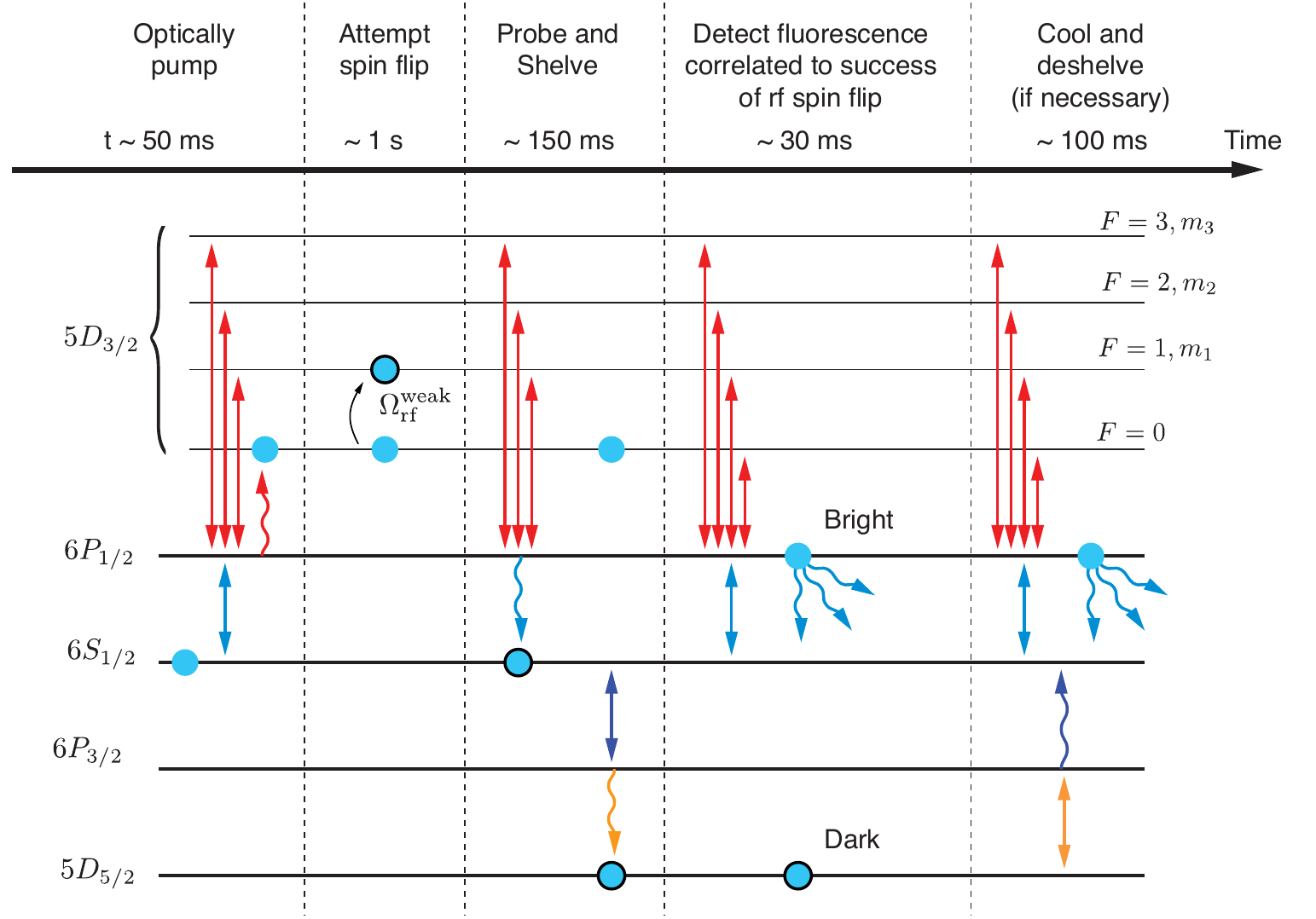}
\caption[Hyperfine splitting measurement sequence: $5D_{3/2}, F = 0 \leftrightarrow F =1$]{A possible hyperfine splitting measurement sequence: $5D_{3/2}, F = 0 \leftrightarrow F =1$ if the \emph{only} optical pumping mechanism available places the ion into the $F = 0$ state.  Refer to the text for a full description, and to Chapter~\ref{sec:lightShiftChapter} for a fuller description of the single ion spin resonance detection technique.  Briefly, we arrange through optical pumping and electron shelving that a successful rf-spin flip can be detected through a correlation in ion fluorescence.}
\label{fig:hyperfineMeasurementSequence}
\end{figure}

\begin{figure}
\centering
\includegraphics{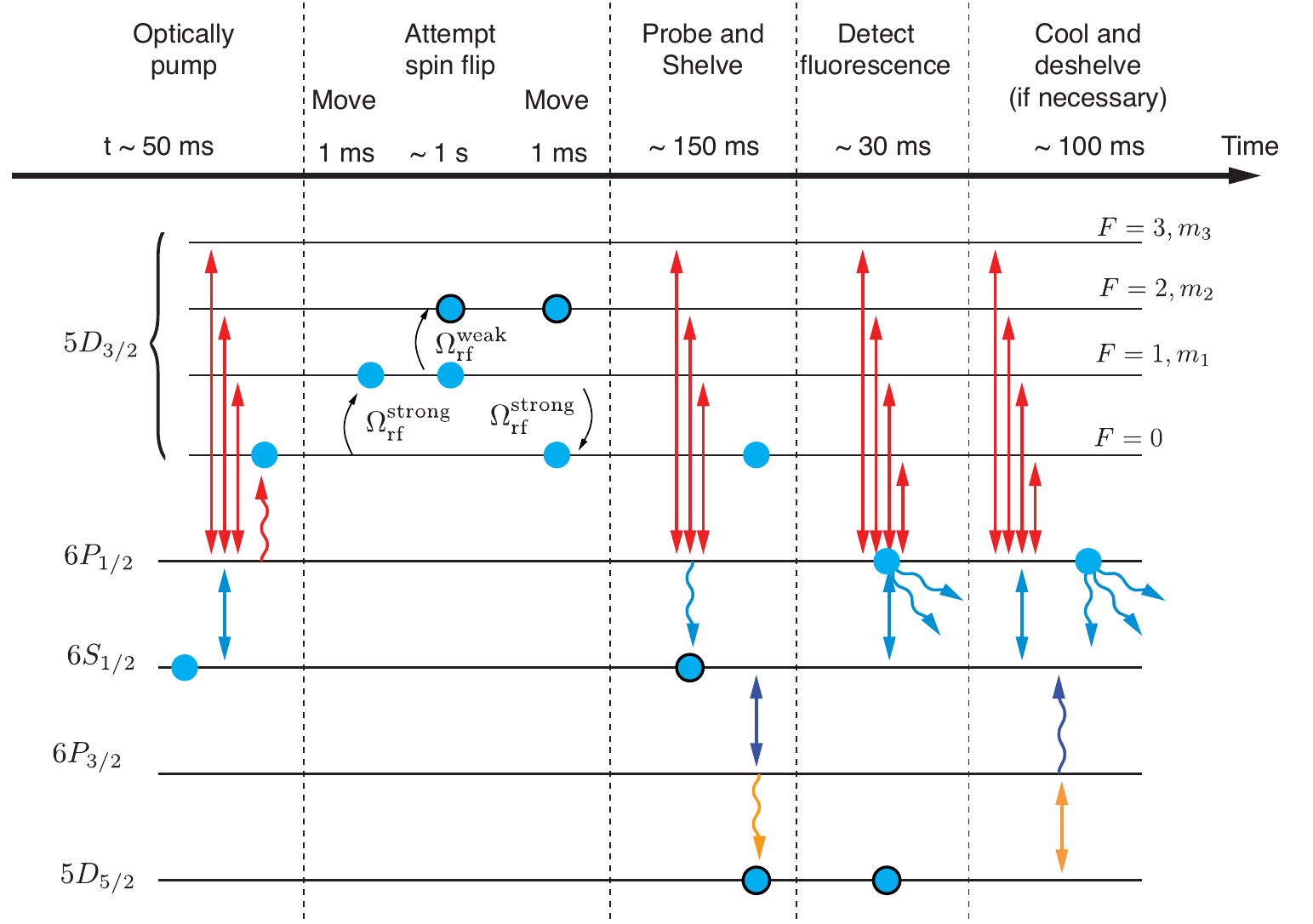}
\caption[Hyperfine splitting measurement sequence: $5D_{3/2}, F = 1\leftrightarrow F =2$]{A possible hyperfine splitting measurement sequence: $5D_{3/2}, F = 1 \leftrightarrow F =2$ if the \emph{only} optical pumping mechanism available places the ion into the $F = 0$ state.  Refer to Figure~\ref{fig:hyperfineMeasurementSequence} to see how the $F = 0 \leftrightarrow F= 1$ splitting is established.  Then, high power `population movement' $\pi$-pulses $\Omega^\text{strong}_\text{rf}$ manipulate the $5D_{3/2}$ population for a weak `spectroscopy' $\pi$-pulse $\Omega^\text{weak}_\text{rf}$.  Of course, a more elegant approach would use direct optical pumping into and out of the $F = 1$ state, but this procedure shows that one must only have one optical pumping mode to succeed.}
\label{fig:hyperfineMeasurementSequence2}
\end{figure}

\begin{figure}
\centering
\includegraphics{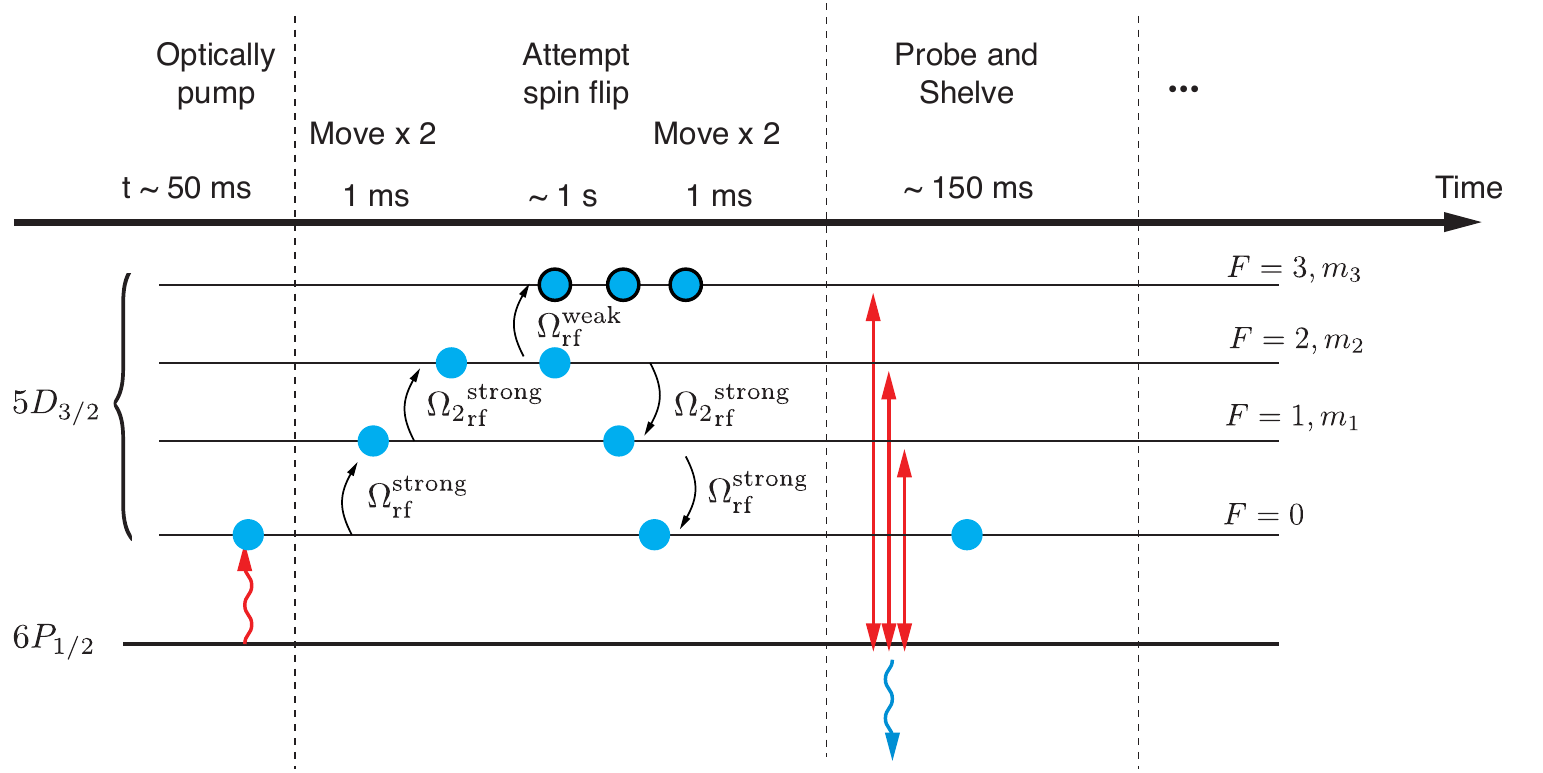}
\caption[Hyperfine splitting measurement sequence: $5D_{3/2}, F = 2\leftrightarrow F =3$]{Part of a possible hyperfine splitting measurement sequence: $5D_{3/2}, F = 2 \leftrightarrow F =3$ if the \emph{only} optical pumping mechanism available places the ion into the $F = 0$ state.  Now, two `population movement' $\pi$-pulses are employed around the spectroscopy pulse. Of course, a more elegant approach would use direct optical pumping into and out of the $F = 2$ state, but this procedure shows that one must only have one optical pumping mode to succeed.}
\label{fig:hyperfineMeasurementSequence3}
\end{figure}

We intend to measure the hyperfine frequency splittings
\begin{align*}
\omega_{01} &\equiv \tfrac{1}{\hbar}(E_{F=1, m_F=0} - E_{F=0,m_F=0}) \\
\omega_{12} &\equiv \tfrac{1}{\hbar}(E_{F=2, m_F=0} - E_{F=1,m_F=0}) \\
\omega_{23} &\equiv \tfrac{1}{\hbar}(E_{F=3, m_F=0} - E_{F=1,m_F=0})
\end{align*}
so that we are insensitive to first-order Zeeman shifts.  We will accomplish the $\Delta_{F=0 \leftrightarrow 1}$ measurement first using the following technique, outlined in Figure~\ref{fig:hyperfineMeasurementSequence}:
\begin{enumerate}
\item The ion is optically pumped into the $5D_{3/2} F=0, m_F =0$ state by maintain cooling on all 493~nm transitions and on 650~nm transitions that couple the $5D_{3/2} F=1,2,3$ states to $6P_{1/2}$.  The ion fluorescence will cease.
\item With all lasers turned off, we apply a weak rf $\pi$-pulse that will drive the ion to $5D_{3/2} F=1, m_F =0$ if it is on resonance.  A magnetic field of $\sim$ 1~G ensures that no other $F=1$ sublevels are populated.
\item To check if the spin flip transition occurred, we again apply weak 650~nm light resonant with the $5D_{3/2} F=1,2,3$ levels.  Any population in the $F = 1$ level is transferred to the ground state $6S_{1/2}$ with high probability while population remaining in $F=0$ is left alone.
\item  A shelving pulse of 455~nm transfers the ground state population to $5D_{5/2}$.
\item We turn on all cooling and repumping transitions and test if the ion is fluorescing or shelved.  This is correlated to whether the rf spin flip is on resonance.
\end{enumerate}
We can precisely measure the second-order Zeeman and electric quadrupole systematic shifts by similarly performing rf probes of the $5D_{3/2} F = 0, m_F = 0 \leftrightarrow F = 1, m_F = \pm 1$ transitions.

Having measured this splitting, $\omega_{01}$, we can now measure the other splittings with no additional optical pumping schemes.  For example, to measure $\omega_{12}$, we follow Figure~\ref{fig:hyperfineMeasurementSequence2}:
\begin{enumerate}
\item Again the ion is initialized in the $5D_{3/2} F=0, m_F =0$ state via optical pumping on the 650~nm transition.
\item Now, since $\omega_{01}$ is known, a \emph{strong}, frequency insensitive $\pi$-pulse (we we call a \emph{movement} pulse) at $\omega_{01}$ moves the population to $F = 1, m_F =0$ quickly and with little sensitivity to any frequency programming errors.  The magnitude of the movement pulse Rabi frequency is perhaps 10~kHz so that the pulse is still not resonant with other $F = 1$ magnetic sublevels.
\item We apply a weak \emph{spectroscopy} $\pi$-pulse that probes the splitting $F = 1, m_F = 0 \leftrightarrow F = 2, m_F = 0$ with a long coherence time.
\item Another (frequency insensitive) movement $\pi$-pulse moves population left in $F=1$ back to $F=0$.
\item Levels $F=1,2,3$ are pumped to $6S_{1/2}$ using resonant 650~nm light, leaving population in $F=0$ alone.
\item A 455~nm shelving pulse moves $6S_{1/2}$ population to $5D_{5/2}$.
\item We detect fluorescence upon application of all cooling and repumping transitions correlated to whether the spectroscopy pulse was successful.
\end{enumerate}

Hopefully one can see that the final splitting, $\omega_{23}$ can be measured in a similar way with two movement rf pulses once the splittings $\omega_{01}$ and $\omega_{12}$ are well known.  See Figure~\ref{fig:hyperfineMeasurementSequence3} for the important portion of the idea.

\subsection{Systematic effects}
\begin{figure}
\centering
\includegraphics{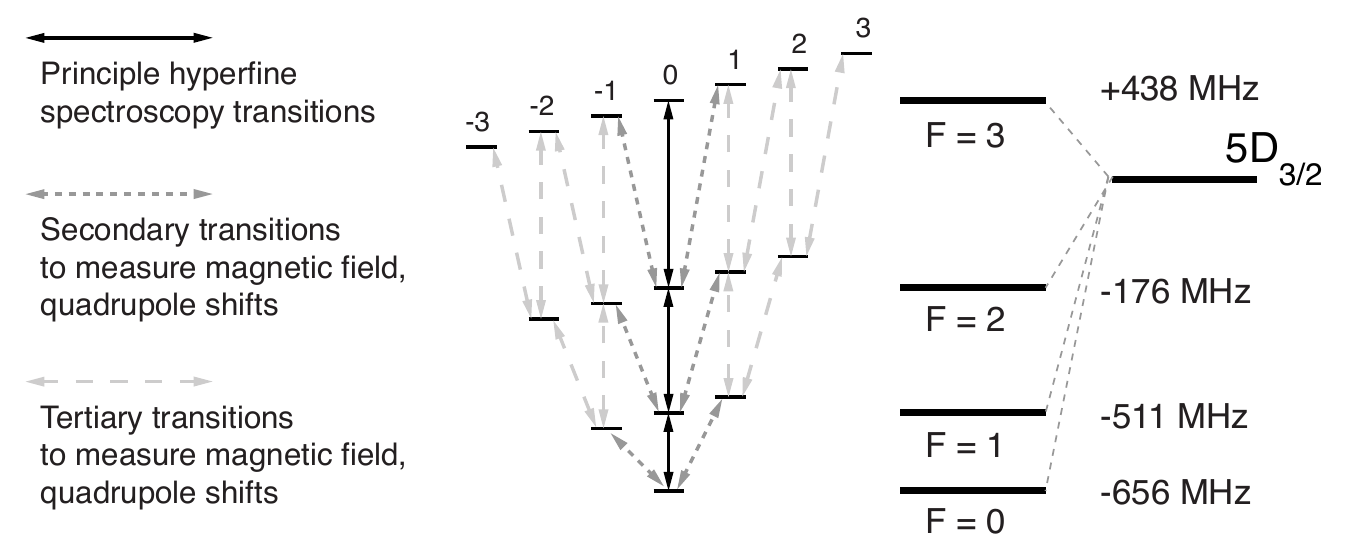}
\caption[Sets of $5D_{3/2}$ hyperfine transitions for isolating systematic effects]{Sets of $5D_{3/2}$ hyperfine transitions for isolating systematic effects.  The principle transitions, $m = 0 \leftrightarrow m'=0$, have small quadratic dependence on the magnetic field.  To correct for this second-order Zeeman effect, the magnetic is sensed on transitions involving $m, m' \ne 0$ which have strong linear dependence on $B$.  Measurements of many transitions allow sensitivity to additional systematic shifts with tensor structure:  dc-quadrupole Stark shifts from electric field gradients near the trap,  light shifts, etc.}
\label{fig:hyperfineRadioTransitions}
\end{figure}

\begin{figure}
\centering
\includegraphics{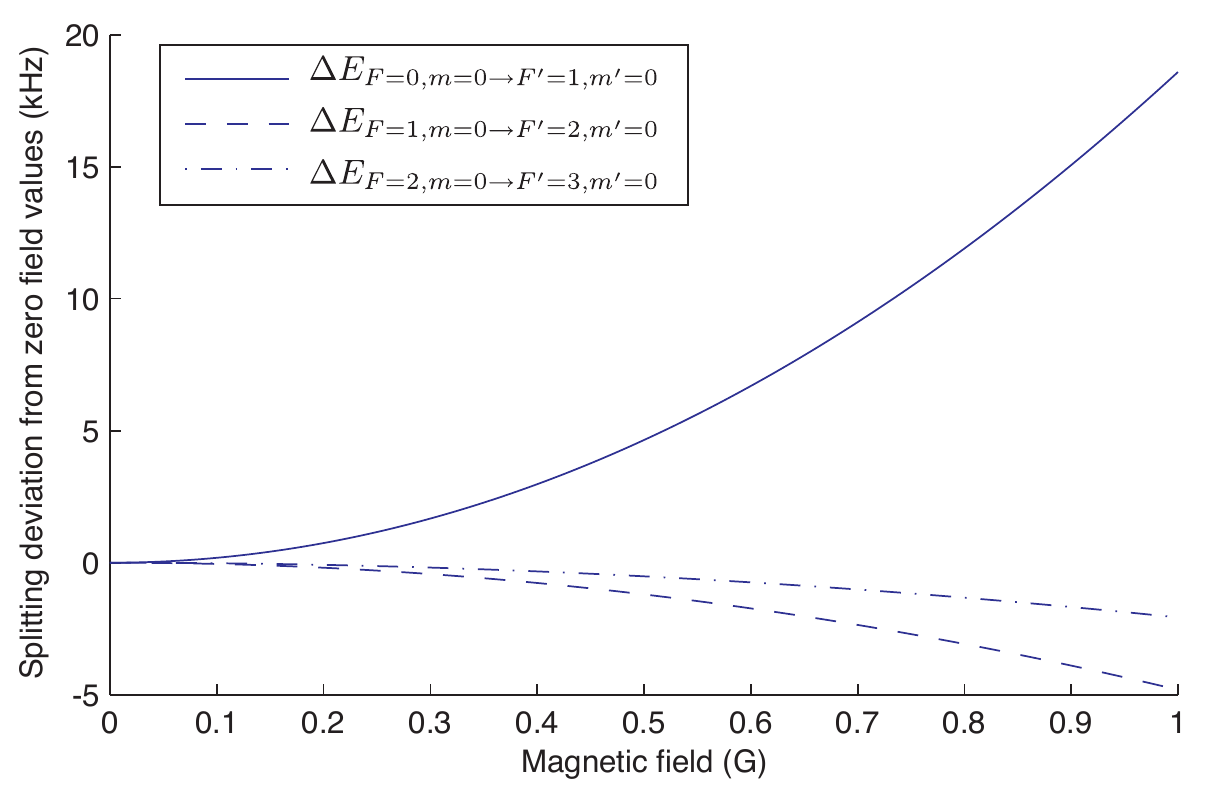}
\caption[Deviations of hyperfine splittings due to the second order Zeeman effect]{Here we show deviations of hyperfine splittings due to the second order Zeeman effect.  These deviations must be removed from measured splittings.  We anticipate using splitting measurements with $m, m' \ne 0$ in order to measure the magnetic field to a precision of $\sim 10^{-5}$.  Then, a lookup table such as this figure allows a precision correction to be applied to splittings with $m, m' = 0$, removing the second-order Zeeman effect at high accuracy.}
\label{fig:zeemanHyperfineDeviations}
\end{figure}
The chief systematic effects to consider are the second-order Zeeman shift (controllable by changing the applied magnetic field) and the electronic quadrupole shift which can vary at the 1~Hz level due to patch-potentials and stray fields.  Other effects to analyze are ac-Zeeman shifts due to stray currents at the trap frequency $\omega_\text{trap rf}$ (see Section~\ref{sec:acZeemanLS}), micromotion induced shifts, and the Bloch-Siegert shift due to the on-resonant spectroscopy pulse.

One strategy, suggested by Figure~\ref{fig:hyperfineRadioTransitions} is to measure hyperfine splittings with $m, m' \ne 0$ in order to sample the first-order Zeeman effect, electric quadrupole shifts, and any other shifts with tensor structure other than scalar.  Then, the `precision' splitting measurements with $m, m' =0$ can be corrected;  for example, the second-order Zeeman shift can be removed using a calculation illustrated in Figure~\ref{fig:zeemanHyperfineDeviations}.

\section{M1/E2 forbidden transition rate ratio at 2051~nm}
\begin{figure}
\centering
\includegraphics{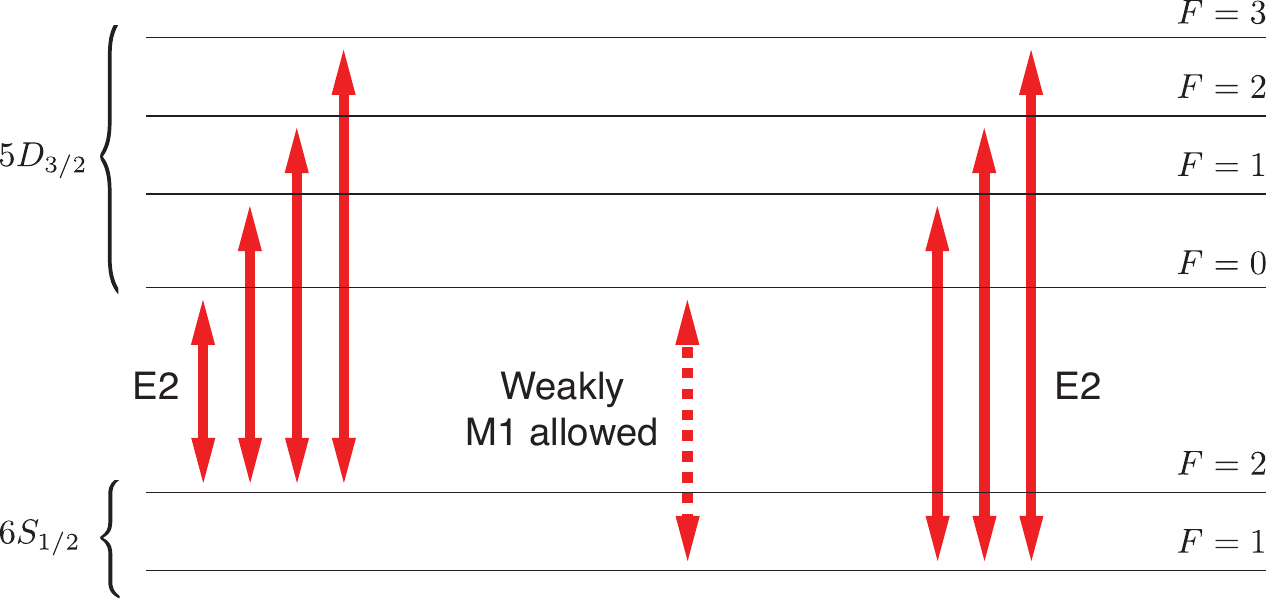}
\caption[Only magnetic dipole $6S_{1/2}, F = 1 \leftrightarrow 5D_{3/2}, F = 0$ transitions are allowed]{Only magnetic dipole $6S_{1/2}, F = 1 \leftrightarrow 5D_{3/2}, F = 0$ transitions are allowed.  Measuring this special transition rate relative to the more relevant E2 coupling for the other transitions would yield information about the $e$-$e$ coupling physics that gives rise to the larger than expected M1 coupling.}
\label{fig:m1E2AllowedTransitions}
\end{figure}
\begin{table}
\centering
\caption[Ba$^+$ 2051~nm resonances verified to be within mode-hop free tuning range]{Barium ion 2051~nm resonances verified to be within mode-hope free tuning range of the Tm,Ho:YLF laser.}
\begin{tabular}{lllllc}
Transition & Utility & Frequency (GHz) & $\lambda_\text{vac}$ (nm) & Wavenumber (cm$^{-1}$)	& Coupling \\ \hline \hline
$^{138}$Ba$^+$ 				    & Parity 	& 146114.408 & 2051.7652 & 4873.852 	    	& E2, M1 \\ \hline
$^{137}$Ba$^+  F=1 \leftrightarrow 0$ & M1 only 	& 146118.491 & 2051.7079 & 4873.988 		& M1 \\ 
$^{137}$Ba$^+  F=1 \leftrightarrow 1$ &  			& 146118.636 & 2051.7058 & 4873.993 		& E2, M1\\
$^{137}$Ba$^+  F=1 \leftrightarrow 2$ &			& 146118.971 & 2051.7011 & 4874.004 		& E2, M1 \\
$^{137}$Ba$^+  F=1 \leftrightarrow 3$ & 			& 146119.585 & 2051.6925 & 4874.024 		& E2 \\ \hline
$^{137}$Ba$^+  F=2 \leftrightarrow 0$ & Clock 	& 146110.454 & 2051.8207 & 4873.720 		& E2 \\
$^{137}$Ba$^+  F=2 \leftrightarrow 1$ & 			& 146110.599 & 2051.8187 & 4873.725 		& E2, M1 \\
$^{137}$Ba$^+  F=2 \leftrightarrow 2$ & 			& 146110.934 & 2051.8140 & 4873.736 		& E2, M1 \\
$^{137}$Ba$^+  F=2 \leftrightarrow 3$ & 			& 146111.548 & 2051.8054 & 4873.757		& E2, M1
\end{tabular}
\label{tab:2micronTransitionsTable}
\end{table}
\begin{table}
\centering
\caption[Calculations~\cite{sahoo2006lms} show enhanced $\langle nS_{1/2} || M1 || n'D_{3/2} \rangle$ matrix elements]{Calculations~\cite{sahoo2006lms} show enhanced $\langle nS_{1/2} || M1 || n'D_{3/2} \rangle$ matrix elements for Ba$^+$ and similar ion species.}
\begin{tabular}{l l l l}
									&  Ba$^+$				&  Sr$^+$	 		& Ca$^+$ \\ \hline \hline
Dominant E2 coupling~\cite{sahoo2006lms} 	&  12.734 $ea_0^2$ 	& 11.332 $ea_0^2$	& 7.973 $ea_0^2$ \\
M1 coupling, coupling~\cite{sahoo2006lms}	&  $8 \times 10^{-4} (\alpha/2)ea_0$ & $5 \times 10^{-4} (\alpha/2)ea_0$ & $7 \times 10^{-4} (\alpha/2)ea_0$ \\ \hline
Naive expected M1 coupling				& \multicolumn{3}{c}{$\sim \alpha^2(\alpha/2)ea_0 \approx 5 \times 10^{-5} (\alpha/2)ea_0$}
\end{tabular}
\label{tab:m1calculations}
\end{table}
Some modern theoretical models that treat the high-order corrections important to non-$s$ wavefunctions report~\cite{sahoo2006lms} that the magnetic dipole matrix element $\langle 6S_{1/2} || M1 || 5D_{3/2} \rangle$ is larger (perhaps by a factor of 10) than expected due to enhanced electron-electron correlation effects, see Table~\ref{tab:m1calculations}.  We can test this assertion by measuring the ratio of the magnetic dipole and the much more well known electric quadrupole transition amplitudes in Ba$^+$.  The odd isotope $^{137}$Ba$^+$ offers a good opportunity for such a measurement because neighboring transitions at 2051~nm have different selection rules (see Table~\ref{tab:selectionRules}) that are illustrated in Table~\ref{tab:2micronTransitionsTable} and Figure~\ref{fig:m1E2AllowedTransitions}.

\subsection{Method 1: Probe M1, probe E2, calibrate laser}
A straightforward way to measure the relative rates is to independently drive the M1 allowed, E2 forbidden $6S_{1/2}, F = 1 \leftrightarrow 5D_{3/2}, F =0$ transition and the E2 allowed $6S_{1/2}, F = 1 \leftrightarrow 5D_{3/2}, F =1$ which is 150~MHz away at comparable rates.  If the laser intensity is kept constant (see, for example, Section~\ref{sec:lightShiftBeamStabilization}) over the measurement of Rabi frequencies on these transitions, the ratio of matrix elements directly results.  Or, the laser intensity can be measured accurately for each transition;  for instance, one might want to heavily attenuate the laser when driving the E2 transition.  To make sure the measurement is immune to jitter and drift in the laser frequency, the driving Rabi frequencies must be made much larger than the presumed linewidth of the laser.

Leaving out angular factors, two Rabi oscillation frequencies scale with the applied laser electric field $E$ as
\begin{align}
\frac{\Omega_{E2}}{E} &\sim \frac{12.742 ea_0}{\hbar} \left(\frac{a_0} {\lambda} \right) \approx 397  \left[\frac{Hz}{V/cm} \right] \\
\frac{\Omega_{E2}}{E} &\sim \frac{8 \times 10^{-4} (\alpha/2) ea_0}{\hbar} \approx 3.7 \left[\frac{Hz}{V/cm} \right]. \\
\end{align}

\subsection{Method 2: On-resonance M1, off-resonance E2}
Another method to measure the relative transition rates is to make the 2051~nm resonant with the weak M1 transition while monitoring for off-resonant transitions on the relatively much stronger E2 transition 150~MHz away.  If the M1 Rabi frequency can be made high enough, say 1 MHz, which is a challenging proposition, then off-resonant excitation of the E2 transition $6S_{1/2}, F = 1 \leftrightarrow 5D_{3/2}, F =1$ occurs with a comparable rate.  One can take advantage of the rf spin-state population movement scheme presented earlier in this Chapter to determine which $F'$ state is excited.  Then, scanning the laser across the $6S_{1/2}, F = 1 \leftrightarrow 5D_{3/2}, F=0$ resonance removes the need to calibrate the laser intensity.
\chapter{Towards a single-ion frequency standard}\label{sec:clockChapter}
\begin{quotation}
\noindent\small An airline suitcase is 9 inches tall.  So, a couple of inches for the atomic standard, and then a plate.  A couple of inches for the femtosecond comb, and then a plate.  A couple of inches for the rf switchyard and electronics.  Now \emph{that} would be \emph{some} suitcase!  \\ \flushright{---John L. (Jan) Hall, paraphrased, during a colloquium at the University of Washington, 5/17/2006.}
\end{quotation}
\begin{figure}
\centering
\includegraphics{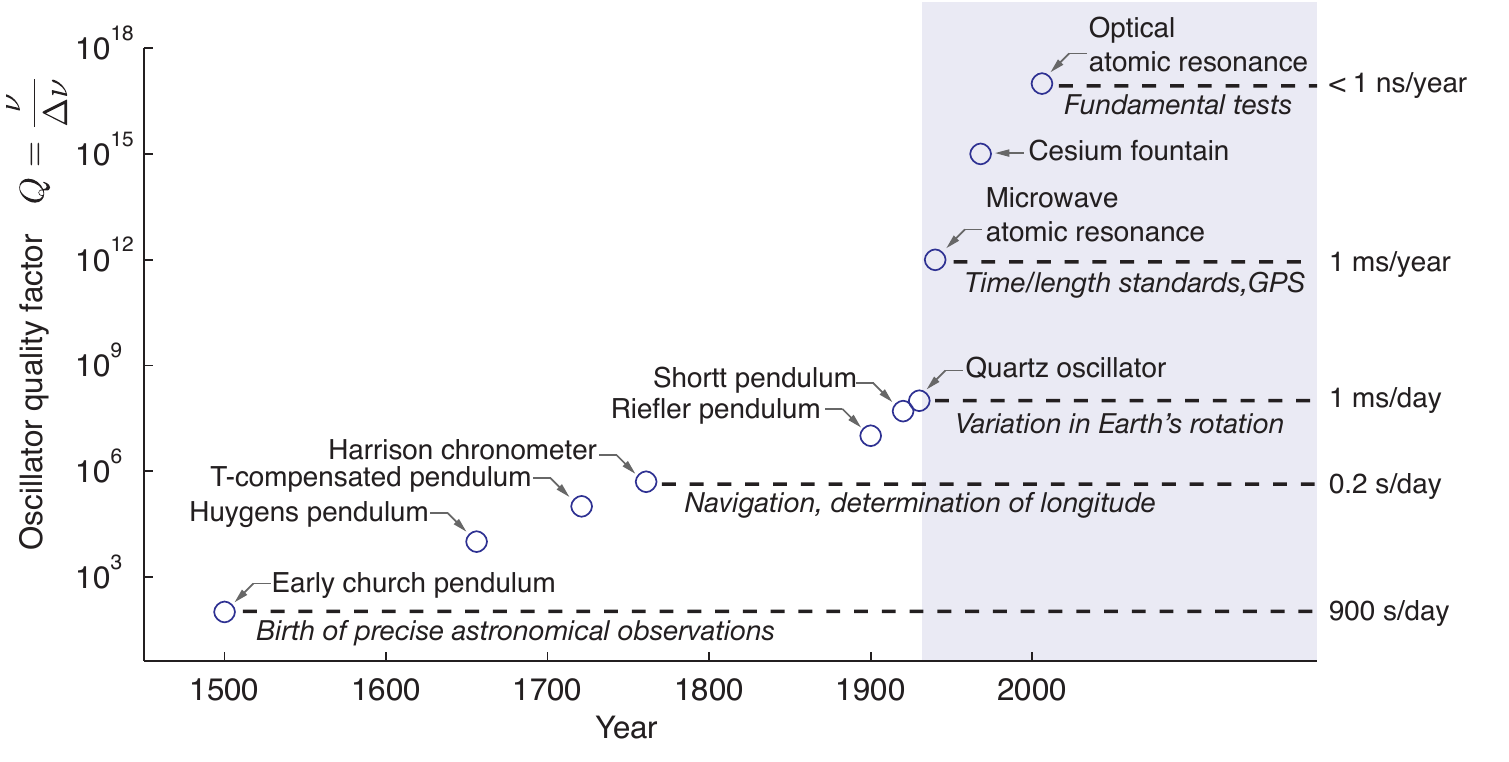}
\caption[Progress of frequency standards over four centuries]{A graphical depiction of the progress in frequency standards over four centuries, largely inspired by~\cite{riehle2004fsb}.  Historically, every incremental improvement of time keeping technology has resulted in new discoveries.  The birth of optical atomic frequency standards may offer new tests of fundamental physics, such as the temporal constancy of quantities such as the fine-structure constant $\alpha$.}
\label{fig:clockProgress}
\end{figure}

\begin{table}
\centering
\caption[Clocks of equal $Q$ are not equally good clocks for terrestrial time keeping.]{Clocks of equal $Q$ are not equally good clocks for terrestrial time keeping.}
\begin{tabular}{m{2 in}ccc}
							& Pulsar		& Cs-fountain	& Optical atomic clock \\ \hline \hline
\raggedright Oscillation frequency $\nu$		& 1 kHz		& 10 GHz		& $10^6$  GHz \\
\raggedright Quality factor $Q = \nu/\Delta \nu$	& $10^{15}$	& $10^{15}$	& $>10^{15}$ \\
\raggedright Time to detect a $\pi$ phase shift	 $t_\pi = \pi Q / \nu$	&  $\sim 10^5$ years & $\sim 1$ day & $\sim 1$ second
\end{tabular}
\label{tab:clockQcompare}
\end{table}

\section{Introduction:  optical frequency metrology}
An oscillator at frequency $\nu \pm \Delta \nu$ is said to have a quality factor
\begin{equation}
Q \equiv \frac{\nu}{\Delta \nu}
\end{equation}
which for classical damped oscillators is a measure of the number of cycles that elapse before the stored energy in the oscillator decreased to $e^{-1}$ of its original value.  Contrary to intuition, clocks of equal quality factor are not necessary equally good clocks for timekeeping.  Table~\ref{tab:clockQcompare} compares three of the best clocks observed in nature or created by man:  they all have similar quality factors $\sim 10^{15}$.  After examining this information, imagine trying to determine if a pulsar's rotation rate were slowing down or speeding up by using another pulsar's rotation rate as a reference.  From the data shown it would take $\sim 10^5$ years to resolve an evolution of $\pi$ phase between two such oscillators.  Note that the same phase shift can be detected in similar quality optical oscillators in just 1 second.

For metrology purposes, faster oscillators are better---and the fastest oscillators in the laboratory are optical. In 2006, the absolute frequency measurement of optical transitions became limited by the stability of the microwave standard that defines the second.  For instance, the absolute frequency measurement of a single Hg ion optical transition has a relative accuracy of $4 \times 10^{-16}$, while the total fractional uncertainty assigned to the Hg ion and apparatus itself is below $7.2 \times 10^{-17}$~\cite{oskay2006sao}.  Further, relative comparisons between different optical frequency standards should reach the same level of accuracy in orders of magnitude less averaging time.  This latest leap forward in timekeeping technology (see Figure~\ref{fig:clockProgress}) resulted from a long chain of advancements in stable lasers, neutral atom and ion trapping techniques, and the development of the femtosecond laser frequency comb, discussed later in this chapter, which netted its key inventors the 2005 Nobel prize in physics.

%MORE DETAIL HERE
%\begin{equation}
%\sigma_y^2(\tau) = \left\langle \sum_{i=1}^2 \left( \bar{y}_i - \frac{1}{2} \sum_{j=1}^2 \bar{y}_j \right)^2 \right\rangle = \frac{1}{2} \langle (\bar{y}_2 - \bar{y}_1)^2 \rangle.
%\end{equation}
%MORE DETAIL HERE

\section{The trapped ion frequency standard scheme} \label{sec:clockScheme}
\subsection{Overview}
\begin{figure}
\centering
\includegraphics[width = 5 in]{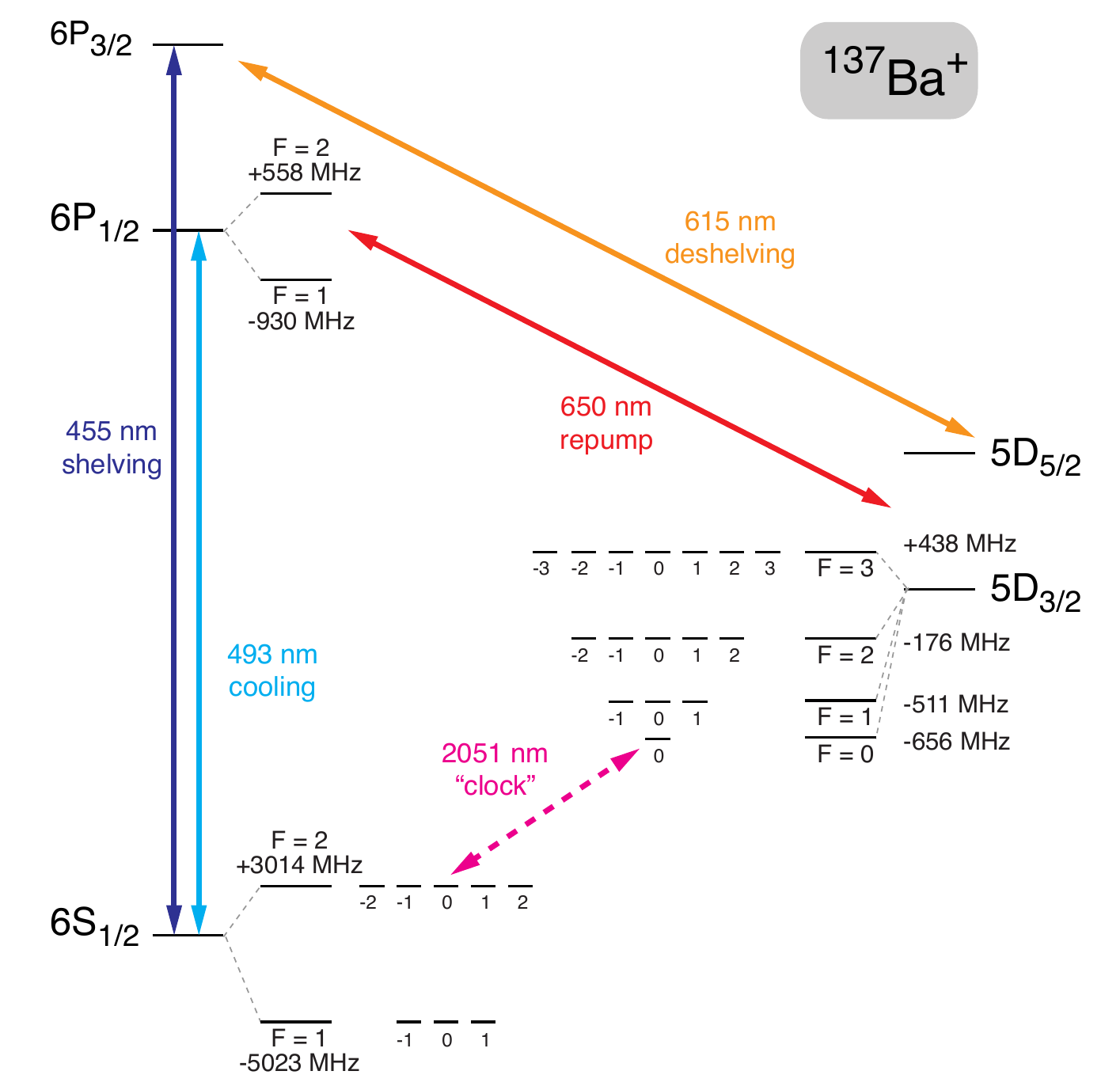}
\caption[$^{137}$Ba$^+$ energy level diagram with emphasis on the 2051~nm clock transition]{This $^{137}$Ba$^+$ energy level diagram shows the hyperfine structure of the $6S_{1/2}$, $6P_{1/2}$ and $5D_{3/2}$ states.  The 2051~nm electric-dipole forbidden ``clock'' transition $6S_{1/2}, F = 2, m = 0 \leftrightarrow 5D_{3/2}, F=0, m = 0$ is also highlighted.  The principle advantage of this transition compared to similar species is the lack of an electric quadrupole (or gradient) Stark shift in the excited state.  The long state lifetime $5D_{3/2}$ also implies a high ultimate quality factor $Q = \nu / \Delta \nu > 10^{16}$ given a sufficiently narrow laser.}
\label{fig:bariumEnergyLevelsClock}
\end{figure}
We view the laser as an oscillator that continuously measures out time with an electric field $E(t) = E_0 \cos(\omega t +\phi)$.  We could rely on these fast metronome beats of frequency $\omega$ to \emph{define} units of time only if we can be sure the laser output frequency is drift-free---that is, free from physical changes to the laser that are too small to control or measure.  Over short times a carefully constructed length standard used in a Fabry-Perot cavity stabilizes the laser:  the linewidth of the laser is narrowed and drift is suppressed.  Over long times the Fabry-Perot cavity itself is unreliable as a reference, due to length variation with temperature changes, vibrations, and material relaxation, so we use a narrow optical atomic resonance to correct it as often as possible.

We will show in this section that the short term \emph{stability} of the system as a timekeeper is dependent on a wide variety of physical effects that require rigorous attention:  thermal fluctuations, vibrations, engineering parameters in the feedback loop, noise on detectors, etc.  However, the ultimate \emph{accuracy} of the clock when averaged over long timescales should have none of these dependencies.  Instead, long-term drifts will only be due to fluctuating systematic energy level shifts in the atom which turn out to be more easily characterized and managed.  Further, we will see that once these atomic energy level shifts are under control, remaining long term drift in the clock frequency can be interpreted as temporal variations in the fundamental constants of nature itself.
 
\begin{figure}
\centering
\includegraphics{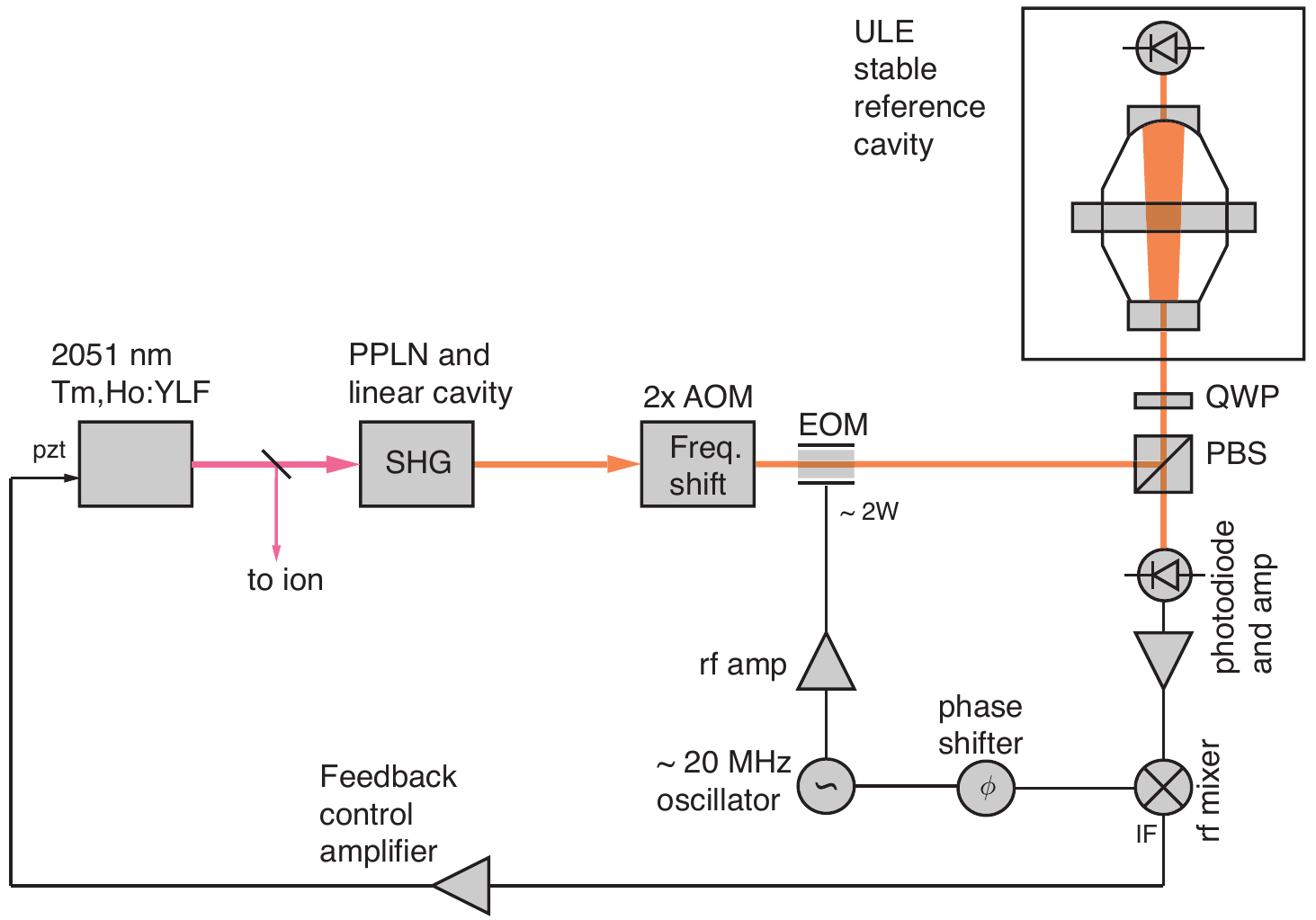}
\caption[An overview of the 2051~nm laser stabilization scheme]{An overview of the 2051~nm laser stabilization scheme.  First, a medium-finesse linear enhancement cavity is servo locked to the 2051~nm beam to increase the efficiency of an intra-cavity PPLN crystal phase matched for second-harmonic generation.  This source of coherently linked 1025~nm light is frequency shifted and sent to a very high finesse stable reference cavity.  The 2051~nm is servo-locked to this cavity using a high bandwidth Pound-Drever-Hall scheme in order to combat laser drift and to narrow the laser linewidth, hopefully to 10~Hz or below after mechanical isolation of the cavity.}
\label{fig:poundDreverHall}
\end{figure}

\begin{figure}
\centering
\includegraphics{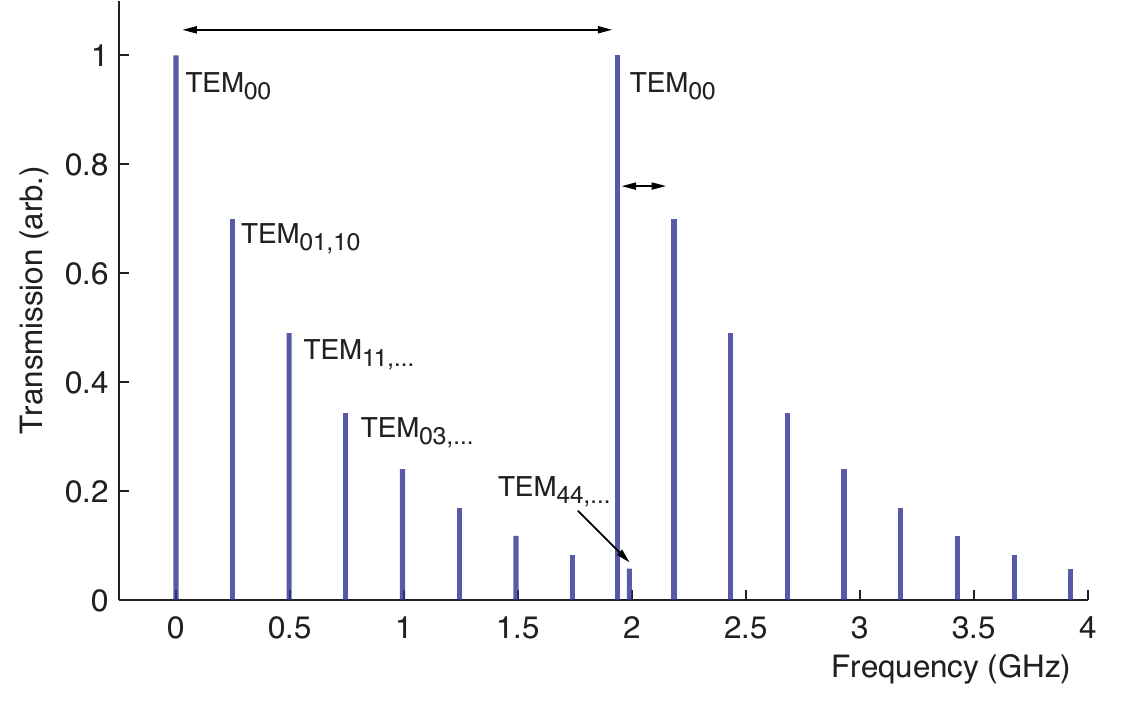}
\caption[Theoretical mode spectrum of a reference cavity at 1025~nm]{The theoretical mode spectrum of a Fabry-Perot reference cavity of length $L$ is dominated by the free-spectral range, or frequency difference between longitudinal TEM$_{n00}$ standing wave modes $\nu_\text{fsr} = c/2nL$.  Since the boundary conditions allow for higher order transverse mode structure, additional resonator modes TEM$_{nxy}$ occur at frequencies given by Eq.~\ref{eq:cavityModes}.}
\end{figure}
The frequency difference between cavity longitudinal modes is called the \emph{free spectral range} and is
\begin{equation}
\nu_\text{fsr} = \frac{c}{2nL}
\end{equation}
for a cavity of length $L$ filled with material assumed to have an index of refraction $n$ which is unity only in perfect vacuum.  As discussed in Appendix~\ref{sec:opticalCavities}, there are additional transverse modes allowed to resonate in the cavity.  These TEM$_{xy}$ modes occur at frequencies~\cite{yariv2006poe}
\begin{align} \label{eq:cavityModes}
\nu_{p,x,y} &= \frac{c}{2nL} \left[p + (x + y + 1) \frac{A}{\pi} \right] \\
\intertext{where}
A &= \cos^{-1} \left( \pm \sqrt{\left(1 - \frac{d}{R_1} \right) \left((1 -\frac{d}{R_2} \right)}\right).
\end{align}
In the expression for $A$, the plus sign is taken with both $(1 - d/R)$ factors are positive, the minus sign when both factors are negative.  For our reference cavity, shown in Figures~\ref{fig:referenceCavityDiagram} and~\ref{fig:referenceCavityPhoto},
\begin{align*}
d &= 77.500 \text{ mm} \\
R_1 &= \infty \quad \text{(flat mirror)} \\
R_2 &= 500 \text{ mm} \pm 3\% \\
\nu_\text{fsr}  &\approx 2.000 \text{ GHz}, \\
A \nu_\text{fsr} &\approx 250 \text{ MHz},
\end{align*}
meaning that useful frequency markers every 250~MHz are easily in range of an AOM frequency shift, and the preferred, dominant TEM$_{00}$ modes are within $\pm 1$~GHz, just within reach of a double-passed, high bandwidth AOM, or a broadband waveguide EOM.  The decision to specify one of mirrors flat is due to the manufacturer's claim that such a choice dramatically decreases the complexity and cost of a high finesse coating run.

\subsection{Reference cavity stability}
\begin{table}
\centering
\caption[Reference cavity material properties~\cite{riehle2004fsb}, figures of merit]{A table of cavity spacer material properties~\cite{riehle2004fsb} and figures of merit.  Ideally, one desires a long thermal time constant, low sensitivity to vibrations, and linear long-term relaxation behavior. Actual spacer design, mounting, and quality of manufacturing are just as important as the material choice.}
\begin{tabular}{m{2 in} l | c c c}
Quantity	& Units & Invar & ULE  & Zerodur \\ \hline \hline
Temperature coefficient $\alpha$ & $10^{-8}$ K$^{-1}$ & 150 & $\lesssim 0.3$ & $\lesssim 1$ \\
Young's modulus $E$	& $10^{9}$ N~m$^{-2}$ & 145	& 68  &  89 \\
Density $\rho$ &  $10^3$ kg m$^{-3}$ & 8.13 	&  2.21	& 2.52 \\
Specific heat $c_\rho$ & J (kg K)$^{-1}$ & 500	& 0.77	& 0.81 \\
Heat conductivity $\lambda$ & W (m K)$^{-1}$ & 10.5	& 1.31 & 1.63 \\ \hline
Thermal time constant figure of merit (per unit of cross-sectional area) & $\displaystyle \frac{\lambda}{\rho c_\rho A}$ & $3 \times 10^{-6}$  & $8 \times 10^{-4}$ & $8 \times 10^{-4}$ \\
Sensitivity to accelerations (per unit length) & $\displaystyle \frac{L \rho}{E}$& $6 \times 10^{-8}$  & $3 \times 10^{-8}$ & $3 \times 10^{-8}$ \\
Long term relaxation characteristics & & \multicolumn{3}{c}{Not quantified here}
\end{tabular}
\label{tab:refCavityMaterials}
\end{table}
How stable are the cavity mode frequencies?  Astonishingly, researchers have constructed reference cavities that are stable to parts in $10^{16}$ over 1~second time intervals~\cite{young1999vls} which roughly translates into fluctuations in the cavity spacer length $L \sim 1$~fm or about the width of a proton.  In this section we discuss the mechanisms for cavity $L$ instability and fluctuation:  material creep, temperature fluctuations, and coupling to mechanical vibrations.  Usually design constraints pull us in opposing directions.  For instance, passive temperature stability is best achieved with a long massive cavity while immunity to likely sources of vibration ($\sim 3 \times 10^{-9}$~m/$\sqrt{\text{Hz}}$ in a bandwidth of 1--30~Hz~\cite{hall2006dmo}) is highest for small light cavities.

Table~\ref{tab:refCavityMaterials} gives the physical properties of several candidate cavity spacers reproduced from~\cite{riehle2004fsb}.  A chief performance metric is the fractional length change given a change in temperature.  Manufacturers design special glasses (such as ULE or Zerodur) to have extremely small linear temperature coefficients (ideally zero at some temperature near room temperature).  Given a cross sectional area $A$, heat conductivity $\lambda$, specific heat $c_\rho$, and density $\rho$, the combination $\lambda / \rho c_\rho A$ is a figure of merit for how quickly the material responds to thermal fluctuations.  Given accelerations $a$, the fractional length change of a cavity is small if it is small, light, and stiff---a figure of merit is the combination $\rho L / E$ where $E$ is the Young's modulus of the material.

The observed stability of reference cavities in practice is often limited by jitters in the cavity length caused by laboratory vibrations.  Researchers spend considerable effort building passive and active vibration canceling systems as well as optimizing the mounting~\cite{notcutt2005sac,nazarova2006vir}, and shape~\cite{chen2006vie,webster2007vio} of the cavity.

\subsection{Frequency doubling of 2051~nm} \label{sec:frequencyDoubling2micron}
\begin{table}[p]
\centering
\caption[Candidate crystal parameters for SHG of 2051~nm]{Candidate crystal parameters for SHG of 2051~nm.  $d_\text{eff}$ is the effective nonlinear Kerr coefficient due to a beam at the phase-matching angle $\theta_\text{pm}$.  $n_{x,y,z}$ are refractive indices, $\delta_\text{walkoff}$ is the angle at which generated light diverges from the fundamental beam, and $\eta_\text{crystal}$ is the figure of merit for each crystal, described in the text.  PPLN stands for periodically poled lithium niobate.  The figure of merit for PPLN differs from Eq.~\ref{eq:SHGfigureOfMerit} by $(2/\pi)$ due to the use of quasi-phase matching~\cite{sutherland2003hno}.}
\begin{tabular}{l|cccc|c}
Crystal & $|d_\text{eff}|$ (pm/V) & $n_x,n_y,n_z$ & $\theta_\text{pm}$ & $\delta_\text{walkoff}$ (mrad) & $\eta_\text{crystal}$ \\ \hline\hline
KNbO$_3$  	& 5.84 & 2.245, 2.245, 2.252 & 42.4$^\circ$ & 62.75 & 2.99 \\
LiNbO$_3$	& 3.99 & 2.263, 2.263, 2.260 & 44.6$^\circ$ & 35.92 & 1.36 \\
AgGaS$_2$	& 13.7 & 2.446, 2.446, 2.525 & 58.4$^\circ$ & 19.17 & 12.32 \\
Ag$_3$AsS$_3$ & 25.3 & 2.798, 2.798, 2.936 & 30.8$^\circ$ & 79.74 & 28.17 \\
GaSe		& 55.2 & 2.783, 2.783, 2.903 & 22.2$^\circ$ & 99.90 & 140.69 \\ \hline \hline
PPLN		& $\approx$ 21.9  & 2.263, 2.263, 2.260 & --- & --- & 26.38
\end{tabular}
\label{tab:2051crystalSummary}
\end{table}
Because high finesse mirror coatings could be made commercially 10 times better at 1025~nm than at 2051~nm, and because the available acousto-optic and electro-optic modulators at 2051~nm were quite limited, we made the decision to solve the second harmonic generation (SHG) problem for our clock laser and specify that high finesse cavity mirrors be coated for the shorter wavelength.  Though we had no experience in the doubling of sources this far in the IR, we also stand to gain another factor of 2 in relative stability since we would interrogate the ion at half the frequency at which we stabilize the laser. We considered several nonlinear crystals, some of which are described in Table~\ref{tab:2051crystalSummary}.  The figure of merit for a candidate crystal is the quantity
\begin{equation} \label{eq:SHGfigureOfMerit}
\eta_\text{crystal} = \frac{d_\text{eff}^2}{n_\omega^2 n_{2 \omega}}
\end{equation}
since the efficiency of the SHG process depends explicitly on exactly these crystal parameters.  The quantity $d_\text{eff}$, often expressed in units of pm/V, is the effective second-order Kerr coefficient experienced by a beam traveling at the phase matching angle through the crystal lattice.  The quantities $n_\omega$ and $n_{2 \omega}$, for our Type-I phase matched application, are the indices of refraction for $o$-polarized fundamental radiation and $e$-polarized second harmonic radiation respectively.

\begin{table}[p]
\centering
\caption[Our PPLN and AgGaS$_2$ crystal properties]{Our PPLN and AgGaS$_2$ crystal specified and measured properties.}
\begin{tabular}{m{1.5 in}|cc}
				& \parbox{1.5 in}{\centering Periodically poled lithium niobate (PPLN)} & Silver thiogallate (AgGaS$_2$) \\ \hline \hline
Length			& 40~mm & 8~mm \\
Width $\times$ Height		& 0.5~mm per poling $\times$ 0.5~mm & 3~mm $\times$ 3~mm \\
Crystal cut		& z-cut	& 58.4$^\circ$ \\
\multirow{3}{1.5 in}{Phase-matching} & 30.50~$\mu$m poling at $\approx$55$^\circ$  & \multirow{3}{1.5 in}{\centering Type-I ($ooe$) angle-tuned} \\
	& 30.25~$\mu$m poling at $\approx$105$^\circ$ & \\
	& 30.00~$\mu$m poling at $\approx$155$^\circ$ & \\
\multirow{2}{1 in}{Coatings} & $<$ 1\% at 2051~nm & 0.07\% at 2051~nm \\
					& $<$ 2\% at 1025~nm & 0.1\% at 1025~nm \\
\raggedright Bulk absorption (2051~nm) & $< 10^{-2}$~cm$^{-1}$ & $\approx 10^{-1}$~cm$^{-1}$ \\
\raggedright Single-pass SHG efficiency $\eta$ & $2.5  \times 10^{-4}$ at 60~mW & $1.3  \times 10^{-6}$ at 60~mW
\end{tabular}
\label{tab:2051nmCrystalProperties}
\end{table}
Due to commercial unavailability of quality cuts of some the strong candidates, such as gallium selenide (GaSe) and proustite (Ag$_3$AsS$_3$), we obtained instead a custom cut of silver thiogallate (AgGaS$_2$) and a commercially mass-produced periodically poled lithium niobate (PPLN, sold as Thorlabs OPO-3).  Table~\ref{tab:2051nmCrystalProperties} lists their properties.

The single-pass SHG efficiencies of both crystals proved too low to provide the $\sim 100$ $\mu$W of 1025~nm necessary to comfortably lock the clock laser to our ULE reference cavity, though some extrapolations from examples of similar SHG techniques in the literature (e.g.~\cite{kim2005sfc}) seemed to indicate that the PPLN should have met our needs.  The unanticipated large bulk absorption of the silver thiogallate made it a poor candidate for use in a resonant buildup cavity and the large size of the PPLN crystal and accompanying oven dissuaded us from using it in a bow-tie enhancement cavity.  In the end, we designed and constructed a linear Fabry-Perot type cavity around the PPLN crystal and oven to successfully achieve sufficient 1025~nm power.

\begin{figure}
\centering
\includegraphics[width=6in]{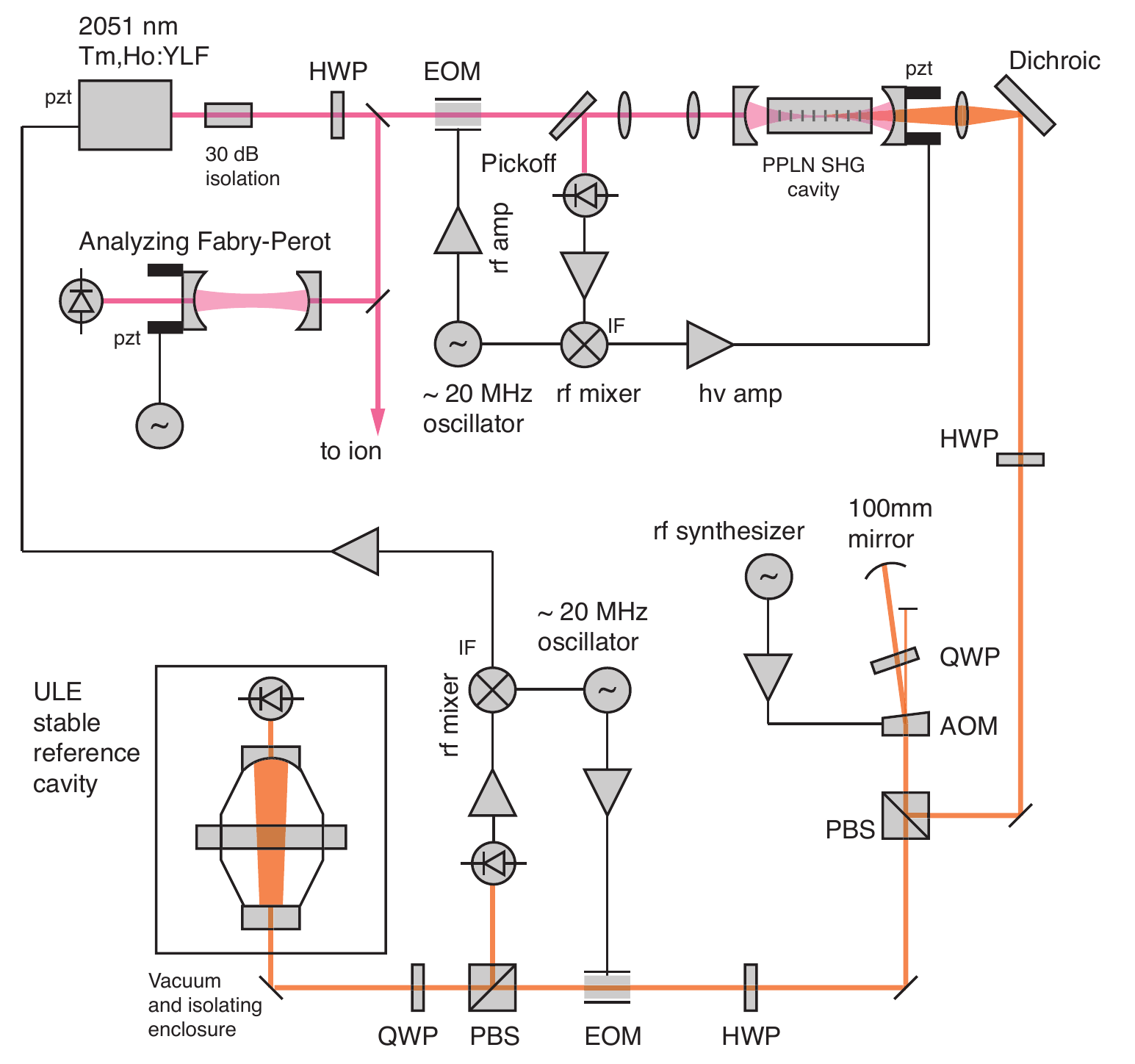}
\caption[Detail of the stabilized 2051~nm clock laser apparatus]{A more detailed layout of Figure~\ref{fig:poundDreverHall}, we show a cavity enhanced frequency doubling stage (see Figure~\ref{fig:2051nmSHGSchematic}) using a PPLN nonlinear crystal to generate 1025~nm light.  This beam is frequency shifted by a wideband AOM in double-pass configuration and sent to the vertically mounted stable reference cavity to derive an error signal which is fed back to the 2051~nm in order to stabilize it.}
\label{fig:clockLaserApparatusClockChapter}
\end{figure}

\begin{figure}
\centering
\includegraphics{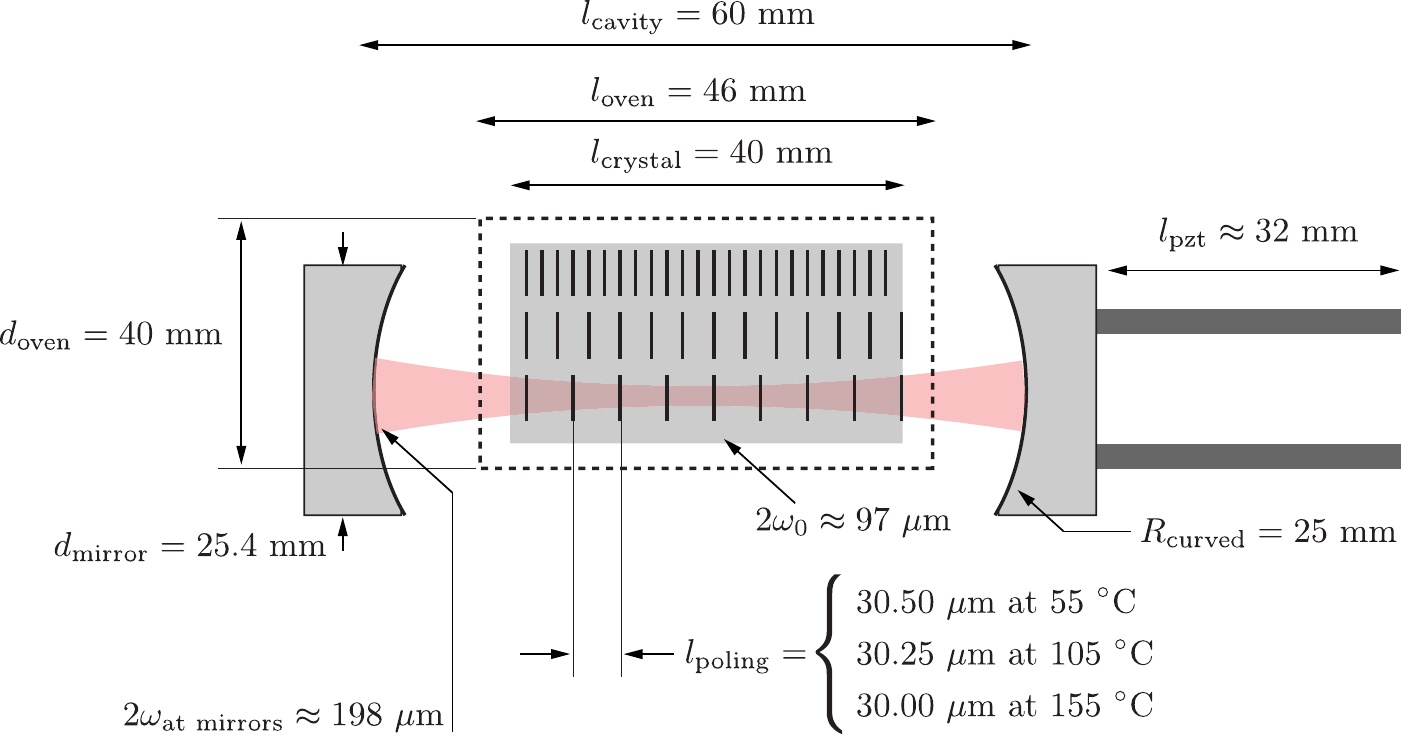}
\caption[Schematic of a linear enhancement cavity for SHG of 1025~nm]{Schematic for a linear enhancement cavity with a periodically-poled lithium niobate (PPLN) crystal for second-harmonic generation (SHG) of 1025~nm light.  The crystal is phase matched along one track of poling by tuning the oven-controlled temperature.}
\label{fig:2051nmSHGSchematic}
\vspace{1.0 in}
\includegraphics{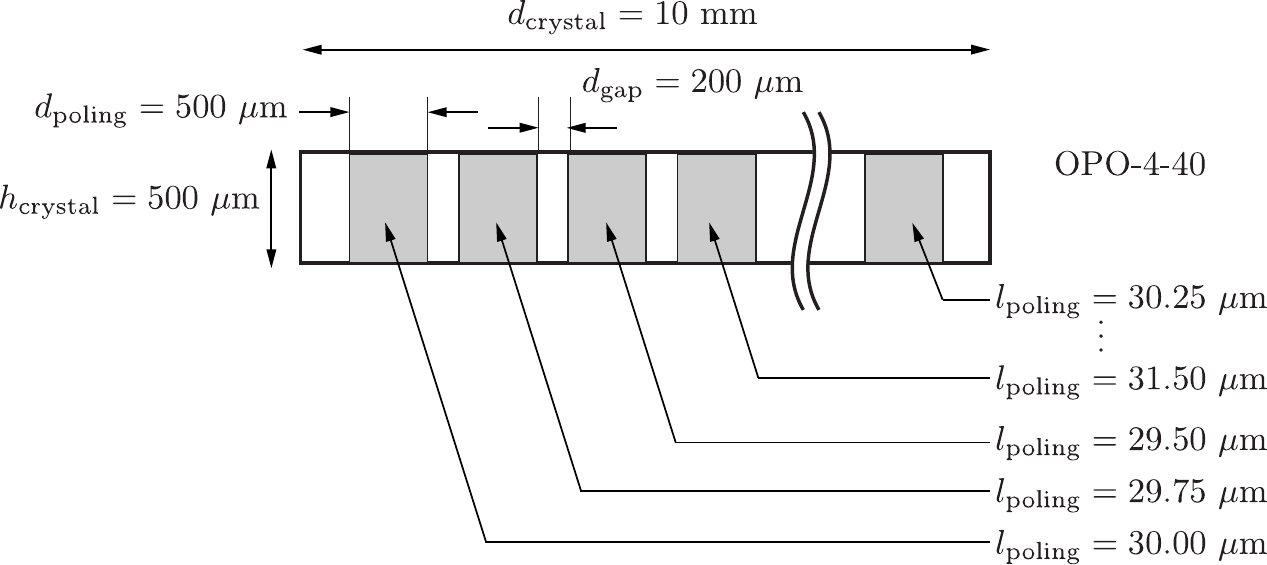}
\caption[Detail view of PPLN crystal for 1025~nm SHG]{Detail view (head-on) of the PPLN crystal for 1025~nm SHG, sold as Thorlabs OPO-4-40.  There are a total of nine poling periods included on the crystal, three of which are useful for temperatures above room temperature but below 200 $^\circ$C.}
\label{fig:2051nmPPLNDetail}
\end{figure}

\begin{figure}
\centering
\includegraphics{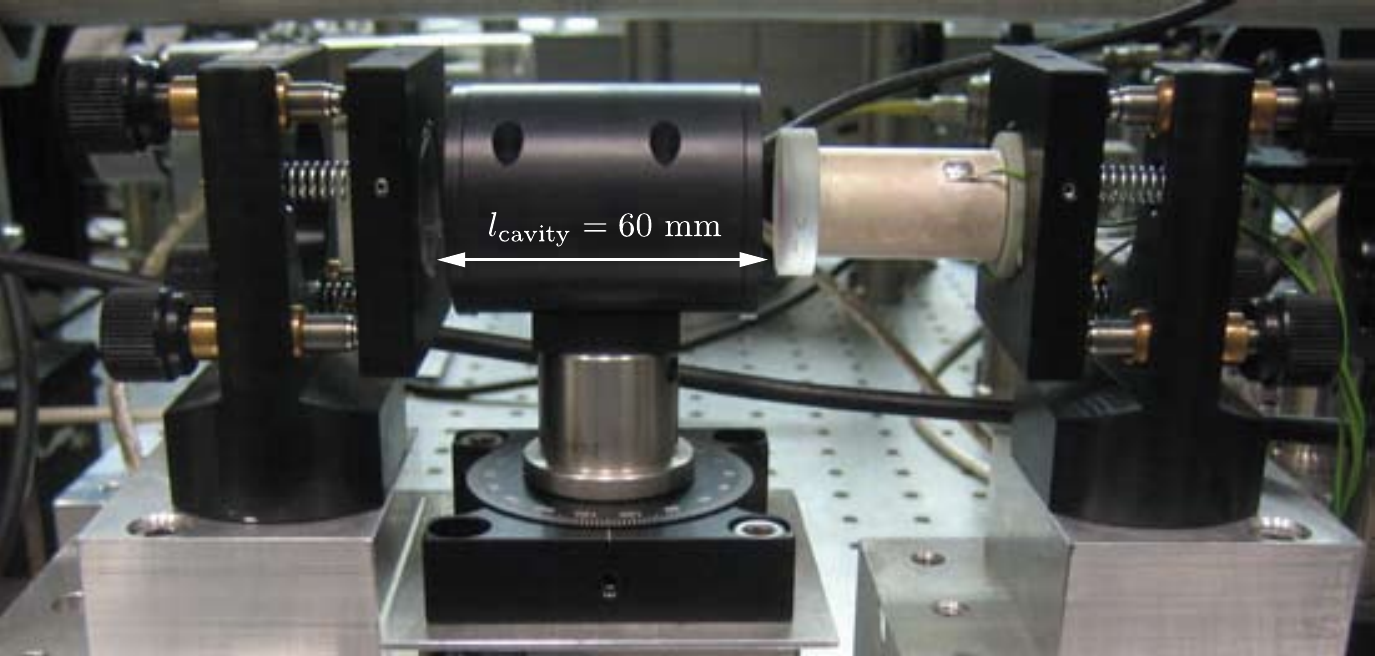}
\caption[Photograph of the linear SHG enhancement cavity for 2051~nm conversion]{Photograph of the linear SHG enhancement cavity for 2051~nm conversion, shown schematically in Figure~\ref{fig:2051nmSHGSchematic}. The cylindrical oven is mounted on a series of stiff translation stages that allow rotation perpendicular to the beam axis, translation in and out of the cavity (to select poling tracks), and adjustment of the crystal height above the table.}
\label{fig:2051nmSHGPhoto}
\end{figure}

\begin{table}
\centering
\caption[Beam parameters for selected 1025~nm SHG cavity lengths]{Beam parameters for selected 1025~nm SHG cavity lengths within the stability regime.  The optimum waist size~\cite{boyd1968pif} for a our crystal is $2 \omega_0 = 96~\mu$m.}
\label{tab:PPLNbeamParameters}
\begin{tabular}{c | c c}
	&					\multicolumn{2}{| c}{Beam sizes~[$\mu$m]} \\
Cavity length $d$~[mm]	& At waist $2\omega_0$ & At mirrors $2\omega$\\ \hline \hline
50	&	105	&  162 \\
55	& 	102	&  177 \\
60	& 	97	&  198 \\
65	&	86	&   247 \\
70	&	55	&  394
\end{tabular}
\end{table}

Figure~\ref{fig:2051nmSHGSchematic} shows the linear enhancement cavity design for SHG of 1025~nm using PPLN.  The mirror radii of curvature are 2.5~cm, and their separation is nominally 60~mm.  The cavity would be an unstable resonator without the presence of the 4~cm PPLN crystal whose index of refraction is 2.263.  The optimum waist size~\cite{boyd1968pif} for a 4~cm crystal is $2 \omega_0 = 96~\mu$m.  The spot sizes at the mirrors is thus $2 \omega_\text{mirrors} \approx 200~\mu$m.  Table~\ref{tab:PPLNbeamParameters} shows beam size information for selected cavity lengths within the stability regime.  One end of a PZT tube (1.25 in.\ long, 0.75 in.\ outer diameter) is affixed to one mirror specified for 99.5\% reflection; the other is glued to an annular optic mount adapter made from G10.  The other mirror serves as an input coupler with $R = 98(1)$\%.  Both mirrors are 1 in.\  in diameter, 0.375 in.\ thick, and were manufactured by CVI.  As shown in Table~\ref{tab:2051nmCrystalProperties}, three of the crystal's poling tracks are available to us at reasonable oven temperatures.  Each seem to give similar power output, but we will likely operate at the lower temperature options to minimize air currents around the oven and evaporation of oil or other contaminants onto the mirror surfaces located only millimeters away from the oven faces, as shown in Figure~\ref{fig:2051nmSHGPhoto}.

Figure~\ref{fig:2051nmSHGSchematic} also details the Pound-Drever-Hall feedback system (see Appendix~\ref{sec:opticalCavities} or~\cite{black2001ipd} for a review) we employ to lock the cavity length to track the laser frequency.  A homemade 1.75~ in.\ long LiNbO$_3$ transverse electro-optic modulator (see Section~\ref{sec:phaseModulation}) puts phase modulation sidebands on the beam at $\omega_m / 2\pi \approx 30$~MHz.  The reflected 2051~nm beam is sampled with a wedged plate instead of a PBS/QWP combination to obtain optimal 1025~nm power out of the doubling stage.  We are currently investigating other types of feedback control options that will do away with the need to modulate the laser beam.  For instance, dithering the linear cavity PZT and phase sensitively detecting the modulated 1025~nm output should allow for stable locking while the resulting amplitude noise can be removed at a later time.

\subsection{Locking to the 1025~nm reference cavity}
The 1025~nm beam is frequency shifted and sent to the stable reference cavity where we generate a cavity/laser frequency error signal using the Pound-Drever Hall technique (again, see Appendix~\ref{sec:opticalCavities} or~\cite{black2001ipd} for a review).  The error signal is amplified through a PI (proportional and integration gain) stage and fed back to the 2051~nm laser's PZT frequency control, closing the loop and locking the laser to the cavity.  We do not expect the presence of the second harmonic generation to affect the lock since the free-running noise bandwidth of the 2051~nm laser is $\sim 10$~kHz (see Section~\ref{sec:clockLaserApparatusChapter}) which is far below the acceptance bandwidth (and response) of the locked doubling cavity.  In this section we discuss the technical challenges in accomplishing the fast photodiode and amplifier chain involved in completing the Pound-Drever-Hall lock.  In summary, these design steps are:
\begin{itemize}
\item Understanding the detector, a photodiode, and its properties which limit performance (SNR and bandwidth),
\item Designing an optimum transimpedance amplifier to convert the detected photocurrent to a voltage for further amplification,
\item Mixing the detected signal with a reference oscillator to form an error signal,
\item Designing a feedback control amplifier that servos the laser frequency to null fluctuations with respect to the optical cavity.
\end{itemize}

\subsubsection*{The detector:  a fast photodiode}
\begin{table}
\centering
\caption[Fast Si and InGaAs photodiode data for signals at 1025~nm and 2051~nm]{Fast photodiode data for a detection of Pound-Drever-Hall signals at 1025~nm and 2051~nm.  Here we choose Hamamatsu models S8223 and G8422-03 for example design constraints.}
\begin{tabular}{p{1.5in} | p{2 in}p{2 in}}
& Si photodiode S8223 & Long-$\lambda$ InGaAs G8422-03 \\ \hline \hline
Photosensitivity	&  $\approx 0.25$~A/W at 1025~nm	& 1.1 A/W at 2051~nm \\
Active size		&  $0.8$~mm dia. & 0.3~mm $\times$ 0.3~mm \\
\multirow{2}{*}{Dark current}		& $<1$~nA at $V_R = 5$~V  & $\sim 30$~nA at $V_R = 0$~V \\
							& $<1$~nA at $V_R = 10$~V & 55--550~nA at $V_R = 1$~V\\
\multirow{3}{*}{Diode capacitance}	& 8~pF at $V_R = 0$~V & 20~pF at $V_R = 0$~V   \\
							& 3~pF at $V_R = 5$~V & 8~pF at $V_R = 1$~V \\
							& 2~pF at $V_R = 10$~V&  \\
Noise equivalent power &	  $3.4 \times 10^{-15} \text{ W}/\sqrt{\text{Hz}}$ at $V_R = 5$~V & $1.5 \times 10^{-13} \text{ W}/\sqrt{\text{Hz}}$ at $V_R = 1$~V \\
Other notes	& Can be highly reversed biased, inexpensive. & Maximum reverse bias voltage is 2V, very expensive, can be cooled with a TEC to achieve better noise performance.
\end{tabular}
\label{tab:photodiodeData}
\end{table}
We cannot send too much optical power to the reference cavity.  Due to its extremely high finesse, the large intracavity powers that build up can cause thermal oscillations.  Other researchers recommend 10--100 $\mu$W of optical power.  Since we need to detect this small light level at several MHz, it is essential that we choose a proper photodiode.  Such a device should have peak sensitivity near 1025~nm, and be physically small to ensure good speed and noise characteristics.  For example design constraints, we refer to the Hamamatsu S8223 data in Table~\ref{tab:photodiodeData}.

\begin{figure}
\centering
\includegraphics{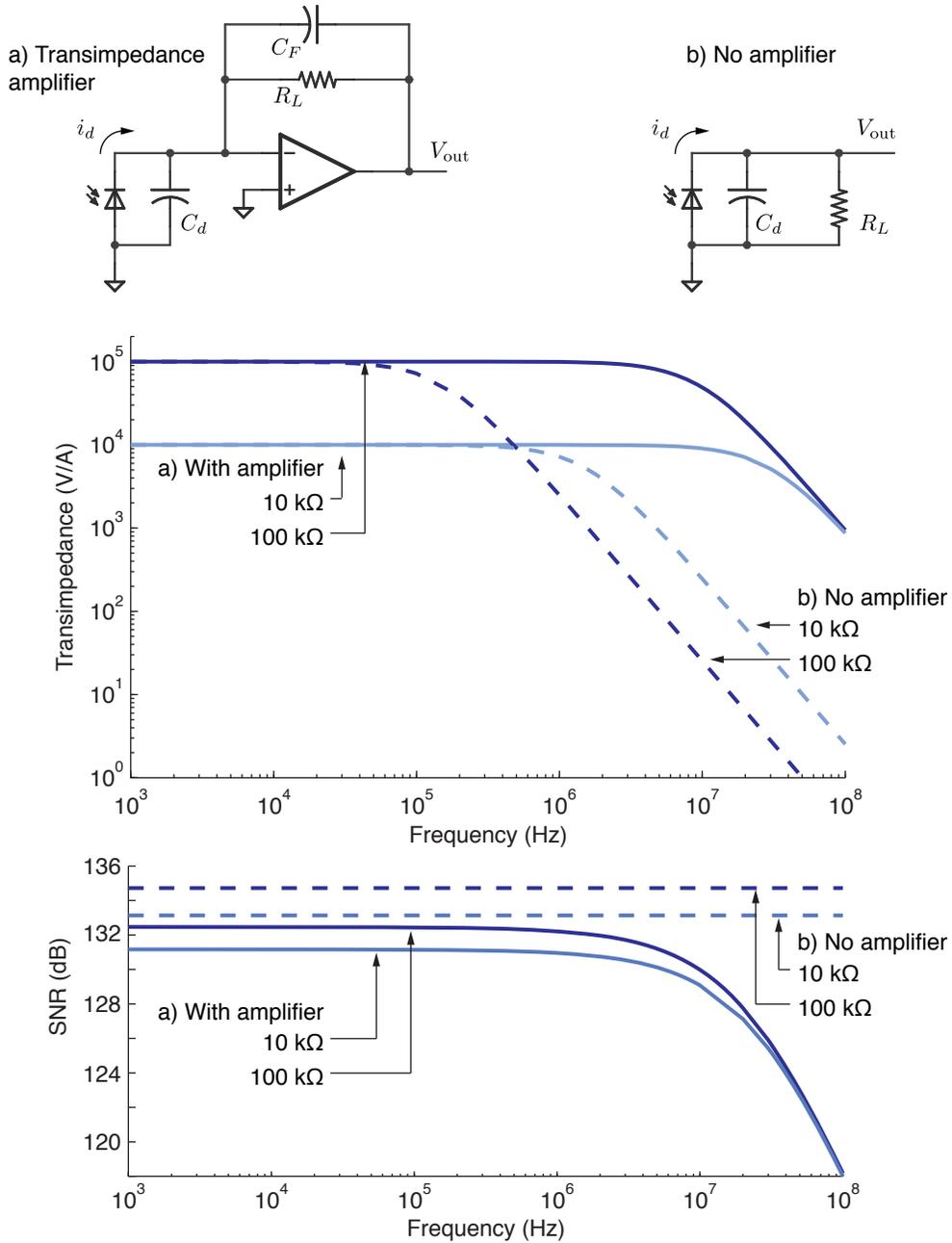}
\caption[Bandwidth and signal to noise tradeoff in the transimpedance problem]{There is always a trade-off between signal-to-noise and bandwidth in the transimpedance problem.  Here we the simulated (transimpedance) signal gain for two load resistors with and without amplifiers.  Though the amplifiers extend the bandwidth considerably, they do so at the cost of noise, especially amplifier voltage noise which increases at high frequency.}
\label{fig:transimpedanceCompare}
\end{figure}

The reflected beam of the cavity is sent to a fast silicon photodiode for detection.  The factor that traditionally limits the performance of photodiode detection schemes is the diode capacitance $C_d$.  For instance, simply connecting a (properly reversed biased) diode with $C_d = 3$~pF to a 1~M$\Omega$ resistor yields a 3~dB bandwidth of 50~kHz.  This isn't acceptable since the signal of interest in the Pound-Drever-Hall scheme is at several MHz.  However, lowering the transimpedance resistor, say to 1~k$\Omega$, throws away signal while \emph{increasing} current noise in the resistor since $i_N = \sqrt{4 kT / R}$.  A compromise~\cite{hobbs2000beo} is to choose a resistor such that the photocurrent drops about 0.2~V;  then the thermal noise power is 1/4 the shot noise, and only $\approx 1$~dB in SNR is lost.

For our application, shown in Figure~\ref{fig:transimpedanceCompare}, we assume an optical power of 10~$\mu$W corresponding to a photocurrent of $I_p \approx 2.5~\mu$A.  The transimpedance load resistor $R_L = 100$~k$\Omega$ succeeds in developing 0.25~V.  However, the bandwidth is still too small:
\begin{equation}\label{eq:rcBandwidth}
f_{RC} = \frac{1}{2 \pi R_L C_d} \sim 500 \text{ kHz} 
\end{equation}
while we need to operate at $\sim$10~MHz.  A transimpedance amplifier (in the form of an operational amplifier) helps tremendously by employing huge intrinsic gain to reduce the voltage swing across $C_d$.  The use of negative feedback with the operational amplifier assures us that the effective gain will be set by reliable, well characterized passive components in the feedback network, rather than the intrinsic gain of the amplifier itself.

\subsubsection*{The transimpedance amplifier:  a fast op amp}
By connecting the load resistor $R_L$ across a transimpedance amplifier as shown in Figure~\ref{fig:transimpedanceCompare}, the photodiode works into a virtual ground, reducing the voltage swing across the diode capacitance $C_d$ and increasing the detection bandwidth.  The price we pay is increased noise from the amplifier, as well as input capacitance that is added to $C_d$.  An additional design requirement emerges:  the amplifier must be kept stable by appropriate feedback compensation.

Operational amplifiers usually feature specified gain-bandwidth products (GBP), approximately equal to the unity-gain crossover frequency $f_T$.  This frequency sets an upper limit on the bandwidth of the amplifier in unity-gain mode for stability.  A rule of thumb~\cite{hobbs2000beo} is that the closed-loop bandwidth in our transimpedance application is extended from Eq.~\ref{eq:rcBandwidth} to
\begin{equation}
f_{-3\text{dB}} \approx \frac{\sqrt{f_T f_{RC}}}{2}.
\end{equation}
If we consider using a AD829 op-amp, for instance, $f_T = 600$~MHz, so
\begin{equation*}
f_{-3\text{dB}} \approx \frac{\sqrt{f_T f_{RC}}}{2} \approx 17 \text{ MHz}
\end{equation*}
which extends to our frequency of interest of several MHz.

One prevents the amplifier from oscillating by robbing it of gain at high frequencies when phase shifts might accumulate to 180$^\circ$.  Two options~\cite{hobbs2000beo} that result in putting the necessary zero in the op-amp transfer functions are placing a feedback capacitor in parallel with $R_L$,
\begin{equation}
C_f = \frac{1}{2\pi R_L \sqrt{f_T f_{RC}}},
\end{equation}
or putting a resistor in series with the photodiode,
\begin{equation}
R_s = \frac{1}{2 \pi C_d \sqrt{f_T f_{RC}}}.
\end{equation}
Since component variation and device construction techniques alter the actual capacitance in the circuit,  oscillations must be searched for and snubbed out empirically.

When choosing amplifiers, one must carefully consider noise properties.  The input current noise of the amplifier $i_{N_\text{amp}}$, Johnson current noise on the feedback resistor $i_{N_\text{th}}$, and shot noise of the photocurrent $i_N$ are all amplified in parallel with the photocurrent $i_d$.  They all receive the same transimpedance gain and therefore don't affect the SNR in a frequency dependent way.  The one noise source that \emph{does} is the amplifier's input voltage noise $e_\text{amp}$.  This noise is transferred to the output with a factor of the closed loop gain~\cite{hobbs2000beo}
\begin{equation}
A_\text{V,cl} = \frac{A_\text{V,ol}}{1 + A_\text{V,ol}/(1 + i \omega C_d Z_F)}
\end{equation}
which begins to rise at the frequency $f_{RC}$.  Notice the falling SNR in \ref{fig:transimpedanceCompare} for the amplified case compared to the (theoretically) constant SNR exhibited by the simple load resistor.

\subsubsection*{Mixing to obtain the Pound-Drever-Hall error signal}
From the review of the Pound-Drever-Hall scheme in Appendix~\ref{sec:opticalCavities}, we recall that the error signal is obtained by detecting the phase of the interference signal detected in the optical reflection.  An rf mixer becomes our phase sensitive detector, as shown in Figure~\ref{fig:clockLaserApparatusClockChapter}.  An optional phase shifter placed between the reference oscillator and the mixer allows us to isolate the dispersive (rather than absorptive) part of the detected signal.  After the transimpedance amplifier, optional low noise radio frequency amplifiers help to bring the detected signal up to the operating point of the mixer, about +7~dBm.  It is useful to amplify before the mixer at high frequencies, rather than after at baseband, since by doing the former one avoids many sources of technical noise.

\subsubsection*{A feedback control system to servo the laser}
After mixing, we low-pass filter the signal to suppress mixer image frequencies.  The result is the detected error signal;  it is nearly ready to be fed back to the laser for closed-loop locking.  At this point, however, it is worthwhile to consider that fluctuations at low frequency will have the largest absolute effect on the laser frequency.  For instance, a 1~mK change in temperature (a slow fluctuation, to be sure) will move the laser frequency by over 1~MHz.  Also, most vibrations in the laboratory only have significant power below 10~kHz.  The ansatz is that we would prefer to have high gain at low frequencies at the expense of smaller gain at high frequencies where noise frequencies aren't as important.

From a practical standpoint, since we intend to servo the laser using a PZT element, our servo bandwidth is constrained by the presence of mechanical resonances which bring with them phase shifts~\cite{bass1994hoo}.  We have measured our laser's first PZT resonance at 112~kHz by constructing an ac-voltage divider with it and a resistor and observing the characteristic high impedance at resonance.  Given this high a resonance frequency, a control bandwidth of 10~kHz seems within reason.  Our estimate of the laser's free-running linewidth is also $\lesssim 10$~kHz, so this control bandwidth should be sufficient to hold the laser to the optical cavity and significantly reduce its short-term linewidth.

\begin{figure}
\centering
\includegraphics{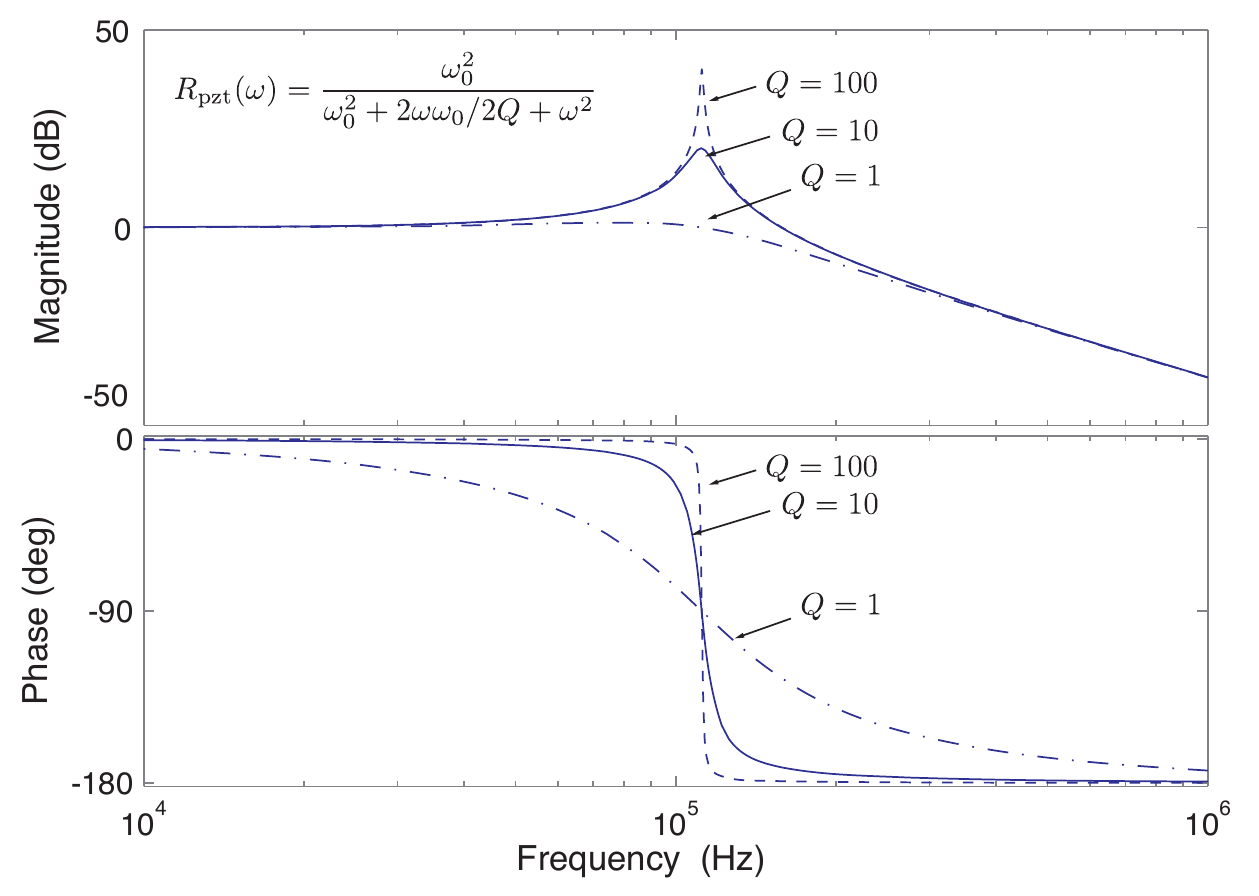}
\caption[The transfer function of a PZT impacts feedback control design]{The transfer function of a PZT impacts feedback control design due to at least one mechanical resonance which increases gain while simultaneously introducing a -180$^\circ$ phase shift.}
\label{fig:pztResonanceBodePlots}
\includegraphics{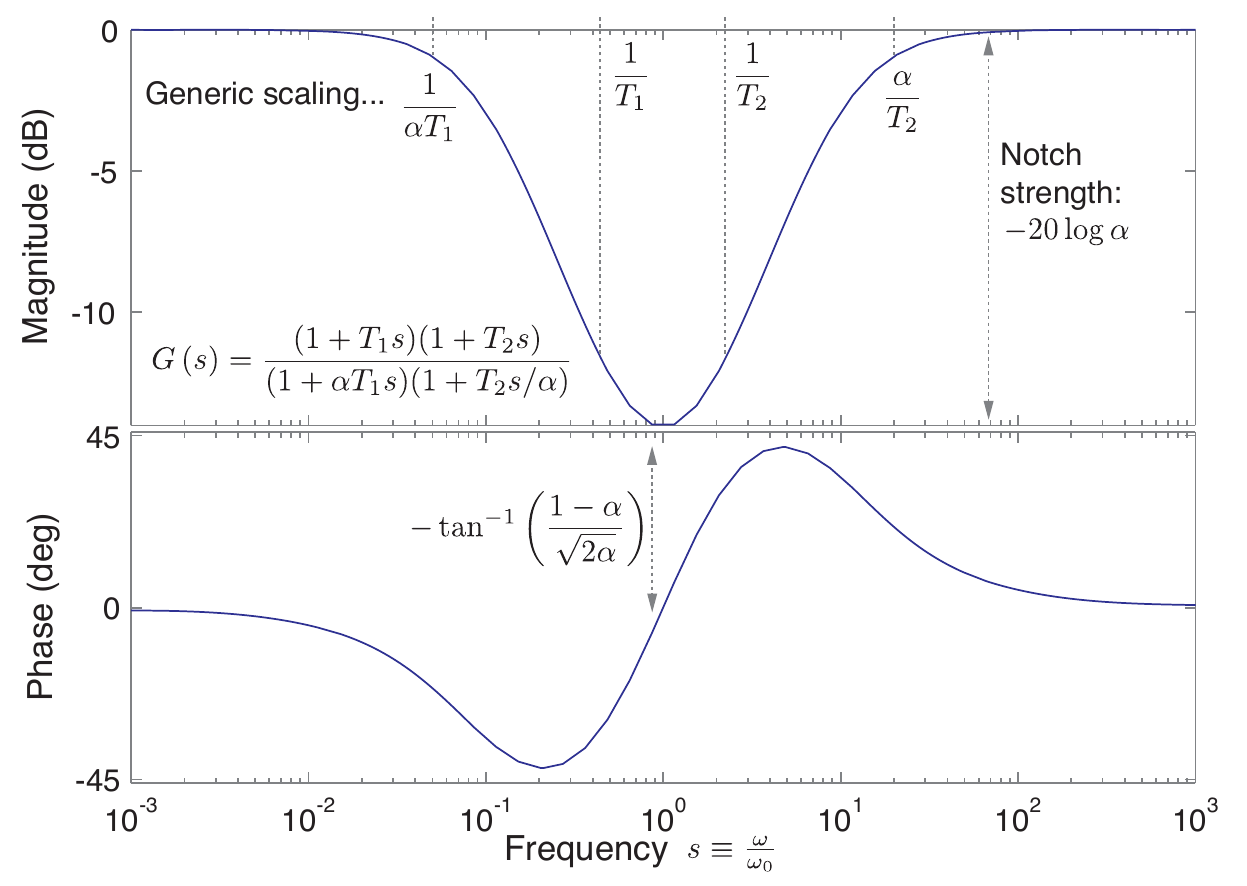}
\caption[The transfer function of a lag-lead compensator]{A lag-lead compensator introduces a notch in the system gain as well as increased phase at high frequencies.  It can be realized as an RC electrical network and is employed to suppress the effect of mechanical resonances such as~\ref{fig:pztResonanceBodePlots}. }
\label{fig:leadlagcomp}
\end{figure}

No perfect guide to laser/cavity stabilization exists, but a great start is Ch.~27 of~\cite{bass1994hoo}, written by John Hall and collaborators.  In the end, these authors advocate a PID feedback loop (independently adjustable proportional, integral, and differential gain) or two PI stages with a notch circuit designed to compensate for the mechanical resonance of typical PZTs.  The transfer function of a generic PZT is
\begin{equation}
R_\text{pzt}(\omega) = \frac{\omega_0^2}{\omega_0^2 + 2 \omega \eta \omega_0 + \omega^2}
\end{equation}
where $\omega_0$ is the first resonance (angular) frequency and $\eta = 1/(2Q)$ describes the width of the resonance.  As shown in Figure~\ref{fig:pztResonanceBodePlots}, the glaring practical problem with $R_\text{pzt}$ is the introduction of -180$^\circ$ phase shift.

A lead-lag compensator, whose response is depicted in Figure~\ref{fig:leadlagcomp}, can be realized as an electrical network of two poles and two zeros with generic transfer function
\begin{equation}
G(s) = \frac{(1+ T_1s)(1 + T_2 s)}{(1 + \alpha T_1s)(1 + T_2 s /\alpha)}, \qquad (\alpha > 1, \quad T_1 > T_2)
\end{equation}
where $s = \omega / \omega_1$ is a scaled frequency variable.  A design heuristic~\cite{bass1994hoo} is to lower the PZT resonance $Q$, if possible, by mounting with light and study mechanical construction and damping materials, such as wax.  Then, a single PI gain stage along with lead-lag compensator are designed to achieve a desired unity gain frequency $f_u$;  this sets an upper limit of the bandwidth of the servo.  Finally, to increase the gain at low frequencies---where it is sorely needed most---one introduces a second PI stage with a corner frequency in the range 0.1$f_u$--$f_u$, trading good step-function response for low frequency gain.  Optionally, the authors recommend an `adaptive clamp' on the second PI stage so it is slowly introduced into the feedback system when lock is achieved;  it therefore won't interfere with reacquisition if the lock is disrupted.  At present our lock system is functional but not optimal.

\section{Ultraprecise frequency comparisons with a femtosecond comb}
A natural goal for study of the 2051~nm transition in Ba$^+$ is the direct, absolute measurement of that transition frequency.  Yet another goal would be a precise comparison of the transition frequency in Ba$^+$ with a similarly narrow transition frequency in another atomic species, perhaps a single In$^+$ ion which is currently trapped in the same laboratory.  Both goals can be satisfied with an instrument called a mode-locked femtosecond laser frequency comb.  In this section we review the mechanics and history of the fs-comb and describe implementations relevant for the narrow transitions in In$^+$ and Ba$^+$.

\subsection{Mode locked lasers and the femtosecond comb}
\begin{figure}
\centering
\includegraphics{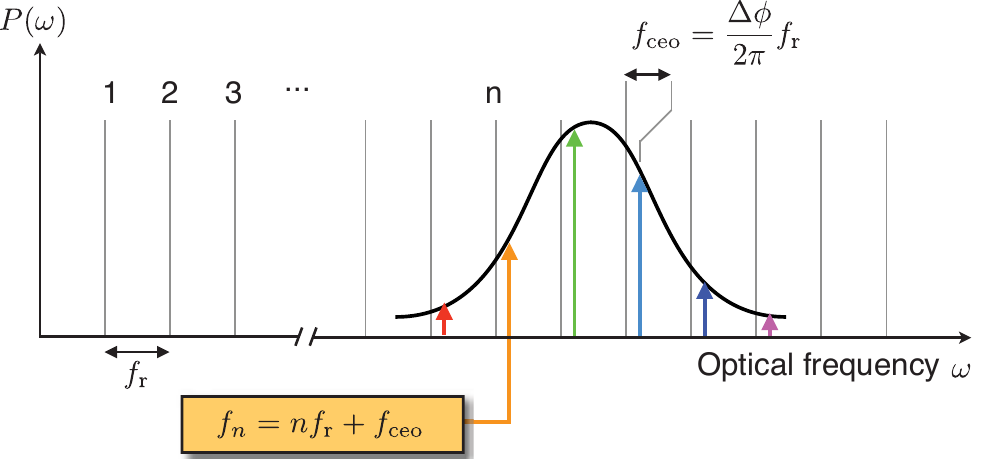}
\caption[Cartoon of a mode-locked laser output spectrum]{Cartoon of a mode-locked laser output spectrum.  A mode-locked laser supports many longitudinal laser modes with a fixed phase relationship.  Here we see depict the spectral output of a mode locked laser as a `comb' of frequencies, spaced by the pulse repetition rate $f_\text{r}$, and offset from a comb extending to zero frequency by the carrier-envelope offset frequency $f_\text{ceo}$.}
\label{fig:modeLockedLaser}
\end{figure}

Stable continuous-wave lasers in the laboratory often produce a single longitudinal mode of the laser cavity.  This desirable feature is brought about by purposefully introducing dispersion into the cavity that allows just one frequency to posses high gain in the laser.  As practical examples, dye laser cavities contain frequency-dependent birefringent tuning elements and intracavity etalons;  some semiconductor lasers incorporate dispersion in the form of an etched intra-cavity diffraction grating.  An ideal mode-locked laser seeks the opposite goal:  broadband operation with very little and tightly controlled dispersion.  Several longitudinal modes are simultaneously supported in the laser cavity with a fixed phase relationship, as implied by Figure~\ref{fig:modeLockedLaser}.

The output of a mode-locked laser, a series of coherent pulses, can be understood by summing the contributions from each laser mode~\cite{cundiff2003cfo}
\begin{equation} \label{eq:modeLockedLaserSum}
E(t) = \sum_m C_m e^{i(\omega_0 + m \nu_\text{fsr})t + \phi_m)},
\end{equation}
where $C_m$ and $\phi_m$ are the relative constant amplitude and phase of mode $m$.  The free-spectral range $f_r = f_r$ sets the mode frequency spacing.  The output is periodic with period $\tau = 2\pi / f_r)$~\cite{yariv2006poe},
\begin{align}
E(t + \tau) &= \sum_m C_m \exp \left[i \left( (\omega_0 + m f_r)(t + \tau) \right) + \phi_m \right] \\
		&= \sum_m C_m \exp \left[i \left( \omega_0 + m f_r)t + \phi_m \right) \right] \exp \left[ i \left( 2 \pi (\omega_0 / f_r + m ) \right) \right] \\
		&= E(t) e^{i 2 \pi \omega_0 / f_r}
\end{align}
up to a constant phase factor sometimes called the \emph{phase slippage}
\begin{equation}
\Delta \phi = 2 \pi \omega_0 / f_r.
\end{equation}

To qualitatively understand the prediction of Eq.~\ref{eq:modeLockedLaserSum}, it useful to assume that there are $N$ laser modes operating at constant amplitude $C_m = 1/\sqrt{N}$ with relative phases $\phi_m = 0$.  Then,
\begin{align}
E(t) &= \frac{1}{\sqrt{N}} \sum_{m=1}^N \frac{\sin\left( \frac{N 2\pi f_\text{r} t}{2}\right)}{\sin\left( \frac{2\pi f_\text{r} t}{2} \right)} e^{i (\omega_0 + (N+1) 2\pi f_r/2)t} \\
\intertext{which implies a laser power $P(t) = E(t)E^*(t)$ strongly peaked at the frequency $f_r$,}
P(t) &\propto \frac{1}{N} \frac{ \sin^2\left( \frac{N 2 \pi f_r t}{2}\right)}{\sin^2\left( \frac{2\pi f_r t}{2}\right)}.
\end{align}
In this last step we have averaged over a period long compared to the optical oscillation frequency $\omega_0$ but short compared to the pulse repetition rate $f_r$ in order to qualitatively determine the pulse shape of the output.  
%% PLOT IF YOU HAVE TIME

\subsection{Characterizing and stabilizing the comb for metrology applications}
The spectral output of a stabilized femtosecond mode-locked lasers, shown in Figure~\ref{fig:modeLockedLaser}, becomes a comb of delta functions at \emph{known} frequencies $f_n$:
\begin{equation*}
f_n = n f_r + \underbrace{\frac{\Delta \phi f_r}{2 \pi}}_{\equiv f_\text{ceo}}
\end{equation*}
where the quantity $f_\text{ceo} = \Delta \phi f_r / 2\pi$ is called the \emph{carrier-envelope offset} frequency.  From a metrology  perspective $f_\text{ceo}$ is the offset from zero frequency a mode would have if the bandwidth of the comb stretched all the way back to dc.  Both $f_r$ and $f_\text{ceo}$ are directly measurable as radio frequencies and can be stabilized.  The picture presented in Figure~\ref{fig:modeLockedLaser}, that the comb modes exist at well defined frequencies, is actually only true when the parameters $f_r$ and $f_\text{ceo}$ are aggressively stabilized.

Simply sending the pulsed comb output onto a fast detector such as a photo-multiplier tube or avalanche photodiode measures the pulse repetition rate $f_r$. The equivalence of $f_r$ as the mode spacing as well as pulse repetition rate has been confirmed to a few parts in $10^{17}$~\cite{udem1999aml}.  The beatnote resulting from $f_r$ being heterodyned with a trusted radio frequency reference (or, better, an optical reference~\cite{jones2001sfl}) is an error signal with which one can servo the laser cavity size (perhaps with a PZT-mounted mirror), stabilizing $f_r$.

\begin{figure}
\centering
\includegraphics{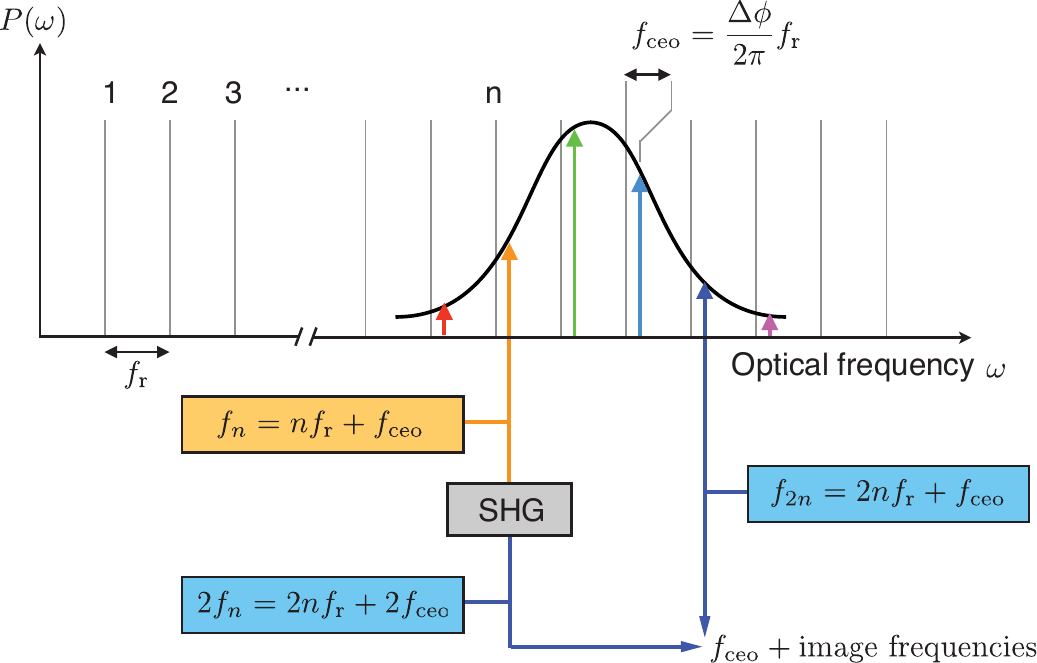}
\caption[A scheme to measure the carrier envelope offset frequency $f_\text{ceo}$]{A scheme to measure the carrier envelope offset frequency $f_\text{ceo}$.  If the mode-locked laser spectrum spans an octave of frequencies, interferometry between the red and blue ends of the comb yields $f_\text{ceo}$ as a beat-note on a photodiode.  Second harmonic generation (SHG) of the red portion of the spectrum in a nonlinear crystal (often single pass is sufficient given the high peak powers in the pulses) converts several laser modes around $f_n$ to $2(f_n) = 2n f_\text{r} + 2 f_\text{ceo}$.  Beat against the blue end of the comb spectrum (near modes $f_{2n} = 2n f_\text{r} + f_\text{ceo}$), $f_\text{ceo}$ and other frequencies such as $f_\text{r}$, $f_\text{r} \pm f_\text{ceo}$, etc. appear on a photodiode. Proper rf filtering yields a signal that can be frequency locked, stabilizing a potentially varying $f_\text{ceo}$.}
\label{fig:f2fInterferometerScheme}
\end{figure}
Measurement of $f_\text{ceo}$ is less straightforward, but can be accomplished is several waves.  Perhaps the easiest to implement is known as the $f$-$2f$ interferometer scheme shown in Figure~\ref{fig:f2fInterferometerScheme}.  Using dichroic optics, the red portion of the comb spectra is split off and frequency doubled using standard second harmonic generation techniques in nonlinear optical crystals.  This spectra now contains frequencies
\begin{equation*}
2f_n = 2n f_r + 2 f_\text{ceo}
\end{equation*}
for several values $n$.  If this doubled spectrum is then heterodyned with the blue (un-doubled) part of the fs-comb spectrum surrounding $f_{2n} = 2n f_r + f_\text{ceo}$, the lowest frequency component of the resulting electronic signal is $f_\text{ceo}$.  More explicitly, the heterodyne detector output is
\begin{align}
E_\text{detect}(t) &\propto \sum_{n} \cos (2n f_r + 2 f_\text{ceo})t \times  \cos (2n f_r + f_\text{ceo})t \\
\begin{split}
&\propto \left[ a_0 \cos f_\text{ceo} t + a_1 \cos f_r t \right. \\
&+ \underbrace{a_2 \cos(f_r - f_\text{ceo})t + a_3 \cos(f_r + f_\text{ceo})t}_\text{intermodulation products} \\
& \left. + \text{ higher harmonics }  \right]
\end{split}
\end{align}
for some amplitudes $a_i$.  A low-pass filter can select just the frequency component $f_\text{ceo}$ (see~\cite{helbing2002ceo}, for instance).

For this scheme to work, the bandwidth of the femtosecond comb must span an octave so that modes at $f_n$ and $f_{2n}$ exist for at least one integer $n$.  If a comb falls short of this width but contains modes $f_n$ and $f_{3n/2}$, $f_\text{ceo}$ can be measured in an analogous $3f$-$2f$ interferometer scheme (see~\cite{ramond2002pcl}, for instance).  Beating a measured $f_\text{ceo}$ against a trusted rf reference produces an error signal with which one can servo the laser cavity.  Because $f_\text{ceo}$ is proportional to both $f_r$ and the intracavity phase slippage $\Delta \phi$, both must be kept stable.  If a direct servo of $f_r$ is in place, the phase slippage $\Delta \phi$ can be locked by carefully controlling the intracavity dispersion of the laser.  One technique successful for Ti:Sapphire fs-combs is to servo the Nd:YAG pump power with an extra-cavity acousto-optic modulator (AOM) preceding the mode-locked laser (\cite{helbing2002ceo} is one example).

With $f_r$ and $f_\text{ceo}$ stabilized, the fs-comb becomes a tool for measuring optical frequencies with same ultimate precision as the microwave standards used to establish $f_r$ and $f_\text{ceo}$.  A laser to be measured $f_\text{laser}$ is heterodyned with the femtosecond comb;  the lowest frequency component of the beat note is the difference between the laser frequency and the nearest comb mode $n^*$:
\begin{align*}
E_\text{detect}(t) &\propto \cos (\omega_\text{laser} t) \times \sum_n C_n \cos(n\omega_\text{r} + \omega_\text{ceo})t \\
& \propto \left[a_0 \cos \left( |\omega_\text{laser} - n^*\omega_\text{r} - \omega_\text{ceo}|t \right) + \text{ higher frequency components } \right].
\end{align*}
One can measure $n^*$ by measuring $\omega_\text{laser}$ to a precision $\pm f_r/2$ with a wavemeter or other interferometer while simultaneously measuring and stabilizing $f_r$ and $f_\text{ceo}$.

\subsection{Suitable fs-combs for use with Ba$^+$ and In$^+$ frequency standards}
\begin{figure}
\centering
\includegraphics[width=6in]{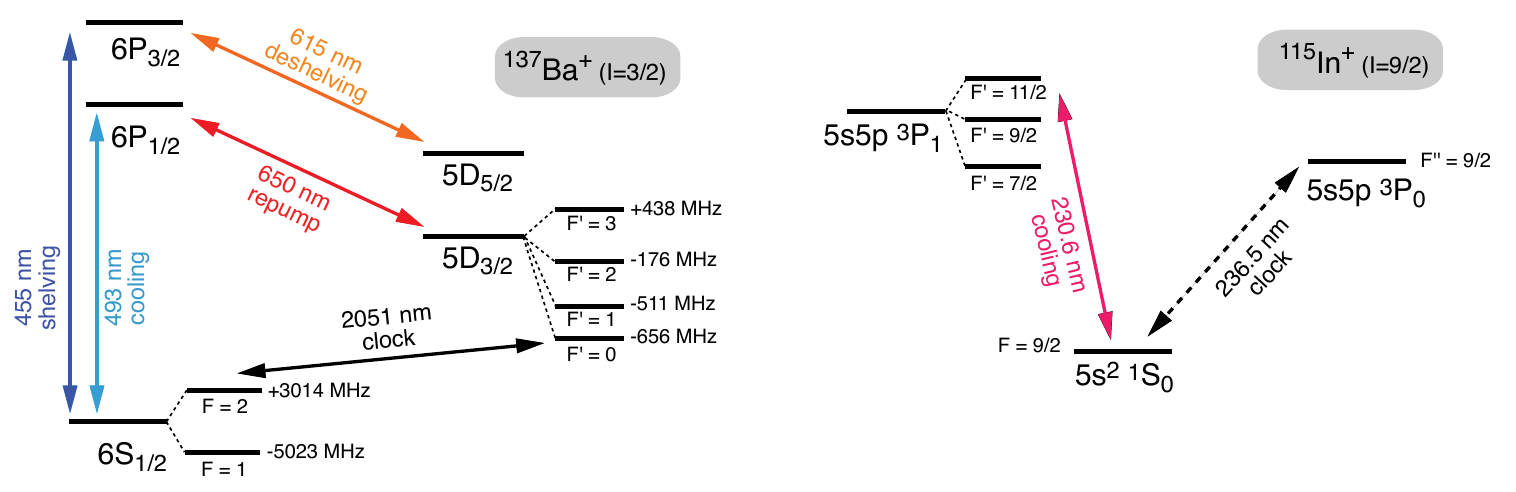}
\caption[In$^+$ and Ba$^+$ energy levels and clock transitions]{In$^+$ and Ba$^+$ energy levels and clock transitions.  Development of both frequency standards is ongoing at the University of Washington.  The addition of a suitable femtosecond laser frequency comb allows a precision comparison of the two clock transition frequencies, shown schematically in Figure~\ref{fig:InBaClocksFsComb}.}
\label{fig:InBaClocks}
\vspace{0.25 in}
\includegraphics{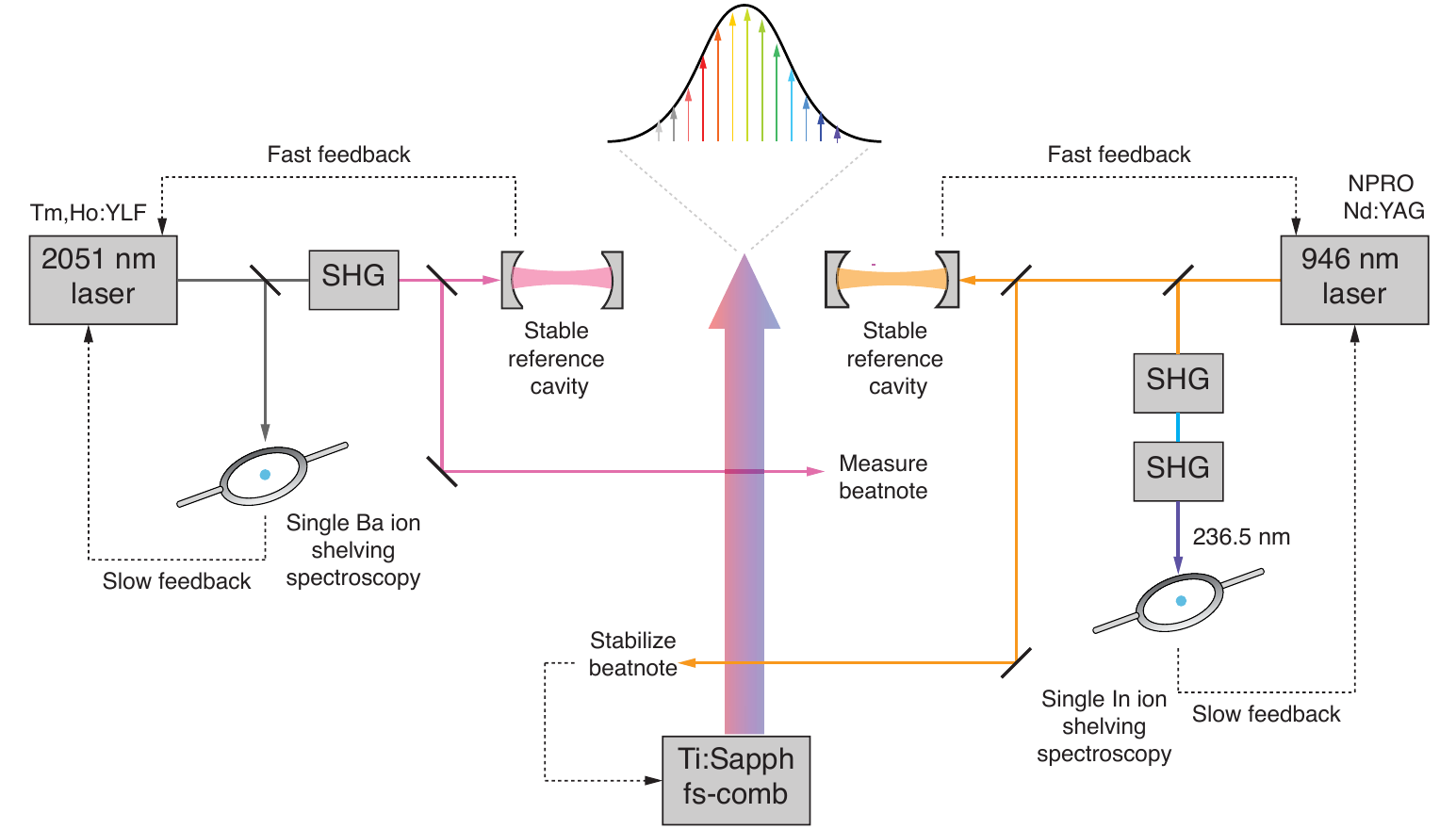}
\caption[Comparison of Ba$^+$ and In$^+$ optical frequency standards using a suitable fs-comb]{Comparison of Ba$^+$ and In$^+$ optical frequency standards using a suitable femtosecond laser frequency comb (fs-comb).  Lasers at 2051~nm and 946~nm are made very narrow and stable by servo-locking them to stable optical reference cavities over short times and to single trapped ions over long times using a shelving interrogation scheme.  Since the Indium transition frequency is faster, and the state lifetime shorter, it makes sense to stabilize the femtosecond frequency comb by locking the difference frequency between the 946~nm laser and a nearby comb mode.  Then, the beatnote between the doubled 2051~nm laser and another comb mode gives the difference in the Ba$^+$ and In$^+$ optical frequencies.  Simpler schemes involving a microwave reference allows the absolute frequency measurement of either transition.} 
\label{fig:InBaClocksFsComb}
\end{figure}
An ideal femtosecond frequency comb useful for absolute and relative frequency measurements between the clock transitions in In$^+$ and Ba$^+$ (see Figure~\ref{fig:InBaClocks}) would:
\begin{enumerate}
\item Yield an output spanning an octave with enough useful power to measure and stabilize $f_\text{ceo}$ with the $f$-$2f$ interferometer technique discussed earlier
\item Posses a bandwidth either covering 2051~nm and 946~nm (which is $4 \lambda_{\text{In}^+ \text{ clock}}$) or 1025~nm (which is $2 \nu_{\text{Ba}^+ \text{ clock}}$) and 946~nm
\item Feature a mode-spacing/pulse repetition rate $f_r > 300$~MHz so that mode numbers can be easily discriminated using a wavemeter
\end{enumerate}
A bow-tie, prism-less, Ti:Sapphire mode-locked laser design~\cite{fortier2005mos} satisfies all these requirements (and would require frequency doubling of the barium clock laser to 1025~nm).  Less ideal solutions include doped Er:fiber fs-combs~\cite{rauschenberger2002cfc} which have low repetition rates and much larger inherent phase noise but almost cover the bandwidth 946~nm--2051~nm.  A bow-tie Cr:Forsterite laser~\cite{nogueira2006egt} may cover a similar IR bandwidth and satisfy all of the requirements. 

\section{Observation of the 2051~nm clock transition}
\subsection{Adiabatic rapid passage} \label{sec:ARPobservation}
\begin{figure}
\centering
\includegraphics[width=6in]{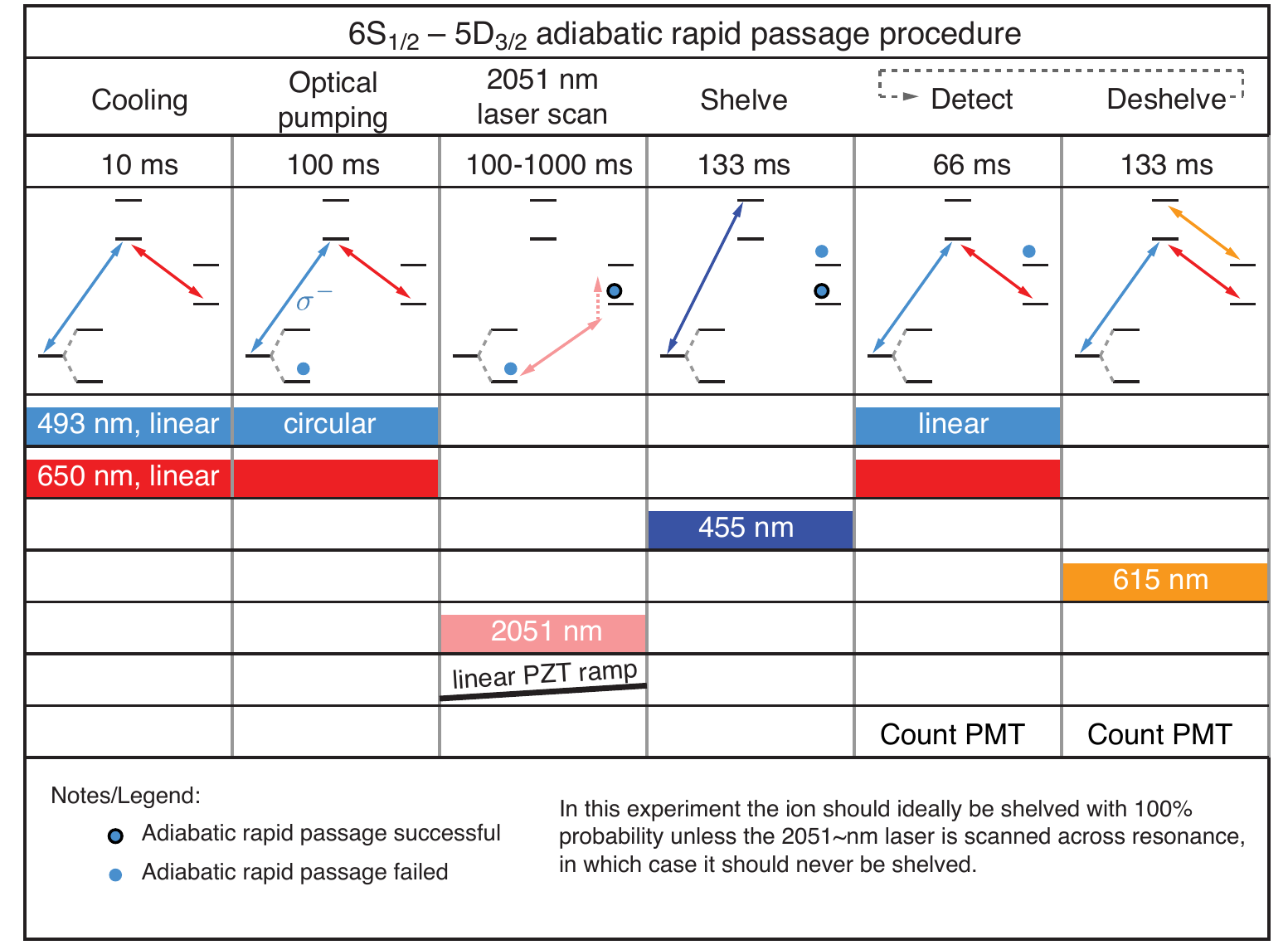}
\caption[Procedure for 2051~nm adiabatic rapid passage]{This graphical schematic outlines the procedure for performing adiabatic rapid passage on the $6S_{1/2} \leftrightarrow 5D_{3/2}$ 2051~nm transition.}
\label{fig:procedureARP}
\end{figure}
\begin{figure}
\centering
\includegraphics{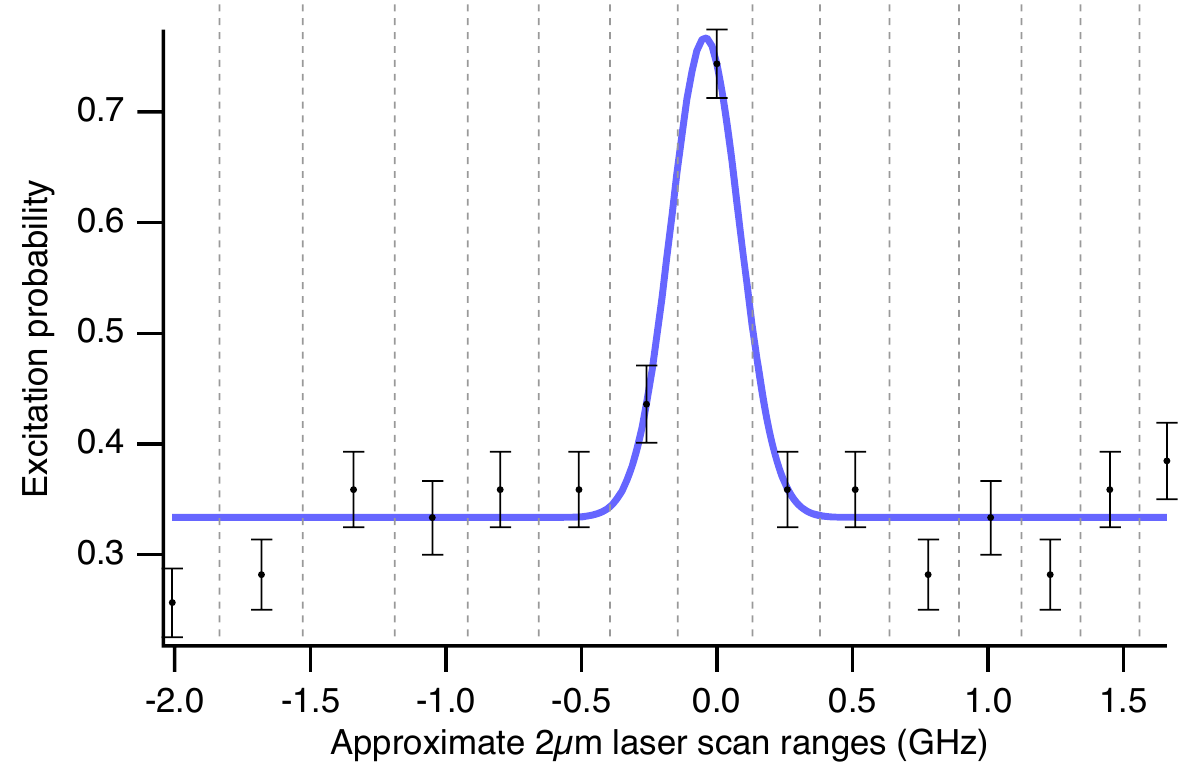}
\caption[Observation of adiabatic rapid passage on the $6S_{1/2} \leftrightarrow 5D_{3/2}$ transition]{Observation of adiabatic rapid passage on the $6S_{1/2} \leftrightarrow 5D_{3/2}$ transition. The measurement procedure is outlined in Figure~\ref{fig:procedureARP}.  The frequency scan ranges differ in size by $\sim$ 10\% due to non-linearity in the laser PZT response.}
\label{fig:2micronARPclockChapter}
\end{figure}
We first observed the electric-dipole forbidden $6S_{1/2} \leftrightarrow 5D_{3/2}$ 2051~nm transition in the even isotope $^{138}$Ba$^+$ using the technique of adiabatic rapid passage.  From the discussion in Section~\ref{sec:atomicARP}, we distill the important result that nearly 100\% efficient transfer of state population is achievable by scanning a laser over a transition as long as
\begin{itemize}
\item The scan time $t_\text{scan}$ is much faster than the excited state lifetime $\Gamma$,
\item The scan time $t_\text{scan}$ is much slower than the Rabi oscillation period $\Omega \propto E \langle g | \text{E2} |e \rangle$,
\end{itemize}
expressed succinctly as $\Gamma \ll t_\text{scan}^{-1} \ll \Omega$.  In an experimental sequence outlined in Figure~\ref{fig:procedureARP}, we programed an amplified function generator to ramp the 2051~nm laser PZT voltage which tunes the frequency by approximately -75~MHz/V (see Table~\ref{tab:2micronPZTtune}).  After cooling and preparing the ion in a definite magnetic sublevel of $6S_{1/2}$ using optical pumping, we expose the ion to the 2051~nm and trigger the frequency sweep.  We test about 15 sweep ranges, each covering approximately $200$~MHz, and expect that only one sweep range contains the 2051~nm resonance since 200~MHz is much larger than the $5D_{3/2}$ Zeeman splittings, and locations of motional sidebands (see Section~\ref{sec:coolingRegime}).  If the adiabatic rapid passage is unsuccessful, a shelving pulse at 455~nm moves the ion with high probability to the $5D_{5/2}$ state (the decay $6P_{3/2} \to 5D_{5/2}$ is about 75 times more likely than $6P_{3/2} \to 5D_{3/2}$).  We then shine 493~nm and 650~nm cooling and repumping beams on the ion and detect scattered fluorescence.  Positive detection is correlated with the adiabatic rapid passage being successful.  A typical data run, Figure~\ref{fig:2micronARPclockChapter}, shows high excitation probability for just one laser frequency scan range.

\subsection{Electron shelving}
\begin{figure}
\centering
\includegraphics{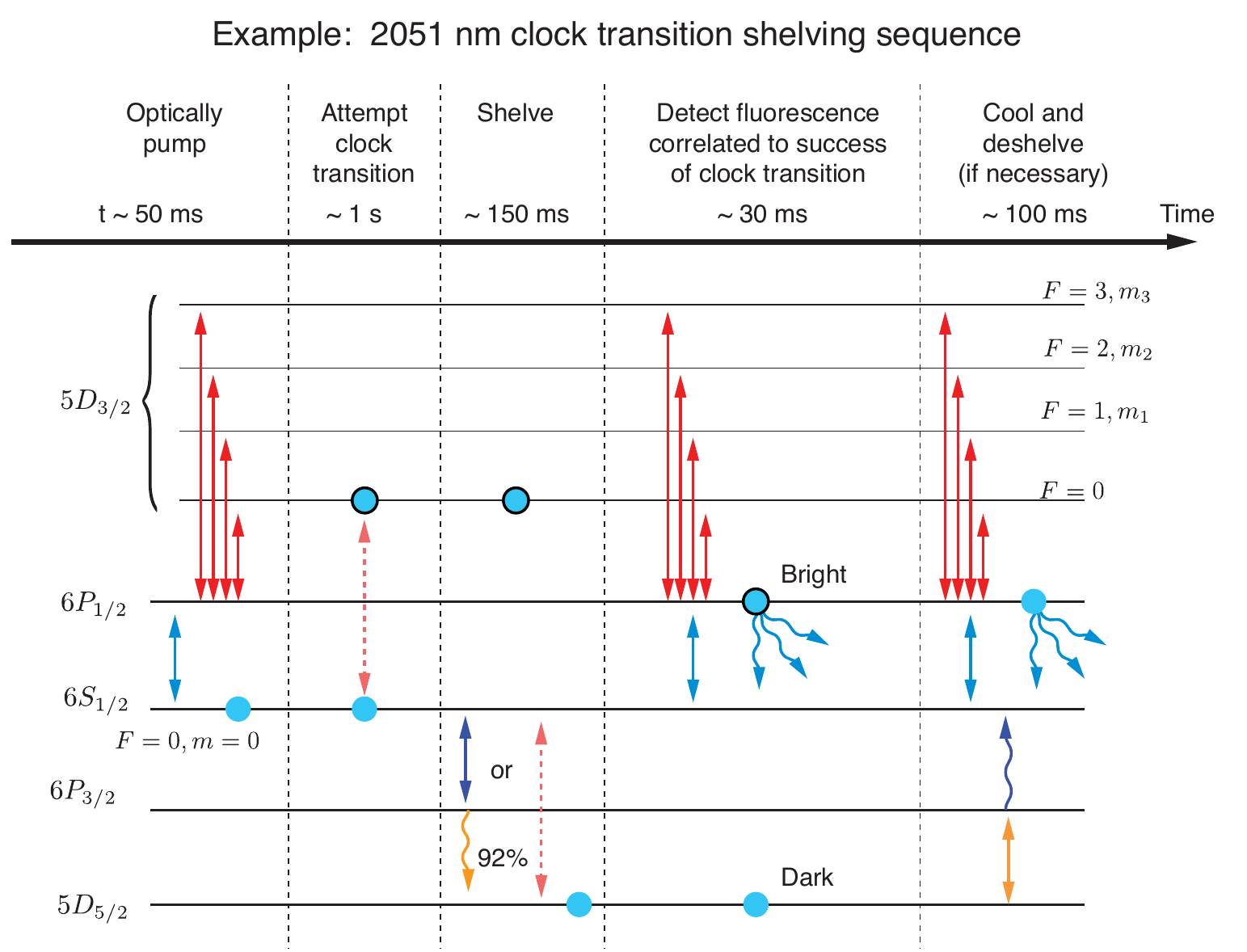}
\caption[An example scheme for shelving approach to the 2051~nm transition]{An example scheme for shelving approach to the 2051~nm transition. In this scheme, success of an excitation of an ion to $5D_{3/2}$ following a pulse of 2051~nm light is correlated to whether the ion is found shelved in the $5D_{5/2}$ state.  An ion successfully excited by the clock pulse will exhibit macroscopic fluorescence when resonant 493~nm and 650~nm lasers are applied.}
\label{fig:electronShelvingClock}
\end{figure}
The next step for future work is to attempt electron shelving on the 2051~nm $6S_{1/2} \leftrightarrow 5D_{3/2}$ transition as shown in Figure~\ref{fig:electronShelvingClock}.  Because both the ground and excited states are also used in the cooling/repumping cycle, a direct electron shelving approach outlined in Section~\ref{sec:electronShelving} must be modified.  As in the single ion rf spectroscopy measurements in Section~\ref{sec:spinResonanceLS}, indirect shelving to the $5D_{5/2}$ state is possible.  A procedure might be
\begin{enumerate}
\item Ground state preparation in $6S_{1/2}, F = 2, m =0$ using optical pumping.
\item Execute a timed, shaped pulse of resonant 2051~nm radiation.
\item Shelve any remaining population in $6S_{1/2}$ to $5D_{5/2}$ using the 455~nm excitation to $6P_{3/2}$ and decay ($\sim$70\% efficient) or an adiabatic rapid passage over the 1762~nm $6S_{1/2} \leftrightarrow 5D_{5/2}$ transition (recently demonstrated, nearly 100~\% efficient).
\item Test whether the clock pulse moved the ion to $5D_{3/2}$ by applying blue and red lasers.
\end{enumerate}

\section{Application:  search for drift in fundamental constants}
As we will show in Section~\ref{sec:clockSystematicEffects}, the frequency of the proposed clock splitting is very robust.  Systematic shifts to the transition due to uncontrolled magnetic fields,  electric fields associated with the trapping voltages and blackbody radiations, and uncertainty in the ion's second order Doppler shift might all add up to a total relative clock frequency uncertainty of less than $10^{-17}$.  At this level of precision, new physics involving drifting fundamental `constants'~\cite{karshenboim2003tas} might be directly observed by measuring the absolute frequency of the clock transition over many years, or by measuring the relative drift between optical clock frequencies in different atomic and ion optical clock species\footnote{These two methods are not identical.  Absolute measurements of optical standards can be made in perennially different labs around the world and compared.  A relative measurement between optical standards require that both standards plus an appropriate femtosecond laser frequency comb be present at the same laboratory but yields far higher precision.}.

Though the case for drifting fundamental constants is not established (see~\cite{uzan2003fca} for a consummate review current in 2003), studies of distant quasar spectra hint of $\sim 10^{-6}$  drift over 1 billion years in the fine structure constant $\alpha$~\cite{webb2001fec,murphy2003fev}.  In these observations, the spectral absorption features from many  elements are shown to have relative shifts among them when the common-mode red shifts are accounted for.  Specifically, \cite{murphy2003fev} observes evidence for a relative shift in $\alpha$ of
\begin{equation*}
\frac{\Delta \alpha}{\alpha} = (-0.543 \pm 0.116) \times 10^{-5}
\end{equation*}
at high statistical significance (4.7 $\sigma$) over the observational red-shift range $0.2 < z_\text{abs} < 3.7$.  Linearly extrapolated, this implies a current drift rate of $\alpha$
\begin{equation*}
\frac{\dot{\alpha}}{\alpha} = (6.40 \pm 1.35) \times 10^{-16} \text{ yr}^{-1}
\end{equation*}
which ought to be resolved by relative comparisons of optical atomic frequency standards in just a few years of observation time.  On the other hand, a similar study~\cite{srianand2004ltv} with comparable accuracy of a different portion of the sky shows no resolved temporal drift in $\alpha$.  

Also, recent studies~\cite{lamoreaux2004nmo} of a unique and fascinating terrestrial phenomenon, the Oklo natural nuclear reactor, demonstrate that a different value of $\alpha$ during the activity some 2 billion years ago might leave a signal in the relative abundances of isotopes nearby $^{149}$Sm which has a low-energy neutron absorption resonance.  This study concludes a positive result, a change in $\alpha$ at the $10^{-8}$ level (over a 1~billion year timescale) and extrapolates that the current rate of change is bounded by
\begin{equation*}
\frac{| \dot{\alpha} |}{\alpha} < 3.8 \times 10^{-17} \text{ yr}^{-1}
\end{equation*}
at 95\% confidence.

\begin{table}
\centering
\caption[Expected shifts in selected optical frequencies with a time varying $\alpha$]{Expected absolute and relative shifts in optical frequency standards with a changing fine-structure constant $\alpha$. The present day transition frequency is $\omega_0$.  Parameters $q_1$ and $q_2$ are found in theoretical work~\cite{dzuba1999cre,dzuba2000aoc,angstmann2004ret}.  Hg$^+$ is an especially good candidate given its high sensitivity an opposite sign.  The proposed Ba$^+$ clock transition has comparable relative sensitivity.}
\begin{tabular}{l l c l  l l l l}
	&	 	 &			&			&		&	       & \multicolumn{2}{c}{Shift with $\dot{\alpha}/\alpha = 10^{-16} \text{ yr}^{-1}$} \\ 
Z  	& Species & Transition	 & $\omega_0$ (cm$^{-1}$) & $q_1$ & $q_2$ & $\Delta \omega$ (Hz yr$^{-1}$) & Relative \\ \hline \hline
20	& Ca		& $4 {}^{1}S_0 \leftrightarrow {}^{3}P_1$	& 15210	& 230 & 0 & 0.014 &$ 3.0 \times 10^{-17}$ \\
38	& Sr$^+$	& $5S_{1/2} \leftrightarrow 5D_{5/2}$ & 14836.24 & 2852 & 160 & 0.19 & $4.3 \times 10^{-17}$ \\
49	& In$^+$	& $5 {}^{1}S_0 \leftrightarrow {}^{3}P_0$ & 42275	 &	2502 &956  & 0.26 & $2.1 \times 10^{-16}$ \\	
56	&  Ba$^+$	 & $6S_{1/2} \leftrightarrow 5D_{3/2}$ & 4843.850 &	5402	 &221 & 0.35 & $2.4 \times 10^{-15}$ \\	
56	&  Ba$^+$	 & $6S_{1/2} \leftrightarrow 5D_{5/2}$ & 5674.824 &	6872	 &-448 & 0.36 & $2.1 \times 10^{-15}$ \\	
70	& Yb$^+$	& $6S_{1/2} \leftrightarrow 5D_{3/2}$ & 24332 & 9898	&1342& 0.76 & $1.0 \times 10^{-15}$ \\
80	& Hg$^+$& $ {}^2S_{1/2} \leftrightarrow  {}^2D_{5/2} $ &35514& -36785 &-9943& -3.40&  $-3.2 \times 10^{-15}$ \\ 
81	& Tl$^+$ & $6 {}^{1}S_{0} \leftrightarrow {}^{3}P_0$ &  49451 & 1661 & 9042 & 1.18 & $8.0 \times 10^{-16}$ \\
88	& Ra$^+$ & $7S_{1/2} \leftrightarrow 6D_{5/2}$ & 13743.11 & 19669 & -864 &1.08 & $2.6 \times 10^{-15}$	
\end{tabular}
\label{tab:alphaDotTransitions}
\end{table}

Laboratory searches for temporal drifts in fundamental constants using optical atomic frequency standards~\cite{bize2003tsf}, microwave standards~\cite{marion2003svf}, and accidentally degenerate transitions~\cite{cingoz2006ltv} are ongoing.  When studies incorporate several atomic species, one gains sensitivity to drift in more than one `constant'~\cite{peik2004lpt}.

The proposed clock transition in Ba$^+$ is an excellent candidate for such a search.  While the expected absolute shift (in units of Hz) given a particular $\dot{\alpha}$ is not as large as other candidate species, the clock transition itself is deeper in the infrared than all other candidate species making the \emph{relative} clock shift in barium large and easy to detect. Theoretical studies~\cite{dzuba1999cre,dzuba2000aoc,angstmann2004ret} estimate what the relative transition drift rate might be under a given time changing $\dot{\alpha} = d\alpha / d t$
\begin{equation}
\left. \frac{\dot{\omega}}{\omega_0} \right|_{\alpha_0} = \frac{(2 q_1 + 4 q_2)}{\omega_0} \frac{\dot{\alpha}}{\alpha}
\end{equation}
where $\omega_0$ is the present day transition frequency, and the model parameters $q_1$ and $q_2$ are tabulated for selected transitions in Table~\ref{tab:alphaDotTransitions}.  This table also estimates the absolute and relative clock frequency shifts associated with a hypothetical yearly change of $\dot{\alpha}/\alpha = 10^{-16} \text{ yr}^{-1}$.

\section{Systematic effects}\label{sec:clockSystematicEffects}
The ultimate accuracy of a single-ion optical frequency standard is limited by the degree to which one can measure and control systematic shifts to the clock transition.  We estimate in this section that the total relative magnitude of systematic shifts on the proposed clock transition is approximately $6 \times 10^{-17}$ with a relative uncertainty of $\pm 6 \times 10^{-18}$ at cryogenic temperatures.  We will discuss our estimates for all known systematic shifts:
\begin{itemize}
\item Shifts due to the fundamental motion of the trapped ion:  the first and second-order Doppler shifts
\item Shifts due to applied magnetic fields:  the first and second-order Zeeman shifts
\item Shifts due to background and applied electric fields:  the blackbody Stark shift, gradient or quadrupole Stark shifts due to stray electric field gradients, and Stark shifts due to the rf trapping field
\item Shifts due to the clock laser itself:  the ac-Stark effect (light-shifts)
\end{itemize}
We will demonstrate in this section a very special feature of the proposed $^{137}$Ba$^+$ frequency standard:  immunity to first order from shifts due to stray electric field gradients.  In many other ion clock transitions such effects can lead to uncertain shifts of up to 1~Hz, corresponding to a relative systematic error of $\sim 1 \times 10^{-15}$.  Though there are measurement techniques for nulling this effect~\cite{dube2005eqs,oskay2005mhq}, it still may be worthwhile to begin with a system completely free of the effect.  We will also see that a chief disadvantage of the proposed standard is a very large black-body Stark shift;  cryogenic cooling will be mandatory for ultimate accuracy.

\subsection{The first-order Doppler shift}
Confining a particle in the Lamb-Dicke regime ($k \sqrt{\langle x^2 \rangle} \ll 1$) eliminates the first-order Doppler shift (see \cite{wineland1987lcl}, for instance).  Consider a particle trapped in a harmonic potential with energy level spacing $\hbar \nu$, with the secular frequency $\nu$ chosen much larger than a transition's natural linewidth $\nu \gg \Gamma$.  The trapped particle experiences a frequency modulated probe laser;  the Lamb-Dicke limit guarantees that the modulation depth is very small compared to the modulation frequency $\nu$.  Thus, almost all of the spectral energy is at the ``rest-frame'' carrier frequency and first sidebands (which are far off resonance).

However, there is another opportunity for our trapped ion to acquire a first-order Doppler shift:  a lack of rigidity or relaxation in optical mounts or the ion trap itself over the 1 second timescale that comprises the clock excitation (e.g.\ a $\pi$-pulse).  We must put a limit on the possible \emph{systematic} motion of the optics and trap and later verify the performance of the real system. Two techniques one might employ to reduce the effect are:
\begin{itemize}
\item  Excite the clock transition from two anti-parallel directions on alternating cycles.
\item  Verify the trap and optical system rigidity interferometrically with a co-propagating HeNe beam or homodyne the clock laser itself.
\end{itemize}

If apparatus motion with average speed $v$ is present, then a worst case observed Doppler shift is
\begin{equation}
\frac{\delta_\text{Doppler}}{\omega_0} = \frac{v}{c}.
\end{equation}
Suppose one can constrain movement to 1~nm over 1~s.  This leads to a maximum relative Doppler shift 
\begin{equation*}
\frac{\delta_\text{Doppler}}{\omega_0} = (1/3) \times 10^{-17} \left(\frac{v}{1 \text{ nm/s}} \right).
\end{equation*}
This could be fairly serious if one cannot rule out systematic motion of the system during a clock transition interrogation.

\subsection{The second-order Doppler shift} \label{sec:secondOrderDopplerShift}
\begin{table}
\centering
\caption{Expected clock transition second-order Doppler shifts}
\begin{tabular}{l|lcl}
Cooling scheme & $\langle n \rangle$ & $\delta_{2D} / \omega_0$ & Absolute shift \\ \hline \hline
Sideband cooling & $\approx 0$ & $-4.7  \times 10^{-20}$ & -6.9 $\mu$Hz \\
Optimal Doppler cooling & 9.5 & $-9.4  \times 10^{-19}$ & -0.14 mHz \\
Poor cooling & 100 & $-9.5 \times 10^{-18}$ & -1.4 mHz
\end{tabular}
\label{tab:dopperShiftTable}
\end{table}

Confining the ion in the Lamb-Dicke regime does not eliminate the second-order (relativistic) Doppler shift
\begin{equation}
\frac{\delta_{2D}}{\omega_0} = - \frac{\langle v^2 \rangle}{2c^2}.
\end{equation}
In terms of the trap secular frequency $\nu$ and mean vibrational occupation number $\langle n \rangle$ after cooling, this is
\begin{align*}
\frac{\delta_{2D}}{\omega_0} = - \frac{\langle v^2 \rangle}{2c^2} &= -\frac{\nu}{2} (2 \langle n \rangle + 1) \frac{\hbar}{2 M c^2} \\ 
&= -4.7 \times 10^{-20} \left( \frac{\nu / 2 \pi}{1 \text{ MHz}} \right) (2 \langle n \rangle +1) \label{eq:barium2Dshift}
\end{align*}
for a secular frequency $\nu / 2 \pi = 1$~MHz and the barium mass $Mc^2 \approx 137$~GeV.  For optimal Doppler cooling on a strong transition (i.e.\ $\Gamma \gg \nu$) the mean occupation number is given by \cite{wineland1987lcl}:
\begin{equation}
\langle n \rangle \approx \frac{1}{2}\left( \frac{\Gamma}{\nu} - 1 \right).
\end{equation}
For barium, assuming $\Gamma/2 \pi = 20$~MHz and $\nu / 2 \pi = 1$~MHz, this yields $\langle n \rangle \approx 9.5$.  The Table \ref{tab:dopperShiftTable} computes the shift for different cooling regimes.

So far this treatment has not considered the effects of micromotion which can turn out to be dominant.  See Section~\ref{sec:clockMicromotion} for their treatment.  Estimates \cite{wineland1987lcl} show that small uncompensated asymmetries in the trapping field lead to orders of magnitude more kinetic energy than the zero-point secular motion of the harmonic trap.  A robust minimization scheme \cite{berkeland1998mim} will be necessary.

\subsection{Static magnetic fields: the linear and quadratic Zeeman shifts}\label{sec:clockZeeman}
The linear Zeeman effect gives energy level shifts of
\begin{align}
\Delta E_{\gamma,F,m_F} = g_F(\gamma) \mu_B m_F B.
\end{align}
Since our clock states are both $m_F = 0$, there is no linear Zeeman shift on the clock transition.  For the other states, the $g_F$ factor is
\begin{equation}
\begin{split}
g_F(\gamma) = & g_J(\gamma) \frac{F(F+1) + J(J+1) - I(I+1)}{2F(F+1)} \\
& - g_I'(\gamma) \frac{F(F+1) - J(J+1) + I(I+1)}{2F(F+1)}
\end{split}
\end{equation} 
where $g_I'$ is the nuclear g-factor which is much smaller $(< 10^{-3})$ than the electronic factor $g_J$.  Notice that because $I = J = 3/2$ in the $5D_{3/2}$ state, $g_F(5D_{3/2})$ is independent of $F$:
\begin{align*}
g_F(5D_{3/2}; F) &= \frac{g_J - g_I'}{2} \approx 0.3995 \\
g_F(6S_{1/2}; F=2) &= \frac{g_J - 3g_I'}{4} \approx 0.5003 \\
g_F(6S_{1/2}; F=1) &= \frac{-g_J - 5g_I'}{4} \approx -0.5010
\end{align*}
with \cite{marx1998pgf,knoll1996egf},
\begin{align*}
g_J(5D_{3/2}) &= 0.799 327 8(3) \\
g_J(6D_{3/2}) &= 2.002 492 2 \times 10 \\
g_I(^{137}\text{Ba}) &= 0.623 876(3) \\
\Rightarrow g_I'(^{137}\text{Ba}) &= g_I(^{137}\text{Ba}) \frac{\mu_N}{\mu_B} \approx 0.624 \left(\frac{1}{1836}\right) = 3.399 \times 10^{-4}
\end{align*}

\begin{figure}
\begin{center}
\includegraphics[scale=0.75]{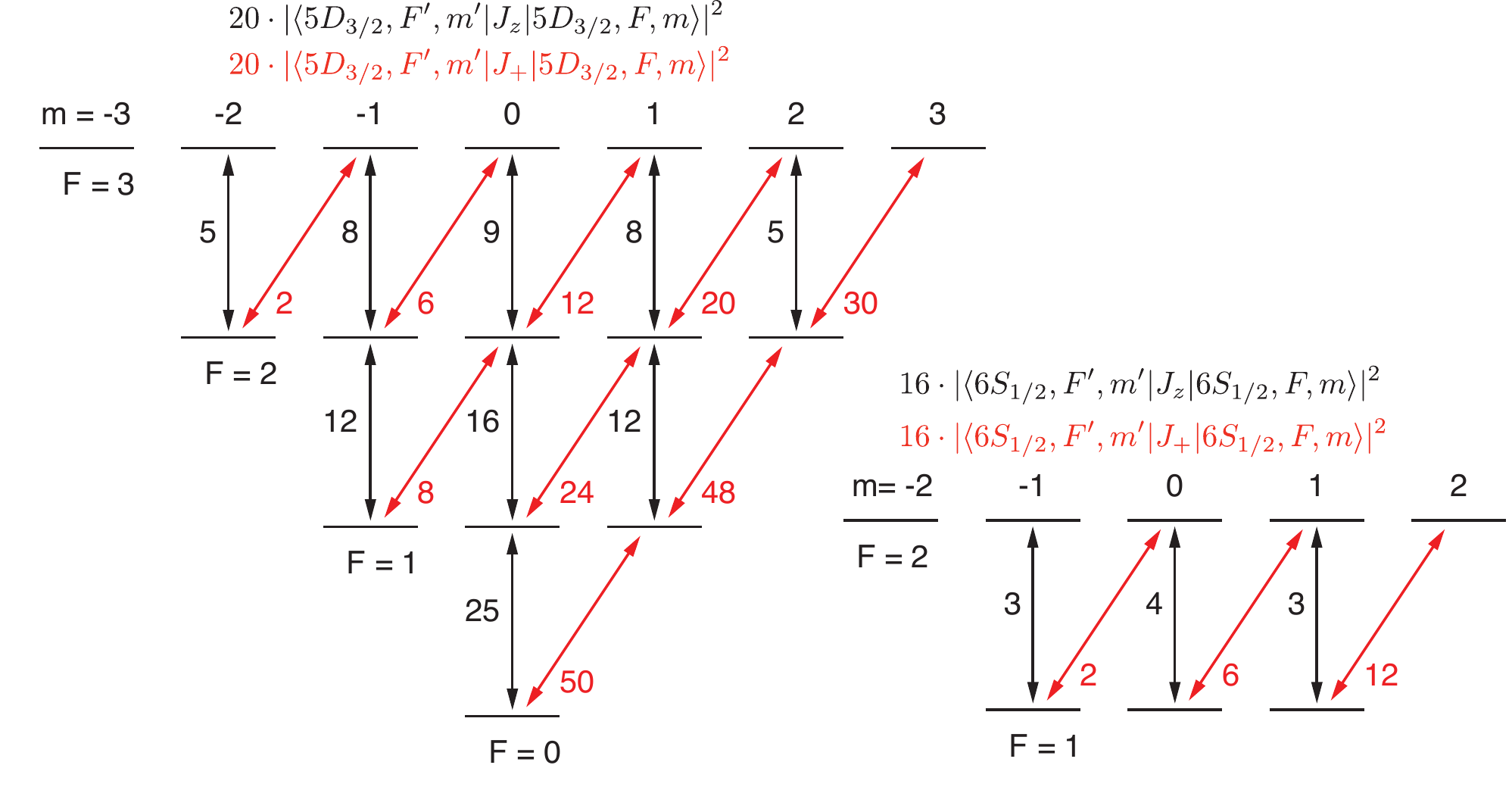}
\end{center}
\caption[$J_z$ and $J_+$ matrix elements in the $6S_{1/2}$ and $5D_{3/2}$ states]{$J_z$ and $J_+$ matrix elements in the $6S_{1/2}$ and $5D_{3/2}$ states.  These factors are useful in this context for calculating second-order Zeeman shifts.}
\label{fig:oddIsotopeJmatrixElements}
\end{figure}

Using the relevant $J_z$ matrix elements, listed in Figure \ref{fig:oddIsotopeJmatrixElements}, the second-order Zeeman shift (neglecting the nuclear term) is given by
\begin{equation}
\Delta_{\gamma,F,m_F} = \sum_{F',m_F'} \frac{| \langle \gamma, F, m_F | J_z | \gamma, F',m_F' \rangle|^2 g_J^2 \mu_B^2 B^2}{E_{F,m_F} - E_{F',m_F'}}
\end{equation}
Given the matrix elements for $J_z$ above, we see that the only relevant non zero couplings are 
\begin{align*}
|6S_{1/2}, F=2, m_F=0 \rangle \quad &\text{with } |6S_{1/2}, F=1, m_F=0 \rangle, \quad \text{and,} \\
|5D_{3/2}, F=0, m_F=0 \rangle \quad &\text{with } |5D_{3/2}, F=1, m_F=0 \rangle.
\end{align*}
The shifts are
\begin{align}
\Delta_{6S_{1/2}, F=2,m_F =0} &= \frac{| \langle 6S_{1/2}, 2,0 | J_z | 6S_{1/2}, 1,0 \rangle|^2 g_J^2 \mu_B^2 B^2}{E_{F = 2} - E_{F = 1}} \notag \\
&= \frac{(4/16)(2.0025)^2(1.3996 \text{ MHz/G})^2 B^2}{8037.7 \text{ MHz}} \notag \\
&= (+ 244 \text{ Hz/G$^2$} ) B^2 \\
\Delta_{5D_{3/2}, F=0,m_F =0} &= \frac{| \langle 5D_{3/2}, 0,0 | J_z | 5D_{3/2}, 1,0 \rangle|^2 g_J^2 \mu_B^2 B^2}{E_{F = 0} - E_{F = 1}} \notag \\
&= \frac{(25/20)(0.7993)^2(1.3996 \text{ MHz/G})^2 B^2}{-145 \text{ MHz}} \notag \\
&= (-10.78 \text{ kHz/G$^2$} ) B^2.
\end{align}

\begin{table}
\let\PBS=\PreserveBackslash
\centering
\caption{First and second-order Zeeman shifts for the $6S_{1/2}$ and $5D_{3/2}$ states.}
\begin{tabular}{c|>{\PBS\centering}m{0.9 in}>{\PBS\centering}m{0.9 in}} 
State & Second-order shift for $m=0$ (kHz/G$^{2}$) & First-order shift for $m \ne 0$ (kHz/G) \\ \hline \hline
$6S_{1/2}, F = 1$  & \multicolumn{1}{c}{-0.244} & \multicolumn{1}{c}{-701 $m_F$}\\
$6S_{1/2}, F = 2$  & \multicolumn{1}{c}{+0.244} & \multicolumn{1}{c}{+702 $m_F$}\\
$5D_{3/2}, F= 0$  & \multicolumn{1}{c}{-10.789} & \multicolumn{1}{c}{---}\\
$5D_{3/2}, F= 1$  & \multicolumn{1}{c}{+7.800}  &\multicolumn{1}{c}{+559 $m_F$} \\
$5D_{3/2}, F= 2$  & \multicolumn{1}{c}{+2.989}  &\multicolumn{1}{c}{+559 $m_F$} \\
$5D_{3/2}, F= 3$  & \multicolumn{1}{c}{+0.917} & \multicolumn{1}{c}{+559 $m_F$}
\end{tabular}
\label{tab:oddIsotopeZeemanShifts}
\end{table}

\begin{table}
\centering
\caption{Expected size/uncertainty of the Zeeman shift on the clock transition.}
\begin{tabular}{m{1.75 in}l|cc}
Conditions & Operating field & Absolute shift & Relative uncertainty \\ \hline \hline
\raggedright Current configuration & $\sim$1 G & -11.024(1) kHz & $\sim 8 \times 10^{-15}$ \\
\raggedright Optimized configuration, cycled cooling beam polarization & 1.00(1) mG & -11(10) mHz & $\sim 8 \times 10^{-19}$
\end{tabular}
\label{tab:zeemanShiftsTable}
\end{table}

Table \ref{tab:oddIsotopeZeemanShifts} summarizes all the first and second-order Zeeman shifts.  The shift to the clock transition frequency is
\begin{equation}
\Delta f_0^\text{Zeeman} = \Delta_{5D_{3/2}, F=0,m_F =0}- \Delta_{6S_{1/2}, F=2,m_F =0} = (-11.024 \text{ kHz/G$^2$}) B^2
\end{equation}
Luckily, the operating field will be small and well controlled.  We've demonstrated 10~$\mu$G stability on a $\sim 1$G background field using two layers of mu-metal shielding and a precision current supply over several hours during our rf-spectroscopy measurements \cite{sherman2005pml}.  Therefore it is reasonable to expect much better than 0.1~mG stability on a 1~mG field used to break the degeneracy of magnetic sub-levels in the $6S_{1/2}, F=2$ state.  As shown in Table \ref{tab:zeemanShiftsTable}, this puts the relative systematic uncertainty of the effect below $10^{-18}$.

Note that the magnetic field can be precisely measured by Zeeman transitions in either the $6S_{1/2}, F=2$ or $5D_{3/2}, F=2$ manifolds;  both are completely free of systematics due to electric field gradients (shown later).  Note also that the shift in the $5D_{3/2}, F=2$ state is considerably smaller.  This might make a better clock transition since the second-order zeeman shift is smaller by a factor of 5.

\subsection{The blackbody Stark shift}
The Plank radiation law states that the electric field energy density at temperature $T$ due to black-body radiation is \cite{itano1982shs}
\begin{equation}
E^2(\omega) = \frac{8 \alpha}{\pi} \frac{\omega^3}{e^{\omega/k_B T} -1} d\omega .
\end{equation}
Since we are considering temperatures below room temperature, most of the radiation is at frequencies far below the electric-dipole transitions in Ba$^+$.  Therefore, we can treat the energy level shifts due to this radiation as a time-averaged quadratic Stark shift.  The time averaged electric field strength of the Blackbody field obtained by integrating the equation above scales strongly with temperature:
\begin{equation}
\langle E^2 \rangle  = \left(8.319 \text{ (V/cm)}\right)^2 \left( \frac{T}{300 \text{ (K)}} \right)^4
\end{equation}

In general, the energy level shift of each state has a scalar and tensor part.  For isotropic blackbody radiation, the tensor part is zero, but it is interesting that the tensor term is identically zero for our clock states notwithstanding.
\begin{align}
\Delta E^\text{blackbody} &= -\frac{1}{2} \alpha_\text{scalar}(\gamma,J,F) E^2 \\
	&+ -\frac{1}{4} \alpha_\text{tensor}(\gamma,J,F) \frac{3m_F^2 - F(F+1)}{F(2F-1)}(3E_z^2 - E^2).
\end{align}
The scalar polarizability $\alpha_\text{scalar}$ is expressible as
\begin{equation}
\alpha_\text{scalar}(\gamma,J,F) = \alpha_\text{scalar}(\gamma,J) = 
\frac{8 \pi \epsilon_0}{3 (2J+1)} \sum_{\gamma',J'} \frac{| \langle \gamma J|| ez || \gamma' J' \rangle|^2}{E_{\gamma',J'} - E_{\gamma,J}}
\end{equation}
The tensor polarizability is a similar summation but includes a $6j$-symbol which vanishes for our states:
\begin{align*}
\alpha_\text{tensor}(\gamma,J,F) &\propto \left\{\begin{array}{ccc}  F & J & I  \\ J & F & 2 \end{array} \right\}  \alpha_\text{tensor}(\gamma,J) \\
 \left\{\begin{array}{ccc}  0 & 3/2 & 3/2  \\ 3/2 & 0 & 2 \end{array} \right\}   &= 0 \qquad 5D_{3/2}, F=0 \text{ state} \\
  \left\{\begin{array}{ccc}  2 & 1/2 & 3/2  \\ 1/2 & 2 & 2 \end{array} \right\}   &=0 \qquad 6S_{1/2}, F = 2 \text{ state}
\end{align*}

With the dipole matrix elements given in Table~\ref{tab:dipoleMatrixElementTable}, one can calculate the expected level shifts:
\begin{align*}
\begin{split}
\Delta_{6S_{1/2}}^\text{BB} &= -\frac{E^2e^2}{3c} \left[
\frac{|\langle 6S_{1/2} ||r|| 6P_{1/2} \rangle|^2}{W_{6P_{1/2}}} +
\frac{|\langle 6S_{1/2} ||r|| 6P_{3/2} \rangle|^2}{W_{6P_{3/2}}} \right. \\
&\left. \qquad + \frac{|\langle 6S_{1/2} ||r|| 7P_{1/2} \rangle|^2}{W_{7P_{1/2}}} + 
\frac{|\langle 6S_{1/2} ||r|| 7P_{3/2} \rangle|^2}{W_{7P_{3/2}}} + \cdots \right]
\end{split} \\
&= -\frac{E^2}{3} \left[0.0298 + 0.0549 + 10^{-5} + 10^{-4} + \cdots \text{ (Hz/(V/cm)$^2$)} \right] \\
&= -E^2 (28.2 \text{ (mHz/(V/cm)$^2$)}, \\
\begin{split}
\Delta_{5D_{3/2}}^\text{BB} &= -\frac{E^2e^2}{3c} \left[
\frac{|\langle 5D_{3/2} ||r|| 6P_{1/2} \rangle|^2}{W_{6P_{1/2}} - W_{5D_{3/2}}} +
\frac{|\langle 5D_{3/2} ||r|| 6P_{3/2} \rangle|^2}{W_{6P_{3/2}} - W_{5D_{3/2}}} \right. \\
&\left. \qquad + \frac{|\langle 5D_{3/2} ||r|| 4F_{5/2} \rangle|^2}{W_{7P_{1/2}} - W_{5D_{3/2}}} +
\frac{|\langle 5D_{3/2} ||r|| 4F_{7/2} \rangle|^2}{W_{7P_{3/2}} - W_{5D_{3/2}}} + \cdots \right]
\end{split} \\
&= -\frac{E^2}{3} \left[0.0309 + 0.0053 + 0.0171 + 0.0032 + \cdots \text{ (Hz/(V/cm)$^2$)} \right] \\
&= -E^2 (18.9 \text{ (mHz/(V/cm)$^2$).}
\end{align*}

Therefore the shift in the clock transition is
\begin{equation}
\Delta f_0^\text{BB} = \Delta_{5D_{3/2}}^\text{BB} - \Delta_{6S_{1/2}}^\text{BB}  = E^2 (9.3  \text{ (mHz/(V/cm)$^2$)}).
\end{equation}

\begin{table}
\centering
\caption{Estimated size/uncertainty of the blackbody shift on the clock transition.}
\begin{tabular}{rc|ccc}
Temperature & $\langle E^2 \rangle$ (V/cm)$^2$ & $\Delta f_0^\text{BB}$ (mHz) & Relative shift & Relative shift uncertainty \\ \hline \hline
300 K & $(8.319)^2$ & 643 & $4 \times 10^{-15}$ & $\sim 1 \times  10^{-16}$ \\
77 K & $(0.548)^2$ & 2.79 & $2 \times 10^{-17}$ &  $\sim 1 \times  10^{-18}$ \\
4 K  & $(0.002)^2$ & $3 \times 10^{-5}$ & $2 \times 10^{-22}$ & $\sim 1 \times 10^{-22}$
\end{tabular} 
\label{tab:blackbodyShifts}
\end{table}

The results are summarized in Table \ref{tab:blackbodyShifts} for a range of operating temperatures.  Note that the relative shift uncertainty is largely due to $< 5$\% uncertainties in the important matrix elements in the calculation, a justification in itself for the light shift work presented in Chapter~\ref{sec:lightShiftChapter}.  For extremely cold temperatures, the coupling between hyperfine levels needs to be considered since the peak of the black-body spectral energy spectrum enters the radio-frequency regime.

The magnetic field in blackbody radiation contains an equal amount of radiation energy as the electric field.  For near-zero background magnetic field and at temperatures such that the blackbody peak is at much higher frequencies than hyperfine splittings, the shift in hyperfine splittings is \cite{itano1982shs}
\begin{align}
\frac{\delta \omega_\text{hfs}}{\omega_\text{hfs}} &\approx - \frac{\alpha^2}{g_J - g_I}^2 \int_0^\infty \frac{B^2(\omega)}{\omega^2} d \omega \\
& = -\frac{\pi}{18} (g_J - g_I)^2 \alpha^5 (kT)^2
\end{align}

With $g_J = 2$ and ignoring the nuclear g-factor, this relative shift is 
\begin{equation}
\frac{\Delta E_{6S_{1/2}}^\text{Blackbody Zeeman}}{\hbar \omega_\text{clock}} \approx -1.304  \times 10^{-17} [T(\text{K})/300]^2
\end{equation}
which is a good estimate for the $6S_{1/2}$ state.  The $5D_{3/2}$ state shift should be considered separately, but this magnetic term is likely to be smaller than the electric black-body component and similarly negligible at cryogenic temperatures.

\subsection{Stray field gradients: the quadrupole dc-field Stark shift is zero}
One of the special features of the proposed barium ion clock transition is that it is free to first order of Stark shifts due to electric field \emph{gradients}.  In other trapped ion systems these shifts can be in the 1~Hz regime \cite{itano2000efs}.  Techniques for dealing with these shifts have been developed---either averaging the clock transition under three orthogonal magnetic fields or interrogating many magnetic sublevels to obtain the shift directly (see~\cite{itano2000efs, margolis2004hlm,oskay2005mhq, dube2005eqs}).  Our system is completely free of this shift to first order so no such cancellation techniques need to be employed.
\begin{align}
\Delta E &= \langle \gamma J F m_F | H_Q | \gamma J F mF \rangle \nonumber \\
&= \frac{\left( 3m_F^2 - F(F+1) \right)}{\left( (2F+3)(2F+2)(2F+1)2F(2F-1)\right)^{1/2}} f(A,\epsilon, \alpha, \beta) \langle \gamma J F || \Theta^{(2)} || \gamma J F \rangle \\
\intertext{where} 
f(A,\epsilon, \alpha, \beta) &= A \left( (3 \cos^2 \beta - 1) - \epsilon \sin^2 \beta (\cos^2 \alpha - \sin^2 \alpha) \right)
\end{align}
describes the strength, orientation, and alignment of the electric field gradient and the quadrupole matrix element is expressible in terms of the electronic quadrupole moment $\Theta(\gamma, J)$ in the IJ-coupling approximation as
\begin{equation}
\langle \gamma J F || \Theta^{(2)} || \gamma J F \rangle =  (-1)^{I+J+F} (2F+1) \left\{\begin{array}{ccc}  J & 2 & J  \\ F & I & F \end{array} \right\}  \left( \begin{array}{ccc}  J & 2 & J  \\ -J & 0 & J \end{array} \right)^{-1} \Theta(\gamma, J). \label{eq:quadShift}
\end{equation}

\begin{table}
\centering
\caption{Table of $6j$-symbols and angular factors relevant for quadrupole shifts.}
\begin{tabular}{c|c c}
$F,m_F$ & $3m_F^2 - F(F+1)$ & $\left\{\begin{array}{ccc}  J & 2 & J  \\ F & I & F \end{array} \right\}$ \\ \hline \hline
$0,0$ & 0 		& 0 \\ \hline
$1,0$& -2 		& \multirow{2}{*}{ $\displaystyle -\frac{1}{5}\sqrt{\frac{2}{3}}$ }\\
$1,\pm1$ & 1 	& \\ \hline 
$2,0$& -6  	& \multirow{3}{*}{0}\\
$2,\pm1$ & -3	& \\
$2,\pm2$ & 6 	& \\ \hline
$3,0$& -12 	& \multirow{4}{*}{ $\displaystyle \frac{1}{5}\sqrt{\frac{3}{7}}$}\\
$3,\pm1$ & -9 \\
$3,\pm2$ & 0 \\ 
 $3,\pm3$ & 15
\end{tabular}
\label{tab:quadrupoleAngularFactors}
\end{table}

The $6S_{1/2}$ state of barium has less electronic angular momentum than is required to couple to tensor operators ($J<1$), and therefore no electronic quadrupole moment $\Theta(6S, 1/2) = 0$.  The ground state is therefore only coupled to electric field gradients through a nuclear interaction ($I > 1$) which is many orders of magnitude weaker.  Though the excited clock state $5D_{3/2}$ has a finite quadrupole moment $\Theta(5D, 3/2) = 2.297 ea_0^2$ \cite{itano2006qmh}, the hyperfine state $F = 0$ has no total angular momentum and is thus free from any shift.  In other words, the 6$j$-symbol present in Eq.~\ref{eq:quadShift} evaluates to zero for $F=0$.  Interestingly, this is also true for the $F=2$ state meaning that it is a strategically important manifold for measuring other systematic shifts or perhaps a more viable clock transition (for instance, the second-order Zeeman shift is a factor of 6 smaller). Further analysis reveals that the factor $\left( 3m_F^2 - F(F+1) \right)$ is 0 for the $F = 3, m_F = \pm 2$ sub-states.  Table \ref{tab:quadrupoleAngularFactors} contains all the $5D_{3/2}$ angular factors.

%\begin{equation}
% \left( \begin{array}{ccc}  3/2 & 2 & 3/2  \\ -3/2 & 0 & 3/2 \end{array} \right)^{-1} = 2 \sqrt{5}
%\end{equation}

\subsection{The rf trapping field: the quadrupole ac-field Stark shift}
We just showed that both the ground and excited states are free from static field dc quadrupole Stark shifts to first order.  However, the next order term might be of considerable size since the large oscillating  rf trap field contributes.  The $6S_{1/2}$ still has a negligible shift at this order of perturbation theory  because the only nearby states are hyperfine components of $6S_{1/2}$ which has no electronic quadrupole moment ($J < 1$).

The $5D_{3/2}, F=0$ state does show a substantial second-order shift, however, due to coupling with the nearby $F'=2$ state (coupling to $F=1$ and $F=3$ are not allowed by the electric quadrupole selection rules:  $F + F' \ge 2$).  The size of the shift depends critically on the orientation of the trap field with respect to the quantization axis:  there may be a particular orientation that makes the shift 0.  However, to naively estimate the size of the effect, we first derive the electric field gradient beginning with the trap potential:
\begin{align*}
\phi(x,y,z) &= \frac{x^2 + y^2 - 2z^2}{r_0^2} \frac{V_0}{\eta} \cos \Omega t \\
\boldsymbol{E} = - \nabla \phi &= -\frac{x \bhat{x} + y \bhat{y} - 2z \bhat{z}}{r_0^2} \frac{2V_0}{\eta} \cos \Omega t \\
\nabla \boldsymbol{E} &= - \frac{\bhat{x}\bhat{x} + \bhat{y}\bhat{y} - 2\bhat{z}\bhat{z}}{r_0^2} \frac{2V_0}{\eta} \cos \Omega t = \frac{2V_0}{\eta} \cos \Omega t  (\bhat{r}\bhat{r} - 3\bhat{z}\bhat{z}) \\
\Rightarrow | \nabla \boldsymbol{E} | &\equiv \sqrt{ \langle (\nabla \boldsymbol{E})^2 \rangle} = \frac{2V_0}{\eta r_0^2} \langle \cos^2 \Omega t \rangle = \frac{V_0}{\eta r_0^2}
\end{align*}
where $\eta \sim 10$ represents an `efficiency factor' that accounts for the non-hyperbolic electrode shape in the Paul-Straubel trap (see the discussion in Section~\ref{sec:realWorldTrapShapes}).

An estimate for the shift in the $5D_{3/2}$ state is then: 
\begin{equation}
\Delta_{5D_{3/2}} \sim \frac{1}{2} \sum_{F',m',\pm} \frac{|\langle 5D_{3/2}, F=0 \left| e(\nabla E) r^2 \right|5D_{3/2}, F' \rangle |^2}{(\omega_{F=0} - \omega_{F'}) \pm \omega_\text{rf}} \\
\end{equation}
Because $\omega_\text{rf}/2 \pi = 10$~MHz is much smaller than the relevant hyperfine splitting $(\omega_{F=0} - \omega_{F'=2})/2 \pi = 480$~MHz, we can drop the `light-shift detuning' term from the denominator and write
\begin{align*}
\Delta_{5D_{3/2}} &\sim \frac{|\langle 5D_{3/2}, F=0 \left| e(\nabla E) r^2 \right|5D_{3/2}, F'=2 \rangle |^2}{(\omega_{F=0} - \omega_{F'})} \\
&\sim \frac{|\langle 5D_{3/2} \left| e r^2 \right|5D_{3/2},\rangle |^2 (V_0 / \eta)^2}{r_0^4(\omega_{F=0} - \omega_{F'=2})} \\
&= \frac{(2.3 ea_0^2)^2 (V_0 / \eta)^2}{r_0^4 (480 \text{ MHz})}
\end{align*}
For a $r_0 = 0.5$~mm trap with $V_0 / \eta = 100$ V rf voltage, and using the factors $a_0 = 0.053$~nm and 1 eV = $2.42 \times 10^{14}$~Hz, we have:
\begin{align*}
\Delta_{5D_{3/2}} &\sim \frac{(2.3)^2 (0.053 \text{ nm})^4 (100)^2 (2.42 \times 10^{14} \text{ Hz})^2}{(500  \times 10^3 \text{ nm})^4 (480  \times 10^6 \text{ Hz})} \\
&= 0.8 \text{ mHz}.
\end{align*}
This shift is potentially serious only if the trap rf amplitude varies wildly.  Patch potentials, modulated by the trap rf will have a totally insignificant effect.

\subsection{The (micromotion-enhanced) dipole ac-field Stark shift}\label{sec:clockMicromotion}
A single ion trapped in the center of an rf Paul trap experiences an oscillating electric field at $\omega_\text{rf}/2\pi \approx 10$~MHz.  In the best case this field is due only to finite secular motion bringing the ion out from its equilibrium position where the rf-field is zero.  Any patch potential or uncompensated dc-field pushes the ion to an equilibrium position where $\langle E^2_\text{rf} \rangle$ becomes large---micromotion results.  Following \cite{berkeland1998mim}, we will estimate the likely size of this effect and calculate its impact on the clock transition;  the formalism will be identical to the blackbody shift since both assume a second-order Stark shift due to a time averaged field.

Recall from Section~\ref{sec:paulTrap} that the rf-driving voltage $V_0 \cos(\omega_\text{rf} t)$ and dc-ring voltage $U_0$ can be written as  dimensionless parameters describing the trap modes and stability:
\begin{align*}
q_z = -2q_r &= \frac{-4eV_0}{mr_0^2 \omega_\text{rf}^2} \qquad \text{(rf voltage term)} \\
a_z = -2a_r &= \frac{-8eU_0}{mr_0^2 \omega_\text{rf}^2} \qquad \text{(dc voltage term)}
\end{align*}
where $r_0$ is the dimension of the trap, $e$ and $m$ are the charge and mass of the ion.  Also recall that the typical operating conditions are $|q_i| \ll 1$ and $|a_i| \ll 1$ (the pseudo-potential approximation) which makes the trap secular frequencies $\nu \ll \omega_\text{rf}$:
\begin{equation}
\nu \cong \frac{1}{2} \omega_\text{rf} \sqrt{a_i + \frac{1}{2} q_i^2}
\end{equation}
We often work with $a_i \ll q_i$ and a secular frequency $\omega_s \approx \omega_\text{rf} /10$.  In this case, the above equation gives $q_i \approx 0.28$.  

To first order, the solution to the Mathieu equations gives the orbits of the ion:
\begin{equation*}
u_i(t) \approx u_{1i} \cos (\nu t + \phi_i) \left[1 + \frac{q_i}{2} \cos  \omega_\text{rf} t \right].
\end{equation*}
Here we see an important feature: the motion is dominated by sinusoidal secular motion at $\nu$ but is subject to micromotion at $\omega_\text{rf}$ that increases with $q_i$, roughly the strength of the rf-field.  This micromotion is unavoidable:  cooling the secular motion will reduce it by reducing the orbit amplitude $u_{1i}$ but cannot eliminate it.  While side-band cooling can cool to the ground state of secular motion, $\langle n_i \rangle \approx 0$, this cannot be done to the micromotion since it is driven motion.

However, much larger \emph{excess} micromotion can be observed if the ion's equilibrium position is displaced from the rf-symmetry point in the trap by some stray field $\boldsymbol{E}_\text{dc}$.  Micromotion can also be induced by a phase-imbalance in the electrodes comprising the trap, but for simplicity this effect will be ignored here.  For a small perturbing field, the secular orbit amplitude $u_{1i}$ is unchanged to first order, making the orbit equation
\begin{equation*}
u_i(t) \approx \left( \frac{Q \boldsymbol{E}_\text{dc} \cdot \bhat{u}_i}{m \nu^2} +  u_{1i} \cos (\nu t + \phi_i) \right) \left[1 + \frac{q_i}{2} \cos  \omega_\text{rf} t \right].
\end{equation*}

The kinetic energy along each direction can be greatly increased by the presence of such excess micromotion \cite{berkeland1998mim}:
\begin{equation*}
E_{\text{kin},i} =\frac{1}{4} m u_{1i}^2 \left( \nu^2 + \frac{1}{8} q_i^2  \omega_\text{rf}^2 \right) +  \underbrace{\frac{4}{m} \left( \frac{Q q_i \boldsymbol{E}_\text{dc} \cdot \bhat{u}_i}{(2a_i + q_i^2)  \omega_\text{rf}}\right)^2}_\text{Excess micromotion} .
\end{equation*}
The first term is from the secular and unavoidable micromotion while the second is induced by the excess micromotion. With some typical parameters, $q_i = 0.28$, $Q = e$, $mc^2 = 137$~GeV, and $ \omega_\text{rf} = 2 \pi \cdot 10$~MHz one can show that a $\boldsymbol{E}_\text{dc} \cdot \bhat{u}_i = 1$~V/cm field can yield a kinetic energy component of $8.4  \times 10^{-5}$~eV which is of order 1~K.  This far exceeds the $\sim 1$~mK one expects from just the secular energy and clearly has an impact on the second-order Doppler shift.

The time-averaged electric field can also be computed under these conditions \cite{berkeland1998mim}:
\begin{equation}\label{eq:eFieldDueToMicromotion}
\langle E_i^2 \rangle \cong \frac{m  \omega_\text{rf}^2 k_B T_i}{2 Q^2} \frac{a_i + 2q_i^2}{2a_i + q_i^2} + \underbrace{8\left( \frac{q_i \boldsymbol{E}_\text{dc} \cdot \bhat{u}_i}{2a_i + q_i^2}\right)^2}_\text{Excess micromotion}.
\end{equation}
Again, the first term is due to unavoidable micromotion that scales with the ion temperature.  The second term, due to excess micromotion, can be much larger than $|\boldsymbol{E}_\text{dc}|^2$ because we often operate with $a_i \ll q_i \ll 1$.  Notice that the term due to excess micromotion actually grows as $q_i \to 0$.  This means that a stiff trap with moderate $q_i$ is favorable, even though this corresponds to increasing the amount of rf on the trap electrode.  The intuitive reason is that a stronger pseudo-potential means that a given stray electric field results in a smaller displacement of the ion from the rf symmetry point of the trap.

To set some limits, consider the first term under the same trap conditions posed above assuming an ion cooled to 1~mK:
\begin{align*}
\langle E_i^2 \rangle_\text{1 mK}^\text{(0)} & \approx \frac{m  \omega_\text{rf}^2 k_B T_i}{2 Q^2} \frac{a_i + 2q_i^2}{2a_i + q_i^2} \\
& = \frac{(138 \times 10^9 \text{ eV})(2\pi \cdot 10 \times 10^6 \text{ Hz})^2 (8.3 \times 10^{-8} \text{ eV})}{2 e^2 (3 \times 10^{10} \text{ cm/s})^2}(2) \\
& = (7.1 \text{ mV/cm})^2
\end{align*}
This is orders of magnitude smaller than the time-averaged field due to blackbody radiation at room temperature $(8.3 \text{ V/cm})^2$ or at 77~K $(550 \text{ mV/cm})^2$.  However, note that this term scales linearly with ion temperature so poor cooling rates or large ion heating rates can increase the amount of rf seen dramatically.

The term due to excess micromotion in Eq.~\ref{eq:eFieldDueToMicromotion}, under the proposed conditions with a 1~V/mm stray field, is
\begin{align*}
\langle E_i^2 \rangle_\text{1 mK}^\text{(1)} & \approx 8\left( \frac{q_i \boldsymbol{E}_\text{dc} \cdot \bhat{u}_i}{2a_i + q_i^2}\right)^2 \\
&= 8 \frac{(1 \text{ V/cm})^2}{0.28} = (5.4 \text{ V/cm})^2
\end{align*}
which is comparable to the room temperature blackbody shift and dominates at lower temperatures.  This term can in principle be made 0 by applying a compensation field $\boldsymbol{E}_\text{applied} = - \boldsymbol{E}_\text{dc}$.  Again, one must compensate for the excess micromotion due to stray fields in all three directions.

For the purposes of systematic shift estimation, we will assume we can compensate a stray field of 1V/cm to 1\% and maintain an ion secular motion temperature $T <$ 1 mK.
\begin{equation*}
\Delta f_0^\text{micromotion} = \Delta_{5D_{3/2}} - \Delta_{6S_{1/2}}  = \langle E^2 \rangle_\text{micromotion} (9.3  \text{ (mHz/(V/cm)$^2$)})
\end{equation*}
with $ \langle E^2 \rangle_\text{micromotion} = (55 \text{ mV/cm})^2$, the relative shift is $2 \times 10^{-19}$.

\subsection{The clock laser ac-Stark shift}
The 2051~nm laser, while resonant with the $6S_{1/2}, F=0, m=2 \leftrightarrow 5D_{3/2}, F=0, m=0$ will cause off-resonant ac-Stark shifts due to dipole coupling of $6S_{1/2}$ to $nP$ levels and $5D_{3/2}$ to $nP$ and $nF$ levels.  Furthermore, off-resonant electric quadrupole coupling between the clock states and nearby hyperfine components are also present.  The relative importance of these couplings will depend on the clock laser intensity as well as the magnitude of the magnetic field which splits the hyperfine degeneracy.

First, we need to estimate what the intensity of the 2051~nm clock laser needs to be to achieve a sufficient Rabi-flopping rate $\Omega$.  For the electric quadrupole transition:
\begin{equation}
\Omega = \frac{1}{2h} \langle 6S_{1/2} || er^2 || 5D_{3/2} \rangle \sum_{q} \frac{(j'm'|2q,jm)}{\sqrt{2j' +1}} \nabla E_q
\end{equation}
Neglecting angular factors (and the $F \to J$ decomposition) to get an order of magnitude estimate, $|\nabla E| \sim 2E / \lambda = 2E / 2051$~nm.  We will also use the factors:
\begin{align*}
I \text{ [mW/cm$^2$]} &\approx 1.33(E \text{ [V/cm]})^2 \\
\frac{e a_0}{h} &= 1.28 \left[\frac{\text{MHz}}{\text{V/cm}} \right] \\
\langle 6S_{1/2} || er^2 || 5D_{3/2} \rangle &= 12 ea_0^2 
\end{align*}
where the quadrupole matrix element is computed in \cite{gopakumar2002edq}.  Therefore,
\begin{align*}
\Omega \text{ [Hz]} &\approx (12) \sqrt{\frac{I}{1.33 \text{ [mw/cm$^2$]}}} \frac{0.05 \text{ nm}}{2051 \text{ nm}} (1.28 \times 10^6) \\
& \approx (325 \text{ [Hz]})  \sqrt{\frac{I}{1 \text{ [mw/cm$^2$]}}}
\end{align*}

This assumes a laser linewidth equal to or less than the state lifetime which isn't likely in the near future given the extreme narrowness of the transition, $\Gamma = 13$~mHz.  In practice, more laser power will have to be used to achieve a given Rabi flopping rate $\Omega$;  this extra power goes to linearly increasing the estimates for light shift systematics.

Leaving out angular factors that will depend on the polarization and alignment of the 2051~nm laser, the dipole light-shift of the $6S_{1/2}$ state can be estimated as:
\begin{align*}
\Delta_{6S_{1/2}} &=  -\frac{E^2}{4c} \sum_{\gamma'} \frac{|\langle 6S_{1/2} || r || \gamma' \rangle|^2}{W_{\gamma} - W_{6S_{1/2}}  - k_\text{clock}} \\
\begin{split}
			   &= -\frac{E^2}{4c} \left[
\frac{|\langle 6S_{1/2} ||r|| 6P_{1/2} \rangle|^2}{W_{6P_{1/2}} - k_\text{clock}} +
\frac{| \langle 6S_{1/2} ||r|| 6P_{3/2} \rangle|^2}{W_{6P_{3/2}} - k_\text{clock}} \right. \\
&\left. \qquad +
\frac{| \langle 6S_{1/2} ||r|| 7P_{1/2} \rangle|^2}{W_{7P_{1/2}} - k_\text{clock}} +
\frac{| \langle 6S_{1/2} ||r|| 7P_{3/2} \rangle|^2}{W_{7P_{3/2}} - k_\text{clock}} + \cdots \right]
\end{split} \\
& = -\frac{E^2}{4} \left[ 0.0393 + 0.0710 + 2 \times 10^{-5} + 0.0002 + \cdots \text{ [Hz/(V/cm)$^2$]}\right] \\
&= -\frac{E^2}{4} (0.1104 \text{ [Hz/ (V/cm)$^2$]}).
\end{align*}
The shifts on the $5D_{3/2}$ state are:
\begin{align*}
\Delta_{5D_{3/2}} &=  -\frac{E^2}{4c} \sum_{\gamma'} \frac{|\langle 5D_{3/2} || r || \gamma' \rangle|^2}{W_{\gamma} - W_{5D_{3/2}}  - k_\text{clock}} \\
\begin{split}
			   &= -\frac{E^2}{4c} \left[
\frac{|\langle 5D_{3/2} ||r|| 6P_{1/2} \rangle|^2}{W_{6P_{1/2}} - 2k_\text{clock}} +
\frac{| \langle 5D_{3/2} ||r|| 6P_{3/2} \rangle|^2}{W_{6P_{3/2}} - 2k_\text{clock}} \right. \\
&\left. \qquad + \frac{| \langle 5D_{3/2} ||r|| 7P_{1/2} \rangle|^2}{W_{4F_{5/2}} - 2k_\text{clock}} +
\frac{| \langle 5D_{3/2} ||r|| 7P_{3/2} \rangle|^2}{W_{4F_{7/2}} - 2k_\text{clock}} + \cdots \right]
\end{split} \\
& = -\frac{E^2}{4} \left[ 0.0450 + 0.0074 + 0.0193 + 0.0036 + \cdots \text{ [Hz/(V/cm)$^2$]}\right] \\
&= -\frac{E^2}{4} (0.0753 \text{ [Hz/ (V/cm)$^2$]}).
\end{align*}

Therefore, the dipole light shift on the clock transition is of order
\begin{equation}
\Delta f_0^\text{(dipole LS)} = \Delta_{5D_{3/2}}- \Delta_{6S_{1/2}} = E^2 (8.8  \text{ [mHz/(V/cm)$^2$}]).
\end{equation}

\begin{table}
\centering
\caption{Estimated sizes/uncertainties of the clock laser light shifts.}
\begin{tabular}{rl|ccc}
$\Omega_\text{clock}$ (Hz) & $I_\text{clock}$ (mW/cm$^2$) & $\Delta f_0^\text{(dipole LS)}$ (mHz) & $\Delta f_0^\text{(E2 LS)}$ (mHz) & $\Delta f_\text{tot}/f_0$ \\ \hline \hline
1 & $10^{-5}$  & $7 \times 10^{-5}$  & $-5 \times 10^{-6}$ & $ 5 \times 10^{-22}$ \\
10 & $10^{-3}$ & $7 \times 10^{-3}$ & $-5 \times 10^{-4}$ & $ 5 \times 10^{-20}$ \\
100 & $0.1$ &  0.7 & $-5 \times 10^{-2}$ & $ 5 \times 10^{-18}$ \\
1000 & $10$	   & 70		    & -5   &  $5 \times 10^{-16}$
\end{tabular}
\label{tab:clockLightShifts}
\end{table}

The quadrupole light shifts coming from off-resonant coupling to nearby hyperfine components have complicated dependence on alignment and polarization of the clock laser beam with respect to the magnetic field.  There will be configurations that minimize the shifts.  We can establish the order of magnitude of such shifts, however.

The nearest quadrupole couplings to the clock transition are $6S_{1/2}, F=2 \leftrightarrow 5D_{3/2}, F =1,2,3$.  An estimate for these shifts, in terms of E2 Rabi frequencies $\Omega$ and detunings $\delta$ is:
\begin{align*}
\Delta f_0^\text{(E2 LS)} &\sim \left( \frac{\Omega_{F=2 \leftrightarrow 1}^2}{2 \delta_{F=2 \leftrightarrow 1}} + \frac{\Omega_{F=2 \leftrightarrow 2}^2}{2 \delta_{F=2 \leftrightarrow 2}}
+ \frac{\Omega_{F=2 \leftrightarrow 3}^2}{2 \delta_{F=2 \leftrightarrow 3}} \right) \\
&\sim \frac{\Omega^2}{2} \left(\frac{1}{-145 \text{ MHz}} + \frac{1}{-480 \text{ MHz}} +\frac{1}{-1094 \text{ MHz}} \right)\\
&= \Omega^2 (-5  \times 10^{-9} \text{ Hz}^{-1}).
\end{align*}
As shown in Table \ref{tab:clockLightShifts}, this shift only becomes important for $\Omega > 1$kHz Rabi frequencies.

If we were to choose the  $6S_{1/2}, F=2, m=0 \leftrightarrow 5D_{3/2}, F'=2, m'=0$ transition instead, the dominant quadrupole light shifts would come from neighboring Zeeman components.  In general the shift will decrease with increasing magnetic field.  Notice also that the Zeeman structure is almost balanced: the light shifts due to the $m'=+1,+2$ states are almost exactly canceled by the shifts due to the $m'=-1,-2$ states.  This cancellation is not perfect due to both the quadratic Zeeman shift and excess laser broadening.  For an order of magnitude estimate, we assume $B = 1.0$~mG and see that
\begin{align*}
\Delta_{F=2, m=0}^\text{Zeeman} &= 3 \text{ mHz}, \\
\Delta_{F=2, m=\pm1}^\text{Zeeman} &= \pm560 \text{ Hz}, \\
\Delta_{F=2, m=\pm2}^\text{Zeeman} &= \pm1120 \text{ Hz}.
\end{align*}
Therefore, the net quadrupole shift to this alternate clock transition is
\begin{align}
\begin{split}
\Delta f_\text{alt. clock.}^\text{(E2 LS)} &\sim \frac{\Omega^2}{2} \left(\frac{1}{560.003} - \frac{1}{559.997} \right. \\
&\left. \qquad+ \frac{1}{1120.003} - \frac{1}{1119.997} + \text{ (other hfs couplings)} \right)
\end{split} \\
& = \frac{\Omega^2}{2} (-1  \times 10^{-8} \text{ Hz}^{-1}).
\end{align}

\subsection{Systematic shift summary}
\begin{table}[p]
\centering
\caption{A summary of estimated sizes/uncertainties of clock systematic shifts}
\small
\setlength{\extrarowheight}{5pt}
\begin{tabular}{m{1.0 in}m{1.25 in} |m{1.25 in}|cc}
\multicolumn{2}{c}{Effect} & \multicolumn{1}{c}{Conditions} & $\Delta f_0 / f_\text{clock}$ & Uncertainty \\ \hline \hline
First-order Doppler 		& \includegraphics[width=1.25in]{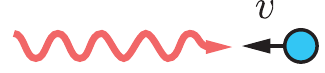} &$<1$~nm/s systematic motion & 0 & $\pm 3 \times 10^{-18}$ \\
Second-order Doppler 	& \includegraphics[width=1.25in]{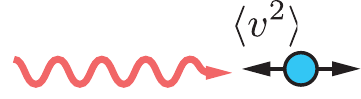} &$\langle n \rangle < 9.5$, $T < 1$~mK, $\omega_s/2\pi = 1$~MHz & $-1 \times 10^{-18}$ & $\pm 1 \times 10^{-18}$ \\
Second-order Zeeman 	& \includegraphics[width=1.25in]{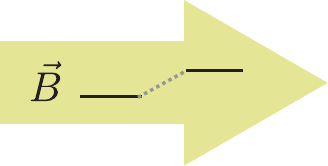} & 1.00(1)~mG, measured in $5D_{3/2}, F=2$ manifold&$ -8 \times 10^{-17} $& $\pm 8 \times 10^{-19}$ \\
\multirow{2}{1.0 in}{Black-body} 			&  \multirow{2}{1.25 in}{\includegraphics[width=1.25in]{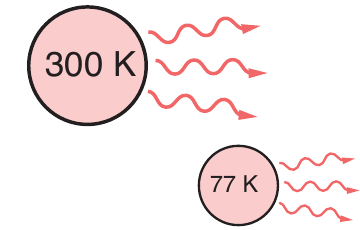}}  & 300 K & ($+7 \times 10^{-15}$) & ($\pm 4 \times 10^{-16}$) \\
		  & &77 K & $+3 \times 10^{-17}$ & $\pm 2 \times 10^{-18}$ \\
Quadrupole Stark 				& \includegraphics{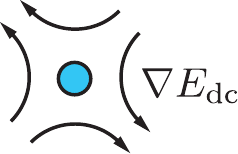} &  Stray dc gradients & 0 & 0 \\
Second-order quadrupole Stark 	& \includegraphics{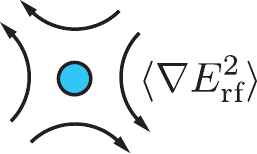} & Central gradient field $| \nabla E_\text{rf}| = 400$V/mm$^2$ &  $-5 \times 10^{-18}$ & $\pm 5 \times 10^{-18}$ \\
Trap-field dipole Stark			& \includegraphics{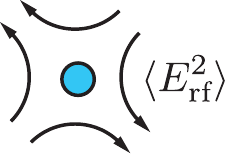} & $T < 1$~mK & $+5 \times 10^{-21}$ & $\pm 5 \times 10^{-21}$ \\
Micromotion enhanced Stark 		& \includegraphics{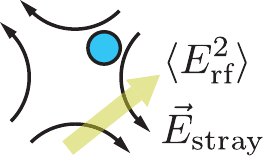} & 1\% compensation of 1V/cm stray field & $+2 \times 10^{-19}$ & $\pm 3 \times 10^{-19}$ \\
2051~nm clock ac-Stark  			& \includegraphics{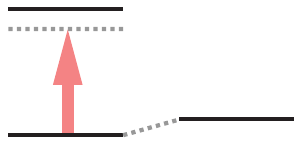} & $I_\text{clock} = 1 \mu$W/cm$^2$ & $+5 \times 10^{-20}$ & $\pm 1 \times 10^{-20}$ \\ \hline
Total & \multicolumn{2}{c}{}& $-5.6 \times 10^{-17}$ & $\pm6 \times 10^{-18}$
\end{tabular}
\label{tab:clockShiftSummary}
\end{table}

Table~\ref{tab:clockShiftSummary} summarizes all the systematic shift estimates from this section.  We estimate that the total relative systematic shift budget on the proposed clock transition is approximately $6 \times 10^{-17}$ with a relative uncertainty of $\pm 6 \times 10^{-18}$ at cryogenic temperatures. From a practical standpoint, the most worrying is the blackbody-shift at room temperature:  cryogenic cooling seems necessary to be competitive with other optical standards.  In summary, because of the unique immunity to dc-quadrupole (gradient) Stark shifts, and because of the high relative sensitivity to temporal drifts in fundamental constants, we feel that the 2051~nm $6S_{1/2}, F = 2 \leftrightarrow 5D_{3/2}, F = 0$ transition in $^{137}$Ba$^+$ is an optical frequency reference worth further development.
\chapter{Towards a single-ion parity violation measurement} \label{sec:ParityChapter}
\begin{quotation}
\noindent\small There are things you can learn by swinging a cat around by its tail that you simply cannot learn any other way. \\ \flushright{---Dubiously attributed to Mark Twain}
\end{quotation}
This research effort was originally launched not to measure off-resonant light shifts, hyperfine structure, or optical atomic clock transitions, the main topics of this thesis.  Instead, the original intent was to precisely measure atomic parity violation in a single trapped barium ion~\cite{fortson1993pmp}.  Subsequent investigation into systematic effects and the feasibility of overcoming many technical challenges has stalled progress on the realization of this goal.  Nevertheless, we wish to document recent conceptual advancements which might lead to a pilot experiment and eventual precision measurement.  Moreover, over the years both theoretical and experimental researches have paid attention to the effort---researchers at KVI may attempt a similar experiment in singly ionized radium.

In order to benefit researchers who are interested in this effort, we decided to document some new ideas concerning the experiment in both Ba$^+$ and Ra$^+$.  This chapter is not intended to be a complete summary of the experiment and heavily relies on previous work.

\section{Atomic parity violation}
The standard model predicts, and all experimental tests confirm, that our physical laws are invariant under the transformation $CPT$,  the combined application of three discrete transformations
\begin{center}
\begin{tabular}{lll}
$C$:  & Charge conjugation & $e \to -e$ \\
$P$:  & Parity inversion & $\boldsymbol{r} \to -\boldsymbol{r}$ \\
$T$:  & Time reversal  & $t \to -t$.
\end{tabular}
\end{center}
Individually, these symmetries are \emph{not} preserved by nature.  A striking example is exhibited in kaon decays.  Decay modes of the kaon lead to products of different parity states:
\begin{center}
\begin{tabular}{lll}
$K_0 \to \pi^+ + \pi^-$ & or $\quad \pi^0 + \pi^0 $& Parity even \\
$K_0 \to \pi^+ + \pi^- + \pi^0$ & or $\quad \pi^0 + \pi^0 + \pi^0$  &  Parity odd
\end{tabular}
\end{center}
Clearly parity cannot be a conserved quantity when a given particle can decay into either a parity even or odd state.  Further dramatic illustration of parity violation is in the handedness present in $^{60}$Co and $\mu$ decays.

Though most experiments aim to measure $CP$ and $T$ symmetry violation at high energy scales, their effects must be present at low atomic energies and at some level contribute to the structure of each and every atom and molecule around us.  More than just a laudable experimental challenge, the precise measurement of parity violation in atoms is uniquely poised to detect as yet undiscovered particles, such as heavy $Z_0$-like bosons, or perhaps particles predicted by certain variations of supersymmetry.

Gross atomic structure is defined by an electromagnetic Hamiltonian that is strictly parity conserving.  For instance, the Coulomb potential on a single electron is
\begin{equation}
V_\text{c}(r_e) = \frac{Z e^2}{r_e}
\end{equation}
where $Z$ is the atomic number, or number of protons in the nucleus and $r_e$ is the radial coordinate of the electron.  Under the parity transformation $\boldsymbol{r} \to -\boldsymbol{r}$, the quantity $r_e$ is unchanged so $V_c$ conserves parity---it is sometimes called \emph{parity even}.  The Standard Model predicts an interaction between the weak $Z_0$ boson and atomic electrons that produces a \emph{parity odd} potential \cite{bouchiat1997pva} (in the non-relativistic limit)
\begin{align}
V_\text{pv}(r_e) &= \frac{Q_W g_{Z_0}^2}{2} \frac{\exp(-M_{Z_0} c r_e/\hbar)}{r_e} \boldsymbol{\sigma_e} \cdot \boldsymbol{v_e}/c + h.c., \\
&\approx \frac{Q_W G_F}{4 \sqrt{2}} \left[ \delta^3(r_e) \boldsymbol{\sigma}_e \cdot \boldsymbol{v}_e + h.c. \right],
\end{align}
where $Q_W$ is quantity called the weak charge, discussed later, $G_F$ is a point contact coupling constant proportional to the underlying physical weak force coupling constant $g_{Z_0}$, $M_{Z_0}$ is the mass of the $Z_0$ boson, and the pseudo-scalar product of the electron spin and velocity $\boldsymbol{\sigma}_e \cdot \boldsymbol{v}_e$ is called the helicity operator.  The notation $h.c.$ stands for the Hermitian conjugate of the first term.  The helicity operator changes signs under parity transformation.  When $\boldsymbol{r} \to - \boldsymbol{r}$, the velocity of the electron also changes sign $\boldsymbol{v}_e \to -\boldsymbol{v}_e$ while the spin $\boldsymbol{\sigma}_e$ is unchanged.

\begin{figure}
\centering
\includegraphics{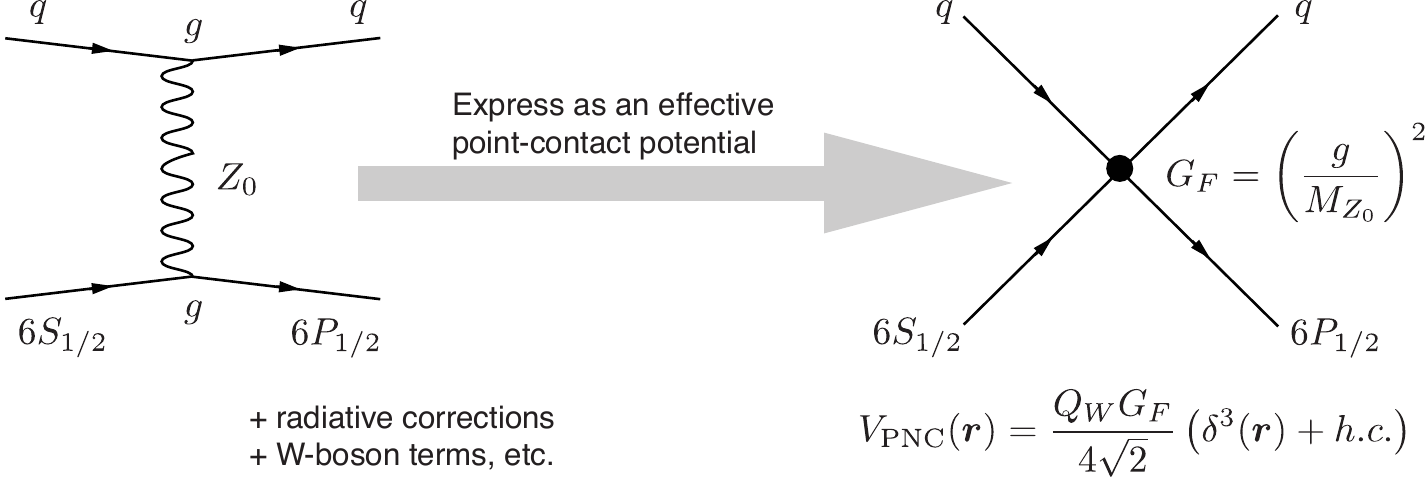}
\caption[Atomic parity violation arises from electron-quark weak interactions]{Atomic parity violation arises from electron-quark weak interactions.  Here we depict in a Feynman diagram the parity violation of an electron due to a t-channel interaction with a $Z^0$ boson inside the nucleus.  This and other weak interactions can be treated in the contact potential approximation given in the text.  Precision measurement of the weak charge $Q_W$ through an atomic parity violation measurement combined with theoretical interpretation can rule out or give evidence to additional $Z'_0$ particles~\cite{bouchiat1997pva}.}
\label{fig:pncFeynman}
\end{figure}

The approximation of the radial form of the potential as a delta function is appropriate given the extremely short range of the weak force compared to the typical atomic scale.  Formally, taking the mass of the heavy $Z_0$ boson to infinity makes the approximation an equality:
\begin{equation*}
\lim_{M_{Z_0} \to \infty} \left(\frac{M_{Z_0} c}{\hbar} \right)^2 \frac{\exp(-M_{Z_0} c r_e /\hbar)}{r_e} = \delta^3(r_e)
\end{equation*}
The weak charge $Q_W$ is the sum of the weak charges of the constituents of atomic nuclei:
\begin{equation}
Q_W = (2Z + N) Q_W(u) + (Z +2N)Q_W(d)
\end{equation}
where $Q_W(u)$ and $Q_W(d)$ are the weak charges of the up and down quarks, $Z$ and $N$ are the number of protons and neutrons in the nucleus.  Recall that a proton is composed of $uud$ quarks whereas neutrons are composed of $udd$.  The standard model predicts
\begin{equation}
Q_W = -N - Z(4 \sin^2 \theta_W - 1) \sim -N
\end{equation}
where $\theta_W$ is the weak mixing angle, a free parameter measured experimentally to be $\sin^2 \theta_W = 0.22(6)$~\cite{arroyo1994pmw}.

How does one go about looking for parity violation in atoms? Remember that because electromagnetic interactions absolutely conserve parity, the existence of atomic transitions forbidden by electromagnetism is evidence of parity violation in the atom arising from the weak force.  In general, matrix elements that enable such parity violating transitions scale quickly with increasing atomic number $Z$---a so-called $Z^3$ scaling law guides our intuition.  From~\cite{bouchiat1997pva} we first have that the nuclear weak charge charge $Q_W$ scales with $N$ which often is larger than $Z$ in many stable nuclei.  Second, in the vicinity of the nucleus, the electron density is known to grow with increasing $Z$, making matrix elements of the parity violating potential $V_\text{pv}$ larger.  Finally, near the nucleus the electron velocity increases with $Z$ making the helicity term $\boldsymbol{\sigma_e} \cdot \boldsymbol{v_e}/c$ larger.

Still, parity violating matrix elements (and therefore transition rates) remain much smaller than typical parity conserving electric dipole matrix elements, by factors of $\sim 10^{-10}$ in even the best cases.  Therefore, the straightforward approach of measuring the absolute strengths of parity violating transitions is bound to fail as weakly allowed parity conserving transitions completely dominate the experiment.  Success follows, however, if we take advantage of the unique transformation properties of the parity violating matrix elements.  For instance measurement of an absolute transition rate
\begin{align*}
R &= \left| \Omega_\text{allowed} + \Omega_\text{pv} \right|^2 \\
    &= \left| \Omega_\text{allowed} \right|^2 + \left| \Omega_\text{pv} \right|^2 + 2 \text{Re}\left( \Omega^\dag_\text{pv} \Omega_\text{allowed} \right)
\end{align*}
is bound to fail to give a value for $\Omega_\text{pv}$ since it is at least quadratically suppressed by a much larger $\Omega_\text{allowed}$.  However, the interference term $2 \text{Re}\left( \Omega^\dag_\text{pv} \Omega_\text{allowed} \right)$ is shown to be a pseudoscalar and will therefore change sign upon a coordinate system reversal.  Further, the absolute size of the interference term grows with $\Omega_\text{allowed}$ so it gains from the strength of the allowed transition.

\begin{figure}
\centering
\includegraphics[width = 5 in]{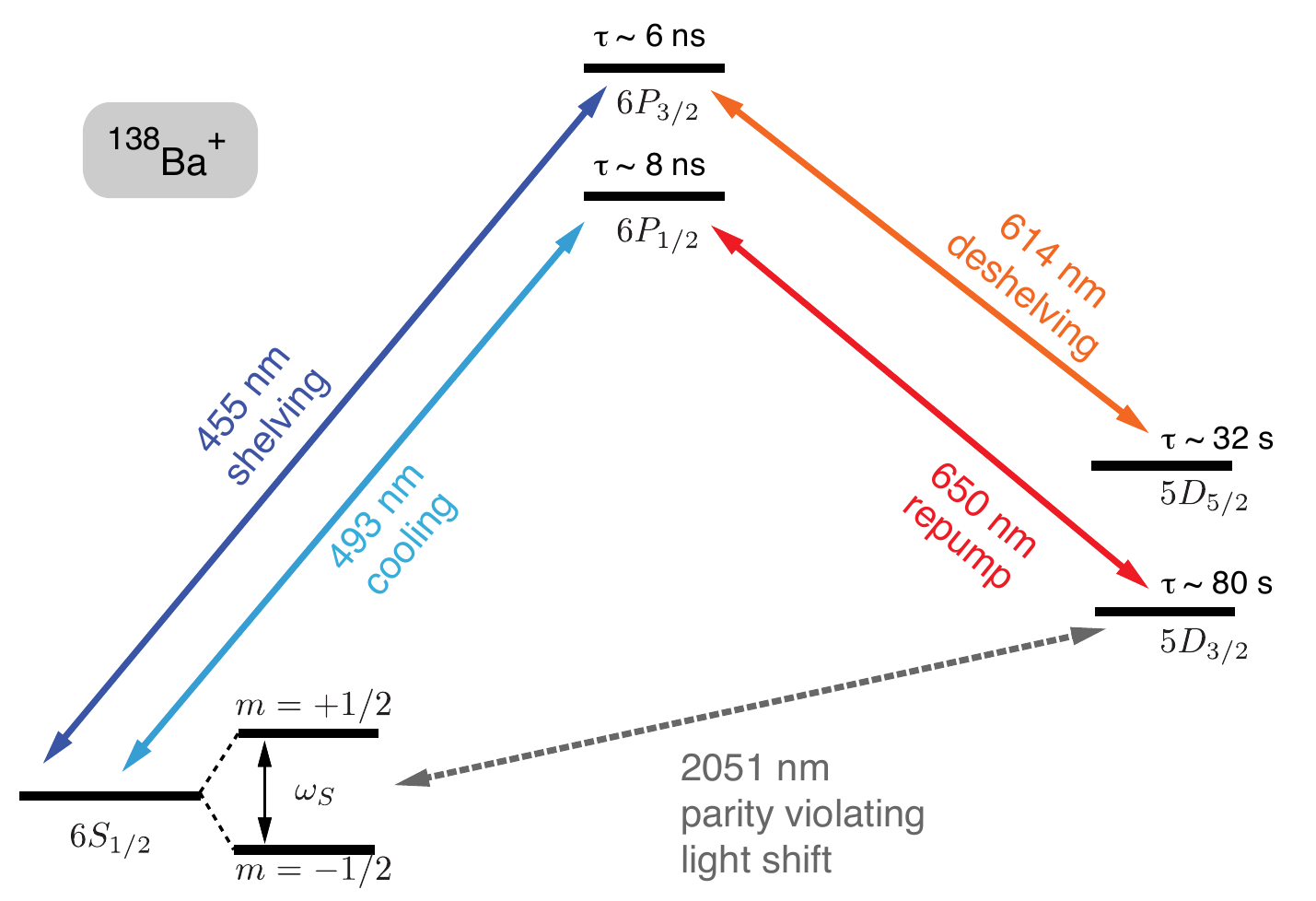}
\caption[Barium level diagram for a parity violation experiment]{The $6S_{1/2}$ and $5D_{3/2}$ are coupled with resonant 2051~nm light using carefully controlled fields such that a parity violating (vector) light shift alters the $6S_{1/2}, m=\pm 1/2$ splitting $\omega_S$.}
\label{fig:bariumEnergyLevelsParity}
\includegraphics{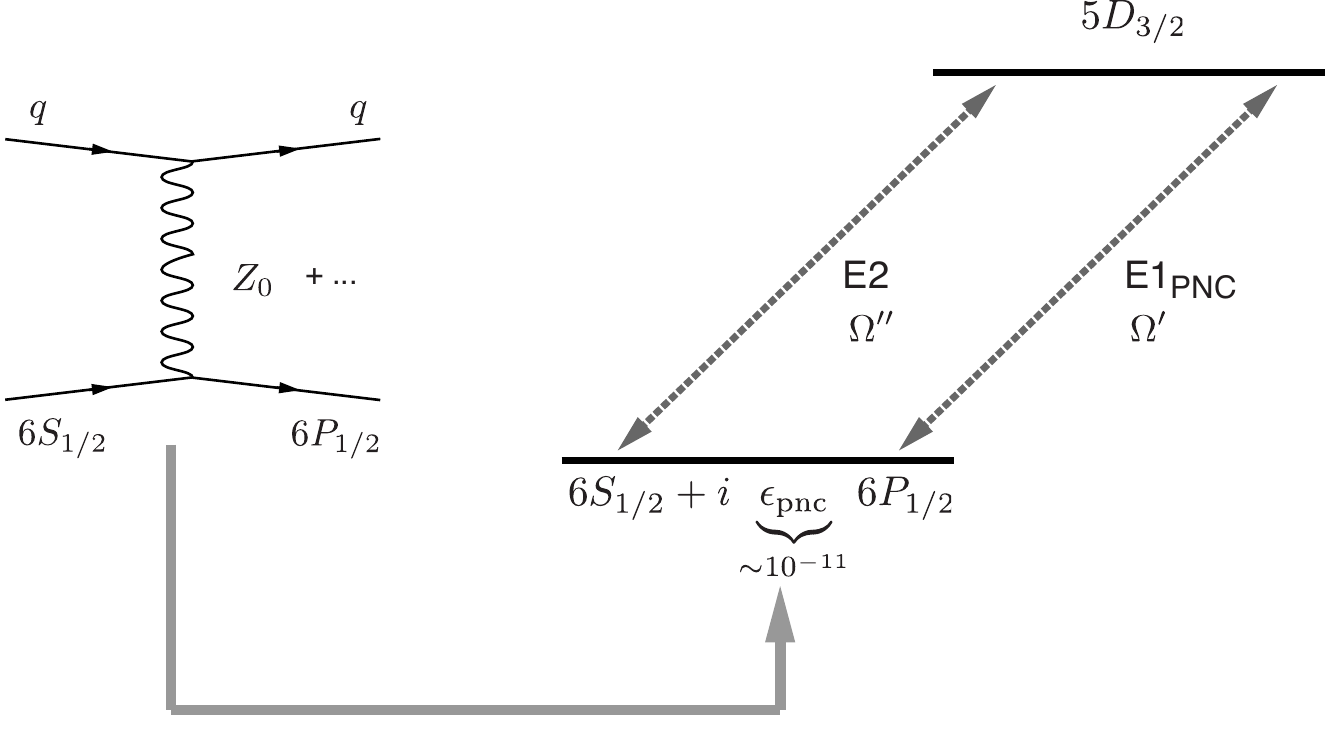}
\caption[Schematic of an E2-E1$_\text{PNC}$ interference]{Schematic of an E2-E1$_\text{PNC}$ interference. Parity violation is manifest in a small admixture $(\sim 10^{-11})$ of $6P_{1/2}$ in the ground state $6S_{1/2}$.  A coupling $\Omega''$ of $6S_{1/2}$ to $5D_{3/2}$ via the electric quadrupole (E2) interaction can interfere with another coupling $\Omega'$ meant to drive the small electric dipole (E1$_\text{PNC}$) interaction allowed by parity violation.}
\label{fig:bariumEnergyLevelsParityCloseup}
\end{figure}

\section{Some discussion of the original proposal}
\begin{figure}
\centering
\includegraphics[width=6in]{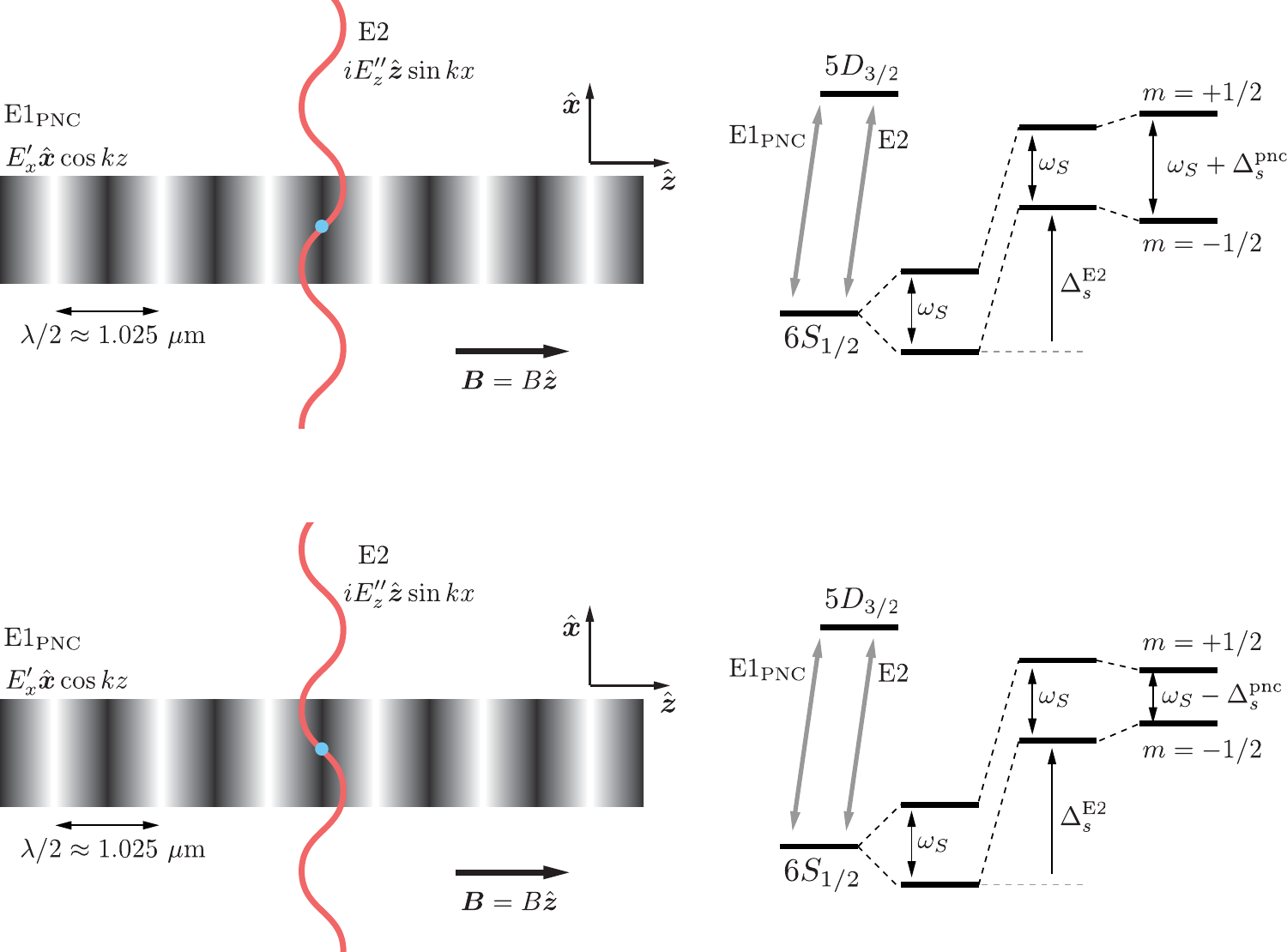}
\caption[Ideal standing wave geometry and light shifts in the parity violation experiment]{Ideal standing wave geometry and light shifts for a single ion parity violation experiment.  We depict the trapped ion located at the anti-node of a standing wave of light, E1$_\text{PNC}$, resonant with the 2051~nm $6S_{1/2} \leftrightarrow 5D_{3/2}$ transition.  In a perpendicular direction, a perpendicular standing wave E2, derived from the same laser, is engineered to have a node at the location of the ion.  We show that the E2 field results in a \emph{scalar} shift $\Delta_s^\text{E2} \sim 10^4$~Hz common to both sublevels of $5D_{3/2}$.  Meanwhile, the interference of E2 and E1$_\text{PNC}$ results in a \emph{vector} shift that changes the Larmor precession frequency $\omega_S \sim 100$~Hz by $\Delta_s^\text{PNC} \approx 0.3$~Hz.  A key signal of parity violation is the \emph{reversal} of $\Delta_s^\text{PNC} $ when the phasing between the two fields are changed by 180$^\circ$, as depicted in the bottom half of the figure.}
\label{fig:parityStandingWavesAndShifts}
\end{figure}
Substantial previous work~\cite{schacht2000thesis, koerber2003rfs,koerber2003thesis} details a method of measuring an on-resonant, parity-violating, vector light shift in the $6S_{1/2}$ state of $^{138}$Ba$^+$ with the application of standing waves of 2051~nm light.  We will briefly summarize the technique, its principle challenges, and new ideas.  

A single trapped ion is located at the anti-node of very intense electric field $\boldsymbol{E}'$ resonant with the 2051~nm $6S_{1/2} \leftrightarrow 5D_{3/2} $ transition.  It is simultaneously placed at an anti-node of a much smaller field $\boldsymbol{E}'' $ of identical frequency.  For now, we define the geometry
\begin{align}
\boldsymbol{E}' &= E'_x \bhat{x} \cos kz, \label{eq:parityIdealGeometry} \\
\boldsymbol{E}'' &= iE''_z \bhat{z} \sin kx \notag
\end{align}
shown in Figure~\ref{fig:parityStandingWavesAndShifts}.  Since the ion sees no spatial electric field gradient due to $\boldsymbol{E}'$, it cannot drive electric quadrupole (E2) transitions;  instead it only couples via a parity violating electric dipole interaction (E1$_\text{PNC}$).  Similarly, the standing wave $\boldsymbol{E}''$ couples only via E2 and not at all through E1$_\text{PNC}$.  We can write down interaction Rabi frequencies
\begin{align}
\Omega_{m',m}' &= \frac{1}{\hbar} \langle \gamma' j' || er^{(1)}, H_\text{pv} || \gamma j \rangle \times \sum_q \frac{\langle j', m' ; 1 q | j m \rangle E_q'^{(1)}}{\sqrt{2j' + 1}}, \\
\Omega_{m',m}'' &= \frac{1}{\hbar} \langle \gamma' j' || e r^{(2)} || \gamma j \rangle \times \sum_q \frac{\langle j' m' ; 2 q | j m \rangle \nabla E_q''^{(2)}}{\sqrt{2 j' + 1}},
\end{align}
where the electric fields $E_q'^{(1)}$ and $E_q''^{(2)}$ are written as rank-1 and rank-2 spherical vectors, and the electric dipole parity violating reduced matrix element is
\begin{equation}
\begin{split}
\langle \gamma' j' || er^{(1)}, H_{pv} || \gamma j \rangle &=
\left( \sum_{n,j_n} \frac{\langle \gamma' j' || e r^{(1)} || n j_n \rangle \langle n j_n || H_1 || \gamma j \rangle}{W_n - W} \right. \\
& \left. \qquad \qquad +  \sum_{n,j_n} \frac{\langle \gamma' j' || H_1 || n j_n \rangle \langle n j_n || e r^{(1)}  || \gamma j \rangle}{W_n - W'} \right)
\end{split}
\end{equation}
where $H_1$ is the nuclear-spin independent part of the parity violating Hamiltonian, the only piece needing treatment for the even isotopes of barium ($I=0$).

The two interactions add coherently
\begin{equation}
\Omega = \Omega' + \Omega'',
\end{equation}
and we might ask what on-resonance light shifts $\Omega^\dag \Omega$ are observable in the ground state $6S_{1/2}$.  Since $J  = 1/2$, we expect a scalar and a vector component,
\begin{equation}
\Omega^\dag \Omega = \underbrace{\Theta}_\text{scalar} + \underbrace{\boldsymbol{\Theta} \cdot \boldsymbol{J}}_\text{vector}.
\end{equation}
The scalar component is independent of sublevel $m$, while the vector piece changes sign for $\pm m$ akin to a magnetic field.  Figure~\ref{fig:parityStandingWavesAndShifts} offers a sketch of the differential shift due to an interference term, along with the sign change upon coordinate reversal typical of a parity violation experiment.  Examining the combination of our E1$_\text{PNC}$ and E2 interactions, we find
\begin{align}
\Omega^\dag \Omega &= (\Omega'^\dag + \Omega''^\dag)(\Omega' + \Omega'') \notag \\
				    &= \Omega''^\dag \Omega + (\Omega'^\dag \Omega'' + \Omega''^\dag \Omega') + \Omega'^\dag \Omega' \notag \\
				    &\simeq \underbrace{\Omega''^\dag \Omega''}_\text{E2-E2} + \underbrace{(\Omega'^\dag \Omega'' + \Omega''^\dag \Omega')}_\text{E2-E1$_\text{PNC}$ interference} \label{eq:parityInterference}
\end{align}
by dropping the quadratic dependence on the very small E1$_\text{PNC}$ interaction.  The first term, just the on-resonant light shift due to the (relatively) strong E2 coupling might itself have both scalar and vector structure,
\begin{align}
(\Omega''^\dag \Omega'')_{m_1, m_2}  &\propto \langle E2 \rangle \langle E2 \rangle \sum_{q,m'} \frac{\langle \tfrac{1}{2} m_1 ; 2,q | \tfrac{3}{2} m' \rangle \langle \tfrac{3}{2} m' ; 2 q | \tfrac{1}{2} m_2 \rangle}{2 (3/2 + 1)} E_q^{(2)} E_q^{(2)} \\
\intertext{or}
\begin{split}
\Omega''^\dag \Omega'' &= \frac{\langle E2 \rangle \langle E2 \rangle}{20 \hbar^2} (\boldsymbol{\nabla} \times \boldsymbol{E}'') \cdot (\boldsymbol{\nabla} \times \boldsymbol{E}'') \\
&+ \frac{\langle E2 \rangle \langle E2 \rangle}{20 \hbar^2} \left[ 2((\boldsymbol{\nabla} \times \boldsymbol{E}'') \cdot \boldsymbol{\nabla})\boldsymbol{E}'' + (\boldsymbol{\nabla} \times \boldsymbol{E}'') \times (\boldsymbol{\nabla} \times \boldsymbol{E}'')  \right] \cdot \boldsymbol{J}
\end{split}
\end{align}
from~\cite{koerber2003thesis}.  However, given the field geometry in Eq.~\ref{eq:parityIdealGeometry}, the vector piece of this interaction vanishes leaving us with just a scalar shift. Therefore, $\Omega''^\dag \Omega''$ does not couple the ground states $m = -1/2$ to $m = +1/2$ so any shift to them is common mode.  The scalar shift due to the E2 field is~\cite{koerber2003thesis}
\begin{equation}
\Delta_S^{E2} = \frac{1}{4 \sqrt{5} \hbar} k E_z'' \langle E2 \rangle.
\end{equation}

Meanwhile, the E2-E1$_\text{PNC}$ interference terms in Eq.~\ref{eq:parityInterference} do yield pure vector shifts~\cite{koerber2003thesis,schacht2000thesis}
\begin{equation}
(\Omega'^\dag \Omega'')_{m_1, m_2} = \frac{\langle E1_\text{PNC} \rangle \langle E2 \rangle}{2 \sqrt{10} \hbar^2} \left[2(\boldsymbol{E}' \cdot \boldsymbol{\nabla}) \boldsymbol{E}'' + \boldsymbol{E}' \times (\boldsymbol{\nabla} \times \boldsymbol{E}'')) \cdot \boldsymbol{J} \right].
\end{equation}

\begin{table}
\centering
\caption[Optimal field strengths and calculations of the parity non-conserving matrix element]{This table reports estimates necessary for calculating the expected size of observed parity non-conserving light shifts:  optimal field strengths and calculations of parity non-conserving matrix elements.  Also, transition wavelengths and state lifetimes are reported to aid in comparison.}
\label{tab:parityCalculations}
\begin{tabular}{l l l l}
						&		&	$^{138}$Ba$^+$		& $^{226}$Ra$^+$ ($\tau_{1/2} = 1600$ yr) \\ \hline \hline
\multicolumn{2}{l}{$S_{1/2} \to D_{3/2}$ wavelength}		&	2051~nm				& 828~nm \\
\multicolumn{2}{l}{Optimal $E'_x$ size~\cite{koerber2003thesis}} 	&  14900 V/cm			& 6800 V/cm \\
\multicolumn{2}{l}{Calculations of $\langle E1_\text{PNC} \rangle$} 	& 2.17~\cite{dzuba2001cpn} & 42.9~\cite{dzuba2001cpn}  \\
\multicolumn{2}{l}{(units of $10^{-11} ea_0 (-Q_W / N)$)} 	& 2.34 ~\cite{dzuba2001cpn} & 45.9  ~\cite{dzuba2001cpn}  \\
&	& 2.35~\cite{geetha2002thesis} & \\
&	& 2.46(2)~\cite{sahoo2006rccpnc} & \\ \hline
\multicolumn{2}{l}{Calculation of $\langle E2 \rangle$} 	& 12.63 $ea_0^2$~\cite{gopakumar2002edq} & \multicolumn{1}{c}{---} \\
\multicolumn{2}{l}{Lifetimes of $P_{1/2}$, $P_{3/2}$}	& 7.8~ns, 6.3~ns	& 8.8~ns, 4.7~ns \\
\multicolumn{2}{l}{Lifetimes of $D_{3/2}$, $D_{5/2}$}	& $\sim 80$~s, 30~s	& 0.6~s, 0.3~s \\  \hline
Cooling  & $S_{1/2} \to P_{1/2}$		& 493~nm			& 468~nm \\
Repump & $D_{3/2} \to P_{3/2}$	& 650~nm`	`	& 1079~nm \\
Indirect shelve & $S_{1/2} \to P_{3/2}$	& 455~nm			& 382~nm \\
Indirect deshelve & $D_{5/2} \to P_{3/2}$ & 615~nm		& 802~nm \\
Direct shelve  &  $S_{1/2} \to D_{5/2}$	& 1762~nm		& 728~nm
\end{tabular}
\end{table}
Only the diagonal elements of the interference term in Eq.~\ref{eq:parityInterference} are important, because we expect the ion to experience a magnetic field of much larger size than the parity violating vector shift.  Therefore, all off-diagonal components act like effective perpendicular magnetic fields and don't contribute to first order.  The remaining diagonal terms,
\begin{align*}
(\Omega'^\dag \Omega'' + \Omega''^\dag \Omega') &\simeq 2 \text{Re } (\Omega'^\dag \Omega'' )_{mm} \\
	&= \frac{\langle E1_\text{PNC} \rangle \langle E2 \rangle}{\sqrt{10} \hbar^2} k E_x' E_z'' j_z
\end{align*}
are proportional to $j_z$, a classic vector operator which leads to an additional splitting, our measurement,~\cite{koerber2003thesis}
\begin{align}
\Delta_S^\text{PNC} &= \Delta_{m=1/2}^\text{PNC} - \Delta_{m=-1/2}^\text{PNC} \\
	&\simeq \frac{1}{2 \sqrt{2} \hbar} E_x' \langle E1_\text{PNC} \rangle.
\end{align}
How large can we make $\Delta_S^\text{PNC}$?  It scales with the size of the $E'_x$ field, but this field cannot increase without limit.  Eventually, through off-resonant allowed E1 interactions, $E'_x$ will shorten the lifetime of both the $6S_{1/2}$ and $5D_{3/2}$ state, principally by admixture with the excited $nP_{1/2,3/2}$ states.  Estimate for the optimal field strengths for both Ba$^+$ and Ra$^+$, along with calculations of the parity nonconserving matrix elements are given in Table~\ref{tab:parityCalculations}.  Using these data, we find that the maximum optimal size of the parity violating light shifts are~\cite{koerber2003thesis}
\begin{align}
\Delta_S^\text{PNC}/2\pi &=  0.37 \text{ Hz} \qquad (\text{Ba}^+), \\
\Delta_S^\text{PNC}/2\pi &=  3.70 \text{ Hz} \qquad (\text{Ra}^+).
\end{align}
Though small, the shifts are larger than expected shifts caused by instabilities in the small, $\sim 1$~mG magnetic field that accompanies the light shifts.  Most importantly, their sign is reversed when the phasing between the $E'_x$ and $E''_z$ fields are altered by 180$^\circ$, equivalent to a parity transformation.

\subsection{Systematic effects}
\begin{table}
\centering
\caption[Phase, symmetry factors, and selection rules for systematic field shifts, from~\cite{koerber2003thesis}]{Phase, symmetry factors, and selection rules for systematic field shifts, from~\cite{koerber2003thesis}.}
\label{tab:parityPhaseFactors}
\begin{tabular}{l c c c c}
Interaction & Field & $P[ \boldsymbol{f}(\boldsymbol{E})]$  & $S[ \boldsymbol{f}(\boldsymbol{E})]$  & $\Delta m$ \\ \hline \hline
\multirow{3}{*}{Electric dipole} 	& $E_z$	&  1	 	& 1		& 0 \\
						& $E_y$	& $i$		& 1		& $\pm 1$ \\
						& $E_x$	& 1		& -1		& $\pm 1$ \\ \hline
Interaction & Field & $P[ \boldsymbol{f}(\boldsymbol{B})]$  & $S[ \boldsymbol{f}(\boldsymbol{B})]$  & $\Delta m$ \\ \hline \hline
\multirow{3}{*}{Magnetic dipole} & $B_z$	&  1	 	& 1		& 0 \\
						& $B_y$	& $i$		& 1		& $\pm 1$ \\
						& $B_x$	& 1		& -1		& $\pm 1$ \\ \hline
Interaction & Field & $P[ \boldsymbol{f}(\boldsymbol{\nabla E})]$  & $S[ \boldsymbol{f}(\boldsymbol{\nabla E})]$  & $\Delta m$ \\ \hline \hline
\multirow{5}{*}{Electric quadrupole} & $\partial_z E_z $				&  1	 	& 1		& 0 \\
							& $\partial_z E_x + \partial_x E_z$	& 1		& -1		& $\pm 1$ \\
							& $\partial_z E_y + \partial_y E_z$	& $i$		& 1		& $\pm 1$ \\											& $\partial_x E_x + \partial_y E_y$	& 1		& 1		& $\pm 2$ \\
							& $\partial_x E_y + \partial_y E_x$	& $i$		& -1		& $\pm 2$				
\end{tabular}
\end{table}
When the experiment deviates from the ideal beam geometry, polarization, and ion location, systematic level shifts may result that mimic the parity violating light shift.  Only \emph{vector} shifts are problematic.  Further, no parity conserving vector shift will reverse in sign when the phasing between $E'_x$ and $E''_Z$ is altered by $180^\circ$, as long as the phase change can be made with high fidelity.  Also, systematic shifts due to just one of the fields ought to be present when the other field is removed, unlike the genuine effect.  Finally, we expect that a background magnetic field will split the $6S_{1/2}$ sublevels $10^2$--$10^3$ times larger than the expected parity violating vector shift.  Therefore, systematic vector shifts perpendicular to $\boldsymbol{B}$ only enter quadratically.   Nonetheless, the systematic effects for this experiment are daunting, and largely the reason it has not been attempted yet.  

We will adopt a consistent theme promulgated by other colleagues working on this problem.  Each systematic effect will be shown to be proportional to one or more `small factors' such as misalignment angles, fractional polarization errors, or displacements.  In practice, one cannot hope to minimize each of these below $\sim 10^{-3}$ without special techniques.  Therefore, one must design the experiment such that likely systematic effects are proportional to many small factors along with a procedure for guaranteeing their `smallness.'  We often refer to the relative strengths of the interactions coupling $6S_{1/2} \leftrightarrow 5D_{3/2}$;  the following order of magnitude estimates of interaction strengths, in consistent units, are useful:
\begin{center}
\begin{tabular}{c c c c c}
$|\langle 6S_{1/2} || \text{E1}_\text{PNC} || 5D_{3/2} \rangle|$ &:& $\langle 6S_{1/2} || \text{M1} || 5D_{3/2} \rangle$ &:&  $\langle 6S_{1/2} || \text{E2} || 5D_{3/2} \rangle$ \\
$ \sim 2 \times 10^{-11} ea_0$	&:&	$\sim 8 \times 10^{-4} (\alpha/2) e a_0$  &:&	$\sim 12 (a_0 / \lambda) e a_0$ \\
$\sim 1$	&:& $\sim 10^5$  &:& $\sim 10^7$
\end{tabular} 
\end{center}

To systematically evaluate deviations from the ideal field construction, we will develop a system~\cite{koerber2003thesis} to distinguish what interactions can mimic the PNC shift.  Three conditions must be meant by products of interactions in order to create spin-dependent shifts in the $6S_{1/2}$ state:  the selection rules of the product must connect states to themselves,  the total phase of the product has to be real, and the product must be antisymmetric with respect to $m$.

\begin{figure}
\centering
\includegraphics{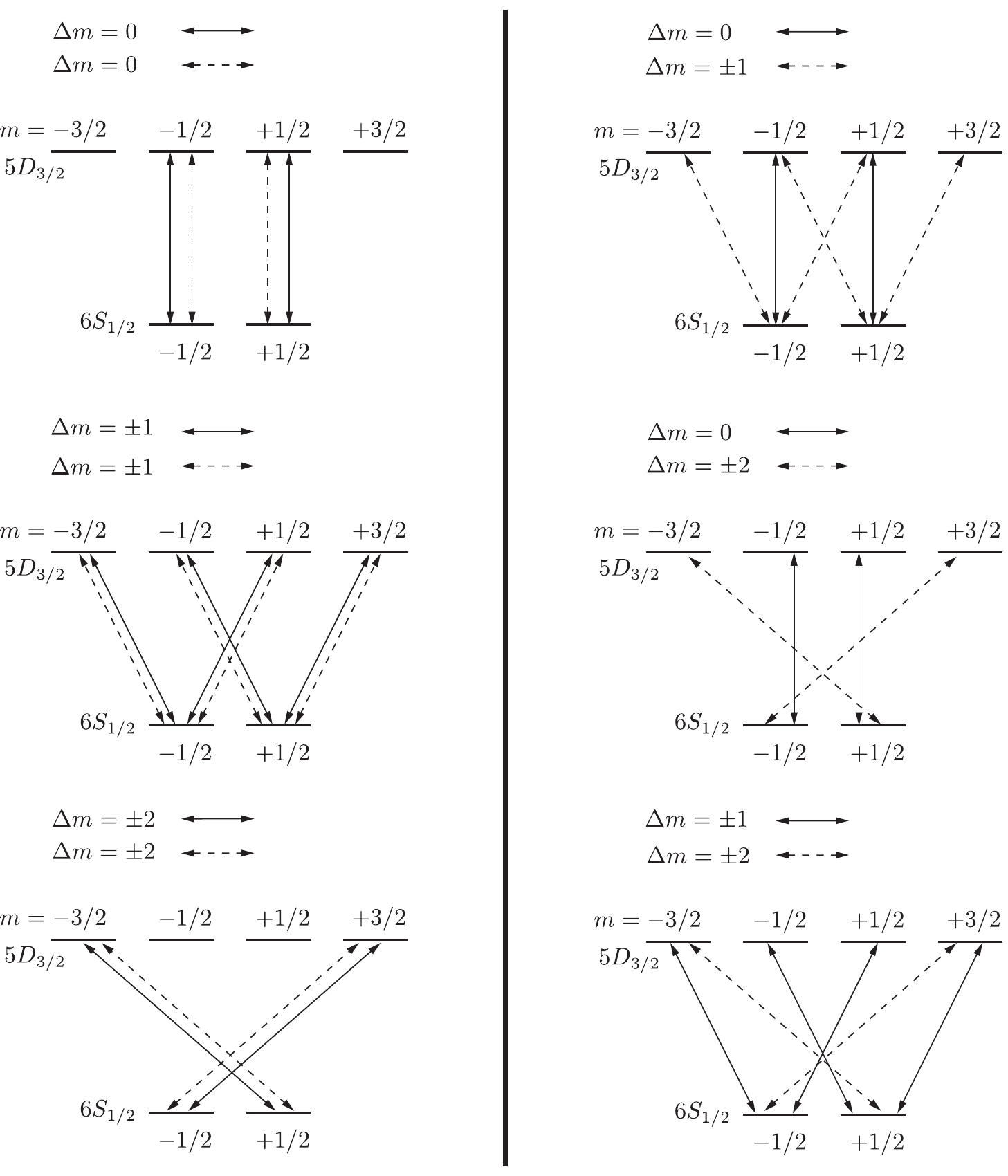}
\caption[Selection rule diagrams for spin-dependent shifts due to product interactions]{Selection rule diagrams for spin-dependent shifts due to product interactions.  Systematic effects due to the product of two interactions can only mimic parity if the selection rules of the product allow spin states to connect to themselves, such as $\{ \Delta m = \pm 1, \Delta m = \pm 1 \}$ and $\{ \Delta m = \pm 2, \Delta m = \pm 2\}$. Selection rules for arbitrary field configurations are tabulated in Table~\ref{tab:parityPhaseFactors}.} 
\label{fig:paritySelectionRuleDiagrams}
\end{figure}

First, we examine the selection rules of a generic product of interactions $\Omega_1^\dag \Omega_2$.  In order to cause a level shift (at first order in perturbation theory), the product of interactions must connect states to themselves.  For instance, the selection rules $\{ \Delta m = \pm 1, \Delta m = \pm 1 \}$ and $\{ \Delta m = \pm 0, \Delta m = \pm 0 \}$ are examples which connect spin states to themselves through both interactions.  Sets such as $\{ \Delta m = \pm 1, \Delta m = \pm 0 \}$ do not.  Graphical representations in Figure~\ref{fig:paritySelectionRuleDiagrams} of this are extremely useful, and the selection rules derived from Clebsch-Gordan rules are tabulated in Table~\ref{tab:parityPhaseFactors}.

Next, we consider the phase of interaction products like $\Omega_1^\dag \Omega_2$.  Following~\cite{koerber2003thesis}, we define a functional $P[\Omega]$ that evaluates the phase of an interaction.  For products of interactions
\begin{equation}
P[\Omega^\dag_1 \Omega_2] = P[\Omega_1] P[\Omega_2],
\end{equation}
which the individual phase terms can be separated into terms that account for the phase of the reduced matrix element (imaginary only for E1$_\text{PNC}$ in this discussion), the geometrical configuration of the field, and the temporal phasing of the field itself:
\begin{align}
P[\Omega_1] &= P[\langle M \rangle] P[\boldsymbol{f}(\boldsymbol{E})] P[\boldsymbol{E}] \qquad (\text{dipole interactions}), \\
P[\Omega_1] &= P[\langle M \rangle] P[\boldsymbol{f}(\boldsymbol{\nabla E})] P[\boldsymbol{\nabla  E}] \qquad (\text{quadrupole interactions}).
\end{align}
Geometrical phasing factors $P[\boldsymbol{f}(\boldsymbol{E})]$  are calculated~\cite{koerber2003thesis} in Table~\ref{tab:parityPhaseFactors}.  The total phase of a product interaction must be real to yield a spin-dependent shift.

Finally, we define the symmetry of product interactions:  a functional $S[\Omega] = 1$ if the matrix elements are symmetric $\Omega_{m,m} = \Omega_{-m,-m}$ and $S[\Omega] = -1$ otherwise.  Like the phase functional, we can separate product interactions, and further, the symmetry of the reduced matrix elements and field configuration:
\begin{align}
S[ \Omega^\dag_1 \Omega_2] &= S[\Omega_1] S[\Omega_2], \\
S[\Omega_1] &= S[ \langle M \rangle] S[\boldsymbol{f}(\boldsymbol{E})] \qquad (\text{dipole interactions}), \\
S[\Omega_1] &= S[ \langle M \rangle] S[\boldsymbol{f}(\boldsymbol{\nabla E})] \qquad (\text{quadrupole  interactions}).
\end{align}
The parity violating interaction is a pseudo-scalar, so $S[ \langle \text{E1}_\text{PNC} \rangle] = -1$.  A rank-2 tensor operator like the E2 interaction is parity-even, so $S[ \langle \text{E2} \rangle] = 1$.  Geometrical factors $S[\boldsymbol{f}(\boldsymbol{E})]$ that rely on whether or not the relevant Clebsch-Gordan coefficients reverse sign when $m \to -m$ are tabulated in Table~\ref{tab:parityPhaseFactors}.

An illustrative example of these three criteria is the analysis of the E2-E1$_\text{PNC}$ interaction in the ideal geometry.  From the fields
\begin{align*}
\boldsymbol{E}' &= E'_x \bhat{x} \cos kz, \\
\boldsymbol{E}'' &= iE''_z \bhat{z} \sin kx,
\end{align*}
and Table~\ref{tab:parityPhaseFactors}, we see that the selection rules are $\{ \Delta_m = \pm 1, \Delta_m = \pm 1 \}$ which do connect spin-states to themselves.  

The total phase of the interaction is real:
\begin{align*}
P[\Omega'^\dag \Omega''] &= \left( P[\langle \text{E1}_\text{PNC} \rangle]  P[\langle \text{E2} \rangle] \right) \left( P[\boldsymbol{f}(\boldsymbol{E}')] P[\boldsymbol{f}(\boldsymbol{E}'')] \right) \left( P[\boldsymbol{E}')] P[\boldsymbol{E}'')]\right) \\
&= \left(i \cdot 1\right) \left(1 \cdot 1 \right) \left(1 \cdot i \right) \\
&= -1.
\end{align*}
The first factor of $i$ arises from the pseudo-scalar nature of the E1$_\text{PNC}$ interaction while the last factor of $i$ arises since the $\boldsymbol{E}''$ field is temporally 90$^\circ$ out of phase with the $\boldsymbol{E}'$ field.  

The symmetry factor of the interaction is
\begin{align*}
S[\Omega'^\dag \Omega''] &= \left( S[\langle \text{E1}_\text{PNC} \rangle] S[\langle \text{E2} \rangle] \right) \left( S[\boldsymbol{f}(\boldsymbol{E}')] S[\boldsymbol{f}(\boldsymbol{E}'')] \right)  \\
&= \left(-1 \cdot 1\right) \left(-1 \cdot -1 \right) \\
&= -1,
\end{align*}
which is antisymmetric.  All three criteria are satisfied---the selection rules connect spin states to themselves, the phase of the interaction is real, and antisymmetric.  As promised, $\Omega'^\dag \Omega''$ yields a vector shift.  Now we will explore some scenarios of systematic effect which manifest themselves as vector shifts.

\subsubsection{Ion displacement, beam misalignment}
An obvious source of trouble is displacement of the ion along the $E'_x$ standing wave set up to drive only the E1$_\text{PNC}$ interaction.  Suppose the ion is displaced along $\bhat{z}$ by a small fraction $\delta$ of the standing wave spacing $\lambda/2 \approx 1025$~nm.  As an initial estimate, we note that recent work with trapped ions in the presence of optical cavities have realized ion wavefunction confinement to 42~nm~\cite{keller2003dcs} which gives $\delta \sim 0.04$.  An electric field gradient appears due to $E'_x$ that drives an additional E2 interaction we call $\delta \Omega'_\text{E2}$.  Including this coupling in Eq.~\ref{eq:parityInterference},
\begin{equation} \label{eq:displacementParitySytematics}
\Omega^{\dag} \Omega = \underbrace{\Omega''^{\dag} \Omega'' + \delta^2 \Omega'^{\dag} _\text{E2} \Omega'_\text{E2}}_\text{E2 single field terms} +  \underbrace{2 \text{Re}(\delta  \Omega'^{\dag}_\text{E2}  \Omega'')}_\text{E2 interference term} + \underbrace{2 \text{Re}(\Omega'^{\dag} \Omega'')}_\text{PNC term} + \cdots
\end{equation}
where have we not yet included a term $2 \text{Re}(\delta  \Omega'^\dag_\text{E2} \Omega')$ that accounts for a small error in the measured PNC effect but is isolated experimentally since it does not depend on the presence of the $\Omega''$ interaction.

Of the extra terms in Eq.~\ref{eq:displacementParitySytematics}, which can lead to a false vector shift?  Applying our criteria to the term that depends on the presence of both fields,
\begin{align}
\delta  \Omega'^{\dag}_\text{E2}  \Omega'':  \qquad &\{ \Delta m = \pm 1, \Delta m = \pm 1 \}, \\
P[\delta  \Omega'^{\dag}_\text{E2}  \Omega''] &= P[\boldsymbol{f}(\boldsymbol{\nabla E'})] P[\boldsymbol{f}(\boldsymbol{\nabla E''})], \\
S[\delta  \Omega'^{\dag}_\text{E2}  \Omega''] &= S[\boldsymbol{f}(\boldsymbol{\nabla E'})] S[\boldsymbol{f}(\boldsymbol{\nabla E''})].
\end{align}
The $\delta  \Omega'^{\dag}_\text{E2}  \Omega''$ term could lead to a shift if the product phase is real and field configuration is antisymmetric.  An analysis~\cite{koerber2003thesis} shows that this is the case if a small amount of circular polarization $\alpha$ is present, creating a component of $\boldsymbol{E}'$ along $\bhat{y}$ which we call $\alpha i \delta_z E'_y$.  Now, though this term is suppressed by two small quantities, the displacement from the anti-node $\delta_z$ and the error in polarization (and phase) $\alpha$, we find using the analysis of phase and symmetry factors above that the resulting shifts mimics the parity violating effect.  Worse, because the electric quadrupole coupling is some $10^{7}$ times stronger than the parity violating effect, even miniscule $\delta, \alpha < 10^{-3}$ are very problematic.

The solution comes about from the \emph{tensor} component of the ordinary off-resonant dipole light shift on the $5D_{3/2}$ state due to couplings with $n P_{1/2}$, $n P_{3/2}$, and $n F_{5/2}$ states.  The large linearly polarized $E'_x$ beam, at the proposed strength of $\sim 1.4 \times 10^{4}$~V/cm, shift the $m_x = \pm 3/2$ states away from the $m_x = \pm 1/2$ states by a large amount, roughly $300$~kHz.   By $m_x$, we mean the magnetic sub-levels defined by the electric field vector $E'_x$, not the magnetic field $B_z$ because one finds that this ordinary dipole light shift is the biggest dipole shift in the problem.   Rotation of state basis.  From Eq.~\ref{eq:d32rotationmatrix}, we can derive the transformation of states in the Zeeman basis to states in the light shift basis, rotated $90^\circ$:
\begin{equation}
d^{(3/2)}_{m' m}(\pi/2) = \frac{1}{(\sqrt{2})^3} \begin{pmatrix}
1		& -\sqrt{3}		& \sqrt{3}		& -1 \\
\sqrt{3}	& -1			& -1			& \sqrt{3} \\
\sqrt{3}	& 1			& -1			& -\sqrt{3} \\
1		& \sqrt{3}		& \sqrt{3}		& 1 \end{pmatrix}
\end{equation}
so the magnetic states transform as
\begin{align*}
| m= -3/2 \rangle_z &\to \frac{1}{(\sqrt{2})^3} \begin{pmatrix} -1 & \sqrt{3} & -\sqrt{3} & 1 \end{pmatrix}, \qquad
| m= -1/2 \rangle_z \to \frac{1}{(\sqrt{2})^3} \begin{pmatrix} \sqrt{3} & -1 & -1 & \sqrt{3} \end{pmatrix}, \\
| m= +1/2 \rangle_z &\to \frac{1}{(\sqrt{2})^3} \begin{pmatrix} -\sqrt{3} & -1 & 1 & \sqrt{3} \end{pmatrix}, \qquad
| m= +3/2 \rangle_z \to \frac{1}{(\sqrt{2})^3} \begin{pmatrix} 1 & \sqrt{3} & \sqrt{3} & 1 \end{pmatrix}.
\end{align*}

\begin{figure}
\centering
\includegraphics{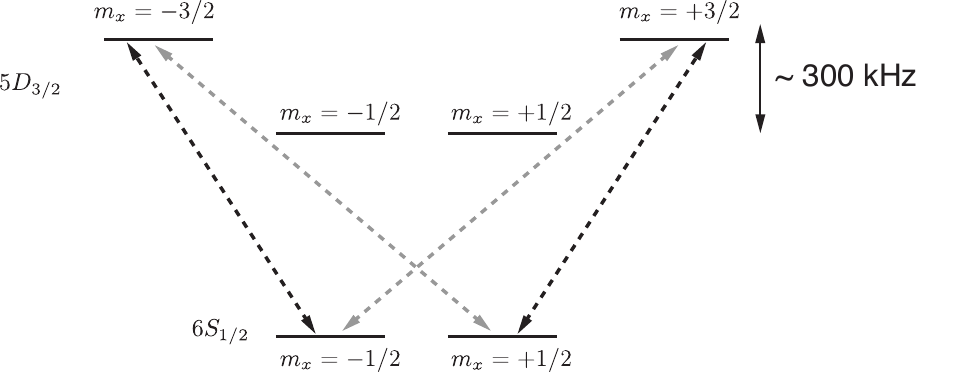}
\caption[Ordinary 2051~nm dipole light shifts split the $m = \pm 3/2$ states away]{Ordinary 2051~nm dipole light shifts split the $m_x = \pm 3/2$ states away from the $m_x = \pm 1/2$ states in $5D_{3/2}$, where the sublevel basis is now defined by the off-resonant dipole light shift, the biggest vector shift in the problem.  The tensor component of this shift, which separates the inner from the outer sublevels, leads to an additional suppression of on-resonant couplings requiring $\Delta m = \pm 2$.  This effect turns out to be a necessary source of suppression from systematic shifts arising from an ion (slightly) displaced in the strong $E'_x$ field.}
\label{fig:parityDipoleLS}
\end{figure}

This is highly relevant because the quadrupole coupling that comes about from an ion displaced from the anti-node of $E'_x$ has the operator structure $\partial_z E_y + \partial_y E_z$ which in the basis defined by the dipole light shift only allows $\Delta m = \pm 2$.  But couplings to these states are reduced due to the off-resoant light shift by a factor of
\begin{equation*}
R \sim  \frac{ \Omega''}{| \Delta_{m=3/2} - \Delta_{m = 1/2}|}.
\end{equation*}
Under certain experimental conditions, the numerator can be made very small, perhaps 10~kHz or smaller, while the denominator can be made very large, perhaps 300~kHz or larger.  This additional suppression factor allows us to be sure that our experiment is free of systematic shifts due to the $E'$ field yielding an E2 coupling as long as ion/beam displacements and polarization errors are each controlled to $10^{-3}$ levels

\subsubsection{Circular polarization}
So far we have treated both the $E'$ and $E''$ fields as being perfectly linearly polarized.  Small misalignments in the polarization vectors are tenable because they effect the measurement only to second order.  However, small amounts of circularly polarized light can potentially be disastrous.  Even without displacement of the ion, as in the last section, a component of the E1 field $i E'_y$ directly leads to a vector shift.  This isn't a complete disaster; the parity violating shift only appears when both $E'$ and $E''$ are on;  turning $E''$ off might leave a vector shift due to circular polarization in $E'$ that can then be adjusted to zero.  Still, sufficient noise in the polarization will dominate the parity violating effect, so this must be studied in earnest.  Also, $i E'_y$ can be directly detected by tuning the laser $\sim 1$~MHz from resonance and looking for \emph{vector} dipole light shifts in either the $6S_{1/2}$ state or the $5D_{3/2}$ state in a similar manner to what is described in Chapter~\ref{sec:lightShiftChapter}.

However, any out of phase $i E'_y$ running-wave field brings with it a magnetic field $i B'_x$ which is definitely problematic, and treated in the next section. 

\subsubsection{Magnetic dipole transitions}
Among the most troubling source of systematic effects is the magnetic field that must accompany the large driving field $E'_x$ through Maxwell's equation
\begin{equation}
\boldsymbol{\nabla} \times \boldsymbol{E} = -\frac{\partial \boldsymbol{B}}{\partial t}.
\end{equation}
Since we intend to put the ion on an anti-node of the electric field standing wave, where the spatial derivative is zero, the curl in the above equation means that the magnetic field strength will ideally be zero.  The time derivative gives a relative temporal phase of $i$ between the two waves, unlike the case for running waves.  In summary, we have
\begin{align}
\boldsymbol{B}' &= -i \nabla \times \boldsymbol{E}' \\
			 &= -i B'_y \bhat{y} \sin kz.
\end{align}
If the ion is displaced along $\bhat{z}$ by a small wavelength fraction $\delta$, the ion experiences a magnetic field of $-i \delta B'_y$.  We can now modify the interference picture in Eq.~\ref{eq:parityInterference} and we did in Eq.~\ref{eq:displacementParitySytematics},
\begin{align}
\Omega^{\dag} \Omega &=  (\Omega'' + \Omega' + \delta \Omega'_\text{M1} )^\dag (\Omega'' + \Omega' + \delta \Omega'_\text{M1} ) \\
				&=  \underbrace{(\Omega''^{\dag} \Omega'' +  \delta^2 \Omega'^{\dag}_\text{M1}\Omega'_\text{M1}}_\text{single field terms} + \underbrace{2 \text{Re}(\delta  \Omega'^{\dag}_\text{M1}  \Omega'')}_\text{M1-E2 interference term} + \underbrace{2 \text{Re}(\Omega'^{\dag} \Omega'')}_\text{PNC term} + \cdots
\end{align}
Can the M1-E2 interference term give rise to a systematic shift that mimic the parity violation signal?  Use of the selection rule, phase, and symmetry criteria show that a field $i B'_x$ will give a shift that mimics our parity violating signal.  Such a field can arise from any remnant circularly polarized running wave.  Though the strength of the coupling is 100 times weaker than the electric quadrupole case, there is no additional suppression factor in analog to the one we found in treating ion displacements.

\subsection{New ideas}
\subsubsection*{Parity violation in $^{137}$Ba$^+$}
We have considered~\cite{koerber2003thesis} the possibility of performing a similar parity violation experiment in the odd-isotope, $^{137}$Ba$^+$.  The advantages of these schemes come with different challenges than the approach presented for $^{138}$Ba$^+$.  For instance, given the hyperfine structure in the $6S_{1/2}$ and $5D_{3/2}$ states, there exists a transition that is electric quadrupole (E2) forbidden:
\begin{equation}
6S_{1/2}, F = 1 \leftrightarrow 5D_{3/2}, F' = 0  \qquad (F + F' < 2 \Rightarrow \text{ E2 forbidden}).
\end{equation}
Now the interference partner for the parity violating amplitude $\Omega'$ becomes a weak (compared to the E2 interaction) magnetic dipole transition $\Omega''_\text{M1}$.  This shrinks the size of misalignment systematic effects, but correspondingly demands that the laser linewidth be much smaller.  Further, now the ground state is $F = 1$, so scalar, vector, and \emph{tensor} shifts are now possible.  Also, we have not experimentally developed rf-spectroscopy in a $F= 1$~\cite{koerber2003thesis} as we have for $J = 1/2$ and $J = 3/2$ states.

Another wrinkle in a proposal to measure parity violation in the odd-isotope is that the underlying physics is not as simple when the nuclear spin $I > 0$.  The parity violation may now arise not only due to the fundamental electron-quark interaction (as shown in Fig~\ref{fig:bariumEnergyLevelsParityCloseup}), but also due to parity violating interactions among the quarks themselves.  This form of parity violating nuclear structure is called an anapole moment.  Some theoretical work~\cite{geetha1998nsd,sahoo2006rccpnc} has addressed these issues relevant to $^{137}$Ba$^+$.

\subsubsection*{Speculation}
\begin{figure}
\centering
\includegraphics{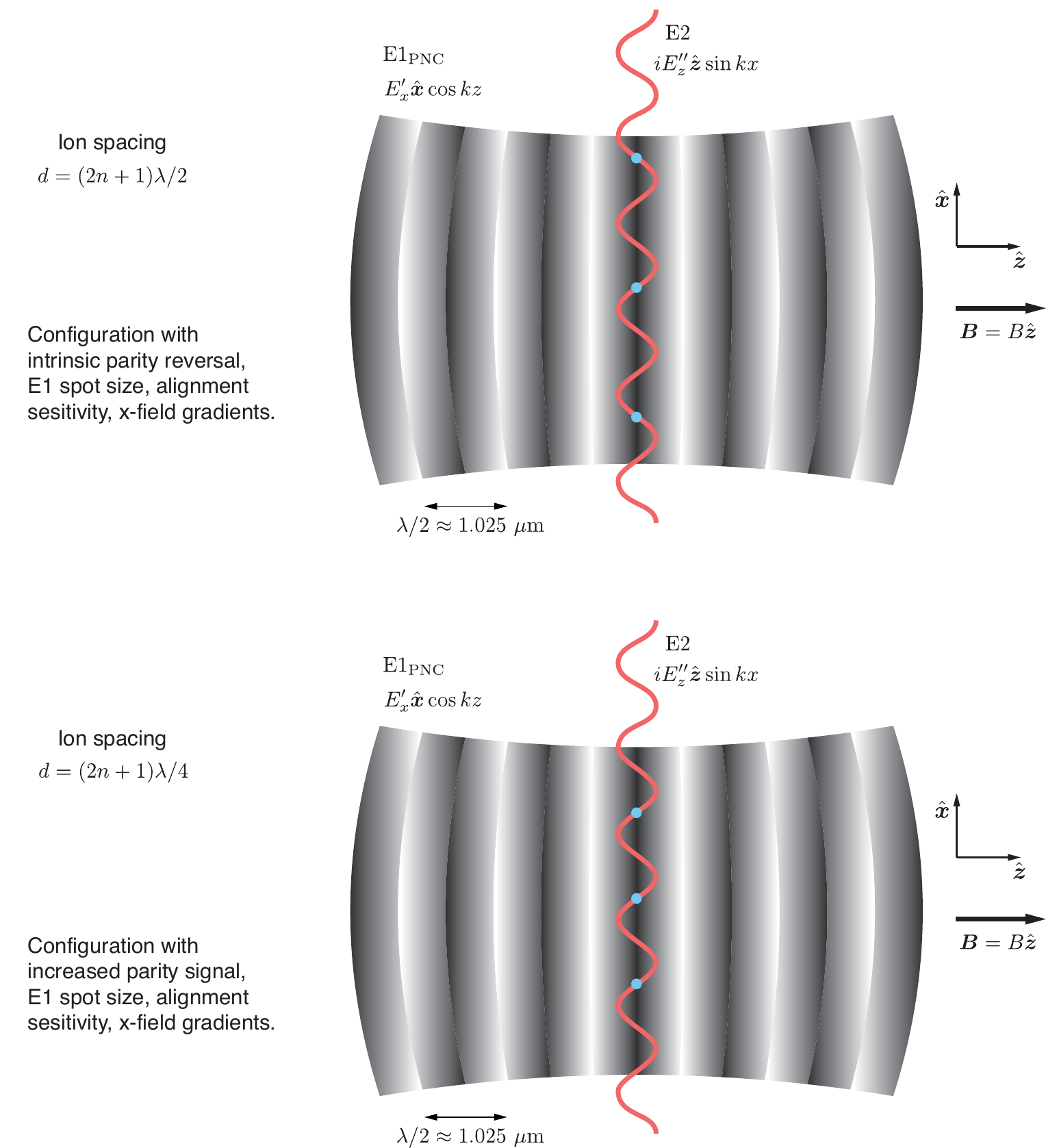}
\caption[Three ions in a standing wave provide extra systematic control]{Transitioning to a linear ion trap allows us to consider placing more than one ion in the standing waves of 2051~nm light.  Since the ion axial spacing is mostly controlled by the end-cap voltages, one could in principle position the ions as shown to increase the signal to noise, sense field gradients, provide real-time `reversal' information, or sense systematics through purposeful misalignment.  Shown here are only two possibilities, one should also consider placing the axial axis along the E1 field in order to measure the displacement of the ions from the anti-nodes, for instance.}
\label{fig:parityThreeIons}
\end{figure}

Future progress on both the optical frequency standard and parity violation projects will demand more exquisite control of ion micro-motion than I believe our current apparatus is capable.  Construction of a linear ion trap, with an appropriate set of optical access ports, seems inevitable.  A linear ion trap also offers great advantages in the parity violation experiment outlook. We recall that in a linear ion trap, axial confinement is mainly provided by dc endcap voltages which are easier to control at a precision level than an rf power.  This means that an ion's position in a 2051~nm standing wave might be manipulated with high resolution and regularity, allowing the experimenter to explore and minimize the displacement systematic effects discussed earlier.

Further, the linear ion trap has proven robust at holding many ions in the ground state of motion for various projects in quantum information.  The ion separation is largely a function of the applied end-cap voltages.  A very interesting arrangement is shown in Figure~\ref{fig:parityThreeIons}:  the end-cap voltages are adjusted to set the ion spacing to precise fractions of 2051~nm.  Now, one has sensitivity to axial field gradients, and adjacent ions are positioned to affect a parity transformation with respect to the standing wave field phasing.  Signal to noise is improved modestly, but now we can imagine using the outside ions as systematic effect sensors.  At this time, we have undertaken no serious study of this design, but it merits future consideration.

Finally, we recall that the reason Ba$^+$ looks so favorable for an atomic parity violation experiment is that the theory is expected to be as tractable as in neutral Cs.  Traditionally, many parity violation experiments worked by observing the parity violation on an aspect of an interrogating laser beam, via optical rotation.  While a single ion has a hopelessly small cross section, series of aligned, segmented ion traps designed to contain many (perhaps millions of) ions might present enough signal.  A magnetic dipole interaction is required as an interference partner for an optical rotation experiment.  While the M1 amplitude in the Ba$^+$ $6S_{1/2} \leftrightarrow 5D_{3/2}$ transition is smaller than those in elements where optical rotation has been successfully measured, the fractional PNC effect is much larger.  Further, calculations~\cite{sahoo2006lms} of the strength of this M1 show it to be roughly 10 times higher than expected due large electron-electron correlation effects (see Table~\ref{tab:m1calculations}).  The E2 coupling presents absorption without the required interference effect, but in the $^{137}$Ba$^+$ isotope we recall that one of the transitions is free of E2 coupling.

A vision for a future experiment might be a series of very large, well machined ion traps, being continuously loaded to achieve some steady state average number of trapped ions.  Perhaps the ion lifetime can be made a large fraction of a second or greater.  Any absorption or genuine optical rotation effect due to the atoms can be distinguished from birefringence in the optics by \emph{shelving} a large fraction of the trapped ions, or by simply shutting of the rf and emptying the trap for a brief time.  This seems like an elegant analog to the `empty dummy tube' approach commonly employed in vapor cell optical rotation experiments.  Preliminary work shows that this approach is likely not fruitful, but speculation is ongoing.

% ==========   Bibliography
%
\nocite{*}   % include everything in the .bib file, regardless of what was explicitly cited
\bibliographystyle{abbrv}
\bibliography{shermanBib}
%
% ==========   Appendices
%
\clearpage
\appendix
\raggedbottom\sloppy
 
 \chapter{Optical cavities} \label{sec:opticalCavities}
\begin{table}
\centering
\caption[Collected facts about Gaussian beams and useful unit conversions]{Collected facts about gaussian beams and useful unit conversions, mostly obtained from \cite{budker2004ape, hobbs2000beo}.  NA stands for numerical aperture and CA for clear area.  The left figure above shows the intensity of a gaussian spot as a function of distance from the center.  The right figure shows the focus of a gaussian beam and denotes many of the quantities listed in this table.}
\begin{tabular}{m{3.0 in}l}
\multicolumn{2}{c}{\includegraphics[scale=0.8]{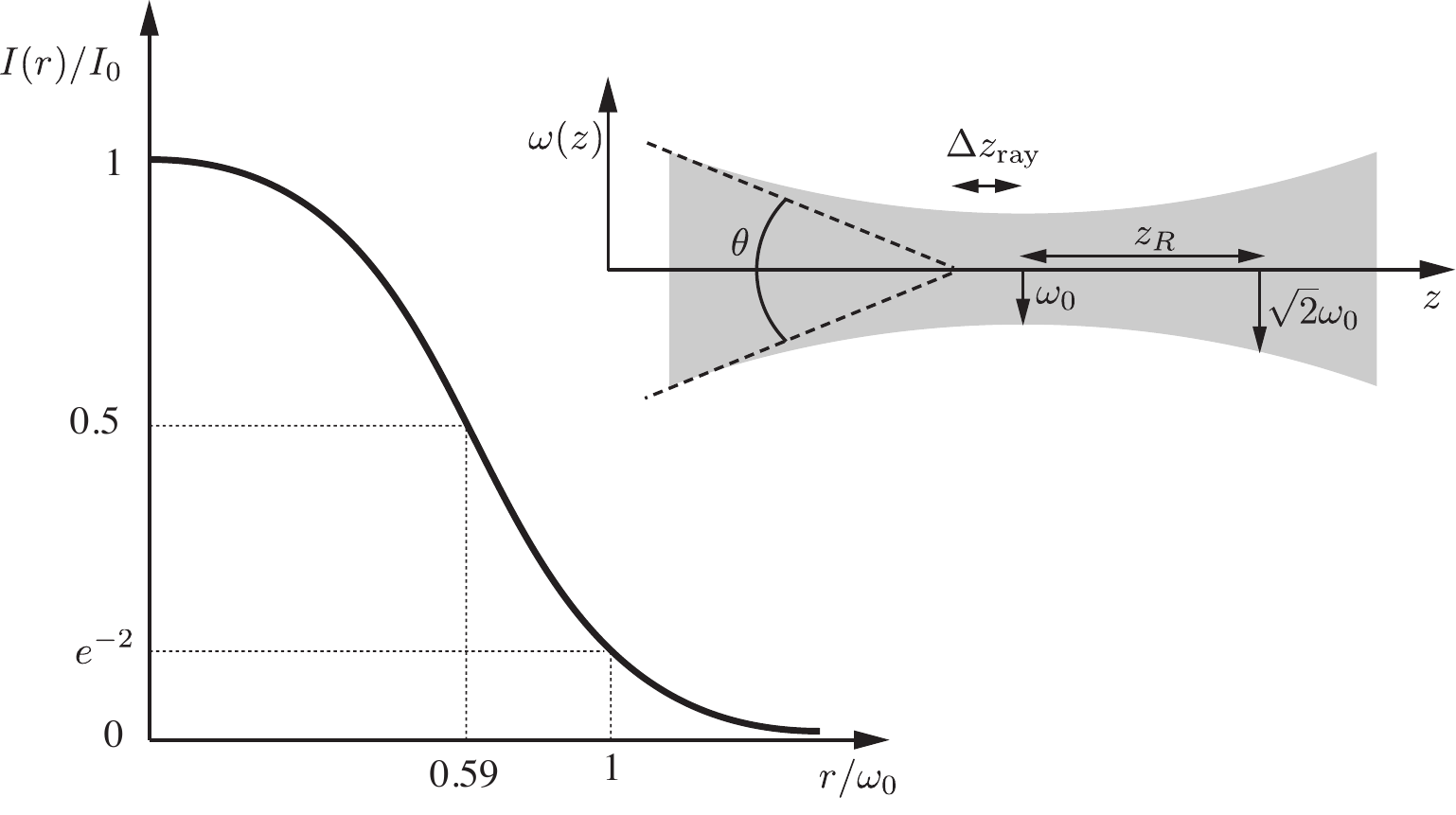}} \\
Electric field distribution 		& $|E(r)| = E_0 e^{-r^2 / \omega_0^2}$ \\
Intensity distribution 			& $I(r)| \sim |E(r)|^2 = I_0 e^{-2r^2/\omega_0^2}$ \\
Power enclosed in radius $r$	& $P(r) = P_0 \left( 1 - e^{-2r^2/\omega_0^2} \right)$\\ \hline
Power $\leftrightarrow$ central intensity & $I_0 = \frac{2P}{\pi \omega_0^2}$ \\
Beam waist $\leftrightarrow$ NA $\leftrightarrow$ f-number & $\omega_0 = \frac{\lambda}{\pi (\text{NA})} \approx \frac{2 \lambda}{\pi}(f/\#)$ \\
Rayleigh range (central intensity halves, waist increases by $\sqrt{2}$) &$ z_R = \pi \omega_0^2 / \lambda = \frac{\lambda}{\pi (\text{NA})^2} = \frac{\omega_0}{(\text{NA})}$ \\
$1/e^2$ power-density radius	& $\omega(z) = \omega_0 \sqrt{1 + (z/z_R)^2}$ \\
Wavefront radius of curvature	& $R(z) = z + \frac{z_R^2}{z}$ \\
Waist displacement from ray optics prediction & $\Delta z = -z_R^2/f$ \\
Equivalent solid angle 		& $\Omega_\text{eq} = \pi(\text{NA})^2 = \frac{\lambda^2}{\pi \omega_0^2}$\\
\'{E}tendue (light gathering efficiency) & $E = n^2 (\text{CA}) \Omega$ \\
Full angular width (far-field)	& $\Theta = \frac{2 \lambda}{\pi \omega_0} \approx (f/\#)^{-1}$ \\ \hline
$E$(V/cm) $\leftrightarrow$ $I$ (mW/cm$^2$) & $ E \approx \sqrt{I / 1.33}$ \\
Atomic dipole moment units	& $ea_0 / h \approx 1.28 \frac{\text{MHz}}{\text{V/cm}}$
\end{tabular}
\label{tab:gaussianBeams}
\end{table}

Resonant optical cavities find so many uses in our experiment:  as laser spectrum analyzers to diagnose laser problems, optical power buildup resonators to enhance second-harmonic generation processes, and as ultra-stable optical frequency references.  In this appendix we describe the resonant optical cavity and summarize these techniques.

\subsection*{Optical modes, characterization of a cavity}
\begin{figure}
\centering
\includegraphics{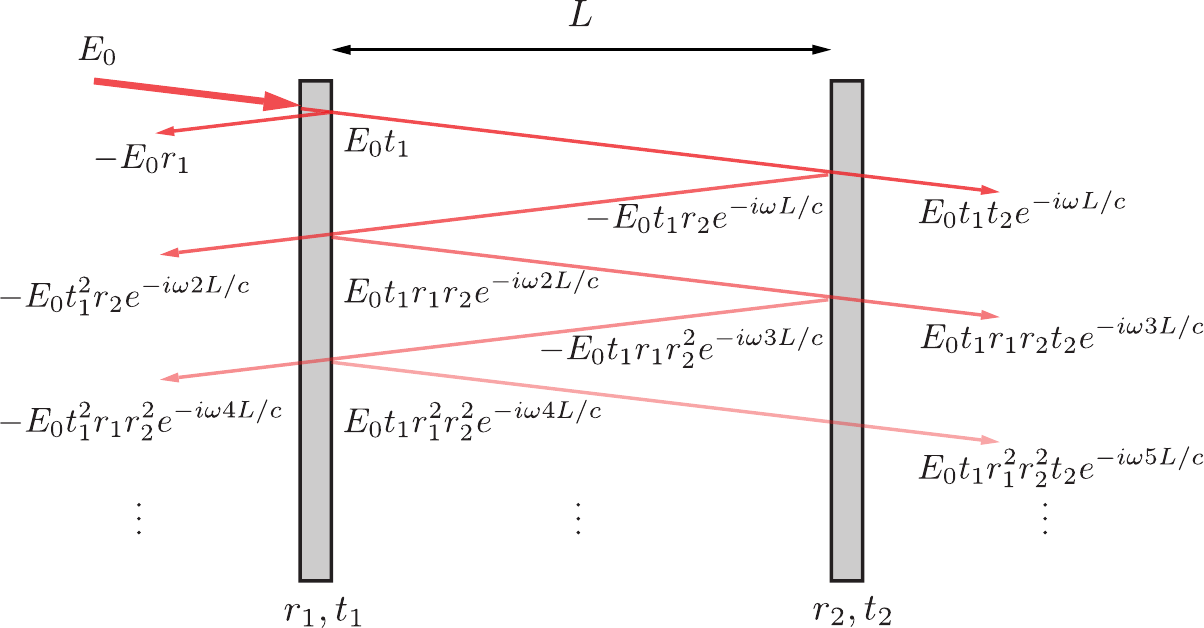}
\caption[Interfering reflections from and transmissions through an optical cavity]{Depiction of the interfering reflections from and transmissions through an optical cavity.}
\label{fig:cavityBounces}
\vspace{1 in}
\includegraphics{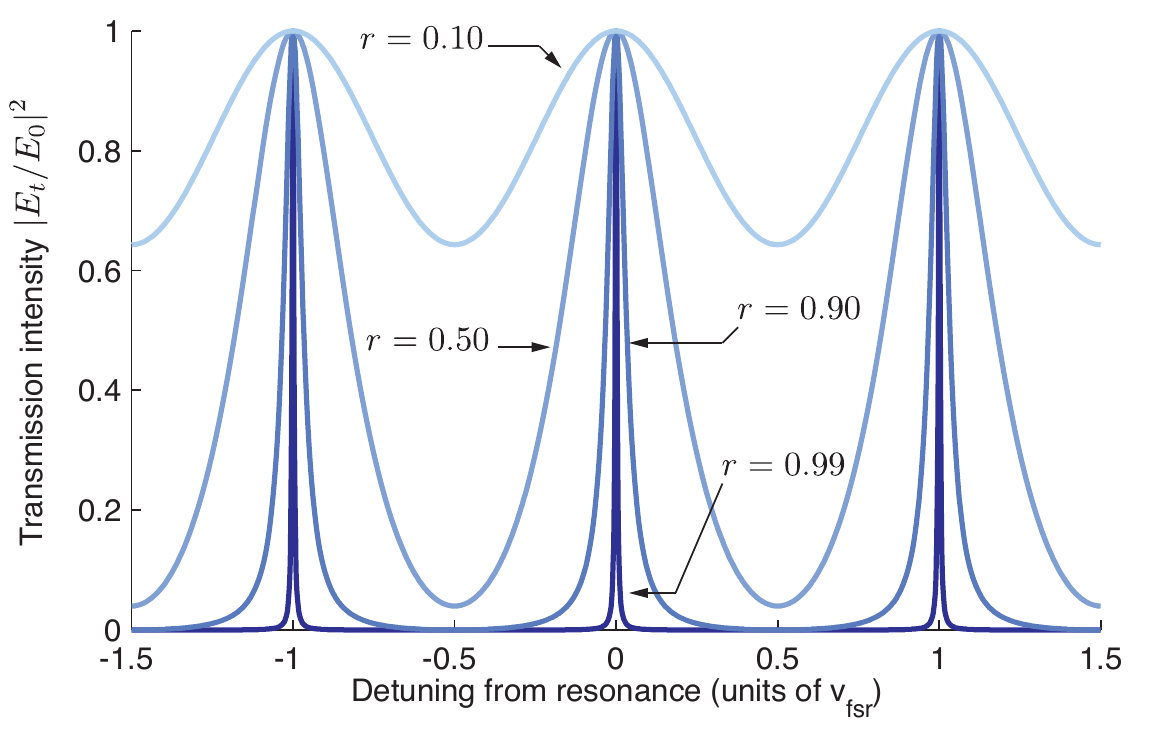}
\caption[Transmission spectra of Fabry-Perot cavities]{Transmission spectra of Fabry-Perot cavities with differing mirror reflectivities.  Here we plot $|E_t/E_0|^2$ as a function of detuning, measured in units of the free spectral range $\nu_\text{fsr} = c/2L$, from a cavity resonance.  From darkest to lightest curve, we plot the cases $r = 0.99$, $r=0.90$, $r=0.50$, and $r=0.10$.}
\label{fig:cavityTransmissionCurves}
\end{figure}

A typical depiction of the planar Fabry-Perot cavity is shown in Figure~\ref{fig:cavityBounces}.  Two mirrors with inner-surface reflectivity $r_1$ and $r_2$ are separated by length $L$.  If we ignore absorption and assume the exterior surfaces of the mirrors are perfectly transmissive, then we can assign transmission factors $t_1 = 1-r_1$ and $t_2 = 1-r_2$ to the mirrors as well.  Given an incident electromagnetic wave $E_0 e^{-i (\omega t +\phi)}$, we see that the prompt reflection $-E_0 r_1$ is interfered with many terms representing waves that have undergone one, two, three, etc., round-trips in the cavity.  During each round trip these components are slightly attenuated upon reflection and pick up a phase factor:
\begin{equation*}
E_i \to E_i r_2 e^{-i \omega 2L/c}
\end{equation*}
The reflected electric field is a sum of all these amplitudes~\cite{riehle2004fsb},
\begin{align}
E_r &= E_0 \left[r_1 - t_1^2 r_2 e^{-i\omega 2L/c} - t_1^2 r_1 r_2^2 e^{-i\omega 4 L/c} - t_1^2r_1^2r_2^3 e^{-i \omega 6 L/c} + \cdots \right], \\
E_r &= E_0 \left[r_1 - \frac{t_1^2 r_2 e^{-i \omega 2L/c}}{1 - r_1r_2 e^{-i \omega 2L/c}} \right],
\end{align}
where we've made use of the geometric series summation
\begin{equation*}
\sum_{n=0}^\infty q^n = \frac{1}{1-q}  \qquad \text{with } q= r_1r_2 e^{-i \omega 2 L/c}.
\end{equation*}

Meanwhile, the steady-state transmitted wave is also an interference of many waves undergoing successive bounces through the cavity,
\begin{align}
E_t &= E_0 t_1 t_2 e^{-i \omega L/c} + E_0 t_1 t_2 r_1 r_2 e^{-i \omega 3L/c} + E_0 t_1 t_2 r_1^2 r_2^2 e^{-i \omega 5 L/c} + \cdots, \\
E_t &= E_0 t_1 t_2 e^{-i \omega L/c} \left[1 + r_1 r_2 e^{-i \omega 2L/c} +  r_1^2r_2^2 e^{-i \omega 4 L/c} + \cdots \right], \\
E_t &= E_0 \frac{t_1 t_2 e^{-i \omega L/c}}{1 - r_1r_2 e^{-i \omega 2 L/c}},
\end{align}
by application of the same geometric series summation.

Finally, in the absence of losses, the internal electric field inside the cavity increases dramatically in magnitude.  For instance, just inside the first mirror \cite{warrenClassNotes},
\begin{align}
E_c &= E_0 t_1 + E_0 r_1 r_2 t_1 e^{-i \omega 2L/c} + E_0 t_1 r_1^2 r_2^2 e^{-i \omega 4L/c} + \cdots \\
	&= \frac{E_0 t_1}{1 - r_1 r_2 e^{-i \omega 2L/c}}.
\end{align}
An expression for the power and intensity enhancement inside the cavity, is
\begin{equation}
\eta_\text{enhance} \equiv \frac{I_c}{I_0} = \left| \frac{E_c}{E_0} \right|^2 \approx \frac{1-r_1^2}{(1-r_1 r_m)^2}
\end{equation}
where we have assumed $t_1 = 1 - r_1$ and folded intracavity losses $l$ into the second-mirror reflection by defining $r_m = r_2 (1- l)$.  For a cavity with only $r_1 = r_m = 0.99$, this enhancement factor is already above 50.  For $r_1 = r_m = 0.999$ mirrors, $\eta_\text{enhance} \approx 500$.  Because the efficiency of non-linear optical processes (like second-harmonic generation) scale with the power of incident light (meaning that the output power of the process scales \emph{quadratically} with the incident power), optical cavities of even moderate quality are an important practical tool in second-harmonic generation.
 
A natural frequency scale is made evident by each of the expressions for reflected, transmitted, and internal electric fields:
\begin{equation}
\nu_\text{fsr} = \frac{\omega_\text{fsr}}{2\pi}  \equiv \frac{c}{2L}
\end{equation}
is called the \emph{free spectral range} of the optical cavity.  In geometrical terms, $\omega_\text{fsr}$ is the additional frequency one must tune a laser frequency $\omega$ to fit another half-wave $\lambda/2$ in the optical cavity and correspondingly is the period at which reflection and transmission peaks appear in $E_r(\omega)$ and $E_t(\omega)$ (see Figure~\ref{fig:cavityTransmissionCurves}, for instance).  Frequencies resonant with the optical cavity that differ by $\omega_\text{fsr}$ are called successive \emph{longitudinal modes} of the cavity.  The planar-mirror geometry sketched in Figure~\ref{fig:cavityBounces} is actually unstable.  Slight misalignments eventually allow the beam to escape the two finite-sized mirrors.  Even for perfect alignment, diffractive losses of the necessarily diverging optical beam inside the cavity limit the number of bounces that survive.  

\begin{figure}
\centering
\includegraphics{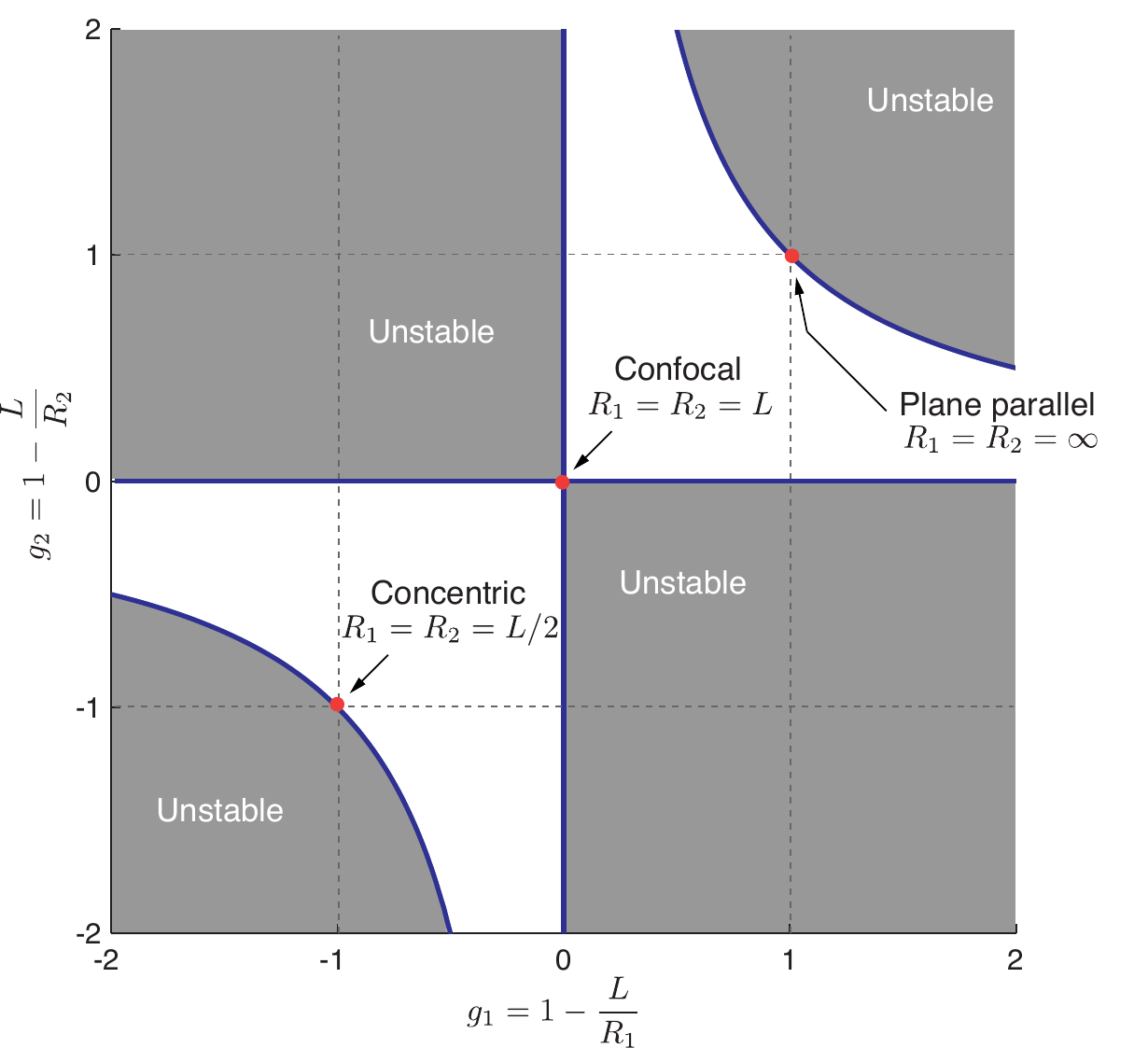}
\caption[Stable optical cavity geometries]{Following classic references (such as \cite{kogelnik1966lbr}) we can define the dimensionless parameters $g_1 = 1 - L/R_1$ and $g_2 = 1 - L/R_2$ from the mirror radii of curvatures $R_1$, $R_2$, and separation $L$.  The confocal cavity geometry $R_1 = R_2 = L$ is at the center of the stability diagram and maximizes the central cavity beam waist.  The concentric cavity $R_1 = R_2 = L/2$ yields the smallest allowed central waist size for a given $L$ but lies on the edge of instability just like the plane parallel mirror configuration $R_1 = R_2 =\infty$.}
\label{fig:opticalCavityStability}
\end{figure}
\begin{figure}
\centering
\includegraphics{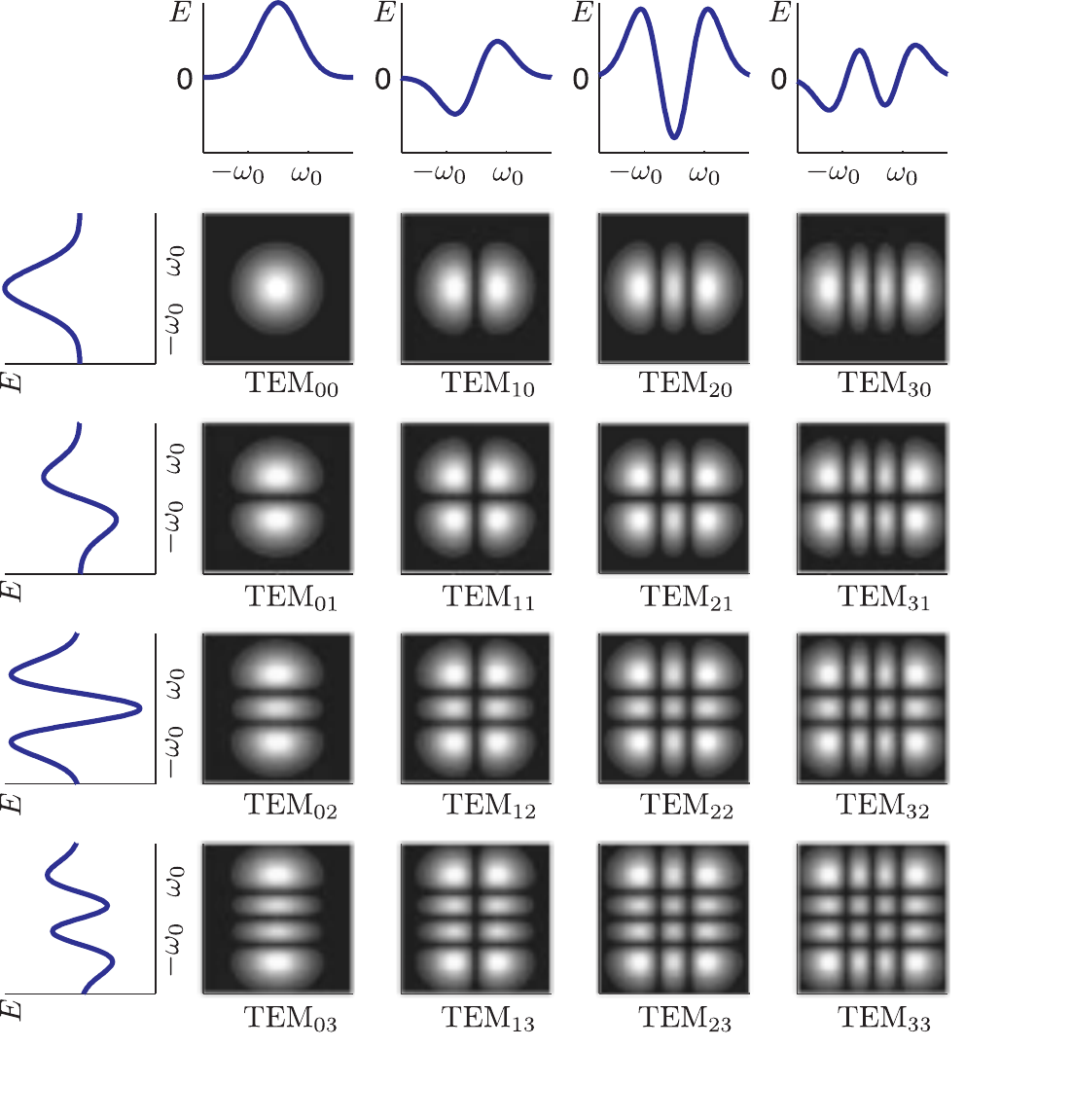}
\caption[A gallery of Hermite-Gaussian modes]{A gallery of Hermite-Gaussian modes that serve as the basis for a resonant optical cavity.}
\label{fig:hermiteGaussModes}
\end{figure}

At least one of the mirrors must be curved.  Several treatments of physical optics (e.g.~\cite{kogelnik1966lbr, yariv2006poe}) give us the choices of mirror radii of curvatures $R_1$, $R_2$, and separation $L$ that lead to a stable resonant cavity:
\begin{equation}
0 < \underbrace{\left(1 - \frac{L}{R_1} \right)}_{\equiv g_1} \underbrace{\left(1 - \frac{L}{R_2} \right)}_{\equiv g_2} < 1.
\end{equation}
The stability condition is graphically summarized in Figure~\ref{fig:opticalCavityStability}.  Inside a stable linear or ring optical cavity is a standing or traveling wave whose transverse profile must also satisfy the boundary conditions of the dielectric resonator.  These mode profiles are in general linear combinations of Hermite-Gaussian modes, some of which are shown in Figure~\ref{fig:hermiteGaussModes}.

The transverse eigenfunctions are indexed according to how many zeros they have along each orthogonal axis.  For instance, the lowest order mode, TEM$_{00}$, has a Gaussian radial profile with no zeros in any direction.  TEM$_{21}$, however, has two zeros along the horizontal axis and one zero along the vertical dimension.  For cavities with rectilinear geometries (and misalignments), the modes of the cavity are exactly the Hermite-Gaussian modes.  For optical cavities and transmission lines with cylindrical geometry (e.g.\ fiber optics), the normal modes are best represented by Laguerre-Gaussian modes which are in general expressible as linear combinations of Hermite-Gaussian modes.

Because higher-order transverse modes acquire additional phase shift during a round trip in the cavity, their resonant frequencies are displaced from the TEM$_{00}$ modes by a factor that depends on the geometry of the cavity.  If we express the frequency of longitudinal mode $q$, transverse mode TEM$_{xy}$ as
\begin{equation}
\nu_\text{q,xy} = \frac{c}{2L} \left[q + \frac{1}{\pi}(x + y + 1) A \right]
\end{equation}
where $A$ is
\begin{equation}
A = \cos^{-1} \left( \pm \sqrt{\left(1 - \frac{d}{R_1} \right) \left((1 -\frac{d}{R_2} \right)}\right).
\end{equation}  
The factor of $A$ present for $x=y=0$ is due to the so-called \emph{Gouy phase shift} of a Gaussian beam with a waist inside the cavity.  A special case occurs when the mirror radii of curvature are equal to the cavity length, $L = R_1 = R_2$.  This arrangement is called a \emph{confocal} cavity.  The transverse phase factor $A$ is equal to one half a free spectral range
\begin{equation*}
A_\text{confocal} = \frac{1}{2} \frac{c}{2 L},
\end{equation*}
which means that all the modes are degenerate at $c/4L$ intervals.  This is an especially good arrangement for an optical cavity used as a scanning optical spectrum analyzer since the simple periodic mode structure at $c/4L$ is insensitive to the input beam alignment and mode-matching.  The arrangement is not good at all for cavities used as optical frequency references since the degeneracy is never perfect in practice;  even if $R_1 = R_2 = L$ is engineered perfectly, slight mirror ellipticity break even the degeneracies between transverse mode groups (e.g.\ TEM$_{02}$, TEM$_{20}$, and TEM$_{11}$).
 
\subsection*{Locking to a cavity}
\begin{figure}
\centering
\includegraphics{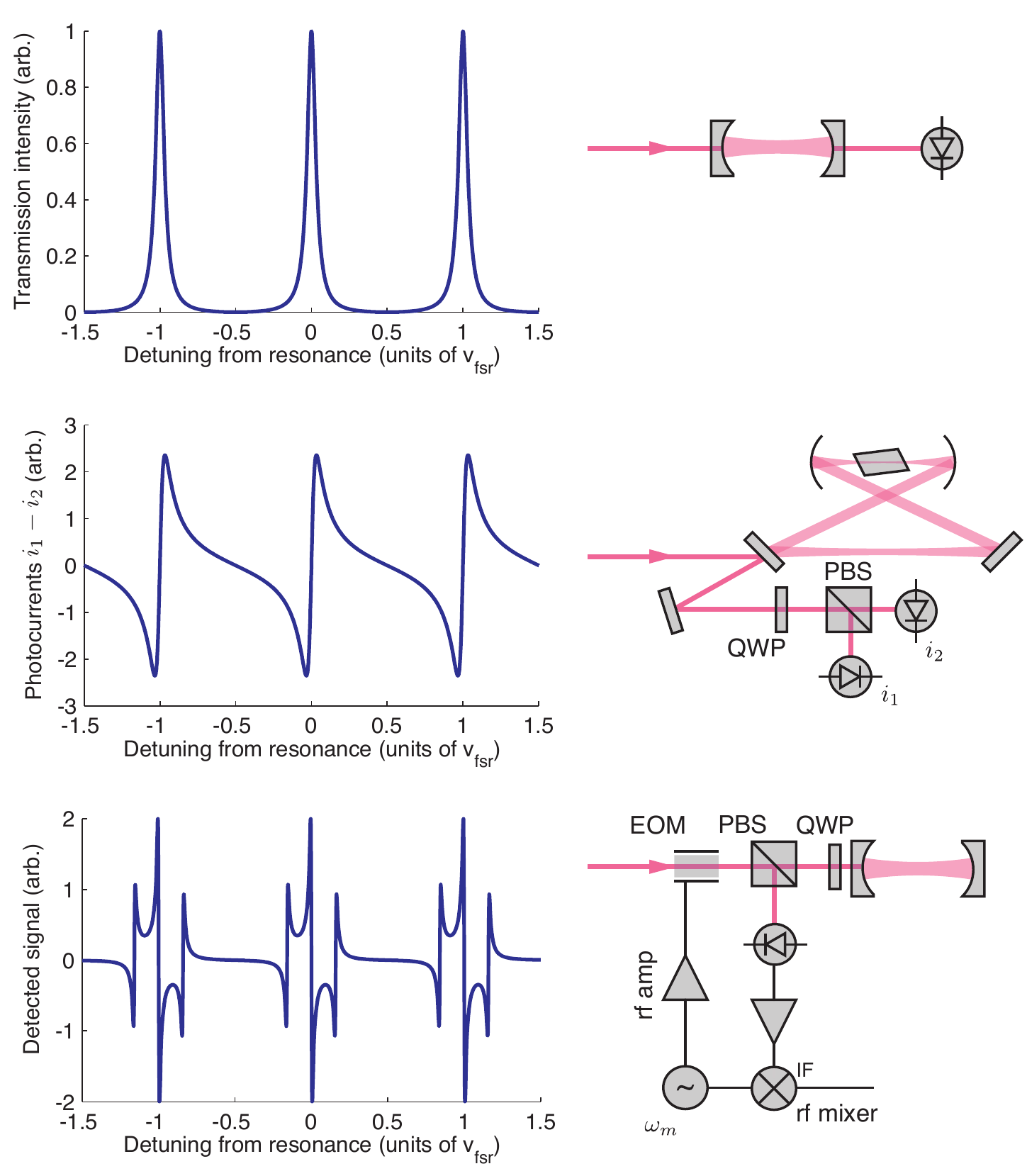}
\caption[Cavity spectra:  transmission, H\"{a}nsch-Couillaud, Pound-Drever-Hall]{Cavity spectra:  transmission peaks, and dispersion lock signals using the H\"{a}nsch-Couillaud~\cite{hansch1980lfs}, and Pound-Drever-Hall~\cite{black2001ipd} schemes.}
\label{fig:cavityResonances}
\end{figure}
If we labor to make the length on an optical cavity very stable, its eigen-frequencies are also stable and the cavity becomes a useful optical frequency reference with frequency markers at period $A$ and strong frequency markers every $\nu_\text{fsr}$.  Even a well insulated laser often has far worse free-running frequency noise and drift than a moderately stabilized cavity, so we can construct an electro-optic feedback loop that servos various laser controls which have an impact on its operating frequency in order to keep it locked to a chosen mode in an optical cavity.  We employ the full toolbox of linear feedback control theory by first considering the ways of constructing an \emph{error signal}, or an electronic signal proportional to the difference between the instantaneous frequency of a laser $\omega$ and a nearby cavity mode $\omega_0$.  Several schemes are shown in Figure~\ref{fig:cavityResonances}.

Perhaps the most straightforward idea is to monitor the transmission through the cavity.  We might choose to place the laser on the side of one of the cavity's transmission features.  If the laser drifts, the transmission signal either increases of decreases;  this information can be amplified and sent back to the laser to correct for any frequency error.  Or, the cavity (or laser) could be frequency-modulated at some rate $\omega_m$ so the center of the transmission feature is turned into dispersion shaped curved by either mixing or a lock-in amplifier.  Either of these methods are fundamentally limited in bandwidth due to the fact that the transmitted intensity of an optical cavity cannot slew faster than the \emph{cavity storage time}
\begin{equation}
\tau_\text{cavity} = \frac{1}{2\pi} \frac{F}{\nu_\text{fsr}},
\end{equation}
where
\begin{equation}
F = \frac{\pi \sqrt{r_1 r_2}}{1 - r_1 r_2} = \frac{v_\text{fsr}}{\Gamma},
\end{equation}
is called the \emph{finesse} of the cavity.  $\Gamma$ is the full-width at half-maximum of the transmission features, also called the \emph{resolution} or \emph{cavity linewidth}.  The use of this language also implies a cavity quality factor
\begin{equation}
Q_\text{cavity} = \frac{\nu_\text{laser}}{\Gamma_\text{cavity}} \qquad (\text{for } \Gamma_\text{cavity} \gg \Gamma_\text{laser}).
\end{equation}
The effective low-pass filter behavior imposed by $\tau_\text{cavity}$ can be quite limiting.  For instance, a $\nu_\text{fsr} =$2~GHz cavity with high finesse of $F = 10^5$ has a transmission bandwidth limit of $\tau_\text{cavity}^{-1} = 20$~kHz.    Most free-running laser linewidths are much larger than this, implying that most of the noise spectral power occurs are frequencies higher than 20~kHz.  So, a servo based on the transmission fluctuations would not help to reduce the linewidths of most laser systems that aren't stabilized in some other fashion.  Thus, the most fruitful technique is to derive an error signal from the optical cavity via the reflectance.  This section covers two schemes known as the H\"{a}nsch-Couillaud and Pound-Drever-Hall techniques.  

\begin{figure}
\centering
\includegraphics{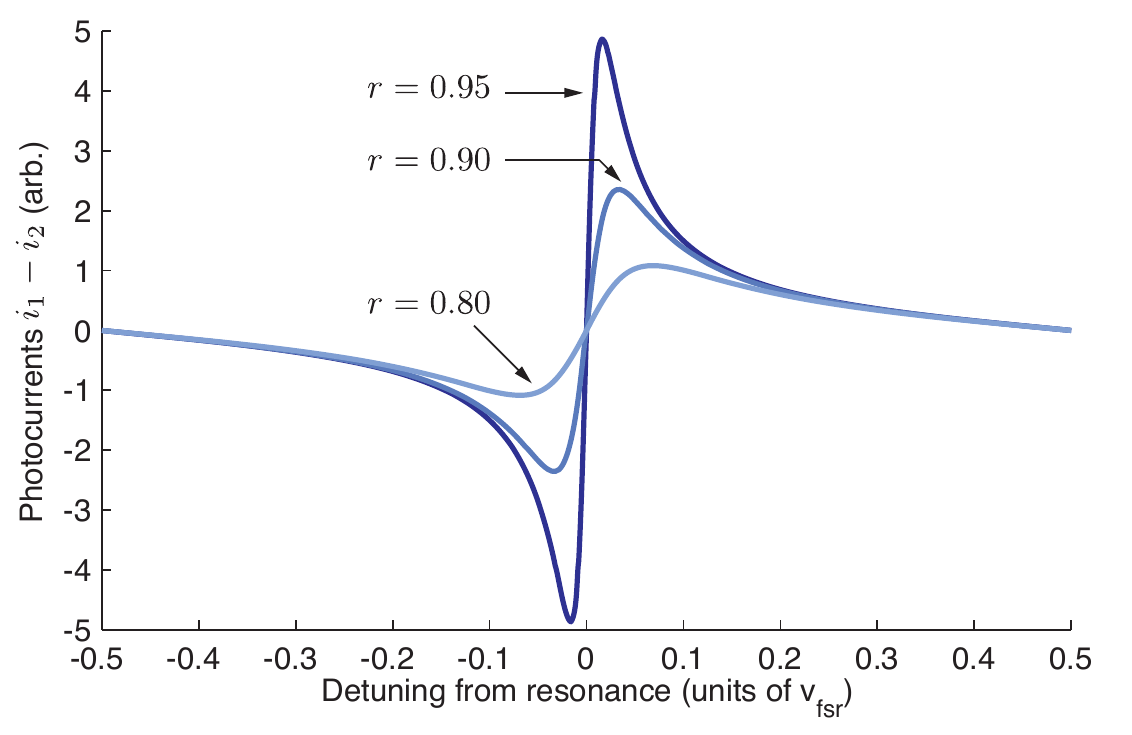}
\caption[The H\"{a}nsch-Couillaud cavity/laser error signal]{The H\"{a}nsch-Couillaud cavity/laser error signal as a function of cavity mirror reflectivity.}
\label{fig:hanschScheme}
\end{figure}
The H\"{a}nsch-Couillaud setup, sometimes called the \emph{polarization scheme}, is to make use of a polarizing or birefringent element inside of the optical cavity.  In practice, this is often a non-linear optical crystal, but a plate set at Brewster's angle also works.  Two orthogonal polarization states suffer different losses in the cavity and therefore the reflected waves exhibit differing phase shifts near resonance:
\begin{align}
E_1^{r} &= E_1 r_1, \\
E_2^{r} &= E_2 \left(r_1 - \frac{t_1^2}{r_1} \frac{r e^{-i \delta 2L/c}}{1- r e^{-i \delta 2L/c}}\right),
\end{align}
where $\delta$ is the detuning from resonance $\omega - \omega_0$ and the reflection coefficient $r$ takes into account internal losses during one round-trip through the cavity.  The two polarization states in the reflected beam are independently measured using a polarizing analyzer (quarter-wave plate and a polarizing beam splitter).  If the resulting photocurrents, $i_1$ and $i_2$, are electronically subtracted, a linear dispersion shaped signal results that features a steep linear slope around cavity resonances
\begin{align}
i_1 - i_2 &\propto |E_1|^2 - |E_2|^2 \\
	       &\propto |E_0|^2 \frac{t_1^2 r^2 \sin \delta}{(1-r^2)^2 + 4 r^2 \sin^2 \delta/2}.
\end{align}
A typical case is plotted in the middle graph in Figure~\ref{fig:hanschScheme}.  An attractive feature of this dispersion signal is the large \emph{capture region} outside the central slope.  Recorded dispersion and transmission curves from a 986~nm doubling cavity used in our experiment are displayed in Figure~\ref{fig:doublingCavitySignals}.

\begin{figure}
\centering
\includegraphics{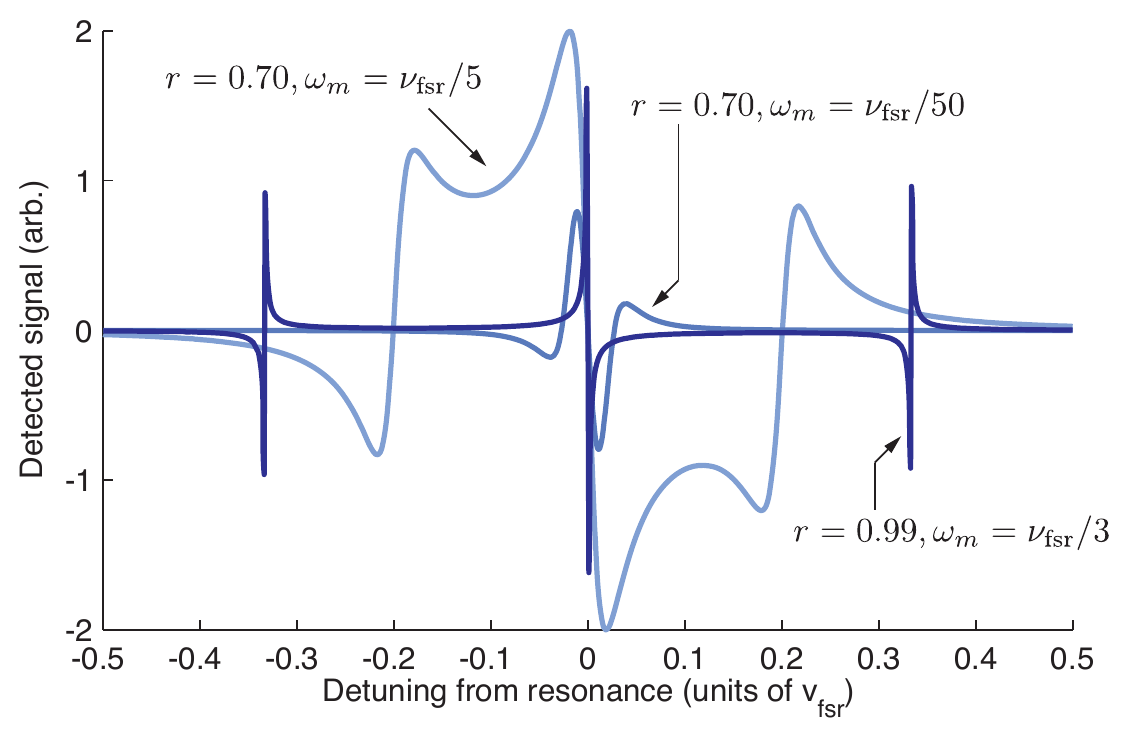}
\caption{The Pound-Drever-Hall cavity/laser locking error signal.}
\label{fig:poundDreverHallScheme}
\end{figure}
Often the best reference cavities cannot afford a polarizing element inside of the cavity.  In these cases a widely used alternative to the H\"{a}nsch-Couillaud scheme is the Pound-Drever-Hall method.  Either using an electro-optic phase modulator, or direct current modulation in the case of diode lasers, the incident beam is given modulation sidebands at $\omega_m$.  With a modulation depth $D \lesssim 1$, the electric field of the incident wave is
\begin{equation}
E(t) = E_0 \left[ J_0(D) e^{i \omega t} + J_1(D) e^{i (\omega + \omega_m)t} - J_1(D) e^{i (\omega - \omega_m)t} + c.c. \right]
\end{equation}
where we have abbreviated the complex conjugate terms and ignored the second and higher order phase modulation sidebands (see Section~\ref{sec:phaseModulation} for a review of phase modulation).  The the carrier (term proportional to $J_0$) is made resonant with the cavity, the sidebands at $\omega_0 \pm \omega_m$ are wholly rejected from the cavity.  The interference of these sidebands with the carrier leaked from the cavity forms a desirable error signal.  In essence, the leaking cavity mode represents a long-term average of the incident electric field since the cavity acts like an optical low-pass filter.  The long-term average leakage beam is interfered with the promptly reflected sidebands that contain in them phase information about recent laser frequency deviations from resonance.  Therefore the response bandwidth is not limited by the cavity storage time.

The reflected beam is routed into a photodiode using a scheme such as Figure~REF.  If we perform phase-sensitive detection by mixing the photocurrent with the modulation frequency $\omega_m$ and low-pass filtering, we can obtain a signal
\begin{equation}
P(\delta) \propto J_0(D) J_1(D) \frac{\omega_m^2 (\Gamma/2) \delta ((\Gamma/2)^2 - \delta^2 + \omega_m^2)}{(\delta^2 + (\Gamma/2)^2)((\delta + \omega_m^2) + (\Gamma/2)^2)((\delta - \omega_m)^2 + (\Gamma/2)^2)}.
\end{equation}
This signal, plotted in Figure~\ref{fig:poundDreverHallScheme}, shows resonances at the carrier and first sideband frequencies and most importantly a very steep linear slope around the carrier frequency at serves as a sensitive error signals.  The strength of the signal, proportional to the product of Bessel functions $J_0(D) J_1(D)$ is maximized with a modulation depth of $D = 1.08$ radians.

How narrow can a laser be made given an optical cavity of quality $Q$?  The ultimate limit with any of the optical reference schemes presented so far is given by the shot-noise limit~\cite{black2001ipd}:
\begin{equation}
\sigma_y(\tau) = \frac{1}{4 Q}\sqrt{\frac{h \nu}{\eta P_d \tau}}
\end{equation}
where $h \nu$ is the energy per laser photon, $\eta$ is the quantum efficiency of a detector upon which power $P_d$ is incident.  This performance limit is in practice never reached;  technical electronic noise, residual amplitude modulation accidentally accompanying the phase modulation in the Pound-Drever-Hall scheme, amplifier noise, and systematic effects in the cavity all degrade performance.

\chapter{Barium atomic data}
\begin{table}[h!]
\let\PBS=\PreserveBackslash
\centering
\caption[Barium isotope data]{Barium isotopic abundances, masses, nuclear spin and magnetic moments \cite{Emsley1995te}, and lifetimes.}
\begin{tabular}{clrc>{\PBS\centering}b{0.5 in}>{\PBS\centering}b{1.0 in}}
Isotope 	& Mass (amu)	& Abundance & Lifetime  & Nuclear spin & Mag. moment ($\mu_N$) \\ \hline \hline
$^{130}$Ba$^+$ & 129.906282 & 0.11 \% & stable  	&  0 		& --- \\
$^{132}$Ba$^+$ & 131.905042 & 0.10 \% & stable  	&  0 		& --- \\
$^{133}$Ba$^+$ & 132.906008 & ---	  & 10.5 y. &  1/2 	& --- \\
$^{134}$Ba$^+$ & 133.904486 & 2.42 \% & stable  	&  0 		& --- \\
$^{135}$Ba$^+$ & 134.905665 & 6.59 \% & stable  	&  3/2 	& +0.8365 \\
$^{136}$Ba$^+$ & 135.904553 & 7.85 \% & stable  	&  0 		& --- \\
$^{137}$Ba$^+$ & 136.905812 & 11.23 \% & stable &  3/2 	& +0.9357  \\
$^{138}$Ba$^+$ & 137.905232 & 71.70 \% & stable &  0 		& --- 
\end{tabular}
\label{tab:bariumIsotopes}
\end{table}

\chapter{Conventions}\label{sec:conventions}
\section*{Circular polarization}
We propagate the conventions in~\cite{koerber2003thesis}. Given an electromagnetic wave propagating along the $\bhat{k} = \bhat{z}$ axis.  Two orthogonal linear polarization states are $\bhat{\epsilon} = \bhat{x}$ and $\bhat{\epsilon} = \bhat{y}$.  In this cartesian basis, we can define left and right circular polarization vectors
\begin{equation}
\bhat{\epsilon}_\pm = \frac{ \bhat{x} \pm i \bhat{y} }{\sqrt{2}}.
\end{equation}
Then any electric field vector can be written in this basis
\begin{equation}
\boldsymbol{E} = E_+ \bhat{\epsilon}_+ + E_- \bhat{\epsilon}_-
\end{equation}
We define the strength of circular polarization $\sigma$ to be the relative asymmetry in these field components
\begin{equation}
\sigma \equiv \frac{E_{+} - E_-}{E_+ + E_-}
\end{equation}
which differs from a conventional definition appropriate when measuring the relative asymmetry in \emph{intensities}
\begin{equation*}
\sigma' = \frac{E_+^2 - E_-^2}{E_+^2 + E_-^2} \qquad \text{(alternative and common definition)}.
\end{equation*}

In a spherical vector basis, the electric components are
\begin{align}
E_0^{(1)} &= 0, \\
E_{\pm 1}^{(1)} &= E_0 \frac{1 \pm \sigma}{\sqrt{2} \sqrt{1 + \sigma^2}},
\end{align}
with $E_0^2 = E_+^2 + E_-^2$.  If the light propagation vector $\bhat{k}$ is projected onto the $xz$-plane and has as finite angle $\theta$ with respect to the $\bhat{z}$ axis, these amplitudes become
\begin{align}
E_0^{(1)} &= E_0 \frac{\sin \theta}{\sqrt{1 + \sigma^2}}, \\
E_{\pm 1}^{(1)} &= E_0 \frac{\cos \theta \pm \sigma}{\sqrt{2} \sqrt{1 + \sigma^2}}.
\end{align}

Measurement of circular polarization quality $\sigma$ in the laboratory is made straightforward by measuring instead the linear polarization quality of the beam intensity
\begin{equation}
Q \equiv \frac{E_x^2 - E_y^2}{E_x^2 + E_y^2}
\end{equation}
where we arrange a polarizing beam cube to maximize, say, the transmitted intensity $E_x^2$ and minimize the reflected intensity $E_y^2$.

%
% ==========   Vita
%
\clearpage
\raggedbottom\sloppy
\chapter*{Vita}
Jeff Aaron Sherman was born to Elaine and James Sherman---an English teacher and jazz musician, who aren't at all sure how their son got interested in science---in the city of bright lights and big buffets, Las Vegas, Nevada.  He obtained his bachelor's degree in Physics in 2001 from the University of Texas---Austin while working at a sonar research facility, a fact that partially explains why he fails to hear most of what his friends say to him.  After spending a summer at Lincoln Lab---MIT in Boston researching what 100 networked electronic ears could do after being scattered about the desert, he began graduate school at the University of Washington in beautiful if damp Seattle.  In June 2007, Jeff anticipates a significant change in longitude, if not climate, when he is due to begin a postdoctoral fellowship at Oxford University.
\end{document}